# RADIO STUDY OF THE ARC AND THE SGR A COMPLEX NEAR THE GALACTIC CENTER

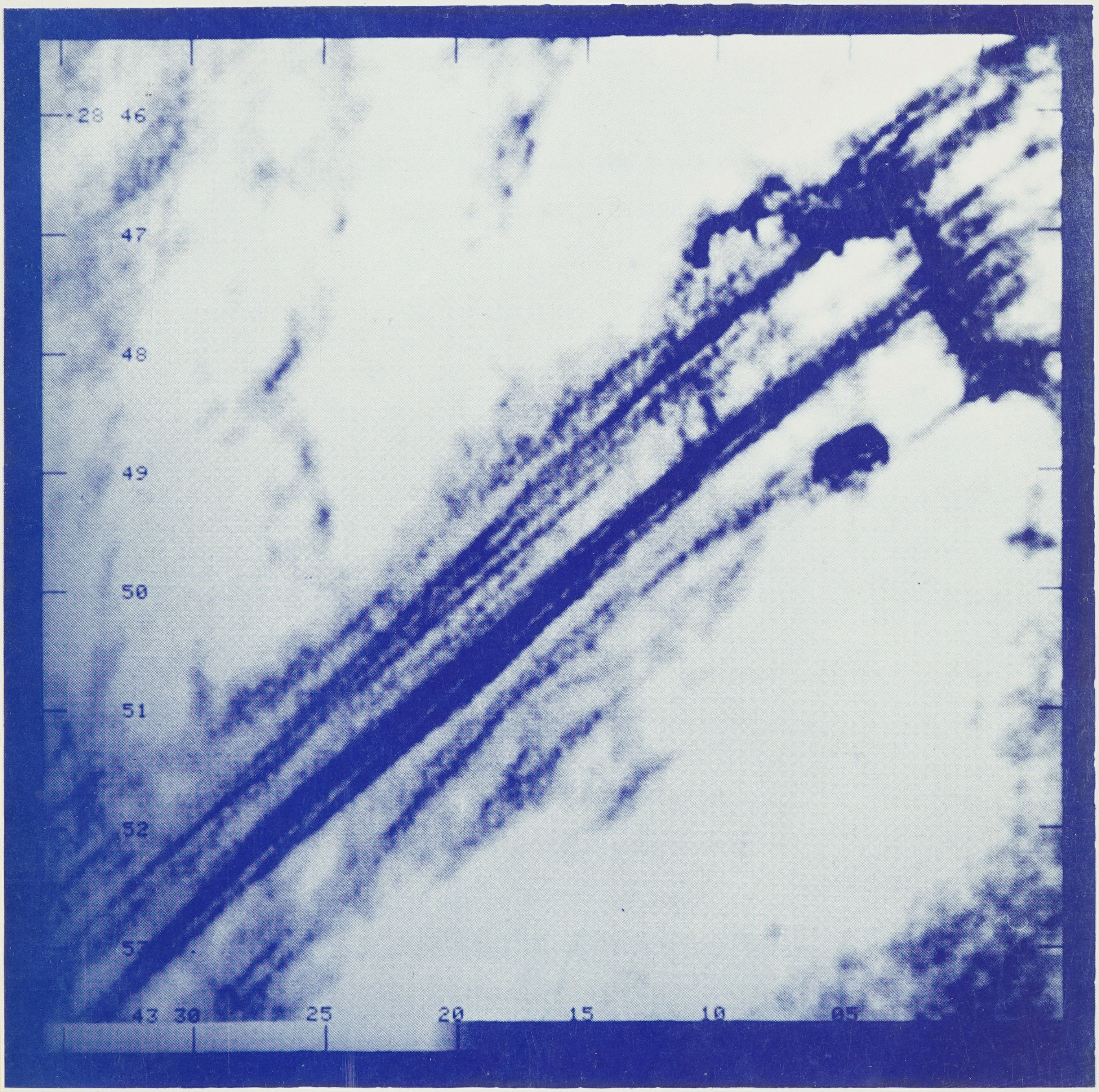

Farhad Yusef-Zadeh

# RADIO STUDY OF THE ARC AND THE SGR A COMPLEX NEAR THE GALACTIC CENTER

Farhad Yusef-Zadeh

ABSTRACT

RADIO STUDY OF THE ARC AND THE SGR A COMPLEX

NEAR THE GALACTIC CENTER

Farhad Yusef-Zadeh

Radio continuum and radio recombination line observations of the inner degree of the galactic center reveal a rich collection of thermal and nonthermal radio structures: (a) A network of linear filaments that are oriented perpendicular to the galactic plane constitute the major portion of the radio Arc at $\ell \sim 0.2°$. These filaments have nonthermal characteristics, show polarized emission at 6, 3 and 2 cm, are organized over a 100 pc scale, and have a flat spectrum. (b) A number of thread-like filaments are situated asymmetrical with respect to the galactic plane and appear to be isolated unlike the linear filaments which are grouped together. The polarization and spectra of these so called "threads" are not yet known. (c) A network of arched filamentary structures that is disorganized in its appearance constitutes the curved portion of the Arc. Radio recombination line emission from these filaments indicates a thermal character for the emission. (d) An interaction between the linear and arched filaments is implied and the strength of the magnetic field along the linear filaments - based on the interaction - is estimated to be between $10^{-3}$ and $10^{-4}$ Gauss. (e) A very steep-spectrum ridge of emission is seen to emerge from Sgr A as it extends perpendicular to the galactic plane. The possibility that this one-sided feature may be a low-energy jet from the galactic nucleus is suggested. (f) the relative location of Sgr A East, West, a cluster of HII regions and the 50 km $s^{-1}$ molecular cloud are discussed. Comparisons of the low and high-frequency maps show clearly that Sgr A East lies behind Sgr A West.

These observations imply that the poloidal component of the magnetic field may dominate in the galactic center region. Curiously, the magnetic field lines appear to be coherent and organized over a large expanse in a region of the Galaxy where the interstellar medium is expected to be characterized by inhomogeneities and violent, disorganized motions.

Acknowledgement

My special thanks goes to Mark Morris, my advisor, who introduced me this fascinating region of the Galaxy. His financial backing from UCLA, his encouragement and his special attention to this project has been extremely helpful during the course of this project.

I like to thank my collaborators D. Chance, E. Fomalont, M. Inoue, U. Klein, A. Lasenby, G. Nelson, J. Seiradakis, J. Van Gorkom, and R. Wielebinski, who helped me out over the years since this work was begun.

It is a great pleasure to thank J. Bally, H. Liszt and Y. Mills who provided me with their unpublished data in advance and K. Prendergast, R. Ekers, B. Elmegreen and D. Helfand for useful discussions and comments on this thesis. I also thank J. Grindlay for sending me a tape including the X-ray data of the Galactic center.

I like to thank numerous members of the astronomical community particularly radio astronomers who gave me lots of encouragement. It was an honor to receive an inspiring letter from Professor J. Oort.

I also like to thank R. Havlen and R. Ekers for allowing me to make numerous copies of the thesis via NRAO.

It is my greatest pleasure to thank Ms. Susan Mescher who typed this thesis in one weekend. This tireless woman has created most valuable atmosphere conducive to the exchange of ideas during my time at Columbia.

Mountains of appreciation go to Kevin Prendergast who did everything he could to provide financial assistance for this very expensive project after M. Morris left Columbia.

This thesis is dedicated to the many members of NRAO whom I got to know during the course of this project.

Table of Contents

# Chapter 1

# An Historical Overview

## I. Introduction

"... An understanding of Sgr A is an
understanding of a phenomenon commonly
found throughout the universe."
Brown and Lo

The proximity of the nucleus of the Milky Way Galaxy gives astronomers an unparalleled opportunity for studying the detailed structure of the core of a spiral galaxy. This advantage, however, is enjoyed mostly by radio and infrared astronomers, since the extinction of light by dust and gas is large enough, i.e., greater than 28 magnitudes, to obscure completely the optical emission. Both radio and infrared astronomers have been accumulating an enormous body of observational data on the galactic center region over the last 35 years, ever since Piddington and Minnet (1951) discovered a strong radio source near the junction of the constellations Sagittarius, Scorpius, and Ophiuchus. During the period since then, the angular resolution with which this source has been observed has increased by a factor of $10^5$. As a result of the present rich body of information, the galactic center region can be differentiated into many complex components whose physical relationships are not yet clear. A careful investigation of the physical relationships between the large and small scale radio features, such as the relationship between the lop-sided distribution of the large-scale molecular emission (Cohen and Few 1978), the 50 km $s^{-1}$ Sgr A molecular cloud (Fukui _et al._ 1977; Güsten _et al._ 1981) and the strong compact radio source at the galactic center (see the reviews by Oort 1977, Brown and Liszt 1984), if any, remain one of the most challenging problems in the study of the galactic center region. A coherent understanding of the diverse phenomena on many scales revealed by the recent flood of papers does not yet exist. We are still at the stage of trying to identify which fundamental phenomena are occurring in the galactic center region. We are also searching to determine conclusively whether these phenomena are found analogous to those on a variety of other scale sizes (i.e., the structures seen on the surface of the Sun or in the nuclei of active external galaxies). Because of the complexity seen in the nucleus of the Galaxy, this stage is dominated by observations rather than theory. Metaphorically, a theorist finds himself (herself) as if he (she) is trapped by a spider's web which becomes more tangled and incomprehensible if any act of struggling is made.

"In nature's infinite book of secrecy
A little I can read."
Shakespeare

Indeed, because of the unique location of the galactic center and because of the constant progress in instrumentation, a "surgical" examination of the properties of the galactic center region can be very important in further understanding the source of activity in galactic nuclei in general. The resemblance of the nonthermal compact radio sources seen in nuclei of galaxies and quasars and in the nucleus of the Galaxy suggests further study of the features which surround the galactic center in order to find out more about the environments in which such compact sources reside. Detailed study of the galactic center region might clarify and perhaps answer a number of questions, and, in so doing, might lead us to better understanding of the nuclei of spiral galaxies. [It is worthy of mention that the galactic center region has also been studied as a potential locale of communication relays/ space vehicles in radio bands (Valleé and Simard-Normadin 1985)].

Throughout this thesis, I will concentrate mostly on the radio study of an "intermediate regime", i.e., the inner 30 arcminutes of the Galaxy, and attempt to discuss the possible link between many of the features uncovered in the intermediate regime and the phenomena occuring on other scales. A good portion of this thesis deals with the phenomenology associated with the galactic region (i.e. astro-geography). It is hoped that a more satisfactory explanation of some of the observed features will be achieved with the participation of experts in the field in the near future. Each chapter is intended to be self contained. This chapter deals with a brief history of the features which have previously been recognized in this region. The interpretations of the observed features will not be stressed, since the picture painted of the galactic center region has been neither permanent nor unambiguous.

## I.A Radio Continuum Observations (1960's)

"There are no truths, only interpretations."

F.W. Nietszche

Radio continuum observations, which were carried out by Drake (1959, see Steinberg and Lequeux 1963) using the 85-foot telescope of the National Radio Astronomy Observatory (NRAO), resolved the galactic center source into four components and obtained the first high-resolution radio map of Sgr A. Drake's map, which has a resolution of 6' at 3.7 cm, is reproduced in figure 1; it shows the Sgr A complex joined by a hook-like structure and two extended sources to the north and south aligned in the direction of the galactic plane. These sources are called "the continuum Arc", "Sgr B2", and "Sgr C", respectively. Thus, the subject of the continuum Arc, with which this thesis deals extensively is as old as a good wine. Drake suggested two models for the nucleus and it is historically interesting to state them almost in their entirety.

> "Two models of the nucleus are proposed. In one, the 'static nucleus' model, there are in the center about $10^9$ solar masses of Population II stars in two small bodies similar to the nucleus of M31. The emitting gas is ejected from these stars, and the blue Population II stars excite the gas. A disk of neutral hydrogen rotates nearly as a solid body around this. In the second model, the 'evolving nucleus' model, gas flows into the central parts of the nucleus, where massive young blue stars and the observed HII regions are formed."

A surge of observations followed Drake's work in order to resolve the complex sources in the galactic center region. Numerous maps of this region were made at 1.4 GHz (Kerr), 3 GHz (Cooper and Price 1964), 5 GHz (Broten et al. 1965), 8 GHz (Downes et al. 1965), and 14.5 GHz (Hollinger 1965) with resolutions of 6!7, 4!1, 4!2, and 5!9, respectively. These were reproduced and compared in a comprehensive review paper by Downes and Maxwell (1966). These maps were used mainly to find a) the spectral index distribution of the galactic center sources, b) the linear size of the resolved sources, and c) the correct position of Sgr A - supposedly the center of our Galaxy. These studies are still the subjects of current investigations.

Downes and Maxwell find the spectral index of Sgr A in the microwave band to be $\alpha = -0.7$ ($F_\nu \propto \nu^\alpha$) and suggest that Sgr A could be a supernova remnant. Early investigations of the radio recombination line emission from Sgr A using the NRAO 140-foot telescope showed a lack of line emission from this region, at least in the radial velocity range -87 to 141 km $s^{-1}$ (Mezger and Hoglund 1966). Polarization measurements of this region at 0.4 and 1.4 GHz (Gardner and Whiteoak 1962), 3 GHz (Cooper and Price 1964), 3.2 and 9.52 GHz (Mayer et al. 1963, 1964) showed that the percentage of both linear and circular polarization is less than 2 percent.

One of the first interferometric measurements of Sgr A was carried out by Biraud et al. (1960). Their measurements were made with aerial separation in the range 41 wavelengths ($\lambda$) to 2080 $\lambda$ at 20 cm. Their visibility function shows that the flux density of Sgr A at the shortest spacing is 0.7 of that of Cas A, which is the strongest radio source in the sky beyond the solar system.

Lunar occultations of Sgr A were first carried out by Maxwell and Taylor (1968), Kerr and Sandqvist (1968), and Thompson et al. (1969). Maxwell and Taylor measured the position of Sgr A to fairly good accuracy to within ± 15", $\alpha \sim 17^h42^m30^s$, $\delta \sim -28°59'14"$, and determined the spectral indices of Sgr A based in their 0.23 and 2.4 GHz observations. In their interpretation Sgr A consists of a flat spectrum core structure ($\alpha \sim -0.25$) surrounded by a region with a steeper spectrum ($\alpha = -0.7$). Although this interpretation has remained roughly unchanged, improvements in the radio picture of Sgr A were needed to progress with our understanding of the nature of this object. These were forthcoming in the following decade, upon which I concentrate next.

## I.B Radio Continuum Observations (1970-1985)

Three different mapping techniques continued to be applied to Sgr A in this period. The first involved using single-dish telescopes with improved sensitivity and higher angular resolution. The main motivations for these observations were to find new sources and to refine the picture of the large-scale radio distributions in this region (Whiteoak and Gardner 1973; Kapitzky and Dent 1974; Little 1974; Pauls et al. 1976; Haynes et al. 1978; Altenhoff et al. 1978; Sofue and Handa 1984; Reich et al. 1984; Seiradakis et al. 1985). The second technique was to use lunar occultations (Gopal-Krishna et al. 1972; Sandqvist 1974). Radio interferometry soon replaced this technique and dominated high-frequency observations of Sgr A thereafter (Downes and Martin 1971; Ekers and Lynden-Bell 1971; Dulk and Slee 1974; Balick and Sanders 1974; Brown and Balick 1976; Ekers et al. 1975; Downes et al. 1980; Brown et al. 1981, 1983; Ekers et al. 1983; Yusef-Zadeh et al. 1984; Mills and Drinkwater 1984; Morris and Yusef-Zadeh 1985).

### I.B.1 Single-Dish Observations

High-frequency observations made by Kapitzky and Dent (1974) at 2 cm with a resolution (full width of half maximum, or FWHM) = 135", as shown in figure 2, added a new complexity to the previously held picture. Their map showed that the large-scale extended emission along the galactic plane, which was revealed in a survey map made by Altenhoff (1970) at 11 cm, consists of several discrete sources mostly aligned in the direction of the galactic plane. This observation indicated that the extended emission which was not seen at higher frequencies must have a non-thermal spectrum. In another study, Pauls et al. (1976) obtained the most detailed single-dish radio map of the Sgr A complex plus the continuum Arc. Their map, which has a resolution (FWHM) of 77", is reproduced in figure 3; it resolves both the Sgr A complex and the continuum Arc into multiple components. They showed that Sgr A has a core-halo structure and argued that the halo is non-thermal (see also Krishna et al. 1972 and Dulk and Slee 1974), since its brightness temperature, as measured by Dulk and Slee (1974) at 160 MHz, is much greater than $10^5$ °K.

As part of a radio survey of the galactic plane (Altenhoff et al. 1978), a large-scale radio feature representing a continuation of the extension the continuum Arc at $\ell$ = 0°.2 to latitudes as high as ~ 1° (figure 4) was brought out and was later recognized by Sofue and Handa (1984). This curious feature can also be seen in the most recent survey at 11 cm by Reich et al. (1984).

### I.B.2 Low-Frequency Observations

Several low-frequency observations were made during the 1970's in order to determine the low-frequency turnover in the radio spectrum of Sgr A. Brezgunov et al. (1971) obtained upper limits to the flux densities at six different frequencies in the range 100-120 MHz using the cross radio telescope of Lebedev Physical Institute. Lunar occultations of Sgr A using the Ooty radio telescope at 327 MHz were made by Gopal-Krishna et al. (1972), who recognized a new extended source surrounding Sgr A which had not been seen in previous high-frequency maps. The highest resolution map at 160 MHz (FWHM ~ 1'.9) was published by Dulk and Slee (1974) and Slee (1977) using the Culgoora radioheliograph. This map, which shows a structure similar to that seen in the 408 MHz map made by Little (1974), is taken from an article by Slee (1977) and is reproduced in figure 5. This figure shows an extension toward the southeast, roughly perpendicular to the galactic plane, and in addition, it indicates that the peak does not coincide with Sgr A West, the thermal component of the Sgr A complex.

Dulk and Slee suggested that the low frequency turnover is due to free-free absorption in the intervening interstellar medium.

### I.B.3 Interferometric Observations

Radio interferometric observations of Sgr A using the Cambridge one mile telescope were carried out by Downes and Martin (1971). They correctly deduced that Sgr A consists of a compact structure -- supposedly coincident with the center of Galaxy -- a thermal feature surrounding the compact source (Sgr A West), and a non-thermal component to the east of the thermal feature (Sgr A East). Similar conclusion was also reached - based on lunar occultation observations by Sandqvist (1974) who showed that the Sgr A complex consists of a number of distinct components.

Because it shows properties similar to those of the nuclei of radio galaxies and quasars and because it is located both at the apparent center of luminosity and at the dynamical center of our Galaxy (Lynden-Bell and Rees 1971; Becklin and Neugebauer 1975; Lacy et al. 1982; Ekers et al. 1983; Lo et al. 1985) much attention has been given to the compact radio source and the infrared source - known as IRS 16 - at the galactic center. A strong infrared source seen at 2.2 $\mu m$, IRS 16, appears to be located at the center of the stellar cluster toward Sgr A (Rieke and Low 1973; Becklin and Neugebauer 1975). This infrared source was shown recently to be separated from the compact radio source by ~ 1" (Ekers [private communication] Henry et al. 1984; Storey and Allen 1983). The non-thermal spectrum of the compact radio source was first recognized by Balick and Brown (1974). VLBI observations of this source indicate that this radio object has a scale size $\leq 3\times10^{14}$ cm (the distance to the galactic center, Ro, is assumed to be 10 kpc, throughout this thesis) and shows an elongation in the direction of the minor axis of the Galaxy (Lo et al. 1981, 1985). Since the brightness distribution of this object has not yet been mapped, the intrinsic elongation of the source is still a tentative result (Lo et al. 1985). Brown and Lo (1982) reported that this compact source is variable on time scale from days to years. This tentative and important result needs to be confirmed.

Radio interferometric observations of Sgr A since 1983 have been extremely revealing. Ekers et al. (1983), Brown and Johnston (1983), and Lo and Claussen (1984) used the VLA in a number of different configurations at 2, 6, and 20 cm and obtained by far the most detailed maps of Sgr A West and East. The observations which were made by Ekers et al. (1983) and Brown and Johnston (1983) showed that Sgr A West consists of a "3-arm spiral-like" feature in which the optically thick compact source is buried. Observations by Ekers et al. (1983) also showed best the shell-like appearance of Sgr A East and the spectral index distribution across Sgr A East and West. Their 20-cm continuum map and spectral index map is reproduced in figure 6. More detailed discussion of these features will be presented in chapter VI.

## I.C Radio Recombination Line Observations

Line emission from the nucleus of the Galaxy has been a subject of much interest for many astronomers over the last two decades. Clear evidence for the thermal nature of Sgr A West was first reported by Pauls, Mezger, and Churchwell (1974) who detected H109α radio recombination line emission. Their observation confirmed the conclusion of the earlier continuum observation of Downes and Martin (1971) that on the basis of its flat spectrum, Sgr A West has a thermal nature. Also, Wollman et al. (1977) found Ne II 12.8 μ emission from Sgr A West. Both H109α and Ne II emission profiles are very broad (>200 km/s) in Sgr A West. Later observations of H65α, H84α and H94α emission from Sgr A West were made by Rodriguez and Chaisson (1979) who explained the dynamical structure of Sgr A West in terms of Keplerian rotation due to the gravitational field of the normal stellar population plus a central mass point of $5\times10^6$ $M_\odot$. These authors also find a very low electron temperature, 5,000 °K. Recent developments in both infrared and radio interferometric spectroscopy toward Sgr A (Lacy et al. 1982; Van Gorkom et al. 1983) have revealed that Sgr A West is hardly a typical spiral-arm HII region. Lo and Claussen (1983) and Serabyn and Lacy (1984) suggest that the velocity structure of Sgr A West can be explained by the kinematics of gas infall toward a compact massive object at the center of the Galaxy. However, their hypothesis is not universally accepted. (Reviews by Oort [1977], Townes et al. [1982], and, specifically, Brown and Liszt [1984] shed considerable light on the pros and cons of both infall and outflow models toward the center of the Galaxy.)

The first comprehensive line studies of the Arc structure, apart from the first positive line detection by Mezger and Hoglund (1965), were carried out by Pauls _et al._ (1976). Later observations by Whiteoak and Gardner (1976) and Pauls and Mezger (1980) showed clearly that a large portion of the Arc, specifically the northern half of the Arc at positive latitudes, has thermal characteristics. Figure 7, which shows the distribution of line emission in the Arc at a resolution of 2ʹ.6, is reproduced from a map made by Pauls (1979). The kinematics of this thermal structure will be addressed in chapter 9.

## II. Motivation for the Study of the Continuum Arc

I briefly describe the main problems which motivated us to study the Arc and the main questions which we raised prior to our theoretical and observational findings in fall 1982.

Low resolution radio continuum observations had revealed an Arc or spur-shaped geometry, (see for example, figure 3) between $\ell = 0.0°$ and $\ell = 0.2°$, extending to negative latitude near the galactic center (Pauls _et al._ 1979; Downes _et al._ 1978; Altenhoff _et al._ 1978; Gardner and Whiteoak 1977). Molecular observations show that the Sgr A cloud, or 50 km $s^{-1}$ cloud, is nestled within the contours of the continuum emission (Fukui _et al._ 1977).

It was generally believed (before our observations described in chapter 2) that compact objects embedded in the Arc were probably HII

regions produced by newly formed massive O stars associated with a molecular cloud (e.g. Mezger et al. 1974, Pauls 1980, Güsten and Downes 1980). This suggestion was based on Westerbork interferometer observations at 5 GHz (Downes et al. 1978) which revealed numerous compact objects embedded within the extended emission of the Arc. These compact objects were shown to be located preferentially near the outer edge of the Arc ($\ell$ = 0°18 between b = 0°0 and b = -0°2). On the assumption that these compact sources were excited by massive O stars, one might have deduced that star formation has been taking place at a large rate in the galactic center region. Although, the presence of molecular clouds in the direction of extended thermal emission was consistent with this idea, several questions arose. First, why did the recombination line observations in the directions of the compact objects show little correlation in velocity and position with molecular emission from the 50 km $s^{-1}$ molecular cloud (Pauls et al. 1980)? Second, why were the compact objects clustered primarily in such a unique position, i.e. one edge of a molecular cloud opposite the center for the Galaxy and aligned roughly perpendicular to the plane of the Galaxy? Third, if star formation has been taking place at the high rate implied by the abundance of apparent HII regions, why were $H_2O$ masers, which are common indicators of star formation in the galactic disk, not found to be commonplace in the galactic center region, apart from the unusual cloud Sgr B2? Recent searches have uncovered a few (possibly background or foreground) $H_2O$ masers (Güsten 1982; Genzel and Downes 1978), far fewer and far weaker than would be expected if star formation in the galactic center were accompanied by maser emission to the same extent as in the disk (Morris, Yusef-Zadeh and Chance 1984). Fourth, why were the compact sources not concentrated near the peaks of the extended 10.7 GHz radiation as they are in Sgr B2, which is thought to be a site of intense star formation activity (Downes and Martin 1972; Elmegreen et al. 1978)? Finally, if HII regions in the galactic center were formed from O stars, as had been suggested to account for the radio continuum flux, then why were their surface brightnesses more characterstic of HII regions around early B stars (Pauls and Mezger 1980).

Having these questions in mind, we investigated the ionization structure produced by high velocity cloud-cloud collisions, as an alternative explanation for the origin of the compact sources in the Arc. Indeed, molecular couds in the galactic center regions have random velocities large enough, in principle, to produce ionizing shocks as they collide with each other. The algorithm that we used for modelling the cloud-cloud collisions will be described elsewhere.

The preliminary conclusions of this investigation indicated that the width of the ionized clouds produced by shocks could not be as large as that implied by the Westerbork map (Downes et al. 1978). Therefore, we (Morris, Chance, and myself) proposed to observe the compact sources in the Arc with the VLA in order to resolve them and thus, to test the cloud-cloud collision and star formation hypotheses. The subsequent VLA observations showed (see Chapter 3) that many of the previously reported compact sources were an artifact of an interferometric observation which had incomplete spatial frequency coverage and was thus insensitive to large-scale features.

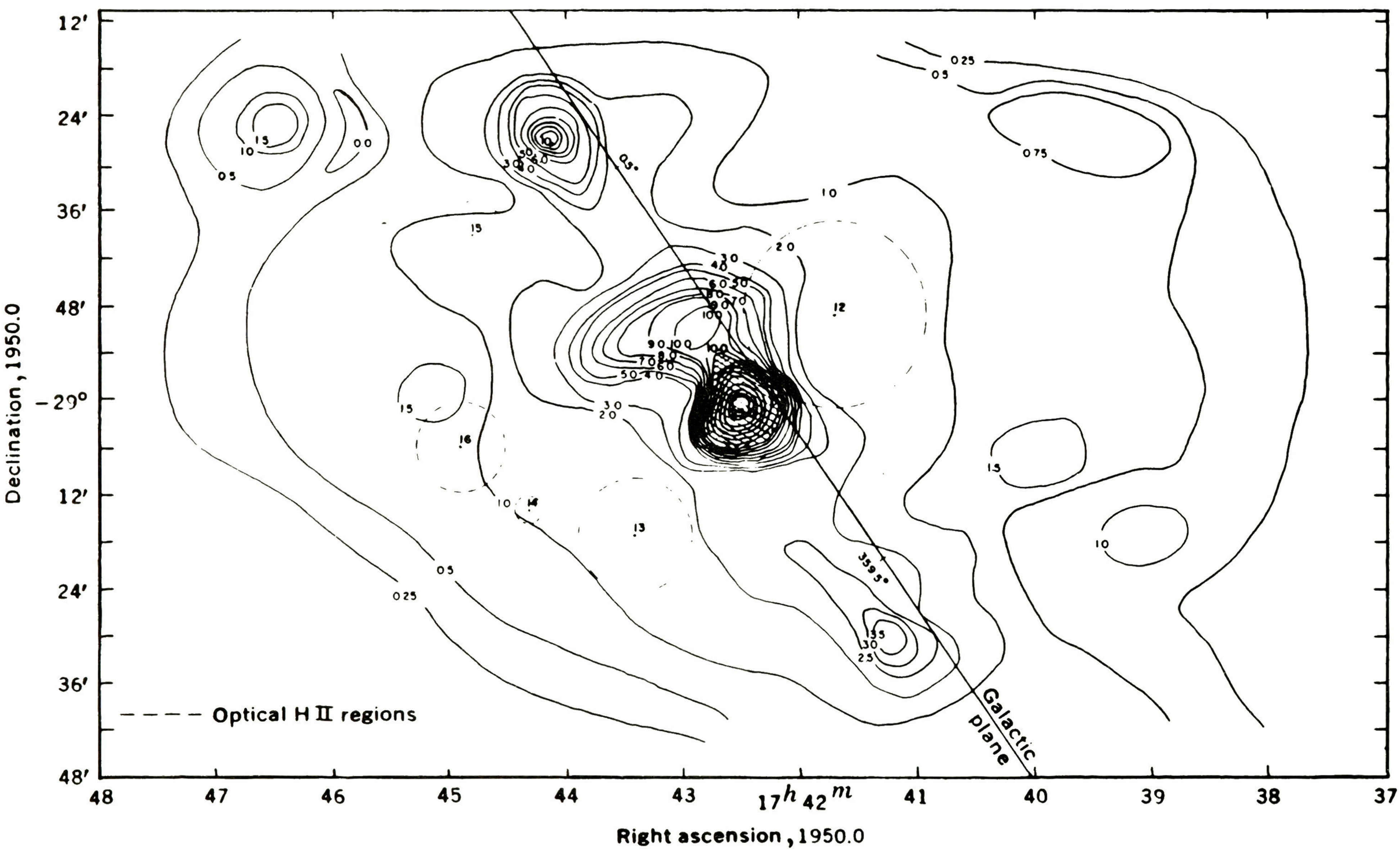

Figure 1: Contour map of the inner 200 pc of the Galaxy is made by Drake (1959) at 8 GHz. The contour unit is 1° K and the resolution is approximately 6 arcminutes.

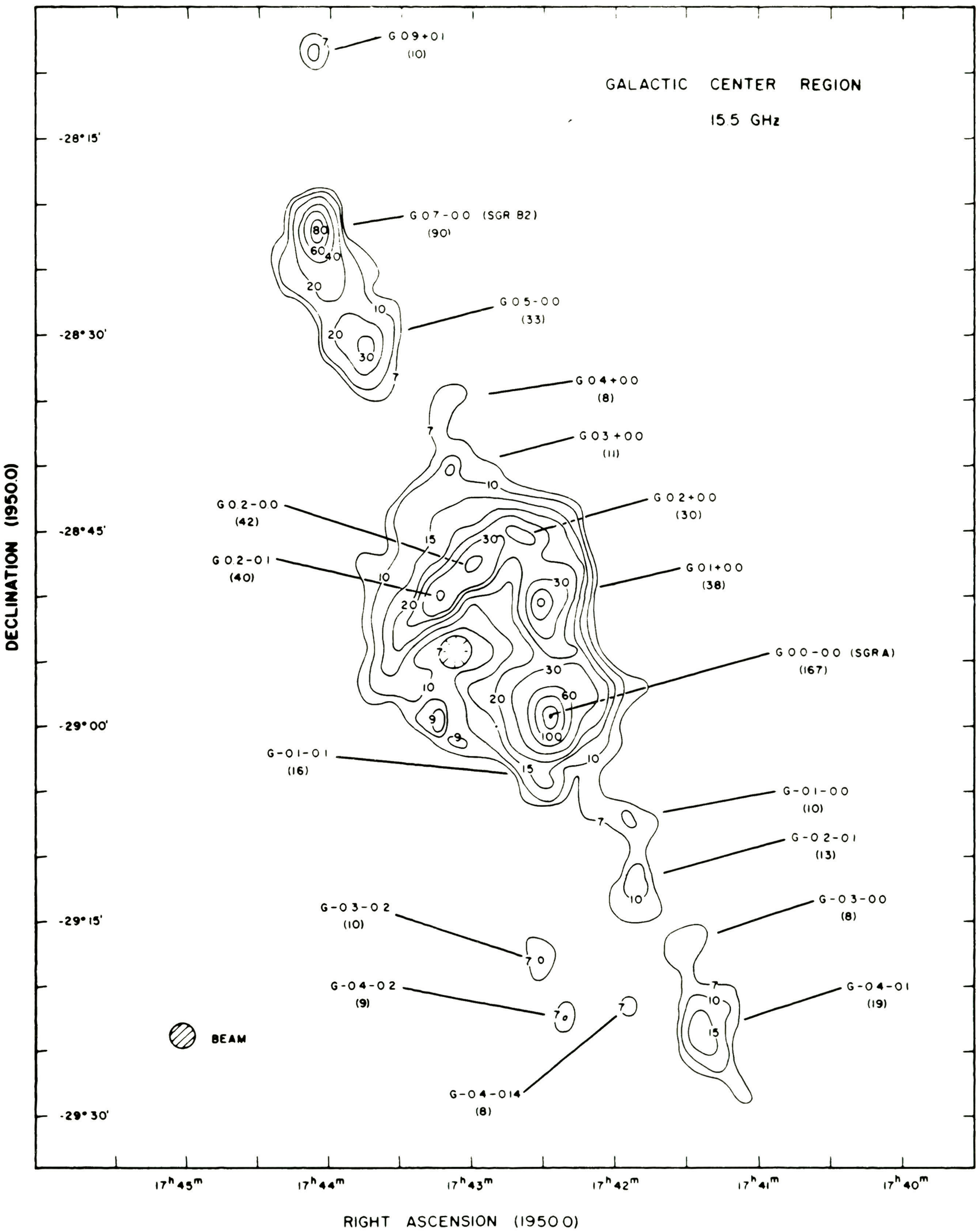

Figure 2: The 15.5 GHz map of the galactic center region made with an effective resolution of 135" by Kapitzky and Dent. The brightness temperature contour are in units of 0°.1 K.

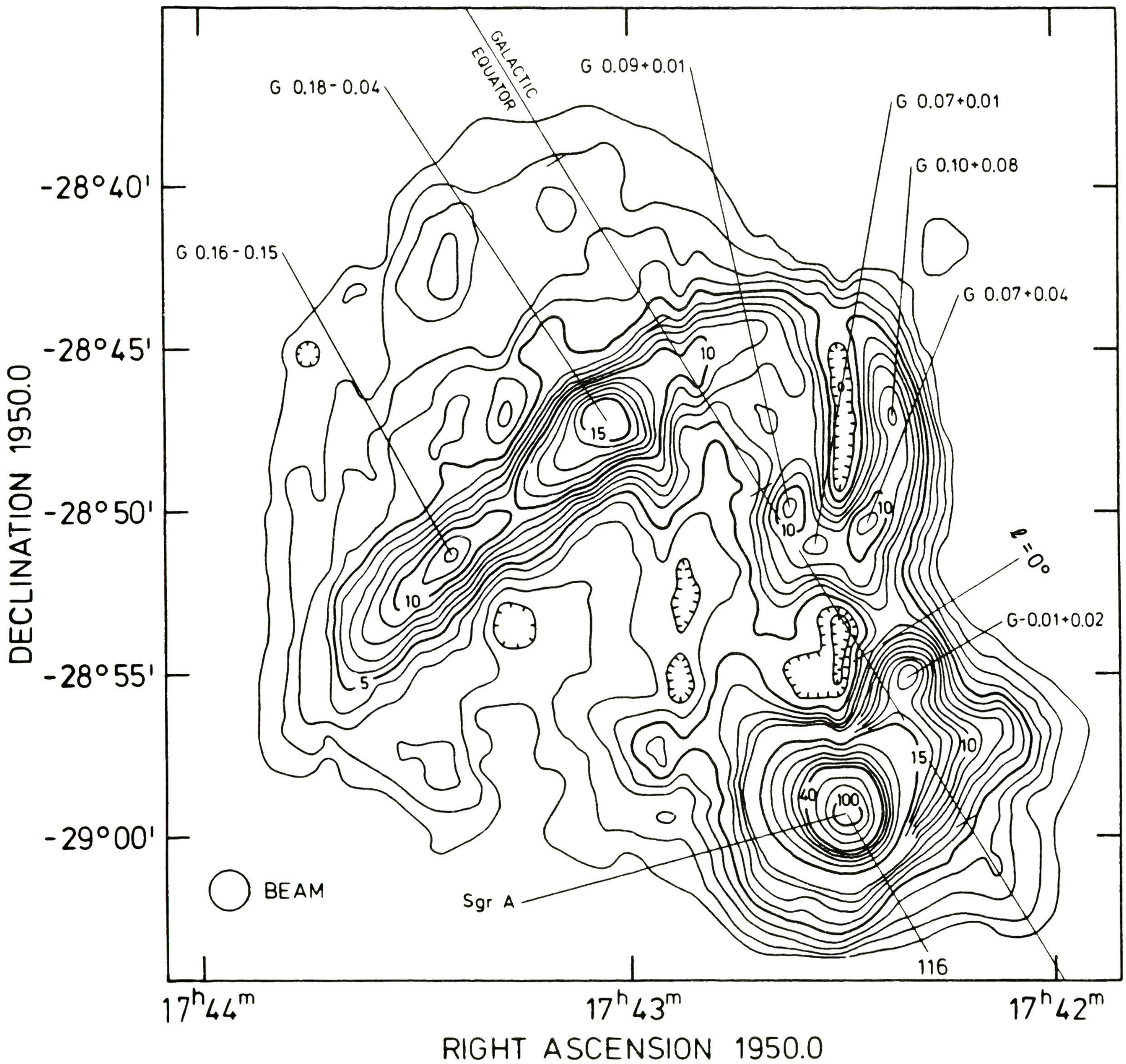

Figure 3: Contour map of the total intensity at 10.7 GHz is made by the 100-m Bonn telescope (Pauls <u>et al.</u> 1976). The spatial resolution (FWHM) is 77".

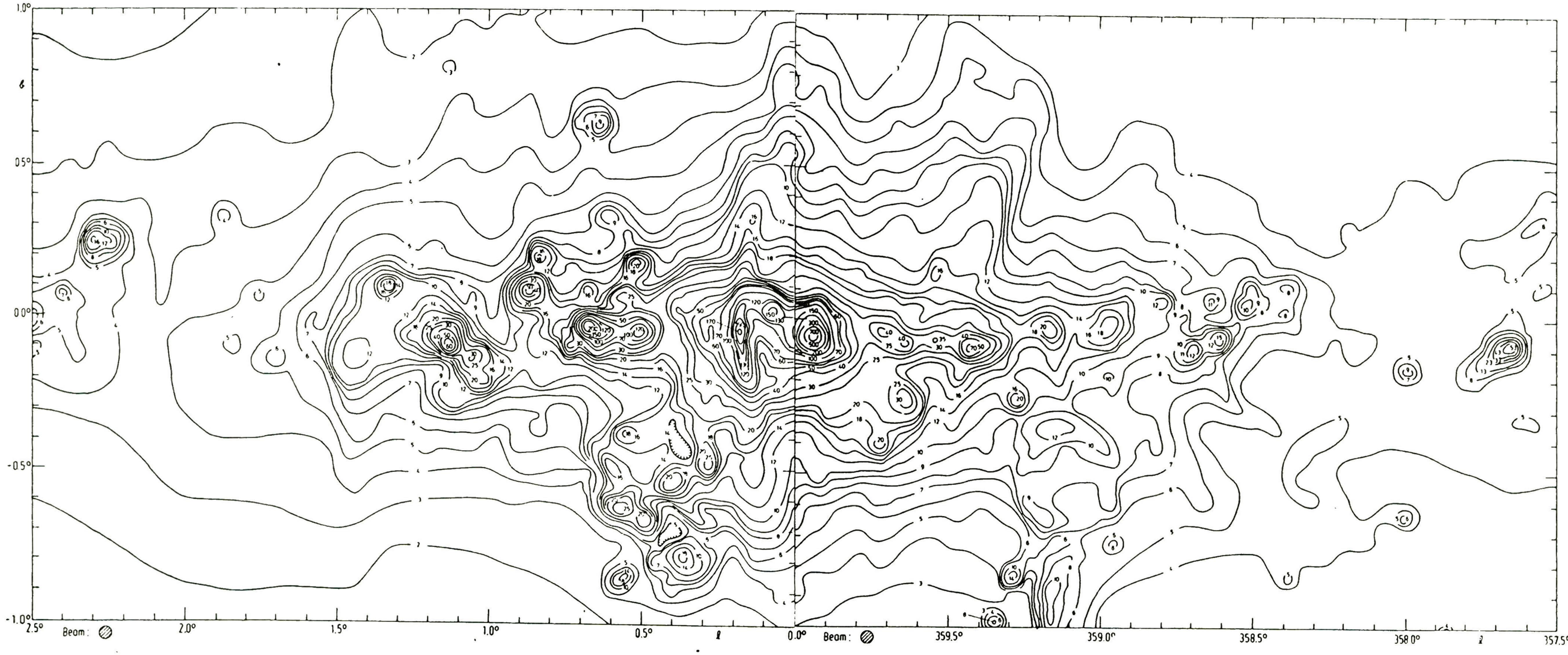

Figure 4: Contour map of the continuum emission from the inner 5°×2° (ℓ × b) of the galactic center at 4.8 GHz made with the 2ʹ.6 beam using the 100-m Bonn telescope of the MPIfR (Altenhoff <u>et al.</u> 1978).

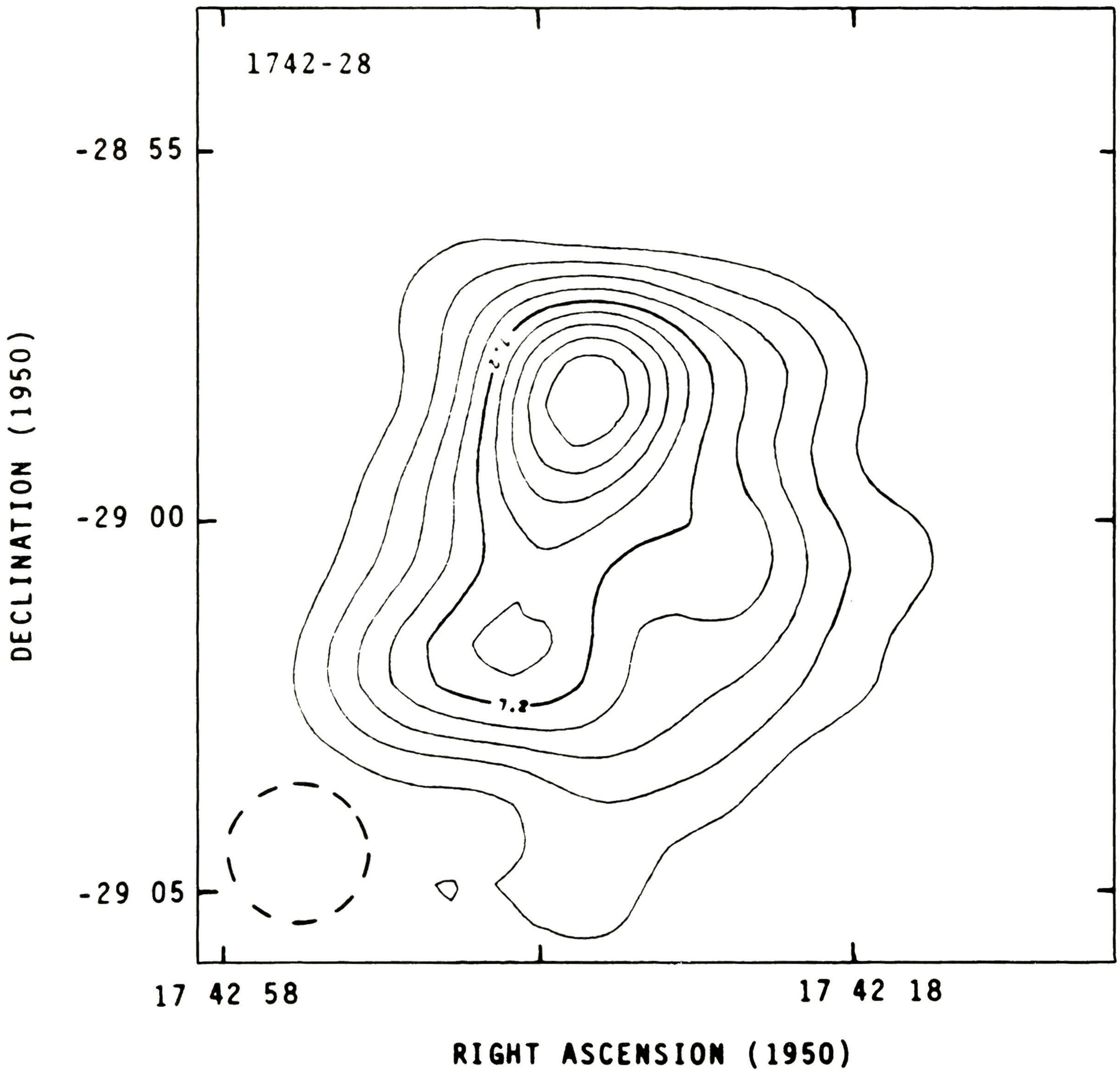

Figure 5: Contour map of Sgr A at 160 MHz observed with the Culgoora radiograph which has a resolution of 2!6 (Slee 1977). The peak brightness is 14.4 Jy/beam and the total flux density is 97.2 Jy. The contour interval and the lowest contour is 5% of the peak brightness.

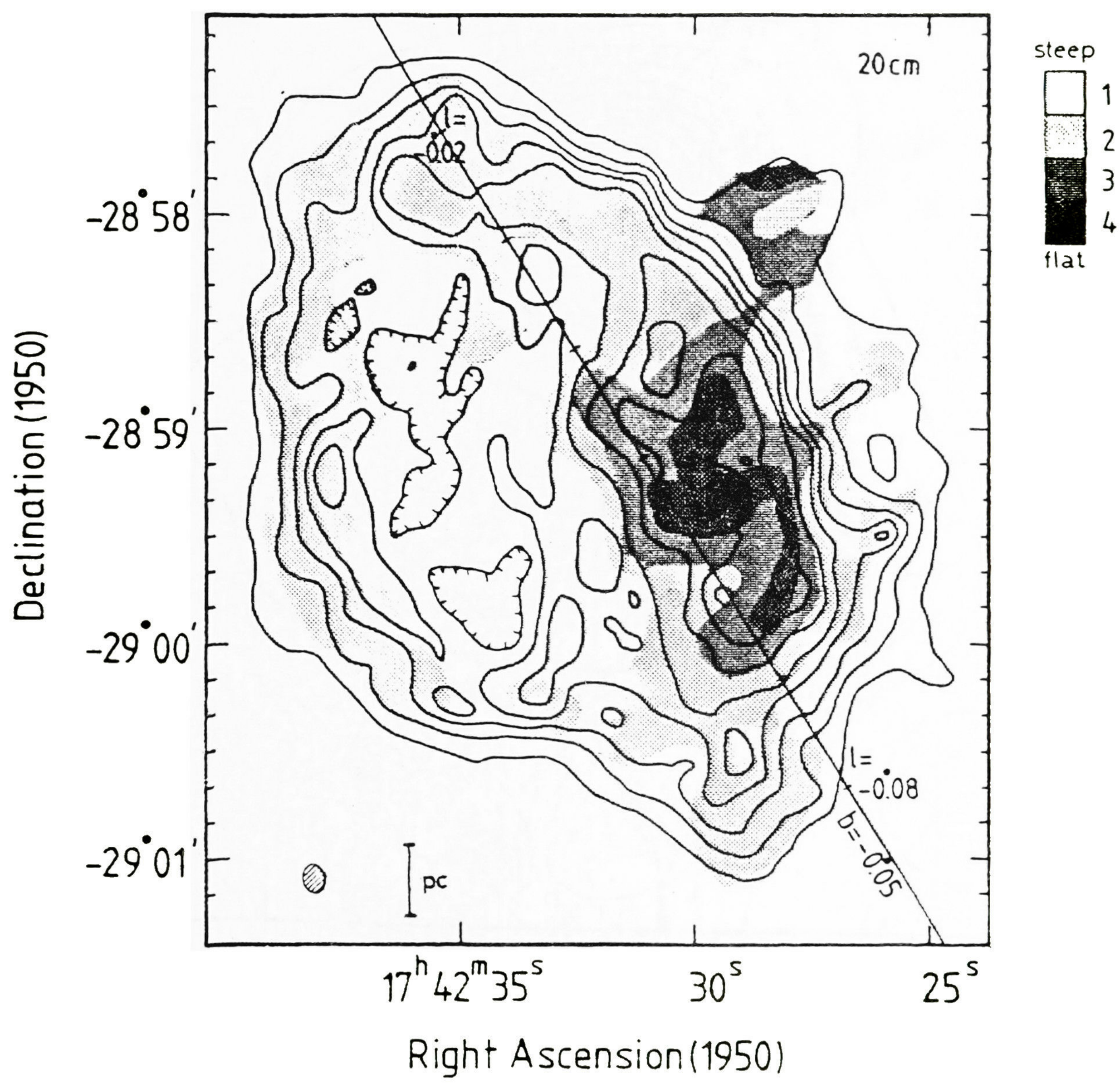

Figure 6: The spectral index (α) distribution is based on the 6 and 20 cm maps using the VLA (Ekers et al. 1983). The 20-cm contours are at 40, 80, 120, 200, 280, ... mJy/beam area (FWHM = 5"×8"). The shading 1, 2, 3 and 4 correspond to $\alpha < -1$, $-1 < \alpha < 0.5$, $-0.5 < \alpha$ 0.0, and $0.0 < \alpha$, respectively.

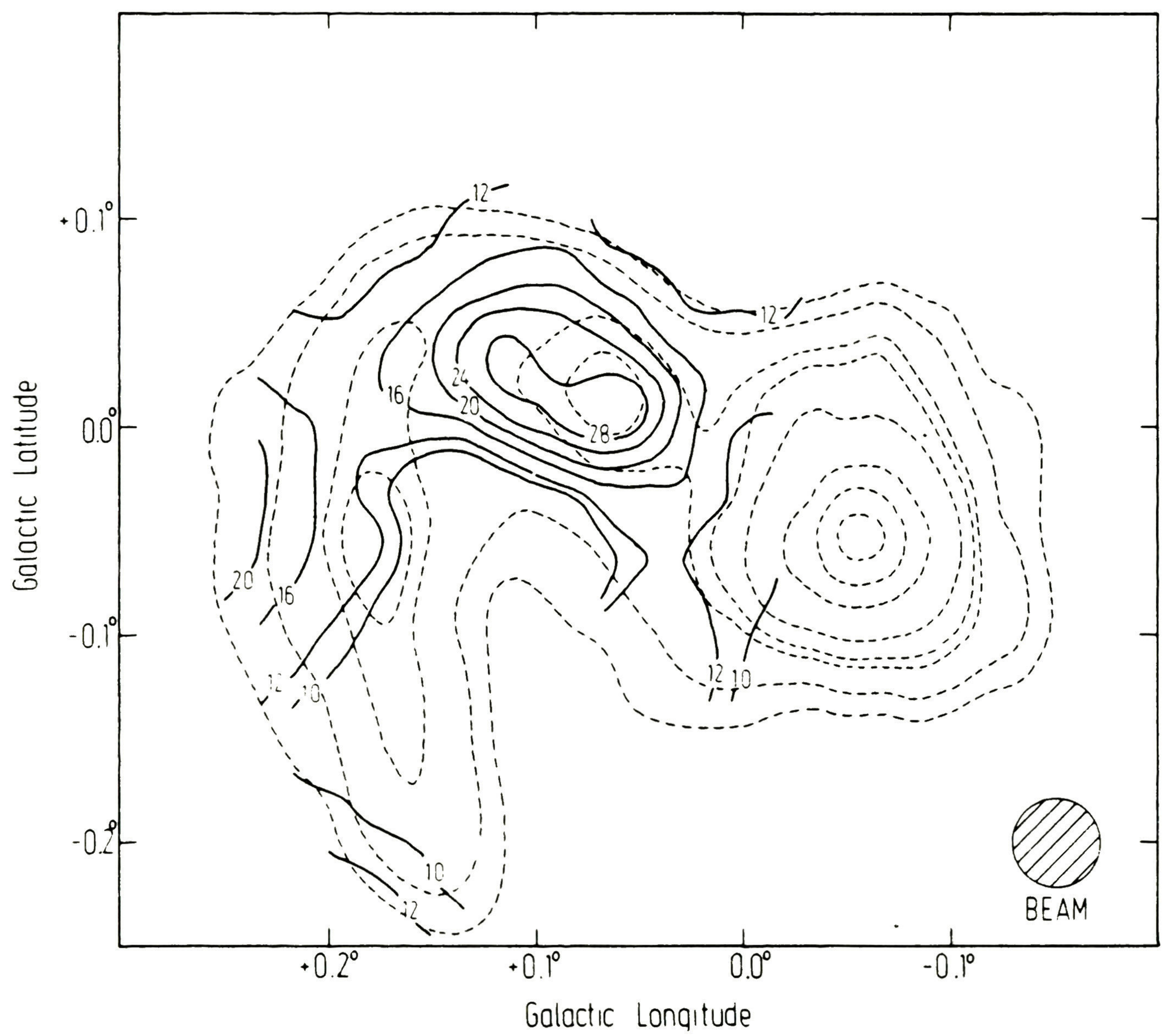

Figure 7: Contour map of the H109α recombination line intensity at a resolution of 2!6 superimposed on the 6-cm continuum map (Pauls 1979). The line intensity is in units of K. km $s^{-1}$. Both maps are based on observations made with the Bonn telescope.

# Chapter 2

## VLA Observations and Data Reduction

### I. Radio Continuum Observations

"... the most important astronomical object within
our Galaxy seems to be one of the most
difficult to observe."
M. A. Gordon

a) Aperture Synthesis

A major portion of our observations of the continuum emission from the galactic center have been carried out with the VLA[1] in four different configurations. There are a number of reviews and lecture notes which deal with the design, performance and operation of the VLA (Fomalont and Wright 1974; Thompson and D'Addario 1982; Hjellming 1982; Napier <u>et al.</u> 1983; Thompson <u>et al.</u> 1980;). Many of the observational parameters can be found in an excellent article by Napier <u>et al.</u> 1983. Here, we briefly describe 60 hours of continuum observations which we devoted to the galactic center. The results of these observations can be found in chapters 3 to 10.

In 1982 September, the telescope was used in the "B" configuration, in which the telescope spacings ranged from 0.2 to 9 km, the resolution was 4".4 9" ($\alpha\times\delta$) at 20 cm (1.45 GHz) and 1.5"×3" at 6 cm (4.8 GHz). We employed this configuration to search for compact sources and thus, observed mostly in short durations (~10-15 minutes) by situating numerous (2 cm) fields along the Arc at wavelengths of 2 and 6 cm. We soon learned that compact sources constitute only a minor portion of the galactic center emission and that the wide fields of view at 6 and 20 cm and the more compact configurations (B/C and C/D arrays) could be more effective in bringing out the relatively extended structures.

---

[1]VLA is a component of the National Radio Astronomy Observatory which is operated by Associated Universities, Inc. under contract with the U.S. National Science Foundation.

In 1983 May, a hybrid C/D configuration was used for the first time (hereafter $C/D^1$), in which the 9 antennas of the northern arm remained in the C configuration (antennas placed up to 2 km from the array center), while the 18 antennas in the other two arms were placed in the D configuration (at distances up to 0.7 km from the array center). The long northern arm compensates for the large southerly declination of the galactic center ($\delta = -29°$) and the consequent foreshortening of the north-south component of the baselines. Therefore, the synthesized beam of the telescope in the C/D configuration was almost circular, having a half-power beam width (HPBW) of 31"×29" at 20 cm and 10.9"×9.4" at 6 cm (similar resolutions are obtained when a high declination source is observed with the D array).

In 1984 March we reobserved the same region using the hybrid B/C array hereafter ($B/C^1$), in which the northern arm was extended to 6.4 km from the array center, in order to improve our spatial frequency coverage and to confirm our earlier findings. The synthesized HPBW of the B/C array was 11"×10" at 20 cm and 3.1"×2.9" at 6 cm. The B/C observations were carried out at two neighboring frequencies simultaneously using two sets of IF's (AC and BD) separated by 150 and 184 MHz at 6 and 20 cm, respectively. This technique has, in principle, the multiple advantage of 1) improving slightly the (u,v) coverage when the two sets in a given wavelength band are combined, 2) offering the potential of measuring the Faraday rotation for polarized sources, and 3) yielding a determination of the spectral index distribution.

In 1984 August a hybrid C/D configuration was again used (hereafter $C/D^2$), in order to add short (u,v) spacings to the data which had been obtained at 4 different frequencies (4.872, 4.722, 1.63, 1.44 GHz) using the B/C array (the technique of using two sets of IF's had not yet become available for earlier observations with the $C/D^1$ array).

In 1985 June (October) a hybrid B/C (C/D) configuration was again used (hereafter $B/C^2$ [$C/D^3$]) in order to study radio extensions of the Arc away from the galactic plane (b = ± 0.3°).[2] We chose a similar set of IF values to those of earlier observations ($B/C^1$ and $C/D^2$) at both 6 and 20 cm. We also studied a selected portion of the Arc at 2 cm.[3] All antennas had the new 2 cm FET amplifiers with system temperatures ~100°K. Two sets of IF's were centered at 14.565 and 15.015 GHz. The frequency bandwidth was 50 MHz in the 2-cm observations.

[2] Observations of the southern extension of the Arc (southern lobe) were performed in collaboration with Drs. Seiradakis, Lasenby, Wielebinski, and Klein.

[3] The 2-cm observations were done in collaboration with Drs. Inoue, Fomalont and co-investigators.

The frequency bandwidths employed were (12.5), 12.5 (25), and 25 (50) MHz at 20 (6 cm) for the B, B/C[1,2] and the C/D[1,2,3] arrays, respectively. These bandwidths were chosen to ensure that there was no significant radial smearing of the radio images over most of the fields of view (The width of the smeared image is proportional to fractional bandpass [$\Delta\nu/\nu$]). The intensity loss of the beam at the half power beam widths (the primary beam widths at half power are listed in Table 1) due to these choices of bandwidths were 90 to 95% of the peak intensity at the phase center.

In order to combine data from different arrays, the observations were carried out with the same phase centers throughout this study. The central positions and diameters of the observed fields of view are shown in figure 1 and are listed in Table 1. The 20 cm field of view upon which we spent most of our observing time (GC20 in Table 1) at 20 cm, was offset from the galactic center in order to include not only the Sgr A complex, but also the entirety of the prominent radio Arc lying to the northeast of the galactic center.

The data were calibrated in all six observing sessions at 6 and 20 cm (2 cm) by measuring a standard calibration source, 1748-253 (1741-312). The scales of flux density were determined by reference to 3C286, which was assumed to have a flux density of 14.6 (13.6) Jy at 1.446 (1.63) GHz, 7.42 (7.57) Jy at 4.872 (4.722) GHz and 3.52 (3.44) Jy at 14.565 (15.015) GHz. The polarization calibrator NRAO 530 was used during the observations with the C/D[1,2,3] and B/C[1,2] arrays. The source 1748-253 also served as a polarization calibrator during the observations with the B/C[1,2] array at 20 cm. In order to remove the instrumental polarization, the polarization calibrator was observed intermittently throughout the observations. The parallactic angle (the angle between the meridian of the polarization calibrator and the elevation great circle of the VLA antennas) coverage for NRAO 530 extended $\sim|70°|$. Because we had a number of fields of view at 2, 6 and 20 cm and because we did not have enough observing time to make full synthesis observations for each of the sources, each source was observed for ~1.5 to 2.5 hours in a 7-hour observing period. An optimum (u,v) coverage was obtained by observing a source 5-7 times throughout the observing run.

The strength of the correlated flux from Sgr A using the C/D configuration was so high that the on-line computers flagged most of the 20-cm data (these data, however, were recovered from off-line computers) because of an increase in the system temperature. Essentially, the on-line computers assumes that the strong correlated flux is interference. There is now an on-line checking program available at the VLA which deals especially with observations of strong sources such as Sgr A. It is noteworthy that a very small portion of data from the calibrators are flagged. This occurs when the system temperature ($T_s$) adjusts itself to a strong source during the actual on-source observations and then the antennas are moved toward the weak calibrators. Such an abrupt change from a strong source to a weak calibrator located near the source causes the incoming data from the calibrators to be flagged for a short period until $T_s$ adjusts itself.

The 20-cm data (both AC and BC IF's) from the four (two) observing runs corresponding to the field designated GC20 (Sgr A Halo), as indicated in Table 1, were combined in order to produce maps based upon sampling over a large range of spatial frequencies. A single configuration can cover a range of 43 in angular scales. This was done by successive iterative applications of a self-calibration procedure (Schwab 1981), which essentially removes antenna-based phase errors, to data from each of the configurations until the solution for phase converged. This required 10(4) iterations for the data set from each observing run. Then the data sets were combined one at a time, followed by a global self-calibration of the data base at each step. For purposes of self-calibration, the data were restricted to spatial frequencies less than 30 k$\lambda$, in order to obtain the fastest convergence of the solutions. We also compared the peak flux density of the Sgr A point source before and after application of self-calibration and showed an increase in the flux density of Sgr A source. (The peak flux of a strong point source should increase after applying such a procedure if the solution is converging).

The 6-cm fields of view are much smaller than the scale of the Arc and do not contain a strong point source -- consequently, self-calibration has not been successful, except for two fields: the one centered on the halo of Sgr A and Arc No. 7 (Table 1).

Because of the complexity of the radio structure in Sgr A and the Arc, and because of the large extent of the maps, in terms of resolution elements along each axis, a large number of clean com-ponents (see the CLEAN algorithm of Hogbom 1974) were computed in order to remove sidelobe features. The number of clean components were also increased by the choice of a small gain factor (~0.05) which was used in CLEAN deconvolution algorithm (the depth to which CLEAN does its deconvolution is roughly proportional to the number of clean components multipled by the gain factor). The small gain factor is used to avoid numerical instabilities in the deconvolution process.

In this study, we have used both the compact and the wide arrays of the VLA and therefore we are sensitive to weaker, finer and larger-scale structures than have appeared in previous interferometric maps of Sgr A and the Arc (Ekers et al. 1983; Downes et al. 1978). Effects of primary beam is corrected in almost all figures presented in this thesis. Furthermore, the extended polarized features and Faraday rotation measurements are verified by a number of adjacent and overlapping fields of view. Since the degree of polarized emission is generally > 15%, the effects of instrumental polarization away from the pointing center of antennas are not significant (i.e. $\lesssim$ 2-3%).

Because radio structure is present on all scales in the galactic center region, and because we have not attempted to account for flux present at scales larger than that which corresponds to the spatial frequency of the shortest antenna baseline, our maps do not accurately represent the extended flux. At 20 (6) cm, features which are extended over more than ~12(5) arcminutes are under-represented or missing from our maps. Since Sgr A and its halo are in fact at the core of a very extended distribution of emission (Altenhoff et al. 1979; Schmidt 1978; Sofue and Handa 1984), an artificial annulus of negative intensity having a diameter of ~15 arcminutes is produced in the VLA maps because of the lack of any information at baselines shorter than 25 m. (These values could be obtained by a single-dish telescope whose diameter is ≥35 m.) Inclusion of zero-spacing flux can improve the quality of a given map. However, the flux values of single-dish measurements of Sgr A at 20 cm (Kerr 1965 [see Downes and Maxwell 1966]; Biraud et al. 1960) is ~960 Jy, which is much greater than the flux seen at the shortest antenna separation (~130 λ) at 20 cm (i.e. ~400 Jy). The single-dish (zero-spacing) fluxes at 6 cm are also 2 to 3 times greater than the observed flux values at the shortest spacing of the VLA. The inclusion of the correct zero-spacing flux in the VLA maps would not bring out the extended features, since the solution to a deconvolution algorithm such as CLEAN (Hogbom 1974; Clark 1980) or the Maximum Entropy Method (Cornwell and Evans 1985) would not converge properly if the zero-spacing flux is ≥ 20 to 30% of the observed short spacing correlated flux. Because of the wrong interpolation of the zero-spacing flux for extended structures as large as the size of the primary beam, the measured values of the degree of the polarized intensity are upper limits.

b) Bandwidth Synthesis

In 1985 January we used the A-array configuration (antennas are placed up to 21 km from the array center) to fill in the long (u,v)

spacings of the 20-cm data set (GC20 in Table 1). Because bandwidth smearing is very significant for an extended source such as the radio Arc in this configuration (at 20 cm, the intensity at the half power points of the primary beam is reduced to 40% of the central intensity due to the bandwidth smearing when a bandwidth of 6.25 MHz is used) and because the continuum mode can not be used at bandwidths narrower than 6.25 MHz, the line mode with 16 channels of 1.5 MHz bandwidth were synthesized at 20 cm. This technique is essentially equivalent to a continuum mode observation which uses 25-MHz bandwidth but which leads to an effective bandwidth smearing characteristic of a 1.5 MHz bandwidth.

The averaging time of incoming data is inversely proportional to the size of the source and to the longest baseline, and thus we chose the averaging time of 5 seconds in this observation (as compared to 20-30 seconds in our observations with smaller configurations of the VLA). Such a short averaging time minimizes the distortion of the source 20' away from the field center. [The results of this observation is not included in this thesis because of the present lack of access to a supercomputer, which is essential in producing large radio images ($>$ 4096 × 4096 pixels). However, we hope to make such images in the near future.]

Table 1

The Observed Fields

| Field Designation | Arrays | Frequency (GHz) AC IF | BD IF | Phase Center α (1950) h | m | s | δ (1950) ° | ' | " | Full Primary Beam Width at Half Power |
|---|---|---|---|---|---|---|---|---|---|---|
| GC20 | B, C/D[1] | 1.44 | — | 17 | 43 | 15.0 | -28 | 52 | 00 | 30 |
| | B/C[1], C/D[2] | 1.446 | 1.630 | | " | | | " | | " |
| | A | " | — | | " | | | " | | " |
| Arc No. 1 | B, C/D[1] | 4.872 | — | 17 | 43 | 35.2 | -28 | 53 | 00 | 9 |
| | B/C[1], C/D[2] | 4.872 | 4.723 | | " | | | " | | " |
| Arc No. 2 | B | 4.872 | — | 17 | 43 | 21.2 | -28 | 50 | 00 | 9 |
| | B/C[1], C/D[2] | 4.872 | 4.723 | | " | | | " | | " |
| Arc No. 3 | B, C/D[1] | 4.872 | — | 17 | 43 | 04.0 | -28 | 48 | 00 | 9 |
| | B/C[1], C/D[2] | 4.872 | 4.723 | | " | | | " | | " |
| Arc No. 4 | B, C/D[1] | 4.872 | — | 17 | 42 | 30 | -28 | 49 | 40 | 9 |
| | B/C[1], C/D[2] | 4.872 | 4.723 | | " | | | " | | " |
| Arc No. 5 | C/D[2],B/C[2],C/D[3] | 1.446 | 1.63 | 17 | 43 | 10 | -28 | 43 | 00 | 30 |
| Arc No. 6 | B/C[2],C/D[3] | 4.872 | 4.723 | 17 | 42 | 25.0 | -28 | 42 | 15 | 9 |
| Arc No. 7 | B/C[2],C/D[3] | 4.872 | 4.723 | 17 | 42 | 35 | -28 | 52 | 40 | 9 |
| Arc No. 8 | C/D[3] | 4.872 | 4.723 | 17 | 43 | 50 | -28 | 56 | 00 | 9 |
| Arc No. 9 | B/C[2] | 1.446 | 1.630 | 17 | 41 | 30 | -28 | 50 | 00 | 30 |
| Arc No. 10 | C/D[3] | 1.446 | 1.630 | 17 | 41 | 37 | -28 | 40 | 00 | 30 |
| Arc No. 11 | C/D[3] | 1.446 | 1.630 | 17 | 44 | 30 | -29 | 00 | 00 | 30 |
| Arc No. 12 | C/D[3] | 14.56 | 15.01 | 17 | 43 | 22.5 | -28 | 50 | 54 | 3 |
| Arc No. 12 | C/D[3] | 14.56 | 15.01 | 17 | 43 | 22.5 | -28 | 50 | 54 | 3 |
| Arc No. 13 | C/D[3] | 14.56 | 15.01 | 17 | 43 | 16 | -28 | 49 | 35 | 3 |
| Sgr A Halo | B/C[1],C/D[2],B/C[2] | 1.446 | 1.630 | 17 | 42 | 25.0 | -28 | 57 | 30 | 30 |
| | C/D[2],B/C[2],C/D[3] | 4.872 | 4.723 | | " | | | " | | 9 |

## II. Radio Recombination Line Observations[4]

In 1984 July, radio recombination line emission (H110α) from the Arc was observed using the VLA in the C/D configuration. Two 6-cm fields of view were centered on the arched filaments (G0.1+0.08) and the sickle-shaped structure (see chapter 3) crossing or wrapping around the linear filaments, (G0.18-0.04). The chosen phase centers for these two fields were identical to those used earlier in the radio continuum observations (Arc No. 3 and 4 in Table 1). We spent ~6 hours of observing time on each of these two sources. The center velocity, $V_{LSR}$, for G0.1+0.08 (G0.18-0.04) was -20(+20) km $s^{-1}$ .

A Hanning smoothing function was applied to the cross-correlation function before it was Fourier transformed. Every other channel from a 64-channel bandpass was then discarded (this was done to optimize the product of the total number of baselines and the total number of frequency channels [i.e. this product is currently limited to 9600 at the VLA]). The total bandwidth used was 3.125 MHz (97 KHz resolution) and the channel spacing was equal to the velocity resolution after smoothing (6 km $s^{-1}$) in this technique. The instrumental parameters for each of the two sources are listed in Table 2.

A bandpass calibration to correct for variation of gain and phase across the 32-frequency channels was obtained from a 30-minute observation of 2134+004. The phase calibration employed in these observations was 1748-253. Standard calibration was then followed.

[4] The line observations were done in collaboration with Dr. J. van Gorkom.

The maps which will be presented in chapter 7 are made with a natural weighting function. Because each visibility point in the (u,v) plane gets an equal weight when this function is applied, the optimum sensitivity is achieved. However, a broad and often elongated synthesized beam is produced. A mean continuum map was formed from averaging the lowest and the highest frequency channels of the spectrum. These channel maps of G0.1+0.08 (G0.18-0.04) had velocities from -110(-52) km $s^{-1}$ to -86 (-28) km $s^{-1}$ and from 46(28) km $s^{-1}$ to 70 (46) km $s^{-1}$. The mean continuum map was then subtracted from all channels and the difference maps -- those showing line emission -- were CLEANed (Clark 1980). The mean continuum map was also CLEANed.

Table 2

Observational Parameters

| | G0.1+0.08 | G0.18-0.04 |
|---|---|---|
| observing data | July 23, 1984 | July 24, 1984 |
| total no. of antennas | 25 | 25 |
| duration of observation | 6h | 6h |
| shortest, longest baseline (m) | 40, 2200 | 40, 2200 |
| half power of synthesized beam (major axis × minor axis, PA) | $22''.5 \times 11''.3$, $-4°$ | $15'' \times 12''$, $4°$ |
| field center (α, δ) | $17^h42^m30.0^s$, $-28°49'40''$ | $17^h43^m04^s$, $-28°48'00''$ |
| half power primary beam | 9' | 9' |
| number of channels | 32 | 32 |
| velocity channel spacing | 6 km $s^{-1}$ | 6 km $s^{-1}$ |
| velocity resolution | " | " |
| rest frequency H110α line | 4874.157 MHz | 4874.157 MHz |
| center velocity (LSR, radio definition) | −20 km $s^{-1}$ | +20 km $s^{-1}$ |

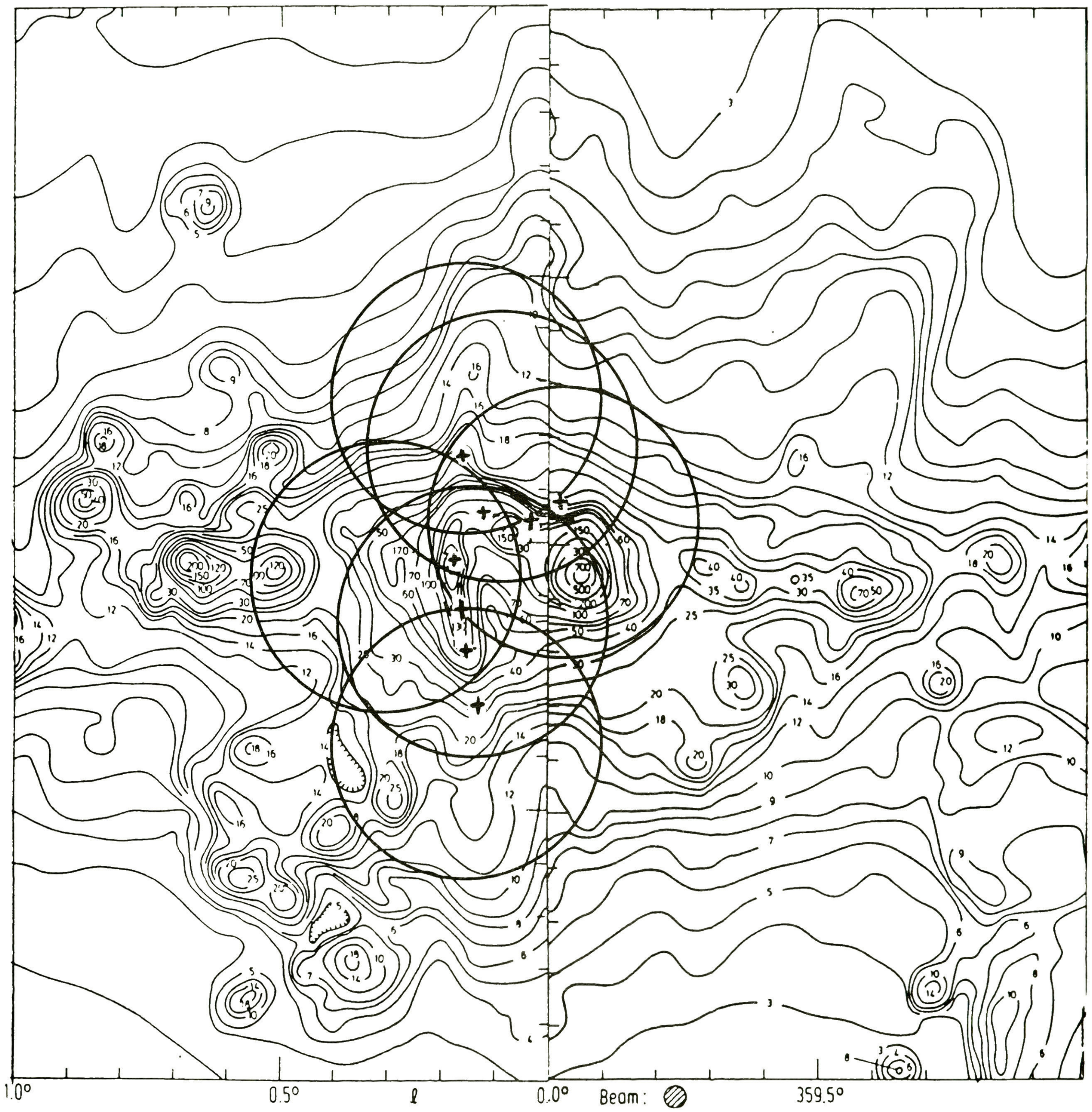

Figure 1: The circles show the observed field of view at 20 cm, i.e. the FWHM of the beam. The phase center of the 6-cm fields are indicated by crosses.

# Chapter 3

## The Discovery of Highly Organized, Large-scale Radio Structures Near the Galactic Center

"... that you can't say it is bigger or brighter or more something than something else. We barely know where to start."

R.L. Brown

## I. Introduction

Much of the past works have been concentrated on the small-scale radio and infrared studies of Sgr A. Recent studies are beginning to shed light on the large-scale structures and the spatial relationship between the ionized, atomic, molecular and dust components of the interstellar environment near the galactic nucleus (Fukui et al. 1977; Liszt and Burton 1978; Güsten and Downes 1980; Güsten et al. 1983; Pauls et al. 1976; Cohen and Few 1980; Gardner and Whiteoak 1977; Pauls and and Mezger 1980; Gatley et al. 1977; Hildebrandt et al. 1978; Dent et al. 1983; Odenwald and Fazio 1984; Stier et al. 1983). However, many fundamental problems on the origin and structure of small, intermediate, and large scale sizes remain unsolved. For example, what is responsible for the ionization, what is the relationship between the ionized gas and dust, and what is the physical relationship, if any, between the intermediate scale features (those that are located within the inner 60 pc of the Galaxy) and the small-scale features (those that are located within the inner 5 pc of the Galaxy).

Population I stars have been considered as the source of ionizing photons in the intermediate regime, (e.g. Pauls and Mezger 1980; Rieke 1981; Downes <u>et al.</u> 1979), but direct evidence for the presence of an adequate number of OB stars is lacking (see chapter 1; Güsten and Henkel 1983). The results of our VLA observations, which will be described below, illustrate that the Arc consists of a highly ordered set of large-scale features which appear to be related strongly to a galactic-scale phenomenon rather than to localized star formation in pockets near the galactic center. Therefore, the case for newly-formed OB stars as the primary source in the region of the Arc is substantially reduced. Furthermore, these results indicate that the idea that cloud collisions are the source of ionization can also be ruled out (see chapter 1).

Because the Arc consists of both large and small scale features and because many of the low surface brightness features in the Arc are not recognizable in traditional contour plots, we present numerous radiographs and gray scale maps in order to clarify their structural details. Indeed, there is a significant overlap between the display of radio images of the galactic center region and analysis. The general procedure used to construct each figure was described in Chapter 2.

# II. Results

"It is a riddle wrapped in a mystery inside an enigma."
W. Churchill [October 1, 1939]

A great deal of information is obtained in the Arc, the most interesting piece of which is the recognition that the Arc breaks up into narrow filamentary structure and a variety of non-filamentary structures surrounding the filaments. Figure 1 (a-d), which exhibits four different aspects of the 20-cm distribution in a 30' field of view, is made at 20-cm wavelength and shows the region of the filamentary Arc and the Sgr A complex. Indeed, multitudes of radio structures with a wide range of scale sizes constitute the Arc and the Sgr A complex (the phenomenology associated with the Sgr A complex is described in chapter 6). The filaments in the Arc are classified into 2 categories, namely, the linear and the arched filaments, each of which is described below, followed by a description of the non-filamentary features, radio shadows, discrete compact sources, the polarization characteristics and the spectral index measurements.

## 1. The Linear Filaments

A system of narrow filamentary structures having typical widths of ~20 arcsecond (1 pc) comprises the linear segment of the Arc as seen in single-dish maps (see figures 1-5 in chapter 1). This system of linear filaments is seen in figure 1, and appears to consist of northern and southern filaments (i.e. bundles or sets of filaments) located between $\ell$ = 0.16° and $\ell$ = 0.18°. They are perpendicular to the galactic plane, parallel to each other, and have a slight curvature, convex toward increasing longitude. The brightest and the longest of these filaments extends at least as much as 15' (i.e. the southern filament) and appears coherent, regular, unbroken, and homogeneous. Its brightness and its width decrease slightly as it crosses the galactic plane in a direction of increasing latitudes. Higher resolution pictures of the linear filaments show substructures which are described next.

a) Thin Strands of Radio Emission

On the finest scale which we can discern in our high-resolution 6-cm maps, the southern edge of the Arc (near G0.16-0.15) is comprised of a network of about a dozen thin strands of radio emission. These features are best shown in figures 3 and 4(a-b), which have resolutions of 7.8"×7.3" and 3.2"×2.8", respectively. These rather remarkable arrangements of radio structures appear to be parallel, organized in their scale sizes, and oriented perpendicular to the galactic plane. A slice cut across these strands, as shown in figure 5, reveals that their spacing is quite regular, with about 10" (or 0.5 pc) separating adjacent strands. Also they appear either to change their brightness abruptly at $\alpha \sim 17^h43^m40^s$, $\delta \sim 28°54'$ along a well-defined line or to abruptly merge into a diffuse, extended medium as they extend toward negative latitudes. Figure 2, which is a 20-cm map with a resolution of 17"×16.4", shows best that the southern filament, perhaps the longest filament in the Arc, appears to extend as much as 0.3° below the galactic plane. The extension of this structure will be discussed in Chapters 8 and 10.

b) Twisting of the Filaments?

The parallel set of thin strands of radio emission, as viewed in figures 3 and 4 (a-b), appear to converge in the direction of increasing latitudes and form a number of distinct filaments which constitute the southern and northern filaments (see figure 2). These filaments can best be seen at 6 cm in figures 6a and 6b which are based on the same data set but are shown with a different transfer function. Note the appearance of distinct filaments which might be due either to the superposition of two separate filaments, or alternatively, to a gentle twist of the filaments about each other with a small pitch angle. The separation between the northern and southern group of filaments appears to widen by ~30" in the direction to the north. A close-up view of the merging filament can also be seen in figure 7 (i.e. $\alpha \sim 17^h43^m10^s$, $\delta \sim -28°49'30"$). Figure 8 shows a slice cut across several linear filaments whose separations by a number of dark lines on the positive radiographs are represented by several dips in the brightness of the filaments. The brightness temperature of these filaments at 6 cm wavelength, as seen in figures 7 and 8 with a resolution of 3.3"×3.3", is ~ 30 °K.

Another aspect of the structure of the linear filaments as they cross the galactic plane can be seen at 6 cm in figures 9 (a-b). These filaments maintain their linearity as they cross a thermal non-filamentary structure (G0.18-0.04), described in §II.3. We note that the brightnesses of the southern group of the filaments decline substantially as they are followed from the southeast to northwest.

2) The Arched Filaments

The arched filaments are those that connect to the western portion of the linear filaments at positive latitudes and arch southeasterly toward the galactic plane. These filaments, which are best exhibited in our high-resolution 6-cm map (FWHM = 3.8"×3.1") -- as seen in figure 10 -- consist of two groups of components: eastern and western, each of which brightens as it nears the galactic plane. These two groups appear to have both a strikingly similar curvature and a broken or patchy structure but they keep their globally filamentary nature as they continue toward the galactic plane. The two groups of arched filaments, unlike the linear filaments, appear to be physically connected by a non-uniform and clumpy structure in the region to the south of $\delta$ = -28°50'. This clumpy structure is best seen in the contours and the radiograph of total intensity at 20 and 6 cm in figures 11a and 11b, respectively. Because of their disorganized and clumpy nature - unlike the linear filaments - we suggest that the arched structures have thermal characteristics (see chapter 9).

The western group of arched filaments appear to consist of two sets of filaments (W1 and W2 in figure 10) separated by a gap at $\alpha \sim 17^h42^m25^s$, $\delta$ = -28°49'. A rather peculiar thread of radio emission crossing the western filaments at $\alpha = 17^h42^m20^s$, $\delta$ = -28°49' can also be discerned in this figure. A full description of this feature is postponed until chapter 5. The eastern arched filaments branch out into two components (E1 and E2 in figure 10) as they extend northward. Figure 12 shows the 6-cm contour map of the bright southeastern filament which splits into two components at $\alpha \sim 17^h36^m$, $\delta \sim -28°48'30''$ as it is followed northward. The detailed kinematic information of the arched filaments is described in chapter 9.

We note that the two systems of linear and arched filaments appear to intersect with each other. It appears that the western arched filaments (W1 and W2) cross the southern linear filament group and merge smoothly with the northern filaments (see figures 1 and 2; chapter 10). The brightnesses of the linear filaments, however, do not increase substantially at the positions where the linear and arched filaments meet. We also note in figure 2 that the angular distance between the two sets of arched filaments is larger by ~1' than those of the linear filaments.

3) Non-filamentary Features

A considerable amount of radio emission from the Arc arises from non-filamentary features which include: features that appear isolated and have a collimated structure (chapter 5); features which are diffuse, non-uniform in their brightness distribution and have weak surface brightness. We describe the latter features next.

a) Diffuse Halo

A rather broad and diffuse structure appears to envelope the bright linear filaments. The 6-cm map shown in figure 13 presents the halo feature. It constitutes a major component of the Arc whose significance will be discussed in detail in chapter 4.

b) Helical Structure

One of the most intriguing characteristics of the linear components is the appearance of a segment of a helical structure winding about the southern end of the twisted-looking filaments with a constant pitch angle. The 20-cm view of this low surface brightness emission, which has a typical brightness temperature, $T_b$, of 120 °K, is presented in figures 14 (a-b) and 15.

This structure consists in part of three helical segments whose axis coincides with the mean position of the linear filaments. The largest segment of the helical structure has a radius of curvature of ~3' (10 pc) and a width of ~1' (3 pc). The "helical" structure is spatially asymmetrical with respect to the galactic plane and can only be seen near the southern edge of the linear filaments.

If our perception of helical geometry is in accord with reality the structure seen in figures 14 (a-b) and 15 adds a third dimension and thus alludes to the geometry of the Arc as whole with respect to the line of sight. One could then argue that the linear filaments are inclined by ~40°-60° with respect to the line of sight -- the southern end of the Arc will be located on the far side -- with the assumption that the linear filaments are aligned along the symmetry axis of the helical structure whose pitch angle is assumed to be ~ 90°. This inclination angle is based on the ratios of major to minor axis of the helical segments. The uncertainty associated with the pitch angle of the helical structure makes it very difficult to interpret correctly the geometry of the Arc without any ambiguities.

c) Large-scale "Arch"

A large-scale "arch" appears to join the two ends of the linear filaments. The photographic representation of the Arc, the "arch", the Sgr A complex, Sgr B2 and G0.52-0.04 can be seen at 20 cm in figure 16. The "arch" structure appears to be anchored at the two ends of the linear filaments. This feature has a brightness temperature of ~100 °K at 20 cm and has a projected diameter of ~30 pc. The "arch" can also be recognized as a diffuse structure in the low-resolution maps at 15.5 GHz and 10.7 GHz made by Kapitzky and Dent (1974) and Pauls et al. (1976), (see figures 2 and 3 in chapter 1, respectively).

d) Sickle-shaped Feature ($\alpha = 17^h43^m03.2^s$, $\delta = -28°47'03''$)

At about the position where the vertical filaments cross the true galactic plane, a radio structure (referred to in single-dish studies as G0.18-0.04) is seen superimposed upon the filaments. As can be seen in figure 9 (a-b), its apparent form is crudely that of a sickle (or a question mark) oriented transverse to the long axis of the filamentary structure.

Figures 9 (a-b), 17 and 18 (a-b) all show the 6-cm maps of the sickle-shaped feature with spatial resolutions of 2.6"×1.9", 4.7"×4.7" and 7.1"×6.3", respectively. These figures strongly suggest an interaction between the filamentary structure and G0.18-0.04: first, the overall brightness of the longest southern linear filament, as seen in figure 17, is stronger on one side of G0.18-0.04 (negative latitudes) than the other (positive latitudes). This effect might be due to dissipation of some of the energy of the longest filament as it crosses through the thermal sickle-shaped feature (see chapter 9). Second, figure 17 shows clearly that some of the northern group of linear filaments brighten substantially very near the sickle-shaped feature. In fact, the northernmost filament appears to coincide with a series of knot-like structures aligned in the direction of the filament [see figures 9 (a-b)]. We note that the linearity of the filaments is maintained as they pass through G0.18-0.04, but with a slight sign of a structural discontinuity as they approach the semi-circular portion of G0.18-0.04. This subtle effect which is perhaps the most convincing argument for the physical interaction of the sickle-shaped feature and the linear filaments can best be seen along the southernmost filament in figure 9a as it crosses $\alpha \sim 17^h42^m58^s$, $\delta \sim -28°48'30''$ and bends by about 5° southward. Third, the brightness of the sickle-shaped feature decreases and it becomes diffuse as it is followed southward along the galactic plane (see figure 18a) where no linear filaments are seen. However, the northern segment of the sickle-shaped structure is discontinued abruptly and appears to be bounded by the northern group of filaments (see fig. 18a). We note that the southern and diffuse segment of G0.18-0.04 is somewhat organized and shows some degree of substructure, particularly at $\alpha = 17^h42^m55^s$, $\delta = -28° 49'$ (cf. figure 9). This diffuse structure appears to show the same curvature as those of the arched filaments (see § II.2) as it gets connected to the northern tip of the Sgr A complex (see chapter 6).

e) Multiple Hot Spot Feature ($\alpha = 17^h43^m04.7^s$, $\delta = -28°48'42"$)

Another non-filamentary component of the Arc which appears to be physically connected to the sickle-shaped structure by a set of very narrow filaments can be seen in Figures 9, 17 and 18 (a-b). This source (G0.16-0.05) consists of at least 5 components which constitute an extended structure with a 30" scale size. The 20 and 6-cm maps of this feature are shown in figures 19 and 23j. The interesting velocity structure of this feature will be addressed in chapter 9.

f) Counter-arch Feature

Figures 1d (SW corner) and 2 show clearly two sets of curved features crossing the southern edge of the Arc. These features, which are located at negative latitudes, appear to constitute a negative-latitude counterpart to the prominent arched filaments at positive latitudes (§II.2). Indeed, the counter-arch features and the arched filaments appear to show similar morphology and interestingly, they both cross the linear filaments at the locations at which the uniform brightness of the linear filaments drops substantially away from the galactic plane. Furthermore, they both appear to interact with the linear filaments. The interaction between the counter-arched filaments and the linear filaments can be noted in figures 4a and 13 where the northern portion of the curved filament

which crosses the linear filaments is displaced by 20" at $\alpha \sim 17^h43^m28^s$, $\delta \sim -28°53'40"$ as it is continued southward.

The evidence for the interaction between the arched and linear filaments is discussed in chapter 10. However, the arched filaments and their counterparts are different in two respects: (1) the counter-arch features are ~ ten times weaker in their surface brightness, are more diffuse, more uniform, less broken, and less filamentary in their appearance than their northwestern counterparts; (2) the counter-arch features appear to be directed toward the nucleus -- though, their surface brightness fades out before reaching the Sgr A nucleus -- whereas the arched filaments display much more complexity as they are directed toward the galactic plane and away from the nucleus of Sgr A.

4) Radio Shadow

At the positions where the large-scale helical structure and the linear filaments cross (see figure 17), the brightness of the northern filament drops by a factor of 1.5 - 2 at $\alpha = 17^h43^m17^s$, $\delta = -28°49'21"$ at both 6 and 20-cm wavelengths. The slices cut along the linear filaments which are shown in figures 20 and 21 illustrate this decrease in surface brightness along the northern filaments at 6 cm and 20 cm.

There are at least four other positions at which a radio shadow can be seen, one of which is located at $\alpha = 17^h42^m51.8^s$, $\delta = -28°46'18"$ along the southern linear filaments as they cross the galactic plane (cf. figure 17). The other can be seen as a dark line of ~0.25 pc thickness crossing the parallel set of radio strands which comprise the southern edge of the Arc (cf. figure 4). A slice cut across this dark feature is presented in figure 22 which shows a reduction by a factor of 2 in the surface brightness of the strands. This dark line at $\alpha \sim 17^h43^m32^s$, -28°52'08" can also be seen at 20 cm in figure 14b where the easternmost segment of the helical structure crosses the linear filaments. The radio shadows can be explained in terms of free-free absoprtion of photons as they pass through cold and dense plasma. If optical depth is $\geq 0.7$, the emission measure has to be $\geq 2.6 \times 10^2\ T_e^{1.33}$, where $T_e$ is the electron temperature of thermal plasma. There is, of course, an ambiguity in determining both the temperature and the electron density of absorbing features.

5) Discrete Sources

The compact sources recognized in the Arc and its vicinity are listed in Table 1 at both 6 and 20 cm. The discrete sources identified in the Sgr A complex (A-M) are reported in chapter 6. We note that many of the sources reported here are either projected against an extended background of radio emission or they are associated physically with an extended structure. The more prominent compact sources are described below.

Sources N1, N2 and N3 (sources 18, 17, 31 by Downes et al. 1978) whose intensity contours are shown in figures 23 (a - c), appear to be associated with sickle-shaped feature (G0.18-0.04) in an unusual manner. Indeed, source N3, which is seen in figures 9 and 17, appears to be linked to the northern segment of the semi-circular feature (i.e. sickle-shaped feature) by a rather diffuse but collimated structure (see figure 12 in chapter 10). This compact source has a spectral index $\alpha^{4.8}_{14.5} = -0.9$ based on the peak brightness temperatures at 6 and 20 cm (see the 2-cm results in the next section). Curiously, the brightness and the physical size of this isolated source (cf. figure 23c) is roughly similar to those of radio knots within the sickle-shaped feature. Sources N1 and N2 appear also to be associated with the southern extension of the diffused segment of the sickle feature (see figure 18a). The compact continuum sources 17, 18, 23, 25, 28, 29, 31, 32, 33, 35, 38, and 40 listed by Downes et al. are part of the sickle and the multiple hot spot features described in §II.3.

Source O1 lies very near the galactic equator (G0.208-0.002) and was listed as source 42 by Downes et al. (see figure 9b). Bonn observations with a 40" resolution (Güsten and Downes 1983) identified an $H_2O$ maser source 20" away from source 42. These authors suggested that this source is associated with a molecular condensation (M0.25+0.01) reported by Güsten et al. (1981). Curiously unlike typical $H_2O$ masers which are embedded within molecular condensations this $H_2O$ maser is not centered on this molecular condensation. Our 20-cm maps shown in figures 1 (bottom left corner) and 2 and 16 suggest that source O1 could be linked by a diffuse structure to the eastern segment of the arched filaments at $\alpha = 17^h42^m40^s$, $\delta = -28°48'$ (see E2 in figure 10). The contour map shown in figure 23e illustrates a steep intensity gradient and an elongation toward southern and northern portions of this source, respectively. If source O1 and its diffuse structure are connected physically to each other, one might infer that the Arc is physically interacting with at least one component of the molecular clouds (M0.25+0.01) known to lie near the galactic center (Güsten et al. 1981). We note that the maser reported by Güsten and Downes appears, within the uncertainties, to be imbedded within the diffuse emission. This compact source was not detected in a far infrared survey made by Odenwald and Fazio (1984).

Sources P1 and P2 which are isolated and appear not to be associated wth the extended features in the Arc, coincide with sources 21 and 19 of Downes et al. The intensity contours of these two sources at 6 cm are shown in figure 23f. Source P2 was also reported by Güsten and Downes (1980) to be an $H_2O$ maser source. The location of this maser - based on its $V_{LSR} \sim 0$ km $s^{-1}$ - was suggested to be in the spiral arms along the line of sight. Their spectra based on their peak brightness temperatures at 6 and 20 cm indicate also that source P2 has a much steeper spectrum ($\alpha > 3.1$) than that of source P1 ($\alpha \sim 0.9$) (see Table 1). Thus, sources P1 and P2 are probably not associated with each other.

Sources Q1, Q2 and Q3 - like P1 and P2 - appear not to be associated with the Arc and are shown at 20 cm in figure 23 g. These sources coincide with a 6-cm source identified by Altenhoff et al. (1978) using the Bonn telescope. A far infrared source (FIR-19) identified by Odenwald and Fazio (1984) appears to be closest to source Q3. These authors derive HII region parameters for FIR-19 and assume that it is a foreground source at a distance of 2 kpc.

Because of their tail-like appearance, we speculate that Q1 and Q2 are not associated with Q3 and are probably extragalactic radio sources.

Sources R1 and R2, which show similar brightness temperatures at 20 cm in figure 23 h are surrounded by an elongated halo structure. These two sources coincide with a radio source (35W60) seen by Isaacman (1981) who lists it as a possible candidate for a planetary nebula in the galactic center region. Sources R1 and R2 are also coincident with source 47 of Downes et al.

Sources R3 and R4 are shown at 20 cm in figure 23i, one of which (R4) appears to have a nearby far infrared counterpart (source number 13 in Odenwald and Fazio). These two sources are coincident with sources 46 and 45 of Downes et al. Sources R1 - R4 all appear to be embedded within an extended source identified recently by LaRosa and Kassim (1985) at 80 MHz with a resolution of 5'×8.6'. The position at which the peak brightness has been seen at 80 MHz ($\alpha = 17^h43^m11^s$, $\delta = -28°31'40^s$) does not coincide with any of the 20-cm sources seen in figures 23 h and 23 i.

Sources S1 and S5 show the peak flux densities of a pistol-like structure (see figure 23j) which is discussed in §II.3f. Two sets of linear filaments appear to cross this structure, one of which is lined up precisely along the extensions of the peaks (S2 - S5) as shown in figure 23j. Such an enhancement of radio emission along the linear filaments suggest an interaction between this structure (G0.16 - 0.05) and the linear filaments. The nearest far-infrared (FIR) source is located at G0.135-0.041 (FIR 32 in Odenwald and Fazio 1984). The positional uncertainties for FIR sources given by Odenwald and Fazio are $3^s$ in $\alpha$ and 48" in $\delta$. The pistol-like structure appears to be displaced by ~3' in R.A. from the peaks noted in FIR map but coincides with FIR source in declination within positional uncertainties. Thus, it is possible that this radio source has an infrared counterpart. Figure 14 of chapter 10 shows clearly the relative positions of these sources in an overlay of the FIR map onto the 20-cm image.

Figure 23k shows the 20-cm contour map of two giant HII regions: G0.67-0.04 (Sgr B2) and G0.51-0.05. A considerable detail can be noted in G0.51-0.05 where it breaks up into a number of compact components surrounded by a rather diffuse structure. The compact sources coincide with sources 49, 51, 52, 53 and 54 of Downes _et al._ A higher resolution (~20"×20") radiograph and contour map of this source, as seen in figures 16 and 23 l, show a bar-like structure at $\alpha \sim 17^h43^m50^s$, $\delta = -28°30'$ titled by 30° (P.A. 60°) with respect to the galactic plane. Observations of G0.51-0.05 made with the 100-m MPIfR telescope show a broad line width of 75 km $s^{-1}$, radial velocity of 42 km $s^{-1}$ and an electron temperature ~6400 °K (Pauls and Mezger 1975). Based on these parameters, Pauls and Mezger (1975) and Whiteoak and Gardner (1973) argue that this source is typical of normal HII regions. This source appears also to be correlated with a far infrared source (FIR-18) observed by Odenwald and Fazio (1984) who find a dust temperature of 70 °K and a total far infrared luminosity of $8.8\times10^{+6}$ $L_\odot$ assuming that this source is located near the galactic center.

Table 1
Discrete Sources near the Arc

| Name | R.A. (1950) | | | Dec. (1950) | | | Peak Brightness Temperature (°K) | | Reference |
|---|---|---|---|---|---|---|---|---|---|
| Arc | h | m | s | ° | ' | " | 4.8 GHz | 1.44 GHz | |
| N1 | 17 | 42 | 52.44 | -28 | 51 | 36.96 | 192 | 196 | Downes *et al.* |
| N2 | 17 | 42 | 59.38 | -28 | 53 | 57.81 | 145 | — | Downes *et al.* |
| N3 | 17 | 43 | 10.57 | -28 | 48 | 56.38 | 103 | — | Downes *et al.* |
| O1 | 17 | 42 | 56.96 | -28 | 44 | 25.1 | 290 | 827 | Downes *et al.* |
| P1 | 17 | 43 | 20.20 | -28 | 55 | 11.2 | 55 | 198 | Downes *et al.* |
| P2 | 17 | 43 | 19.22 | -28 | 55 | 15.36 | 289 | <100 | Downes *et al.* |
| Q1 | 17 | 43 | 56.76 | -28 | 45 | 57.2 | — | 541 | Altenhoff *et al.* |
| Q2 | 17 | 43 | 59.1 | -28 | 45 | 19.1 | — | 329 | Odenwald & Fazio |
| Q3 | 17 | 44 | 01.4 | -28 | 45 | 10.3 | — | 100 | |
| R1 | 17 | 43 | 17.28 | -28 | 35 | 01.4 | — | 987 | Isaacman |
| R2 | 17 | 43 | 18.3 | -28 | 34 | 58.0 | — | 987 | Downes *et al.* |
| R3 | 17 | 43 | 22.63 | -28 | 38 | 08.4 | — | 448 | Downes *et al.* |
| R4 | 17 | 43 | 17.44 | -28 | 38 | 06.2 | — | 455 | Downes *et al.* |
| S1 | 17 | 43 | 03.89 | -28 | 48 | 45.6 | 118 | — | Odenwald & Fazio |
| S2 | 17 | 43 | 04.75 | -28 | 48 | 42.34 | 160 | — | Altenhoff *et al.* |
| S3 | 17 | 43 | 05.03 | -28 | 48 | 44.06 | 145 | — | Altenhoff *et al.* |
| S4 | 17 | 43 | 05.46 | -28 | 48 | 45.5 | 126 | — | Altenhoff *et al.* |
| S5 | 17 | 43 | 05.75 | -28 | 48 | 49.2 | 94.5 | — | Altenhoff *et al.* |
| T | 17 | 43 | 53.7 | -28 | 32 | 15.56 | — | 907 | |
| U | 17 | 43 | 49.6 | -28 | 31 | 39.64 | — | 1136 | |
| V | 17 | 43 | 00.07 | -28 | 22 | 15.2 | — | 640 | |

6) Polarization Measurements

a) 6 and 20-cm Results

Over the region we are concentrating on in this chapter (0°.2 ≤ b ≤ 0°.12), our 6-cm observations indicated that there was much polarized emission from the negative latitude side of the Arc and a complete lack on the positive-latitude side. This apparent asymmetry of polarized emission from the Arc with respect to the galactic plane is consistent with both a lack of radio recombination line emission from the polarized segment of the Arc (Pauls and Mezger 1980; Gardner and Whiteoak 1978) and the low frequency emission from this region (see next chapter). Polarization measurements at 20 cm showed a total lack of polarized emission from both the Arc and the Sgr A compex. Here, we present the linear polarization maps and the Faraday rotation measurements derived from them.

Figures 24 and 25 show the distribution of both polarized and total emission intensity, respectively, from the southernmost portion of the Arc. The line segments superimposed on both maps represent the orientation of the electric vectors, and their lengths correspond to their degree of polarization. These low-resolution maps are made with a synthesized beam having FWHM ~ 20.7"×17.5" in order to bring out the extended structures. The degree of linear polarization is highest, 15%, to the northwest and it decreases to ~10% toward the south. Furthermore, a high degree of non-uniformity can be seen in the distributions of both polarized intensity and the orientation of electric vectors in the region where the highest degree of polarization is seen. Another representation of the polarization character-istics of this region can be seen in figures 26 and 27 which have a resolution of 4.3" × 3.4". (For demonstrative purposes, the radiograph seen in figure 27 has a slightly higher resolution [~3"×2.6"] than the corresponding polarization angle distribution superimposed on it). Indeed, most of the extended polarized emission is overresolved in figure 26, as can be seen by comparing it with figure 24. The highest percentage polarization appears to be ~60% in the region where the distribution of polarized intensity has a compact appearance (cf. fig. 27). This high percentage polarization is an upper limit since the extended unpolarized continuum emission may be partially resolved out.

Two sets of IF's, which were separated by 150 MHz, were employed in order to measure the Faraday rotation. Figure 28 shows the distribution of the polarized intensity with a resolution 11.3"×10.8" at 6 cm. Here, the position angles of the superimposed line segments correspond to the difference in position angles of the electric vectors at 6.166 and 6.363 cm. The length of the line segments is proportional to the polarized intensity at 6.166 cm. If there were no changes in the position angle (p.a.), of the electric vectors at these two wavelengths, the line segments would have been oriented north-south. Indeed, the Faraday rotations of the polarization angle, $\Delta\phi$, are large and uniform, ~ +30° to ~ +45° and are within uncertainties, ± 1°, to the northwest of figure 28, respectively. There is, of course, an ambiguity of $2n\pi$ ($n = 0,1,2...$) in the polarization angle measurements.

Two other 6-cm fields of view were situated along the negative-latitude portion of the Arc, one of which was centered on the sickle-shaped structure (see §II.3d). No polarized emission was detected from this structure (<0.5%). The other fields was situated to the northwest of the center of the field shown in figures 24 to 28 in order to confirm the polarization characteristics which had been seen near the edge of the fields of view in these figures. The results of this field follow next.

Figures 29 and 30 show the distribution of both polarized and total emission at 6 cm from the region near G0.16-0.15 with a resolution of 10.6"× 16.9". The line segments superimposed on figure 20 represent the position angles of the electric vectors and their lengths correspond to their percentage polarization. The highest degree of polarization is ~17%. The polarization properties seen in figures 29 and 30 confirm those depicted in the northwest positions of figures 24 and 25.

Figure 31 again represents the difference in position angles of the electric vectors at 6.166 and 6.363 cm in the same manner as in figure 28. Indeed, with a resolution of 4.3"×3.4" the Faraday rotations of the polarization angle agree very well with those of figure 28 in the overlapping region. In fact, the largest Faraday rotation of ~80° (±2nπ) occurs in both figures 28 and 31 in a small region centered at $\alpha \sim 17^{h}42^{m}26^{s}$, $\delta \sim -28°52'$.

One of the more interesting aspects of figure 27 is the locations at which the strongest polarized emission at 6 cm occur. The polarized distribution has compact appearance and unlike the distri-bution of the total intensity. The positions of strong linear polarization coincide essentially with the positions where two segments of the so-called helical structures cross the linear filaments at $\alpha_1 \sim 17^h43^m16^s$, $\delta_1 \sim -28°49'30''$ and $\alpha_2 \sim 17^h43^m22^s$, $\delta_2 \sim -28°51'$ (see also figures 14 a,b). Furthermore, we note that the Faraday rotations are more uniform and higher near or at the locations where the helical feature crosses the linear filaments. Figure 31 implies that the rotation measure (R.M.) derives from R.M. $= \Delta\phi/(\lambda_1^2-\lambda_2^2)$ - where $\Delta\phi$ is the Faraday rotation and $\lambda_1$, $\lambda_2$ correspond to adjacent wavelengths - is typically $\sim -3000$ rad $m^{-2}$ at or near the location where a shadow on the northern filament can be seen at $\alpha \sim 17^h43^m15^s$, $\delta = -28°49'30''$ (see figures 17 and 14b). Higher values of rotation measure $\sim -5500$ rad $m^{-2}$ can be noted to the southeast of figure 31 where $\Delta\phi \simeq 80°$ (see also figure 28).

The highest degree of polarized emission coincides generally with the locations where the highest rotation measures is observed. It is conceivable that the polarized emission from the filaments pass through plasma which is threaded by a very uniform and extended magnetic field. This medium causes the electric field vectors to rotate uniformly without attenuating the degree of polarized emission if the transverse component of magnetic field in the medium does not change rapidly across the extended source.

b) 2-cm results

A number of 2-cm fields were situated along the filaments in order to determine the polarization characteristics of the linear filaments at this frequency. Here we give a preliminary and partial account of the 2-cm results using only the C/D array data base. More complete 2-cm results using both the B/C and C/D arrays will be published elsewhere (Fomalont, Inoue, Morris, Tsuboi and Yusef-Zadeh 1986).

Two 2-cm fields were centered on the polarized intensity maxima seen at 6 cm. Figure 32a shows the radiograph of polarized emission from a portion of both the northern and southern linear filaments. The most interesting 2-cm result is the recognition that the polarized emission from the Arc has a filamentary appearance which is, unlike the 6-cm polarization structure, similar to what has been seen in the 6 and 20-cm total intensity distribution (see figure 2). The southern linear filament in figure 32a has a higher polarized surface brightness than that of the northern linear filament similar to what is seen in the total intensity distribution (figure 7). The southern linear filament was shown earlier (figure 7) that it might consist of two twisting filaments. Figure 32b shows the contours of polarized intensity at 2-cm which have been superimposed on the 6-cm image of the filaments which appear to be twisted. This map shows clearly that the linear polarization stems mostly from the linear filaments.

Most of the diffuse polarized and unpolarized emission are located to the north of the southern linear filaments, i.e. on the side opposite to that of the galactic center. We note a number of revealing substructures which appear to cast shadows on the relatively extended emission in polarized intensity maps: 1) A narrow loop-like structure to the northwest of figure 32a appears to be projected along the southern filament. A slice parallel to the southern filament is cut across this loop-like structure and is shown in figure 33. The two strongest absorption features in this figure correspond to the locations where the polarized emission becomes null at the outer edges of this loop. 2) A very sharp and elongated absorption structure is seen to split the continuity of the southern filament in figure 32 at $\alpha \sim 17^h43^m19^s$, $\delta \sim -28°50',51''$. This drop in the polarized intensity of the southern filament which has a width of ~7" (0.35 pc) can be seen in figure 34 showing a slice cut along this filament. We note an asymmetry in the polarized intensity of the southern filament on opposite sides (i.e. NW and SE) of the sharp absorbing feature. 3) A circular absorption feature located in the southwest of figure 32 appears to form a polarization shadow on the region between the northern and southern filaments. Because the absorbing features appear only in the polarization intensity maps, they can best be interpreted as depolarizing features lying between the observer and the linearly polarized filaments. Indeed, most of these depolarizing features appear to have one-dimensional structure and show characteristics similar to the radio shadows seen on the total intensity maps at both 6 and 20 cm. Thus we assume that these absorbing features are located in the vicinity of the linear filaments as they depolarize the background emission from the linear filaments.

Figure 35 shows the distribution of electric field vectors in the region seen in Figure 32. Another characteristic of the depolar-izing features is that the electric field vectors change aburptly by ~45° along the southern filament at the location where the sharp elongated structure splits the southern filament. Such a change in the orientation of electric field vectors can also be noted at the location where the southern foot of the loop-like depolarizing feature intersects the southern filament ($\alpha$ ~$17^h43^m21^s$, $\delta$ ~ -28°51'). This effect can also be seen in figure 36 which shows the polarization structure of the region to the northwest of and adjacent to the field of view seen in figure 35. The electric field vectors are generally aligned in the direction of the filaments at the positions away from the absorbing features, i.e., NW and SE of figures 36 and 37, respectively. Because of a sudden change in the direction of the electric field vectors on either side of the absorbing features and because of the total depolarization of the linear filaments at the locations where the absorbing features block the linear filaments, we infer that the depolarizing features consist of ionized gas features which are maximized at the location where total depolarization is seen and which have a larger dimension than that implied by their appearance in figure 32a. This inference is stressed further below.

Comparisons of the 6 and 2-cm polarized intensity distributions can be seen in figure 37. Both maps have similar spatial resolutions and yet show a very different linearly polarized structure. Indeed, the clumpy nature of the polarized emission at 6 cm can best be explained by the depolarizing features seen in the 2-cm radiograph. This comparison can be used as an evidence in support of the presence of non-uniform but coherent features (see chapter 4) which surround the linear filaments and which essentially depolarize the 6-cm polarized emission more than the 2-cm emission (and depolarize the 20-cm emission completely). This effect is attributed to the quadratic dependence of Faraday rotation with respect to wavelength.

Figure 38 shows a measure of Faraday rotation superimposed on the 2-cm polarized intensity distribution. The direction of the line segments correspond to the difference in position angles of the electric vectors at 14.5649 and 15.0149 GHz. The length of the line segments is proportional to the polarized intensity. Most of the region in this figure shows a Faraday rotation of ~ -5° which corresponds to rotation measure of ~ -3480 rad $m^{-2}$. This is somewhat similar to the rotation measure seen at 6 cm in figure 31 (see figure captions for sign convention). Figure 39 shows the distribution of Faraday rotation at 6 cm, as seen in figure 31, superimposed on the 2-cm polarization intensity. It is evident that largest Faraday rotation emerges from the region which shows a number of depolarizing features. These depolarizing features do not appear in the 2-cm total intensity distribution. Figures 40 and 41 show the total 2-cm intensity distribution of the region where two overlapping 2-cm fields were situated. A lack of short (u,v) spacings has severely suppressed the extended structures in these two figures. Because of this, the degree of linear polarization is not represented accurately for weak and extended structures. However, it should be pointed out that the brightest filament has a comparable surface brightness in both the total and polarized intensity maps, and has comparable dimensions. Thus, this implies that the apparent degree of linear polarization at 2-cm is unusually large (~90 - 100%). Further observations of this region including short (u,v) spacings should be carried out in order to determine accurately the degree of linear polarization along the filaments at 2 cm.

Burn (1966) finds that, as a result of depolarization, the degree of polarization decreases substantially when the rotation measure (RM) is greater than $\lambda^{-2}$, where $\lambda$ is the wavelength in meter. Thus, emission with rotation measures greater than 2500 rad $m^{-2}$ should be depolarized. If we use this lower limit in the rotation measure formula which is 8.1 $n_e B_{11} \ell$, where $B_{11}$, the component of magnetic field along the line of sight, is in units of $10^{-5}$ Gauss and $n_e$, the number density of electrons, is in units of $cm^{-3}$, we find that $n_e B_{11} > 1234$. We assumed that the length in which the ionized gas is distributed, $\ell$, is similar to the width of the elongated depolarizing structure, i.e. ~ 0.25 pc. The large rotation measure of ~ -5500 rad $cm^{-2}$ can be explained by assuming that the ionized gas is distributed immediately in the outskirts of the polarized region having a two dimensional slab geometry. Thus, the R.M. becomes ~ 4.05 $n_e B_{11} \ell$ which is consistent with the depolarization formula found by Burn (1966). Further discussion of the electron density near the filamentary Arc is discussed in the following chapter.

7) Spectral Index Measurements

Spectral index maps have been made by comparing the 6- and 20-cm maps made with data in the same range of spatial frequencies (450 $\lambda$ to $30\times10^3$ $\lambda$), ensuring that emission at one frequency was being compared only with data at other frequency which related to a similar scale size. However, the lack of short spacings affect the 6- and 20-cm data sets differently because of the different primary beam size and because of the large scale size of the Arc and cause much uncertainty in precise spectral index measurements. Therefore, we show only the spectral index maps of G0.16-0.15 and postpone a more accurate high-resolution spectral index measurement which includes the short spacing visibilities taken from single-dish measurements.

Over most of the 6-cm fields, the spectrum is flat, with $\alpha \sim$ -0.2 to 0.2 ($F_\nu \propto \nu^\alpha$). Figures 42 and 43 represent intensity distributions of the Arc near G0.16-0.15 at 6 and 20 cm, respectively. The spectrum of this region inferred from the contours of total intensity in these figures is consistent with low-resolution measurements by Mills and Drinkwater (1983). Similar results are obtained along other segments of the linear and arched filaments. Discussion of such a flat spectrum emission from a polarized region is discussed in chapter 4. Note that the double compact source seen in the 6-cm map has disappeared in the 20-cm map. This is due to the absence of short spacings at < 450 $\lambda$, which suppresses emission surrounding the Sgr A complex (see chapter 6) much more at 20 cm than at 6 cm.

# III. Discussion

"It looks magnetic as Hell"

Kevin H. Prendergast

## 1) Location of the Arc

It must first be established that the Arc lies at the distance of the galactic center, and is thus not an accidentally superimposed foreground object or background. The radio structure in the Arc does not correspond to any phenomenon known to occur in the galactic disk. Filamentary radio structures exist (in supernova remnants, for example) but we know of no cases in which radio filaments are: as nearly linear, as even and unbroken on such a large angular scale, and as perpendicular to the galactic plane. The high degree of linear polarization, the large Faraday rotation, radio shadows at such high frequencies, a complex velocity structure, as revealed by radio recombination line emission (see chapter 9), a possible twisting of the linear filaments, and the helical structure are also unknown elsewhere. The existence of the unique radio Arc is therefore easiest to contemplate if it occurs in a unique environment, such as the galactic center. Moreover, the arched filaments appear to originate (or end) near the halo of Sgr A (see chapter 9). That is, the placement and orientation of the Arc indicates a possible physical relationship between the Arc and the galactic nucleus. It is quite unlikely that this is a mere coincidence of projection.

We argued that the linear filaments are interacting with the radio structure G0.18-0.04 (the sickle-shaped feature). Other evidence (Fukui et al. 1977; Güsten and Downes 1980; Güsten et al. 1981, Güsten and Henkel 1983) indicates that G0.18-0.04 is itself interacting with the edge of the "40 km $s^{-1}$" molecular cloud (M+0.11-0.08). In particular, the cloud velocity, ~50 km $s^{-1}$, is similar to that of recombination line emission from G0.18-0.04 (see chapter 9; Pauls and Mezger 1980). Evidence that this molecular cloud lies near the galactic center is fairly strong, and has been discussed elsewhere (Güsten and Henkel 1983 and references therein).

The preceding arguments for the placement of the Arc at the galactic center are circumstantial. A more direct and more conclusive argument can be made on the basis of low resolution $H_2CO$ absorption studies carried out by Bieging et al. (1980) who infer that the Arc is well within the molecular ring. Recent results of HI absorption studies against the Arc using the VLA also confirm the conclusions made by Bieging et al. 1980 (Lewtas, Lasenby and Yusef-Zadeh 1986).

On the strength of these arguments, we proceed with the assumption that the Arc represents a phenomenon occuring within the region of the galactic center.

2) Structure of the Arc

Because a substantial degree of linear polarization has been observed along the filaments and because the appearance of the linear filaments strongly suggests that they are one-dimensional (a two-dimensional structure would show more diffuse structure on either side of the linear filaments [see §III.3 for further discussion]), we hypothesize that the filaments and perhaps the features surrounding them are controlled by magnetic phenomena. Linearly polarized emission from the southern edge of the Arc is explained in terms of synchrotron radiation from relativistic particles in a highly ordered magnetic field. The equipartition magnetic field in the polarized region is $\sim 10^{-5}$ gauss. The mean orientation of the position angle of the polarized electric vectors where the emission is highly polarized (P.A. ~170° - 190° when the effect of Faraday rotation is corrected) implies that the magnetic field lines are roughly parallel to the long axis of the linear filaments. Recent Faraday rotation measurements by Inoue et al. (1984) show that the magnetic field lines are aligned roughly in the direction of the linear filaments. However, the intrinsic electic field vectors show considerable variation in their orientation - after they are corrected for the Faraday effect - and therefore it is very difficult to assess that the magnetic field vectors are aligned in the diretion of the linear filaments throughout the Arc based on the data presented here. Now, we explore somewhat qualitatively a few possibilities which could explain some aspects of the radio structure in the Arc.

a) Force Free Magnetic Fields

Three of the most interesting aspects of the structures seen in the Arc have been the filamentary structure, the appearance of a helical structure and, perhaps, the twisting of the long filaments. If the helicity and twisting of the filaments can be verified, the above structures would support the hypothesis that not only is the magnetic field important, but also, the geometry of the magnetic field is governed by force-free field configuration. Such a configuration occurs in nature when the magnetic field ($B^2/8\pi$) dominates the kinetic ($1/2\,\rho v^2$), the gas ($nkT$) and the gravitational ($GM\rho/R$) pressures (see chapter 10 for estimates of these quantities). If so, the Lorentz force ($J\times B$) is not balanced by any other forces, and the system quickly evolves to a configuration in which the current flows along the field lines. Of course, a current free structure ($J = 0$) can also satisfy $J\times B = 0$. When the current density J is nonzero, then

$$\nabla\times B = 4\pi J = \alpha B \qquad (1)$$

$$\nabla \cdot B = 0 \qquad (2)$$

where $\alpha$ is a scalar function of the space coordinates. It follows from the above equations that $B \cdot \nabla\alpha = 0$; this implies that $\alpha$ is constant along a field line. Equation 1 also shows that, as one travels in the direction of B along a field line, the neighboring field lines curl in a fixed orientation around the field line. By taking the curl of equation (1), it follows that

$$(\nabla^2 + \alpha^2)\, B = 0 \qquad (3)$$

Explicit solutions for the magnetic field when $\alpha$ is constant take a complicated form (see Chandrasekhar 1956; Chandrasekhar and Kendall 1957; and Schlüster 1957) but in a cylindrical geometry with (r, θ, z) coordinates:

$$\frac{d}{dr}(B_\theta^2 + B_z^2) + \frac{2}{r} B_\theta^2 = 0 \qquad (4)$$

when the twist of the field lines given by the simplest possible form is $\alpha \; r \; B_z = B_\theta$, then

$$B_\theta = B_o \alpha r \left\{(1 + (\alpha r)^2\right\}^{-1} \qquad (5)$$

$$B_z = B_o \left\{1 + (\alpha r)^2\right\}^{-1} \qquad (6)$$

where $B_\theta = B_z$ (r = 0) is the field along the axis of symmetry (Pikel'ner 1964). Equations 5 & 6 show that the longitudinal component of the field lines ($B_z$) dominates near the axis of symmetry whereas the azimuthal component of the field lines dominates as r increases. A helical geometry is manifested in such a field configuration. The current which flows along the field lines is confined and leads to a filamentary one-dimensional appearance only on the axis.

Kaplan (1959) finds that the intensity of the synchrotron radiation emitted from such a field geometry does not correlate with the degree of polarization. In fact, he finds that if the axis of the force-free cylinder subtends an angle $\zeta$ with the plane of the sky, then there exists an inverse correlation between the synchrotron emissivity and the degree of polarization (Kaplan 1959).

In view of the structure of the Arc (i.e. filaments, helical appearance) and its polarization characteristic suggest that the force free magnetic field is an attractive hypothesis, but one which leaves a number of questions: 1) The emergent radio emission is cylindrically symmetric in a force free geometry, but a large number of thin strands located to the southern portion of the Arc (figure 3) seems to contradict such an idealized geometry. It is not clear whether this bundle of thin strands is a product of some form of instability or whether they are independent structures which are evolving independently. One question is that if this parallel network of strands is due to some unspecified instability, what form did these features take before they became unstable?

2) It is often pointed out that no field of finite energy can be force-free everywhere. This could only occur when $J \times B = 0$ everywhere and it then follows that the magnetic energy, which is proportional to the integral of $\vec{r} \cdot (\vec{J} \times \vec{B})$ over volume, is zero which then implies that $B = 0$ everywhere. Thus, the force-free field structure must break down at its boundaries. Indeed we note that the brightness and coherence of the long linear filaments change dramatically at their northwestern and southeastern extremes, where the arched filaments and their mirror counterparts (i.e. counter-arch) emerge symmetrically with respect to the galactic plane (see chapter 8 for a discussion of the large scale radio lobes). These thermal, arched structures differ in their morphology from the non-thermal linear filaments but they are physically related to them (see §II.2). Indeed, the electron temperature and the number densities found in the arched filaments by single-dish observations (Pauls and Mezger 1980) indicate that these features are no longer governed by force free field configurations since their thermal pressure is greater than their magnetic pressure unless a large magnetic field strength is invoked (see chapter 9 and 10). One could, however, speculate that the Lorentz force becomes nonzero at the southern and northern edges of the linear filaments and directs the ionized gas toward the galactic plane. It should be pointed out, however, that there are no theoretical grounds to account for the structure of the arched filament except some speculative suggestions stated in chapter 9.

3) It has been suspected that solar flares connect two regions of different plasma characteristics (Jordan 1981): One of which is dominated by a force-free field geometry (low $\beta$), where $\beta$ is the ratio of gas pressure to magnetic pressure. The other is a region of high $\beta$ (i.e. where the force-free field breaks down) and is located well under the sun's surface. The two feet of solar flares are anchored to the surface of the sun. Indeed, a third major question concerning the structure of the filaments in the Arc is where the field lines are anchored if they are dominated by force-free fields. One might speculate that field lines are associated with and anchored to the nucleus and that the structure of the Arc is determined by the illumination of poloidal magnetic field lines which are centered on the nucleus (see chapter 6). Alternatively, one can speculate that the field lines are anchored to dense molecular clouds in the galactic plane such as the 40 km $s^{-1}$ molecular cloud, which is projected along the galactic plane to the south of the Arc ($\ell <$ 0.18°).

b) Pinch Discharge

The filamentary nature of the Arc suggests that pinch discharge process can be important in forming such a structure. Pinch discharge occurs when a strong current passes through a plasma, then the stress of the magnetic field lines, which are azimuthal and have shapes of coaxial circles, confine the plasma along the axial direction. In this configuration, in which the field is purely azimuthal, the equation of force balance when gravity is ignored becomes

$$\frac{dp}{dr} + \frac{d}{dr}\left(\frac{B_\phi^2}{8\pi}\right) + \frac{B_\phi^2}{4\pi r} = 0 \qquad (7)$$

where p is pressure and the azimuthal field can be written as

$$B_\phi = \frac{\mu I r}{2\pi a^2} \quad r < a$$

$$\frac{\mu I}{2\pi r} \quad r > a$$

where r, $\phi$ are the usual cylindrical coordinates, a is the radius of the current - carrying cylinder, I is the current and $\mu$ is the magnetic permeability. The gas pressure decreases radially outward whereas the azimuthal field increases linearly with increasing r inside the cylinder (see Priest 1984).

The pinch discharge hypothesis has some merits in explaining the linear filaments (strong current) and their helical segments (azimuthal field lines) if they are physically associated. However, there are at least 2 drawbacks to this hypothesis. One is the fact that the polarization measurements indicate the magnetic field lines appear to be aligned roughly along some portions of filaments (see §II.6). The other is the appearance of the helical structure, which consists of both longitudinal and azimuthal components. This property of the Arc can also be used to rule out a situation in which purely axial field lines dominate the structure of the linear segment of the Arc if the helical and linear features are physically associated. It is, however, possible that a purely axial field existed before the onset of a current flow along the field lines. In this scenario, not only does a large axial field stabilize a pinched discharge against the sausage instability (see Priest 1984) but it is also compatible with the orientation of the observed polarization vectors.

c) Dynamical effects

At the projected distance of the linear filaments from the galactic center, the period of circular rotation is $\sim 10^6$ years. Also, the filaments are not within the region of solid body rotation. Therefore, unless the lifetime of the filamentary structure is much shorter than $10^6$ years and (or) unless the filaments are rotating at a constant angular velocity ($\omega$), one might expect differential rotation to deform it and wrap it into a primarily azimuthal structure. If the linear filaments have constant angular velocity, then all the points along the filaments have

$$\omega = \left(\frac{GM \cos}{r(r^2+z^2)}\right)^{1/2} = \left(\frac{GM}{r(r^2+z^2)^{3/2}}\right)^{1/2} \quad (8)$$

where r is the distance between the galactic center and the location at which the linear filaments appear to bend abruptly, z is the distance away from the plane and $\theta$ is the angle between r and z. The above formula assumes that the filaments have circular motion and are influenced only by a central gravitational force. The ratio of the distance of the northwesternmost position of the linear filaments from the galactic center and that at the position at which the linear filaments cross the galactic plane is $\sim 1 \pm 0.1$. Indeed, this ratio is sufficient to keep all points on the linear filaments rotating at a constant angular velocity (based on the mass distribution given by Oort 1977). This characteristic of the Arc would permit the filaments to have an arbitrarily long lifetime.

The arched filaments do not appear to fall on the surfaces of constant angular velocity and therefore the above arguments do not apply to them. The placement and orientation of these filaments, which were described in §II.2, strongly indicate that they are interacting with the linear filaments. Here, we argue that the apparent cause of an abrupt change in the direction of the linear filaments cannot be attributed to gravitational force alone.

The balance of forces acting on the northernmost portion of the linear filaments ($\alpha = 17^h42^m30^s$, $\delta = -28°45'$), ignoring electromagnetic forces, is

$$\frac{1}{\rho}\frac{\Delta p}{\Delta z} = \frac{GMz}{(r^2+z^2)^{3/2}} \qquad (9)$$

The left-hand term of equation (9) is the pressure gradient in the direction perpendicular to the galactic plane and the right hand term is the component of the gravitational force in this direction. It follows that, if $\Delta p$ is less than $6\times10^{-12}$ ergs $cm^{-3}$ then the gravitational force alone could dominate the dynamics of the linear filaments and bend them toward the galactic center at the location where the arched and linear filaments meet ($z \sim 20$ pc); a number density of $\sim 15$ $cm^{-3}$ (Mezger et al. 1974) is assumed in this calculation. However, the total pressure of gas, starlight, far infrared radiation and magnetic pressure in the galactic center region (as estimated by Audouze et al. 1977) far exceeds $6\times10^{-12}$ ergs $cm^{-3}$. Inclusion of centripetal force in the above calculation makes it even less likely for the balance of forces. So, either an additional agent such as electromagnetic forces contributes to shaping the linear filaments into arched ones or, possibly, the arched and linear systems of filament are not physically associated. There is much to be learned from these two systems in the near future (see chapter 10).

3) Origin of the Arc

a) Unlikely Hypothesis

The filamentary radio structures in the Arc and the rich collection of other features described in previous sections reveal phenomena which have not previously been considered for the galactic center region or other parts of our Galaxy. The scale of the Arc phenomenon is large (±150 pc, judging by the apparent extension of the Arc in Altenhoff et al. 1978 seen in figure 4 of chapter 1, figure 1 of chapter 10 and the Bonn maps shown in chapter 8), and the radio emission is continuous across this entire scale, so it is unlikely to be linked to localized occurrences such as star formation, supernova remnants (Mezger and Pauls 1979; Downes et al. 1978), or collisions between clouds, as we had hypothesized prior to our observations (see chapter 1). Furthermore, the radio arc is clearly a physically unified phenomenon, rather than the chance superposition of two or more unrelated sources of emission, as had been previously suggested (Pauls and Mezger 1980; Pauls et al. 1976; Downes et al. 1978).

Could the apparently one-dimensional "filaments" comprising the Arc actually be two-dimensional shock fronts that are limb brightened by being viewed in projection from the side? If this were the case, if the medium into which the shock were supposedly expanding is homogeneous, and if the shock resulted from a spherically symmetric explosion, then the center of curvature of the filaments should coincide with the source of the shock. The center of curvature of the linear filaments lies far from the galactic nucleus on the side opposite the Arc, and may coincide with the radio source Sgr C.

Recent observation by Liszt (1985) shows that Sgr C also has a puzzling filamentary structure on a scale much smaller than the linear filaments in the Arc. However, no evidence exists which would suggest that Sgr C has undergone the kind of activity that would lead to such a shock. Also, the radius of curvature, $R_c \sim 110$ pc, is so large that no known event would be likely to generate ionized shocks at such a large distance over such a large angular scale. If the extension of the Arc above and below the galactic plane and the Arc itself are parts of a unified structure, the center of curvature would be at least 150 pc from the galactic center.

Because the shock velocity would decline with increasing density, relaxing the assumption of homogeneity by invoking a plane-stratified medium in which density declines with height could allow a shock to assume the curvature observed even if the origin of the shock is the galactic nucleus. Still, the shock front would be enormous ($R_c \sim 40$ pc) and the energy requirements correspondingly high, although perhaps not prohibitive. In addition, the regularity and straightness of the linear filaments alluded to above are probably inconsistent with a shock model for the arc structure. The inhomogeneities and clumpiness known to exist in the interstellar medium and, particularly, in the galactic center region (for example, the 40 km $s^{-1}$ molecular cloud near Sgr A seen in figure 4 of chapter 4) would be expected to introduce more breaks, kinks, and general disorder in a shock front passing over them than are observed in the filaments making up the Arc.

Finally, the "helical" and the large-scale "arch" structure,

whether apparent or real, would be very difficult to fit into a picture of the Arc as a shock front.

b) More Likely Hypothesis

One interesting aspect of the Arc is that its network of linear features is symmetrical with respect to the galactic plane. This plus the presence of symmetrical arched features connecting to the vertical filaments both above and below the galactic plane suggest strongly that the structure of the Arc is likely to be associated with a phenomenon which is linked to the galactic plane.

It has been suggested that the differential non-uniform rotation of the Galaxy and the turbulent motion of the interstellar gas, when combined, produce a dynamo effect (Parker 1971 a,b,c). Indeed, these two effecs are stronger near the galactic center than in other parts of the Galaxy (see chapter 6).

The non-uniform rotation of the gaseous disk would shear any preexisting radial magnetic field and enhance the azimuthal component. If the field is purely azimuthal, the rotation stretches the azimuthal field and increases its magnetic pressure by a factor of 4 in each rotation (i.e. magnetic field strength is increased by a factor of 2). The rate of shear in the galactic center region is

$$G \equiv r\frac{d\omega}{dr} = \frac{dV\phi}{dr} - \frac{V\phi}{r} \sim \frac{238.2\ r^{-0.9}}{(3.1\times 10^{16})}\ \sec^{-1}\ . \qquad (10)$$

The rotational velocity $V_\phi \sim 264r^{+0.1}$ (km $s^{-1}$) is taken from Sanders and Lowinger (1972), r (kpc) is the distance from the center of the Galaxy. So, $|G| \sim 1.14\times10^{-13}$ $sec^{-1}$ at $r = 0.05$ kpc which is two orders of magnitude greater than the local rate of shear (Parker 1971 a,b). The characteristic time is $\sim 2.8\times10^{5}$ yrs which is less than the rotation period of $\sim 1.5\times10^{6}$ yrs (Oort 1977).

It is also known that the stellar density in the galactic center is quite high (Sanders 1979; van den Bergh 1983) which might imply a high rate of supernovae. The large random velocity of ~200 km $s^{-1}$ near the galactic center (Loose <u>et al.</u> 1982) is interpreted as a consequence of the turbulent motion caused by supernova explosions. The turbulent velocity in the inner 300 pc of the Galaxy is greater than in the solar neighborhood by an order of magnitude.

There are, of course, other characteristics which distinguish the galactic center from the solar neighborhood. For one thing, the thickness of the neutral and molecular gas layer is smaller in the former region. Oort (1971) reports that the thickness of the HI nuclear disk at distances less than 300 pc from the center is ~100 pc and that its HI density averaged over the inner 100 pc is ~3 to 5 $cm^{-3}$ (Oort 1971; Sanders and Wrixon 1973). The high synchrotron emissivity (Little 1975) and the intensity of high-energy cosmic rays observed toward the galactic center region (Audouze <u>et al.</u> 1979) suggest that the strength of the magnetic field is higher, at least by a factor of 3, in the galactic center region than at the solar circle.

Indeed, the parameters given above suggest that the expansive forces associated with magnetic and turbulent pressures might be very important in destabilizing the balance of forces (Parker 1972):

$$\left(P + \frac{B^2}{8\pi} + p\right) \simeq \rho \langle g \rangle \Lambda \tag{11}$$

where p and P are turbulent and cosmic ray pressures, respectively. $\langle g \rangle \simeq 1.6 \times 10^{-9}$ cm s$^{-1}$ is the mean gravitational acceleration over one scale height $\Lambda$ due to stars. Equation (11) assumes that the pressure on the left hand side are roughly constant over a scale height $\Lambda$. Using $\frac{B^2}{8\pi} = 3.98\times10^{-12}$, $P = 5\times10^{-13}$, $p = 2\times10^{-9}$, $\rho g \Lambda = 1.2\times10^{-11}$, and $\rho = 5\times10^{-23}$ (all in c.g.s. units, Parker 1972), it becomes evident that turbulent pressure dominates strongly over all other terms. This dynamical instability causes the field lines to be inflated and the matter to slide down the field lines. If the azimuthal field lines are raised perpendicular to the galactic plane, matter should be concentrate on either side of the inflated field lines near the plane (i.e. the low places). In other words, magnetic arches are expected to be anchored by the gas concentrations. The characteristic scale for such an instability is a few times the scale height, and the growth time is equal to the ratio of the scale divided by the Alfvén speed, which is 1 - $2\times10^5$ yrs. This time scale is much less than that appropriate for the galactic disk ($\sim 5\times10^7$ yrs, Parker 1976). The fast growth of this instability might be responsible for many features seen perpendicular to the galactic plane in the galactic center region (Liszt 1985; Thaddeus [private communication]), including the vertical portion of the Arc.

It is possible that the field lines grow and can be extended perpendicular to the galactic plane until a rapid reconnection between two regions of opposite field direction takes place. If so, particles can be accelerated by dissipation of the field lines. We speculate that, perhaps, the relativistic particles along the linear filaments are accelerated by such a mechanism. Some drawbacks of the above hypothesis are: the linear filaments are symmetrical with respect to the galactic plane, whereas theoretically, Parker's instability should occur only on one side of the plane; also, the gas concentrations in Parker's instability are expected to lie on either side of the arched field line, however, the molecular gas (i.e. the 50 km $s^{-1}$ molecular cloud) is distributed only on one side of the filamentary Arc (the Sgr A molecular cloud); and finally, the vertical and horizontal dimensions are expected to be about the same with much uncertainty (Parker 1976), whereas, the observations of the linear filaments in the Arc indicate that these dimensions differ by ~2 orders of magnitudes. Further theoretical work is needed to understand why if this idea has any validity, the manifestation of such an instability, is so different from those seen in other parts of the Galaxy.

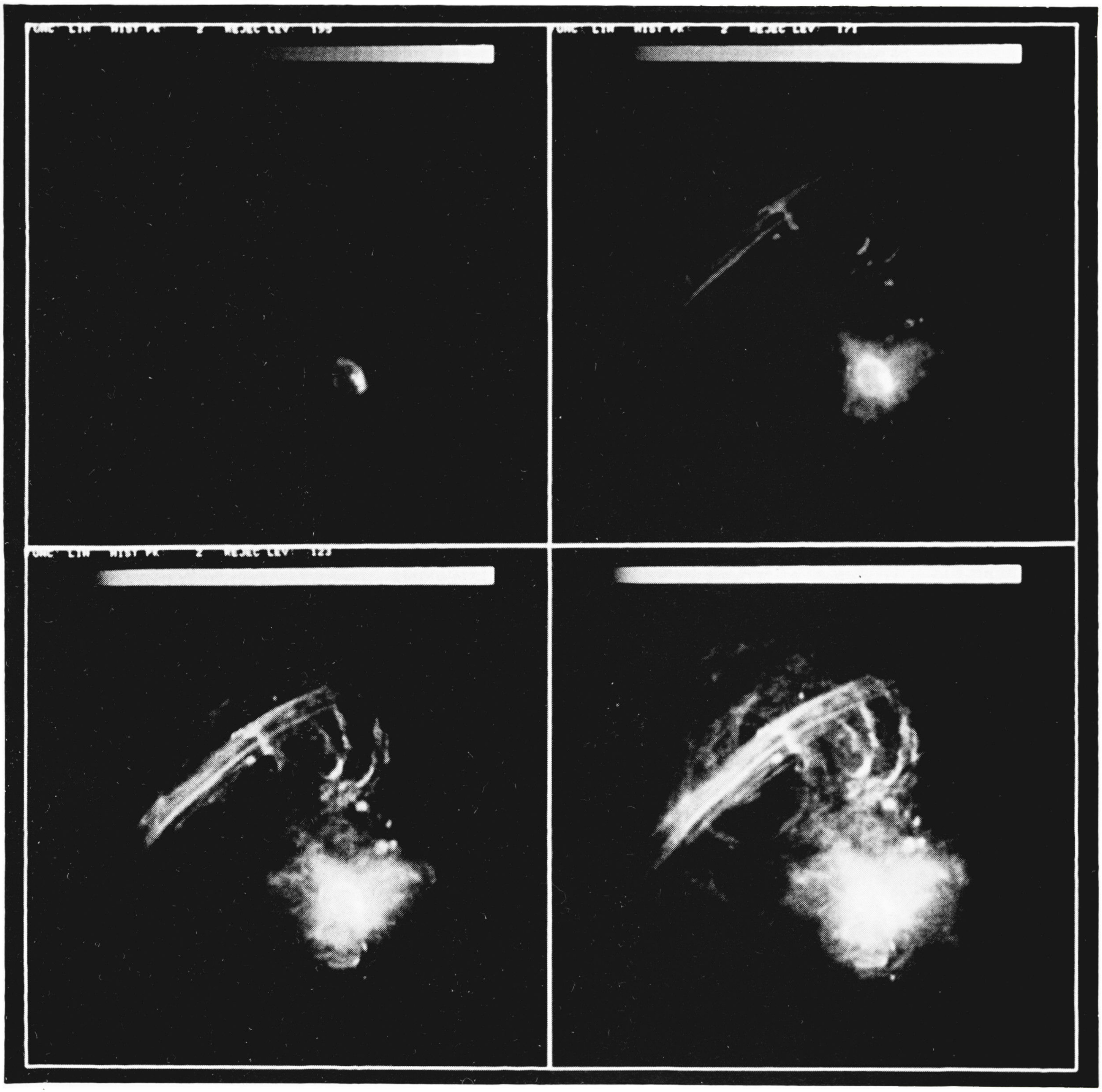

Figure 1 (a-d): In making the 20-cm map shown in figure 1 with four different contrasts a Maximum Entropy (ME) technique was used on the data base corresponding to the designated field GC20 (Table 1 of chapter 2) in order to remove the effects of the beam pattern from the map. This data set, which is based on combining the B, B/C' and the C/D' configuration arrays, was tapered at 25k$\lambda$, Fourier transformed, convolved with a gaussian beam having full width at half maximum (FWHM) = 10.7"$\times$10.1" and subjected to 20 iterations of ME in order to produce the final map. The required noise level for this map was 1.76 mJy/beam area and a total of 450 Jy of flux was required for the solution to converge.

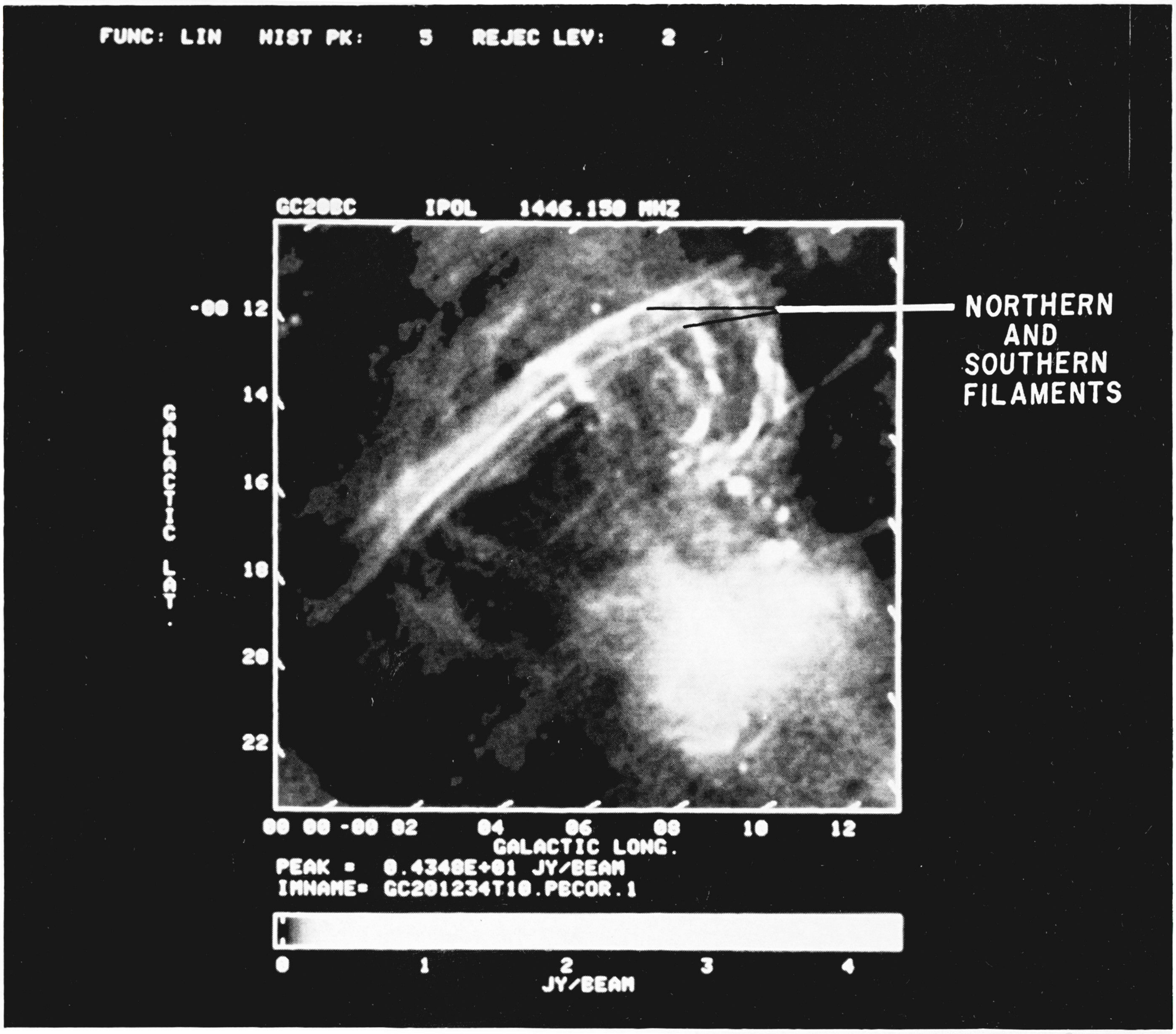

Figure 2: This map is based on the same data as was used for figure 1. The (u,v) data was tapered at 10 kλ, Fourier transformed, CLEANed, and corrected for the response of the primary beam. FWHM = 17.1"×16.4" (P.A. = 35°). The peak residual is 13.8 mJy/beam area. The total CLEANed flux is 455 Jy.

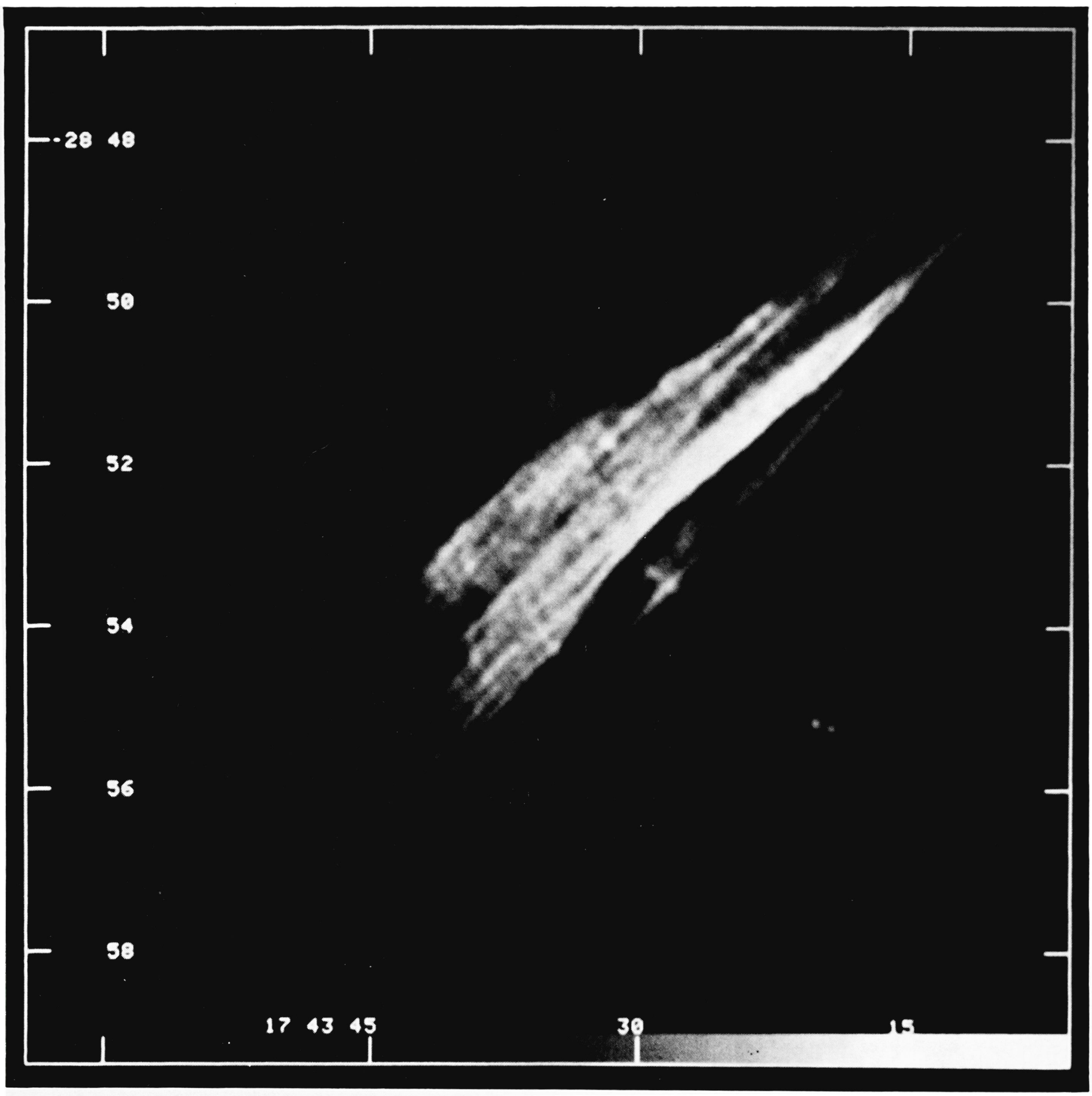

Figure 3: The designated field corresponding to this figure is Arc No. 1 (see Table 1 in chapter 2). The CLEAN components associated with Sgr A West were Fourier transformed and subtracted from the (u,v) data set before this map was CLEANed with a gaussian beam of 7.8" 7.3". The 6-cm data set corresponding to this figure was tapered at 20 k .

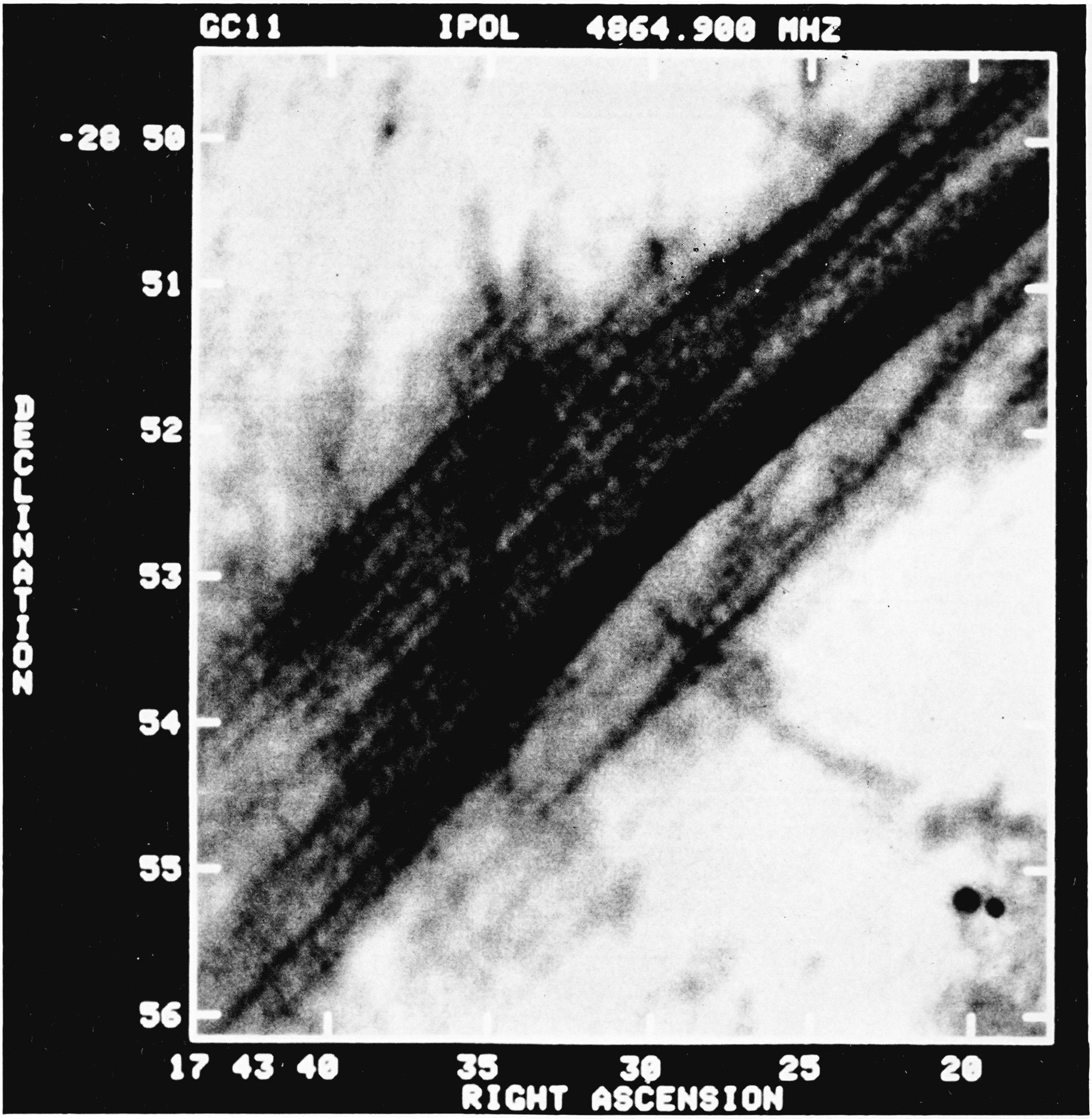

Figure 4 (a-b): Similar to figure 3 except that no taper was applied to this data base. CLEAN beam = 3.2"×2.8". The peak residual flux is 0.75 mJy/beam area. These two figures are different in their transfer functions.

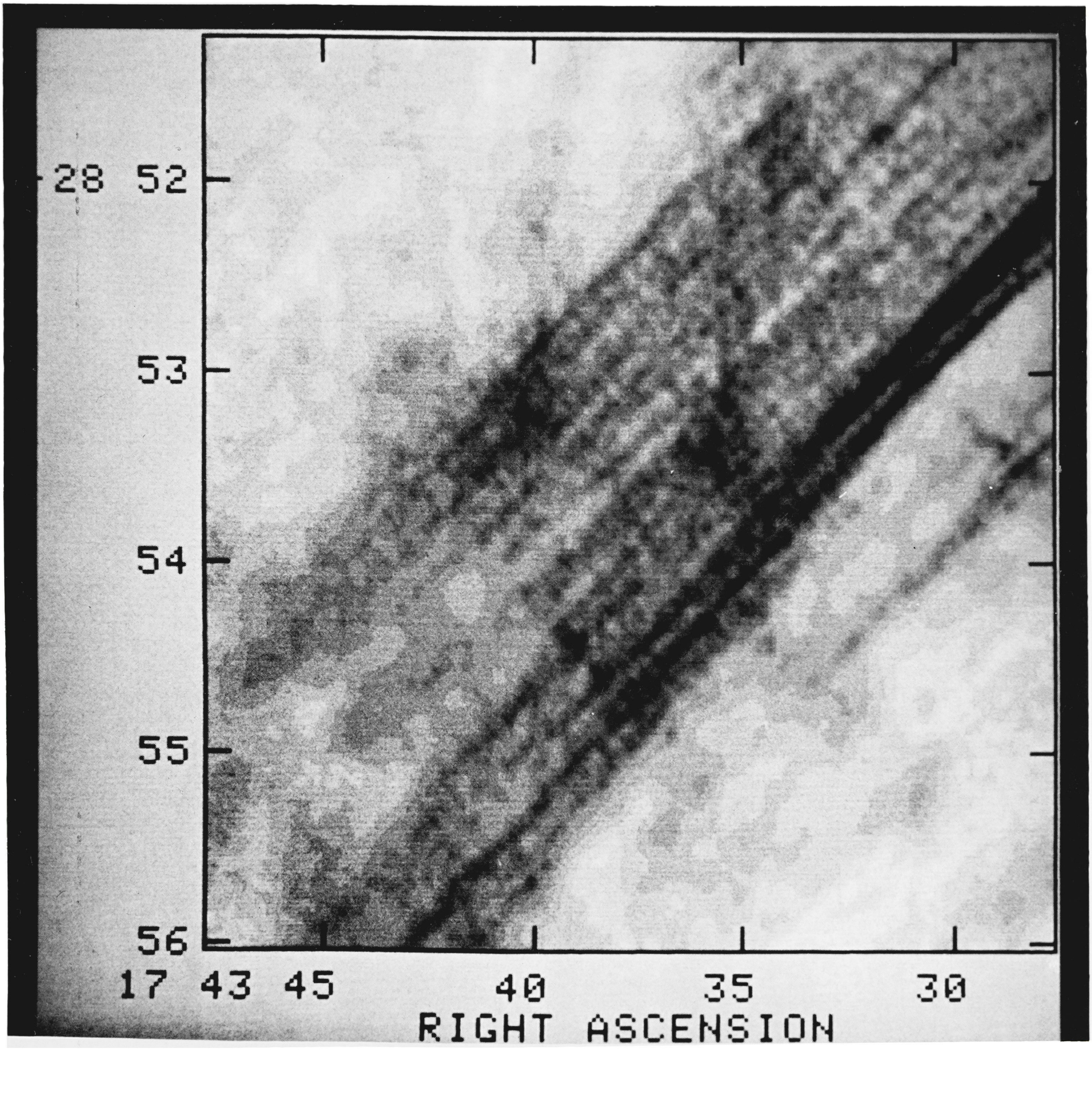

28 52
53
54
55
56
17 43 45
40
35
30
RIGHT ASCENSION

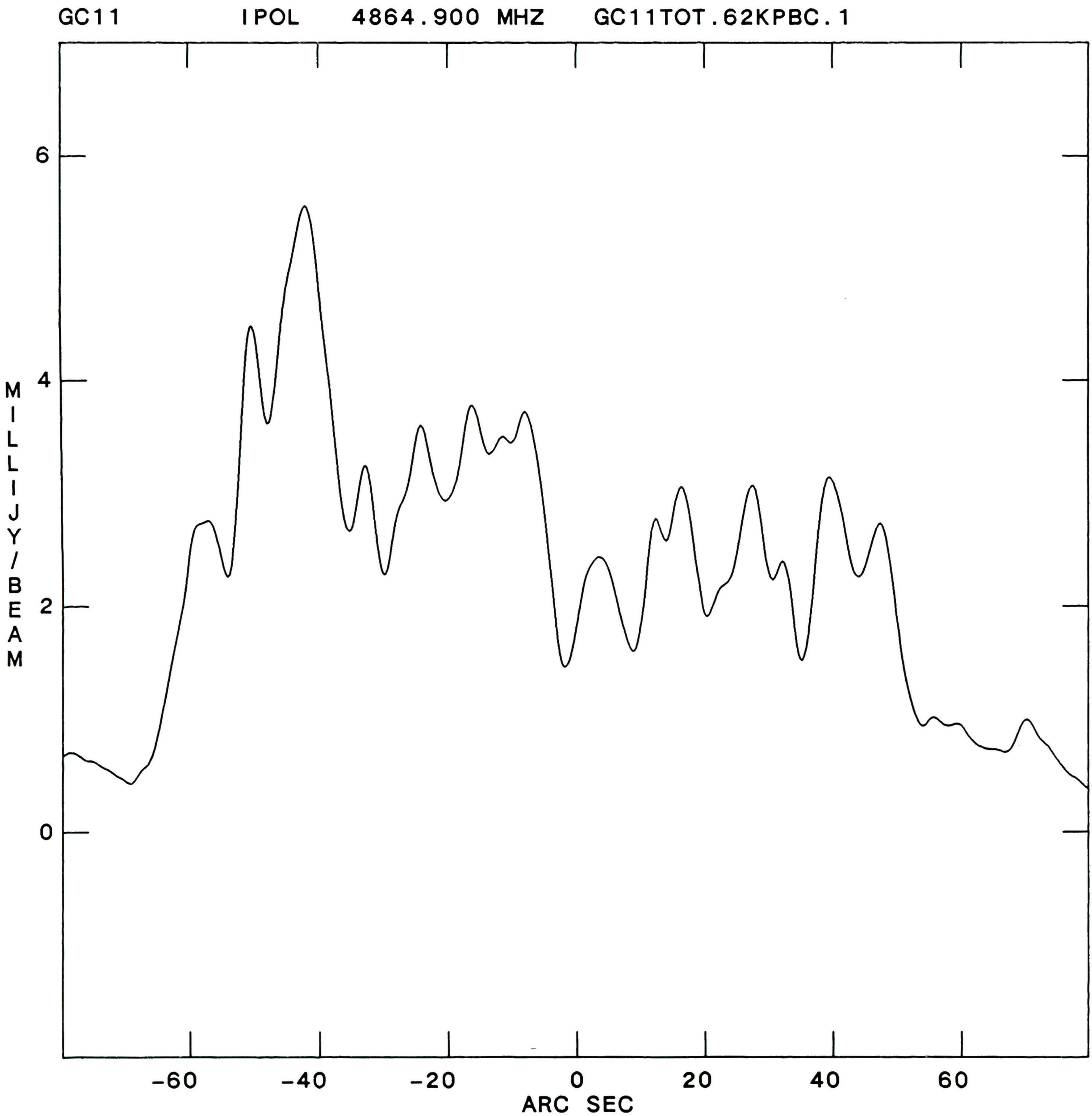

Figure 5: A slice is cut across the strands as seen in figures 4(a-b). The coordinates of this slice are: $\alpha = 17^h43^m36.9^s$, $\delta = -28°53'18.7''$, P.A. = 37°.

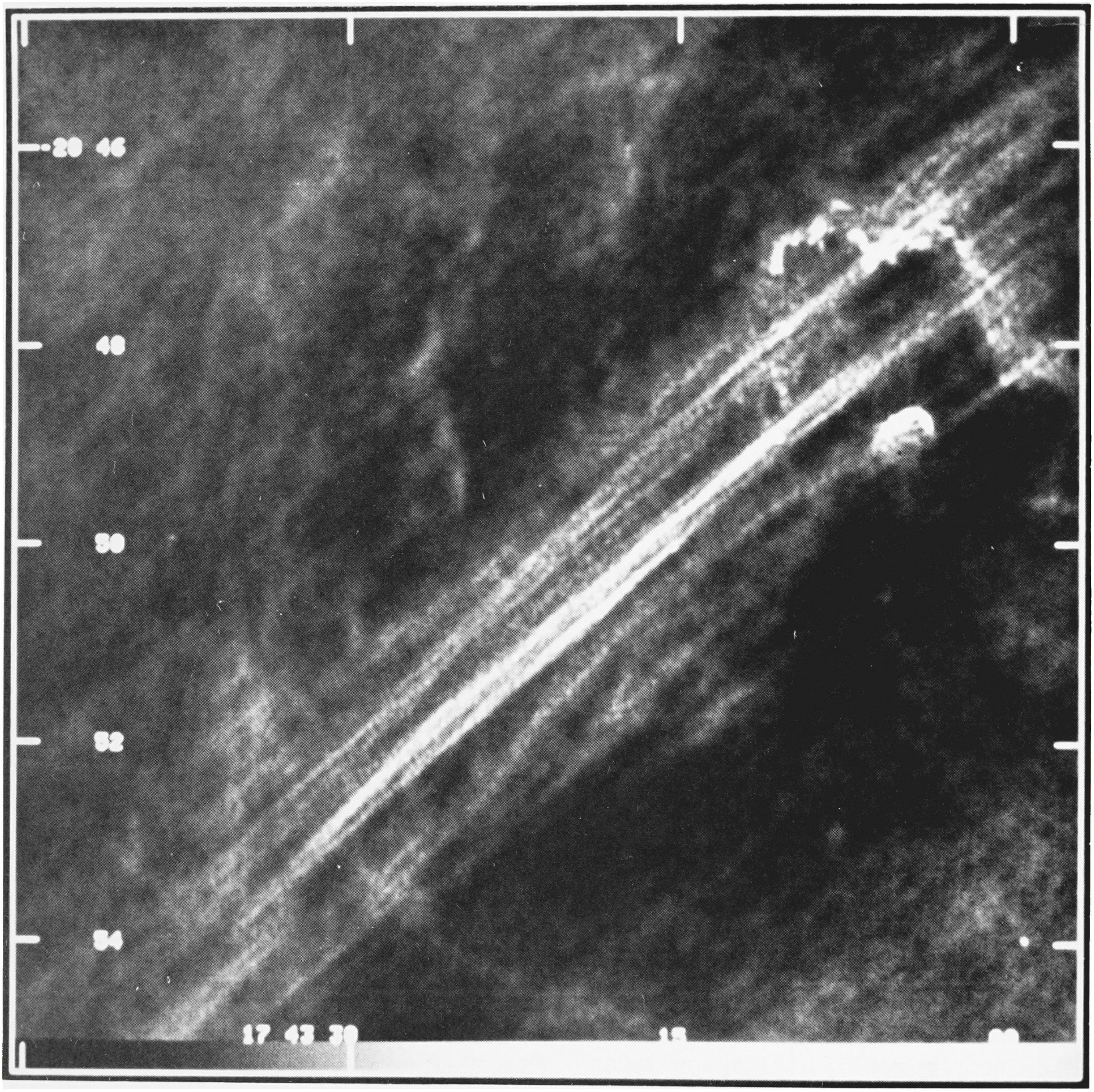

Figure 6(a),(b) and 7: These figures are produced based on the data base designated by field Arc No. 2. The data base is tapered at 75 kλ. The CLEAN beam (FWHM) = 3.3"×3.3". The peak residual flux = 0.8 mJy/beam area.

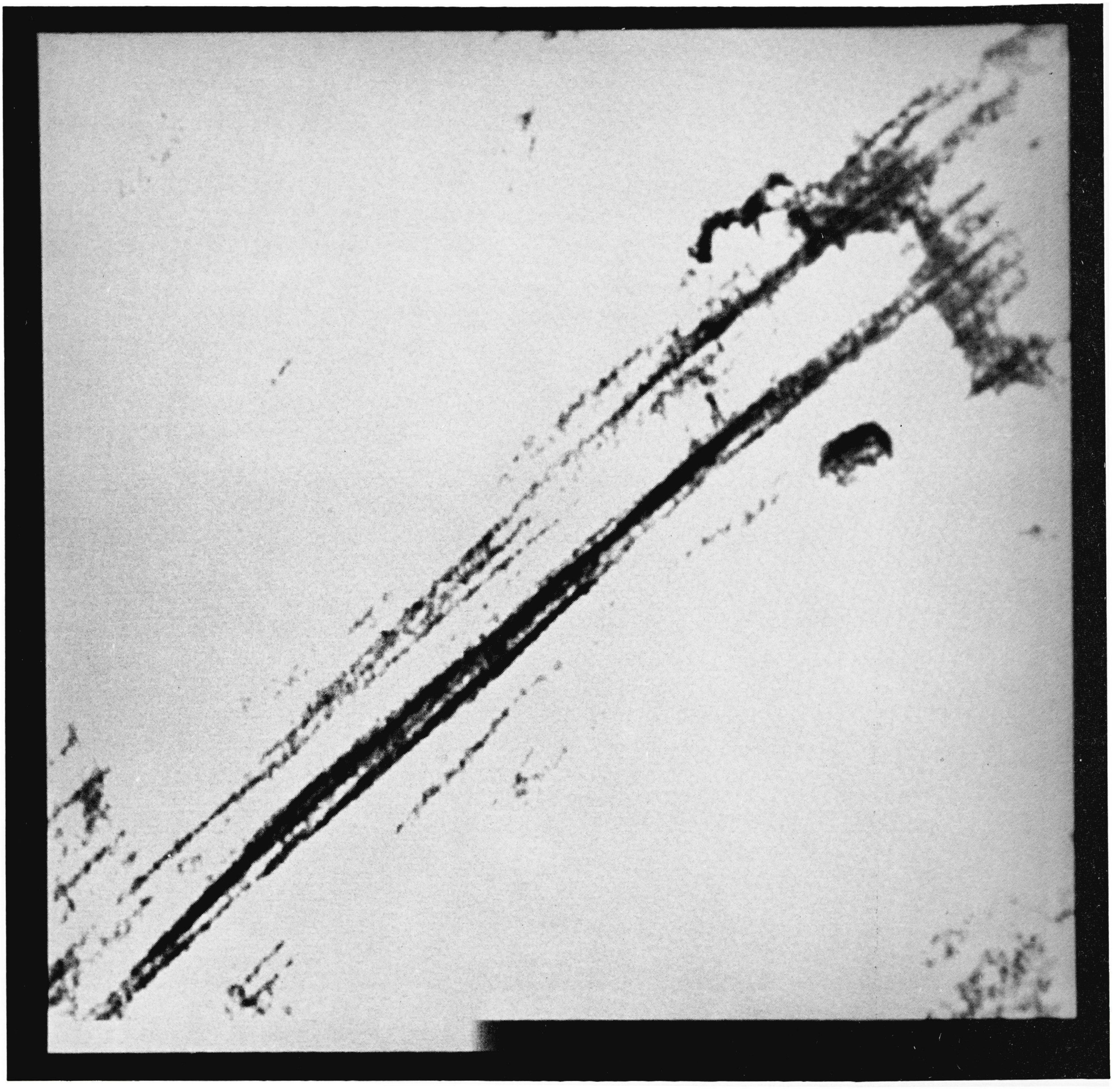

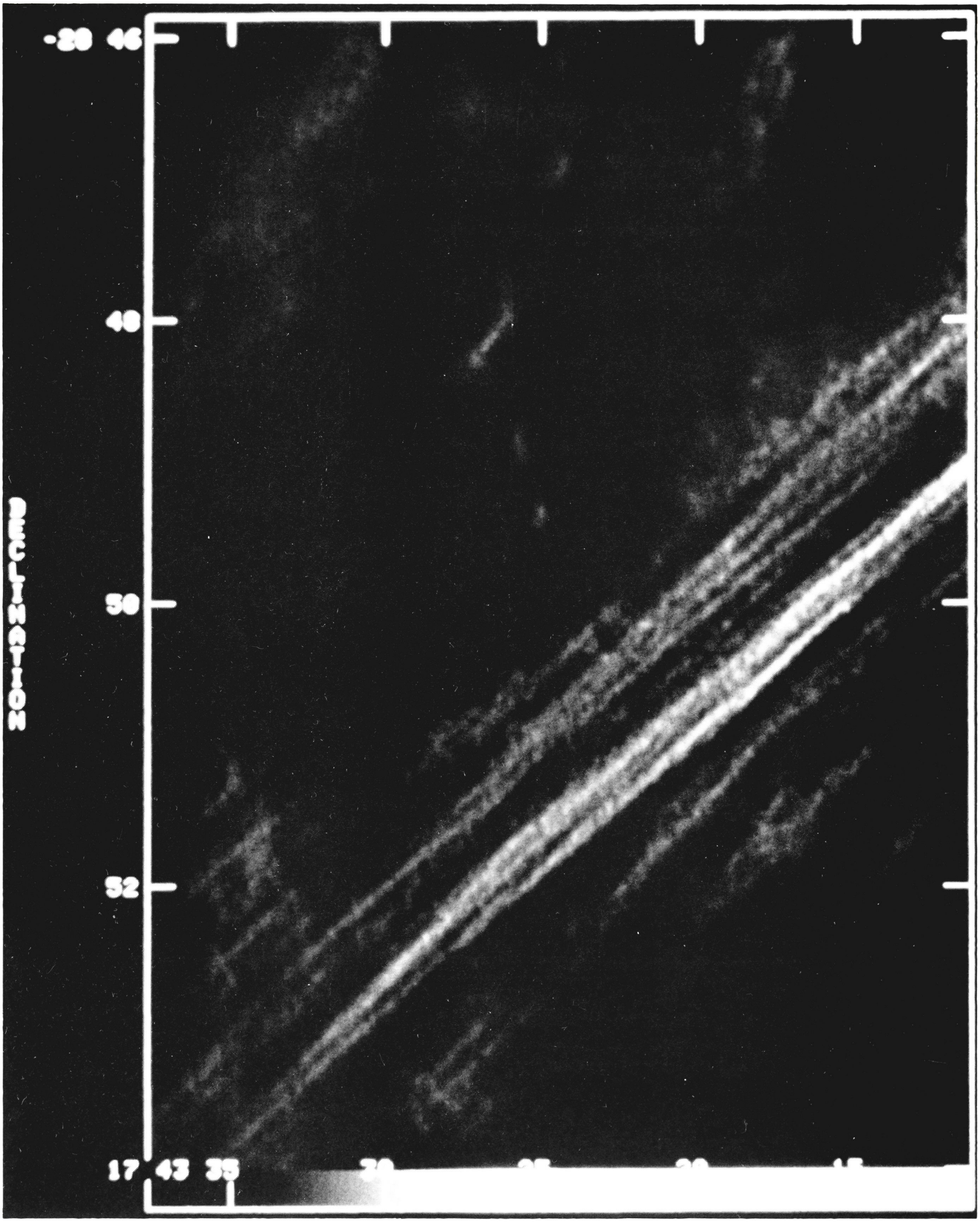
-28 46
48
50
52
DECLINATION
17 43 35
30
25
20
15

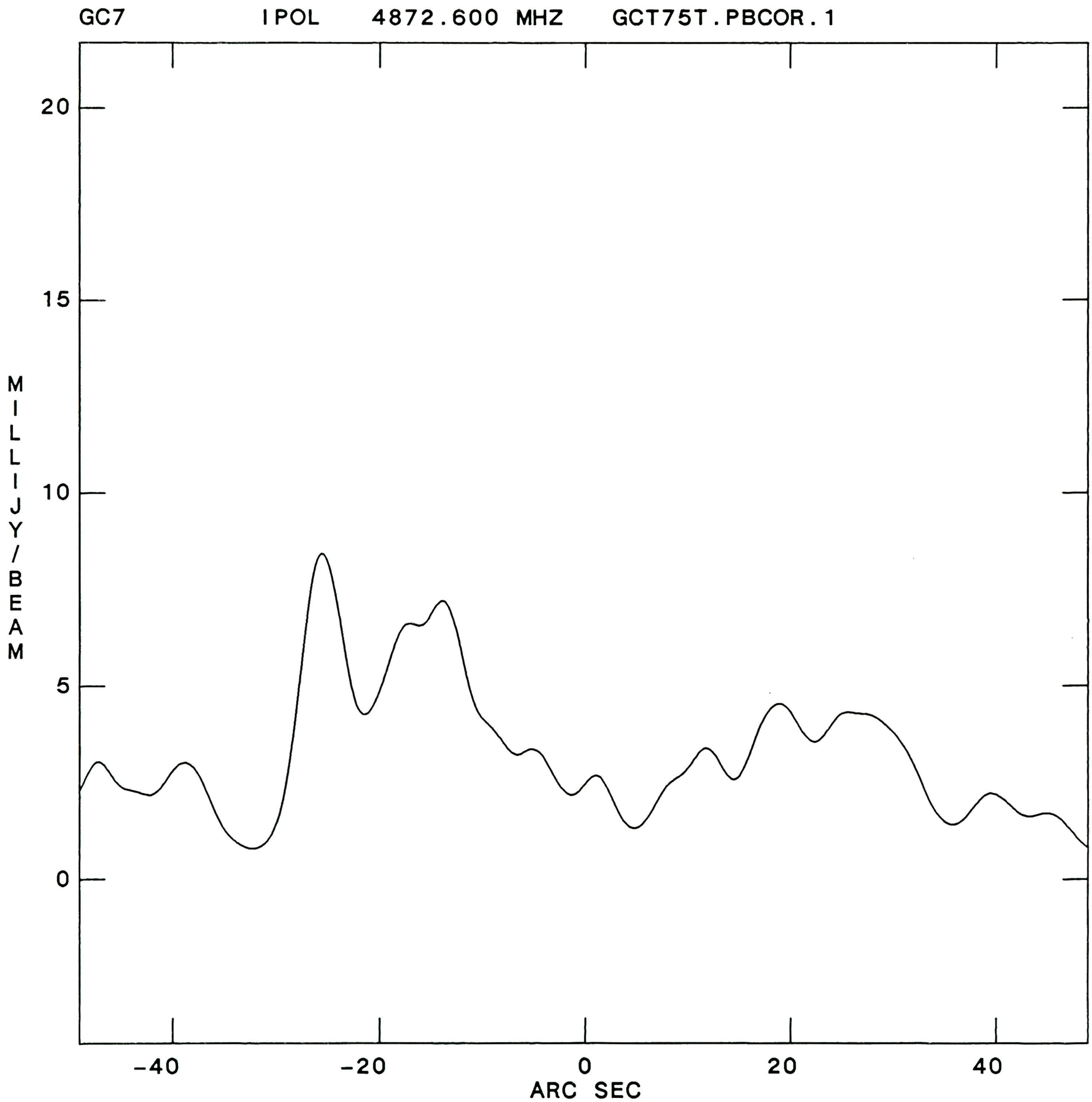

Figure 8: A slice is cut across the linear filaments, as presented in figure 7, in order to show the dark lines which separate the filaments. Coordinates of this slice are $\alpha = 17^h43^m18.4^s$, $\delta = -28°50'03.7''$, P.A. = 52°.

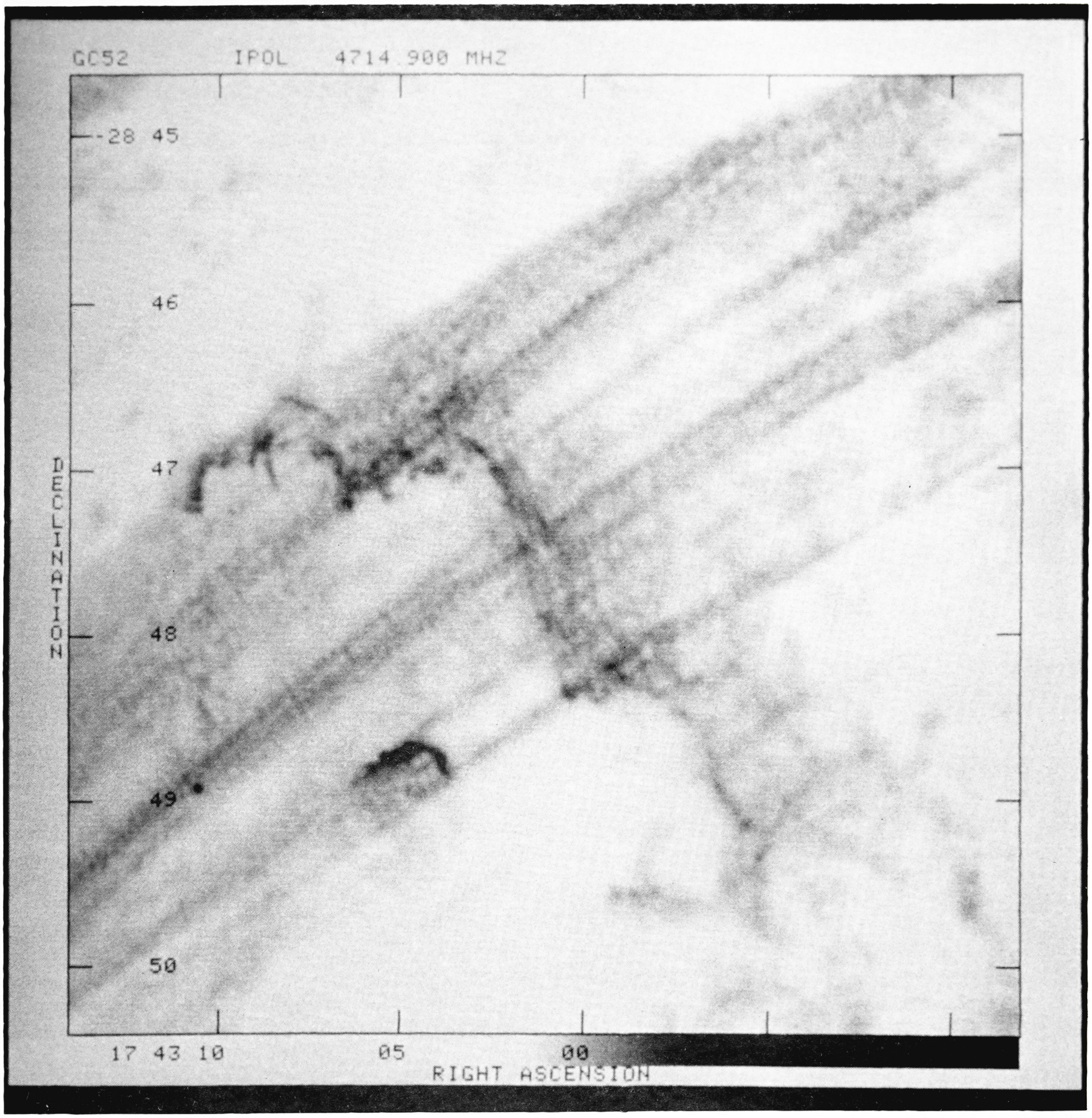

Figure 9 (a-b): Two different aspects of G0.18-0.04 are shown in figures 9a&b based on the data set corresponding to the designated field Arc No. 3. The continuum channel radio recombination line observations are also added to this data set (i.e Table 2 of chapter 2). Figures 9a and 9b were subjected to ME and CLEAN deconvolution algorithms. These maps have spatial resolutions = 2.58"×1.93" and the peak residual flux is 1.29 mJy/beam area for figure 9a.

GC20BC

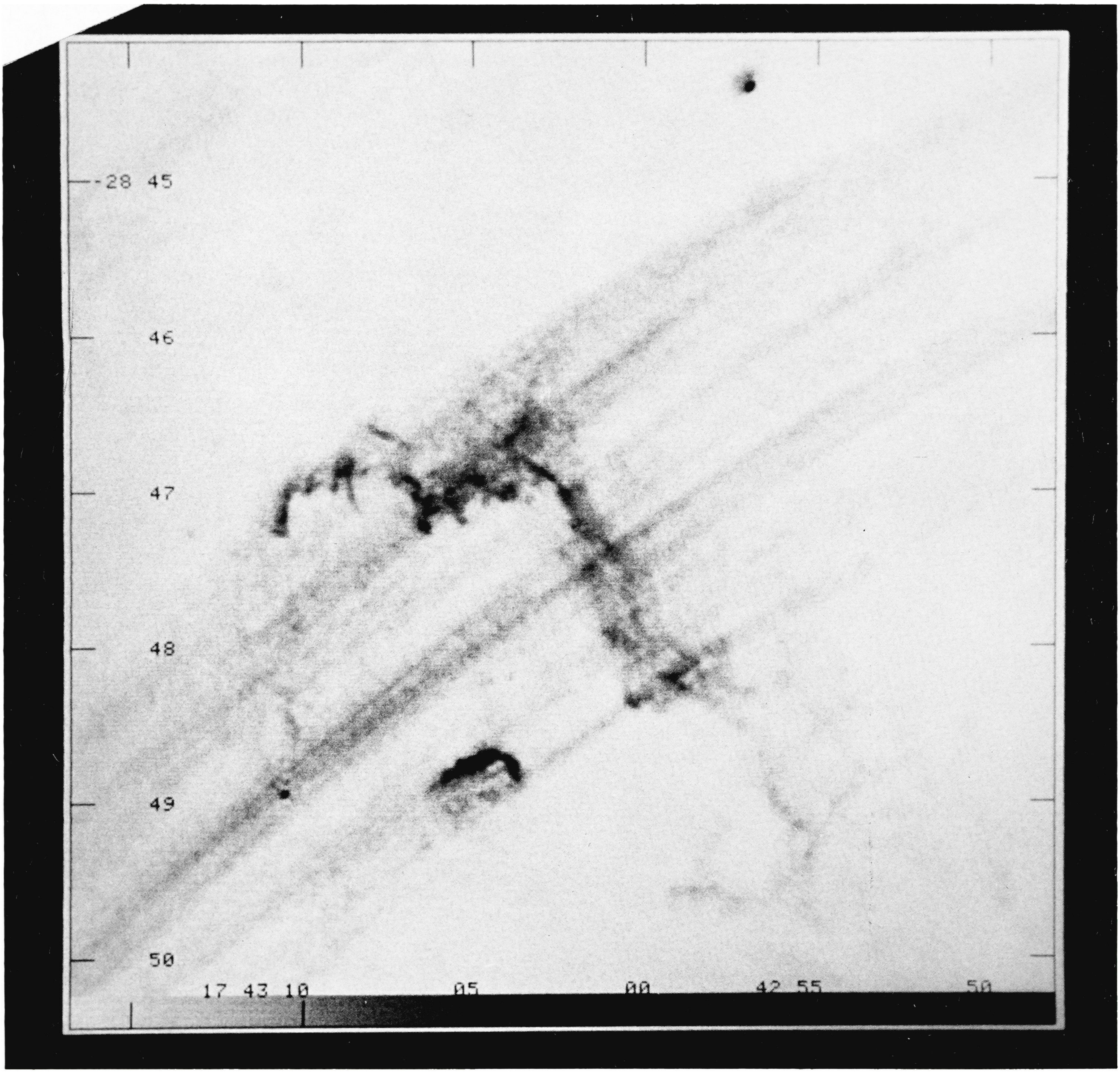

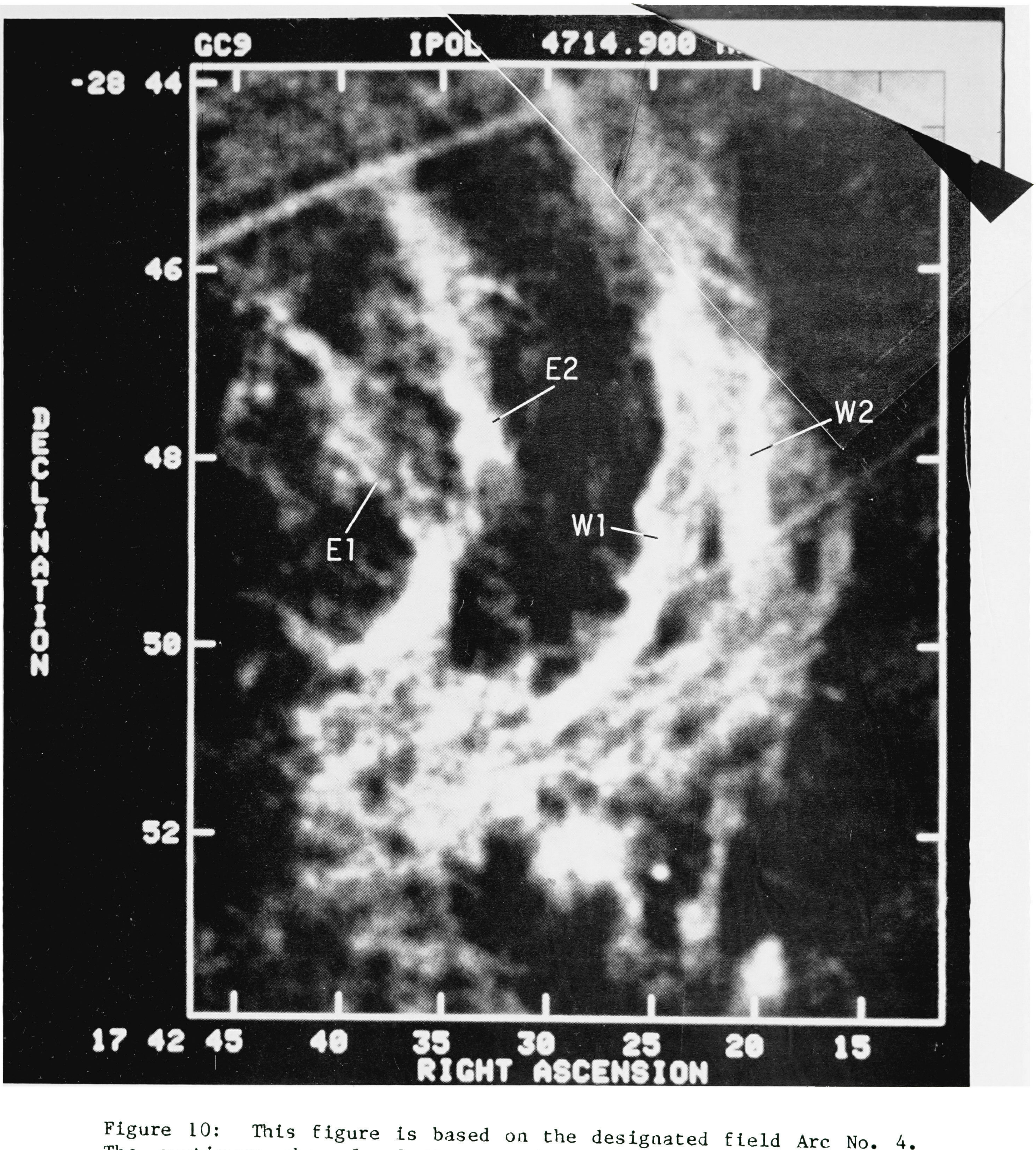

Figure 10: This figure is based on the designated field Arc No. 4. The continuum channel of the recombination line data set is also added to this data base which is tapered at 50 kλ before the final map was CLEANed. FWHM of the gaussian beam = 5.4"×4.5" (P.A. = 80). The peak residual flux = 1.8 mJy/beam area. (See Table 1 of Chapter 2 for the arrays which were used in making this map).

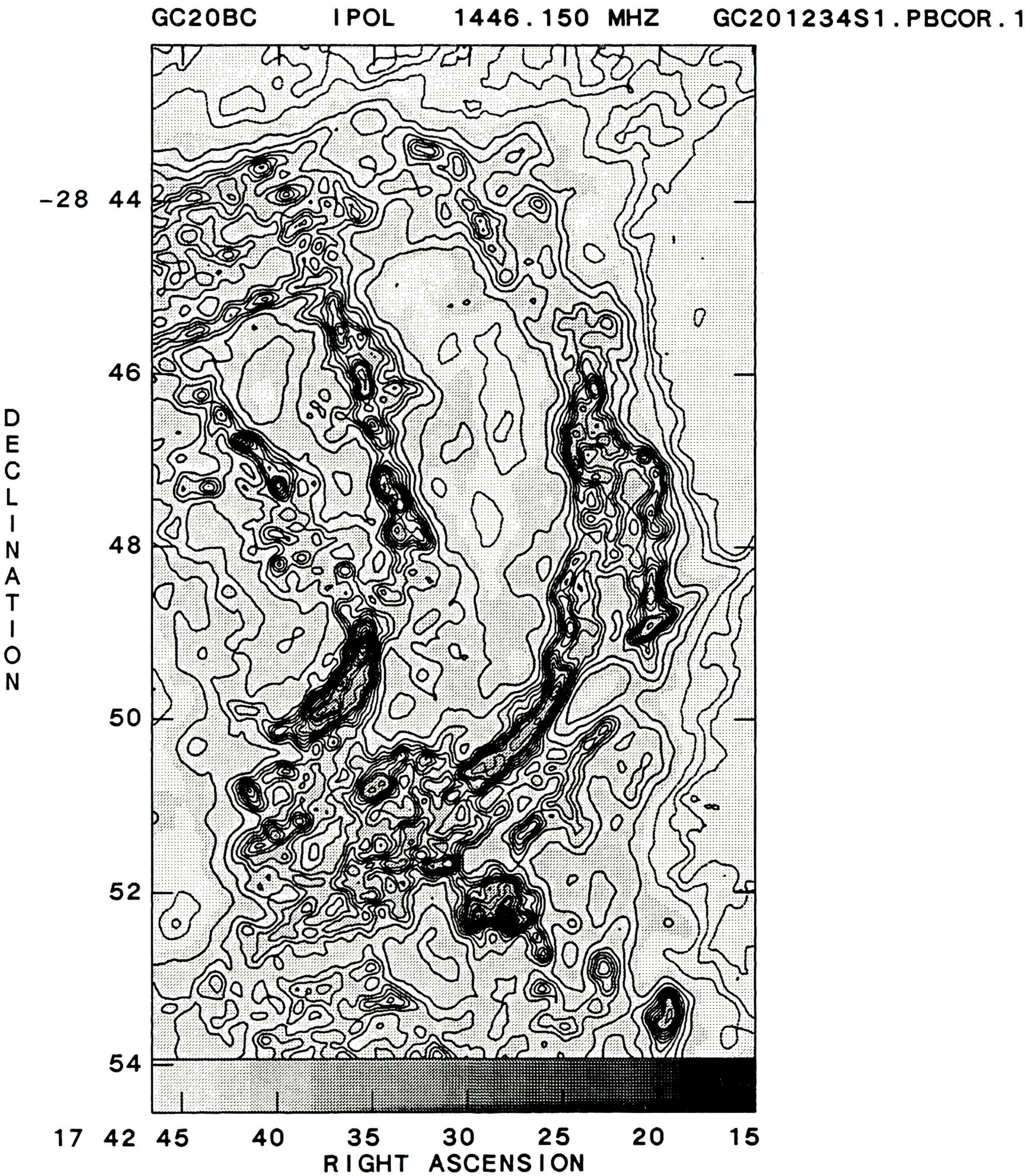

Figure 11a: This is a 20-cm contour map of the arched filaments using CLEAN with a gaussian beam (FWHM) = 8.1×7.1" (P.A. = 30°). The (u,v) data is similar to that of figure 2 except that no tapering was applied but the (u,v) range was limited up to 48 kλ. The peak residual flux is 8.1 mJy/beam area. The contour intensity levels are -15, -0, -5, 5, 10, 15, 20, 25, 30, 35, 40, 45, 50, 60, 70, 75, 80, ..., 135 mJy/beam area.

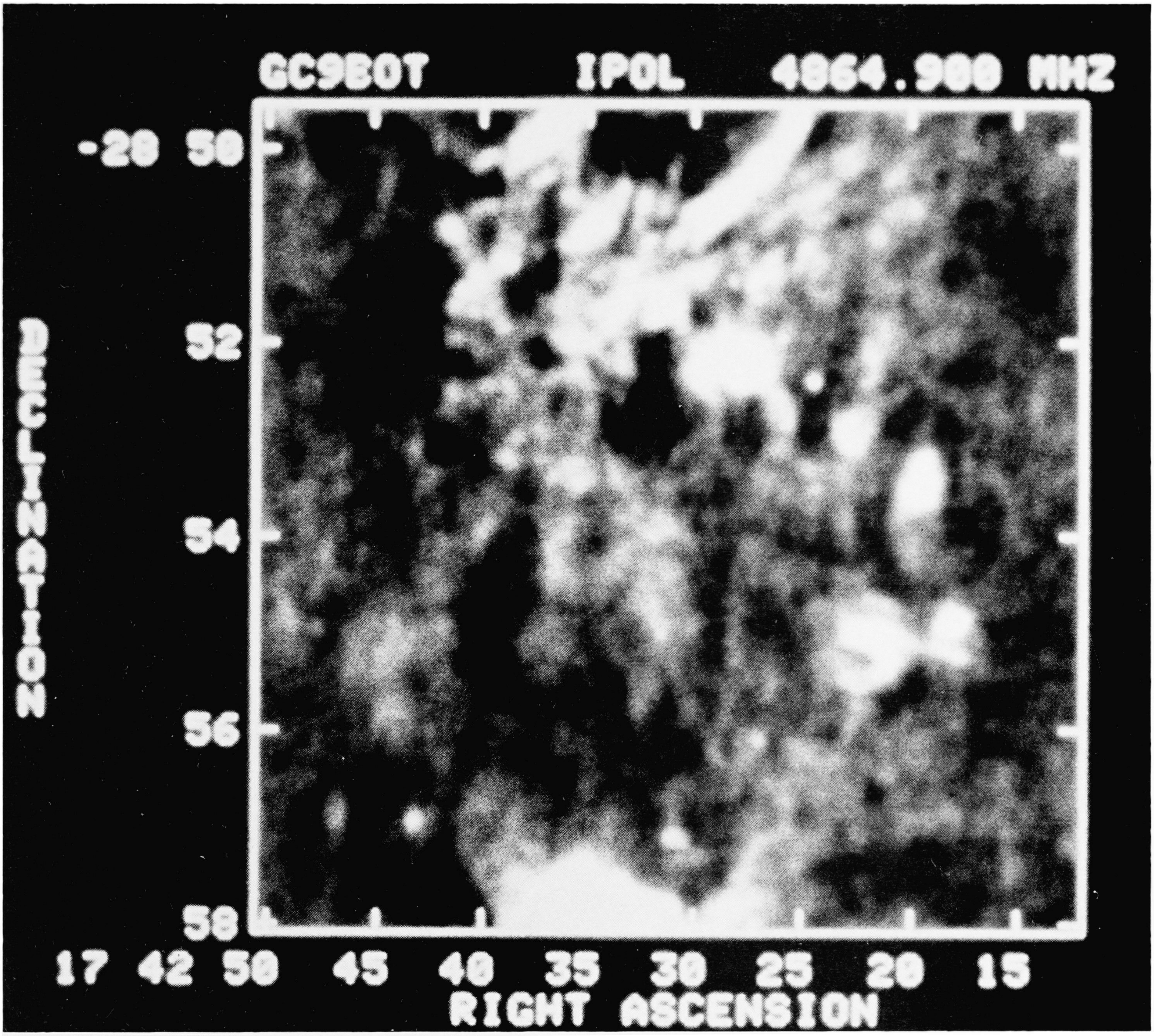

Figure 11b: This 6-cm radiograph is based on a data set which used the B/C[2] and C/D[3] arrays (Arc No. 7 in Table 1 of chapter 2). The data was self-calibrated individually and globally before it was tapered at 50 kλ FWHM = 7"×7".

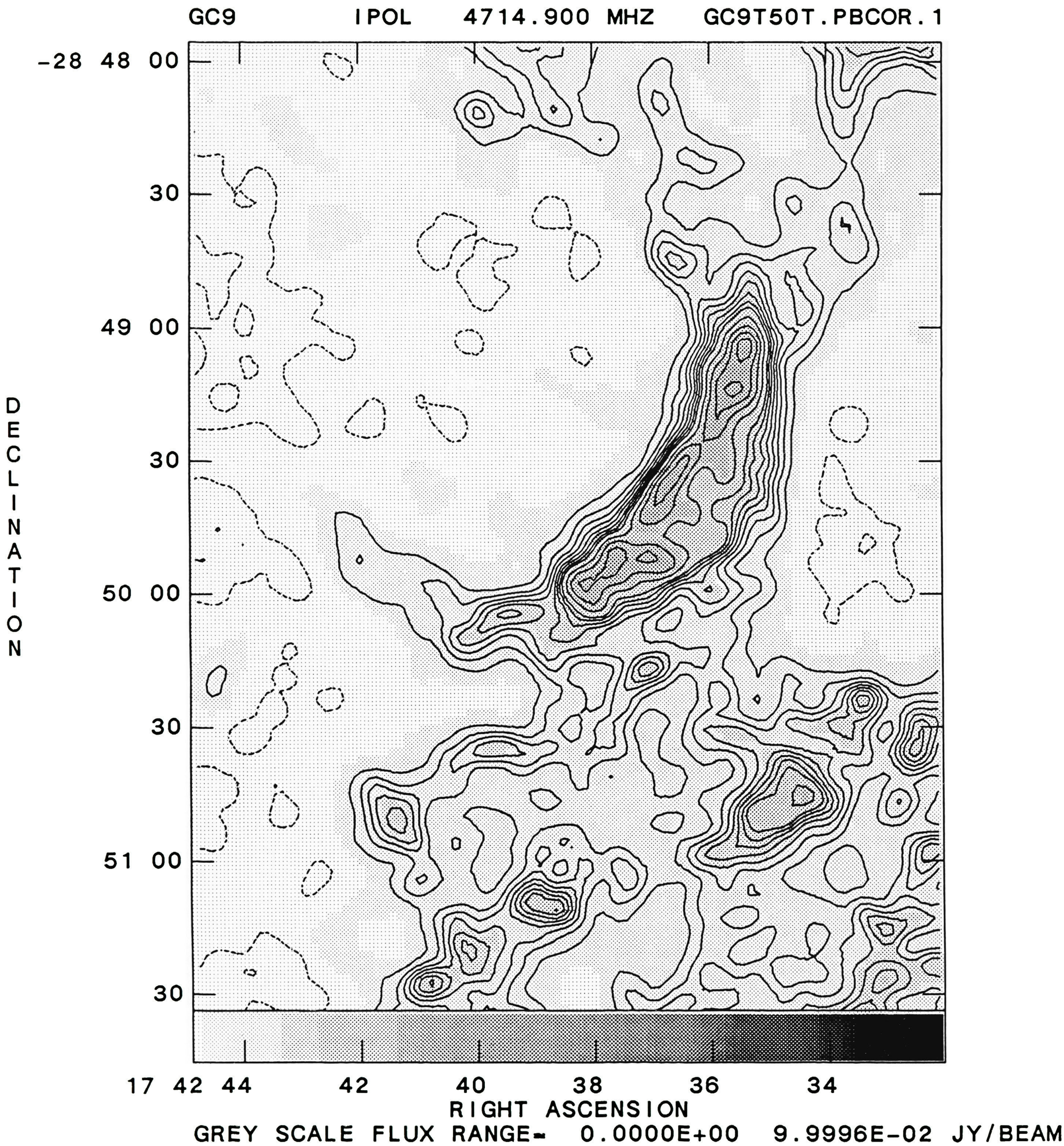

Figure 12: Similar to figure 10. The contour intensity levels are -5, -2.5, 2.5, 5, 7.5, 10, 12.5, 15, 17.5, 20, 25, 30, 35, 40, 45, ..., 100 mJy/beam area.

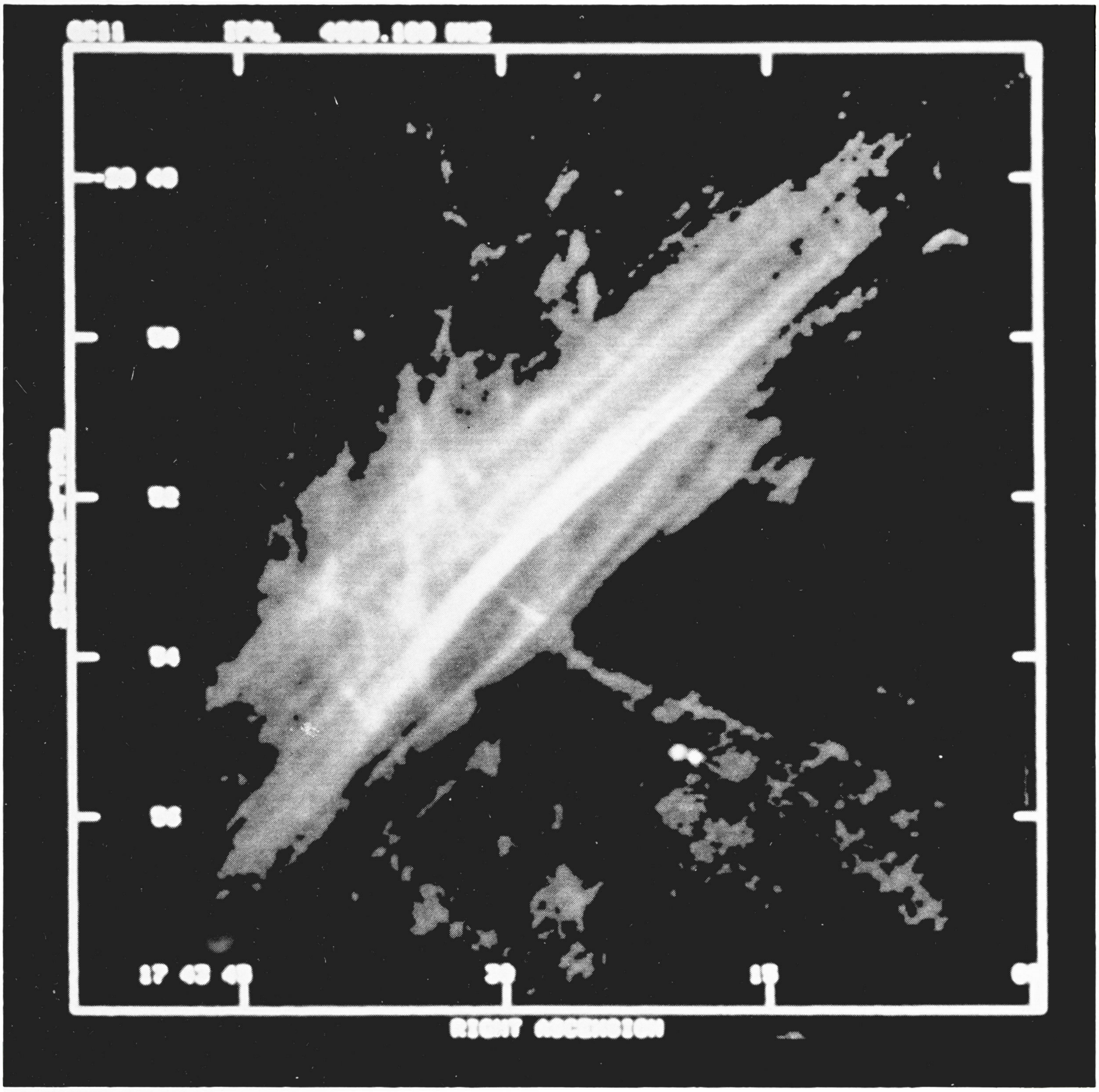

Figure 13: Similar to figure 4 except that ME deconvolution is applied to the data base. The solution converged after 15 iterations requiring 26 Jy of flux with the noise level = 0.2 mJy/beam area.

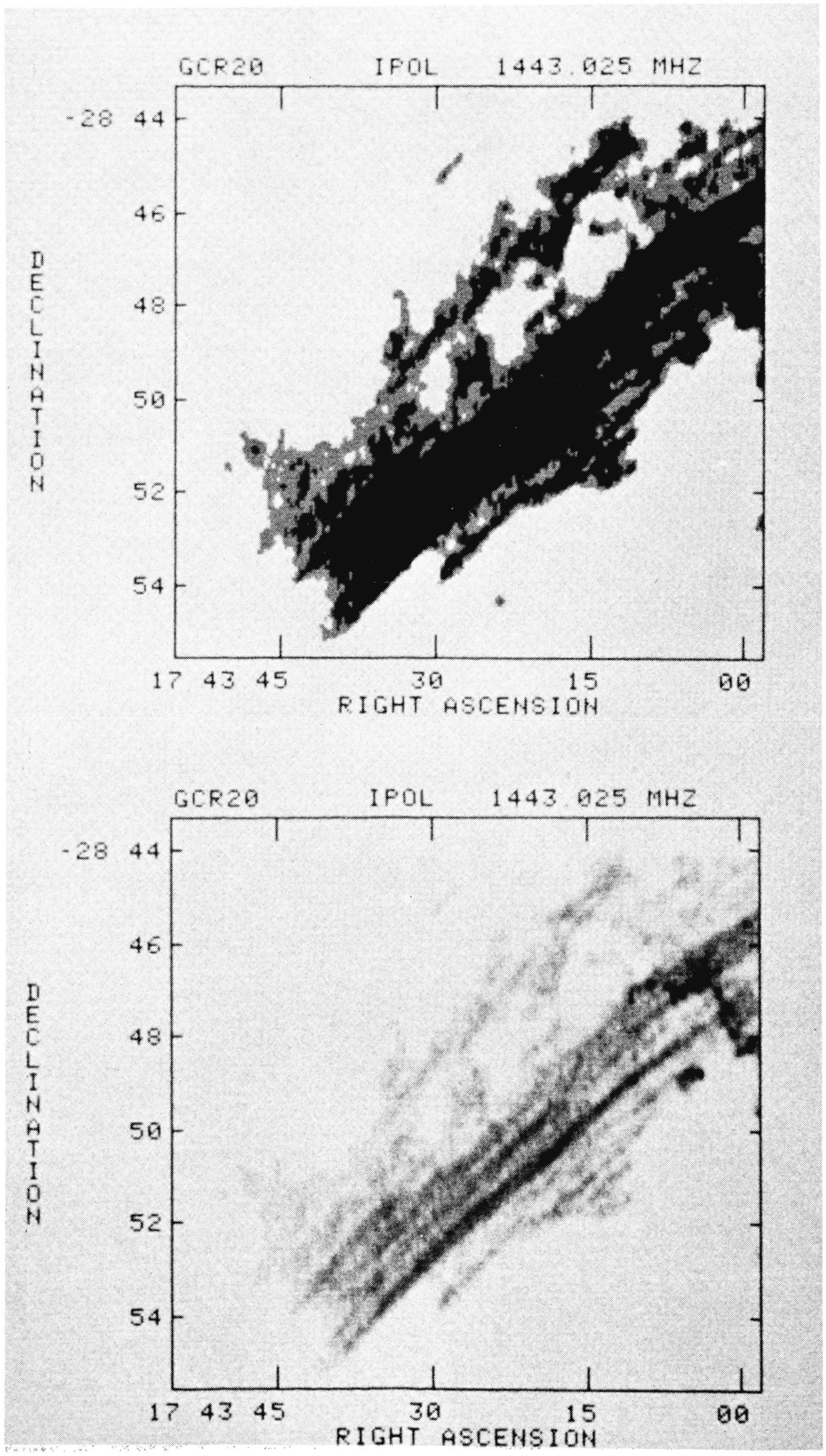

Figure 14 (a-b) and 15: The data set corresponding to these figures is similar to that of figure 1 except that only the data corresponding to B and C/D' were combined. The resolution is 9.2"×4.4" (P.A. = 4°). Figures 14a and 14b have different transfer functions. The intensity contour levels in figure 15 are -30, -20, -10, 10, 20, 30, 40, 50, 60, ... 100 mJy/beam area.

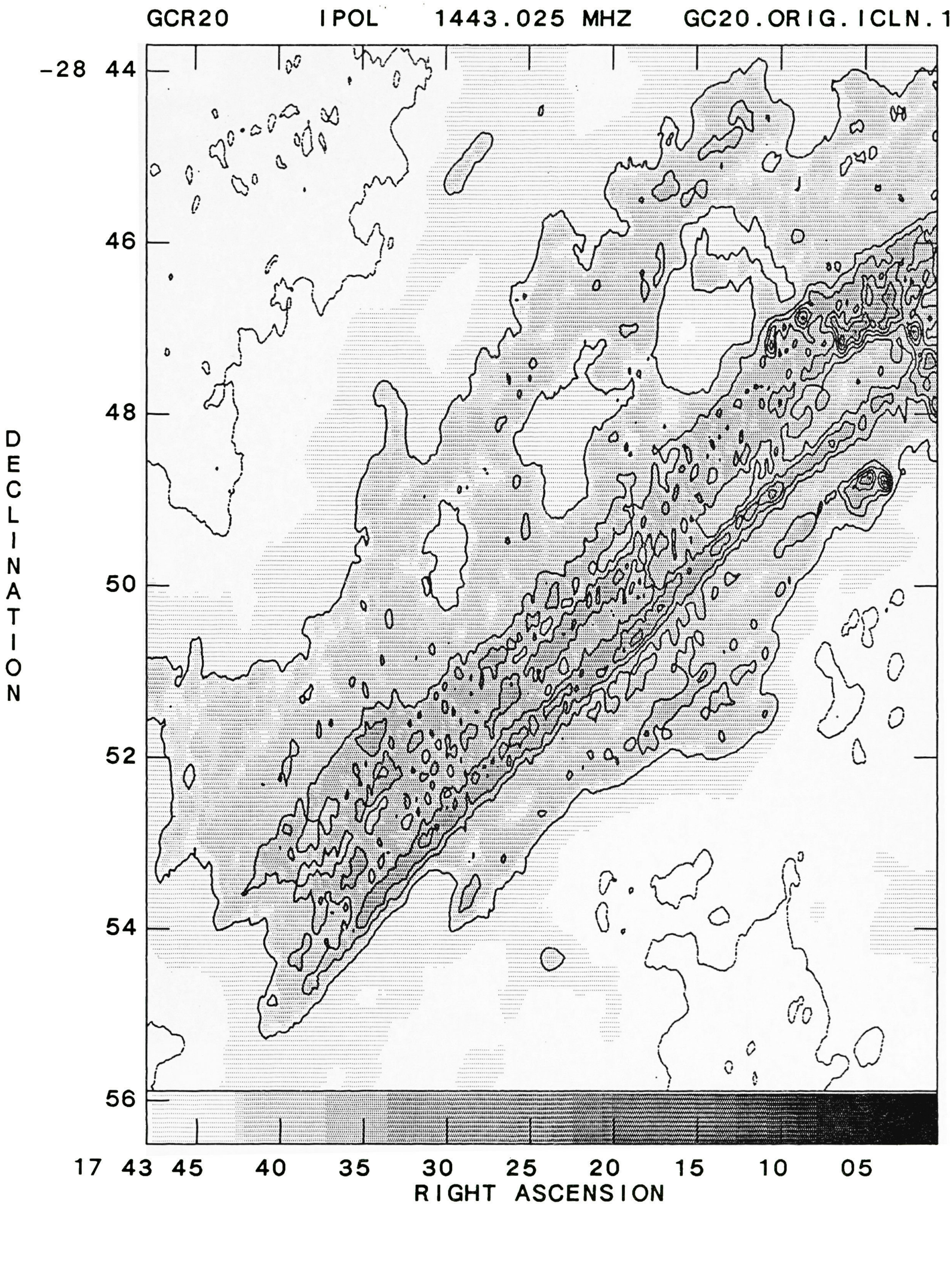
GCR20 IPOL 1443.025 MHZ GC20.ORIG.ICLN.1
-28 44
46
48
50
52
54
56
DECLINATION
17 43 45
40
35
30
25
20
15
10
05
RIGHT ASCENSION

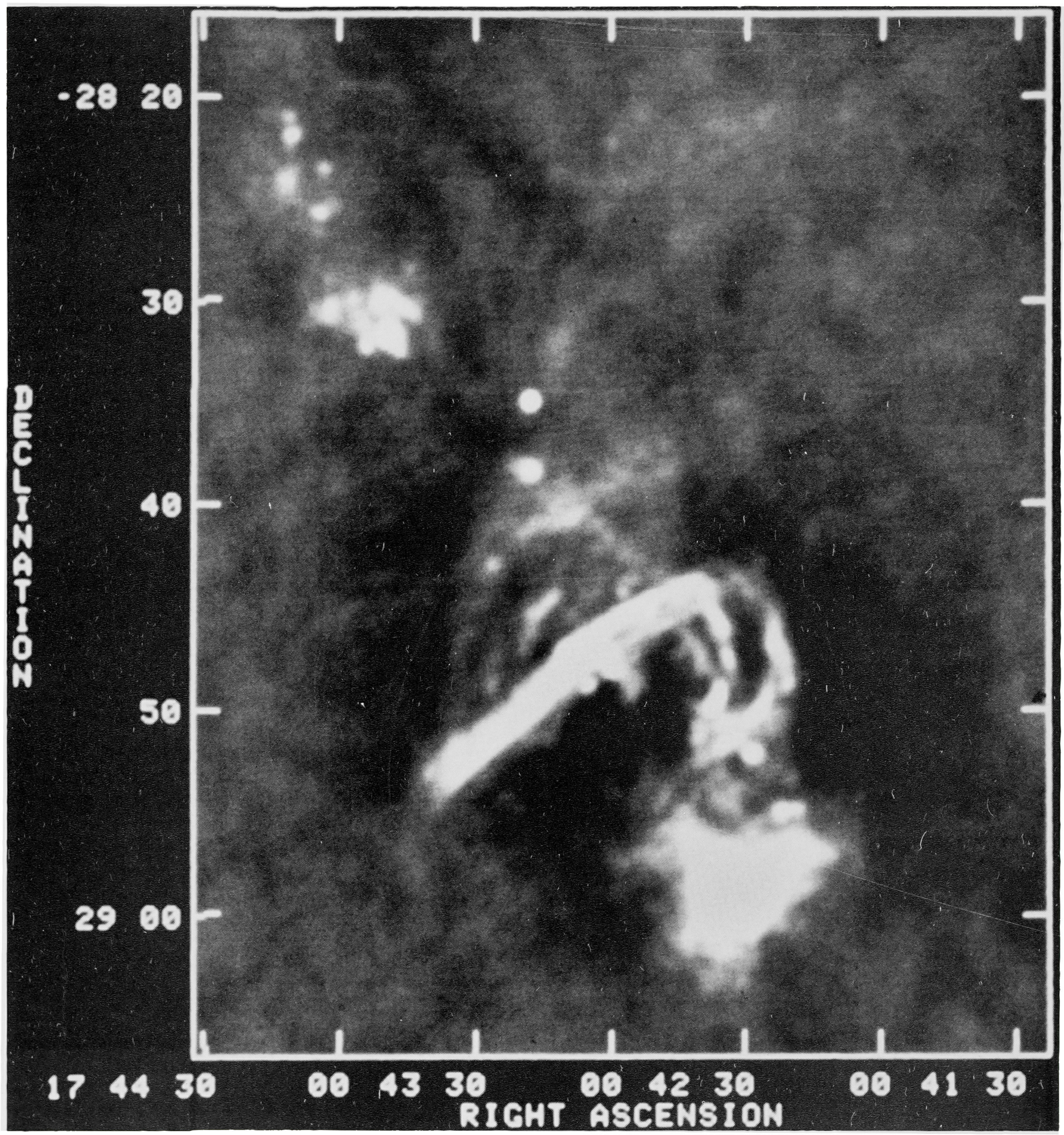

Figure 16: This is based on a data set which used only the C/D' array data and which restricted to (u,v) range between 0.21 to 5 kλ. The CLEAN beam (FWHM) = 37.2"×32.8" (P.A. = -82°). The peak residual flux = 15.1 mJy/beam area.

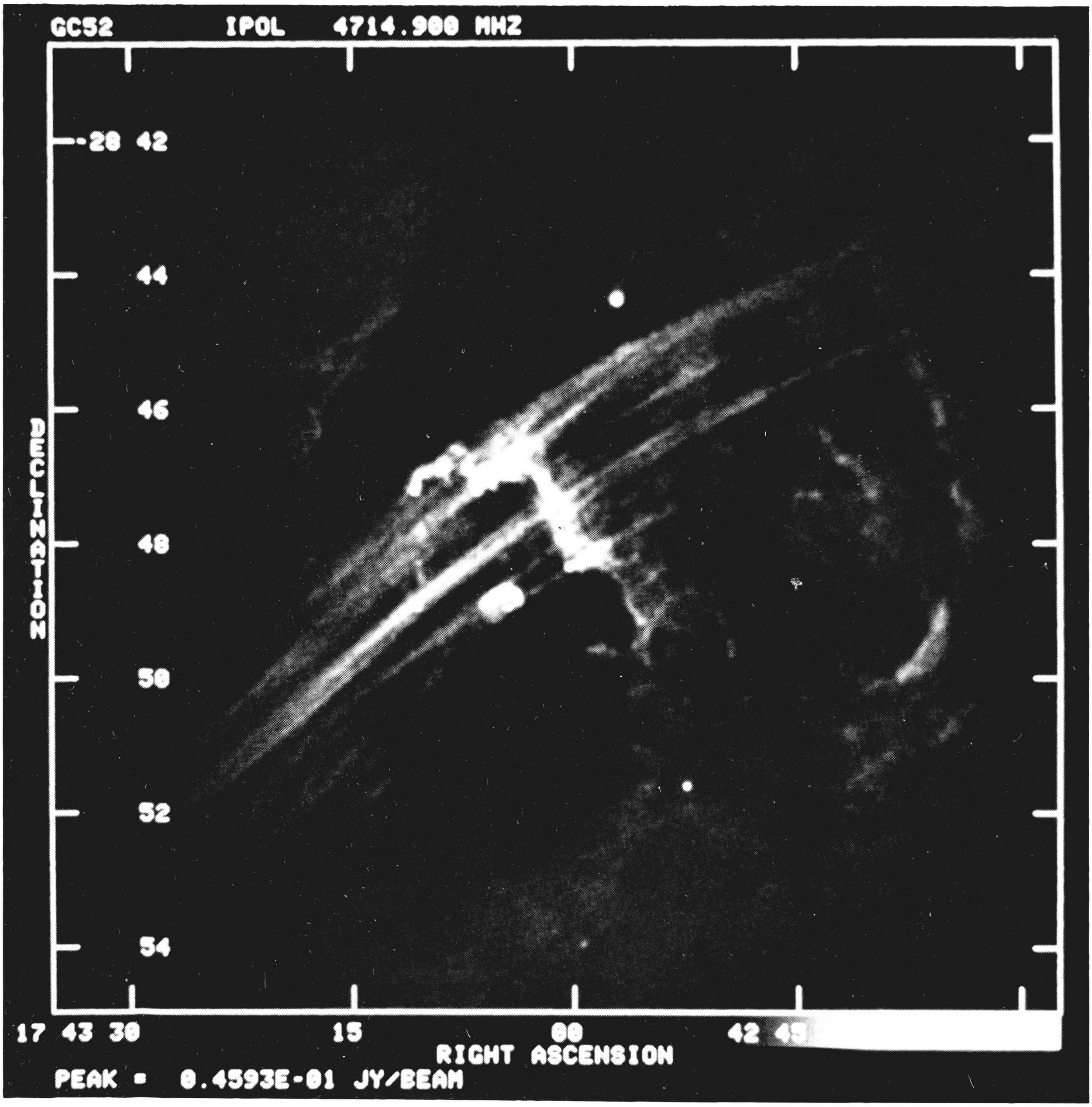

Figure 17, 18 (a-b): These figures are based on the same data base as that of figure 9. Figures 17, 18 (a-b) were constructed by tapering at 50 and 25 kλ, giving resolutions (FWHM) of 4.7"×4.7" and 7.1"×6.2" (P.A. = 9.4°), respectively. The intensity contour levels for figure 18b are -5, -2.5, 2.5, 5, 7.5, 10, 15, 20, 25, ... 65 mJy/beam area. The peak residual flux for figure 17 and 18 (a-b) are 1.4 and 1.32 mJy/beam areas, respectively.

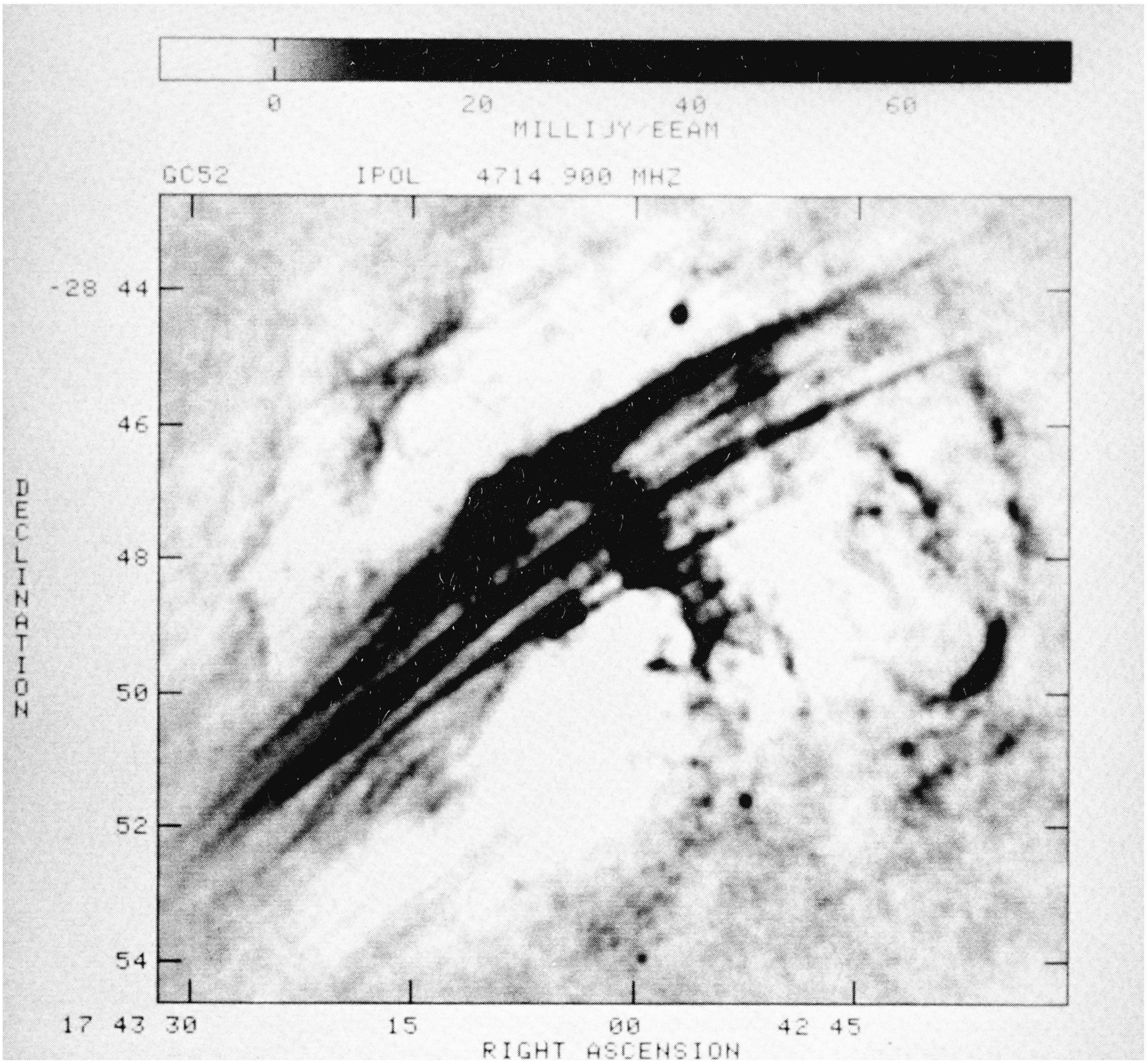

0 20 40 60
MILLIJY/BEAM
GC52 IPOL 4714.900 MHZ
-28 44
46
48
50
52
54
DECLINATION
17 43 30
15
00
42 45
RIGHT ASCENSION

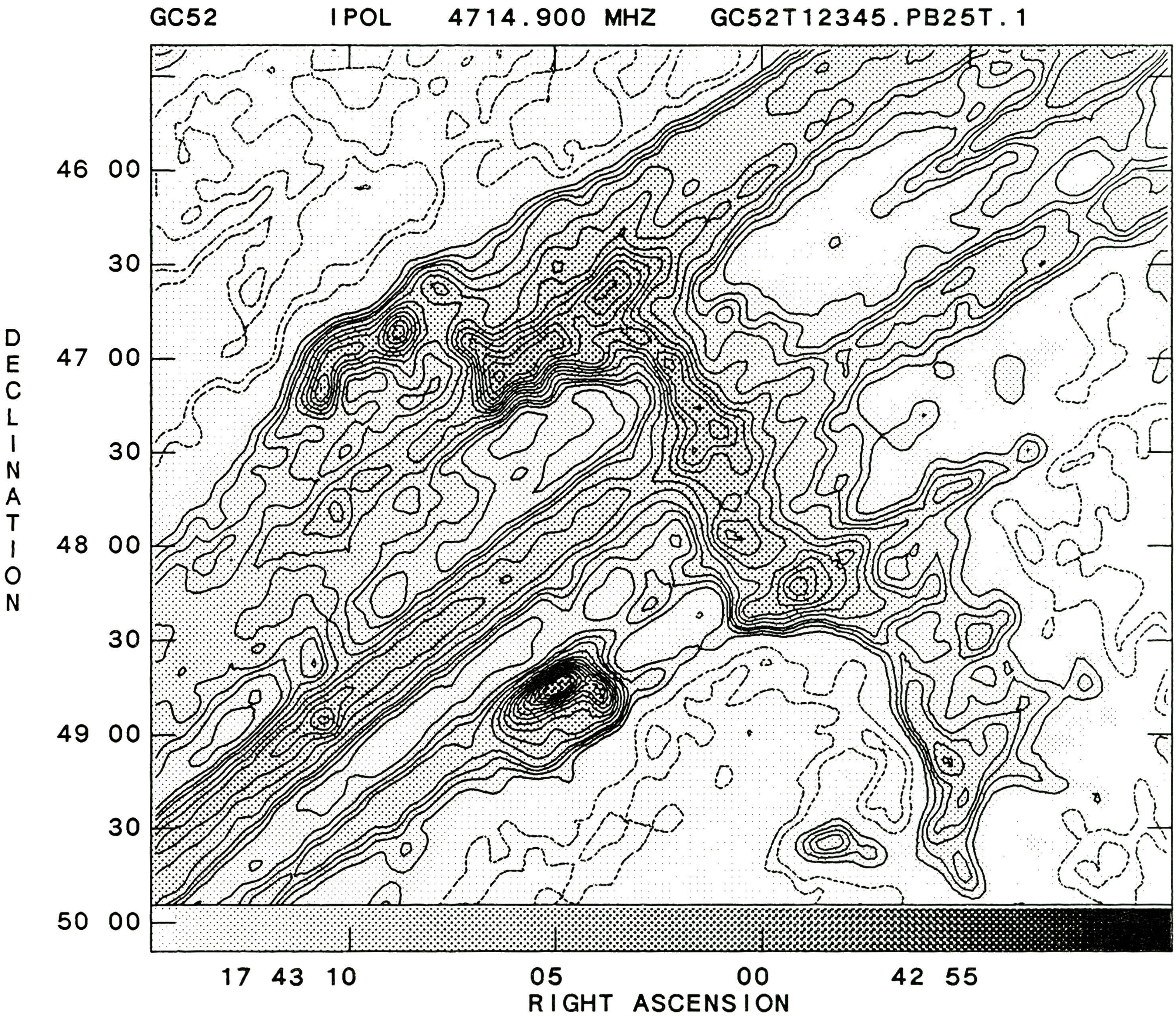
GC52 IPOL 4714.900 MHZ GC52T12345.PB25T.1
DECLINATION
46 00
30
47 00
30
48 00
30
49 00
30
50 00
17 43 10
05
00
42 55
RIGHT ASCENSION

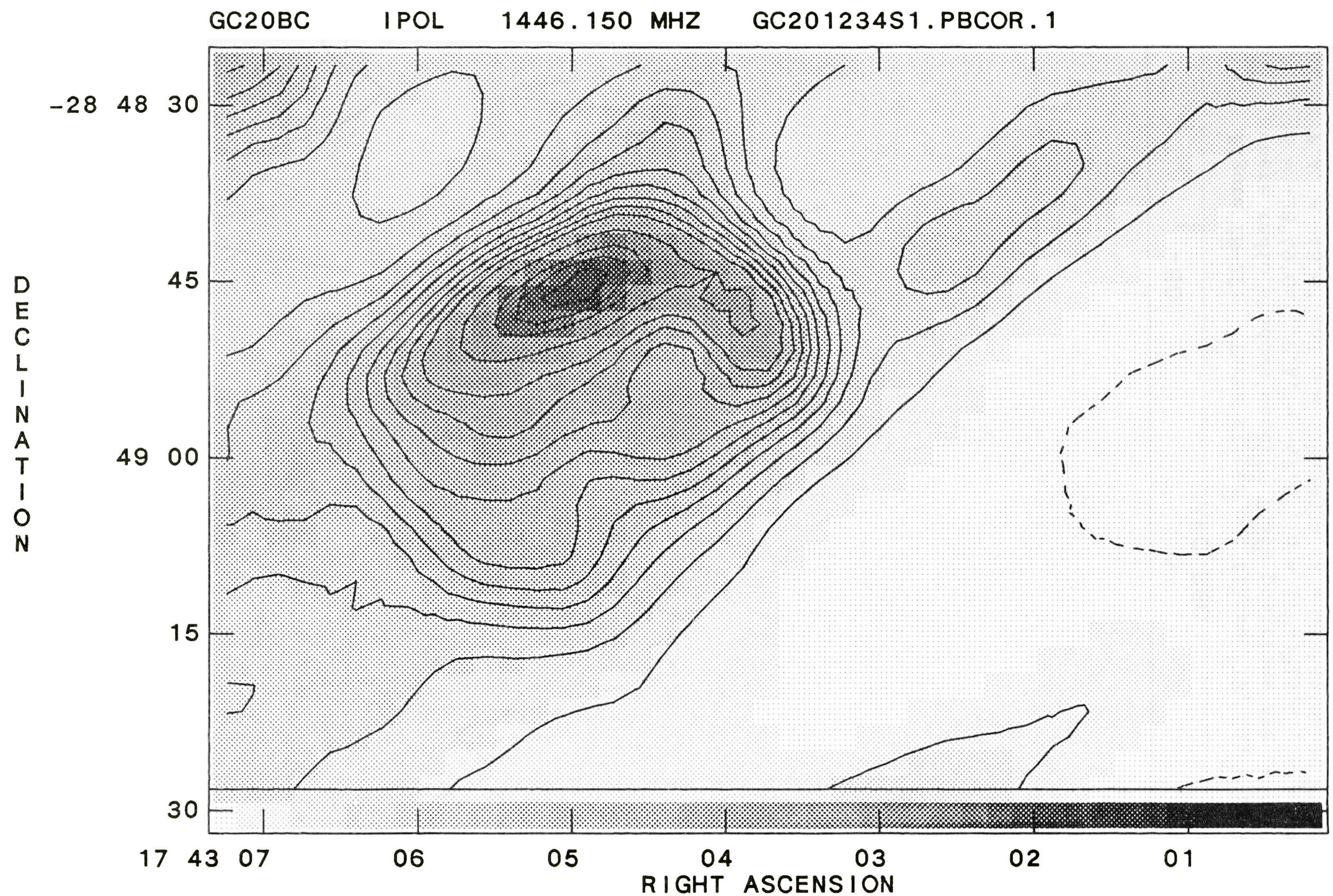

Figure 19: This map shows the multiple-hot feature at 20 cm having identical contour intervals as those seen for figure 11.

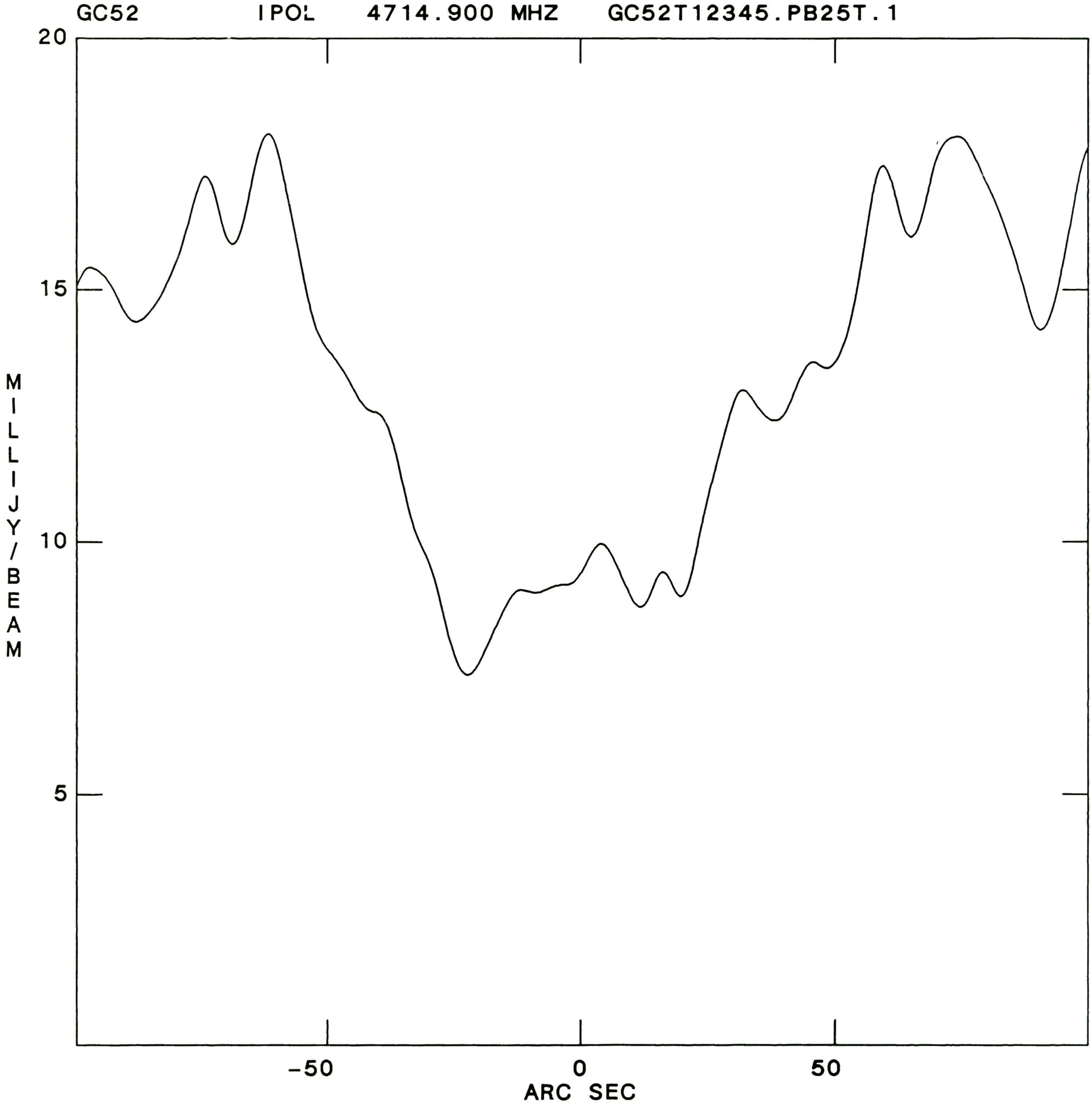

Figure 20-22: Three slices cut figures 18, 11, 4 at the positions where radio shadows occur. These positions are at $\alpha = 17^h43^m15.8^s$, $\delta = -28°49'04''$, P.A. = -47.6°, $\alpha = 17^h43^m15.9^s$, $\delta = -28°49'15''$, P.A. = -48.6° and $\alpha = 17^h43^m32^s$, $\delta = -28°52'08.7''$, P.A. = -51.3° for figures 20, 21 and 22, respectively.

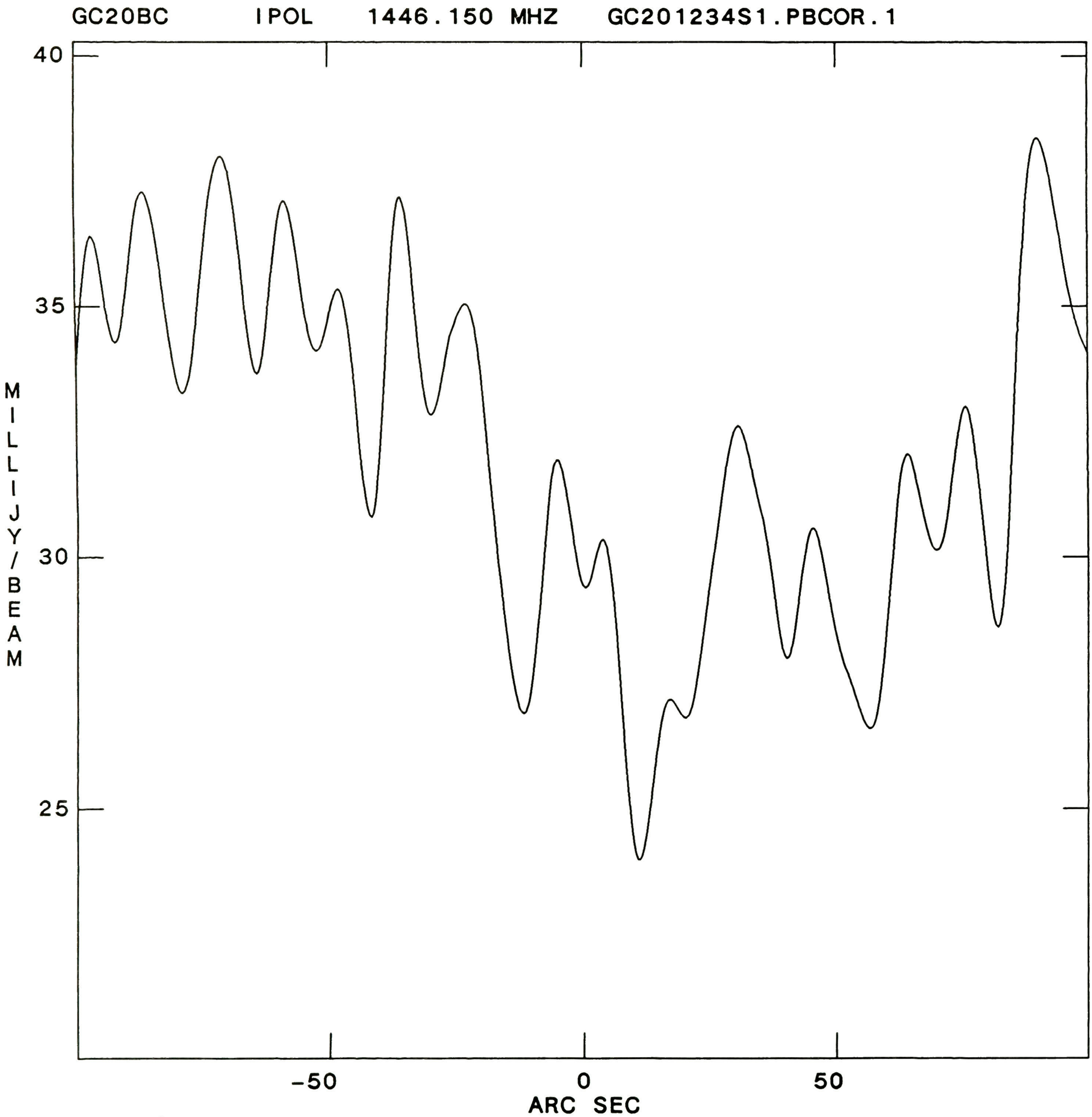
GC20BC IPOL 1446.150 MHZ GC201234S1.PBCOR.1
MILLIJY/BEAM
40
35
30
25
-50
0
50
ARC SEC

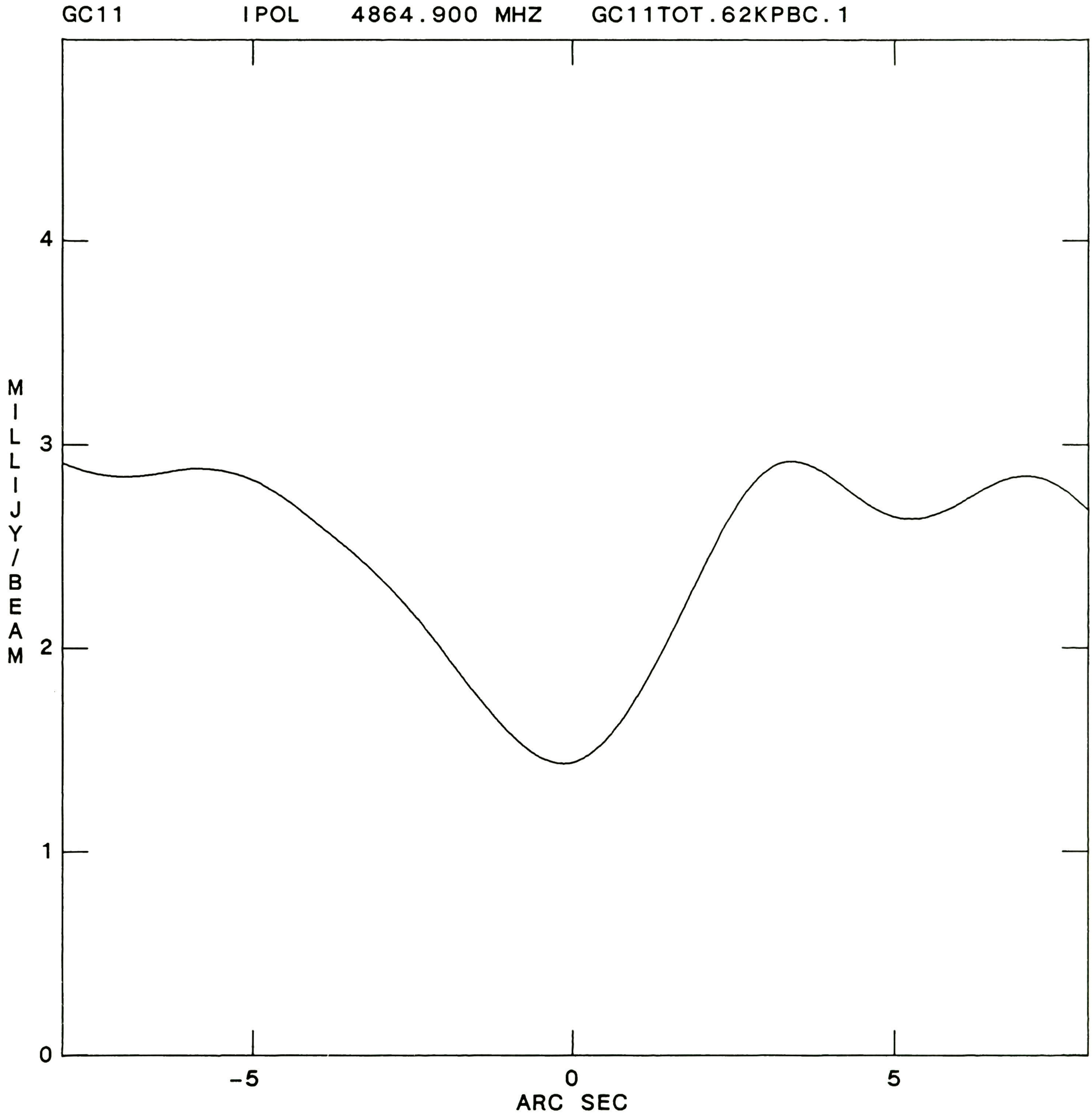
GC11 IPOL 4864.900 MHZ GC11TOT.62KPBC.1
MILLIJY/BEAM
0
1
2
3
4
-5
0
5
ARC SEC

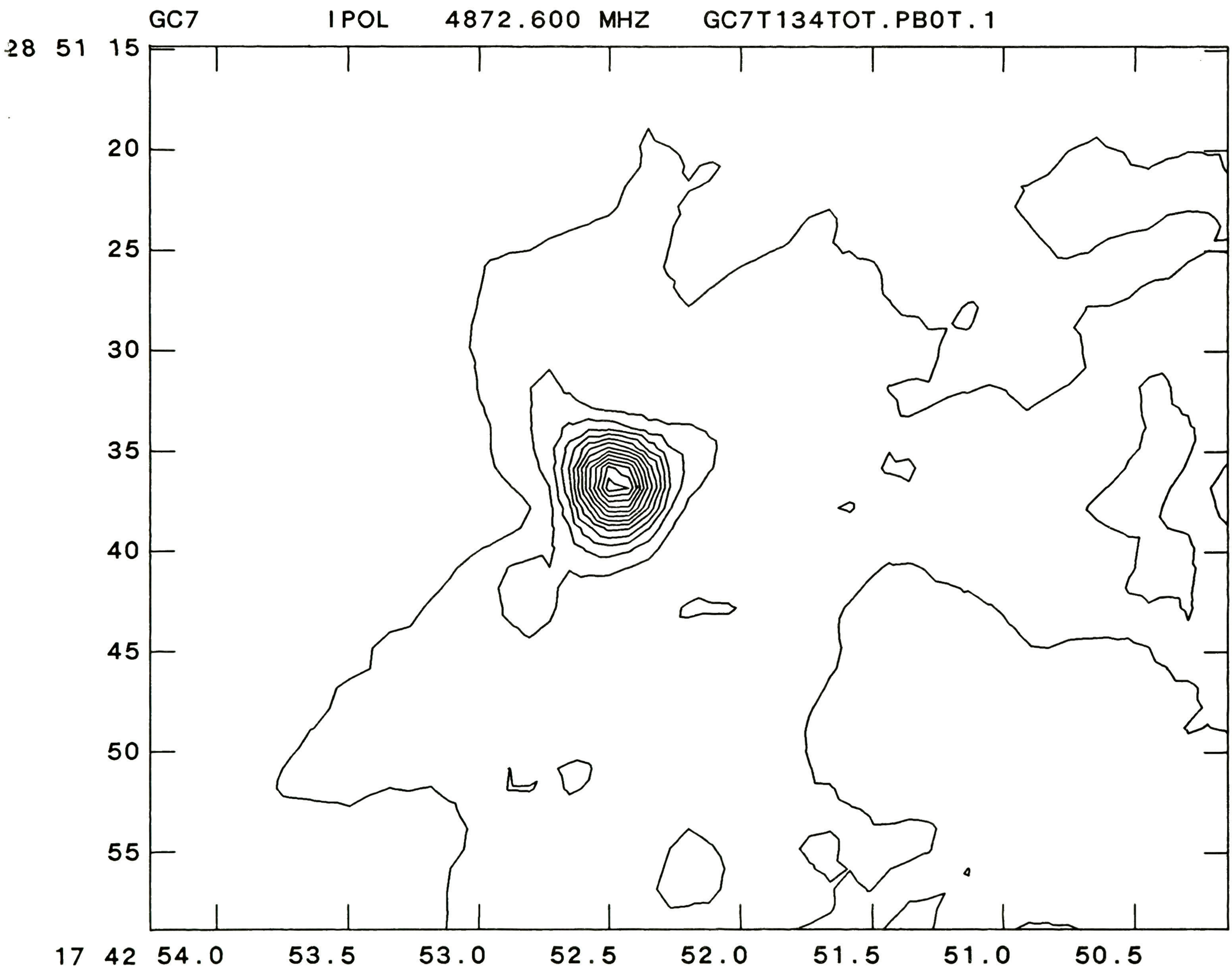

Figure 23a: FWHM = 30.6" x 2.64", the contour intervals 2, 4, 6, 8, 10, 12, ..., 36 mJy/beam area, λ = 6 cm.

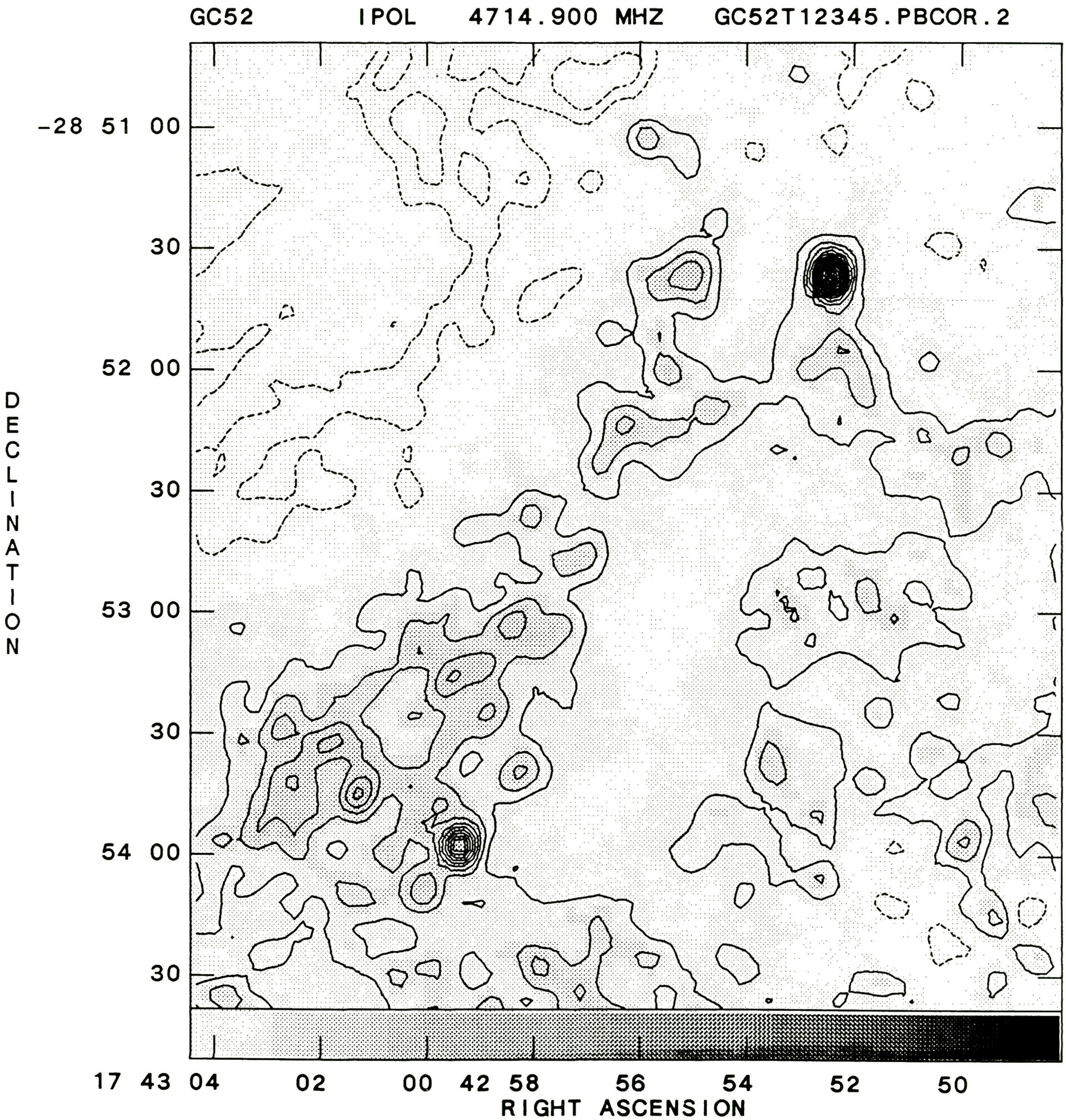

Figure 23b: FWHM = 7.1" x 6.3" (P.A. = 9.5°), the contour intervals are -5, -2.5, 2.5, 5, 7.5, 10, 12.5, 15, 17.5, 20, ..., 45 mJy/beam area, λ = 6 cm.

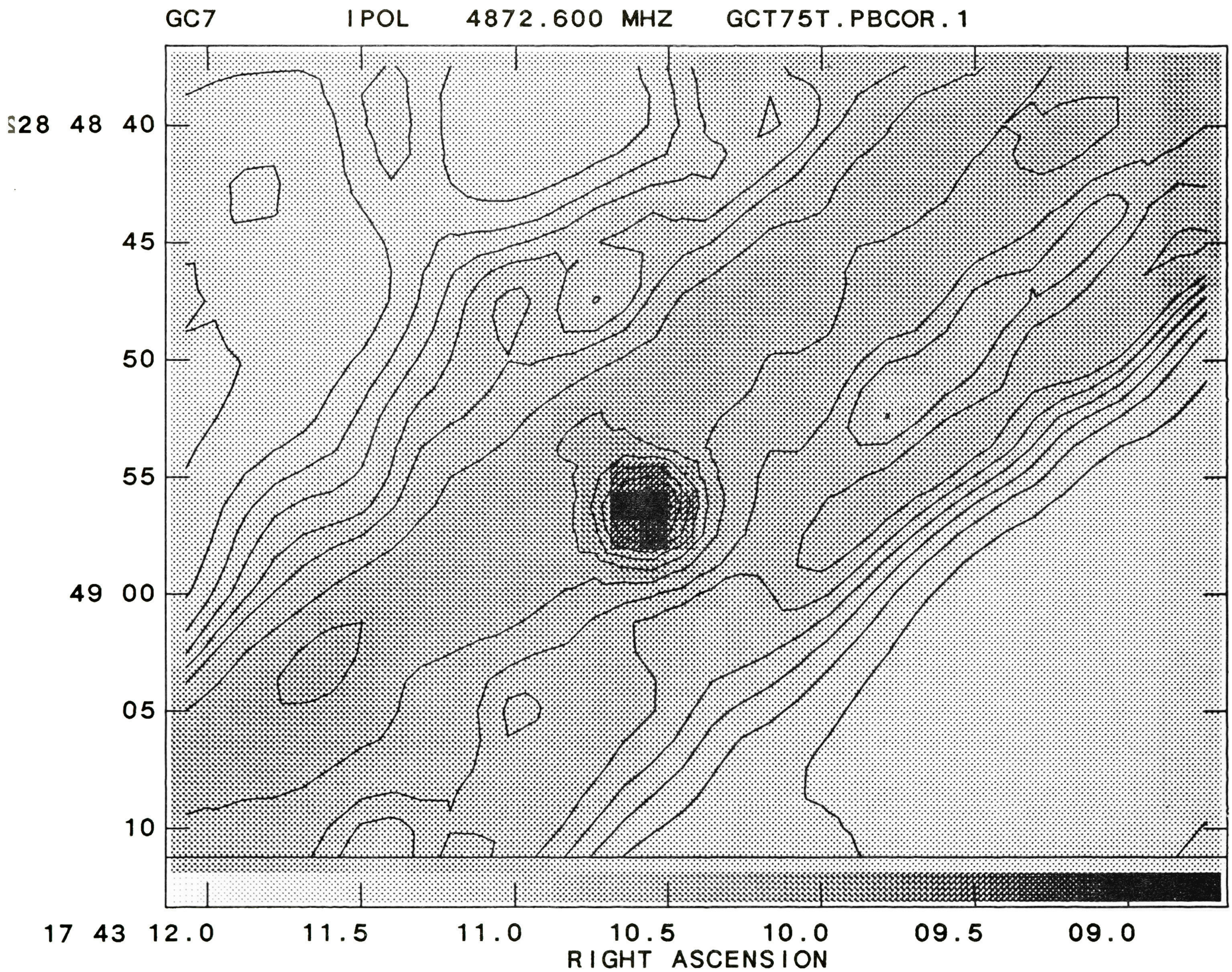

Figure 23c: FWHM = 3.3" x 3.3", the contour intervals are 1, 2, 3, 4, 5, 6, 8, 10, 20 mJy/beam area, λ = 6 cm.

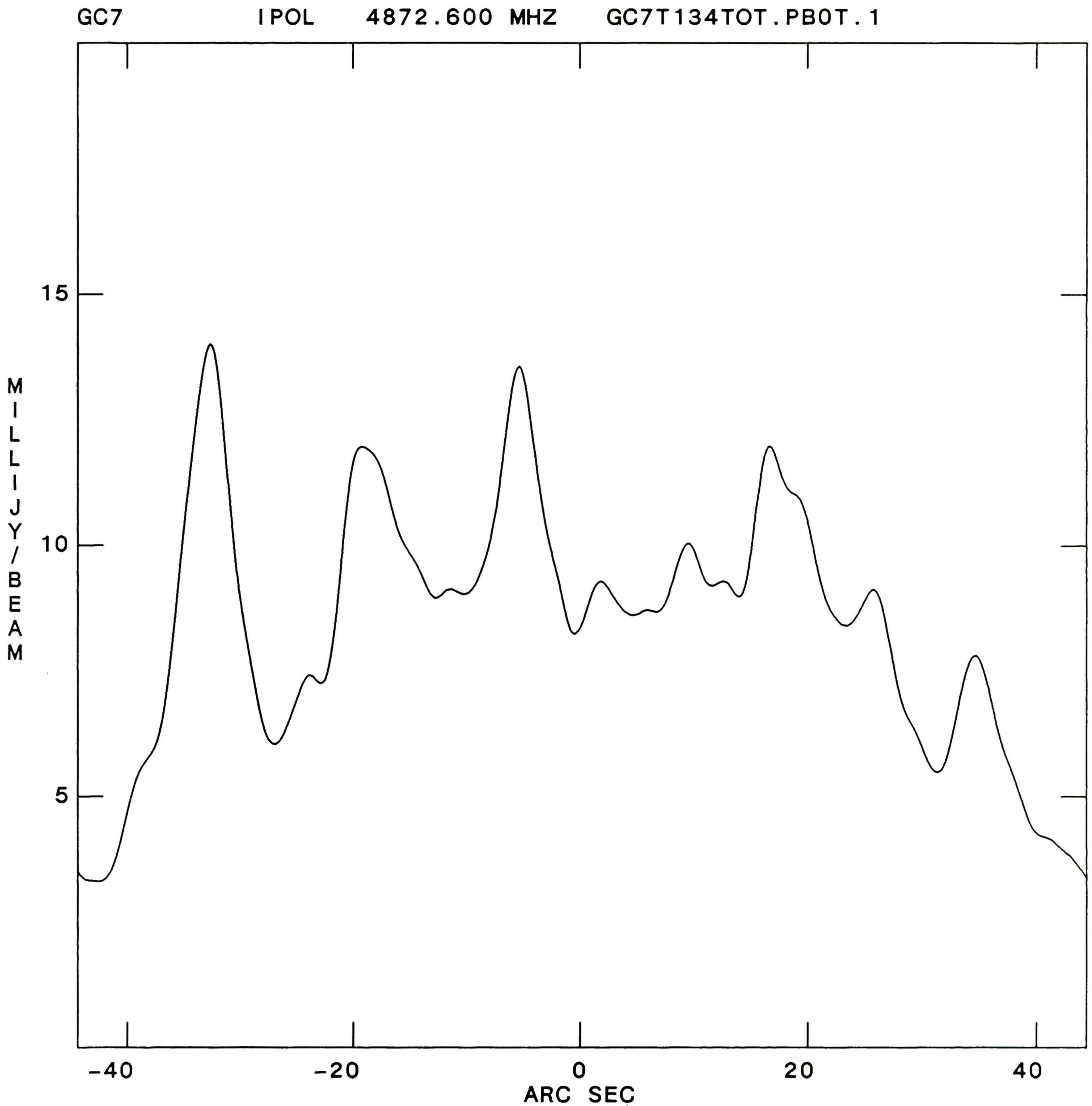

Figure 23d: FWHM = 3.66" x 2.64", the zero position of this slice plot is at $\alpha = 17^h43^m04^s$, $\delta = -28°46'53"$ (P.A. = -50.9°), $\lambda$ = 6 cm.

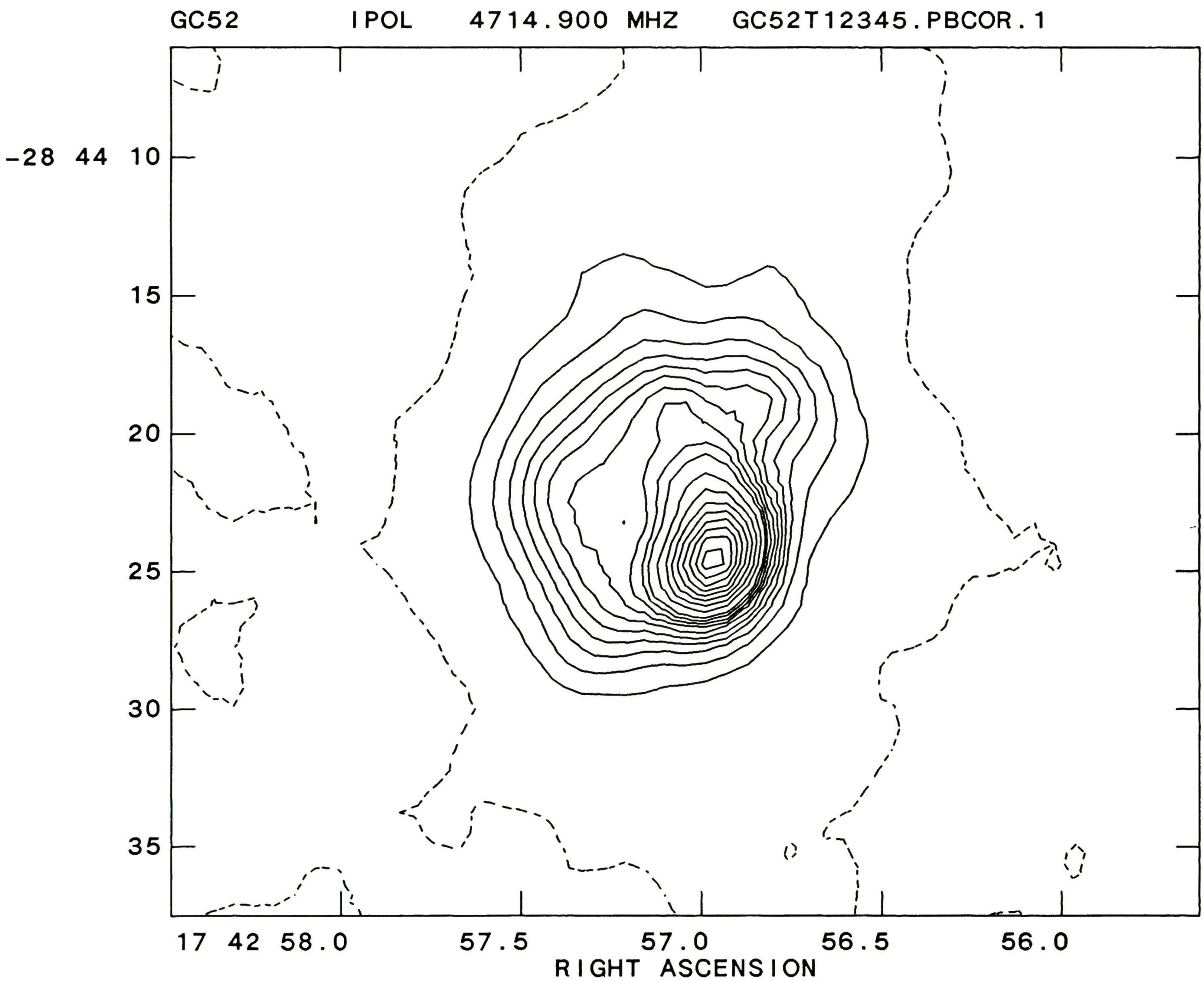

Figure 23e: FWHM = 2.58" x 1.93" (P.A. = 16°), the contour intervals are -1, 1, 2, 3, ... 10, 12, 14, 16, ..., 34 mJy/beam area, λ = 6 cm.

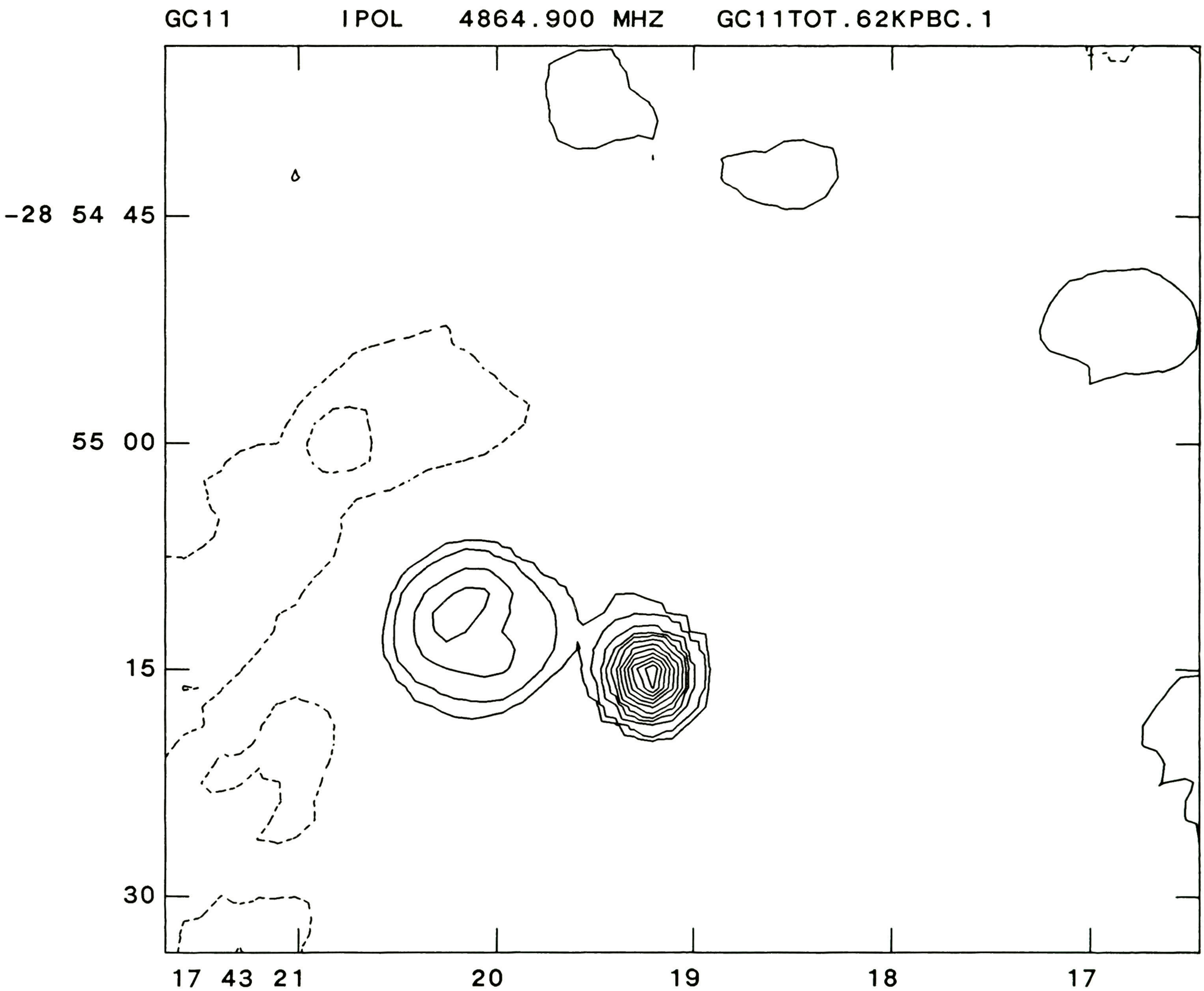

Figure 23f: FWHM = 3.16" x 2.84" (P.A. = 17.5°). The contour intervals are -2, -1, 1, 2, 5, 8, 11, 15, 20, 25, 30, 35, 50, 45 mJy/beam area, $\lambda$ = 6 cm.

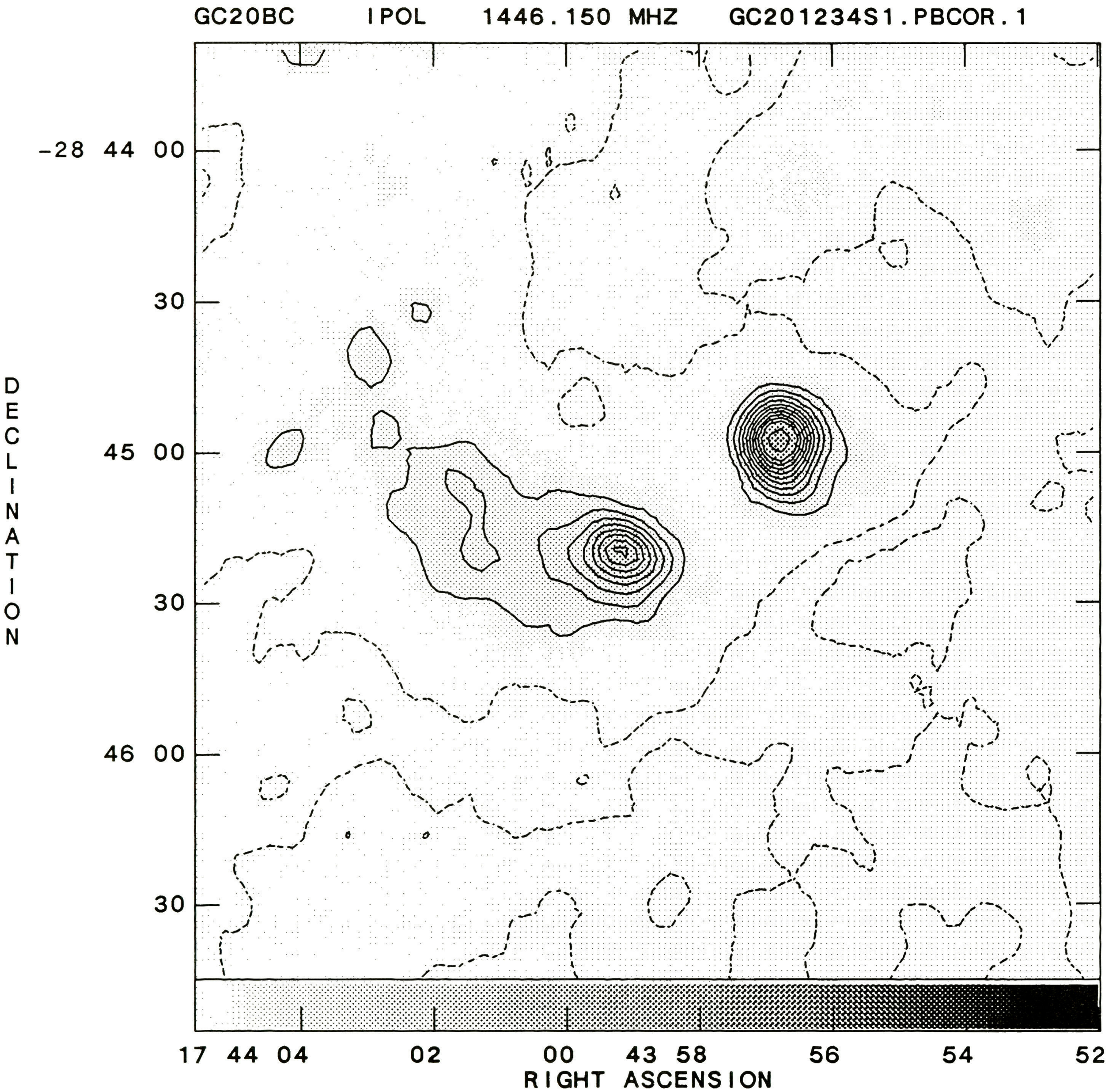

Figure 23 (g-i): FWHM = 8.1" x 7.1" (P.A. = 30.5°), the contour intervals are -3, -2, -1, 1, 2, ..., 10, 12, ..., 20, 23, 26, 29, 33, 37, 42, 48, 55, 63, 72, 82, 93 mJy/beam area, $\lambda$ = 6 cm.

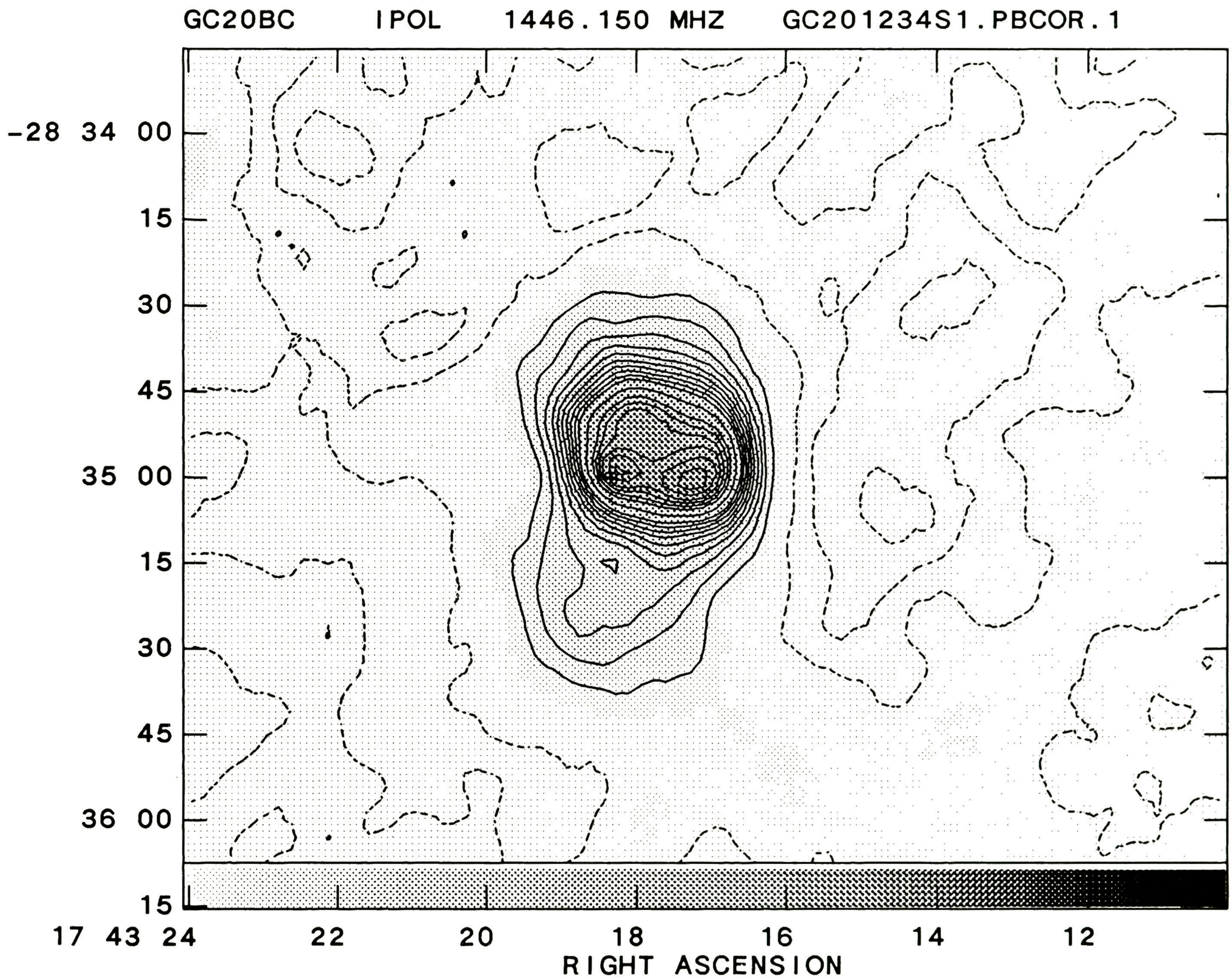
GC20BC IPOL 1446.150 MHZ GC201234S1.PBCOR.1
-28 34 00
15
30
45
35 00
15
30
45
36 00
15
17 43 24
22
20
18
16
14
12
RIGHT ASCENSION

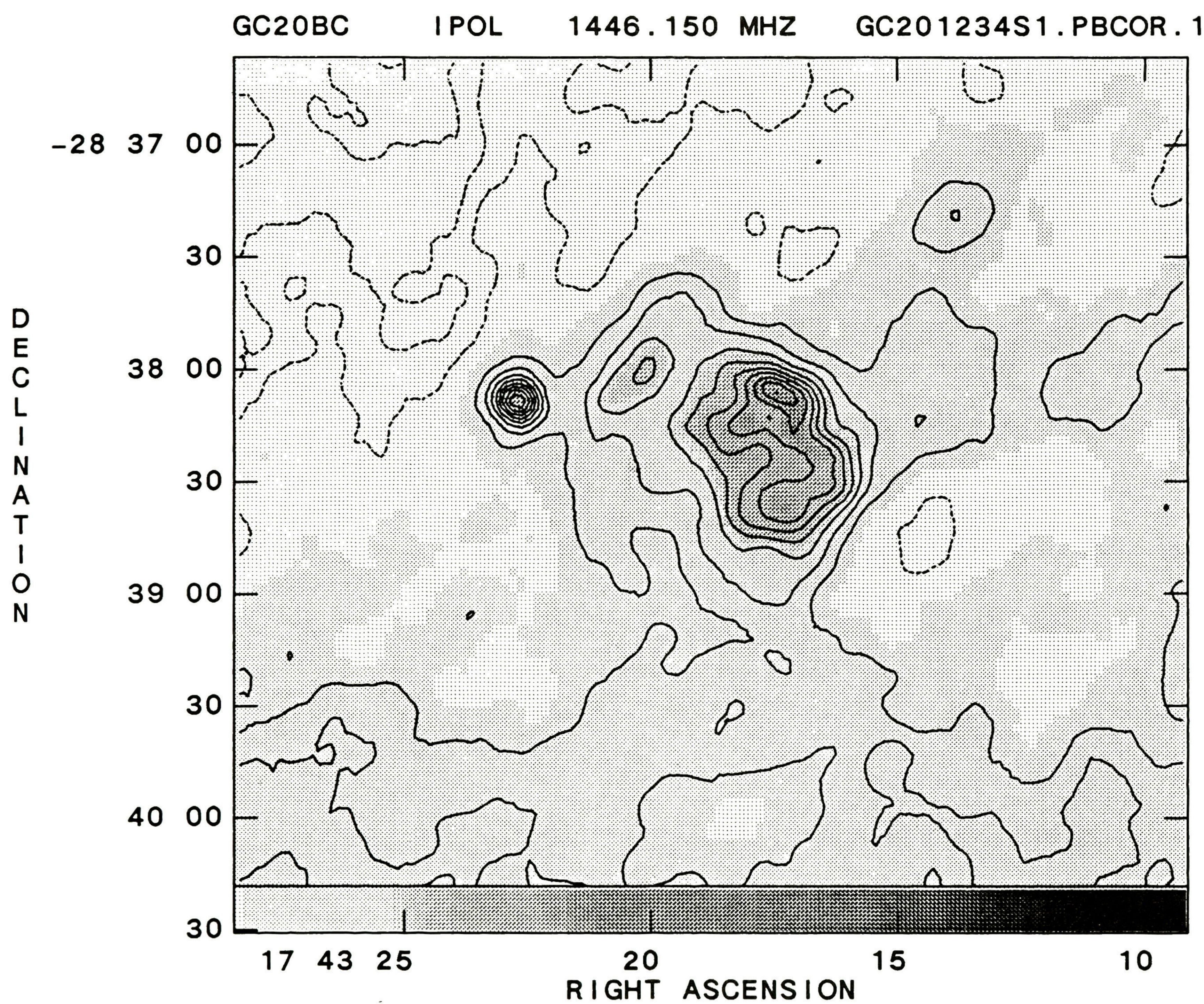
GC20BC IPOL 1446.150 MHZ GC201234S1.PBCOR.1
DECLINATION
-28 37 00
30
38 00
30
39 00
30
40 00
30
17 43 25
20
15
10
RIGHT ASCENSION

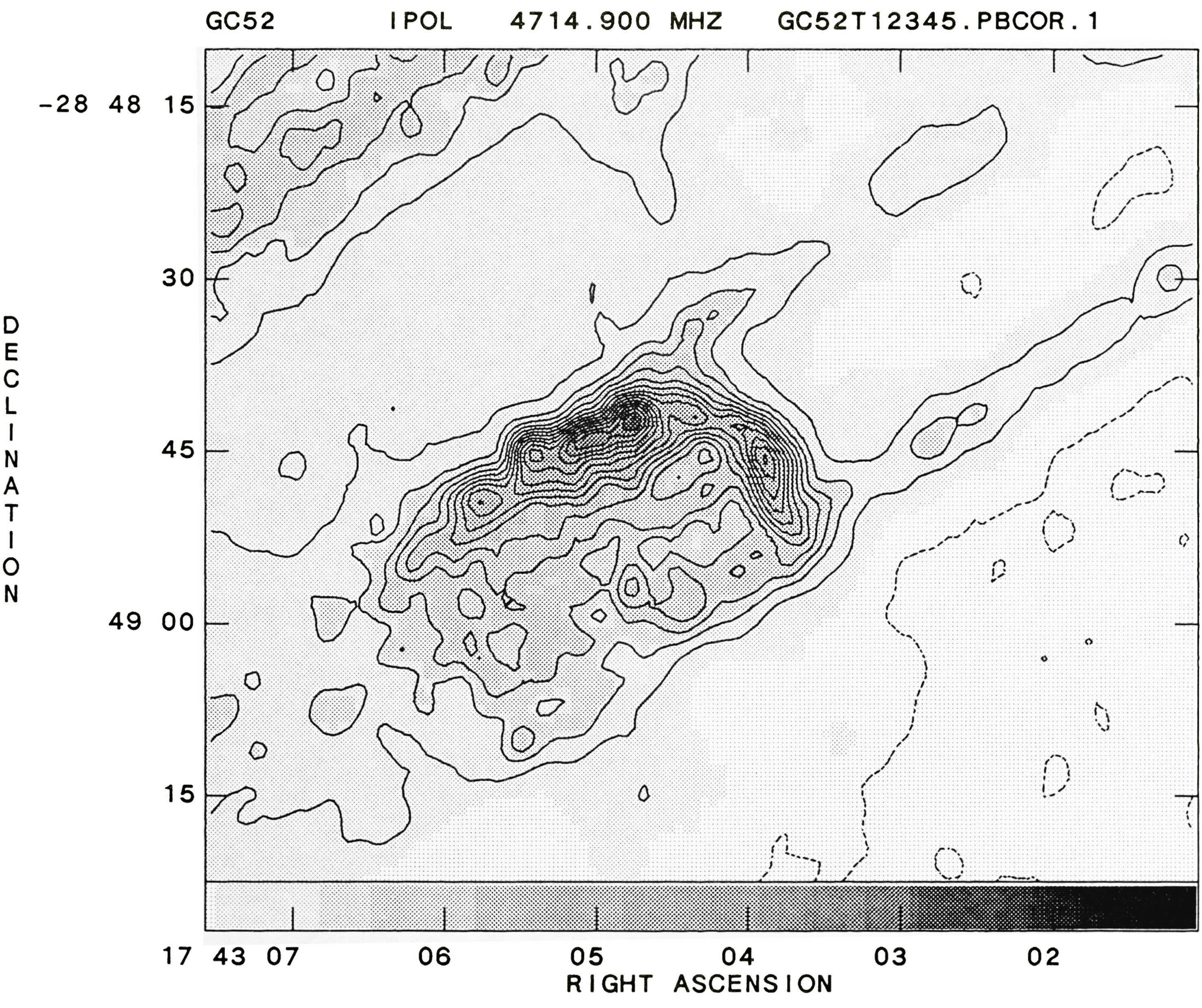

Figure 23j: FWHM = 2.58"×1.93" (P.A. = 16°), the contour intervals are -2, -1, 1, 2, 3, 4, 5, 6, ..., 18 mJy/beam area, λ = 20 cm.

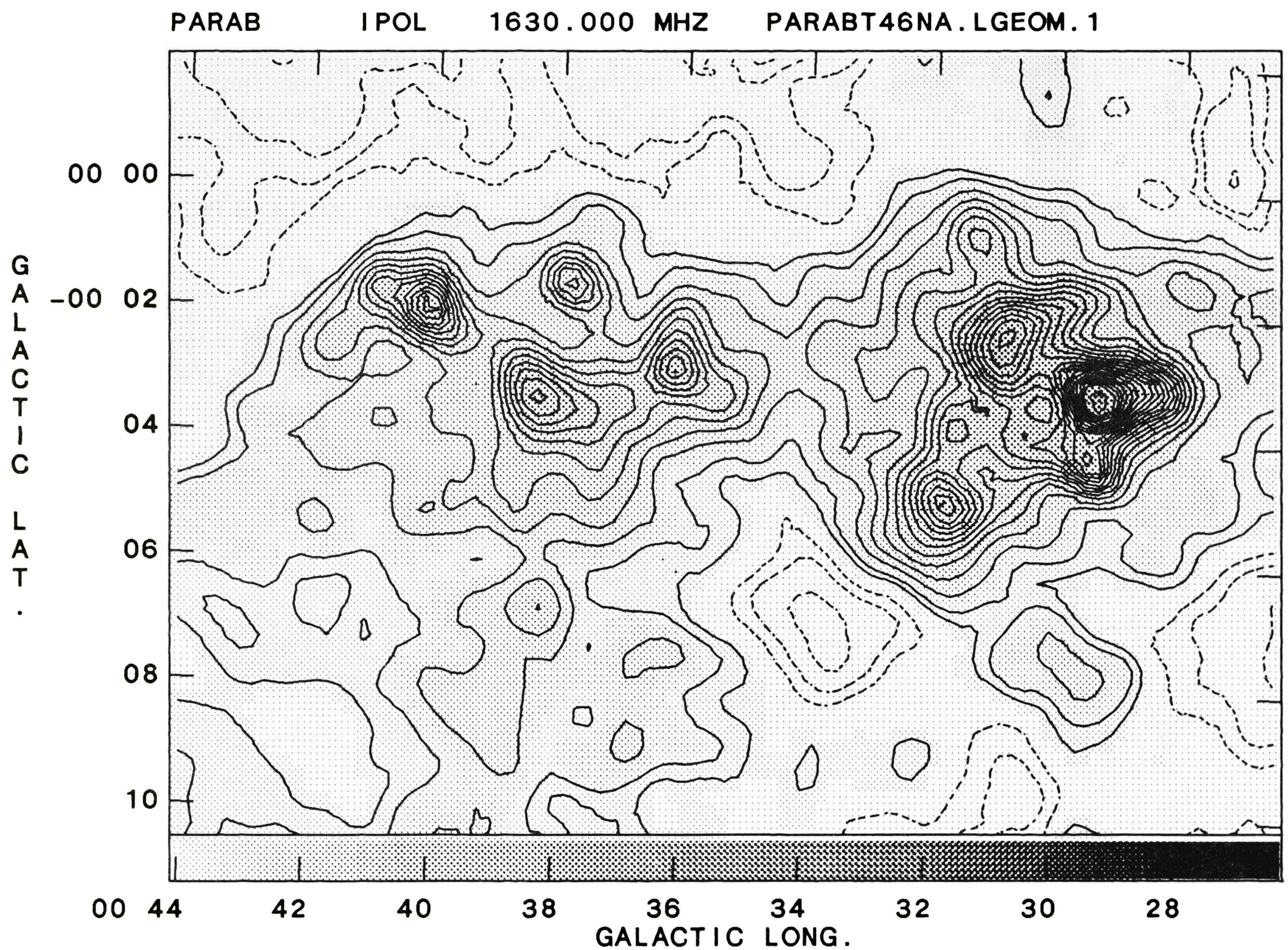

Figure 23k: FWHM = 40.3"×31.4" (P.A. = −18°), the contour intervals are 0.25×(−3, −2, −1, −0.5, 0.5, 1, 2, 3, ..., 20, 22, 24, 26, 28, 30) Jy/beam area. This map is not corrected for the response of the primary beam. The designated field is Arc No. 5 (see table 1 in chapter 2) and λ = 20 cm.

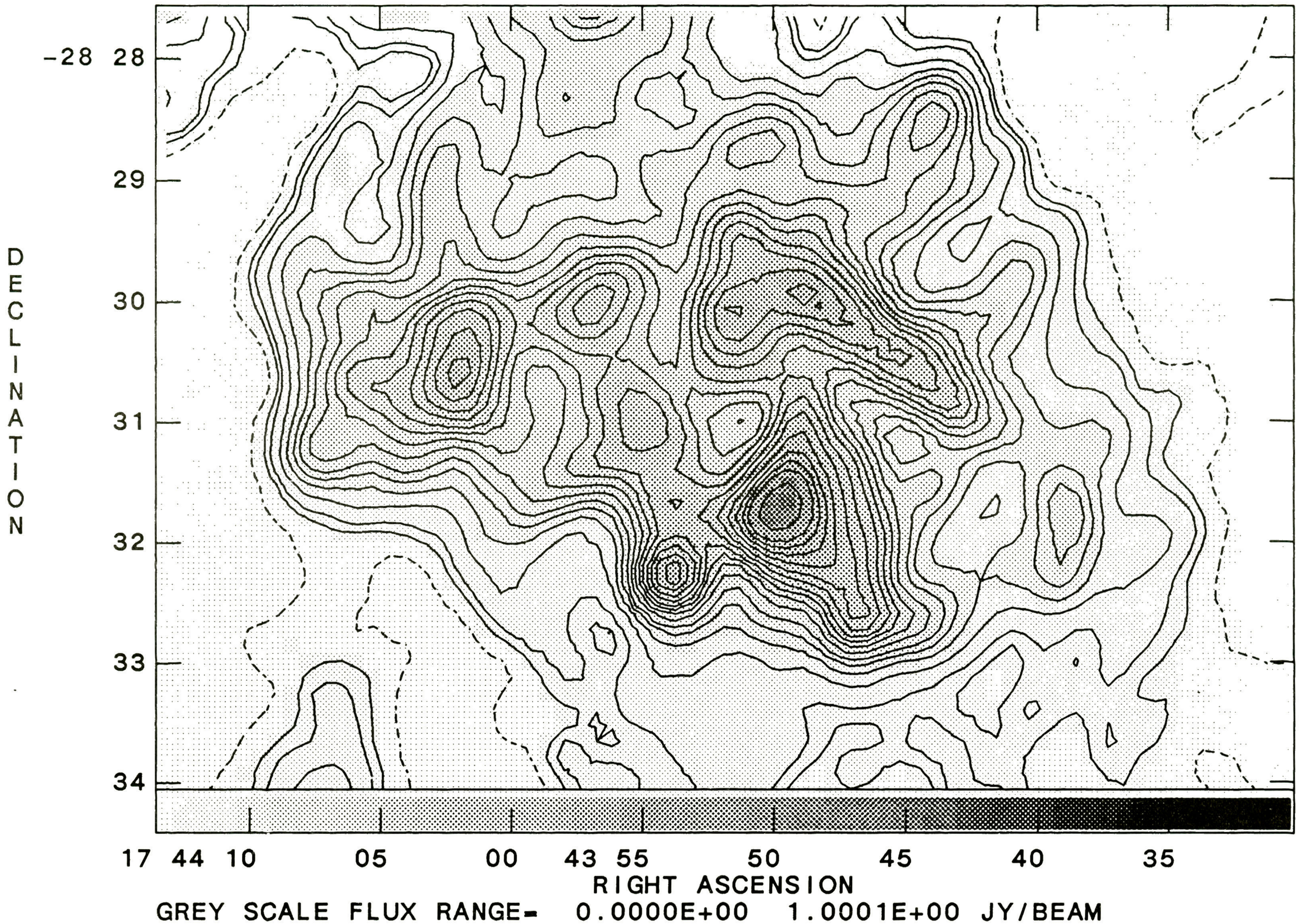

Figure 23 1: FWHM = 21.1" x 21.4" (P.A. = -29°). The contour intervals are -10, 10, 20, 40, 60, 80, 100, 130, 160, 190, 220, 260, 300, 350, 400, 450, 500, 600, 700, 900 mJy/beam area. The designated field is Arc No. 5, and λ = 20 cm.

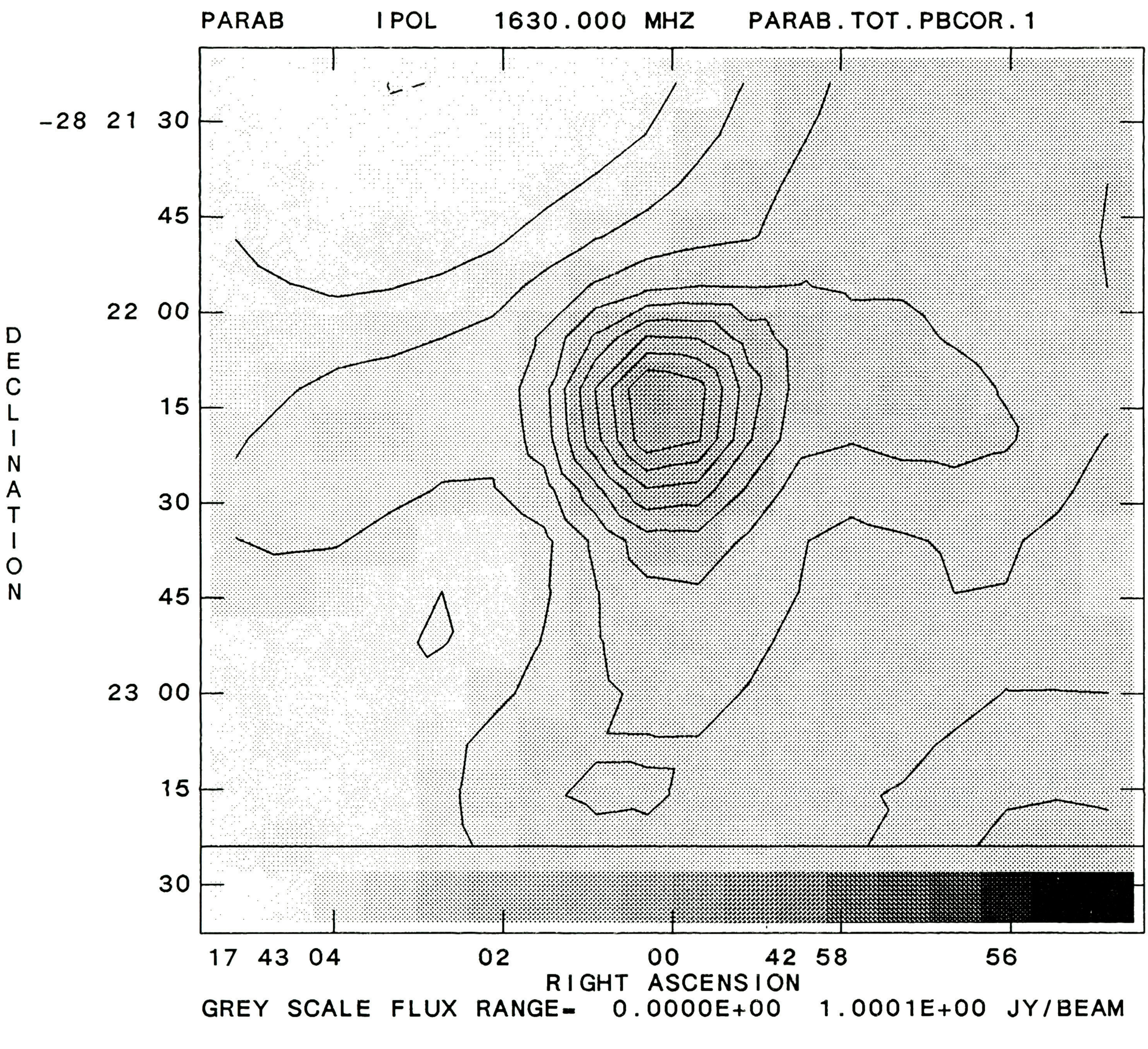

Figure 23m: FWHM = 21.1" × 21.4" (P.A. = −29°), the contour intervals are 100y (−0.5, 0.5, 1, 1.5, 2, 2.5, 3, 3.5, 4, 4.5, 6, 7, 8, 9) mJy/beam area.

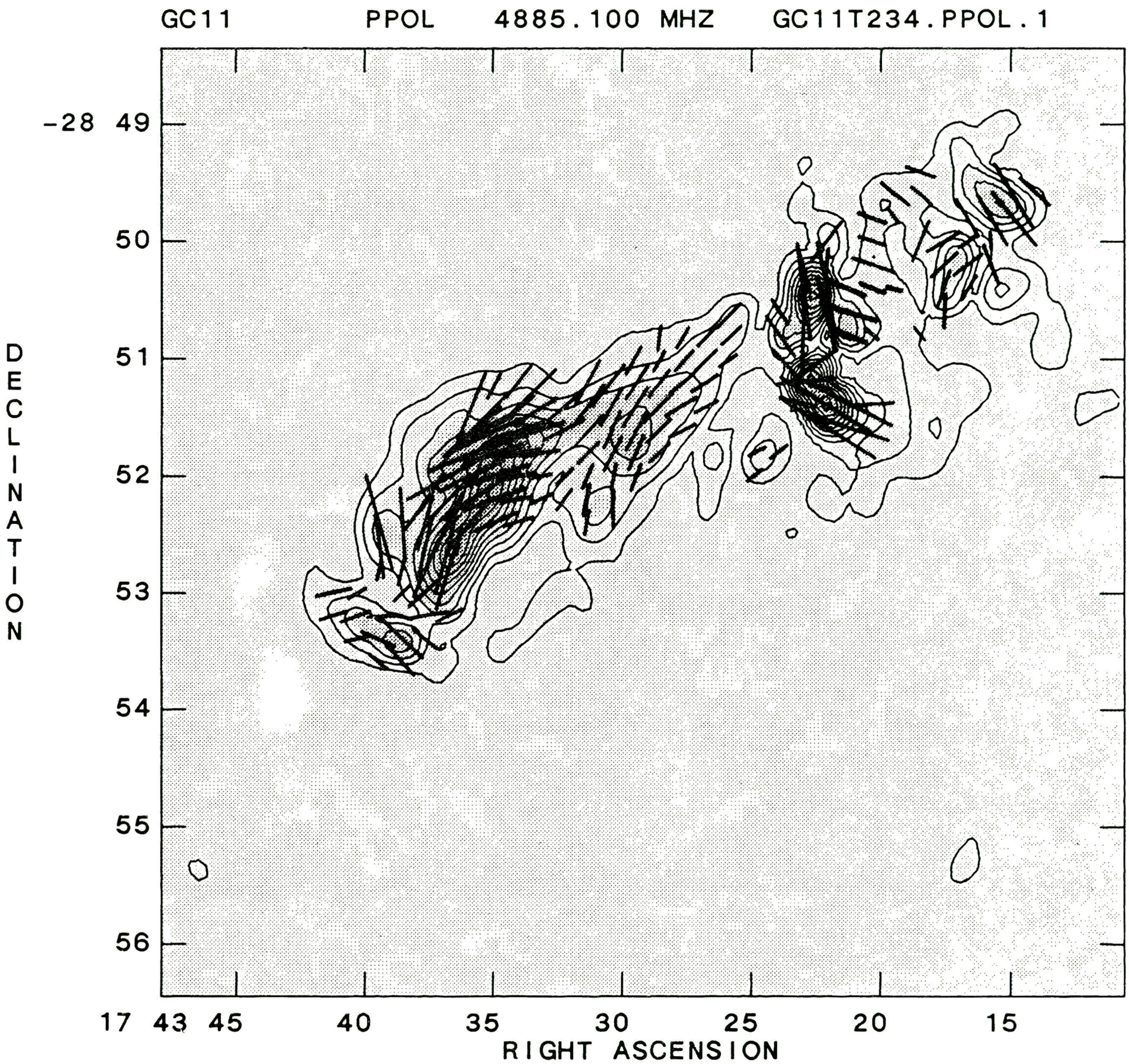

Figure 24: The data set corresponding to this figure is based on combining the AC side of the IF's ($\nu$ = 4.885 GHz) employing B/C[1] and C/D[2] arrays. The (u,v) data is tapered at 7k$\lambda$, the gaussian beam (FWHM) is 20.5"×17.5" (P.A. = -3.4°), and the polarized intensity contour intervals are 1, 2, 3, 4, 5, 6, 7, 8, 9 mJy/beam area. The length and orientations of the line segments represent the degree and direction of the electric field vectors. A line segment of 1' corresponds to 14% linear poalrization. The polarized intensity distribution is not corrected for the response of the primary beam.

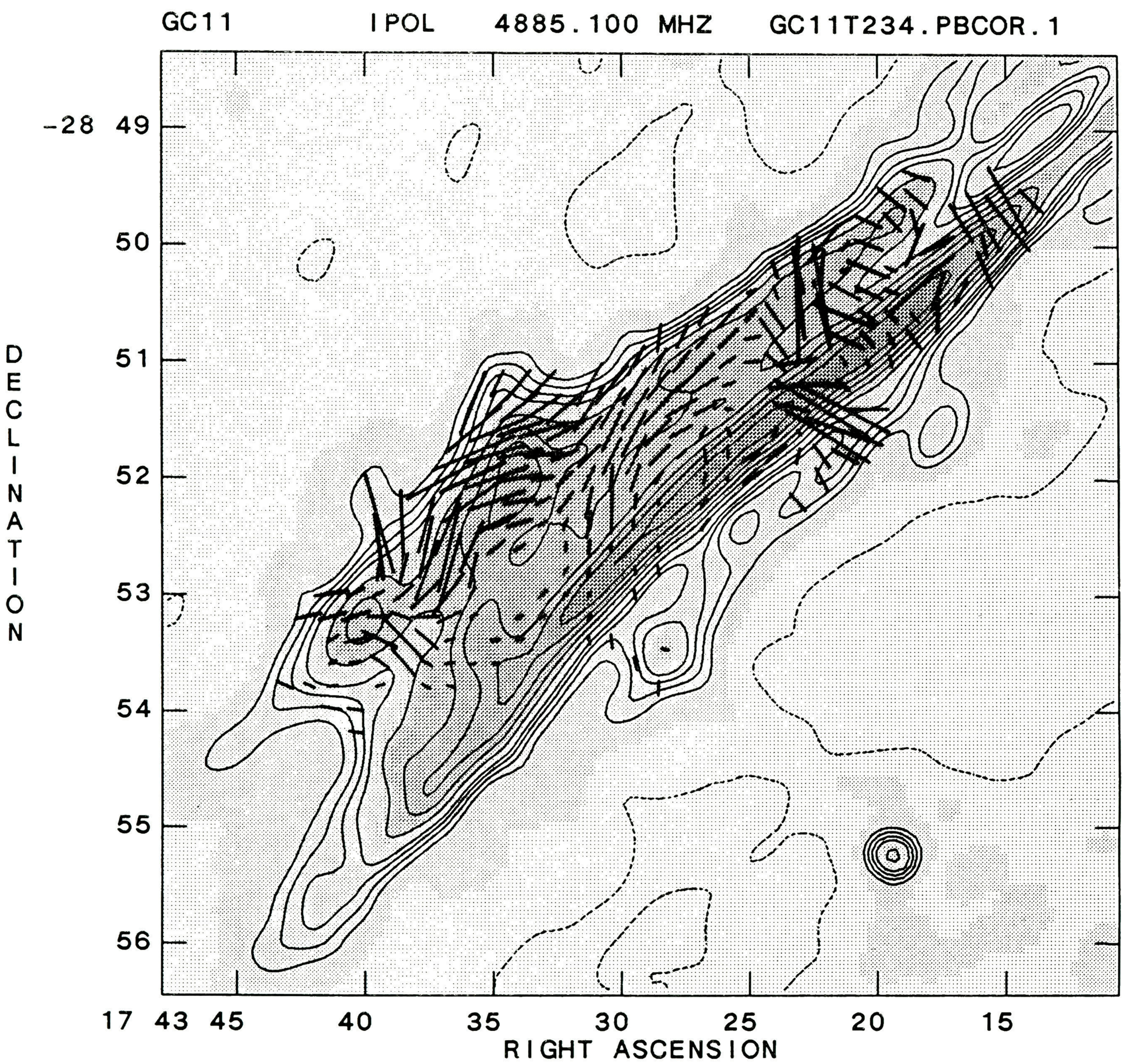

Figure 25: This figure is identical to figure 24 except that the contours correspond to total intensity. The contour intervals are -20, 20, 30, 40, 50 70, 90, 110, 130, 150, 170, 190 mJy/beam area.

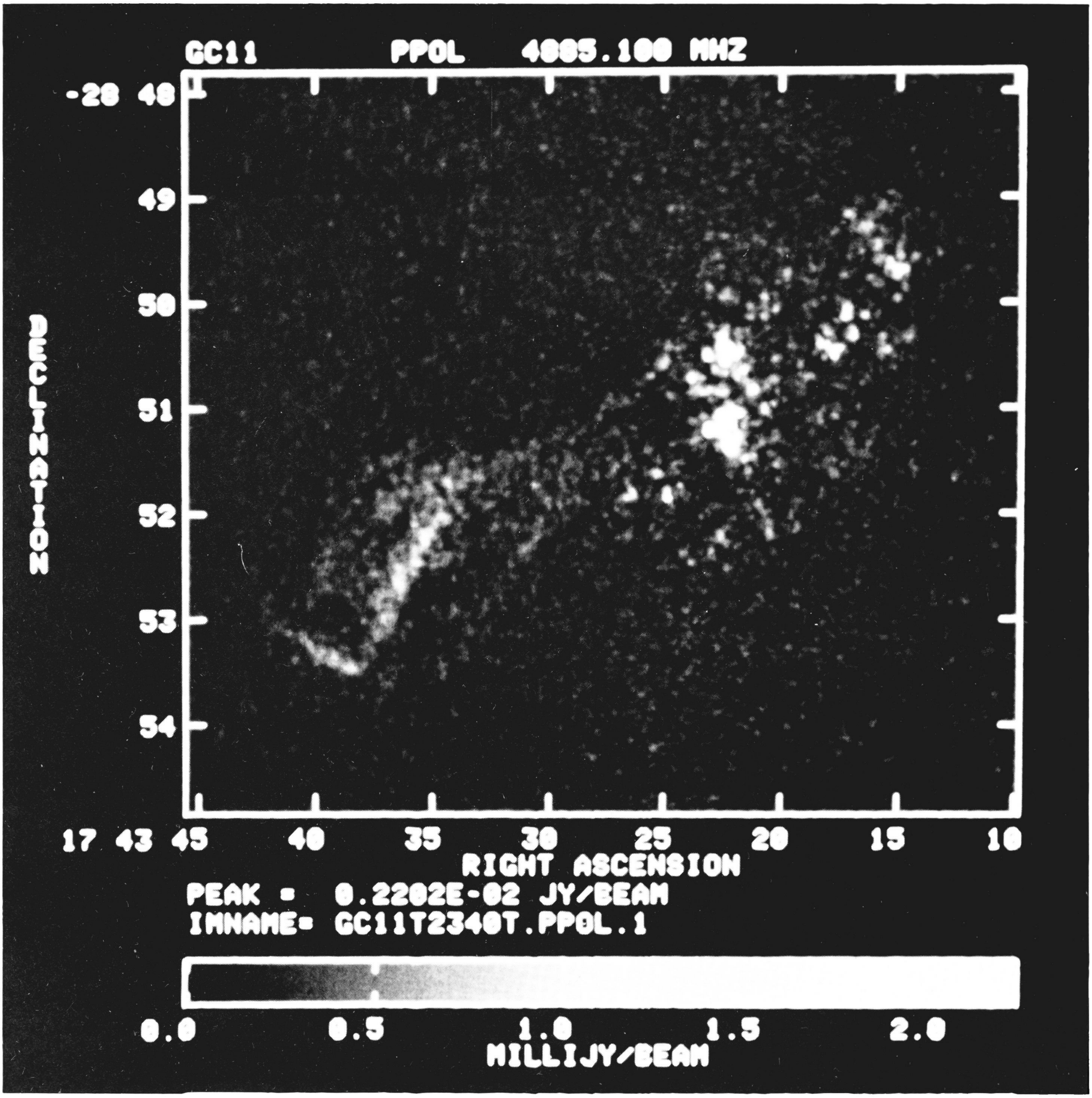

Figure 26: The radiographic presentation of the polarized intensity with the highest available resolution, FWHM = 4.3"×3.4", is shown in this figure which is based on the same data set as that of figure 24 except that the two sets of the AC and BD IF's are also combined.

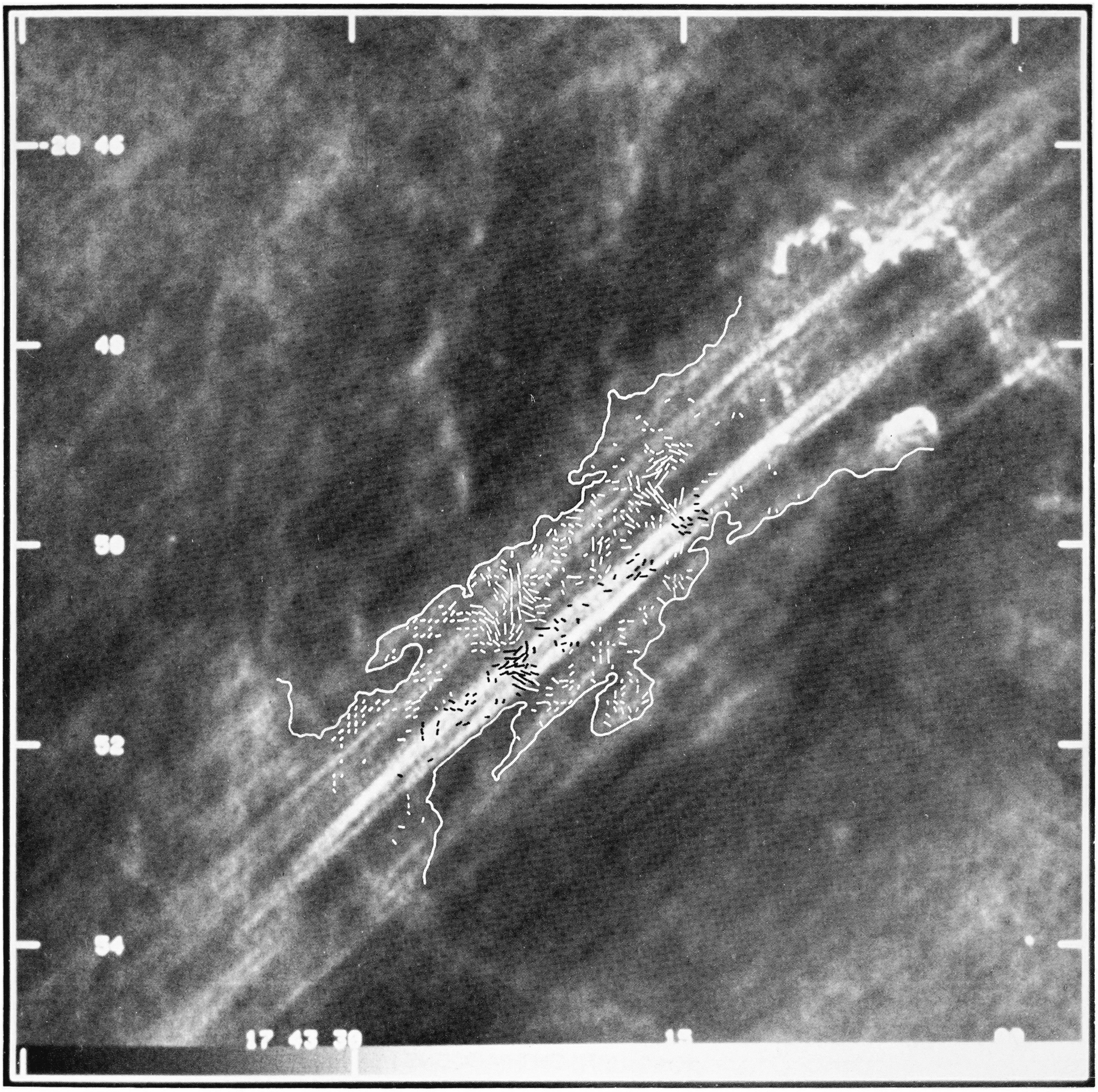

Figure 27: The radiograph shown in this figure is based on the designated field Arc No. 2 and has a resolution (FWHM) of 3"×2.6". A line segment with a length of 1' corresponds to % of linear polarization. The line segments correspond to the orientation of electric field vectors.

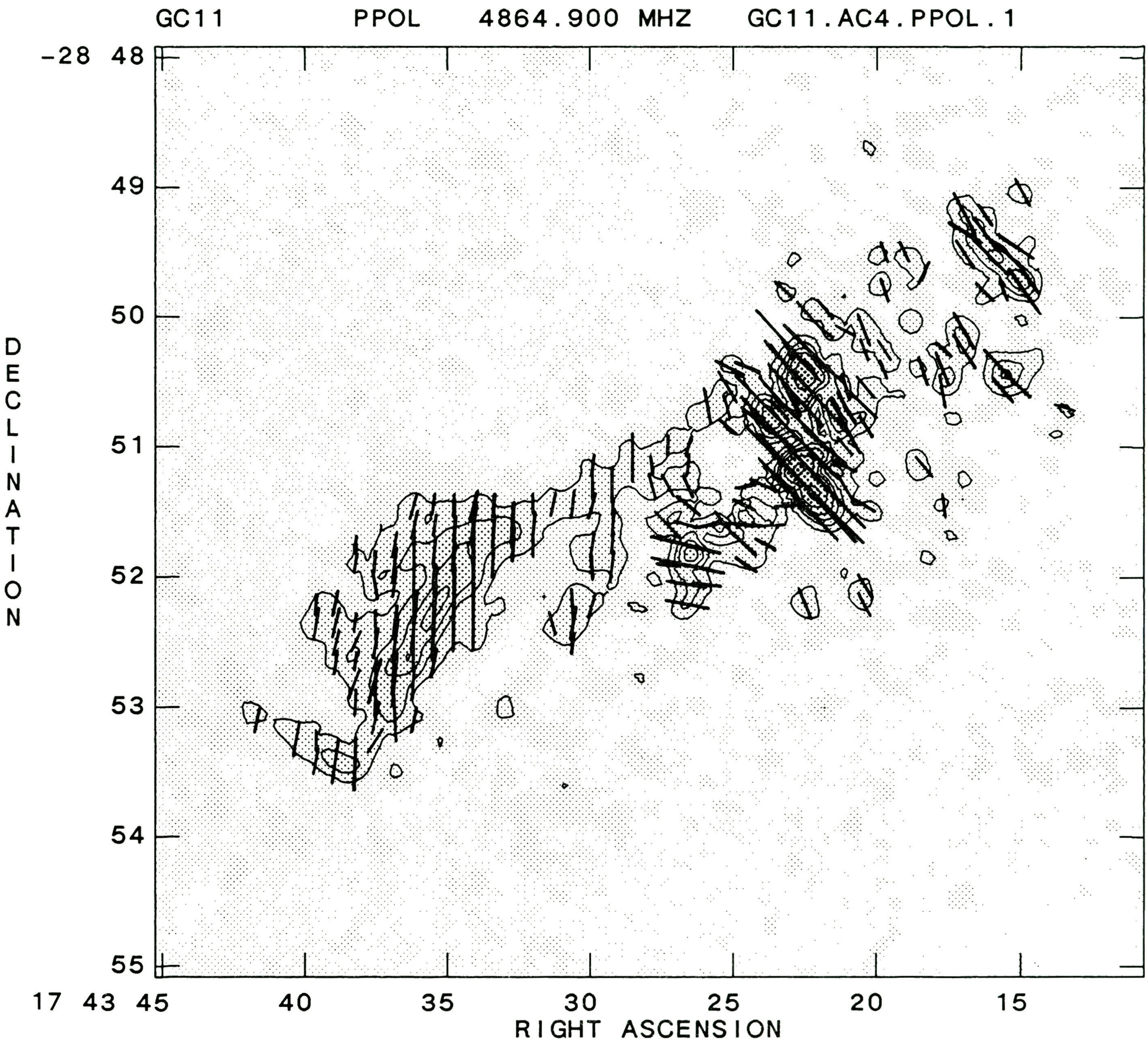

Figure 28: The Faraday rotations of the region to the southeast of the Arc is shown in this figure to be proportional to the angle which the line segments make with respect to vertical axes. The resolution of this map is, FWHM 11.3" x 10.8" (P.A. = -3°) and the polarized intensity contours are 1, 2, 3, 4, 6, 9, 13 mJy beam area. The map is based only on data corresponding to C/D$^2$ array. The two sets of the IF's are separated by 150 MHz.

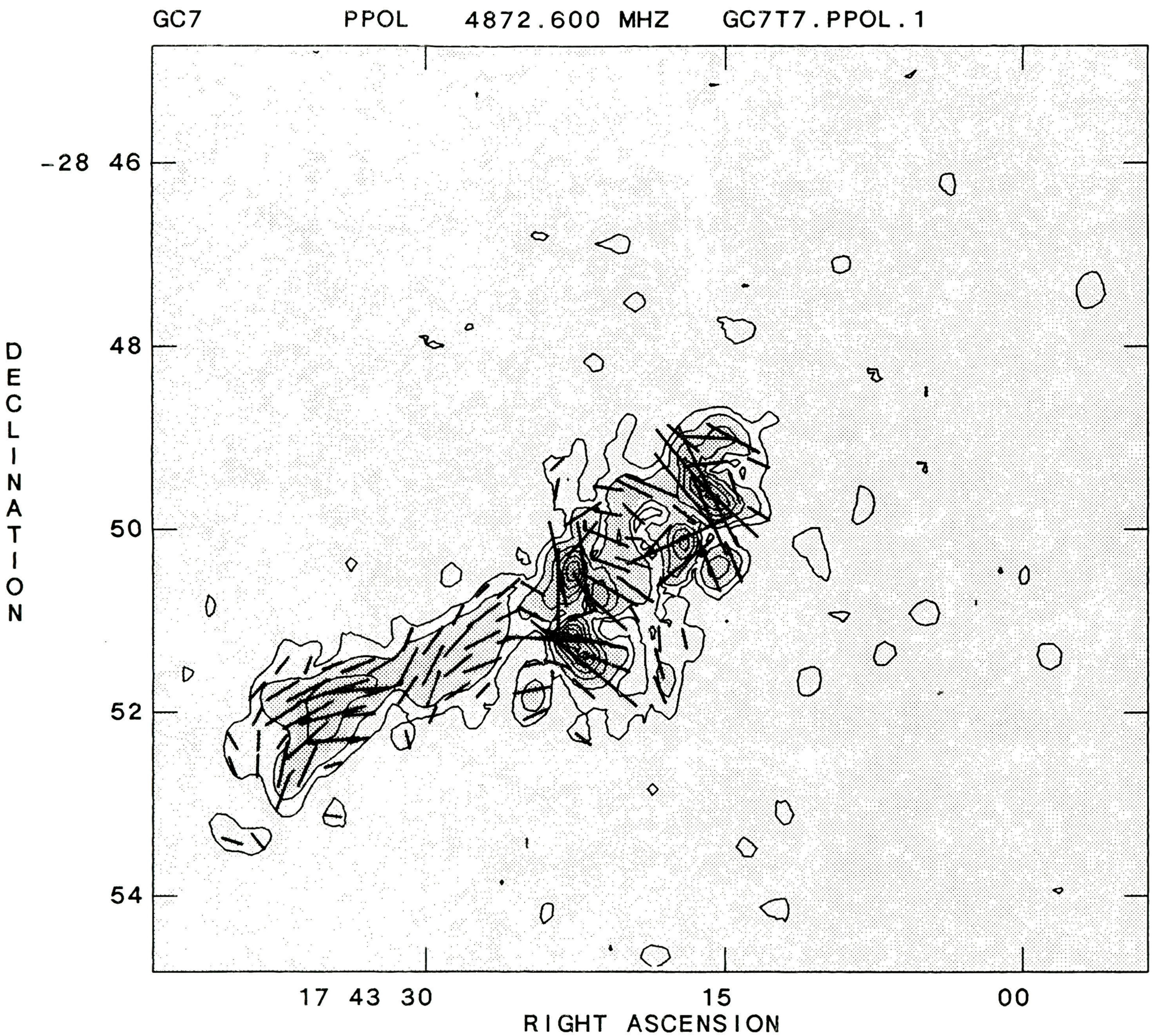

Figure 29-30: The polarization characteristics of the field situated adjacent to that of figures 24 and 25 are shown in figures 29 and 30 (i.e. designated field in Arc No. 2) with resolution of 10.6" x 16.9". The polarized and total intensity contours for figures 29 and 30 are 1, 2, 4, 6, 8, 10, 12, 14, 16 and -20, -10, 10, 20, 40, 60, 90, 120, 160, 200, 250, 300, 360 mJy/beam area, respectively. The length of the line segments are proportional to polarized intensity and their orientations correspond to the direction of electric field vectors.

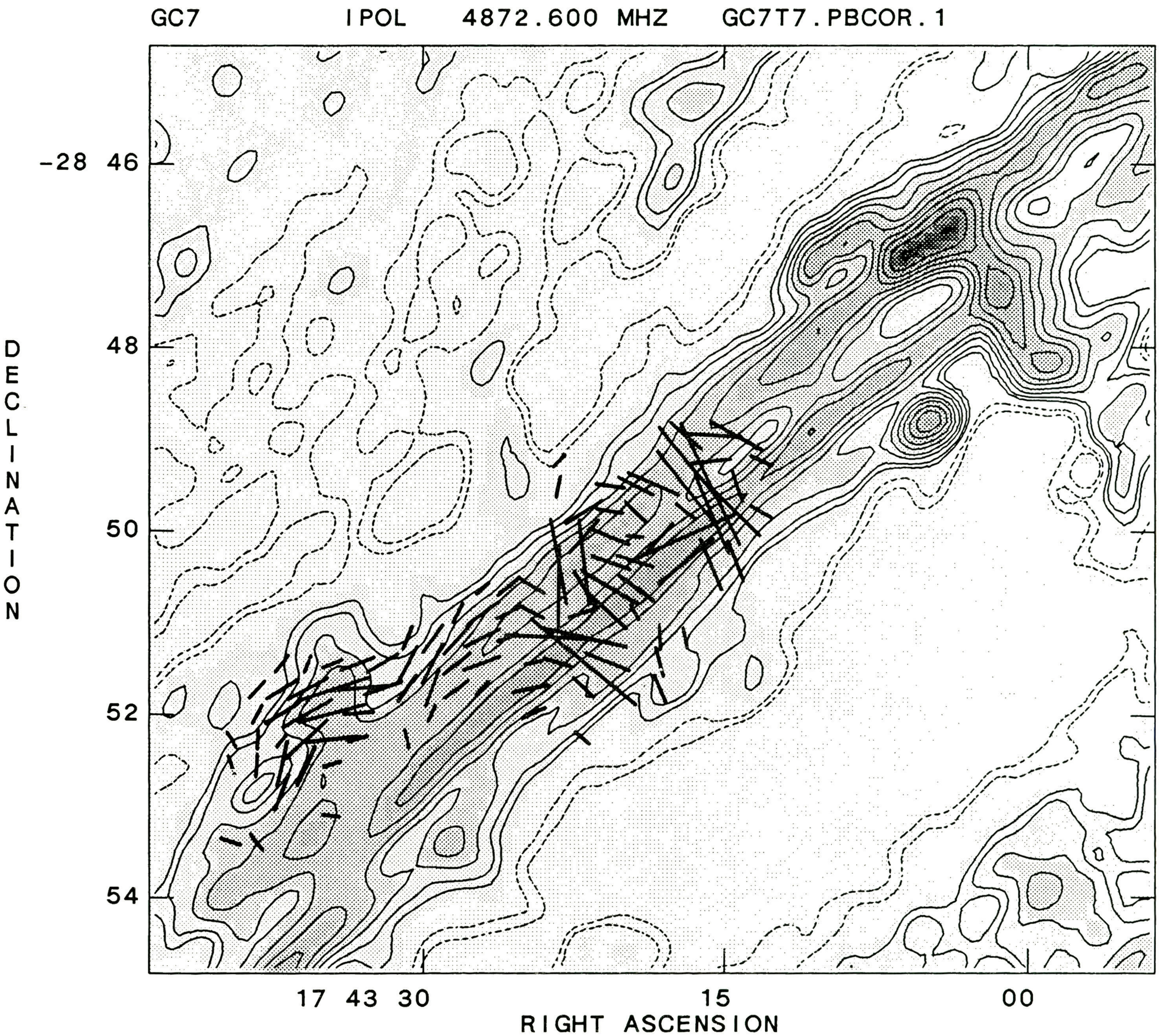
GC7 IPOL 4872.600 MHZ GC7T7.PBCOR.1
DECLINATION
-28 46
48
50
52
54
17 43 30
15
00
RIGHT ASCENSION

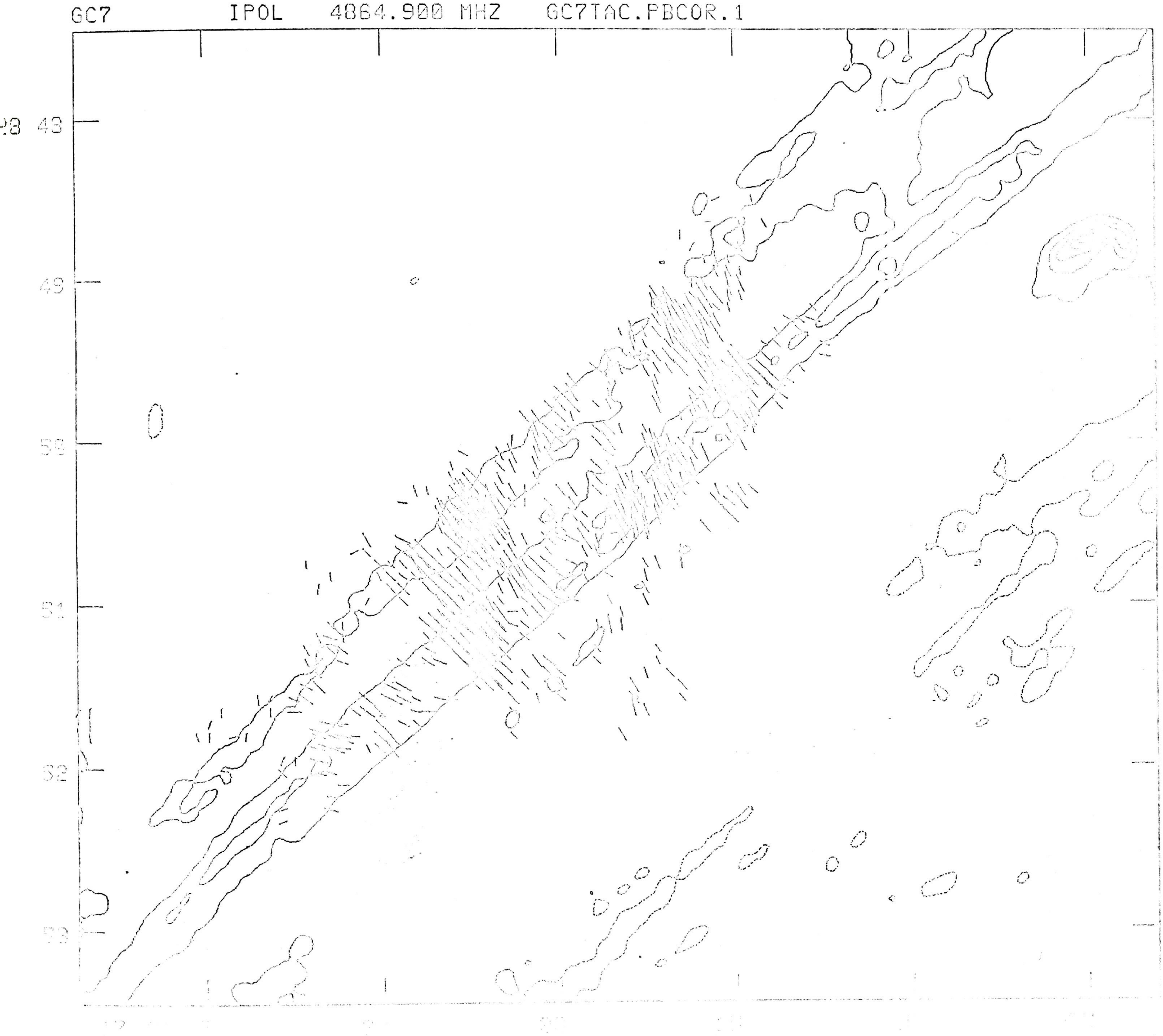

Figure 31: The Faraday rotation of the field corresponding to figures 29 and 30 can be seen in this figure with a resolution of 5.8"×5.6". The intensity contour intervals are -1, 1, 2, 3, 4, 5, 6 mJy/beam. The length of the line segments is proportional to the polarized intensity.

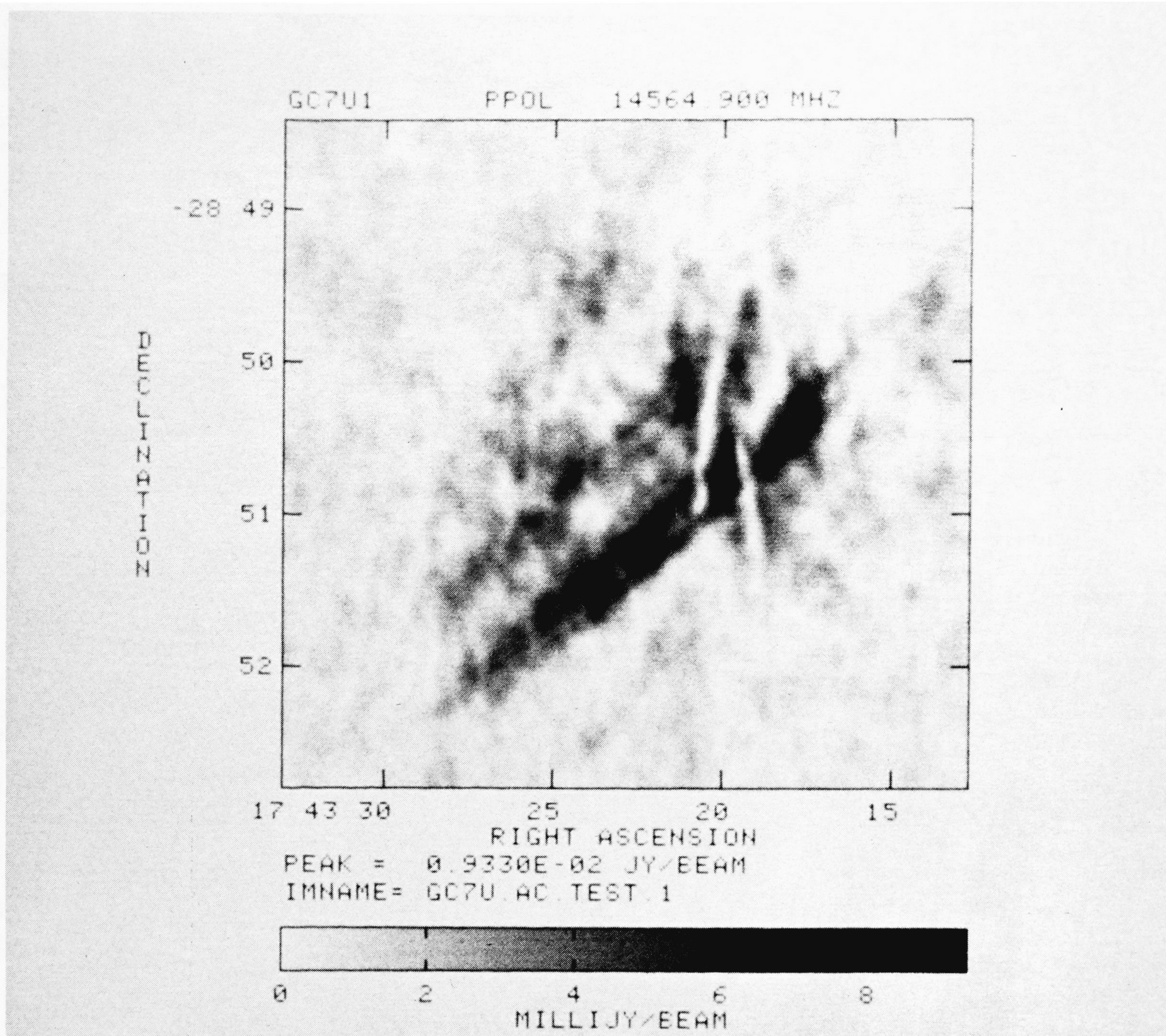

Figure 32a: This is a radiograph of the linearly polarized emission at 14.56 GHz. This figure which is based on the C/D array data base has a resolution of 5"×4.9" (P.A. = 63°) and the peak residual flux is 1.28 mJy/beam area. The total polarized flux density is 3.6 Jy.

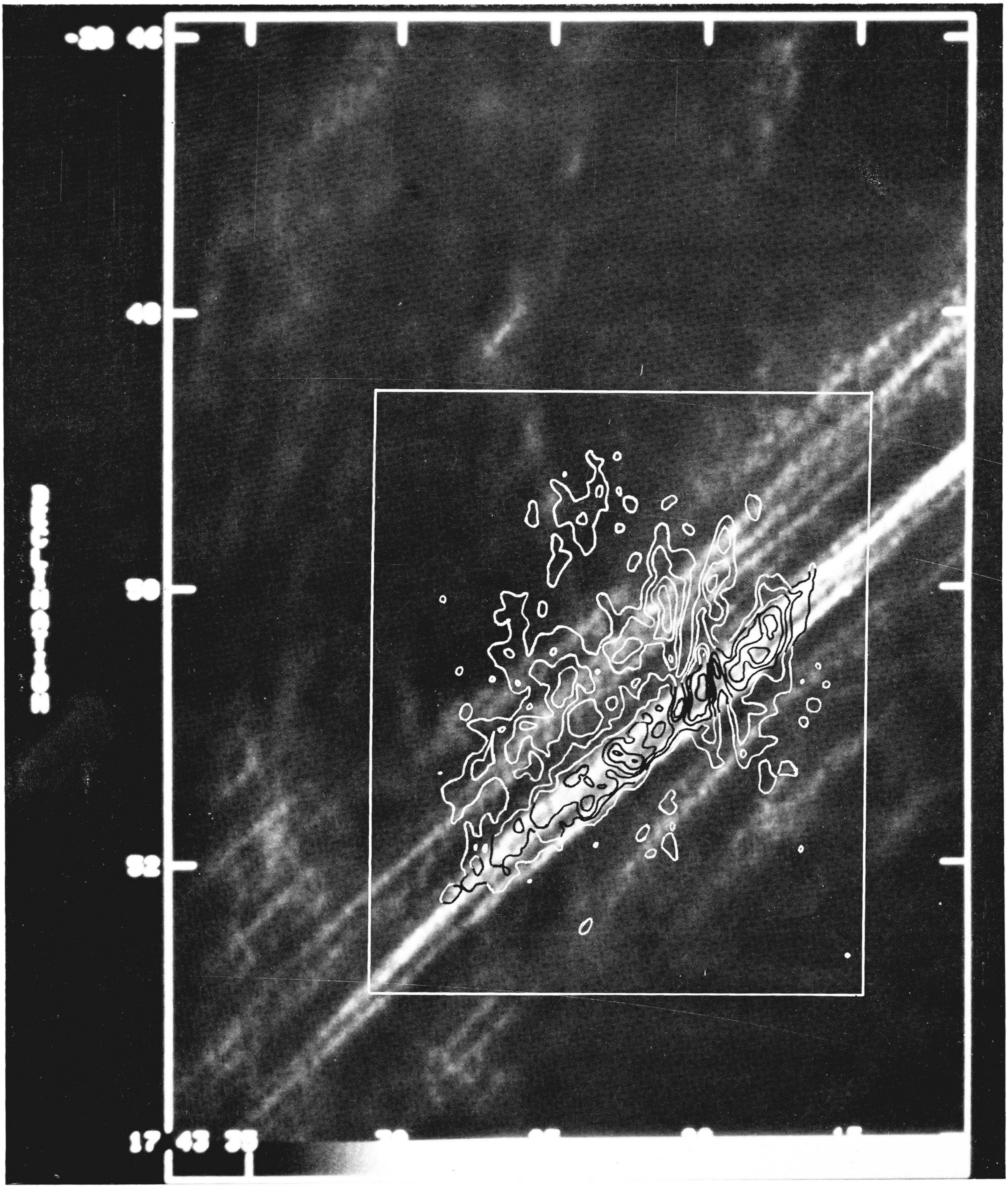

Figure 32b: The contour map of the region shown in figure 32a is superimposed on the 6-cm radiograph of the linear filaments which is identical to figure 7.

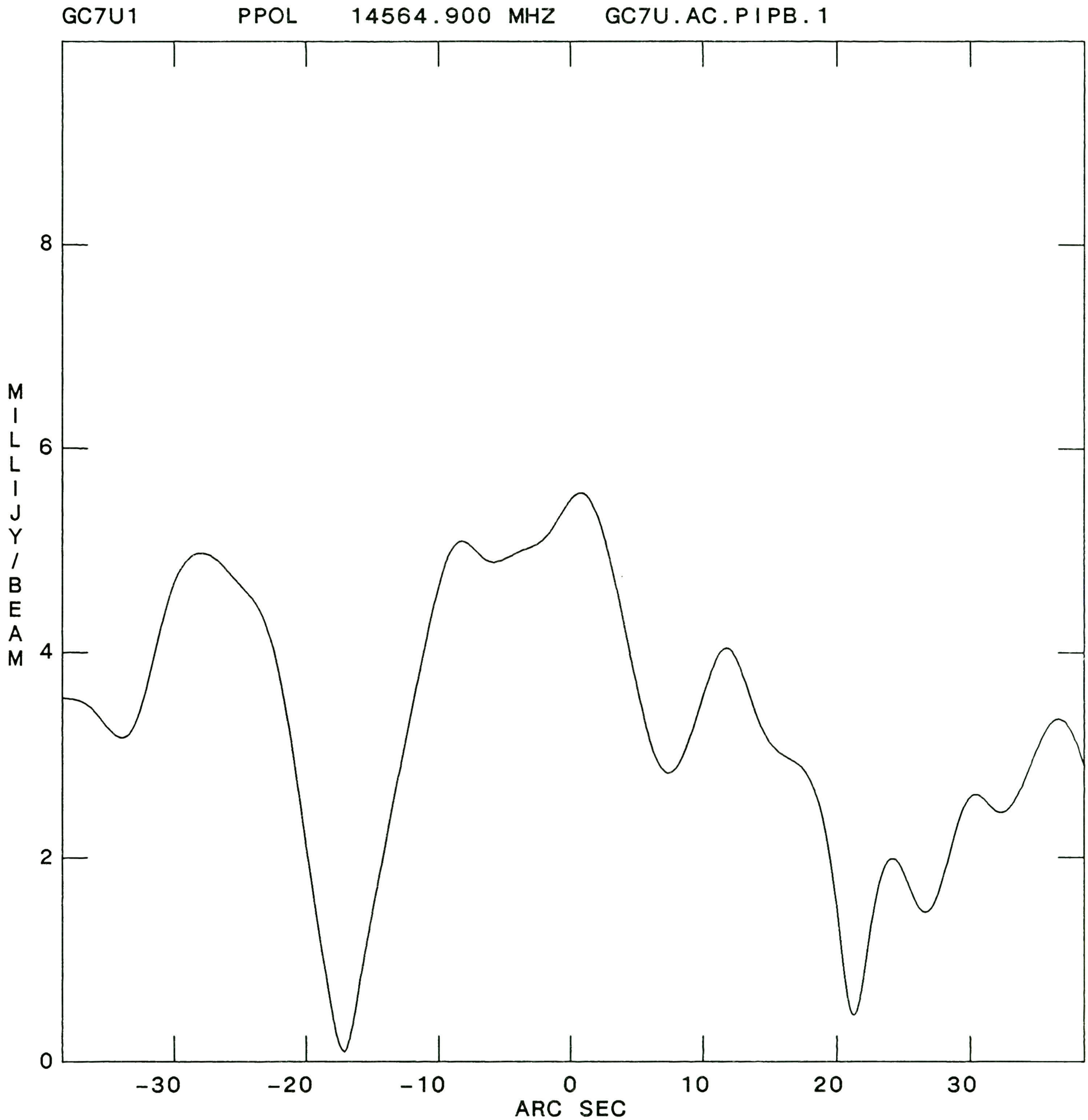

Figure 33: A slice is cut along the loop-like depolarizing feature seen in the polarized intensity radiograph. The center of the cut is at $\alpha = 17^h43^m19.5^s$, $\delta = -28°50'10''$ and its position angle $-44.5°$.

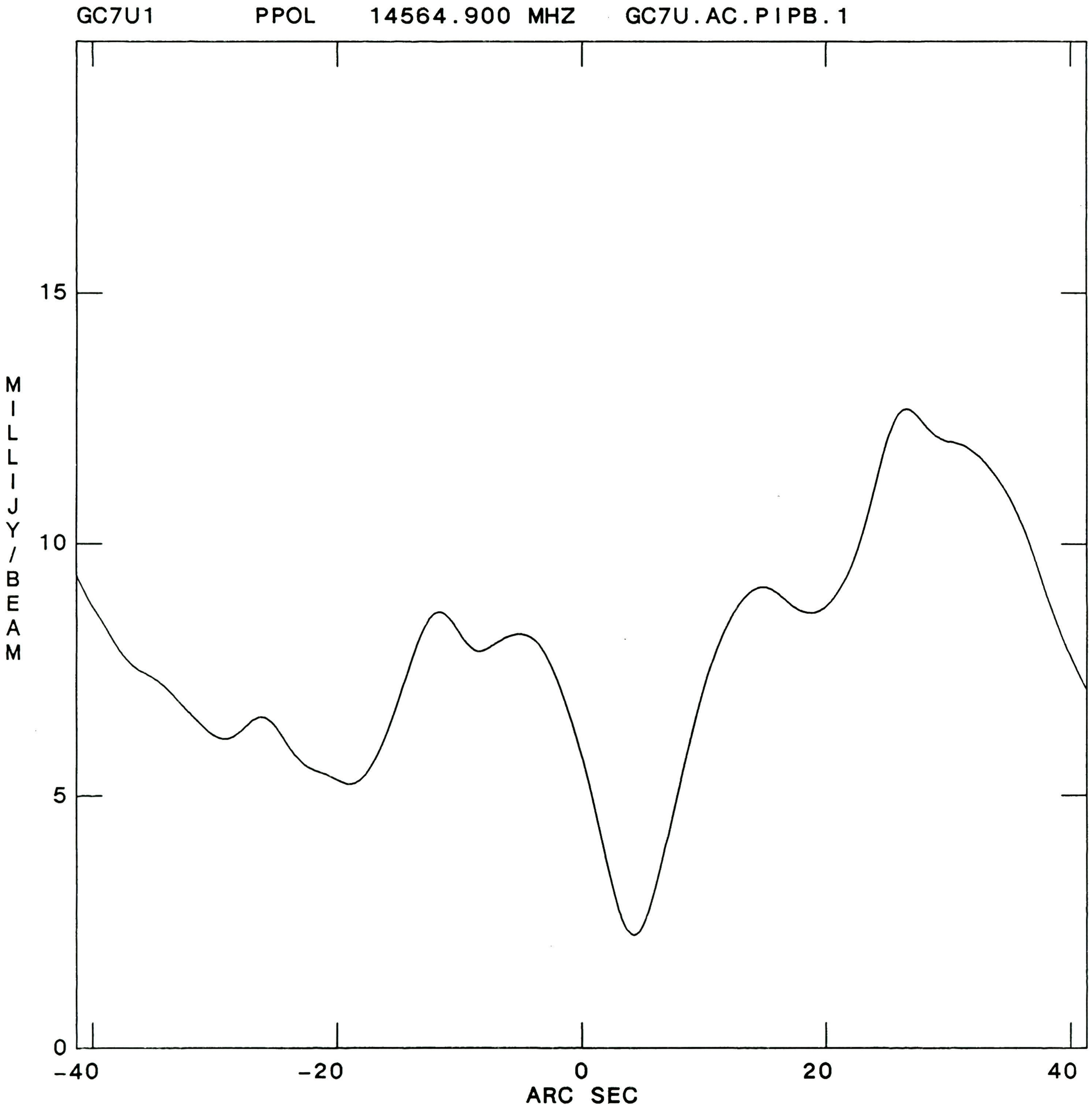

Figure 34: A slice is cut along the elongated depolarizing feature. The center of the cut is at $\alpha = 17^h43^m19.4^s$, $\delta = -28°50'51''$ and its position angle is $-51.9°$.

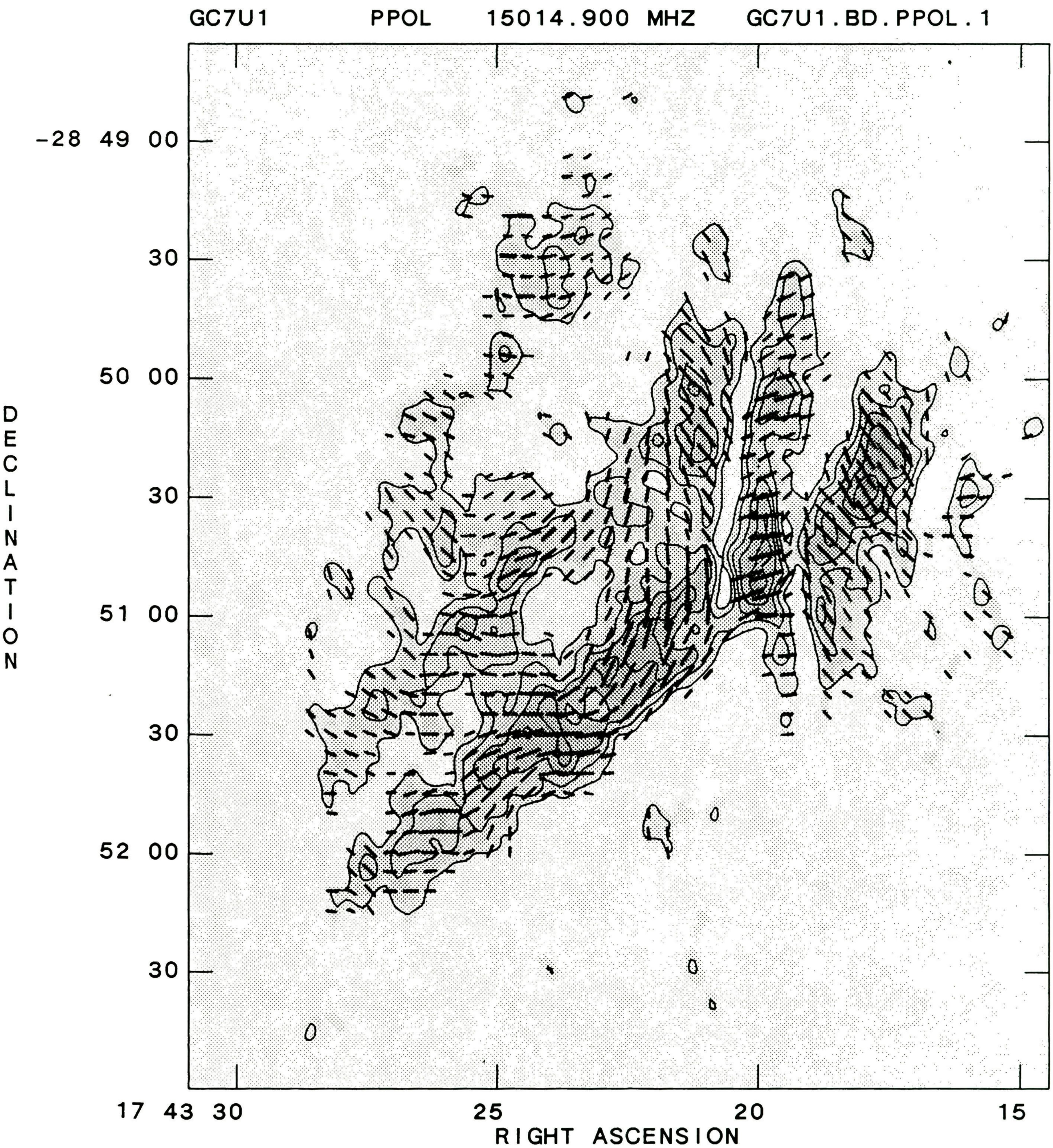

Figure 35: The polarized intensity contour map at 15.015 GHz. The length of the electric vectors is proportional to polarized intensity. This map is not corrected for the response of the primary beam. The contour intervals are 2.5, 4, 4.5, 7, ..., 14.5 mJy/beam area.

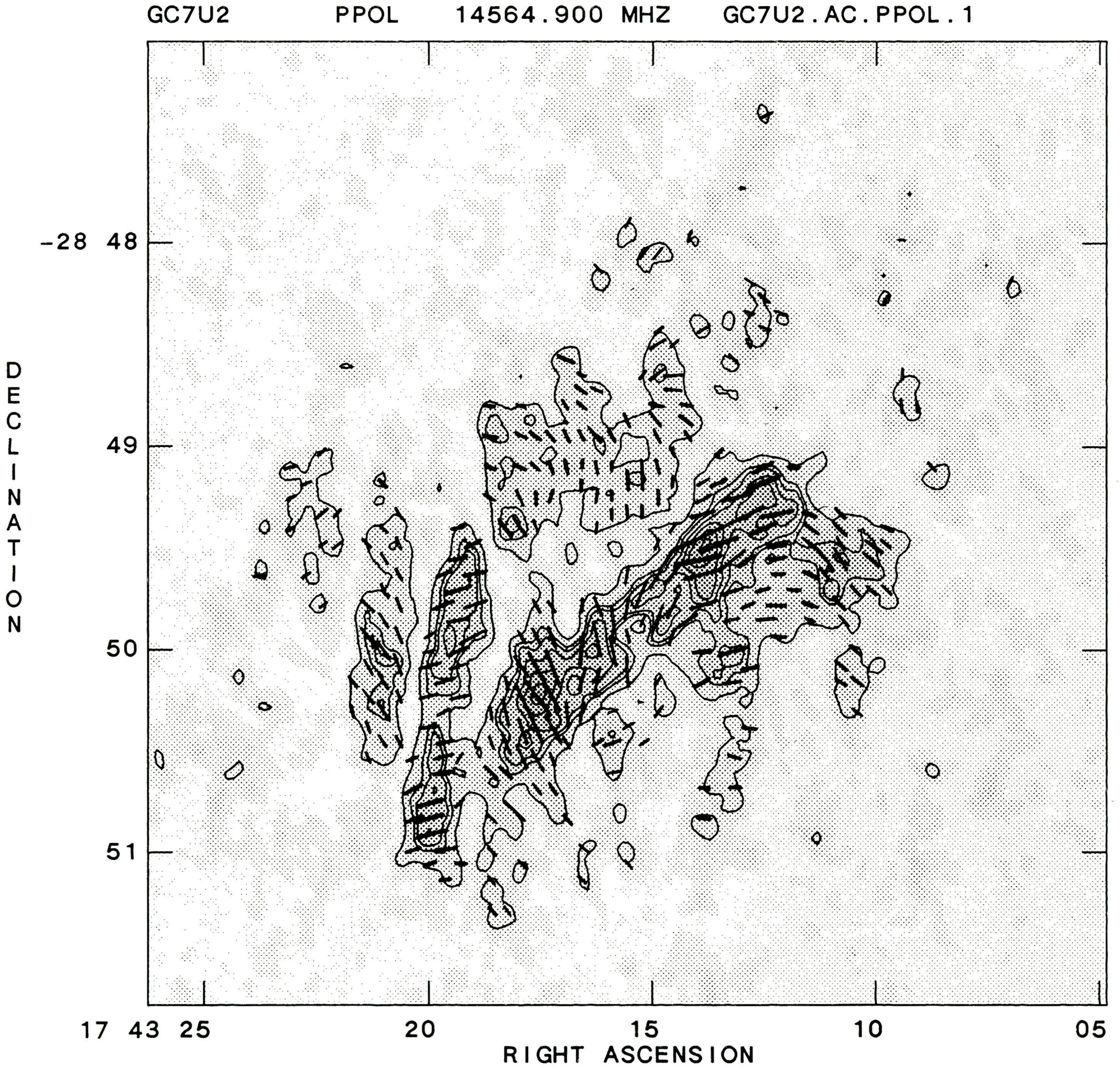

Figure 36: The polarized intensity contour map of the field adjacent to that of figure 35 at 1.456 GHz. The length of the electric vectors is proportional to polarized intensity. The contour intervals are 2.5, 4.5, 5.5, 7, 8.5, 10, 11.5, 13, 14.5 mJy/beam area. The gaussian beam has a FWHM = 5.1" x 4.7" (P.A. = 70°).

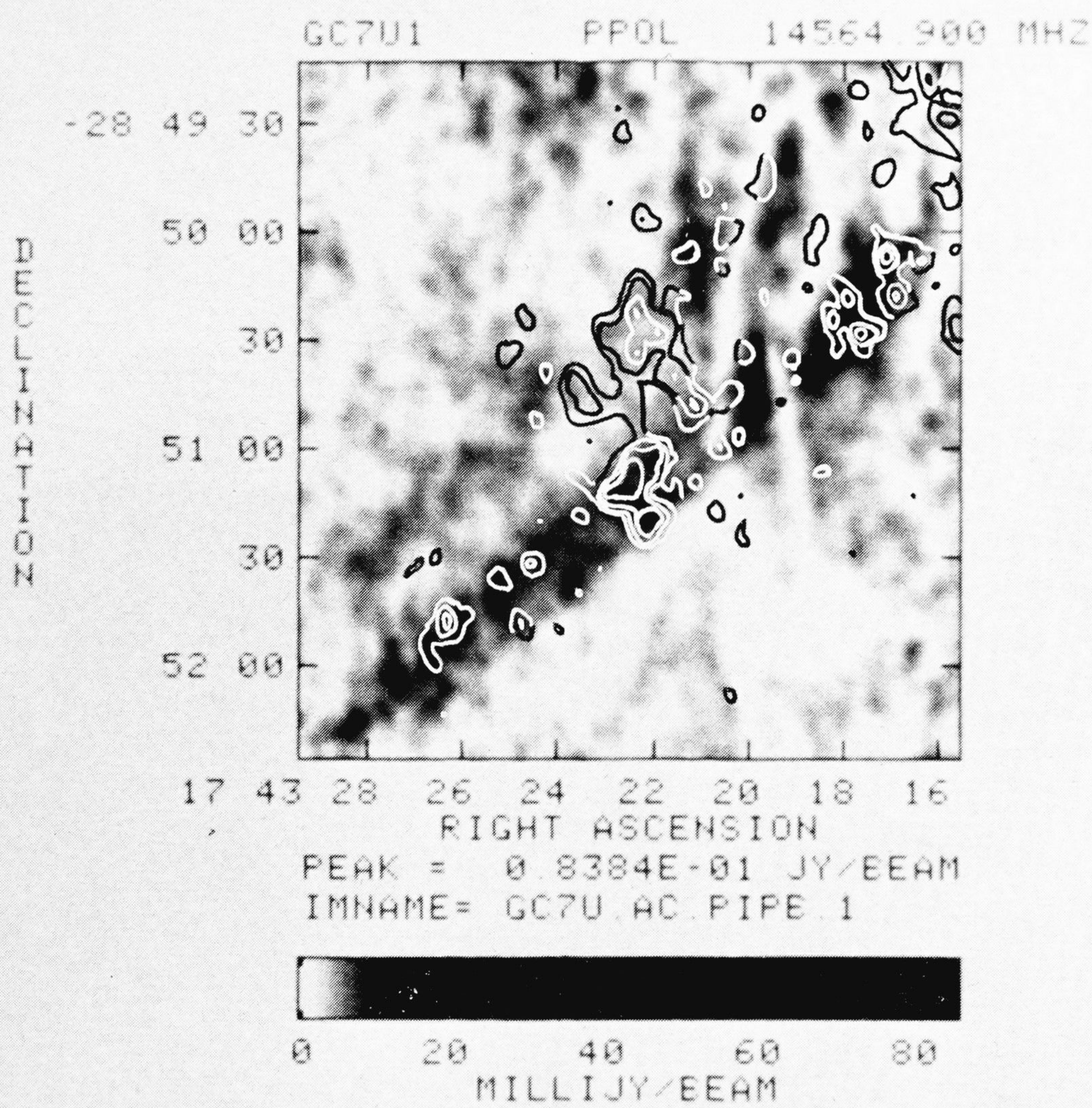

Figure 37: The 6-cm polarized intensity is superimposed on the radiograph of the polarized intensity at 2 cm which is identical to figure 32. The resolution of the 6-cm map is 2.5" x 3".

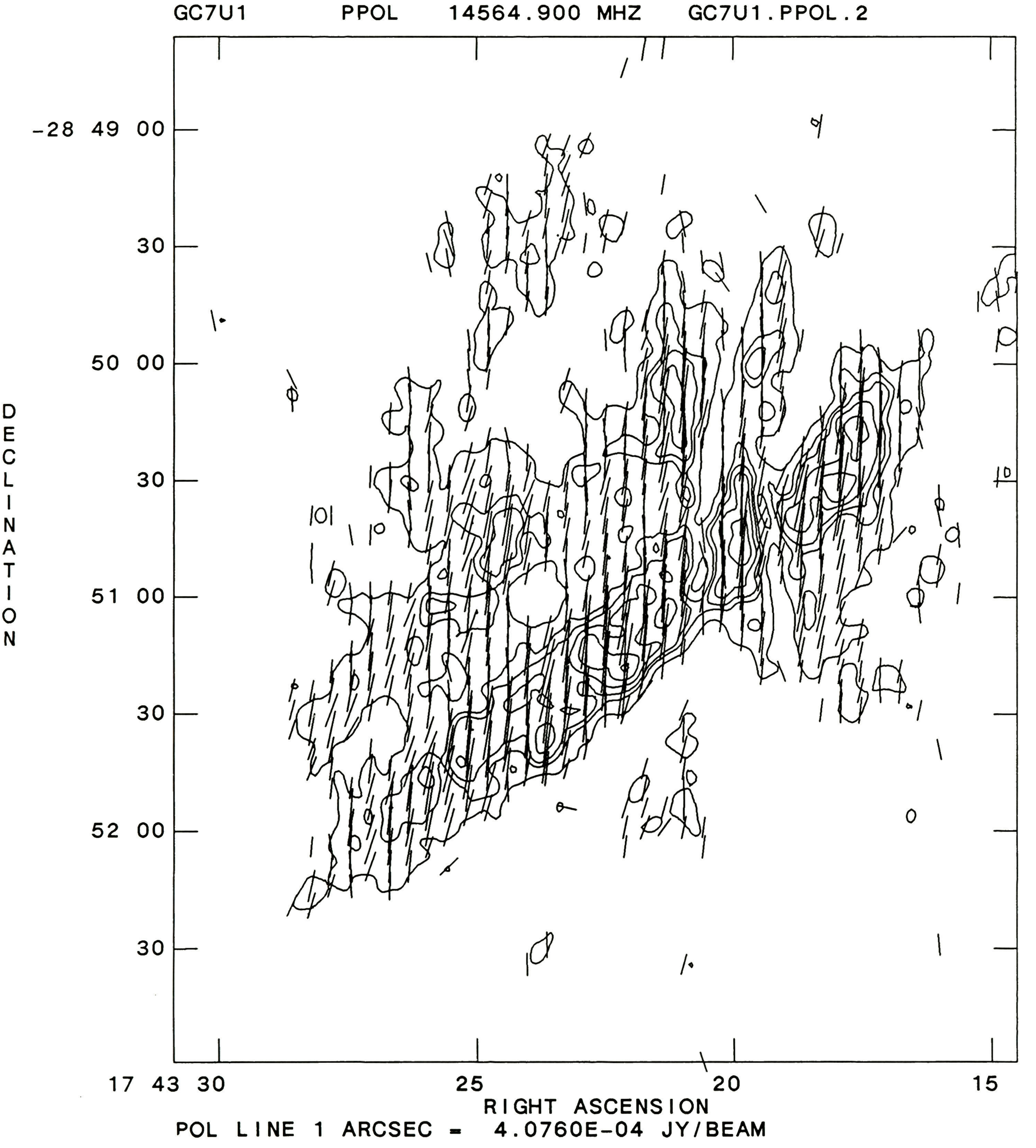

Figure 38: The 2-cm polarized intensity map with contour intervals 2.5, 4.5, 5.5, 7, 8.5, 10, 11.5, 13, 14.5 mJy/beam area. The differential Faraday rotation angles between 1.45649 and 1.50149 GHz are superimposed on this map. The length of the electric vectors is proportional to the total polarized intensity.

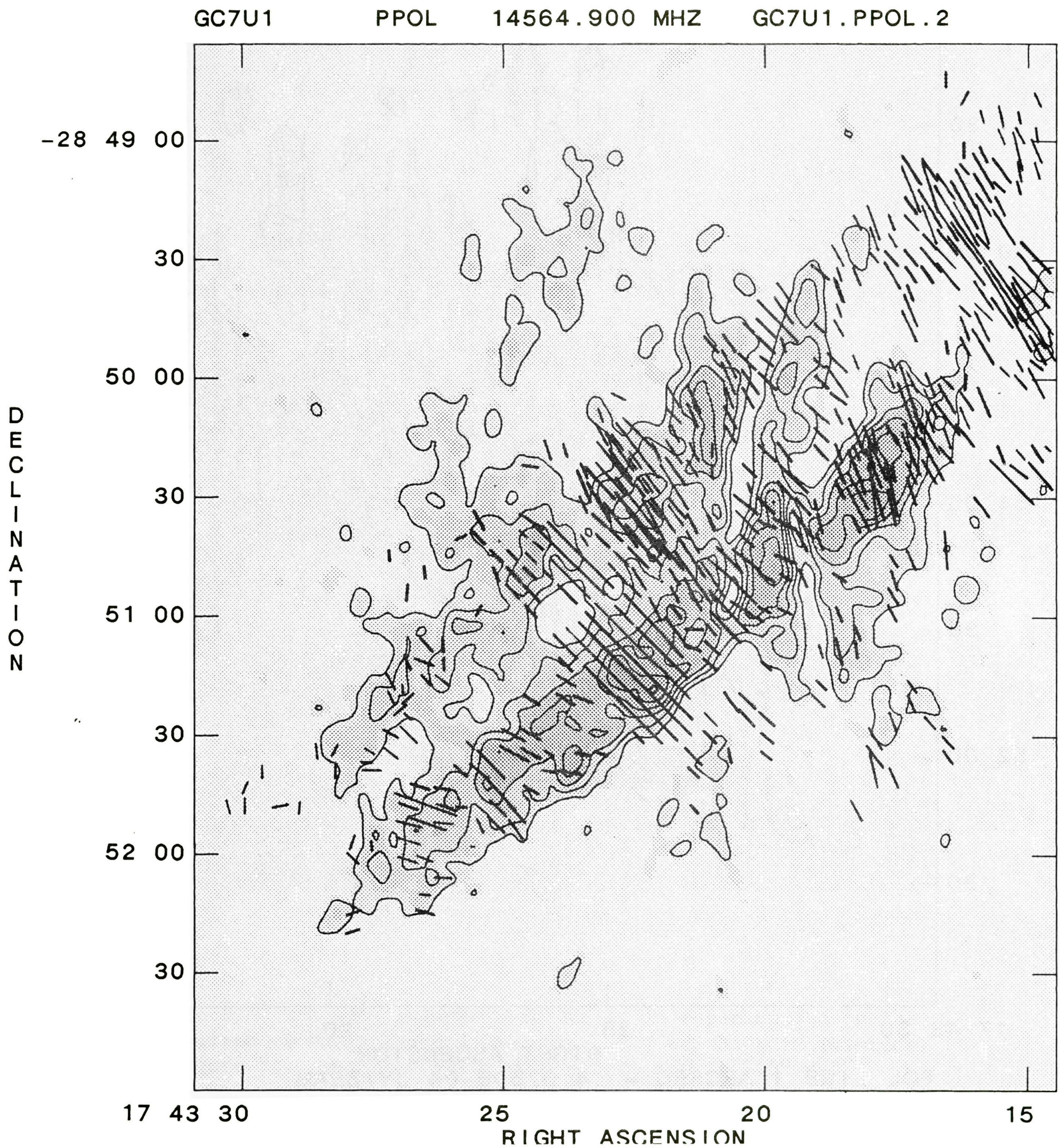

Figure 39: The Faraday rotation shown in figure 31 at 6 cm is superimposed on the 2 cm polarized intensity. The contour intervals are similar to those of figure 35.

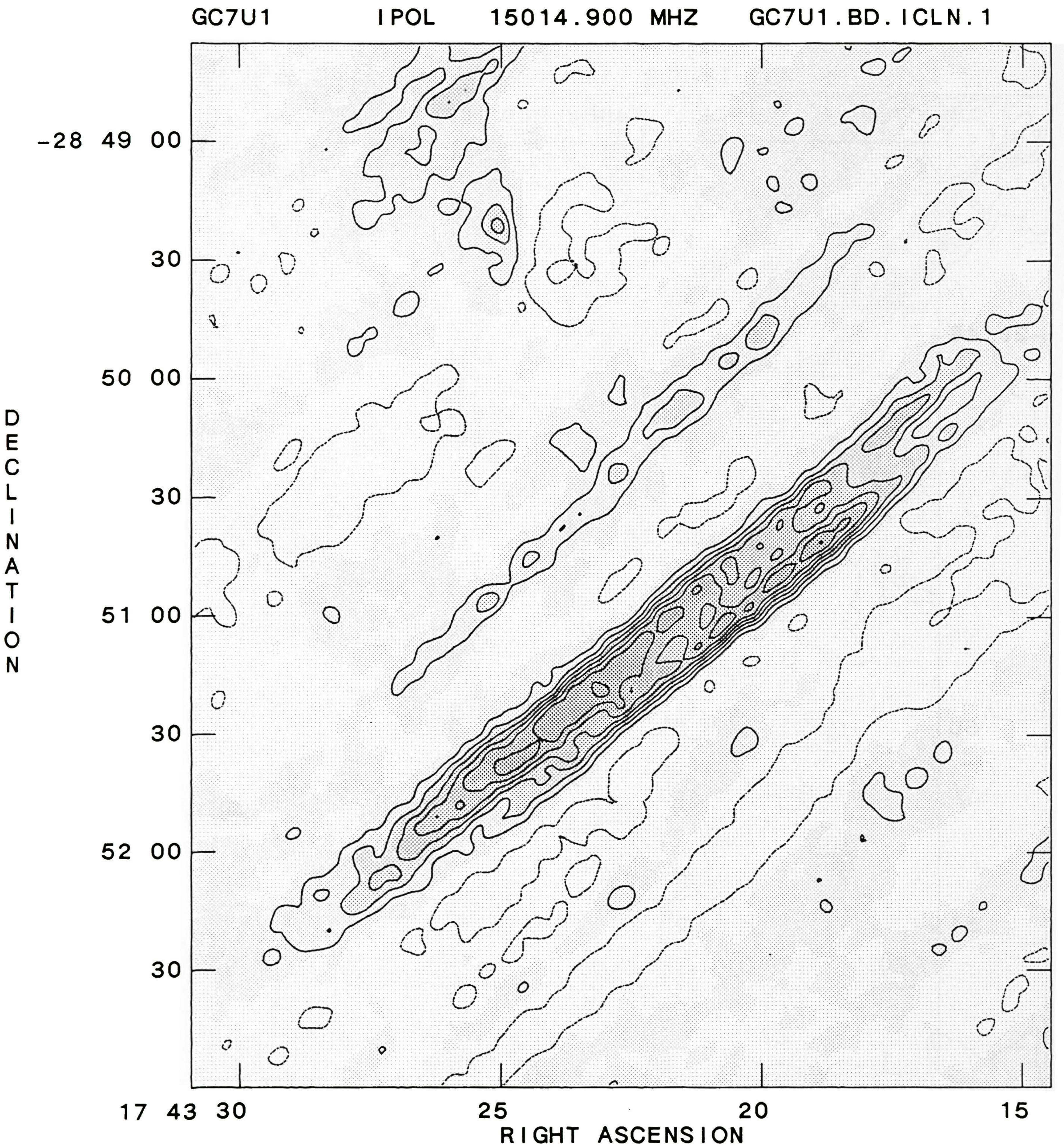

Figure 40-41: The total intensity contour maps of the region shown in figures 35 and 36, respectively. The contour intervals for both figures are -1, 1, 2, ..., 10 mJy/beam area.

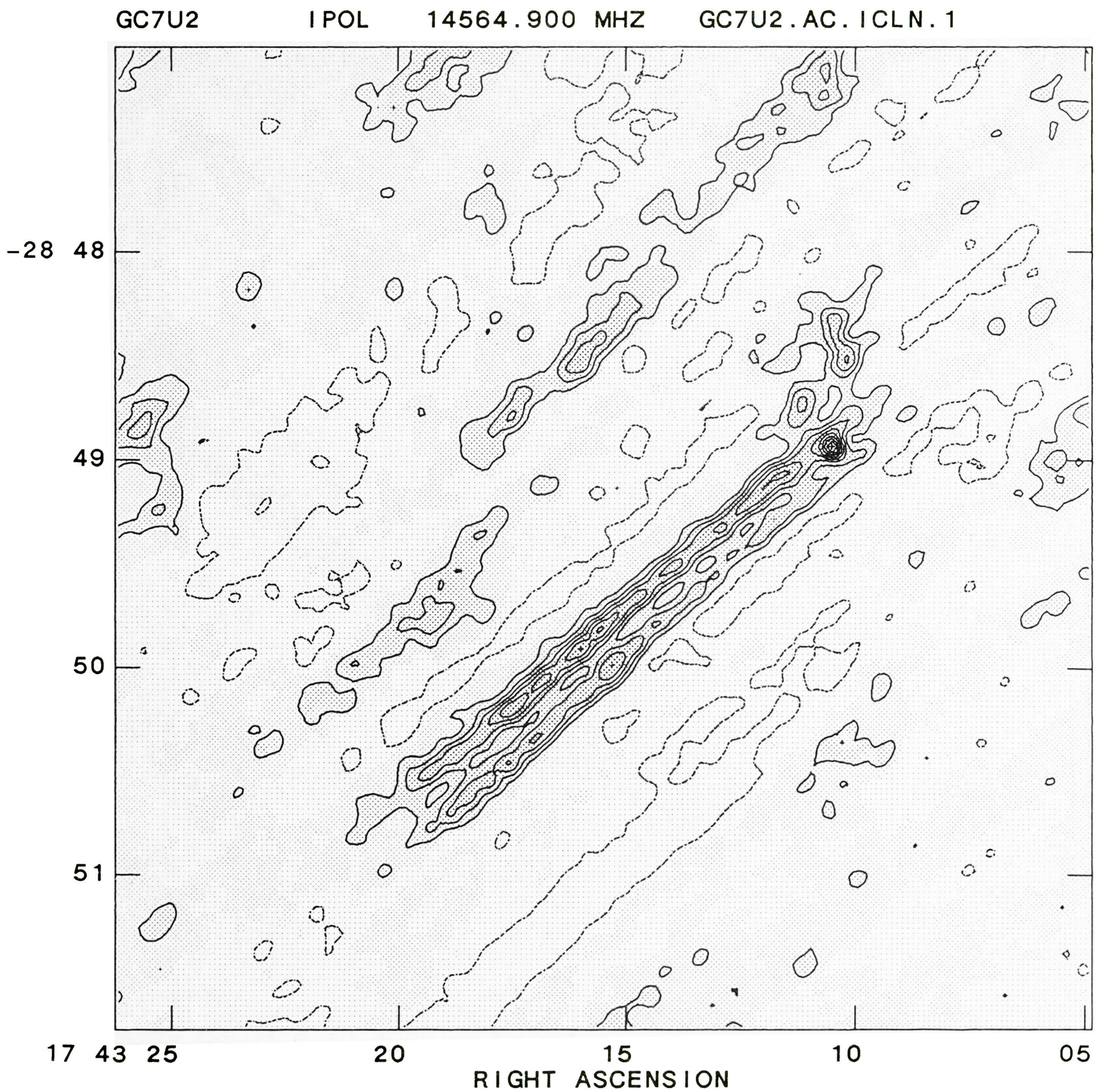
GC7U2 IPOL 14564.900 MHZ GC7U2.AC.ICLN.1
DECLINATION
-28 48
49
50
51
17 43 25
20
15
10
05
RIGHT ASCENSION

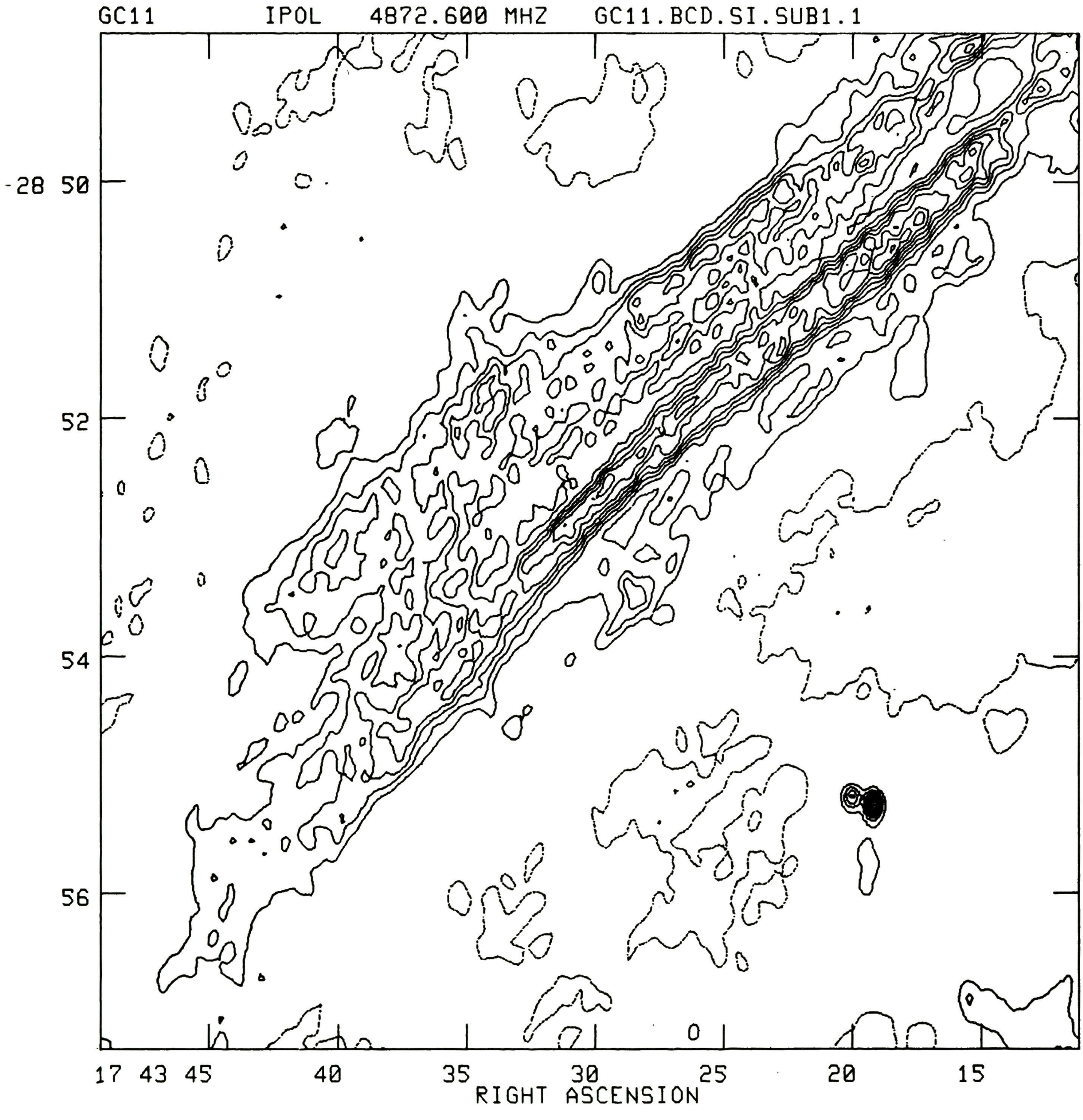

Figure 42: The 6-cm intensity contour intervals are 4.17 (-1, 1, 2, 3, 4, 5, 6, 7, 8, 9) mJy/beam area. This map is convolved with a gaussian beam, FWHM, of ~10" x 10" before it was corrected for the response of the primary beam. The noise level is ~7.5 mJy/beam area.

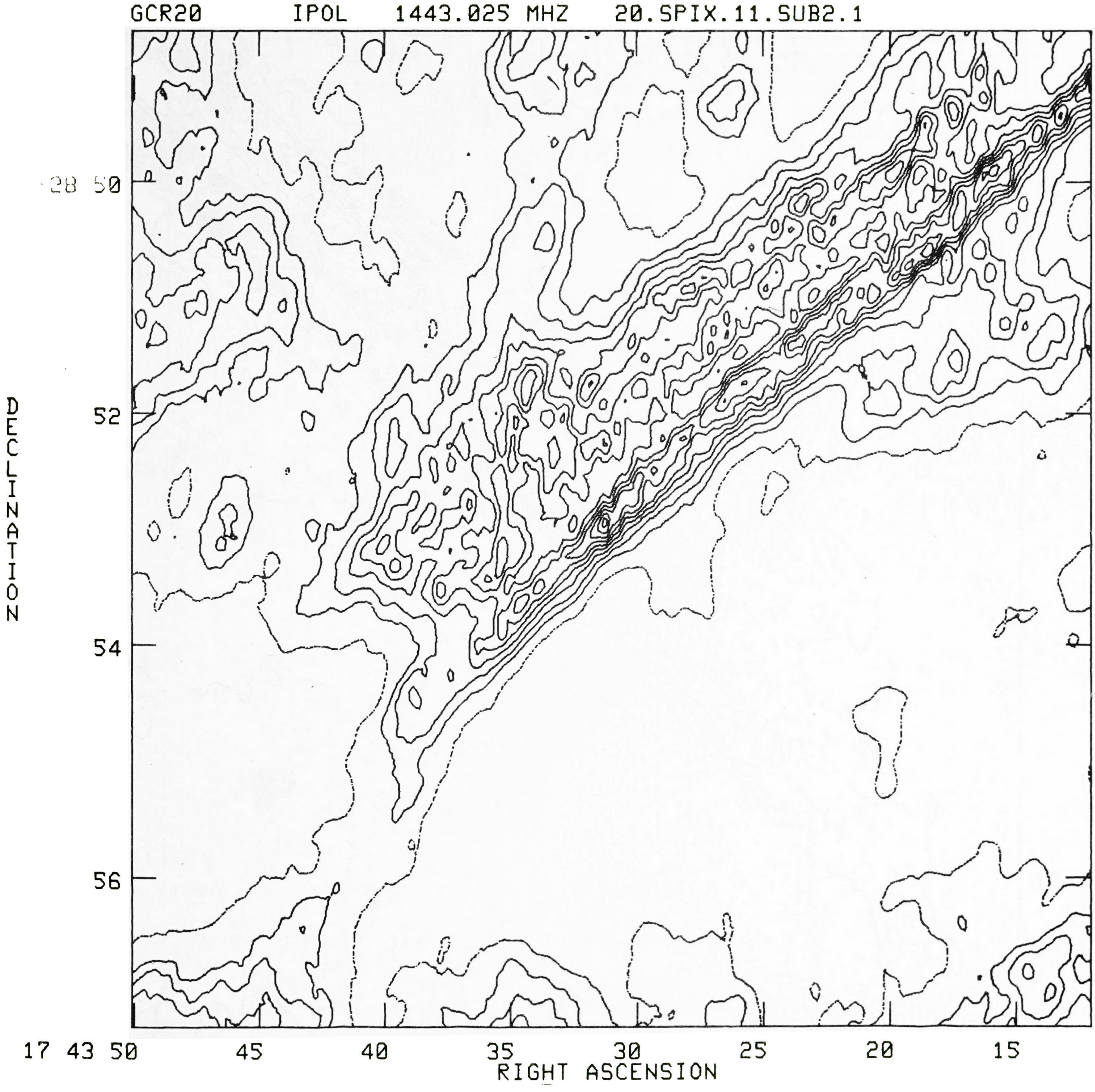

Figure 43: The 20-cm intensity contour intervals are 4.29 (-1, 1, 2, 3, 4, 5, 6, 7, 8, 9) mJy/beam area. This map is convolved with a gaussian beam, FWHM, of ~10" x 10" before it was corrected for the response of the primary beam. The noise level is ~9.5 mJy/beam area.

# Chapter 4

## RADIO EMISSION FROM THE GALACTIC CENTER ARC AT 160 MHz

"Is it possible that absorption is making itself felt at 80 MHz but is hardly significant at 160 MHz and higher?"
O.B. Slee

### I. INTRODUCTION

VLA observations of the radio continuum Arc near the galactic center ($\ell \sim 0^\circ.2$) have shown that this feature consists of a number of long and narrow filaments (see chapter 3), some of which extend as much as 40 pc along a line perpendicular to the galactic plane. In addition, a complex thermal radio structure has been observed at the position where the filaments cross the galactic plane (G0.18-0.04) and apparently at the location where the filaments undergo an abrupt discontinuity (G0.16+0.08) and bend sharply toward the galactic plane (G0.07+0.04).

Radio continuum observations of the Arc at frequencies between 408 MHz and 15.5 GHz (Little 1974; Mills and Drinkwater 1984; Yusef-Zadeh et al. 1984; Pauls et al. 1980; Whiteoak and Gardner 1973; Kapitzky and Dent 1974) all show a spatial asymmetry with respect to the galactic plane; the "vertical" portion of the Arc (i.e., the quasilinear portion oriented perpendicular to the galactic plane) extends slightly more towards negative latitudes than positive ones. Recent observations by Mills and Drinkwater (1984) indicate that the three broad intensity maxima in the Arc (G0.16-0.15, G0.18-0.04 and G0.1+ 0.08) have spectral indices, $\alpha$, between -0.05 and +0.11 ($F_\nu \propto \nu^{+\alpha}$). Previous radio recombination line data suggested that most of the Arc appears to have thermal characteristics except for G0.16-0.15 (Pauls et al. 1976; Pauls and Mezger 1980; Gardner and Whiteoak 1977).

In this chapter we first describe our 160 MHz and 327-MHz observations and present the 160-MHz and 5-GHz maps of the Arc and then argue that two radiation mechanisms operate simultaneously in the enigmatic Arc: On the one hand, the filaments of which the Arc is composed emit synchrotron radiation; on the other hand, a nonuniform distribution of ionized material surrounding the Arc and physically associated with the filaments emits thermal radiation at high frequencies and absorbs the nonthermal emission from the background filamentary structure at low frequencies.

## II. OBSERVATIONS

The 160-MHz observations were made with the Culgoora Circular Array consisting of 96 reflectors spaced around a circle of 3 km diameter. The 32 central beams formed by the array (spaced along the north-south direction) were placed ahead of the field center in a repetitive 128 s cycle. The resulting drift scans were digitized, averaged and recorded on magnetic tape for off-line computer analysis. Contour maps were constructed from 70 drift scans occupying a total observing time of 2.5 h on the evening of 24 September 1984. Observations of a strong flux calibrator (1938-155) directly following those of Sgr A yielded a circular beam of FWHM = 1!9 and side-lobes at the 5% level of peak brightness.

The 327-MHz observations were conducted in two parts due to the halved declination coverage of the 32 beams at this frequency: (i) 128 s drift scans for 2.5 h of the region centered on the vertical segment of the Arc on October 4, 1984; (ii) 64 s drift scans for 2.1 h of the area centered on Sgr A on October 24, 1984. Observations of the strong calibrator source 2012+234 immediately following these measurements showed obvious side lobes at the 20% level of peak brightness (due to our inability to phase the array correctly at 327 MHz). Hence the 327-MHz maps are not well suited for accurate brightness distributions but are useful for estimates of integrated flux density. The width of the main beam to half power points is 56" × 56".

The 160-MHz intensity contours of the Sgr A complex and the Arc are shown in Figure 1 superimposed upon a 1.4-GHz image of the same region made with the Very Large Array (VLA) of the NRAO. The 327-MHz map of the Arc, for which the upper brightness limit is < 1.75 Jy/beam area corresponding to a brightness temperature of $T_B < 6,850$ °K, is not presented here, while the 327-MHz map of Sgr A is being presented in chapter 7.

The observations and data processing procedures for the VLA result are described in full detail in chapter 3. Here we describe briefly the polarization maps of the Arc made from data taken at a wavelength of 6 cm. These maps are based on the observations which were carried out using the VLA in both the hybrid C/D and B/C configurations, in May 1983 and March 1984, respectively. The sources NRAO 530, 1748-253, and 3C 286 were observed for use as a polarization calibrator, a phase calibrator, and a flux calibrator, respectively.

The radiograph shown in Figure 2 displays the distribution of polarized intensity emitted from the tail end of the Arc (G0.16-0.15) at 4.885 GHz, as seen by a beam of FWHM = 17".7 × 20".6 ($\alpha \times \delta$). The complex visibility (u,v) data from the two array configurations were edited, calibrated, combined, tapered at 7 kλ and Fourier transformed before the final map was CLEANed (Högbom 1974; Clark 1980). In this figure, which has a noise level of ~185 μJy/beam area, the total polarized flux is found to be ~390 mJy.

The contour plot shown in Figure 3 represents the 4.8 GHz intensity distribution sampled with a beam having a FWHM of 11"3 × 10"8 (α × δ). This map is made only from data taken with the hybrid C/D array. The noise level in this map is ~3 mJy/beam area. The superimposed line segments represent the predominant direction of the electric vectors and the lengths give the corresponding percentage polarization.

The 125 μm intensity contours of the Arc and Sgr A made using a beam of FWHM =1' (Dent et al. 1982) is superimposed in figure 5 on a 1.4-GHz radiograph with a resolution similar to that of figure 1.

## III. RESULTS

Here we discuss the new 160-MHz details observed in the Arc and compare them with the 5-GHz polarization maps in order to establish the nonthermal character of the Arc and to give a foundation for a proposed explanation for the limited extent of the polarized region.

### a) The 160-MHz Intensity Distribution

At 160 MHz, as at 80 MHz (LaRosa and Kassim 1985), the emission associated with the Arc is asymmetric with respect to the galactic plane. Figure 1 shows that the long, linear geometry which characterizes the Arc at frequencies > 408 MHz, and the long continuous filaments into which it breaks up at high resolution, are no longer identifiable at 160 MHz; only the southeastern portion of the Arc (G0.16-0.15) remains visible.

The brightness temperature at the 160-MHz peak in G0.16-0.15, $T_B$ = 11,850 K, is consistent with a linear extrapolation of the flat ($\alpha$ = +0.01) spectrum found by Mills and Drinkwater (1984) at this position between 0.408 and 10.7 GHz. If the 0.16 GHz result is compared to the 10 GHz result, then the recomputed spectral index $\alpha_{0.16}^{10.7}$ = + 0.11. While this flat spectrum would normally suggest the presence of an optically thin thermal source, such an interpretation is not consistent with the lack of detectable recombination lines in this portion of the arc (Pauls et al. 1976; Pauls and Mezger, 1980) and its polarization characteristics observed at high frequencies. On the other hand, if we compare the 160-MHz and the 80-MHz surface brightness (La Rosa and Kassim 1985), then the spectral index becomes steep ($\alpha$ = 1.6) but with much uncertainty due to the contribution from background emission and foreground absorption at 80 MHz.

b) Polarization Characteristics

The polarized portion of the Arc coincides in position and extent with the 160-MHz emission source at G0.16-0.15. The overall distribution of polarized intensity (Figure 2) is elongated along the filaments, but shows a higher degree of nonuniformity than does the total intensity (Figure 3), which varies surprisingly little through the polarized region.

We note a number of structural differences between the southeastern and northwestern regions of Figures 3 and 4:

1) The discrete, compact, polarized intensity features located to the northwest of the polarized area have the highest degree of linear polarization: ~ 30%, whereas the more extended polarized region to the southeast has a linear polarization of ~ 10%.

2) The position angles of the electric field vectors toward the northwest are oriented roughly perpendicular to the direction of the linear filaments, whereas the position angles of the electric field vectors in the southeast are roughly aligned along the long axis of the linear filaments. The high percentage of linear polarization at 5 GHz implies that the magnetic field is highly ordered in G0.16-0.15 and indicates that the radiation is produced by the synchrotron mechanism.

3) The Faraday rotation in this region is extremely high: ~-1660 rad $m^{-2}$ at 3 cm (Inoue et al. 1984) or -2880 (-200) rad $m^{-2}$ at 6 cm from the northwestern (southeastern) portion of figures 2 and 3. The difference in Faraday rotation at 3 and 6 cm is ascribable to different resolutions.

## IV. DISCUSSION

### a) A Nonuniform Halo Surrounding the Arc

The apparent variation of spectral properties along the vertical filaments might simply be due to an alternation of the predominance of one emission process (i.e., synchrotron radiation) over another (free-free emission). However, this possibility seems quite unlikely in view of the relative uniformity of the intensity and width of the filaments along their long dimension at the higher frequencies; that is, the total intensity maps made at 20 and 6 cm (Figures 1 and 3) give no indication whatsoever that the nonthermal region, G0.16-0.15, has any unusual significance or that the emission mechanism differs between that region and elsewhere along the Arc. Furthermore, the polarized intensity map (Figure 3) shows that the changes in polarization are rather abrupt (the percentage polarization increases by a factor of 30 to 50 at $\alpha = 17^h43^m15^s$, $\delta = -24°49'$), unlike that which one would expect if one emission mechanism merges smoothly into another so that the total intensity doesn't undergo substantial variations. (The abruptness in the polarized emission is especially pronounced in the highest resolution 6-cm maps, as shown in chapter 3, while the total intensity distribution remains uniform in its appearance.)

We therefore suggest an alternative model in which the vertical filaments are taken to be uniform and nonthermal along their full length (~30 pc). An intervening absorbing medium is then held responsible for cutting off the low-frequency emission and for depolarizing (i.e. large Faraday rotation) the radiation from much of the filamentary structure. Under this hypothesis, the variation in spectral characteristics along the Arc is caused by inhomogeneities in the intervening medium.

The required intervening medium appears to actually surround the vertical filaments and thus to be associated with them. Images of the Arc at 20 cm that are tailored to emphasize the faintest details show a nonuniform, 15-pc-diameter halo of emission surrounding the vertical portion of the Arc (see chapter 3). The most prominent features within this halo are the "helical segments" (Chapter 3), which cross in front of the filaments and noticeably absorb some of their higher-frequency emission. (This absorption can be seen in figure 1 at $\alpha = 17^h43^m15^s$ $\delta = -28°49'05"$). Another component of the medium surrounding the nonthermal filaments is the sickle-shaped feature, G0.18-0.04 (see Chapter 3), the appearance of which strongly suggests that it is interacting with the vertical filaments (see chapter 9).

The southernmost component of the Arc ($\alpha$ = $17^h43^m37^s$, $\delta$ = -28°54'30") is a location at which neither emission at 160 MHz nor polarized emission at 5 GHz is seen. These characteristics can also be attributed to an enhancement in the electron column density in the intervening halo surrounding the vertical portion of the Arc which causes both the absorption of low-frequency emission and the depolarization of nonthermal emission from the background filaments.

The $^{12}CO$ intensity map presented by Brown and Liszt (1984, see their Figure 4) and the distribution of $NH_3$ emission shown by Güsten and Henkel (1983) strongly suggest that the 50 km $s^{-1}$ molecular cloud located near the galactic center is interacting with G0.18-0.04 (see chapter 9). Figure 4 shows that the $^{13}CO$ intensity distribution between 30 and 50 km $s^{-1}$ appears to be maximized near the southern edge of the Arc (see also figure 4 of Brown and Liszt 1984). The molecular velocity in here agrees well with the radio recombination line velocities seen in this region (chapter 9; Pauls and Mezger 1980). The HCN peak antenna temperature map made by Fukui et al. (1977, see their figure 11) also indicates that G0.18-0.04 lies near a strong peak in the HCN emission from the 50 km $s^{-1}$ molecular cloud. Formaldehyde studies by Bieging et al. (1980) and Güsten and Downes (1980) show an absorption component against G0.18-0.04 at 50 km $s^{-1}$. High-resolution infrared observations at 55 and 125 $\mu m$ show that the northern half of the radio Arc has infrared counterpart (Dent et al. 1982). Figure 5 exhibits the contours of 125 $\mu m$ intensity distribution superimposed on the 20-cm radiograph. The lack of infrared and molecular emission from the Arc at negative lati-tudes might be accounted for by low molecular and dust column densities or/and low dust temperature in the medium surrounding the radio filaments. Further comparisons of infrared, radio and molecular maps are discussed in chapter 9. On the basis of recombination line measurements, absorption line studies, and the 160-MHz map of the Arc (cf., Figure 1), we infer that the nonuniform halo of thermal gas and possibly dust, the nonthermal filaments and the 50-km $s^{-1}$ molecular cloud are phenomenologically linked, and that at least part of the 50 km $s^{-1}$ cloud intervenes between us and the Arc. Thus, the 50-km $s^{-1}$ molecular cloud with its nonuniform column density projected along the Arc may act as a reservoir which fuels the inhomogeneous, ionized medium surrounding the Arc.

b) Structural Differences Between G0.18-0.04 and G0.16-0.15

The 20-cm map shown in Figure 1 demonstrates that G0.16-0.15 and G0.18-0.04 have structural differences. For one thing, G0.16-0.15 is a highly polarized, 8-pc segment of the 30-pc filamentary system which appears to be surrounded by a helical structure, whereas G0.18-0.04, which has been seen at far IR wavelengths (Odenwald and Fazio 1984), consists not only of a different portion of the same filaments, but also of narrow, shock-like structure oriented roughly perpendicular to the filaments (sickle-shaped feature, chapter 3). It is likely that this structure is a dense portion of the envelope which is surrounding and interacting with the linear filaments. Both the helical and the shock-like structures are assumed to be part of the halo surrounding the filaments. High-resolution radio recombination line observations toward

G0.18-0.04 indicate that, unlike the vertical filaments, the nonfilamentary structures have thermal characteristics. Low-resolution recombination line observations made by Güsten and Downes (1980) also show that the apparent electron temperature increases rapidly toward the intensity peak in G0.18-0.04. These authors argue that nonthermal radiation contributes somewhat to an apparent rise in electron temperatures. Indeed, the high-resolution line observations of G0.18-0.04 strongly support a picture in which the thermal nonfilamentary features and nonthermal filamentary features are interacting with each other in this region (see chapters 3 and 9).

Table 1 lists the peak surface brightnesses of G0.16-0.15 and G0.18-0.04 including their corresponding spectra between 160 MHz and 4.8 GHz. We note that G0.18-0.04 is brighter than G0.16-0.15 by a factor of 1.2-1.4 at frequencies > 843 MHz. However, G0.16-0.15 becomes brighter than G0.18-0.04 by a factor of ~1.2 (> 3) at 408 (160) MHz (Mills and Drinkwater 1984; Pauls <u>et al.</u> 1976). These two peaks appear to have flat spectra at frequencies > 408 MHz, which is consistent with the conclusions of Mills and Drinkwater (1984). However, the spectrum of G0.18-0.04 turns over at 160-MHz whereas that of G0.16-0.15 remains flat. Such a change in surface brightness along the Arc can be accounted for if the nonthermal radiation from the filaments in the Arc is reduced at low frequencies by free-free absorption in the inhomogeneous thermal gas surrounding the Arc, but is enhanced at high frequencies by free-free emission from the thermal plasma.

Assuming that G0.18-0.04 has an intrinsically flat spectrum, the optical depth ($\tau$) toward this peak has to be at least ~2 at 160 MHz in order to reduce the peak flux measured at 1.4 GHz (~7.5 Jy/beam area) by a factor of $\geqslant$ 7.5 to conform with the observed limit at 160 MHz (1 Jy/beam area; the 160-MHz and 1.4 GHz beams have the same areas). The emission measure (E, $cm^{-6}$ pc) and electron temperature are related by (Mezger and Hoglund 1967):

$$E\ T_e^{-1.33} = \frac{12.5\ \tau\ \nu^{2.1}}{a\ (T_e,\nu)} \sim 5.32 \times 10^{-1} \qquad (1)$$

where a is a correction factor of order unity, and, in this case, $\nu$ = 0.16 (GHz). Also, based on their radio recombination line measurements, Pauls <u>et al.</u> (1976) and Pauls and Mezger (1980) find $T_e \sim 10000$ °K toward G0.18-0.04. This value is an upper limit since non-thermal radiation contributes somewhat to the derived $T_e$. However, one can deduce another upper limit to $T_e$ from the failure to detect 160-MHz emission since the limit of ~1 Jy/beam area corresponds to $T_B$ = 5300 °K (the dilution factor is assumed to be unity for this resolved source). Hence, using $T_e = T_B(1-e^{-\tau})^{-1}$, $T_e$ becomes $\leqslant 6.13 \times 10^3$ °K. Thus, from the observed emission measure, $0.532\ T_e^{1.33}$ $cm^{-6}$ pc, one can deduce an electron density of $\leqslant$192 $cm^{-3}$ if the ionized gas near the Arc is uniformly distributed over 3 pc, the width of the helical structure as it crosses the vertical filaments (see chapter 3).

The appearance of several radio features in and around the Arc, such as the helical segments, suggests that a magnetic field plays a strong role in shaping some of the features seen there (chapters 3, 5, & 6) Thus, it is plausible that the strength of the magnetic field in the plasma surrounding the Arc is comparable to that in G0.16-0.15, derived assuming equipartition of energy: $\sim 1\text{-}2 \times 10^{-5}$ gauss. The large column density of ionized gas in G0.18-0.04 implied by the above-derived emission measure coupled with the assumed magnetization, would rotate the plane of polarization by ~15 radians at 6 cm. Thus, the background nonthermal emission from the filaments would be completely depolarized by Faraday rotation in a manner consistent with our observations.

A reduction in the total flux of G0.16-0.15 from 16 Jy at 10 GHz (Pauls <u>et al.</u> 1976) to 12.4 Jy at 160 MHz implies a 160-MHz optical depth > 0.25 in the intervening absorbing medium if $\alpha < 0$. An estimate of the number density of electrons $n_e(\text{cm}^{-3})$, spread over L = 3 pc in front of G0.16-0.15 can be deduced by using the formula for the rotation measure R.M.:

$$n_e = \frac{\text{R.M.}}{(8.1 \times 10^5) B_{\parallel} L} \text{ cm}^{-3} \qquad (2)$$

where $B_{\parallel}$ is the component of the magnetic field along the line of sight taken to be $\sim 10^{-5}$ gauss. The number density of electrons in this region is ~68 $\text{cm}^{-3}$ based on the rotation measure given by Inoue <u>et al.</u> (1984) at 3 cm. High-resolution observations at 6 cm indicate that the rotation measure varies greatly in G0.16-0.15 and therefore $n_e$ could be as low as 8 $\text{cm}^{-3}$ along some lines of sight. One can then find the emission measure, $n_e^2$ L, to be between 50 and $1.4 \times 10^4$ $\text{cm}^{-6}$ pc in the regions in the southeast and northwest of figures 2 and 3, respectively.

c) Optical Depth Toward G0.17+0.0

G0.17+0.0 occupies the northern segment of the vertical filaments. Its location is chosen for comparison because, unlike G0.18-0.04, it appears mostly filamentary and because, unlike G0.16-0.15, it is not observed at 160 MHz. So the halo surrounding the Arc must be the only agent responsible for the factor of $\geq$ 8 which differentiates the 1.4 GHz flux (10 Jy) from the upper limit at 160 MHz; the optical depth at 160 MHz is then $\geq$ 2 (if $\alpha$ at this location is $\leq$ 0). The emission measure of the absorbing medium is then $\gtrsim 0.53 \quad T_e^{1.33}$; using $T_e \sim 5000°$ K, as estimated by Pauls <u>et al.</u> (1976), one finds $n_e \gtrsim 120$ cm$^{-3}$ distributed over 3 pc. Comparison of emission measures from G0.16-0.15 and G0.17+0.0 indicates that the halo which surrounds the Arc has a nonuniform column density.

d) Conclusion

We conclude that the ionized gas lying in front of and presumably surrounding the vertical segment of the Arc is distributed nonuniformly and the measure of its electron density changes by ~ an order of magnitude. Based on estimated electron densities, the total ionized masses in front of G0.16-0.15 (~7' × 3'), G0.18-0.04 (~4' × 3') and G0.17+0.0 (~5' × 2') are all $\geq 320 M_{\odot}$.

e) Summary

We have presented obsevations of the continuum Arc at three different frequencies, 160 MHz, 327 MHz and 5 GHz. A map of the 160-MHz radio continuum and another of the polarized 5-GHz emission indicate clearly that nonthermal processes are responsible for the emission arising from one limited segment of the Arc: that arising from G0.16-0.15, a broad intensity maximum situated on the southern half of the system of parallel filaments oriented perpendicular to the galactic plane. Other portions of this "vertical" system of filaments have radio characteristics consistent with thermal emission. A model is developed in which the filaments forming the core of the Arc are uniformly nonthermal emitters, but are surrounded by a nonuniform distribution of thermal plasma which preferentially absorbs low-frequency radiation, thus flattening the nonthermal spectrum, and depolarizing by Faraday rotation the nonthermal emission from the filaments everywhere except in the vicinity of G0.16-0.15.

Table 1

Spectra of Three Positions Along the Arc

| | Peak Brightness Temperature (k) | | | Spectral Index (α) + | | | | |
|---|---|---|---|---|---|---|---|---|
| | 160 MHz | 1.44 GHz* | 4.8 GHz* | 0.16–1.44 | 0.16–4.8 | 1.44–4.8 | 0.16–0.408 | 0.16–0.843 |
| G0.16–0.15 | $1.18\times10^4$ | 320 | 30 | +0.36 | +0.24 | +0.02 | +0.5 | +0.29 |
| G0.18–0.04 | $<3.95\times10^3$ | 450 | 56 | >1.01 | >0.74 | 0.13 | 1.46 | 0.93 |
| G0.17+0.0 | $<3.95\times10^3$ | 215 | — | >0.67 | — | — | — | — |

* From chapter 3.

\+ The brightness temperatures at 408 MHz and 843 MHz are taken from Mills and Drinkwater (1984)

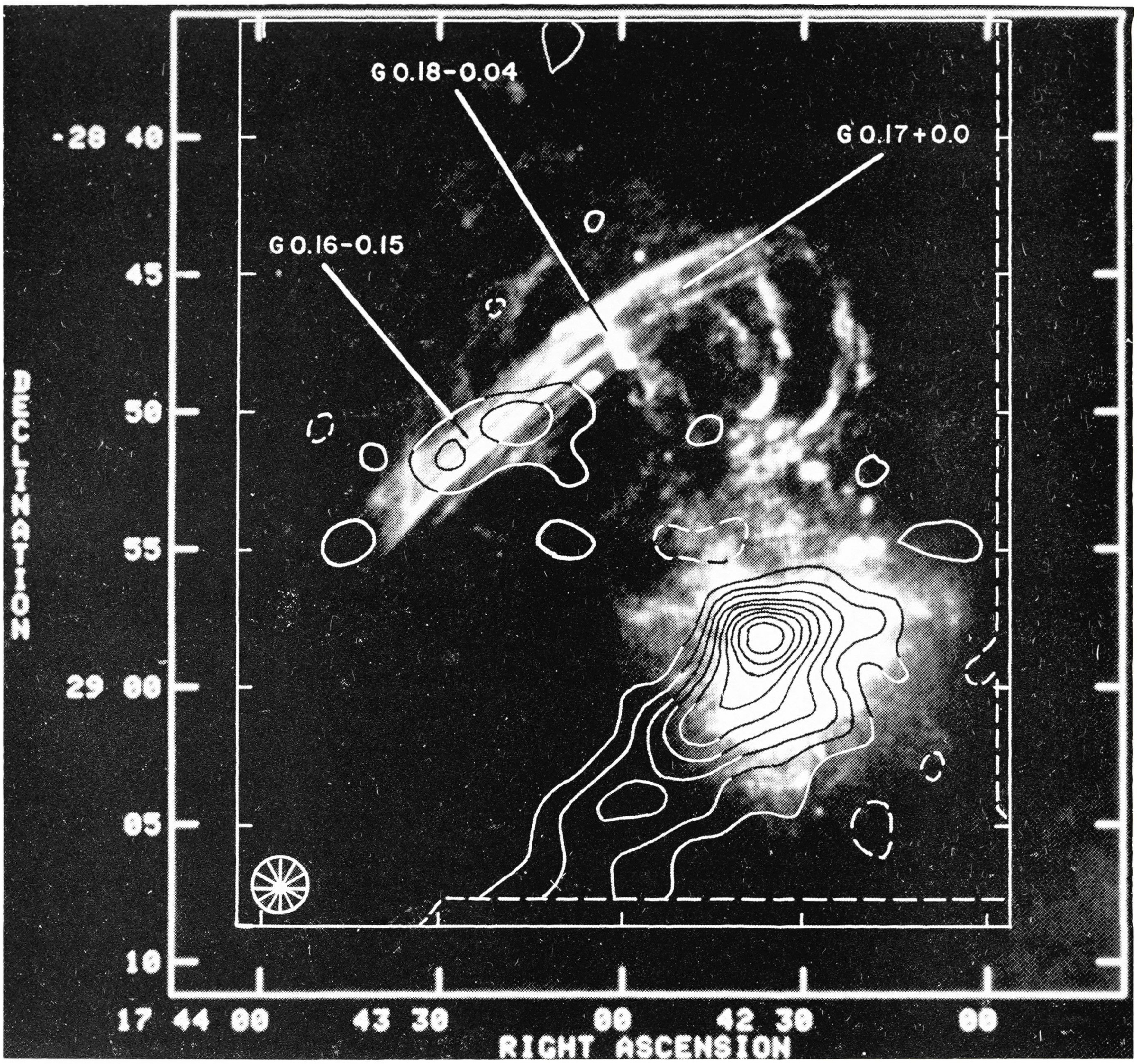

Figure 1: The 160-MHz map of Sgr A and the Arc, is superimposed upon the 1.4-GHz VLA map, made with a beam (FWHM) of 17" × 16" (α × δ). The contour levels are 10%, 20%, ... 90% of the peak flux, which is 12.8 Jy/beam area.

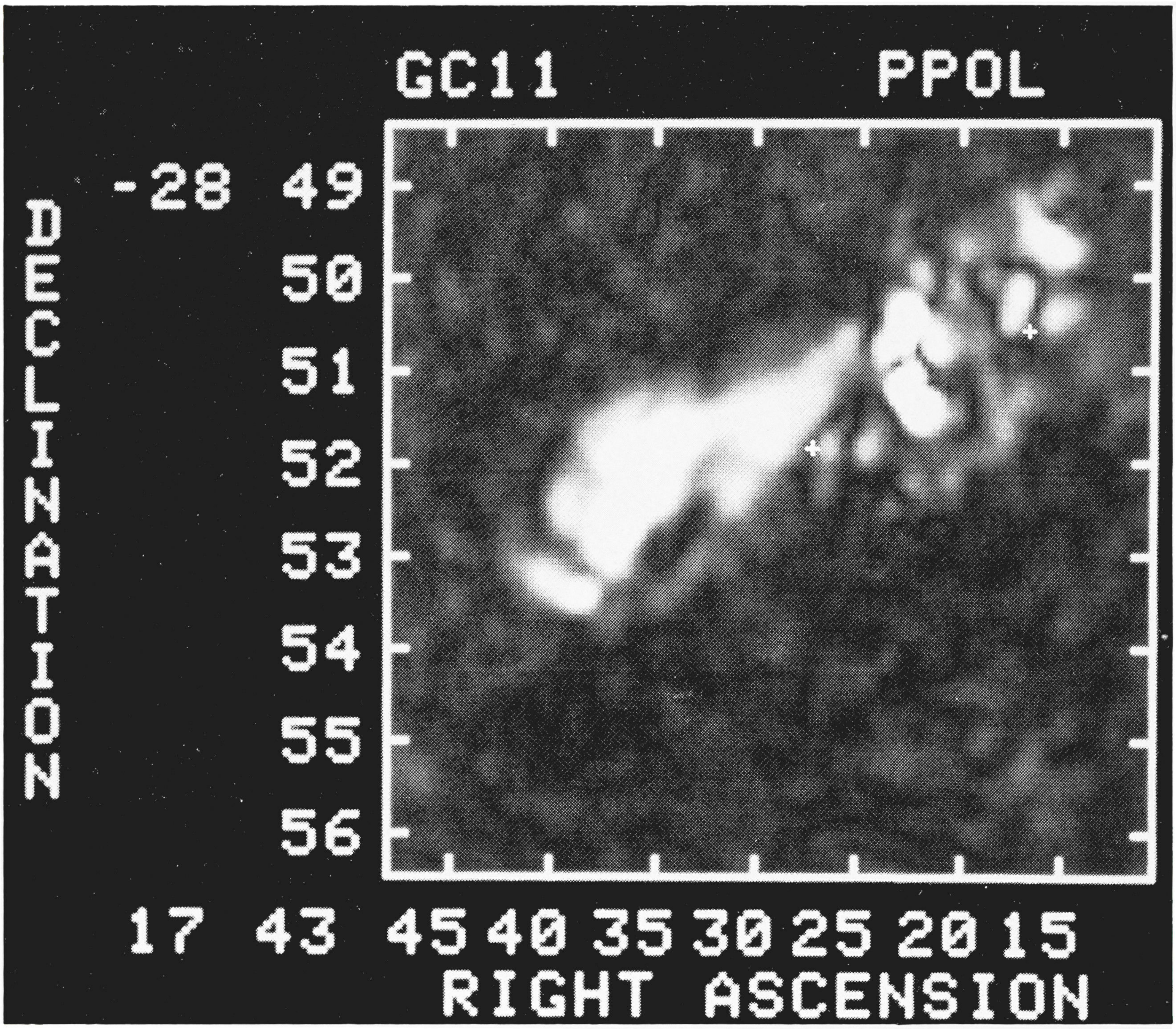

Figure 2: The 4.9-GHz image of the polarized emission from the southeastern segment of the Arc, centered on G0.16-0.15, made with a resolution (FWHM) of 17"7 × 26"6 (α × δ). The two crosses indicate the positions of the peaks seen in the 160-MHz map.

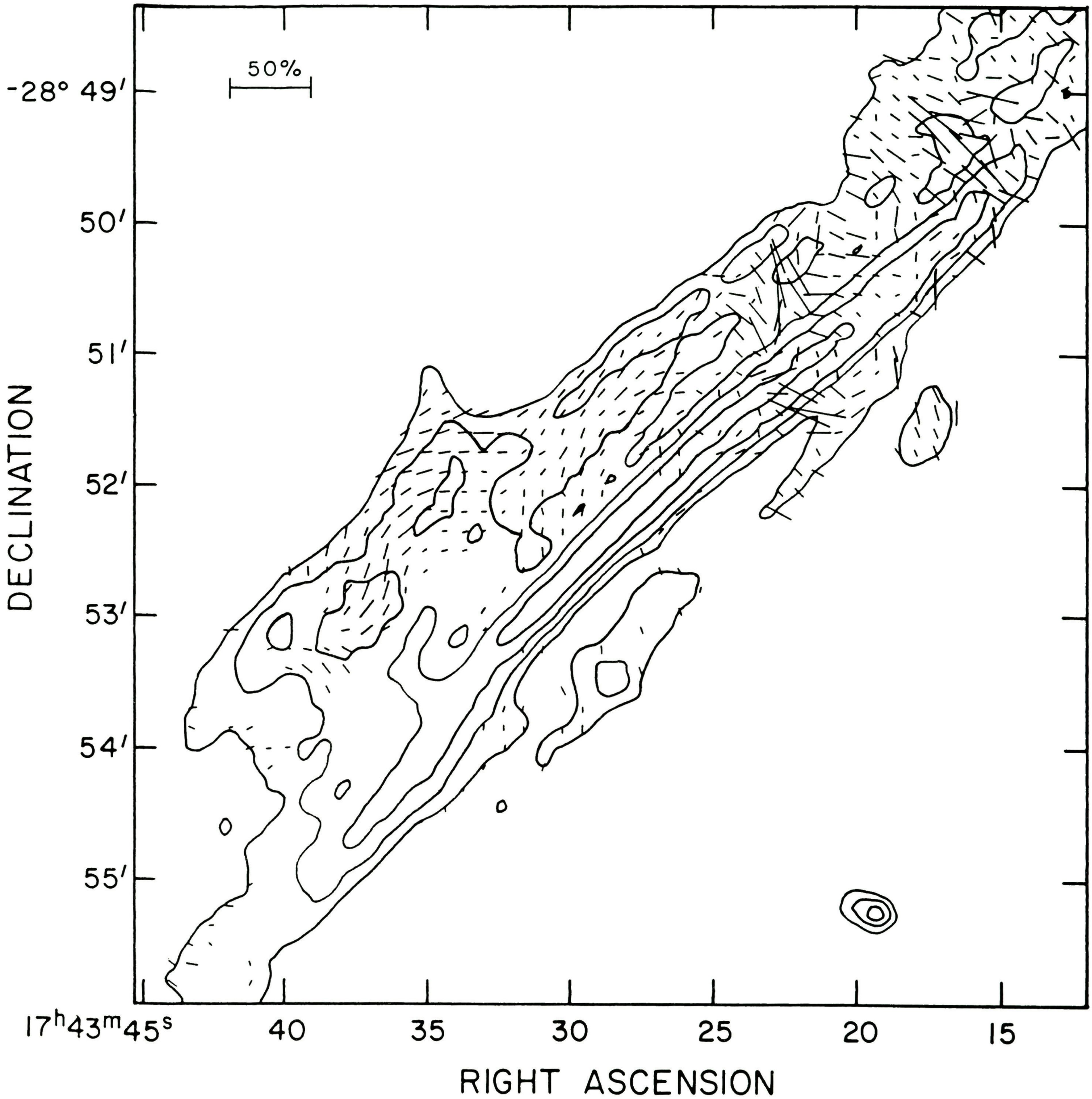

Figure 3: The 4.9-GHz VLA intensity map, made with a beam (FWHM) of 11.″3 x 10.″8 ($\alpha$ x $\delta$). The contour levels are 13, 26, 42.3, 61.9, and 84.8 mJy/beam area. Superimposed upon this are line segments indicating the direction of the electric vectors and the degree of polarization at the same resolution. The length of the superimposed bar in the top left hand corner represents a polarization of 50%.

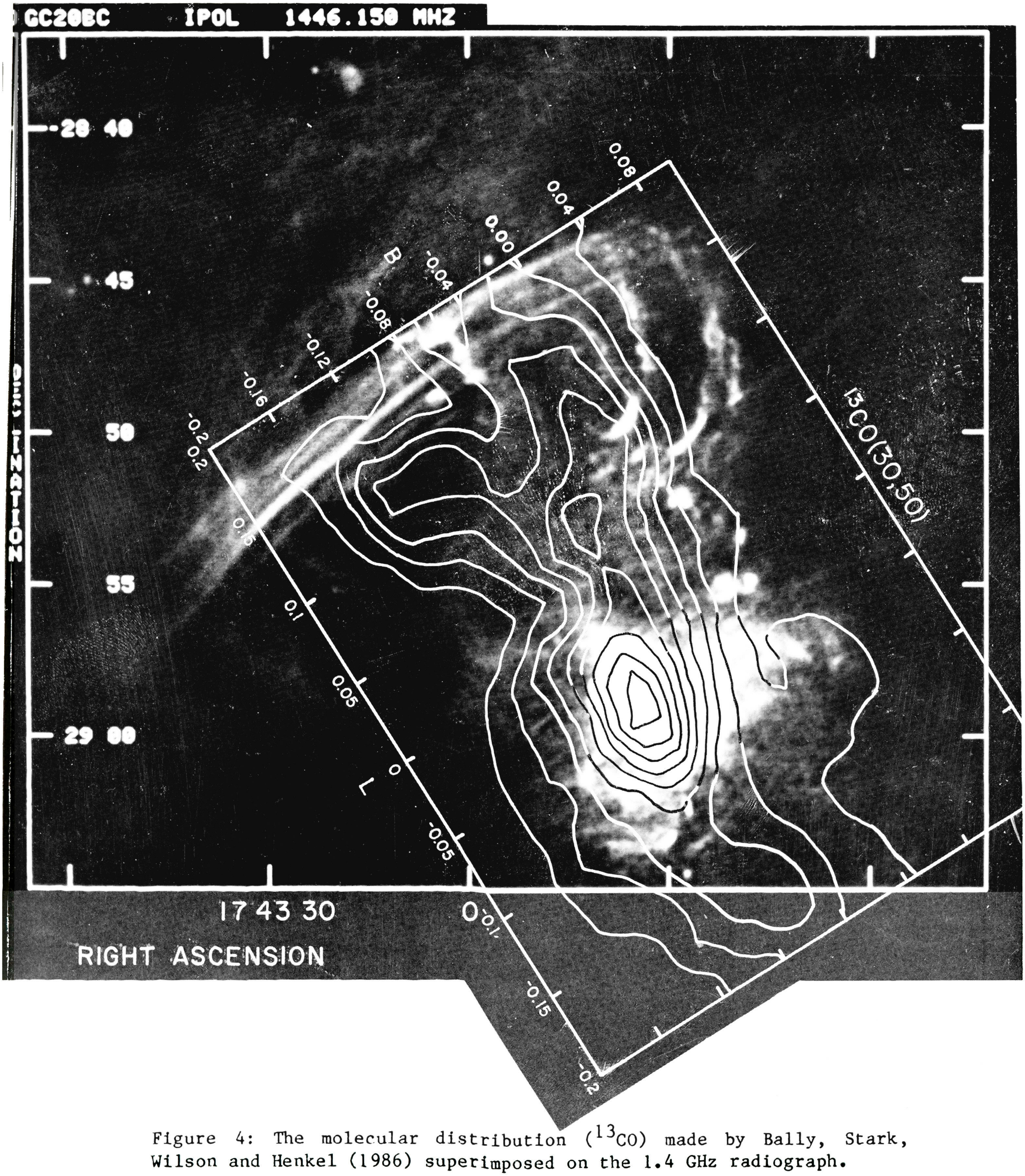

Figure 4: The molecular distribution ($^{13}CO$) made by Bally, Stark, Wilson and Henkel (1986) superimposed on the 1.4 GHz radiograph.

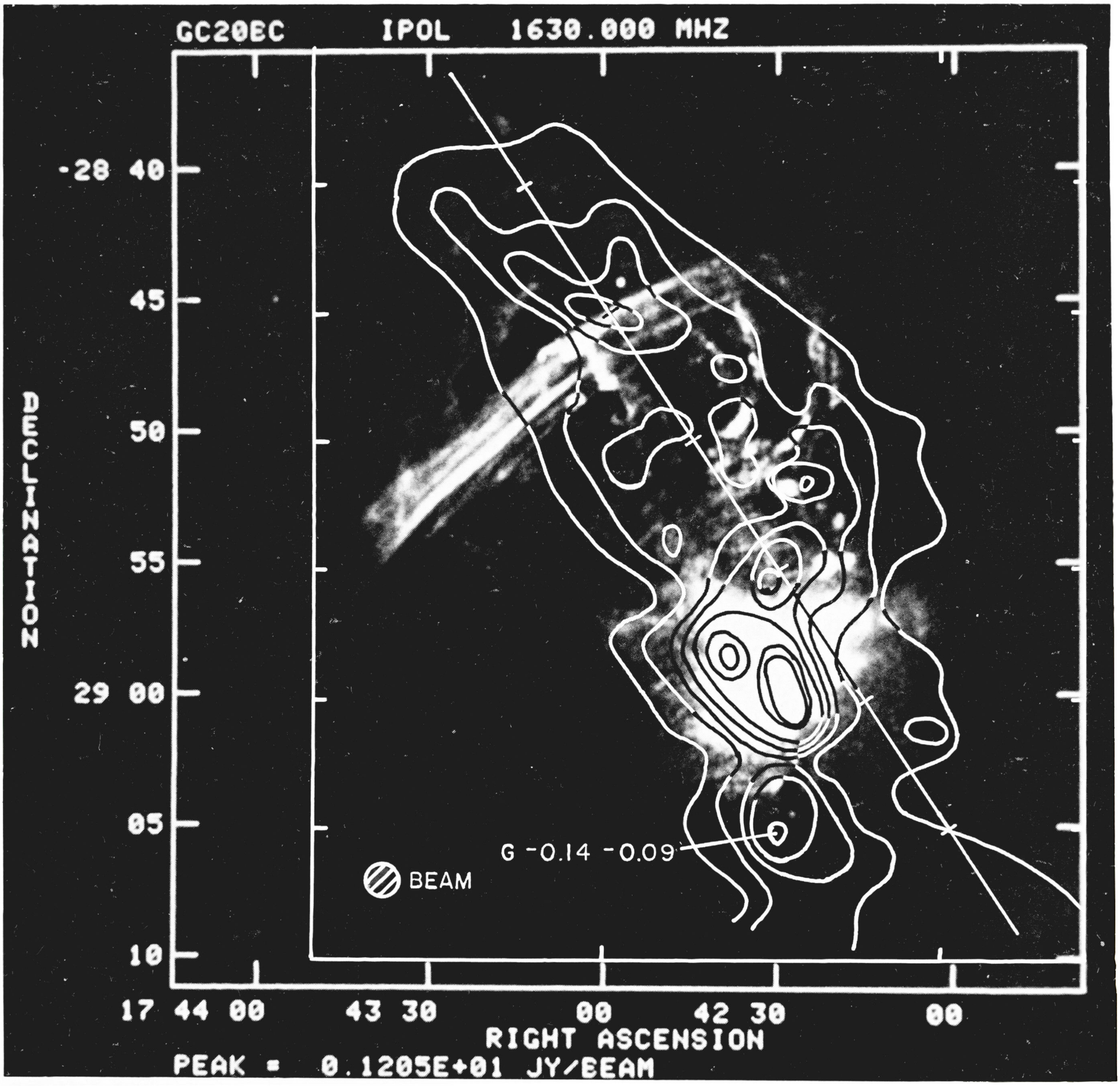

Figure 5: The 125 μ intensity map made by Dent <u>et al.</u> (1982) is superimposed on the 1.4 GHz radiograph. The area of the beam of the infrared map is greater by ~two orders of magnitude than that of the 1.4 GHz radiograph.

# Chapter 5

# PUZZLING THREADS OF RADIO EMISSION NEAR THE GALACTIC CENTER

"The second sweetest set of Three words in
English is "I don't know", ..."
C. Tavris

## I. INTRODUCTION

The radio structure of the central region of our galaxy as revealed by the VLA, and depicted in previous chapters, displays an increasingly rich pheomenology (Brown and Johnston 1983; Ekers et al. 1983; Lo and Cluassen 1983; Yusef-Zadeh, Morris, and Chance 1984; Ho et al. 1985; Liszt 1985). Many of the newly exposed details were quite unexpected, and some have no known counterpart outside this region (see Chapter 3). We report here the detailed structure of further anomalous feature that fit that description.

In VLA radiographs depicting a large field of view (up to 0.5° across) near Sgr A, faint but definite threads of radio emission are seen to traverse the field. These structures are remarkably uniform in width and brightness and have a remarkably smooth and gentle curvature. They are morphologically very similar to the radio filaments reported previously to occur in the radio continuum Arc described in chapter 3, but differ in being isolated from and uncorrelated in position angle with those filaments which compose the Arc. In this chapter, we present pictures of these threads (the term "threads" is employed in this thesis in order to distinguish them from two types of filaments, namely, the linear and arched filaments described in chapter 3). These pictures suggest an interaction between the threads and the arched filaments described in chapter 3. We conclude that the threads are illuminated magnetic field lines and leave open the question of why the threads shown here are singled out among field lines which, presumably, occupy the entire central region of the Galaxy.

## II. OBSERVATIONS

Most of the thread-like features appear at both 6 and 20 cm wavelength. In addition, they were detected at 20 cm using four different phase centers separated by 1/8 to 1/2 degree (see chapter 2 for further observational details). Furthermore, the northern thread can be identified in the 843-MHz map made by Mills and Drinkwater (1984). The reality of these features is therefore not in question. We are unable to extract a reliable measure of the spectral index from our data, first because the strong, extended background emission affects the maps made at the two wavelengths in different ways, and because of the different sampling of the visibility of the (u,v) data at these two wavelengths. For example, in both the 6 and 20 cm maps, Sgr A is surrounded by a shallow negative annulus (i.e., it is sitting in a "bowl") caused by lack of coverage at the lowest spatial frequencies, but this artificial annulus had different characteristics at 6 and 20 cm becuase of the different primary beams and the different frequency coverages. Like numerous features seen near the galactic center, crude estimate of the spectral index of the northern thread (see below) indicates that this feature has a flat spectrum between 6 and 20 cm.

## III. RESULTS

Figure 1 shows the 20-cm image using only the B/C array data with a phase center positioned at $\alpha = 17^h41^m30^s$, $\delta = -28°50'$. The prominent features to the southeast and to the northeast are the Sgr A complex (chapter 6) and the arched filaments (chapter 3). One can recognize a number of gently semi-linear features which strike across the image. These threads of radio emission are characterized by their narrowness, their tight collimation and the uniformity in their surface brightness. Their width is generally ~6 to 20 arcseconds or only 0.3 to 1 pc, and varies by less than 20% over the length of the observed segments. We describe each of two threads located to the north, and to the center of figure 1, separately.

### (A) The Northern Thread:

The brightest of the radio threads has a scimitar-like appearance. It is located in the northern portion of figure 1 and appears to merge with emission from the arched filaments. The northern thread becomes diffuse and weak in its surface brightness as it is directed toward the northwest. Figure 2 shows another view of the northern thread in a 20 cm field having a phase center near Sgr A (see Table 1 in chapters 2 and 3). This image shows clearly that this thread crosses the arched filaments and continues toward the southeast. Figure 3 shows the intensity contours of the westernmost segment of the arched filaments (W2 in figure 10 of chapter 3). The peak brightness temperature of the northern thread is ~250 °K at 20 cm and appears to be located at a position, $\alpha \sim 17^h42^m05^s$, $\delta \sim -28°46'15''$, where there is a slight shift in the curvature along its 10' extent.

At the point where the northern thread, as seen in figure 3, encounters the westernmost arched filament (W2) at $\alpha \sim 17^h42^m20^s$, $\delta \sim -28°49'$, there appears to be sharp discontinuity in the brightness of the arched filament, which would be a rather unlikely occurence if it were mere coincidence. This, perhaps, is more evident in figure 4 which shows the 6-cm detail of the region where the northern thread crosses W2. We also note that the northern thread crosses an adjacent arched filament W1 before it continues its southeasterly direction. The direction in which the surface brightness of W1 and W2 are enhanced at the locations where the northern thread crosses them differ: W1 (W2) becomes brighter (dimmer) by a factor of ~2 as it is followed southward (northward).

The southeastern extension of the northern thread, as seen in figures 2 and 4, appears to coincide with several hot spots between $\alpha = 17^h42^m32^s$, $\delta = -28°50'38''$ to $\alpha = 17^h42^m40^s$, $\delta = -28°51'35''$.

Furthermore, the intensity of the thread appears to increase as it approaches the westernmost arched filament in figure 4. Indeed, these intensity enhancements support the earlier suggestion that the northern thread is physically interacting with the arched filaments.

The arched filaments also appear to change in their geometry as they cross the northern thread. We note that the upper portion of W2 as seen in figure 4 is elongated in the north-south direction. However, south of the filaments, it curves substantially to the east. W1 also shows an increase in curvature but not as marked as that for W2. It appears that this structural change occurs at the location where the northern thread traverses the arched filaments. Both the arched filaments and the northern thread are directed toward or away from the galactic plane. The northern filament appears straight in the region where it travels through the arched filaments, but it has a gentle, but substantial curvature at $b > 0\overset{\circ}{.}17$ (see figure 1). The apparent structural changes that occurs at the intersection of the arched filaments of the Arc lead us to conclude that these two structures are interacting. However, the nature of the interaction is not clear.

(B) The Central Thread

The longest and the most linear thread occupies the central portion of figure 1. This thread merges with the Sgr A complex at its eastern extremity and fades into noise in its western extremity. This thread has a lower surface brightness than that of the northern thread and is also affected more by the artificial annulus, which surrounds the Sgr A complex, than the northern thread. Thus, the surface brightness of the central thread could be more than 50-100 °K at 20 cm. One of the interesting characteristics of this thread is a splitting at $\alpha \sim 17^h43^m$, $\delta \sim -28°54'10''$ in figure 1. This split resembles the apparent twisting of the linear filaments which were shown in figure 7 in chapter 3. There we showed that the twisting filaments were highly polarized, however, no polarization information is presently available for the central thread.

One basic question is whether the central thread is associated with the Sgr A complex and ultimately to the non-thermal radio source in Sgr A West or it is an unrelated feature which is projected upon the Sgr A complex. Figure 5 shows a close up view of the central thread as it merges with a number of radio protrusions which will be discussed in chapter 6. The linear central thread loses its identity at a location near two discrete features (see chapter 6) and near a radio protrusion at $\alpha \sim 17^h13^m$, $\delta \sim -28°54'$. We cannot determine the association of the central thread with any features, conclusively. However, high-resolution observations of this region are in progress and we hope to determine if the central thread is associated with Sgr A West. If so, the central thread could be an opposite counterpart to a one-sided low-energy jet structure which has been seen emanating from the center of the Galaxy (see chapter 7).

(C) Other Features

A wisp-like structure (see figure 12g of chapter 6) is located to the southwestern portion of figure 1. A map of the region to the west of this structure (i.e. beyond the field shown in figure 1) shows three compact sources which appear to have the same curvature as the wisp-like structure (see figure 12h of chapter 6). The nature of association between these compact sources and the wisp-like structure is not known.

Figure 6 which is more sensitive to the weakly emitting features displays a high-contrast radiograph of the region depicted in figure 1. Apart from the extension of the linear filaments which is discussed in chapter 10, we note two weak and thread-like features: one is running almost parallel along the northern thread and the other traverses the region between the northern and central threads. Curiously, the latter feature appears to emerge from or end at a compact source (see source H3 in figure 1 of chapter 6) which is situated to the south of the arched filaments. Both these thread-like structures curve in the same direction as the northern thread.

The 5-GHz maps made by Whiteoak and Gardner (1973) and Altenhoff _et al._ (1978) and the 10-GHz maps made by Whiteoak and Gardner (1973) and Altenhoff _et al._ (1978), the 10-GHz maps made by Sofue _et al._, and Seiradakis _et al._ (1985) all show large scale (0.5° - 1°), possibly unresolved features, located in such a way tha they could represent activity associated with an extrapolation of the threads to a much higher latitude than that revealed in figure 1.

## IV. Discussion

We have no polarization information on the radio threads presented here. However, these features show structures which are very similar to those linear filaments which compose the Arc. Polarization studies of the linear filaments and their northern extensions, which are fully discussed in chapter 3, 8, and 10, show clearly that the linear filaments are magnetic features. Because of the remarkable resemblence between the northern thread and the northern extension of the linear filaments (see chapter 8) and because of the brightening of the northern thread as it is distorted slightly from its linear geometry, we suggest that the threads are produced by synchrotron emission. The latter reasoning arises from the knot-like structure or hot spots noted in astrophysical jets as they are bent; it is suggested that relativistic particles are accelerated and magnetic field is strengthened at these locations where radio emission shows marked enhancement (see Bridle and Perley 1984 for references).

The threads appear to be well collimated and show characteristics, such as the ratio of their lengths to their widths, which are similar to very long astrophysical jets. One major problem, however, is that there are no sources of energy which could account for feeding or producing these threads.

We argued in chapter 3 that the linear filaments are one-dimensional features which are unlikely to be edge-on shock fronts originated from the galactic center. These arguments can also be made for the threads. In fact the orientation of the northern thread is such that its position angle with respect to the galactic center strengthens the argument that the radio filaments associated with the Arc are unrelated to activities (i.e. explosion in the nucleus) at the galactic center. We also reasoned in chapter 3 that the Arc is near the galactic center. Since the northern thread appears to be interacting with a component of the Arc, i.e. the arched filaments, we infer that the threads are located near the galactic center.

It is shown in chapter 9 that the arched filaments are one-dimensional structures which channel the flow of ionized gas. On the other hand, the northern thread appears to be a one-dimensional magnetic structure - based on its appearance - which appears to be interacting with the thermal filaments. The nature of their interaction is extremely puzzling because both of these features have one-dimensional structures and because they are governed by different physical processes. It is possible that the thermal electrons within the arched filaments are feeding the relativistic particles for acceleration in the northern thread. If so, what is the acceleration mechanism and, furthermore, what is it that makes the northern thread - assuming that it is an illuminated preexisting magnetic field lines - so special in its isolated location and its orientation as it bisects the western arched filament? A possible scenario for the latter question is discussed in chapter 9. Indeed, more observational work needs to be carried out in order to deduce the spectra of these new features, as well as the orientation of their magnetic field lines.

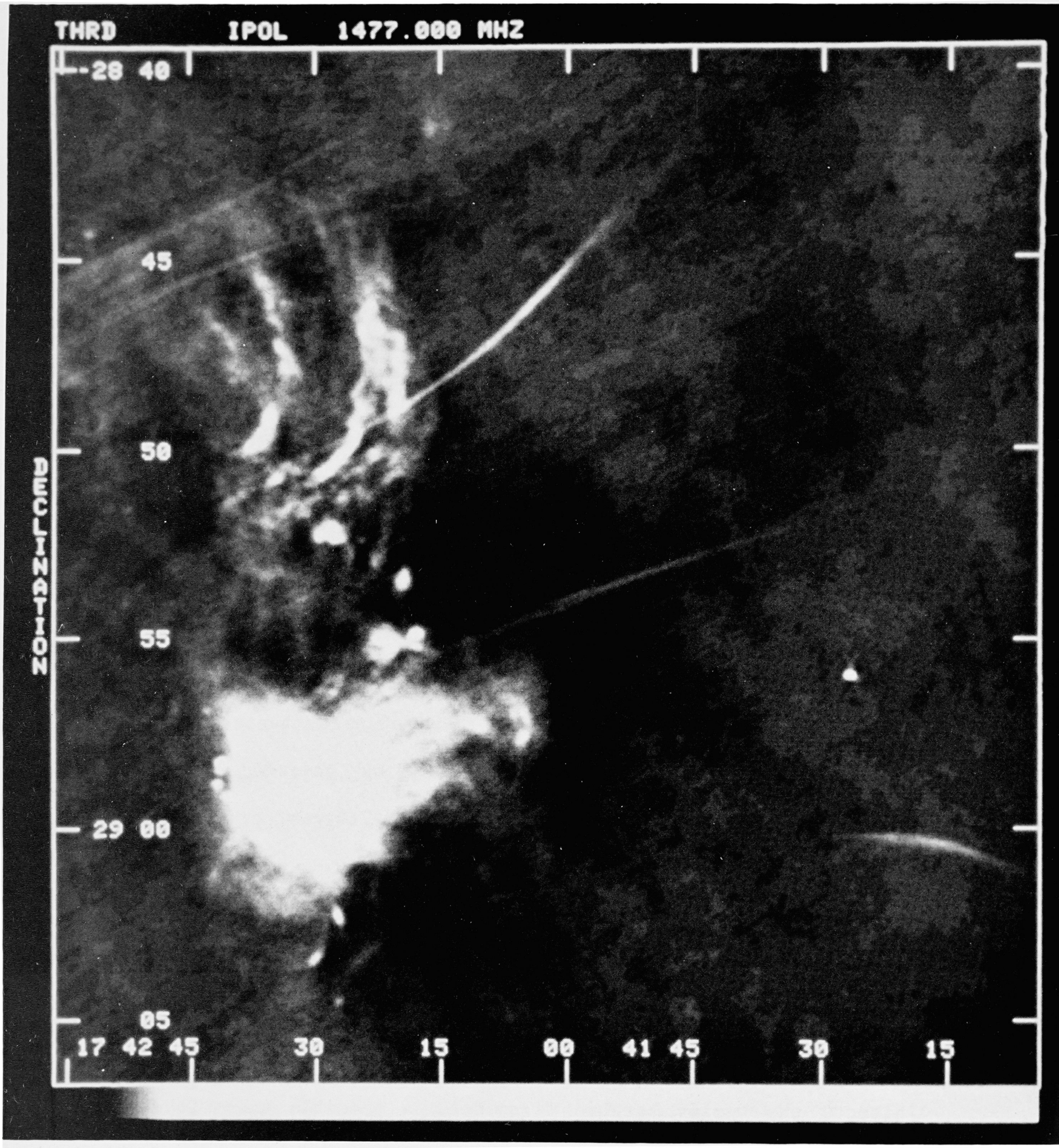

Figure 1: A 20-cm radiograph of the region to the west of the galactic center (positive galactic latitudes). This image was restricted to data taken on baselines ⩾ 400 λ and was based on the data from the B/C[2] array (see chapter 2). The CLEAN beam is 9.16"×8.2" (P.A. = 87.5°).

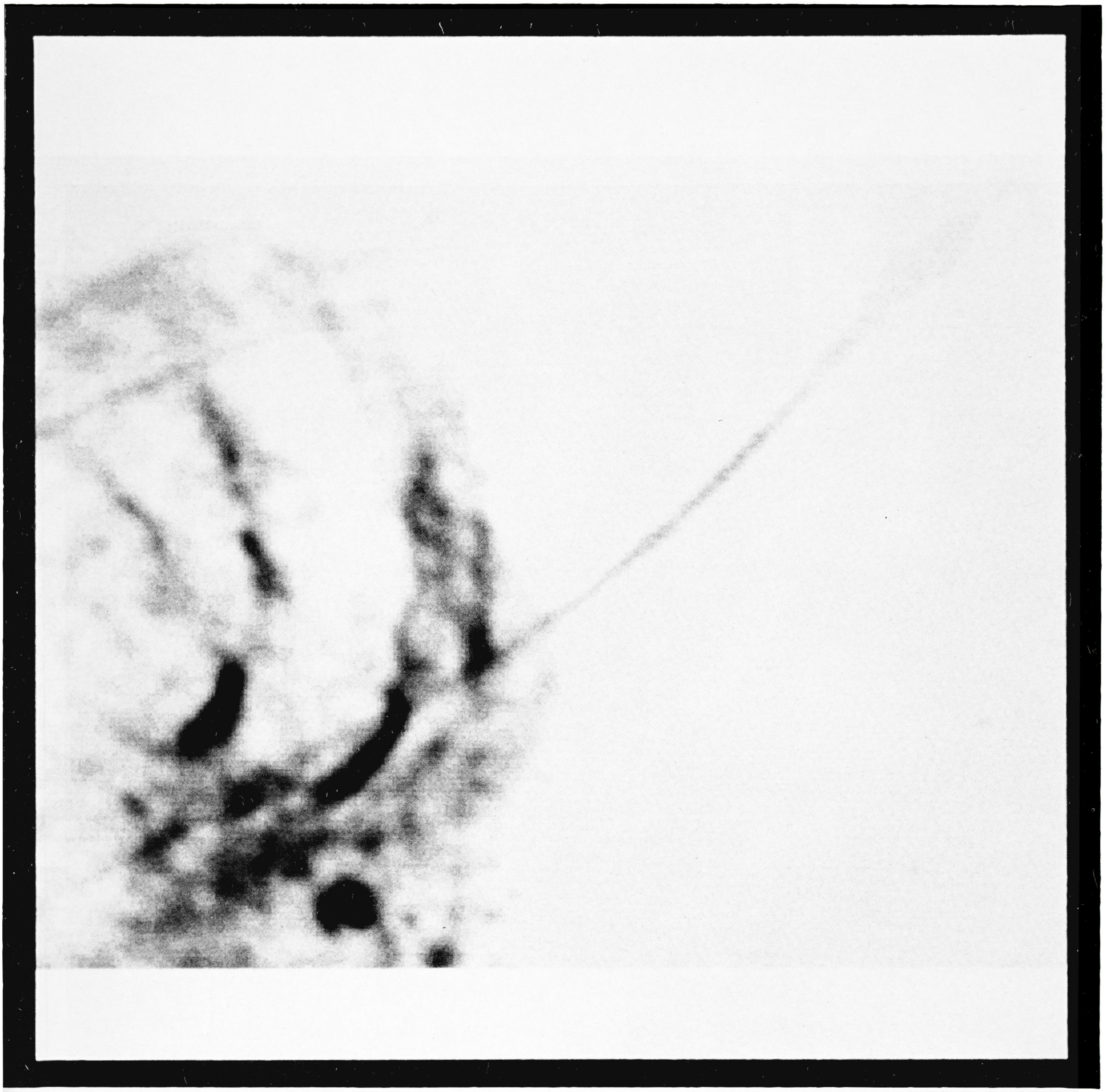

Figure 2: Another 20-cm map showing the northern thread with a phase center near the Sgr A complex (Sgr A Halo in Table 1 of Chapter 2). Maximum Entropy deconvolution is used for this map with the required total flux and required residual error of 440 Jy and 1.5 mJy/beam area. The final map was convolved with a CLEAN beam of ~12.5"×9.7" (P.A. = 62°).

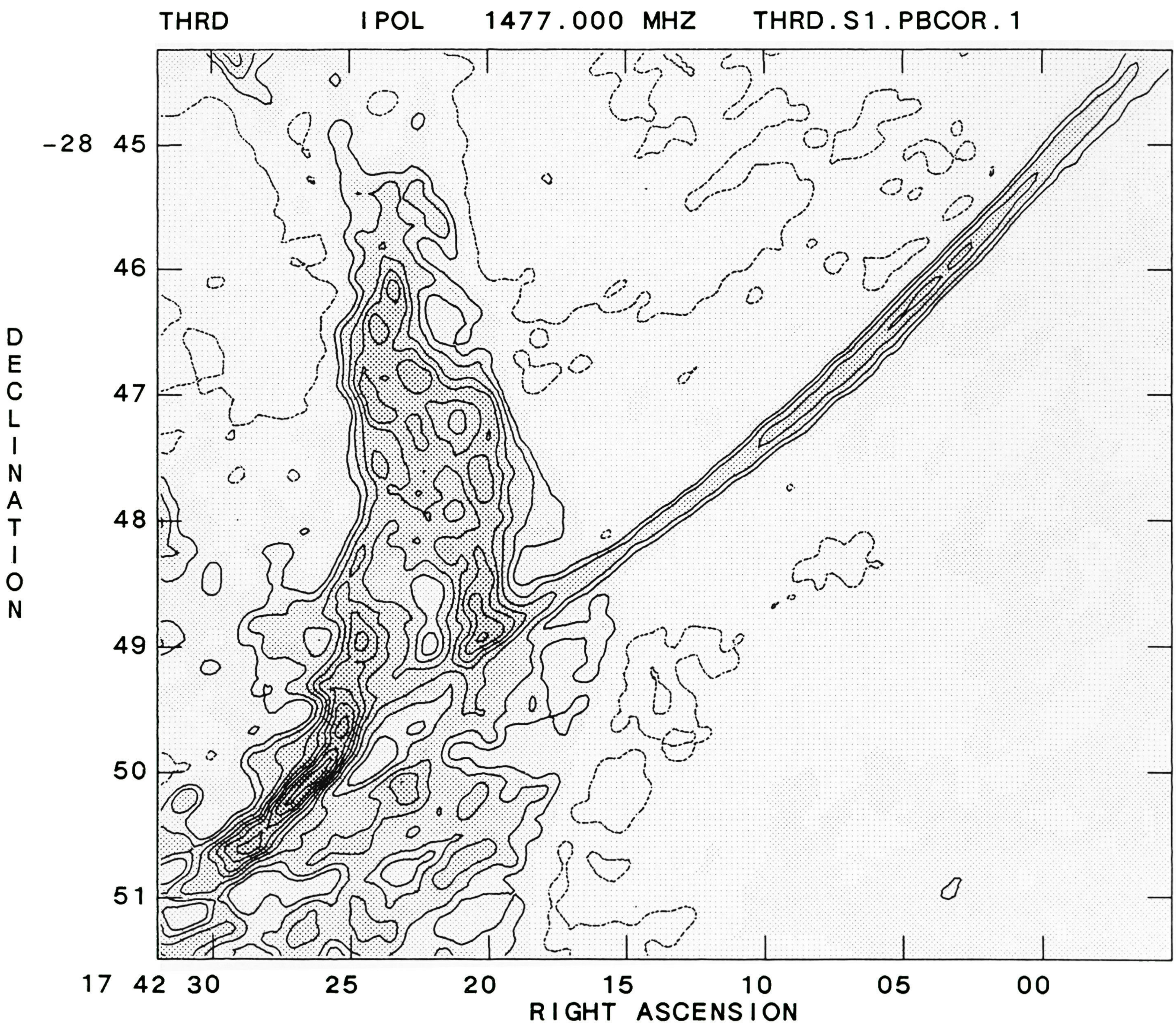

Figure 3: The contours of total intensity at 20 cm with intervals of -5, 5, 10, 20, 30, ..., 100, 120, ..., 200, 240, 280, 320 mJy/beam area. This map is based on the data base described in Figure 1.

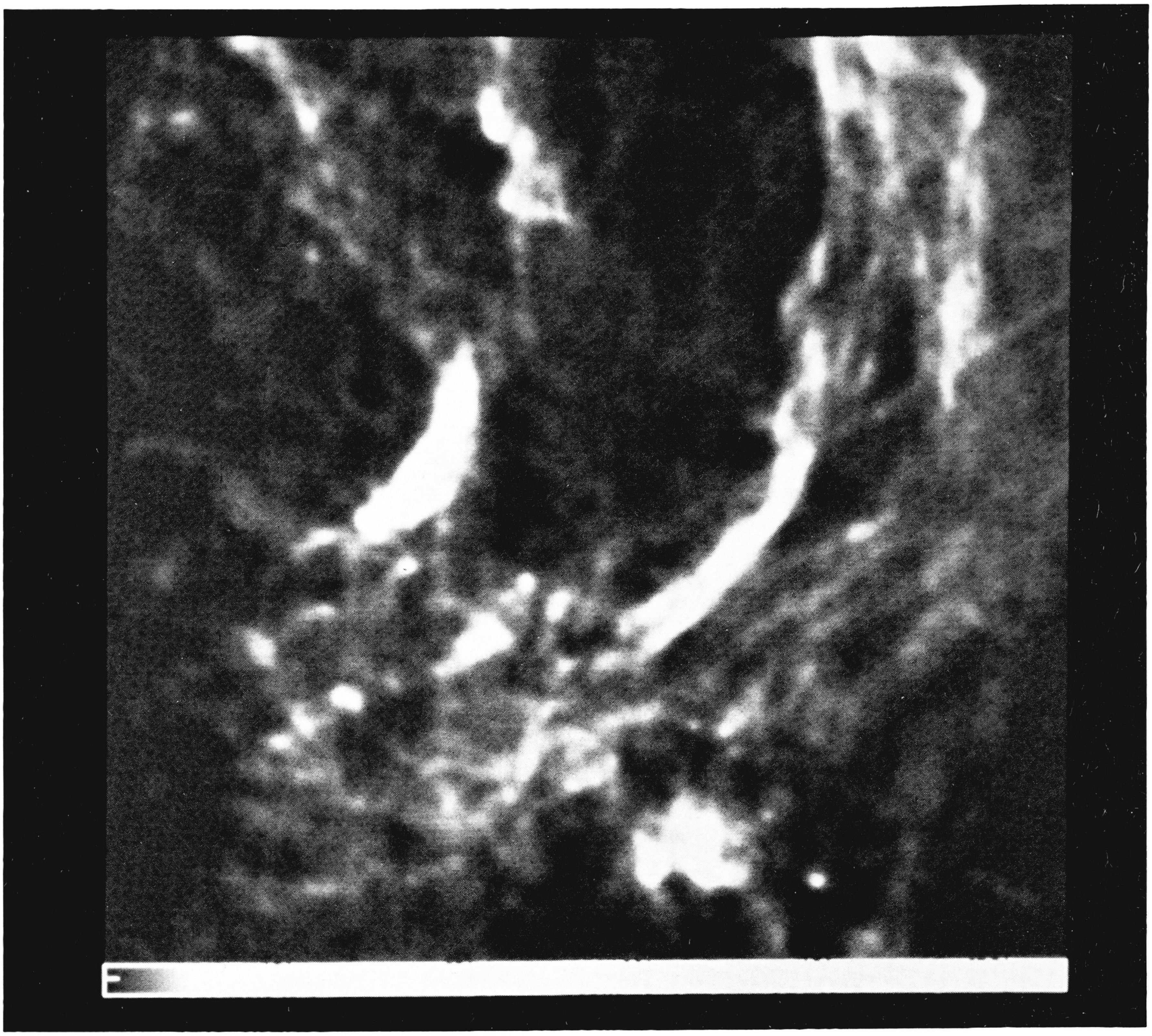

Figure 4: This 6-cm radiograph is a portion of a larger image shown in figure 10 of chapter 3.

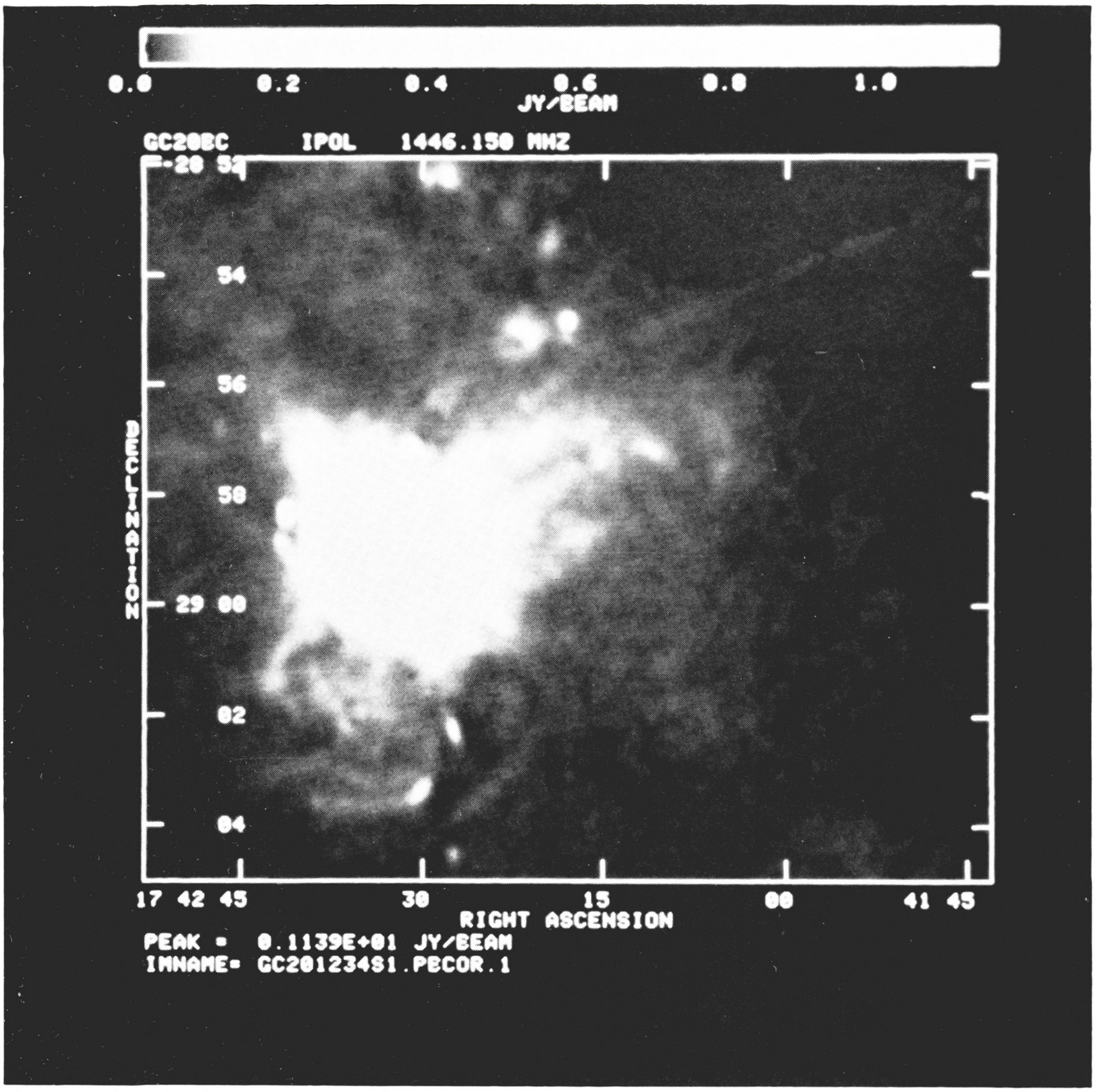

Figure 5: This is a portion of a large image shown in figure 1 of chapter 3. The phase center is located near the polarized portion of the Arc at G0.16-0.15 (GC20 field in table 1 of chapter 2).

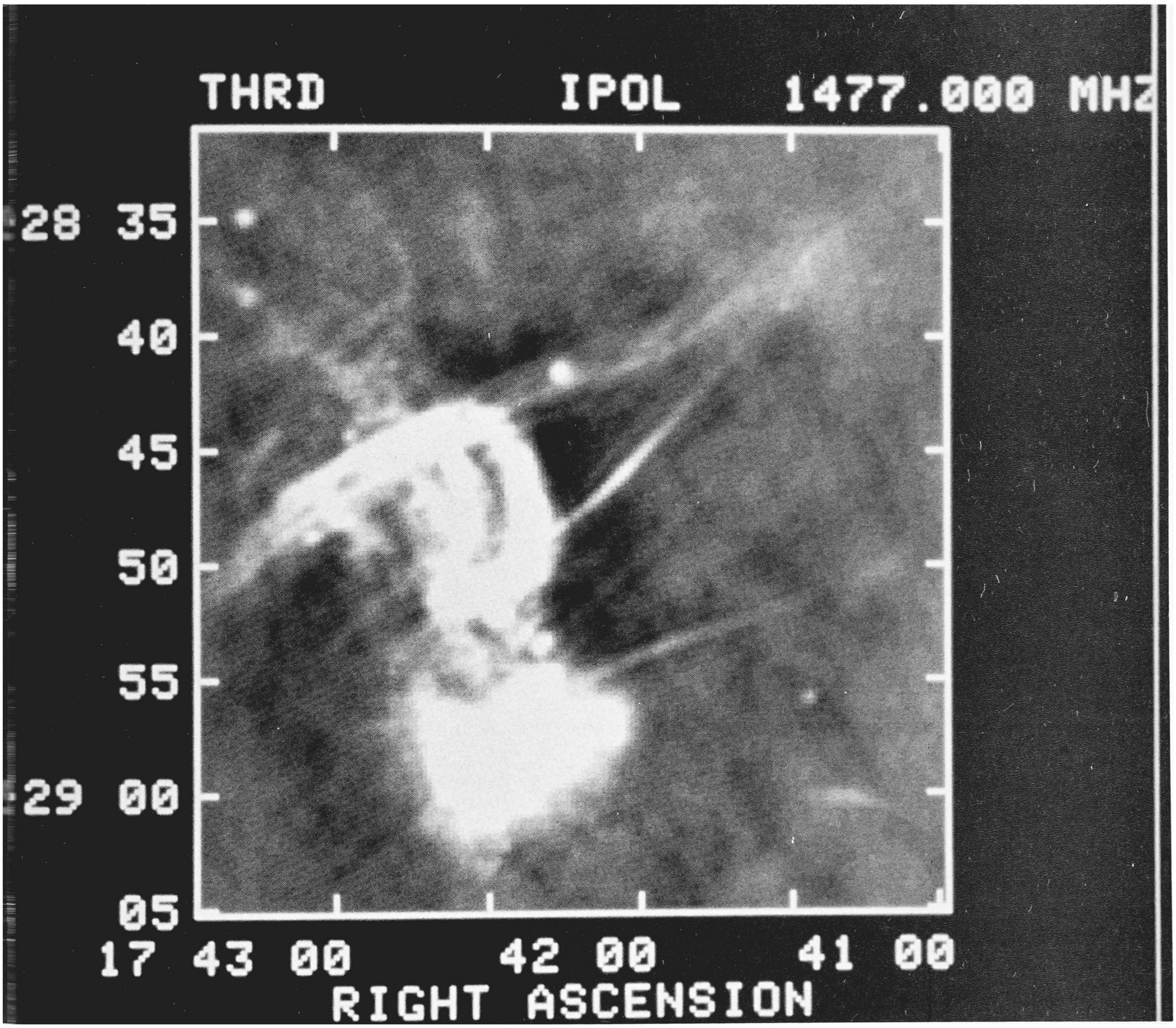

Figure 6: This image is similar to figure 1 of chapter 10 except that a high-contrast transfer function is used. FWHM = 30"×30".

# Chapter 6

## STRUCTURAL DETAILS OF THE Sgr A COMPLEX: POSSIBLE EVIDENCE FOR A LARGE-SCALE POLOIDAL MAGNETIC FIELD IN THE GALACTIC CENTER REGION

"Where is the Life we have lost in living?
Where is the wisdom we have lost in knowledge?
Where is the knowledge we have lost in information?"
T. S. Elliot

### I. Introduction

Although the nucleus of our Milky Way Galaxy is relatively calm in terms of its energy production and radio output, it offers a relatively clear vision of its structure by virtue of its proximity. It may thereby yield essential clues to the nature of activity in galactic nuclei in general.

Radio observations to date show that the structure of the galactic center is complex, but is neither chaotic nor intractable. As can be seen in Chapters 3 and 5, a wealth of new detail is now available. Moderate-resolution radio continuum observations around the galactic center maximum Sgr A - the third brighest radio source outside the solar system in a 30' field of view at 20 cm - reveal a structure consisting of a core surrounded by a halo of 3' radius (Pauls *et al.* 1976). The core itself consists of two components: Sgr A East, which has non-thermal characteristics, and Sgr A West, which has a flat spectrum and is apparently thermal (Downes and Martin 1971). In this chapter we describe new structural details of the Sgr A complex, which consists of Sgr A East, Sgr A West, within which a non-thermal compact radio source is buried, and the halo surrounding them. Furthermore, we attempt to elucidate the physical association between Sgr A East, Sgr A West and the halo component.

Recent observations of Sgr A East (Ekers et al. 1983, hereafter EVSG) clearly show it to be an elliptical shell structure elongated along the galactic plane with a major axis of length 10.5 pc. Its ratio of major to minor axis is approximately 1.3 and its center is displaced by 2.5 pc in projection toward negative galactic latitude from the apparent dynamical nucleus of the Galaxy - Sgr A West. This shell, which appears to enclose the nucleus, shows a characteristically non-thermal spectrum, with spectral index $\alpha$ = -0.7 ($F_\nu \propto \nu^\alpha$) (Downes and Martin 1971; Dulk and Slee 1974; Ekers and Lynden-Bell 1972; Gopal-Krishna and Swarup 1976; EVSG). On the basis of this and its shell structure, Goss et al. (1983) assessed the earlier suggestion that Sgr A East might be a supernova remnant (Downes and Maxwell 1966; Jones 1974; Ekers et al. 1975; Gopal-Krishna and Swarup 1976). On the assumptions that Sgr A East is physically associated with Sgr A West, and that the surrounding density in the interstellar medium near Sgr A East is comparable to that in the galactic disk, Goss et al. (1983) conclude on the basis of the surface brightness-diameter relationship that the supposed supernova must be young, having an age between 140 and 440 years. No significant degree of linear polarization - a characteristic which might be expected to accompany non-thermal radiation from a supernova remnant - has been detected from Sgr A East using single-dish radio telescopes (Gardner and Whiteoak 1962; Mayer et al. 1964). The Westerbork observations of Sgr A East have provided an upper limit of 8% on the linearly polarized flux at 6 cm (Ekers et al. 1975). Recently, Seradakis et al. (1985) found an upper limit ~1% at 3 cm using the Bonn telescope - see chapter 8 for more detail.

The low frequency observations of Sgr A by Dulk and Slee (1974) reveal that the peak flux at 160 MHz does not coincide with Sgr A West and that the Sgr A complex is more extended toward the southeast (negative galactic latitudes) than toward the northwest. Recent 160 and 110 - MHz observations of the Sgr A complex show that this extension has a length of ~30 pc and is directed along the rotation axis of the Galaxy (see Chapter 7; Kassim et al. 1986). The 10-GHz map of Sgr A made by Pauls et al. (1976) shows that the halo of Sgr A has a total flux density of ≃ 40 Jy, a factor of two less than the Sgr A East shell and its interior. They argue that the halo of Sgr A is non-thermal because its brightness temperature is much greater than $10^4$ °K at 160 MHz. Gopal-Krishna and Swarup (1976) determined the spectral index of the halo, which was then called "the background source", and concluded that it had a flatter spectrum than other Sgr A components. More accurate spectral index measurements of the halo were recently made by Mills and Drinkwater (1984) by comparing their 843 MHz observations with the 10-GHz map of Pauls et al. (1976). They concluded that the spectral index of the halo, which ranges from -0.6 to -0.1, has a north-south gradient, and argue that this is indicative of a mixture of thermal and non-thermal radiation.

VLA observations of Sgr A West showed a thermal "3-arm, spiral-like" feature within which an optically thick, compact, non-thermal source dominates the radio emission (EVSG; Brown and Johnston 1983; Lo and Claussen 1984). Becklin and Neugebauer (1975) and Becklin et al. (1978) reported several compact sources at 10 and 20 μm in Sgr A West. Many of these discrete sources were observed to have thermal characteristics and to be associated physically with more extended emission seen in radio maps of Sgr A West (Lacy et al. 1979, 1980; van Gorkom et al. 1984; Brown and Liszt 1984; Serabyn and Lacy 1985; Mezger and Wink 1986). Brown and Liszt (1984), who assembled most of the salient observational information relating to the inner few parsecs of the Galaxy, argued for the rotating ring of gas and dust hypothesized by Sandqvist (1974), Becklin et al. (1982) and Genzel et al. (1984) having an inner radius of 2 pc and an inclination of roughly 70°. Observations made by Lacy et al. (1979, 1980) and EVSG show that a bar-like feature, having a very different kinematic structure than that of the rotating ring, is superimposed on the center of the ring and is oriented along the rotation axis of the Galaxy. (In radio maps, this structure combined with the two northern and southern segments of the ring, has the superficial appearance of a spiral). This "bar" of ionized gas harbors the enigmatic IRS 16 - which marks the highest concentration of stars in the nuclear region - and the nonthermal compact radio source. One of these two objects are presumed to coincide with the dynamical center of the Galaxy (Storey et al. 1983; Henry et al. 1984; Forrest et al. 1986). Based on the most recent kinematic information obtained from NeII observations of Sgr A West, Serabyn and Lacy (1985) argue that the appearance of Sgr A West can be described as the superposition of a number of features orbiting about the nucleus of the Galaxy, although Lo (1985) concludes, on the other hand, that the radio features of Sgr A West appear remarkably continuous and filamentary at high resolution ($0''.3 \times 0''.6$). Quinn and Sussman (1985) have recently modeled the continuous picture of gas in Sgr A West which has resulted from the infall and tidal break up of a molecular cloud in dissipative medium near the galactic center.

Description of the observational technique employed in using the VLA at both 6 and 20 cm can be found in chapter 2. Construction of radio images presented here are similar to the procedure described in chapter 3 (see also chapter 2) In the next section we present our observational results. In the discussion that follows (§III), we suggest that the available evidence is consistent with a large-scale poloidal magnetic field lying at the galactic center. Furthermore, we argue in favor of the suggestion that ionized gas is outflowing isotropically from the nucleus and that Sgr A East is located behind Sgr A West.

## II. Results

The weak, extended emission which surrounds both Sgr A East and West appears to have structure on a variety of scales. In order to delineate the important features seen in this complex region, their descriptions are grouped into 6 subsections followed by presentation of the polarization and spectral index measurements. The "finding chart" seen in figure 1 depicts most of the features which will be described in this section, as well as others which have been noted in the past.

A) Large-Scale Elliptical Halo

The radiograph of 20-cm intensity from the Sgr A complex is shown in figure 2. Sgr A West and the elliptical Sgr A East shell (EVSG) dominate the brightness of this complex. An extended, roughly elliptical emission region which is concentric with, and has the same shape and orientation as, the Sgr A East shell can best be seen in this figure. This elliptical halo component has about the same ratio of major-to-minor axes (~1.5) as Sgr A East, but is about twice as large in each dimension and its mean surface brightness (550 °K at 20 cm) is lower than that of the shell and its interior by a factor of ~3.5. Like the shell structure, the major axis of the elliptical halo is aligned with the galactic plane; both are symmetric about b = -3.'65 which is displaced by 55" toward negative latitudes from the constant latitude plane which passes through the non-thermal point source at the galactic center (indicated as a black dot in figure 2). At positive longitudes the halo shows its greatest extension along the plane. The elliptical halo drops to a surface brightness of 200-300 °K at 20 cm at a distance of 6' from the galactic nucleus - the shape of the halo becomes spherical centered not on Sgr A East but on the nucleus at this distance - and stays at about that level out to 10', where it merges with the emission associated with the Arc (see chapter 3).

B) Large-Scale Protrusions of Sgr A West

One of the major components of the Sgr A complex is the group of several large-scale (2-5 arcminutes) protrusions arising near the center of the complex. The sequence of images shown in figure 3(a-f) displays six different aspects of the structures seen in the complex, using a sequence of contrasts. The protrusions are linear emanations from within the Sgr A East shell and are directed predominantly toward the galactic poles. They are clearly more prevalent toward positive latitudes than negative ones. Some of these linear protrusions display a slight curvature, and at $17^h42^m08^s$, $\delta$ = -28°56'38", 7 arcminutes from Sgr A West where the surface brightness of the northern protrusions decreases to 100 °K, they become parallel with the galactic plane, and possibly even curve back toward the plane (figure 4).

Far infrared observations indicate that a central source has caused heating of the dust ring which surrounds the ionized gas associated with Sgr A West. Contours of 100 μm emission (Becklin et al. 1982), which delineate the limbs of the ring surrounding the 2-pc cavity in the gas and dust distributions, are superimposed upon the 20-cm radiograph in figure 4. Both the northeastern and southwestern radio protrusions appear to be localized and concentrated mostly above or below the inner 10 pc of the Galaxy. This width corresponds to the extent of the 100 μm emission constituting the ring, suggesting that the protrusions may have their origin in or within the ring. In spite of the limited area (4') covered by the 100 μm observations, the comparison of the 100 μm and the 20 cm emission provides evidence in support of the physical association between Sgr A West and the protrusions. Furthermore, we note that the orientation of the continuum bar feature (Brown and Liszt 1984; Lacy et al. 1980) is the same as that of the northern protrusions.

Figure 5 shows a high dynamic range (~1000) radio picture of Sgr A at 6 cm exhibiting the filamentary nature of the northwestern protrusions as they cross the Sgr A East shell. We note that the limb-brightened shell constituting Sgr A East is continuous and is not distorted at the location where the set of filaments which compose the northwestern protrusions cross the shell. The gap that has been seen between the eastern and the northern arms of the "3-arm spiral" (EVGS; Lo and Claussen 1984) is occupied by much of this fine filamentary structure. Some of these protrusions are linked clearly to the northern and eastern arms.

Other new aspects of the structure of Sgr A West can best be seen in figures 6(a-b): the southern arm appears to continue far beyond its presently organized extent with a low surface brightness (EVSG; Lo and Claussen 1984), in the direction of negative galactic latitudes. The southern extension of the arm constitutes the symmetrical counterpart of the northern arm as it merges with northwestern protrusions (figures 1 and 6a). Overall, there is a remarkable symmetry in that respect. Figure 6b shows best a horizontal ridge of weak emission which is located to the south of the ionized bar and which appears to link the southern and northeastern arms ($\delta \sim -28°59'30"$). The kinematic information obtained from spectra 39, 40, 36 and 57 of Serabyn (1984) indicated that this feature is a thermal feature showing both red and blue shift velocities. Its radial velocity changes from 0 to -10 km $s^{-1}$ and then increases to +25 $km^{-1}$ in the direction from west to east. A compact source which coincides with IRS 8 can be recognized at the northernmost tip of the northwestern arm at $\alpha \sim 17^h42^m28\overset{s}{.}5$, $\delta \sim -28°58'45''$). Curiously, a number of weakly emitting features which appear to be associated with the protrusions seem to branch out from this compact source (see figure 5). Indeed, this souce coincides with positions 29 and 30 of Serabyn and Lacy (1985) who show that this is the region where non-circular velocity components are identified.

C) Parabolic Feature

Not all of the northwestern protrusions appear to join the "3-arm spiral" feature evident in figure 5. One of the northwestern protrusions - the "parabola" in figure 1 - appears in both the 6 and 20 cm maps to cross the Sgr A East shell at $\alpha = 17^h42^m30^s$, $\delta = -28°58'20''$ (see figures 7 and 8). To the south of this position, this filamentary feature intersects the extension of the westernmost arm of the "3-arm spiral", and appears to curve gently (possibly as a result of a chance superposition) at $\alpha = 17^h42^m33^s$, $\delta = -28°59'45'$ toward the west (see figure 1). In order to see and appreciate the full extent of this filamentary feature which, if it is continuous, appears to follow roughly a parabolic trajectory, a number of figures with different contrast levels have to be examined and compared (see figures 2, 3d, 5).

D) Radio "Threads"

Curious features which we refer to as "threads" have been discussed in chapter 5 (see also Morris and Yusef-Zadeh 1985). One of these very narrow structures is evident in figure 3f where it merges with the northwestern protrusions (see figure 1 for the feature labelled "thread"). The possibility exists that this feature is continuous with the parabola or is an extension of other protrusions that may connect to the galactic nucleus. The figures presented here do not resolve the above suggestions, but we hope to clarify it with observations now in progress.

E) Other Noteworthy Features

Many fine-scale details which can be seen throughout the Sgr A complex are described next.

1) Linear Striations: In the highest resolution 20-cm map yet made (figure 9), the Sgr A halo appears to be composed of a system of linear striations running roughly perpendicular to the galactic plane. These are most evident throughout the low-surface brightness portions of the halo, particularly to the southwest and northeast. The typical dimensions of a striation are ~3'×10", and the typical separation between them (~30") can be noted in figure 10 which shows a slice cut across the halo along the galactic plane. Further high-resolution 20-cm observations of the Sgr A complex are in progress in order to clarify the nature of this feature.

2) A "Streak" Feature: A thin streak-like feature at $\alpha = 17^h42^m47^s$, $\delta = -28°59'30''$ having a width of a few arcseconds and a length of ~1' at position angle 120°, is one of the more prominent linear features seen in he fibrous halo of Sgr A (see figures 3e, 4). This feature, which is labeled "streak" in figure 1, is reminiscent of, but much shorter than the larger filamentary structures seen in the Arc (see chapter 3), or the radio "threads" seen near the galactic center (§II.D above and chapter 5).

3) Radio Shadows on the Sgr A complex: Several features which can be interpreted as radio shadows are present, presumably caused by free-free absorption within the Sgr A complex. Perhaps the most interesting of these is seen in figure 11a (a low resolution [13"×12"] 6-cm map) where the most prominent northwestern protrusion crosses the Sgr A East shell at $\alpha = 17^h42^m30^s$, $\delta = -29°58'20''$. At this point there is a clear dip (80%) in the intensity of the shell. Figure 8 (20-cm map) illustrates another possible radio shadow at a point where a faint, north-south linear structure crosses the northern part of the Sgr A East shell ($\alpha = 17^h42^m36^s$, $\delta = -28°57'30''$). The dip in the intensity of the shell at this position is 80%. One can therefore conclude that the linear structure and the northeastern protrusion lie in front of the Sgr A East shell.

Another long dark band roughly perpendicular to the galactic plane manifests itself to the south of the halo. This feature, which is best seen in figures 3f and 9 crosses the galactic plane at $\ell = -0.13°$ and has a length of ~3'. It appears to have the same dimen-sion and orientation as the small-scale striations (§II.E.1), which may indicate that at least some, if not all, of the many small-scale striations, may be absorption features attenuating the otherwise relatively uniform background emission from the halo.

4) A "Right-angle" Feature: A linear structure located to the south of the Sgr A halo, as seen in figure 4, extends ~ 2' along the galactic plane and then bends abruptly towards negative latitudes at $\alpha = 17^h42^m20^s$, $\delta = -29°02'30"$.

5) A Gap in the Shell: The Sgr A East shell does not seem to form a completely closed loop. The gap in the shell is best displayed in figures 2, 5, 6, and 7 where it can be seen that the cluster of HII regions reported by EVSG (see §II.E) lie at the position angle of this gap. Figure 11 shows best a link between the largest of these HII regions and the Sgr A East shell.

F) Discrete and Extended Components

Earlier papers by Downes et al. (1978), EVSG, and Ho et al. (1985) list many compact sources in the galactic center region. Here we catalog and present the contour maps of the recognizable emission peaks, including several that have not previously been discussed. The discrete sources near or within the Sgr A complex (the Arc) are listed in Table 1 here (in Chapter 3), very few of which are completely isolated or compact.

Figures 12a,b show the contour maps of the cluster of HII regions (A-D) located to the east of the Sgr A East shell at 20 and 6 cm respectively. These sources, which were first recognized by EVGS, are embedded within the halo. Source C is seen to consist of two components. Sources A, B and C whose maps are also shown at 2 cm by Goss et al. (1985) appear to have shell-like structures. Comparison of the peak surface brightness of source D between 6 and 20 cm indicates that this source has a steeper spectral index (i.e. $\alpha \simeq 1.6$) than those of A, B & C. Figure 12c shows the elongated features located to the south of the halo (E-G). These sources which were recognized by Ho et al. (1985) appear to be parts of two separate semi-circular features. Sources H1-H5 shown in figure 12d are located at the junction between the Arc and the Sgr A halo.

Two weak sources (I1, I2 and I3, see figures 12e and 12f) are located to the immediate north of the shell (see figure 6a); an isolated compact source (J, figure 12g) which is identified by Isaacman (1981) is projected onto the rotation axis of the Galaxy; an elongated (~3') wispy structure with a thread-like characteristic (source K, see figure 12h) and a cluster of 3 weak thermal sources (L1 - L3, see figure 12i) are all located to the southwest of the halo (see figure 1 of chapter 5); an elongated thermal source (M, see figure 12j) which was observed at 12.8 μm (Serabyn 1984) is located near Sgr A West and is clearly part of a large loop-like structure (cf. figure 5). Based on a mixture of red and blue shifts, Serabyn (1984) shows that source M and its loop-like extension (see his spectra 2 and 3) deviates from circular rotation about the galactic center.

The most prominent extended sources in the Sgr A complex are listed in Table 2. The stated 6-cm flux density is a lower limit since source sizes greater than seven arc minutes are resolved out when the VLA is used in the C/D configuration. Figure 12k and 12 l show the low-resolution contour maps of the Sgr A complex at 20 and 6 cm, respectively.

Table 1

The Compact Components of Sgr A

| Name Galactic Center | R.A.(1950) h | m | s | DEC.(1950) ° | ' | " | Peak Brightness Temperature (°K) 6 cm | 20 cm | Reference |
|---|---|---|---|---|---|---|---|---|---|
| A | 17 | 42 | 41.4 | -28 | 58 | 18.5 | 346 | 1370 | EVGS; |
| B | 17 | 42 | 41.5 | -28 | 58 | 29.4 | 165 | 505 | Ho et al. 1985; |
| C1 | 17 | 42 | 41.87 | -28 | 58 | 49.7 | 195 | 622 | Goss et al. 1985 |
| C2 | 17 | 42 | 41.5 | -28 | 58 | 56.1 | 66 | — | Goss et al. 1985 |
| D | 17 | 42 | 40.85 | -28 | 59 | 13.5 | 367 | 600 | Goss et al. 1985 |
| E | 17 | 42 | 30.01 | -29 | 03 | 23.2 | — | 684 | Ho et al. 1985 |
| F | 17 | 42 | 27.3 | -29 | 02 | 19.8 | — | 620 | Ho et al. 1985 |
| G | 17 | 42 | 27.35 | -29 | 04 | 33.1 | — | 446 | Ho et al. 1985 |
| H1 | 17 | 42 | 21.53 | -28 | 55 | 02.5 | 132 | 765 | Downes et al. 1978 |
| H2 | 17 | 42 | 18.01 | -28 | 54 | 53.4 | 606 | 1404 | |
| H3 | 17 | 42 | 19.29 | -28 | 53 | 26.3 | | 410 | |
| H4 | 17 | 42 | 22.81 | -28 | 52 | 57.7 | | 219 | |
| H5 | 17 | 42 | 27.9 | -28 | 52 | 18.0 | 154 | 893 | |
| H6 | 17 | 42 | 26.03 | -28 | 51 | 58.2 | 33 | — | |
| H7 | 17 | 42 | 24.5 | -28 | 52 | 24.5 | 47 | — | |
| H8 | 17 | 42 | 27 | -28 | 56 | 24.5 | 47 | — | |
| I1 | 17 | 42 | 30.99 | -28 | 57 | 06.5 | — | 855 | |
| I2 | 17 | 42 | 30.7 | -28 | 57 | 17.0 | — | 738 | |
| I3 | 17 | 42 | 43.1 | -28 | 56 | 55.9 | 62 | — | |
| J | 17 | 41 | 26.4 | -28 | 55 | 57.5 | — | 945 | Isaacman 1981 |
| K | 17 | 41 | 17.8 | -29 | 00 | 11.0 | — | 233 | |
| L1 | 17 | 42 | 54.4 | -29 | 09 | 02.0 | — | 194 | |
| L2 | 17 | 42 | 55.1 | -29 | 10 | 01.0 | — | 240 | |
| L3 | 17 | 42 | 53.58 | -29 | 10 | 58.0 | — | 344 | |
| M | 17 | 42 | 25.89 | -28 | 59 | 27.9 | 294 | | Serabyn 1984 |

Table 2
The Extended Components of Sgr A

| Name | Flux Density (Jy) 4.86 GHz | 1.446 GHz | Diameter $\theta_{max}$ | $\theta_{min}$ | Reference |
|---|---|---|---|---|---|
| Sgr A Complex | > 91.5 | 328 | 7' | 6' | |
| Sgr A East & West | 67.7 | 137 | 3'.5 | 1'.5 | EVGS |
| Sgr A West & Compact Source | 32.8 | 31.6 | 1.3 | 40" | Brown & Johnston 1983 |
| Northwestern Protrusions (exterior of shell) | >14.5 | 49 | 2'.5 | 2' | |

G) Linear Polarization Measurements

We attempted to measure linearly polarized emission at 6, 18, and 20 cm from the northwest protrusions. No significant polarization was detected from these features to the limit 0.3%. The Sgr A East shell, located a few arcminutes off the phase center, showed a linear polarization of 1% at 6 cm. However, because of an increase in the level of instrumental polarization away from the phase center (Bignell 1982), this value for the polarization is not reliable.

H) Spectral Index Measurements

Comparison of 20-cm and 6-cm maps, centered on a point in the halo of Sgr A and using identical ranges of spatial frequencies, were used to investigate the spectral index of the northwestern protrusions. Our results indicate that this region appears to have a relatively flat spectrum, not greatly different from that of Sgr A West. This result supports our earlier claim based on the comparison between the 100 μm and 20 cm distributions (§II.B) that the northwest protrusion is possibly associated with the northern arm of the "mini-spiral". A comparison of the 2-cm map made by Kapitzky and Dent (1974), which shows an enhancement in the direction of the positive-latitude protrusions extending out to $\alpha = 17^h41^m50^s$, $\delta = -28°57'$, with our 6 and 20 cm maps also suggests that the spectrum of the northwest protrusions must be fairly flat.

Our measurements indicate that the Sgr A East shell has a continuously decreasing spectral index, proceeding from its interior to its exterior. The non-thermal spectra of the Sgr A East shell and its elliptical halo strengthen our argument that they are physically associated with each other.

The spectral index results noted above agree with the higher-resolution map which was made by EVSG. It should be pointed out, however, that there are uncertainties in our measurements which stem from the fact that the short (u,v) spacings, which can only be obtained using single dish observations, are missing in our data sets and therefore the map presented in figure 10 1 misses the flux contributions from structures extended over > 6'-7'. Furthermore, our data sets were not obtained using full synthesis capabilities or scaled array configurations at 6 and 20 cm, and thus our spectral index maps might not be free of systematic errors. Further work to untangle different features at a number of high and low frequencies is required to obtain truly reliable spectral index maps. The spectral index results of the compact sources listed in Table 1 will be published elsewhere.

## III. Discussion

### A) Sgr A East and its Elliptical Halo

On the basis of the above-described properties of Sgr A, we note that the halo of Sgr A appears to be part of the same physical system as Sgr A East. The two features are concentric, are roughly symmetric with respect to the galactic plane, and have roughly the same shape and orientation. Furthermore, the spectral index decreases systematically and continuously outward from the Sgr A East shell through the halo. However, the halo appears to be more extended towards positive than toward negative longitudes. The large-scale molecular and dust distributions in the inner 100 pc of the galactic center are also asymmetric in the same sense, but are much more extended towards positive longitudes than the radio halo of Sgr A (Fukui et al. 1977; Alvarez et al. 1974). The molecular distributions in the galactic center region, as revealed by OH absorption (Cohen and Dent 1983), $H_2CO$ (Scoville 1972), $NH_3$ (Kaifu et al. 1975; Morris et al. 1983), and CO (Bania 1977; Liszt and Burton 1978) indicate that the asymmetry extends even to scales as large as one kpc. Inasmuch as the distributions of ionized and cool gas are similarly asymmetric, they may be physical related (Alvarez et al. 1974; Fukui et al. 1977; Güsten and Downes 1980; Bieging et al. 1980) and if so, this supports the notion that the physical system of Sgr A East and its halo might be located close to the nucleus of the Galaxy based on the evidence that the cool gas is distributed near the nucleus.

Is the halo a secondary manifestation of the supernova explosion which others have suggested to account for the characteristics of the Sgr A East shell (see §I)? Indeed, the core-halo structure of the Sgr A complex does not resemble the structure of a typical supernova. We note that this halo structure resembles somewhat the X-ray halo seen surrounding Cas A (Steward et al. 1983); a number of suggestions have been made to account for this halo (Steward et al. 1983; Morfill et al. 1984). One of the mechanisms that Steward et al. suggest is a leakage of accelerated cosmic ray electrons in the remnant having velocities which exceed the shock speed of the supernova.

This suggestion might be particularly relevant to Sgr A East, assuming that synchrotron radiation from the non-thermal halo of Sgr A is produced by cosmic ray electrons. It is also possible that cosmic ray energy would contribute to the heating of ambient gas in the immediate vicinity of the shocked region (see §III.E).

## B) The Origin of Sgr A East

We consider two alternative hypotheses for the origin of the Sgr A East shell: 1) It is a supernova, as others have suggested (see the references cited in §I), and its positional association with the galactic nucleus is therefore coincidental. The appearance of Sgr A East and its halo indicates that the shell is produced in a special environment such as the galactic center region (§III.A) 2) Sgr A East is phenomenologically coupled with Sgr A West. The difficulty presented by this alternative lies in accounting for the displacement of Sgr A West from the center of the Sgr A East shell and for the radio shadow that the northwestern protrusions from Sgr A West cast on the shell. Even so, it is intriguing that the galactic nucleus lies within the perimeter of the Sgr A East shell (in projection), and it is tempting to consider the possibility that the event which produced the shell is related to an event in the nucleus.

The displacement might be understood if the gaseous medium surrounding the nucleus (the presumed energy source) and the nucleus itself were in roughly uniform motion with respect to each other. In particular, if the circumnuclear medium were moving roughly eastward in the plane of the sky (that is, at about a 45° angle with respect to the galactic plane) at a velocity typical of non-circular motions near the galactic nucleus, say 50 km $s^{-1}$, then the 2.5-pc displacement could be achieved if the energy input came from an impulsive event comparable to a supernova $5\times10^4$ years ago. In this time interval, however, the shell should have entered the snowplow phase and ceased emitting bright, non-thermal radiation (see discussion in Goss _et al._ 1983 and van den Bergh 1983).

A steady nuclear wind might be contemplated as the source of energy. Indeed, observations by Hall, Kleinmann, and Scoville (1983), and more recently by Geballe _et al._ (1984) suggest that a wind of $\sim10^{-3}$ Mo $yr^{-1}$ may emanate from the nucleus. If the Sgr A East shell is an extreme example of an interstellar bubble, then, from Castor, McCray and Weaver (1975) the shell radius is given by

$$R(t) = 28\left(\frac{\dot{M}v_{2000}^2}{n_o}\right)^{1/5} t_6 \ ,$$

where the wind velocity $v_{2000} = \frac{v}{2000}$ km $s^{-1}$, the mass loss rate $\dot{M}_6 = \dot{M}/10^{-6}$ $M_\odot$/yr, and $n_o$ is the ambient density. If $n_0 = 10^2$ $cm^{-3}$ and R = 5 pc, then we find the shell has been $1.4\times10^5$ years in the making. This is greater than the time needed to account for the displacement of Sgr A West from the center of the shell, but in this interval of time, one should expect the shell to have entered the snowplow phase. Despite maps of this region at many wavelengths, the shell has not been detected in 21-cm HI, radio-molecular lines, or IR lines such as those of OI, NeII, or $H_2$. Also, interstellar bubbles do not normally display non-thermal radio emission, so if this scenario is to be applicable, some other process must be added to it, such as the continual generation of relativistic particles within the shell (recent calculation by White [1985] suggest that in the mass outflow of hot stars, electrons can be accelerated to relativistic energies by shocks in the wind and emit synchrotron radiation).

C) Elongation of Sgr A East

What might account for the elongation of Sgr A East along the plane? Since the gas density is maximized in the galactic plane (Mezger and Pauls 1979), one might expect that, in the context of an explosive model for the origin of Sgr A East, the ram pressure in the galactic plane would impede material motions relatively more in directions parallel to the plane than in directions away from the plane. Furthermore, the rotation period of matter 5 pc from the galactic center is $\sim 2\times10^5$ yrs (Oort 1977), and therefore, the Sgr A East shell, whose expansion time scale is much less than this, would be dynamically unaffected by the gravitational field of the galaxy, assuming that Sgr A East is indeed located near the galactic center. Even if these effects were significant, the expected distortions from spherical symmetry would have been contrary to the observed. We therefore consider the possibility that the galactic magnetic field plays a significant role in determining the elongation of Sgr A along the galactic plane.

The high synchrotron emissivity observed in the inner 300 pc of the Galaxy (Little 1974) coupled with the result that the high energy cosmic ray intensity - based on COS-B observations of $\gamma$-rays with energies higher than 100 MeV (Audouze et al. 1979; Blitz et al. 1985) - is not different than the local intensity, imply a larger interstellar magnetic field strength than exists at the solar circle (Audouze et al. 1979). The earlier suggestion that a magnetic field plays a substantial role in confining the linear geometry of the Arc (see chapters 3, 9 and 10), leads us to consider the possibility that such a field is also dynamically important in the Sgr A complex. Such a conjecture is supported by the non-thermal nature of radio emission from the Sgr A East shell.

The piston model of an explosion in a uniform magnetic field in the limits where $V_a \gg V_{exp}$ and $V_a \ll V_{exp}$, where $V_a$ and $V_{exp}$ are the Alfvén and expansion velocities, respectively, was explored analytically by Bernstein and Kulsrud (1965) and Kulsrud et al. (1965). In the case where $V_a \ll V_{exp}$, they find that the shock propagates fastest in the direction perpendicular to the field lines. This effect might account for the observed distortion from spherical symmetry in the Sgr A East shell if the pre-explosion magnetic field is oriented perpendicular to the galactic plane and if it has a strength between $10^{-4}$ - $10^{-6}$ gauss depending on the choices of initial number density, n, the ratio of specific heats, $\gamma$, and $V_{exp}$. A gas density of less than 40 atom $cm^{-3}$, as reported by Watson et al. (1980), $\gamma$ = 1.2 (see figure 5b of Kulsrud et al. (1965), $V_{exp}$ = 160 km $s^{-1}$, and a ratio of semi-major to semi-minor axes ~1.3, as seen in the Sgr A East shell, would imply a magnetic field strength less than $1.1 \times 10^{-4}$ gauss using (see Kulsrud et al. 1965)

$$B = (5.78\times 10^{-2})\ (4\pi\rho\ V_{exp}^{2})^{1/2}\ .$$

This calculation is based on the assumption that the expanding piston has not swept up a quantity of interstellar gas comparable to its own mass.

In other studies by Rimer and Jen (1973), the effect of an interstellar magnetic field on the spherical piston model of a supernova explosion was pursued with no restriction on $e = V_a/V_{exp}$. Their numerical result, which confirms the results of Kulsrud et al., indicates that the shock deviates most from spherical geometry when e = 1; the major axis of the elliptical geometry that is formed is in the direction perpendicular to the magnetic field lines. This rather non-intuitive phenomenon arises from the coupling of gas-dynamical disturbances traveling at the sound speed $C_s$ and magnetic distur-bances traveling at the transverse Alfven speed in the direction perpendicular to the field lines. These effects are uncoupled at the pole and consequently the shock propagates slower along the field lines than in the direction perpendicular to the field lines.

We therefore suggest that the dominant geometry of magnetic field lines in the galactic center region is poloidal and that this field might be essential in shaping many of the radio features occuring in the galactic center region, including the filamentary structure of the Arc (Chapter 3), the very large-scale radio lobe seen by Sofue and Handa (1984) and Uchida _et al._ (1985), the new filamentary structure in Sgr C seen recently by Liszt (1985), and the striations seen in the Sgr A complex region (see §II.E1). Other evidence in support of this suggestion is provided by the infrared polarization of the galactic center sources at 11.5 μm (Capps and Knacke 1976; Knacke and Capps 1977; Maihara _et al._ 1977). The 11.5 μm observations of Sgr A reported by Knacke and Capps (1977) showed that the electric polarization vectors are perpendicular to the galactic plane, and thereby to the direction of polarization vectors of optical light which is presumably caused by extinction of interstellar grains aligned by the magnetic field in the galactic disk. They concluded that the 11.5 μm polarization might be caused by grains aligned by a magnetic field of $10^{-3}$ gauss permeating the innermost portions of the galactic center region and oriented perpendicular to the galactic plane. We now discuss a possible mechanism to maintain or amplify such a field geometry.

D) Dynamo at the Galactic Center?

The turbulent dynamo theory has been considered as a mechanism for the generation of the large-scale galactic magnetic field (Parker 1955, 1970, 1971 (a-c); Steenbeck et al. 1966; Vainshtein and Ruzmaikin 1972). In the "$\alpha$-$\omega$" dynamo process, the toroidal magnetic field is generated by non-uniform rotation ("$\omega$") in a rotating gaseous disk and the poloidal field is accomplished by the presence of turbulence ("$\alpha$" effect) having a definite helicity. A number of authors have attempted to explain the large-scale azimuthal field of the Galaxy inferred by Hiltner 1949, 1951, 1956, Hall 1949, Davis and Greenstein 1951, Vershuur 1974, and Heiles 1976. However universal agreement is far from being reached on how this field is generated. The differences of interpretation arise from the different assumed geometries (Parker 1971; Vainstein and Ruzmaikin 1972; Moffat 1978; Soward 1972; Ruzmaikin et al. 1979; Stix 1975; White 1978).

We attempt to give a plausibility argument for the poloidal magnetic field at the galactic center by exploring qualitatively the dynamo processes. Because of the strong potential well of the gravitational field of the stars, most of the gas is trapped in the central region of the Galaxy, and it might therefore be considered a closed system. It is possible for closed systems such as cloud complexes or the galactic center region to have their own internal differential rotation and cyclonic turbulence and therefore produce their own magnetic field. We discuss two possibilities that could generate poloidal fields in the galactic center region. 1) The dynamos in which the $\alpha$-effect acts as a source of both poloidal and toroidal field are called "$\alpha^2$-dynamos". The $\alpha^2$-approximation can be justified for a solidly rotating disk in which the shear rate is zero. Ruzmaikin et al. (1979) have adopted this approximation and suggested that it might be applied to the inner portion of the galaxy. The rotation curve of the inner portion of the Galaxy illustrated by Oort (1977) supports the adoption of the assumption of solid body rotation for the range of galactocentric distances between 50 and 750 pc. They concluded that in the region $0.28 < \frac{z}{h} < 0.72$, where $\frac{z}{h}$ the ratio of the distance perpendicular to the plane to the half-thickness of the galactic gas disk, the quadrupolar mode of the magnetic field is predominantly excited. Furthermore, in the immediate vicinity of the plane ($z = 0$) and at the boundary ($z/h = 1$) only the dipole is excited. If we assume $h \sim 100$ pc, the dipole generation of magnetic field might indeed be a dominant field structure close to the galactic plane ($h \simeq 30$ pc) and close to the nucleus ($r <$ 1 kpc). Another qualitative difference between the $\alpha^2$-approximation and the $\alpha$-$\omega$ dynamo of the Galaxy is the ratio of the radial component of the poloidal field relative to that of the azimuthal field at an arbitrary point. In the latter, it is expected to be much less than 1, (if $R_\alpha$ the Reynolds number due to cyclonic turbulence is small) and in the former the ratio is $\sim 1$ (see Ruzmaikin et al. 1979). It is noteworthy that the turbulent velocity (i.e. random velocity of clouds) within the inner 300 pc of the Galaxy is at least an order of magnitude greater than in the solar neighborhood (Loose et al. 1982; Sanders 1979) The large turbulent pressure in the galactic center region is presumably caused by a high rate of supernova explosions which stir up the gas. The high rate of supernovae is a natural consequence of the high stellar density in the galactic center region (Sanders 1979). 2) The second possibility is to assume that at the very center of the Galaxy where $R_\omega$, the Reynolds number due to rotation, is 100 times greater than in the solar vicinity, the "$\alpha$-$\omega$" dynamo process in a spherically symmetric geometry maintains the magnetic field. It is possible for the poloidal field to become the dominant mode if the region of interest is far from the central generator (Moffet 1978). This is of course analogous to the dipole field seen far away from the cores of the sun and the earth, where the dynamo generators are presumed to reside. Further detailed work is essential to illuminate and clarify both alternatives that are mentioned here.

E) The Radio Protrusions and Sgr A West

Although numerous features are superimposed throughout the Sgr A complex, the radio protrusions discussed in §II.B are likely to be associated with Sgr A West for the following reasons: first, a few of the protrusions are linked to features such as the "bar" and both the northern and southern arm of the "mini spiral" in the sense that they appear to be extensions of the small-scale linear structures in Sgr A West. Second, like Sgr A West, they have a flat spectrum. Also, most of the high latitude features in the Sgr A complex appear to be located just above and below the ring structure as depicted in figure 4.

We note that the northwest protrusion is coincident with an X-ray feature at $\alpha = 17^h42^m10^s$, $\delta = -28°57'$ which is present in the diffuse X-ray maps made by Watson et al. (1981, figure 2c) using the Einstein Observatory. We obtained the X-ray data of Watson et al. (1981) and convolved their full-resolution map with a gaussian beam of FWHM = $1' \times 1'$ using the CONVolution algorithm of the AIPS software package. The resulting map is reproduced here and is superimposed on our 20 cm radiograph in figure 12. An inter-comparison of the morphology recognized in both X-ray and radio maps suggests that the northwest protrusion has an X-ray counterpart. Watson et al. discuss a number of possibilities that could account for the production of this diffuse X-ray emission. The possibility that the radiation is due to bremsstrahlung in a hot plasma is, perhaps, the most likely one, since the protrusions have a flat radio spectrum. The total X-ray flux emitted from the northwest protrusions is $\sim 10^{35}$ erg/s, which then gives a number density of 1.25 $cm^{-3}$ if the estimated temperature is assumed to be $10^7$ °K. Similar results were obtained by Watson et al. when the number density and the temperature were considered over a larger area in their X-ray maps. Low resolution radio recombination line observations of the Sgr A complex and the Arc, which were carried out by Pauls et al. (1976), indicate that the highest electron temperatures, $T_e$, 9300 to 29 000 °K occur in a region located 7' to the northwest of Sgr A. In fact, the position at which the highest electron temperature was inferred (Pauls et al. 1976) appears to coincide with the northwestern protrusions.

Mills and Drinkwater (1984) calculate the electron temperature of the thermal component in Sgr A by comparing their 843 MHz map with the 10.7 GHz map made by Pauls _et al._ They find $T_e \sim 2 \times 10^4$ °K when Sgr A East is assumed to be located behind the nucleus. Therefore, the line and the continuum observations are consistent with the X-ray observations if both arise in a very hot thermal plasma having a wide range of temperatures but predominantly at 30,000 °K. We now argue that Sgr A must have a similarly hot environment near the non-thermal point source.

The possibility that the X-ray Einstein point source in the direction of Sgr A could be due to a transient A1742-289 has been discussed by Watson _et al._ (1981). We concur with the suggestion, as was stated by Watson _et al._, that the IPC point source might be associated with Sgr A West and not A1742-289 for the following reason: comparison of the diffuse X-ray and radio maps suggests that the overall X-ray distribution, particularly at positive latitudes, follows the radio protrusions perpendicular to the galactic plane (cf. figure 12). It was also suggested above that the radio protrusions are associated physically with Sgr A West and are thermal. Thus, it is plausible to associate the IPC point source with Sgr A West. We also note that the location of peak intensity in the diffuse X-ray map (cf. figure 13) is off set from the non-thermal compact source by 20" to the north coincident with the northern arm of "3-arm spiral" feature. Therefore, we suggest that Sgr A West must also contain a wide range of gas temperatures ($2 \times 10^3$ - $10^7$ °K) to explain the infrared, radio and X-ray emissions (Wollman _et al._

1977; van Gorkom et al. 1983; Watson et al. 1981; see Brown and Liszt for further references 1984).

The double-lobed structure of the dust distribution, as seen in figure 4, suggests that it is heated by a centrally condensed source (Becklin et al. 1982; Gatley et al. 1984), perhaps associated with IRS 16. The infrared distribution at 30, 50, 100 μm shows that the inner 2-pc of the Galaxy is a region of decreasing dust density and increasing dust temperature with decreasing radial distance from the center (Becklin et al. 1982). Kinematic information obtained from NeII emission (Lacy et al. 1979; 1980; Serabyn and Lacy 1985), radio recombination line emission (van Gorkom et al. 1983), 63 μm emission from neutral oxygen (Genzel et al. 1984), and from shocked hydrogen molecules (Gatley 1983; Gatley et al. 1984), combined with the dust distribution yield a picture in which a dust ring surrounds a region of molecular, ionized and neutral gas rotating about the galactic center in the sense of galactic rotation. The total luminosity emitted from the inner 1 pc, based on the infrared distributions, is $1-3\times10^{7}$ $L_{\odot}$ (Becklin et al. 1982). This value is determined by assuming an isotropic luminosity source with a soft spectrum at the center which emits isotropically and that most of the azimuthally-directed radiation is absorbed by the dust ring along the galactic plane. A comparison of the 100 μm dust distribution and the ionized gas distribution at 6 cm, assuming that the protrusions are thermal, shows that (cf. figure 4) the ionized gas extends beyond both Sgr A West and the dust ring in a direction both along and away from the plane. This picture suggests that the gas and dust are mixed within the inner 6' along the plane. Low resolution far-infrared observations at 540 $\mu$m carried out by Hildebrand et al. (1978) suggest that the southern segment of the ring structure is 5' away from IRS 16 along the galactic plane; no radio emission is seen from this region. Figure 7 shows a sharp right angle turn in the spatial distribution of the southern arm before it extends toward negative latitudes and weakens. Furthermore, the diffuse emission to the south of this sharp feature at $\alpha \sim 17^h42^m28^s$, $\delta \sim -28°59'37"$ shows a minimum in the distribution of radio emission when compared to its surroundings. This minimum coincides with the southeastern position of the shocked molecular hydrogen found by Gatley et al. (1984). These new structural details in the distribution of ionized gas in Sgr A West can be used as pictorial evidence in support of an outflow model suggested by Gatley et al. (1984) and Geballe et al. 1984.

The total flux density emitted from Sgr A West, the diffuse component surrounding Sgr A West (Brown and Johnston 1983; EVGS), and the northwest protrusions -40 Jy at 6 cm requires $\sim 4\times 10^5$ of Lyman continuum photons that the radio emission arises thermally from ab isothermal medium at 8000 °K. Most of the power of the central source, however, (i.e. ~80%) is emitted as non-ionizing photons (Gatley 1984). The emission measure in the region of the northwestern protrusions (i.e., exterior to the Sgr A East shell) is $\sim 1.3\times 10^4$ pc cm$^{-6}$ which gives an electron number density 45 cm$^{-3}$ if the depth is equal to 6.4 pc. The total mass of ionized gas from the northwest protrusions is 200 $M_\odot$ assuming that the ionized gas occupies a region of 200 pc$^3$. We note that many of the above physical parameters are uncertain, since the flux density at 6 cm is contaminated by the background radiation from the halo of Sgr A. Spectroscopic observations and high-resolution far infrared observations of the inner 6' of the Galaxy can be fruitful in determining more reliable values of these parameters.

F) The Location of Sgr A East and its Associated HII Regions

The appearance of a gap to the northwest of the Sgr A East shell (see §II.D6) and the location of a cluster of thermal sources at $\ell$ = -0°02, b = -0°07 (A-D in Table 1 and figure 12a), the incomplete shell-like structure of sources A & B (see figure 2a) whose steepest intensity gradients are opposite the Sgr A East shell, the appearance of a diffuse structure linking source A to the Sgr A East shell (see figure 11b) and a lack of much emission from the halo of Sgr A in the direction to the east of the HII regions (see figure 2) all might indicate that the Sgr A East shell and the complex of 5 HII regions are physically associated.

NeII observations of source A shows that it has a radial velocity of ~50 km $s^{-1}$ (Serabyn 1984). Also, radio recombination line emission from the cluster of HII regions was shown by Goss et al. (1985) to have the same radial velocity as M-0.02-0.07 - a molecular condensation in the 50 km $s^{-1}$ molecular cloud (Güsten et al. 1981). Goss et al. (1985) argue that these thermal sources are regions of star formation in the 50 km $s^{-1}$ molecular cloud. It is also known that this molecular cloud lies near the galactic center (Güsten et al. 1981; Güsten and Henkel 1983; the reviews by Oort 1977 and Brown and Liszt 1984; Ho et al. 1985). We also argued in §III.A that the halo surrounding the Sgr A East shell is physically associated with the shell based on their non-thermal spectra and on their concentric elliptical geometry. So, we infer that the Sgr A East shell, the elliptical halo, the cluster of thermal sources, and the 50 km $s^{-1}$ molecular clouds are an integral part of one physical system, and thus, are all located near the galactic center.

Goss et al. (1985) estimated that $1\text{-}6\times10^{48}$ Lyman continuum photons $s^{-1}$ - needed to ionize the HII regions - could have originate from four 07-09 stars in the 50 km $s^{-1}$ molecular cloud. Alternatively, the energy from the supernova explosion could be deposited directly at the edge of molecular cloud M-0.02-0.07 through the gap discussed above. In a relatively high-density medium, it is possible that the relativistic particles, the soft X-rays and the ultraviolet photons from the supernova explosion do not escape to infinity but deposit all their energy in the ambient gas in the immediate vicinity of the shock thus leading to the observed pockets of ionization. Similarly, van den Bergh (1971) suggests that Minkowski's HII region surrounding Cas A could have been formed by the supernova explosion.

We argued earlier that the radio protrusions are physically linked to Sgr A West. Here, we suggest that the Sgr A East shell and its associates lie behind Sgr A West and its protrusions: Figure 10a exhibits subtle radio shadows where the northwestern protrusion crosses the Sgr A East shell. We note that Sgr A East and Sgr A West may not interact with each other is provided by the smooth shape of the Sgr A East shell, which appears not to show any deviations from its elliptical form where the radio protrusions cross the shell. Indeed, if there were any interaction between the system of radio protrusions and the shell, one would expect to see some distortions of the shell at the location where the protrusions cross it. These effects can be explained by absorption of non-thermal emission from the shell as it passes through the thermal ionized gas which constitutes the northwestern protrusions. At lower frequencies the shell is not seen behind the location of Sgr A West itself at 327 MHz, presumably because of absorption. This geometry was also suggested by Güsten et al. (1981) and Güsten and Henkel (1983), who based their arguments on $NH_3$ and $H_2CO$ studies.

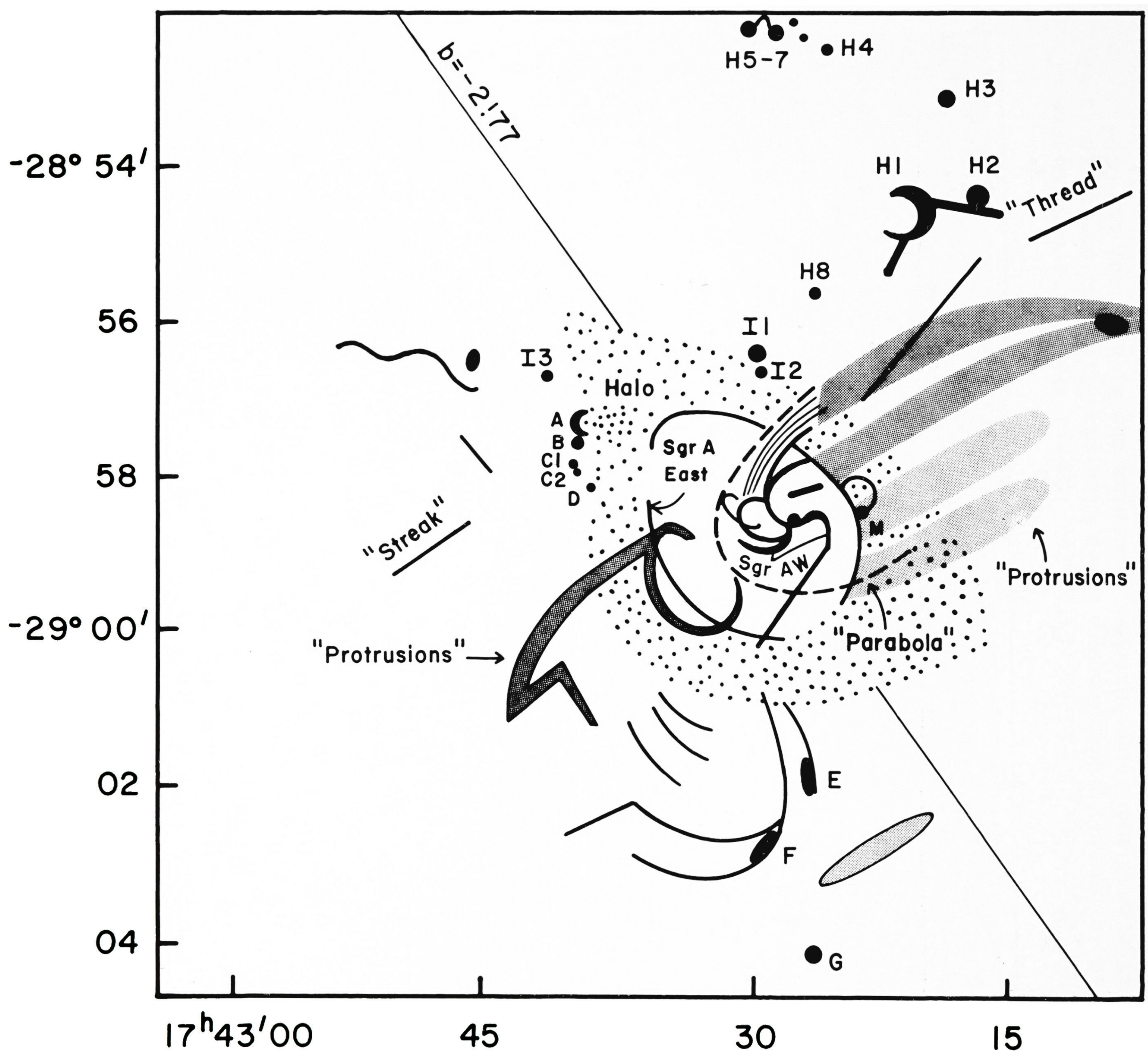

Fig.1: Finding Chart For Sgr A Radio Structures

Figure 1: Schematic view of the radio structure identified in the radio images of the Sgr A complex. The discrete and extended components drawn on this figure are listed in Tables 1 and 2.

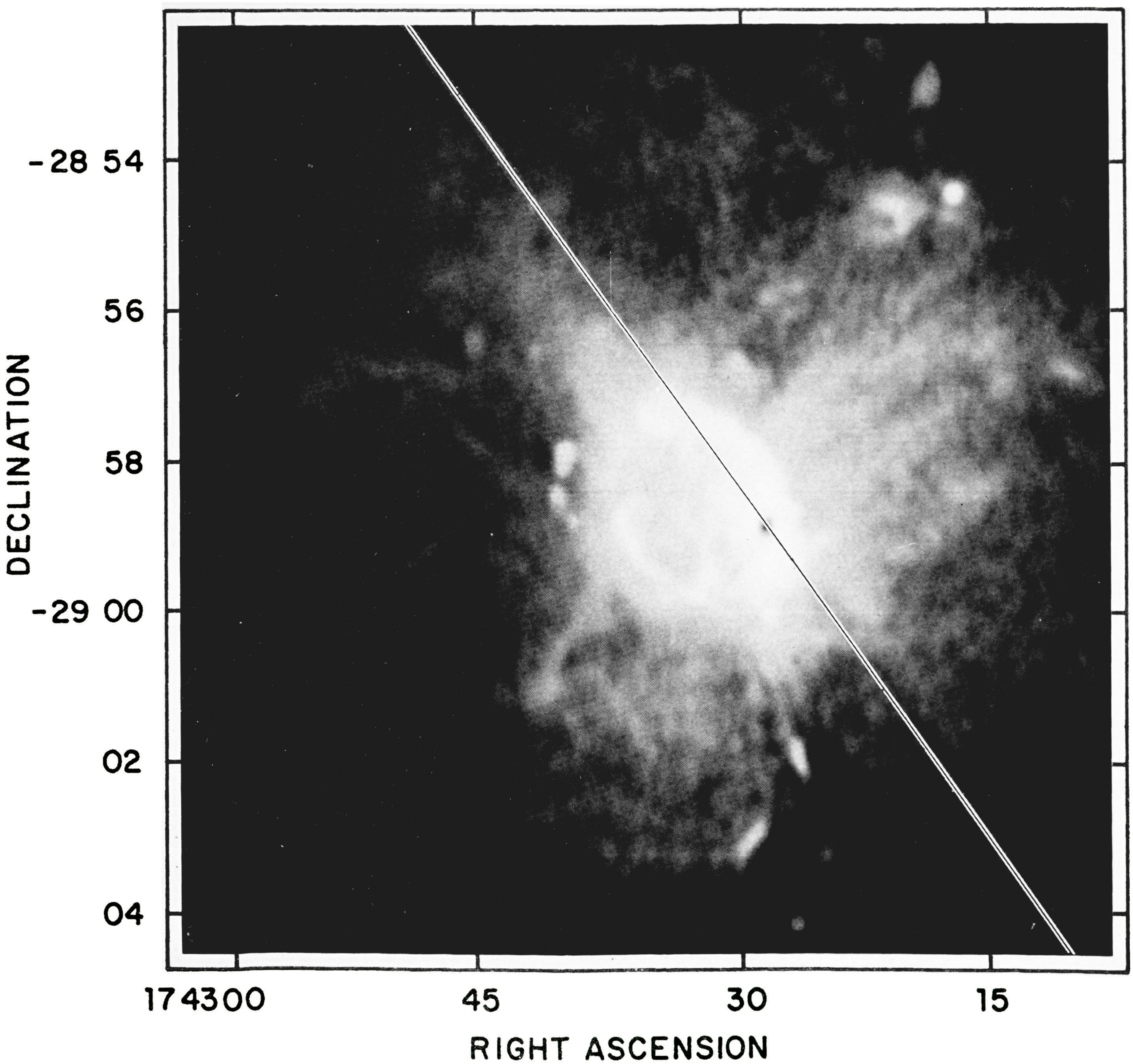

Figure 2: The 20-cm image of the Sgr A complex, including Sgr A West, the Sgr A East shell and its halo. A maximum entropy deconvolution technique was employed in making this figure. The dynamic range in this image, which has a spatial resolution (FWHM) of 5.″8×8.″5, is 800:1.

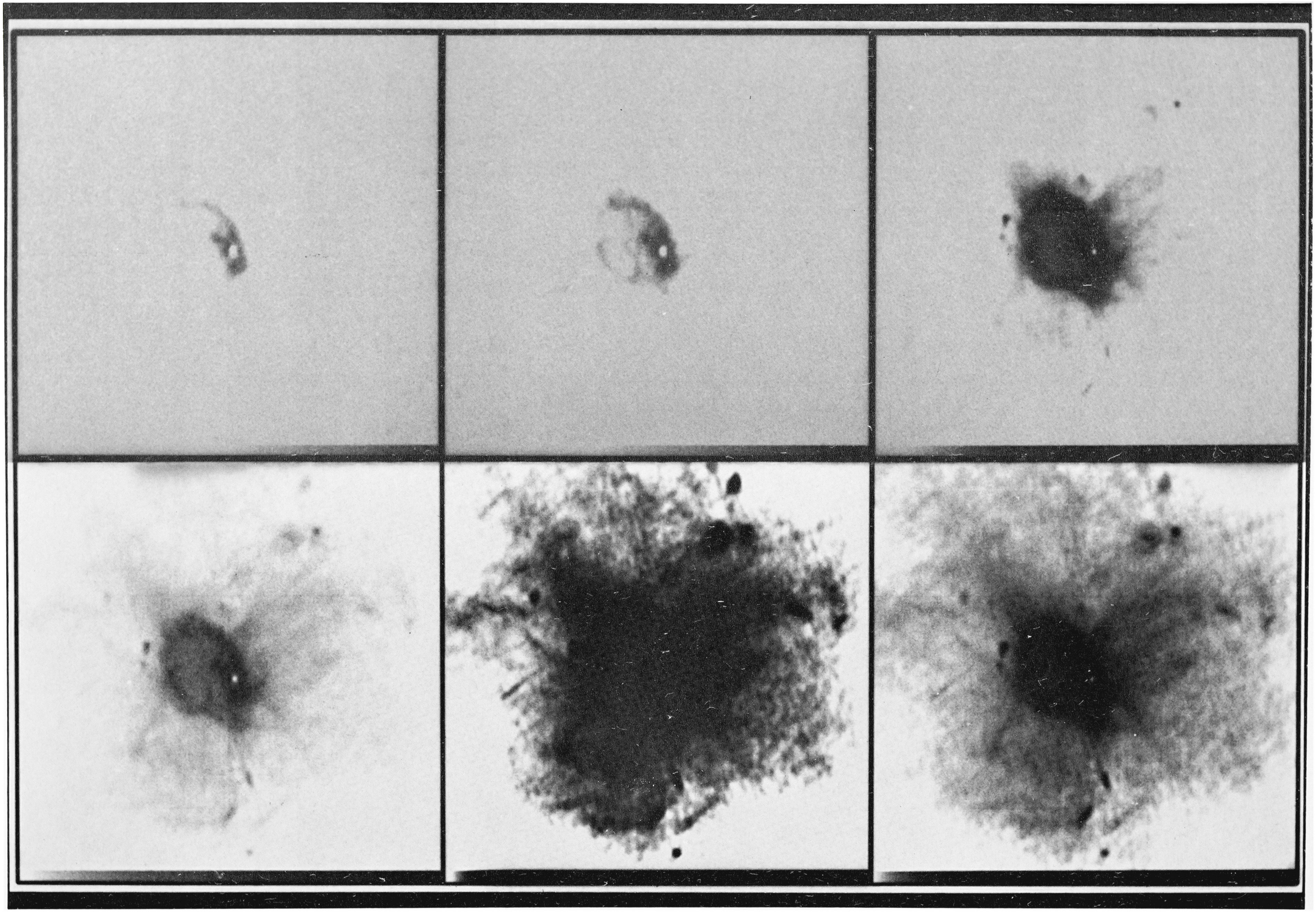

Figure 3 (a-f): This figure is identical to figure 2 except that different aspects of the radio structures in the Sgr A complex are shown with six different contrasts (a-f). The data for both figures 2 and 3 are based on combining the B, $B/C^1$ and the $C/D^2$ configuration arrays.

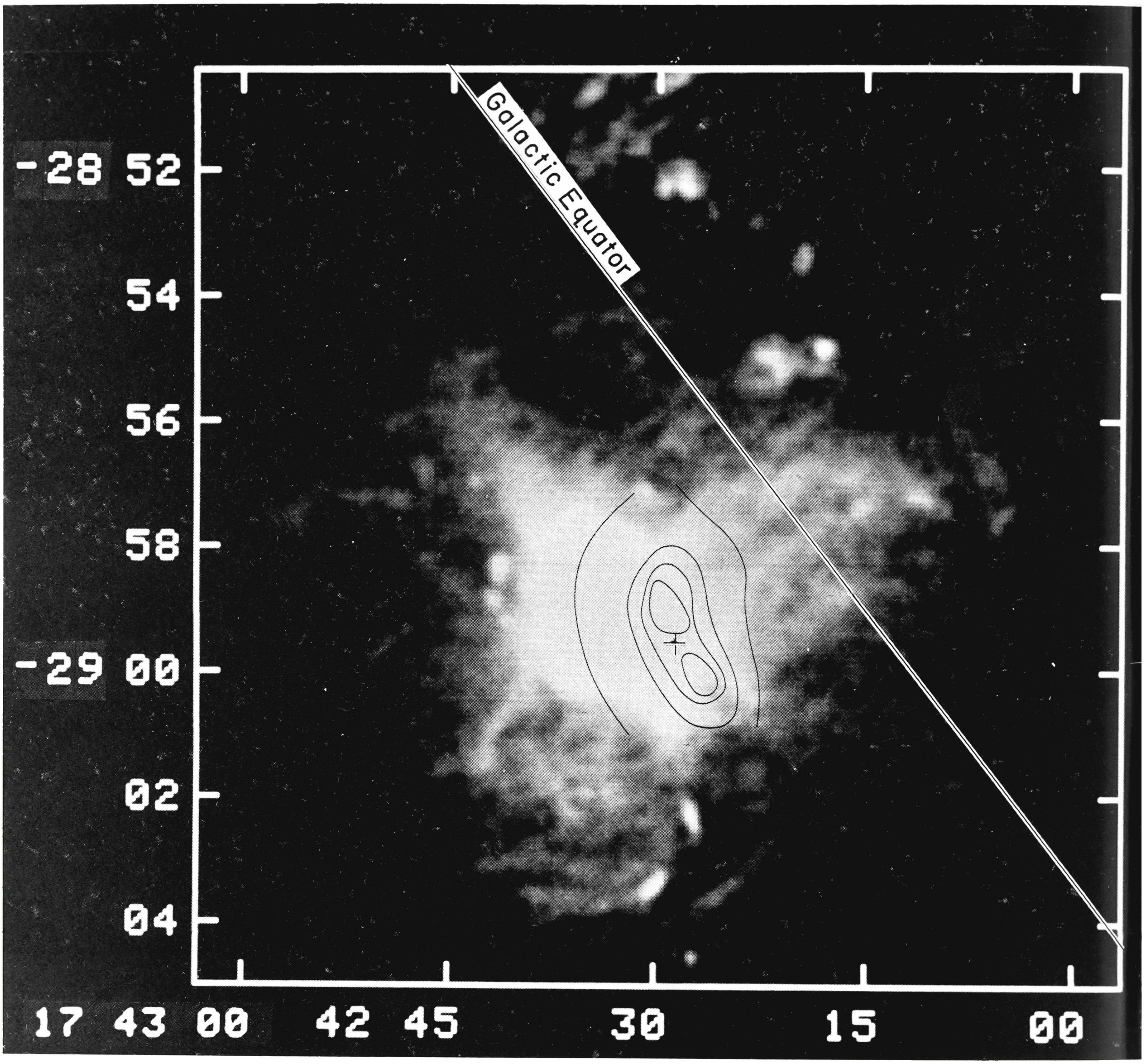

Figure 4: The 100 μm contour map of Sgr A West made by Becklin et al. (1982) is superimposed on the 20-cm map of the Sgr A complex with a resolution, FWHM, of 10.″8×10.″1. The C/D$^2$ array data have been added to those of figure 2 and 3, followed by a 25 kλ taper and application of the CLEAN deconvolution algorithm. The dynamic range in this figure is ~1500 and the peak intensity is 3.66 Jy/beam area.

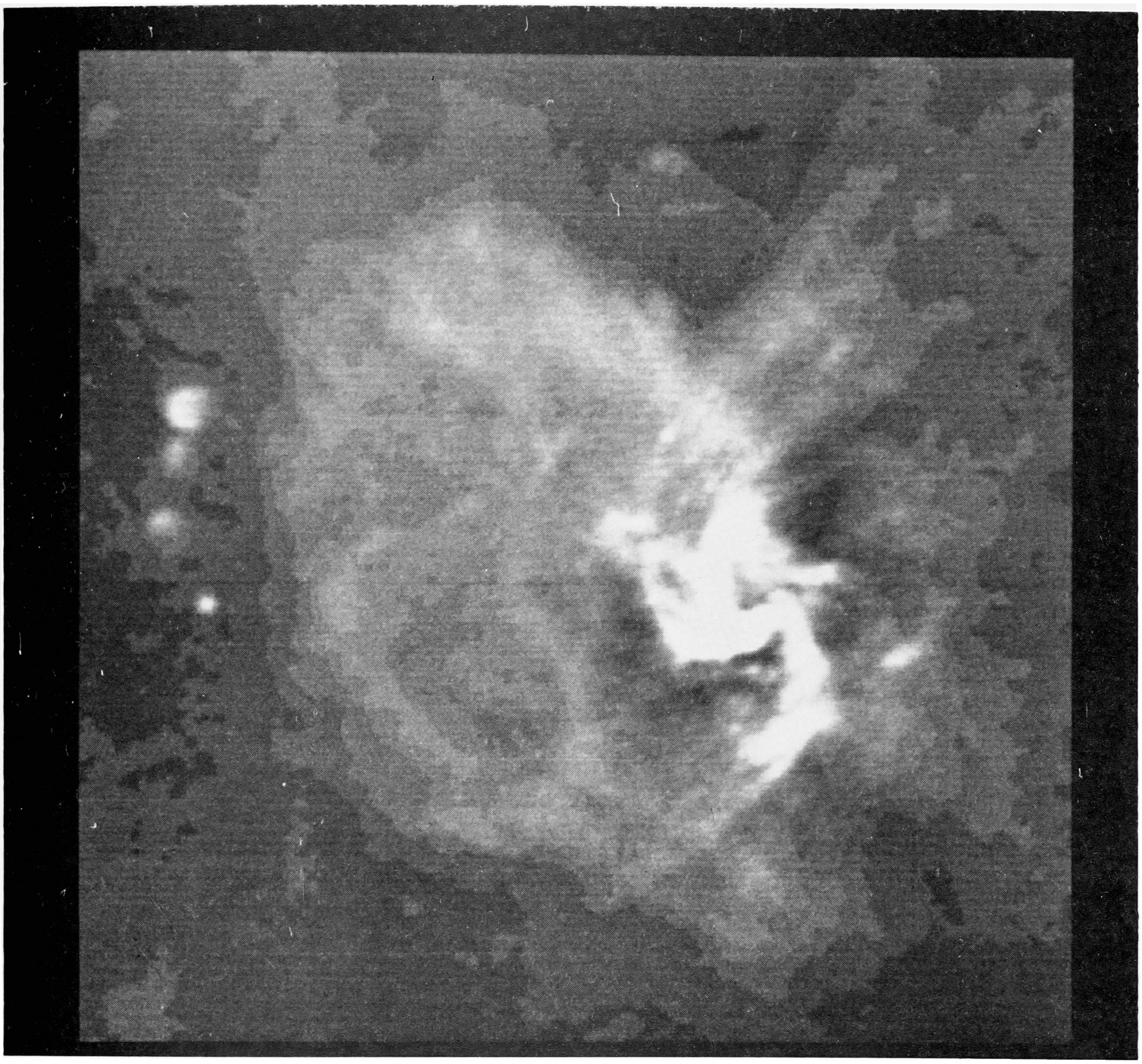

Figure 5, 6a, 6b and 7: These 6-cm maps are based on the $C/D^2$, $B/C^2$ and $C/D^3$ array data which are centered near Sgr A West and are constructed by application of CLEAN algorithm (FWHM = 3".39×2".9). Different transfer functions was used in order to show different aspects of the structures in Sgr A West and its vicinity. The peak flux density and the rms residual flux density are $1.04\times10^3$ and 0.74 mJy/beam area.

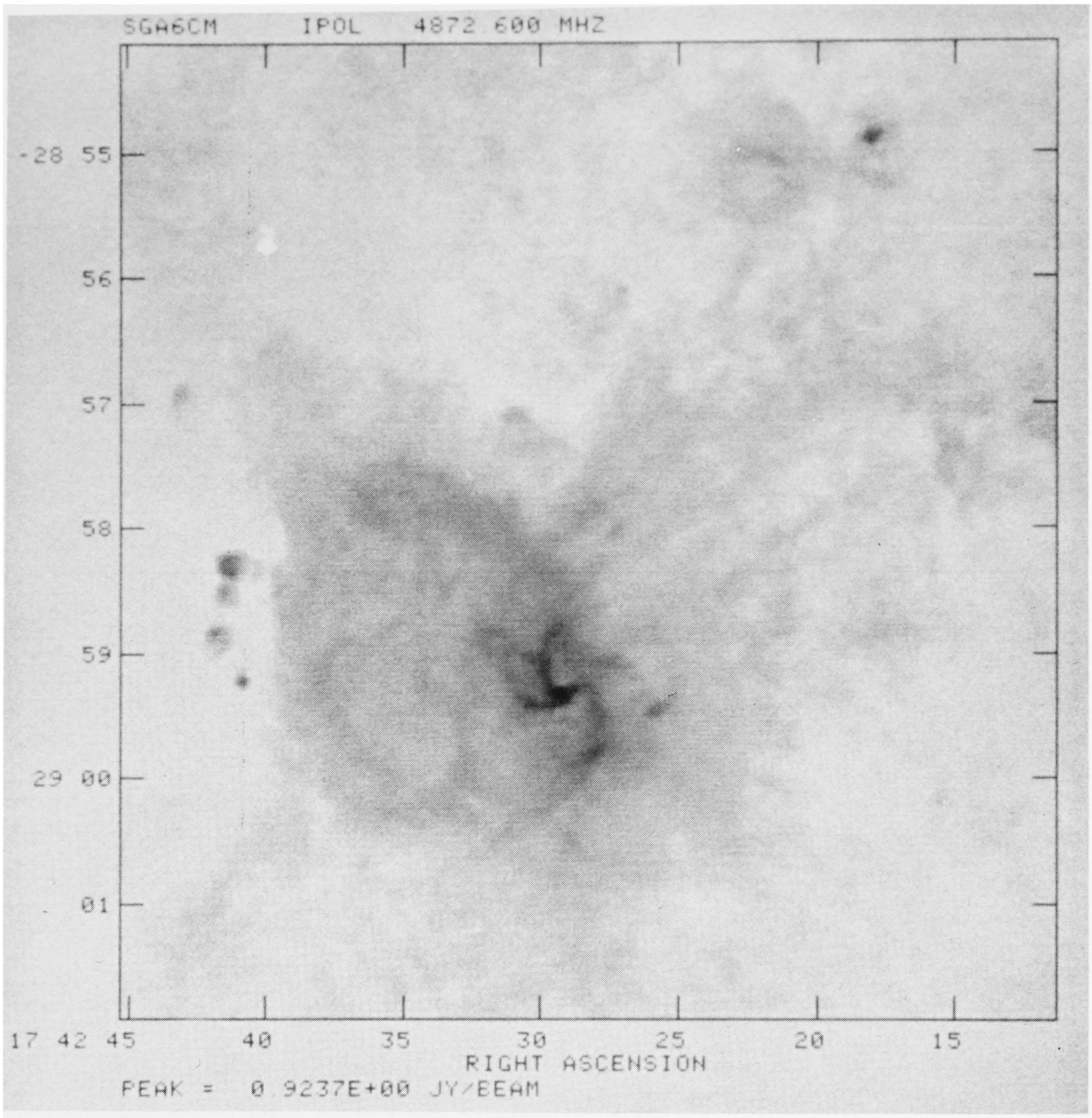
SGA6CM IPOL 4872.600 MHZ
-28 55
56
57
58
59
29 00
01
17 42 45
40
35
30
25
20
15
RIGHT ASCENSION
PEAK = 0.9237E+00 JY/BEAM

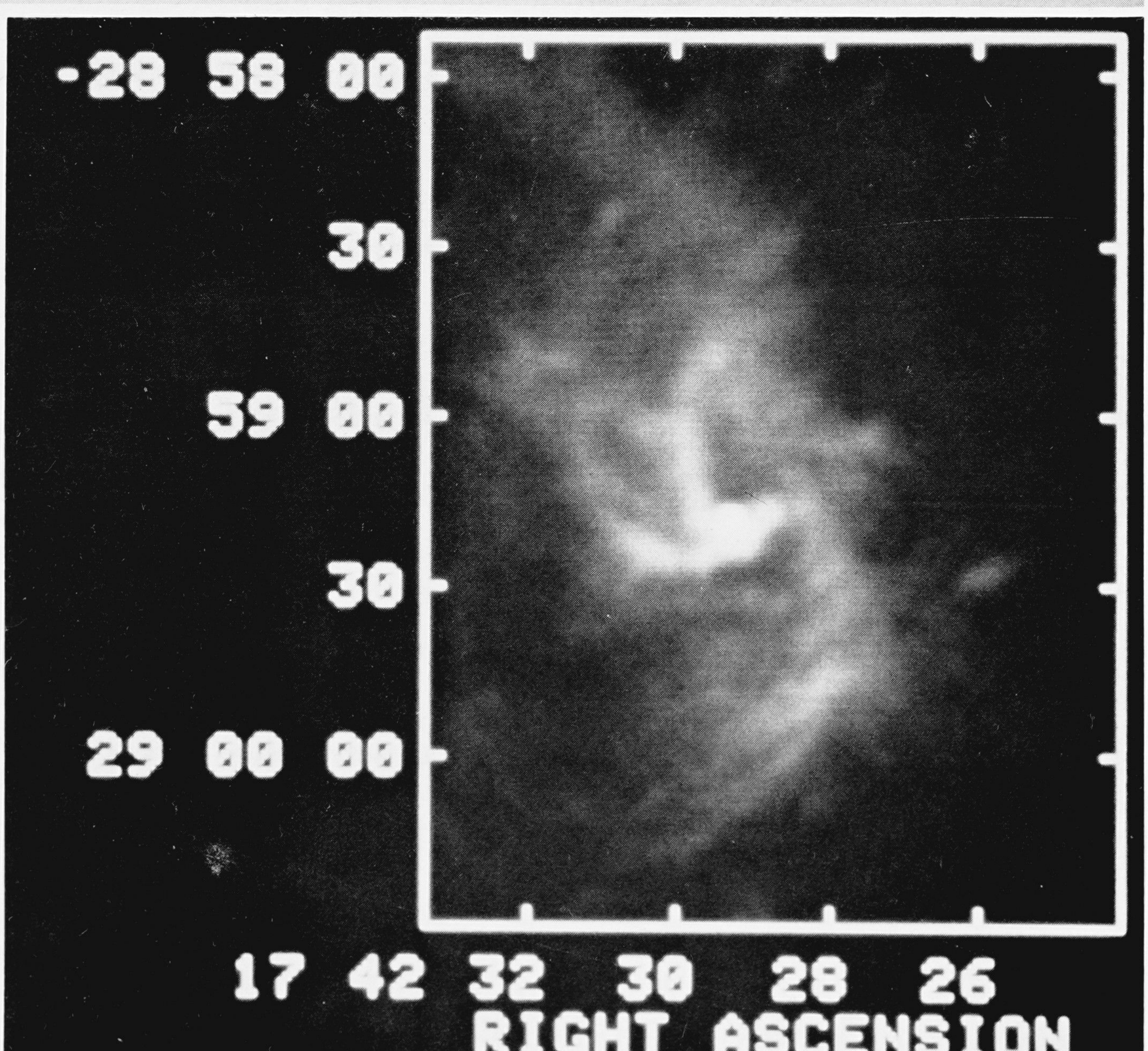
-28 58 00
30
59 00
30
29 00 00
17 42 32
30
28
26
RIGHT ASCENSION

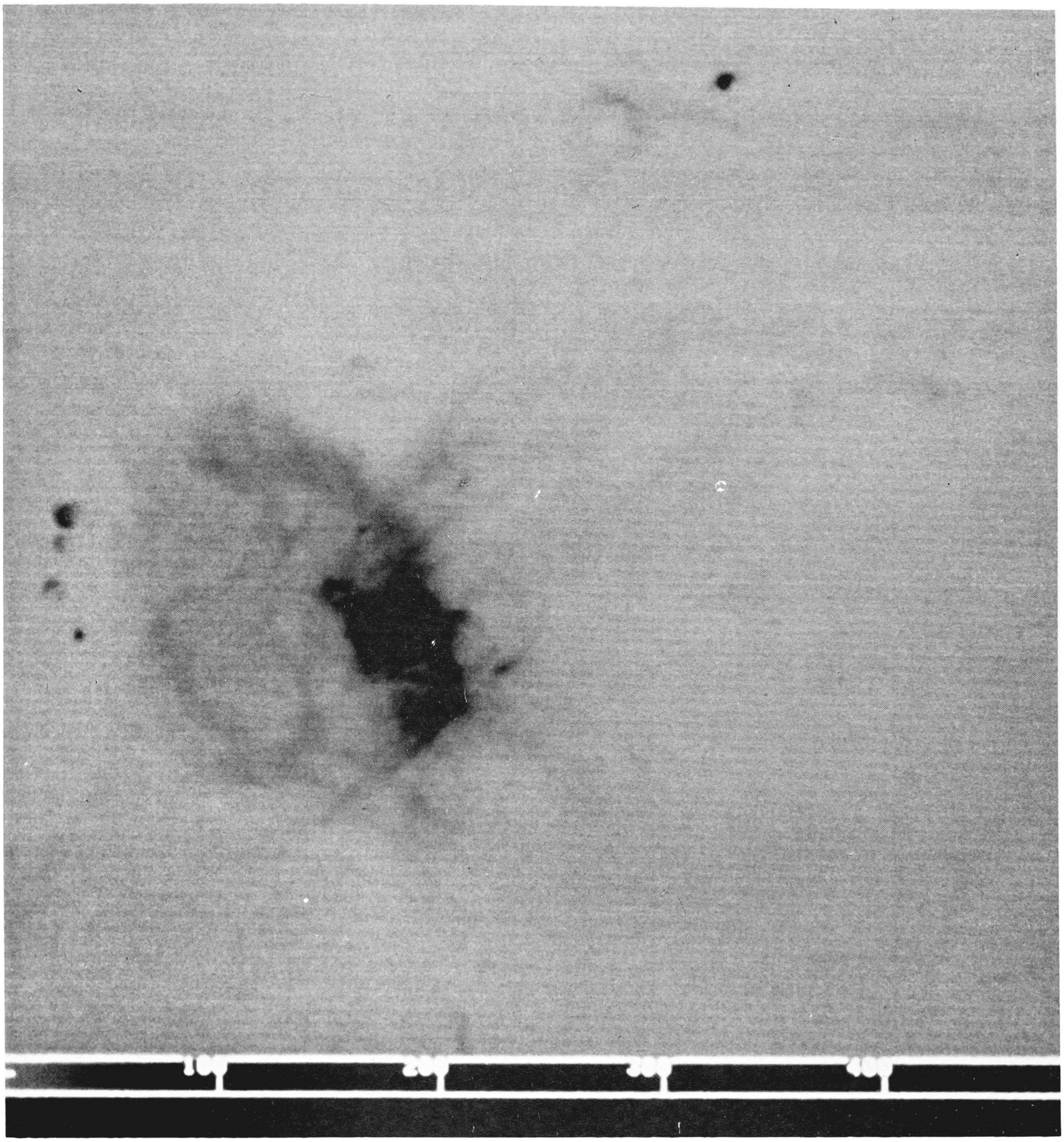
100
200
300
400

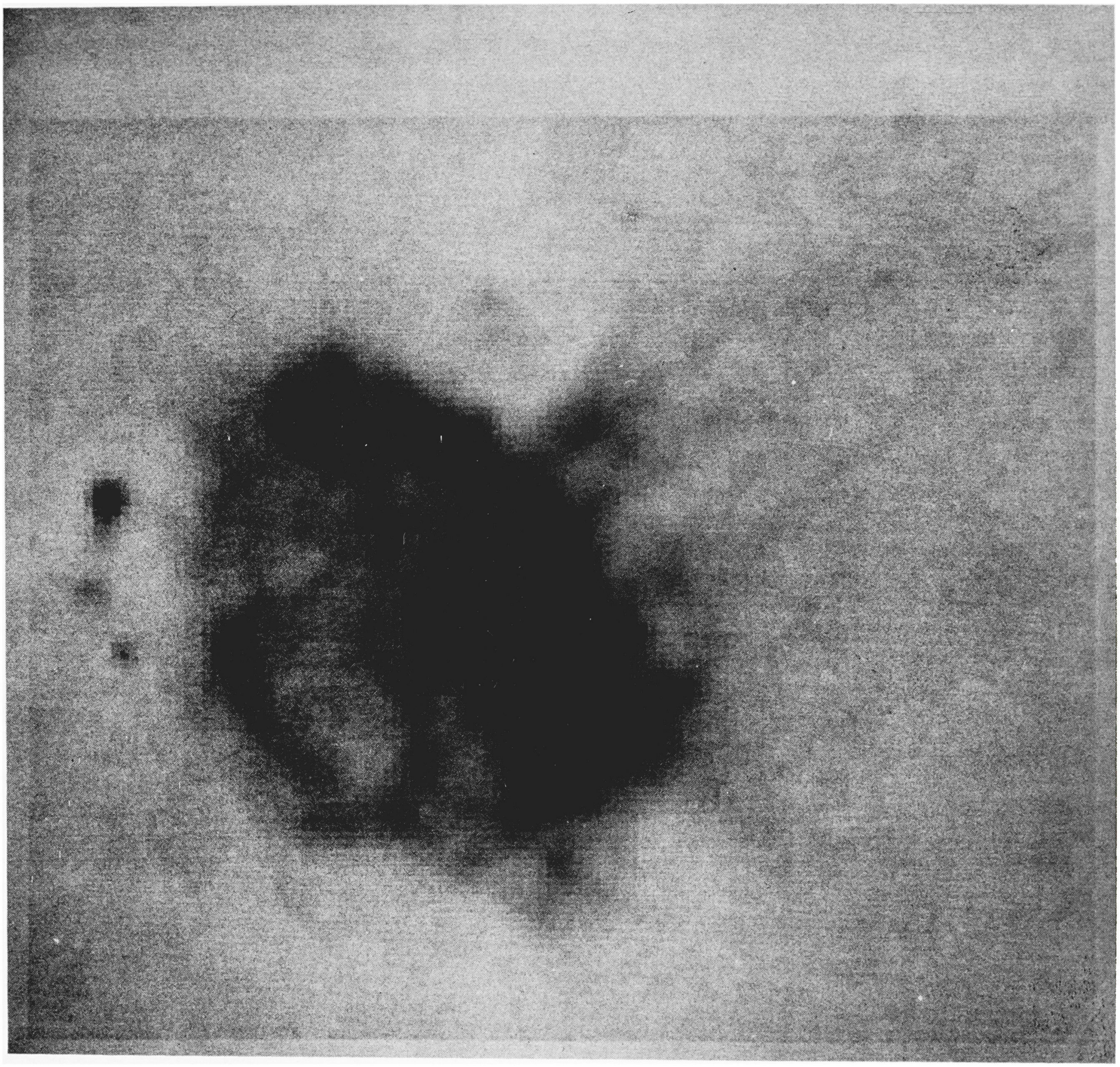

Figure 8: This 20-cm map was centered much closer to the nucleus (the Sgr A Halo field in Table 1 of Chapter 3) than that of figures 2-4. The B/C[1], C/D[2] and B/C[2] were added to construct this image with a resolution 7."7×7."2.

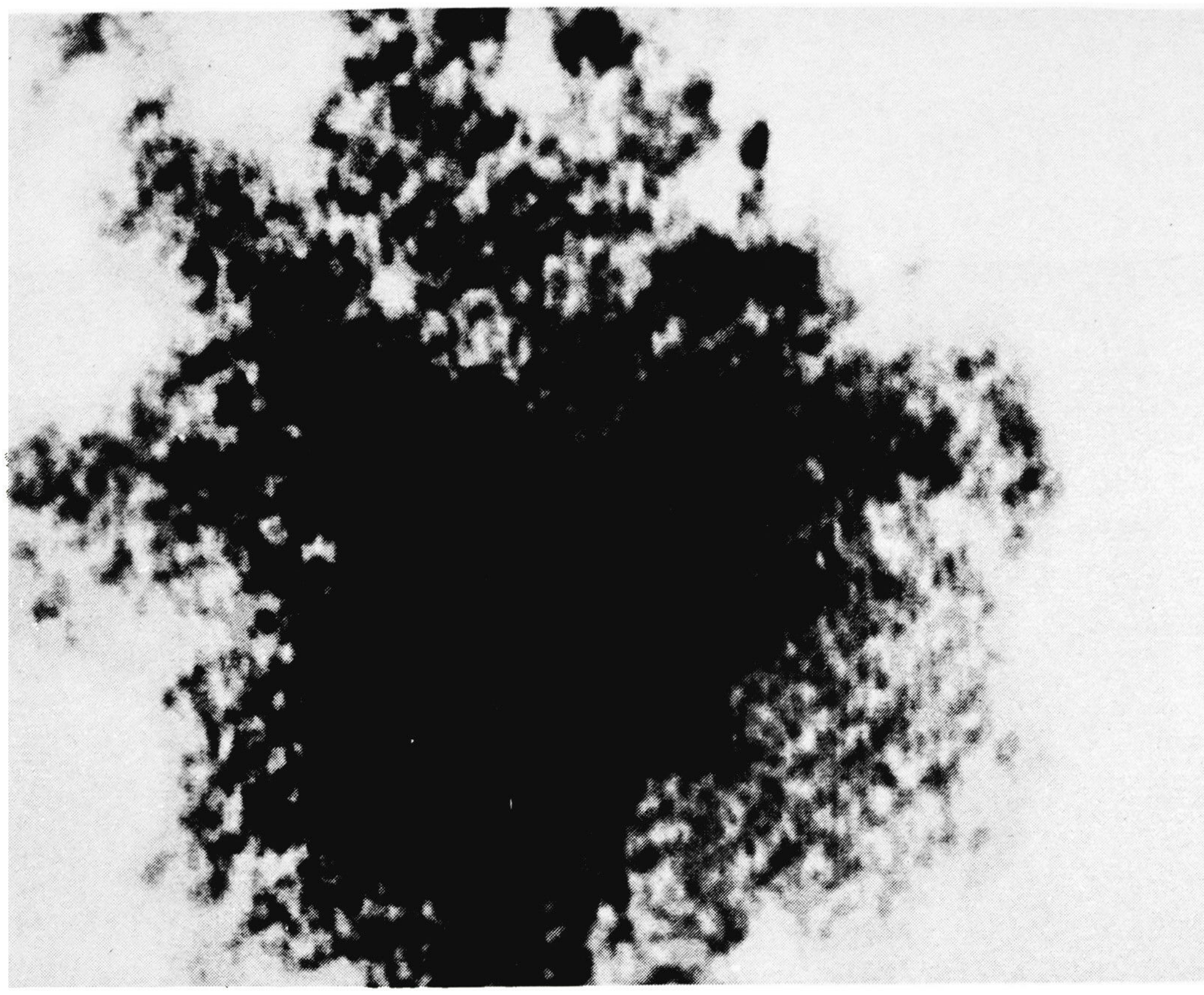

Figure 9: This 20-cm image is produced to show the striations described in the text. This figure, which has a resolution of 4" x 8" was made based on the B and C/D[1] array data sets (the designated field GC20 in Table 1 of Chapter 3).

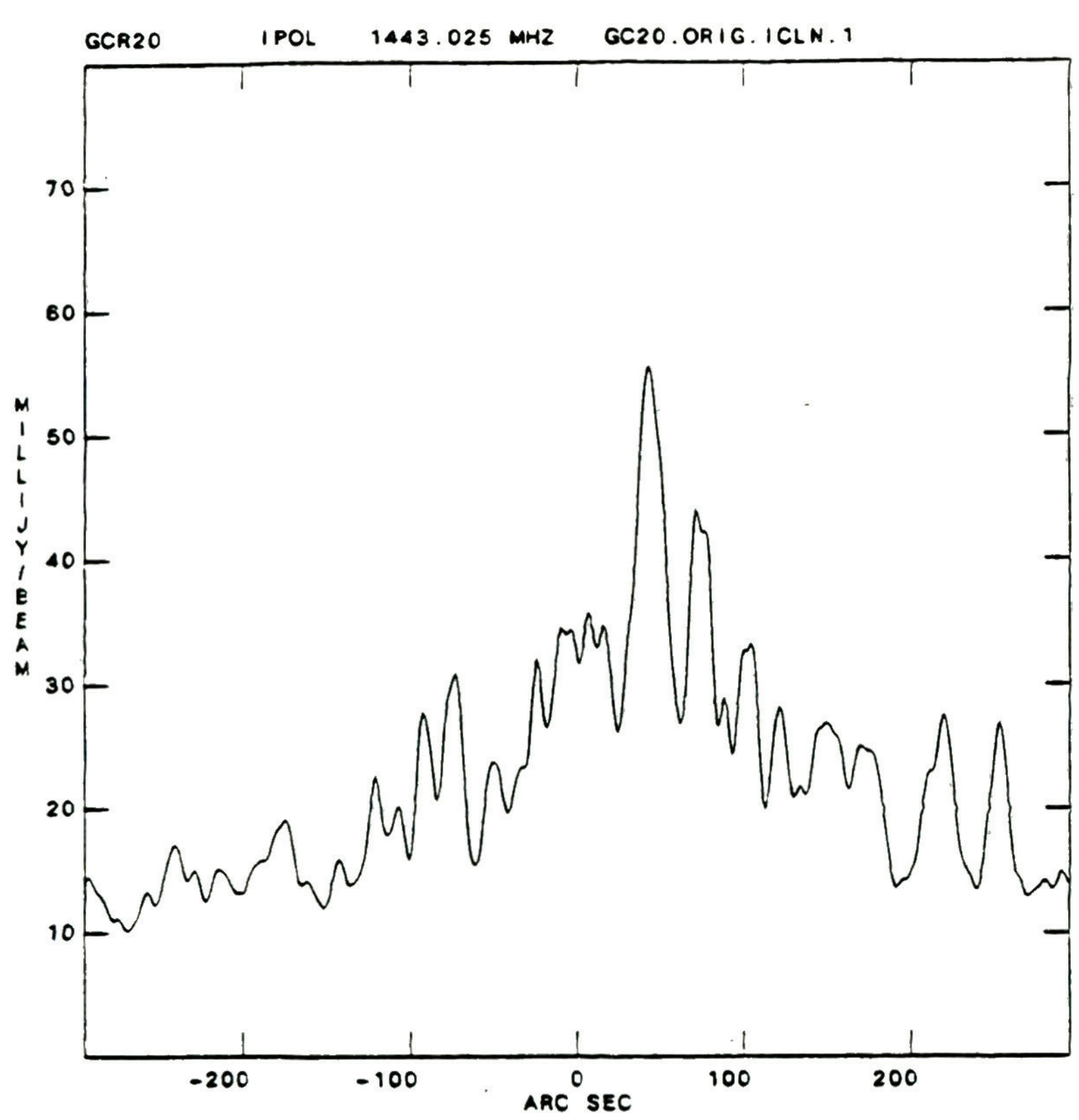

Figure 10: A slice cut across the striations (P.A. = $29\overset{\circ}{.}2$) illustrated are identified in figure 9. The reference position is at $\alpha = 17^h42^m18\overset{s}{.}6$, $\delta = 28°57'27\overset{''}{.}3$.

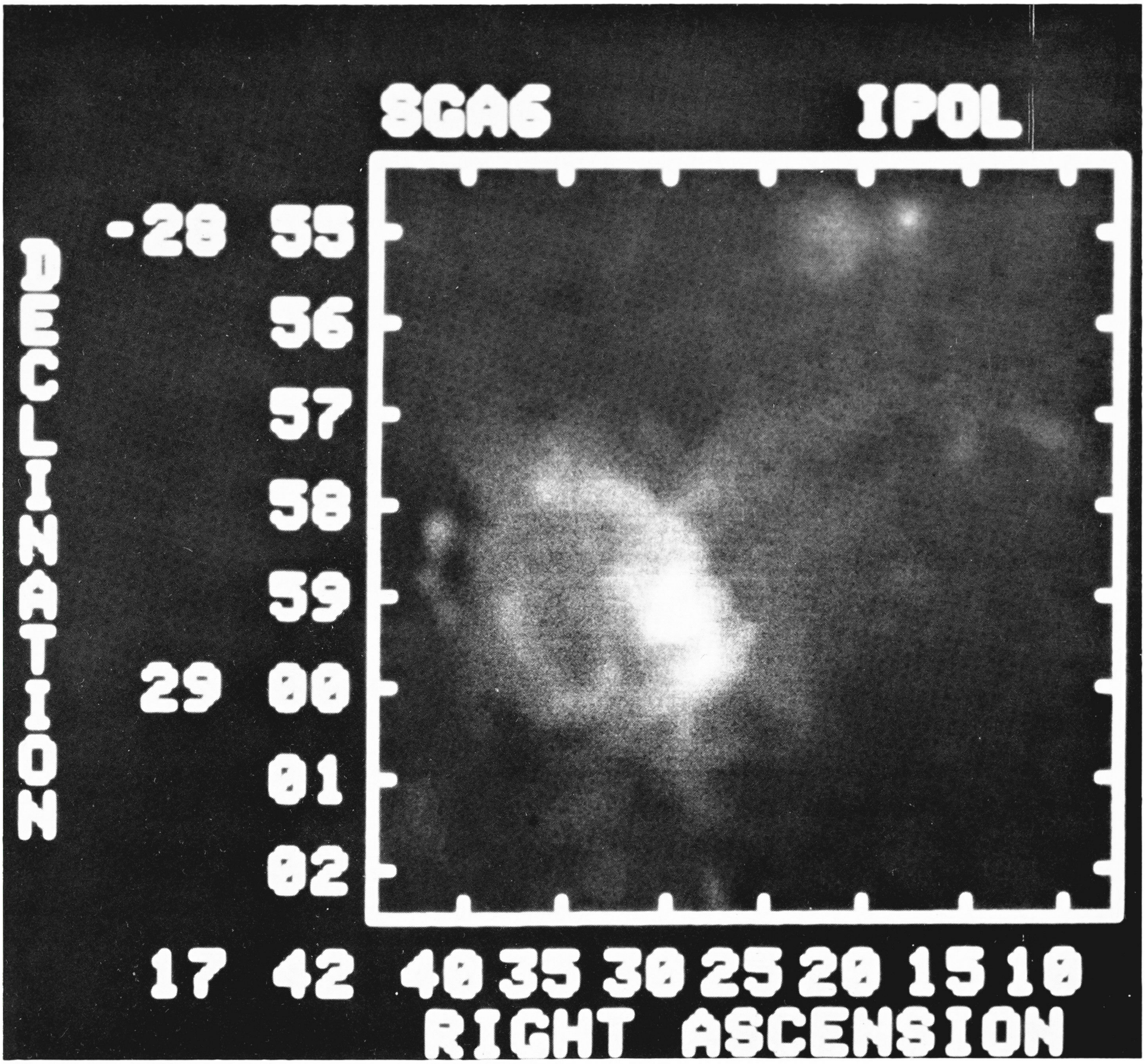

Figure 11a: This is another 6-cm map of Sgr A West and its protrusions in order to show a change in the smooth shell-like geometry of Sgr A East as it crosses the northwestern protrusions. This image is based on the $C/D^2$ array data set which is tapered at 20 kλ and has a resolution of $13''.0 \times 12''.0$.

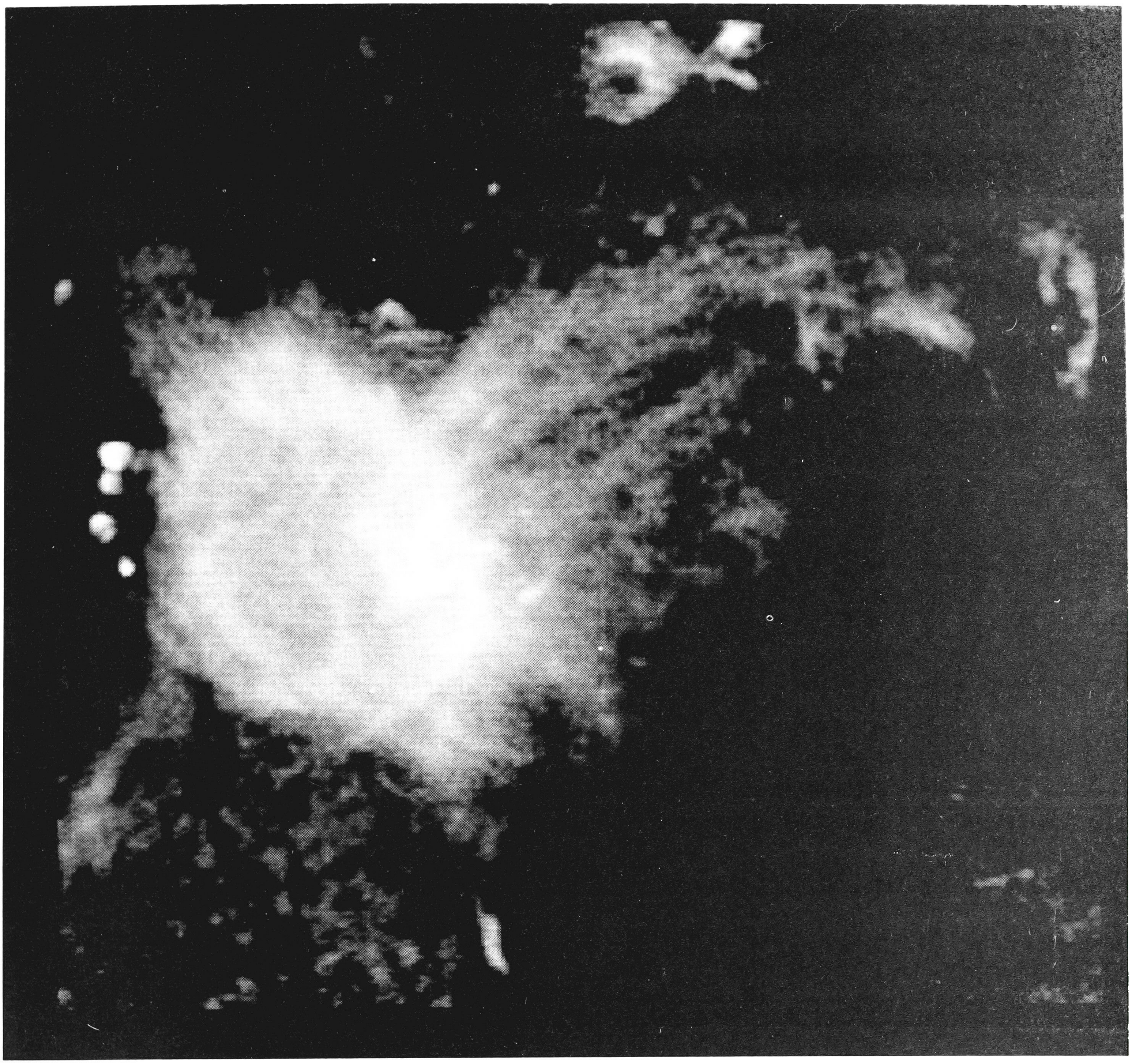

Figure 11b

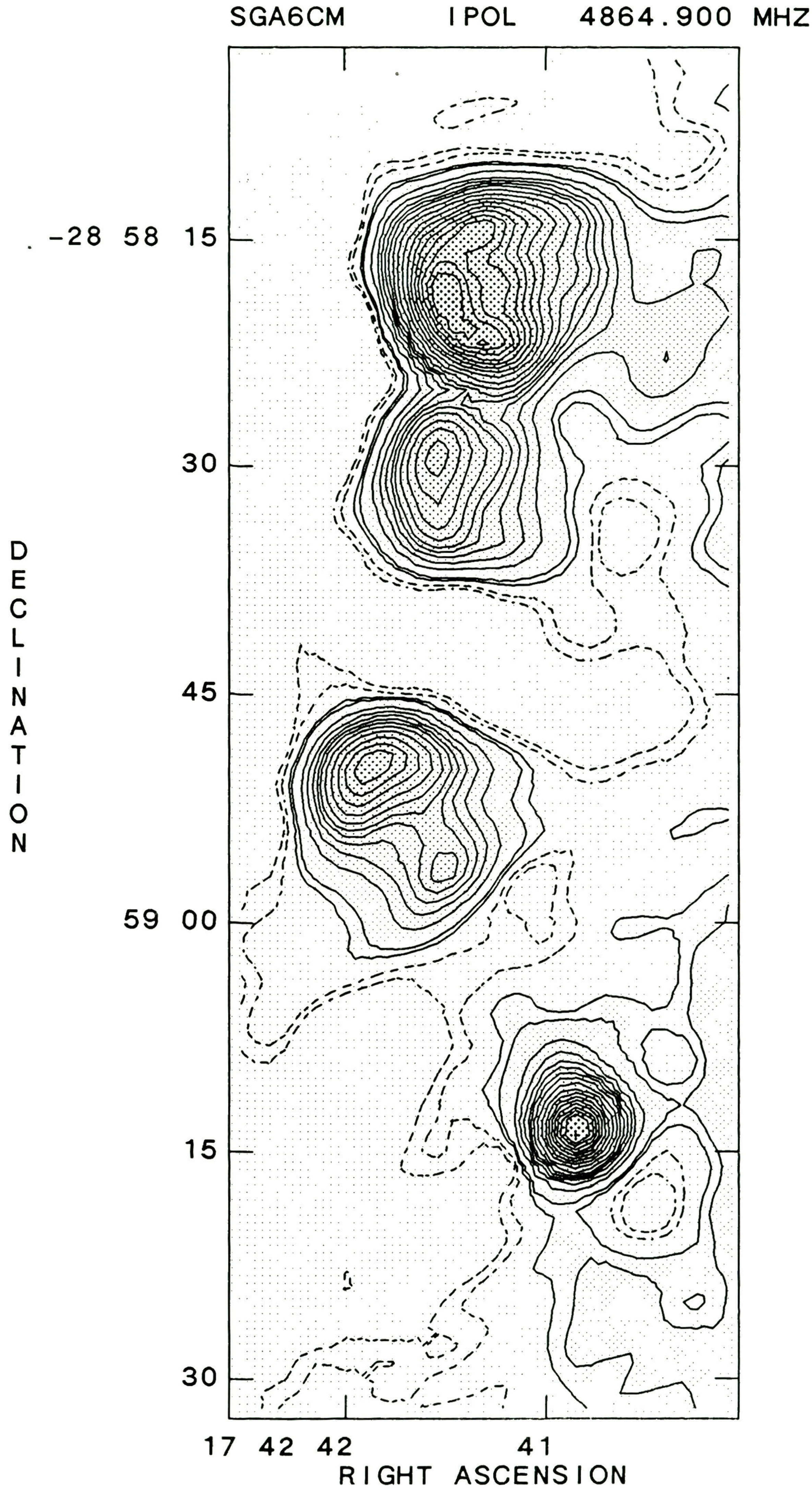

Figure 12(a): The contours of total intensity at 6 cm with intervals of -1, -0.5, 0.5, 1, 3, 5, 7, 8, 11, 14, 18, 23, 28, 33, 38, 43, 48, 53, 58, 68, 78, 88 mJy/beam area (see figures 5-7 for further details).

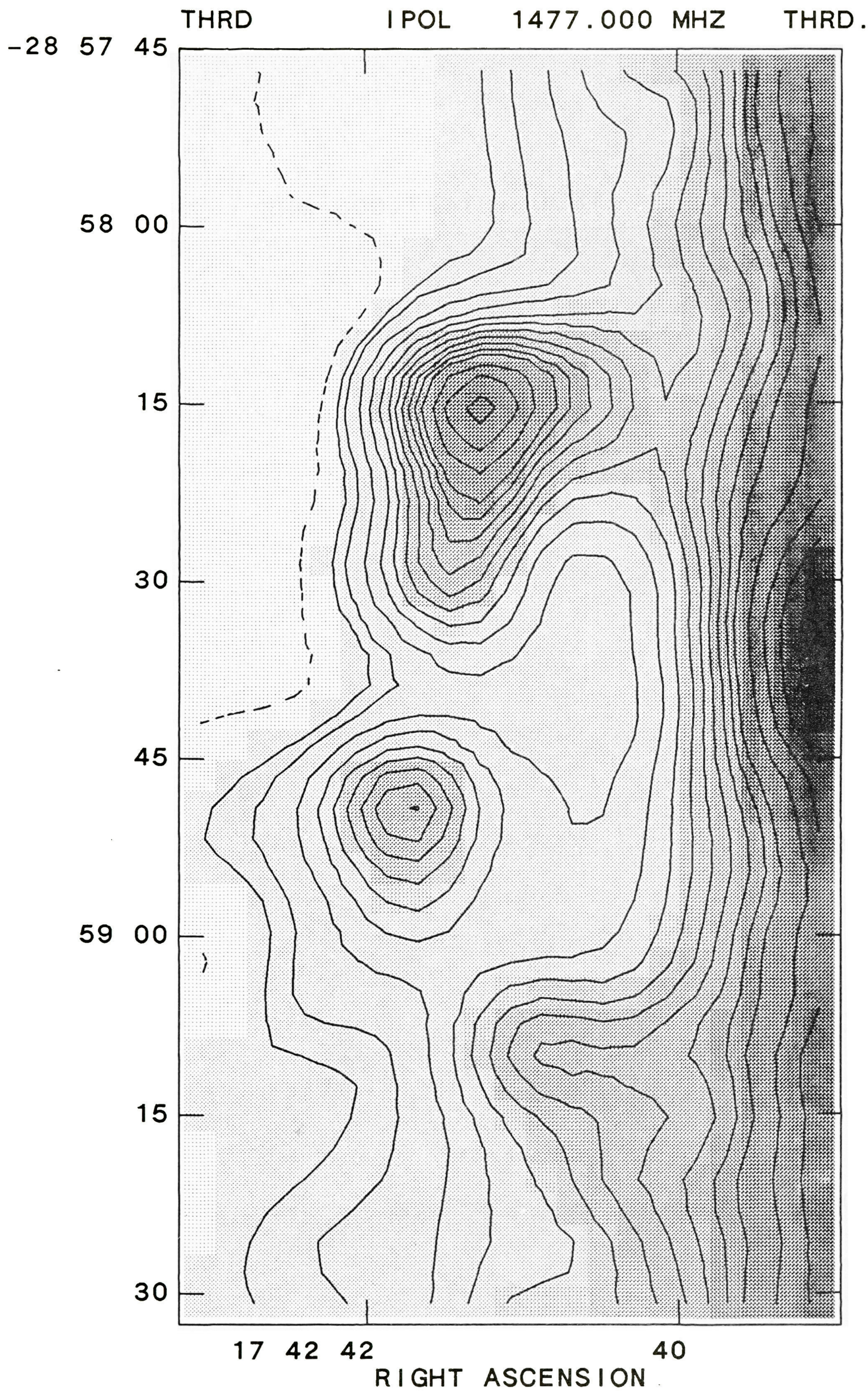

Figure 12(b),(c),(e): The contours of total intensity at 20 cm with intervals of -5, 5, 10, 20, ..., 100, 120, 140, ..., 200, 240, 280, 320, mJy/beam area. This map is constructed using baselines ⩾ 400 λ. FWHM = 9.16" x 8.2" (P.A. = 87.5°).

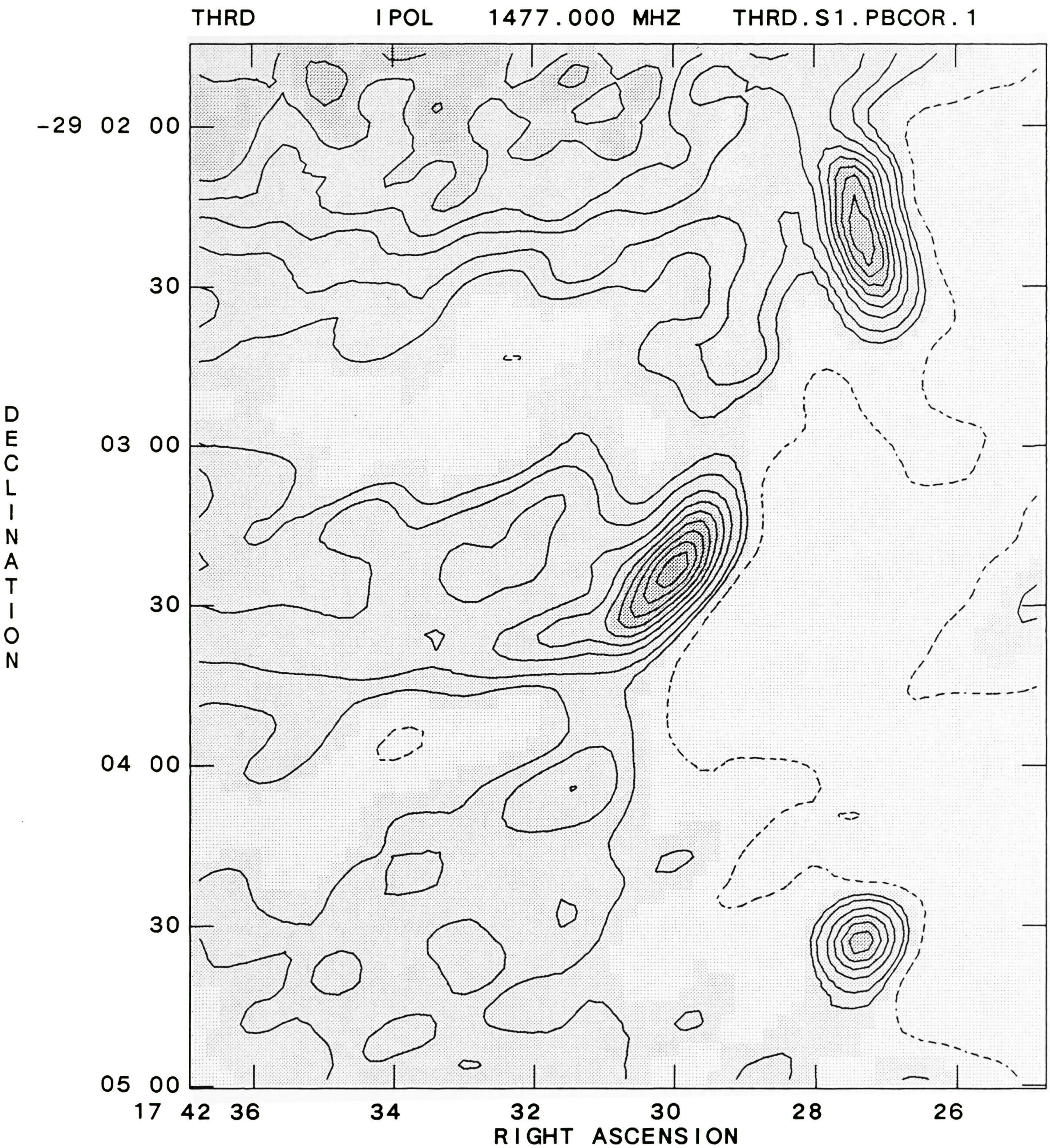
THRD IPOL 1477.000 MHZ THRD.S1.PBCOR.1
-29 02 00
30
03 00
30
04 00
30
05 00
DECLINATION
17 42 36
34
32
30
28
26
RIGHT ASCENSION

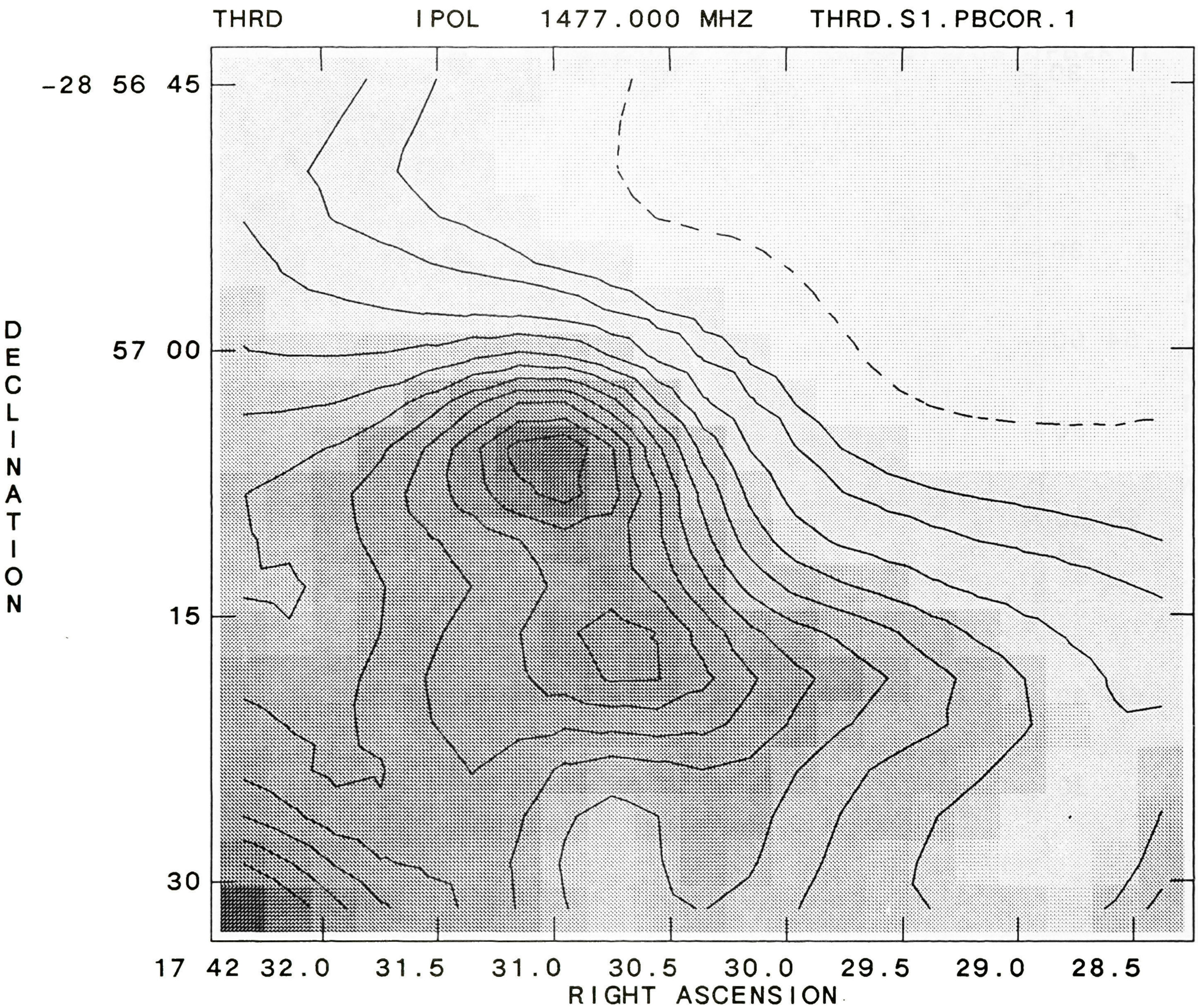

THRD
I POL
1477.000 MHZ
THRD.S1.PBCOR.1
-28 56 45
57 00
15
30
DECLINATION
17 42 32.0
31.5
31.0
30.5
30.0
29.5
29.0
28.5
RIGHT ASCENSION

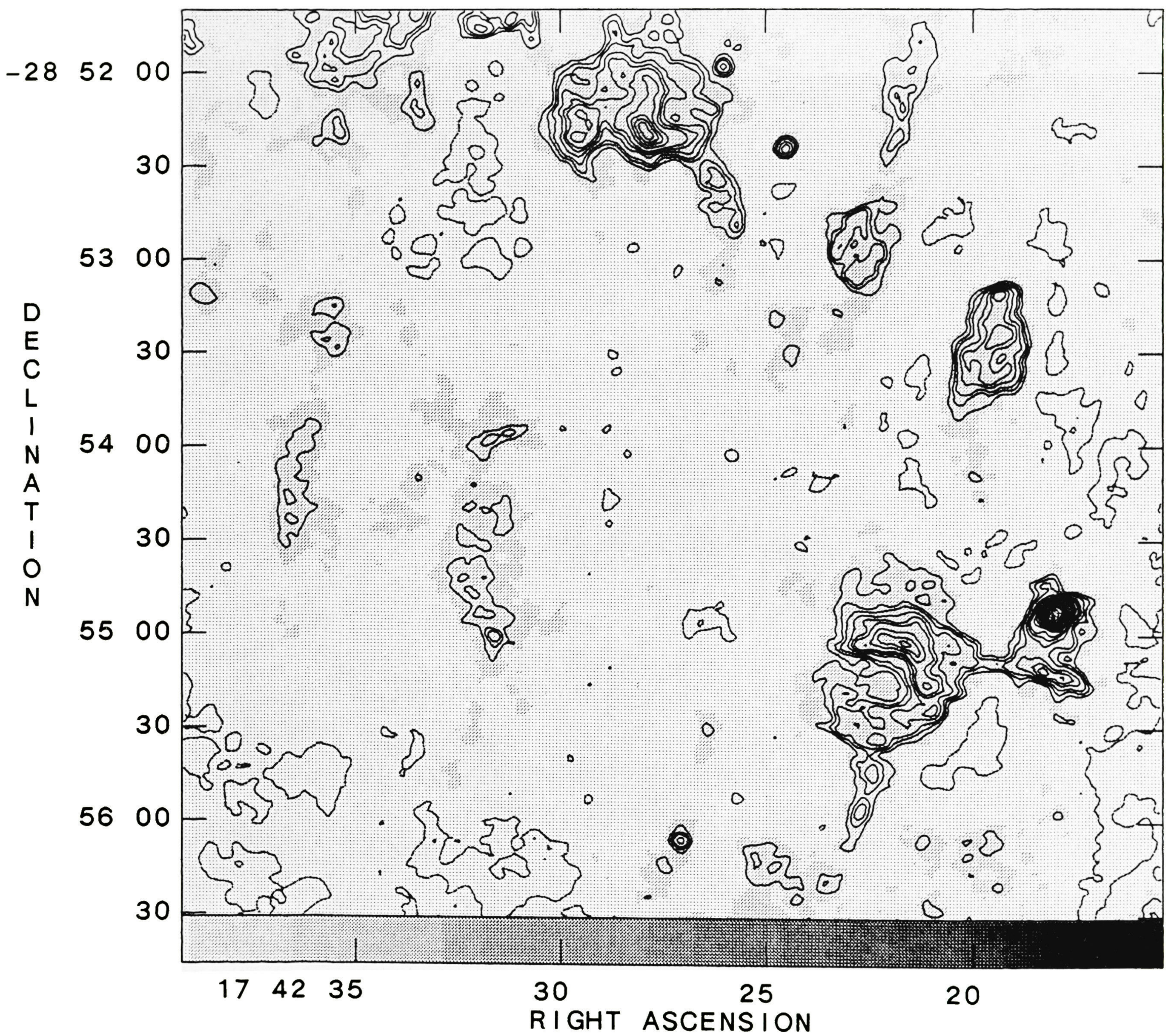

Figure 12(d),(f): The contours of total intensity at 6 cm with intervals of −2, 2, 3, 4, 6, 8, 10, 15, 20, 25, 30, 40, 50, 60, 70, 80, 100, 120, 140, 160 mJy/beam area. FWHM = 3.73"×2..97" (P.A. = 87°).

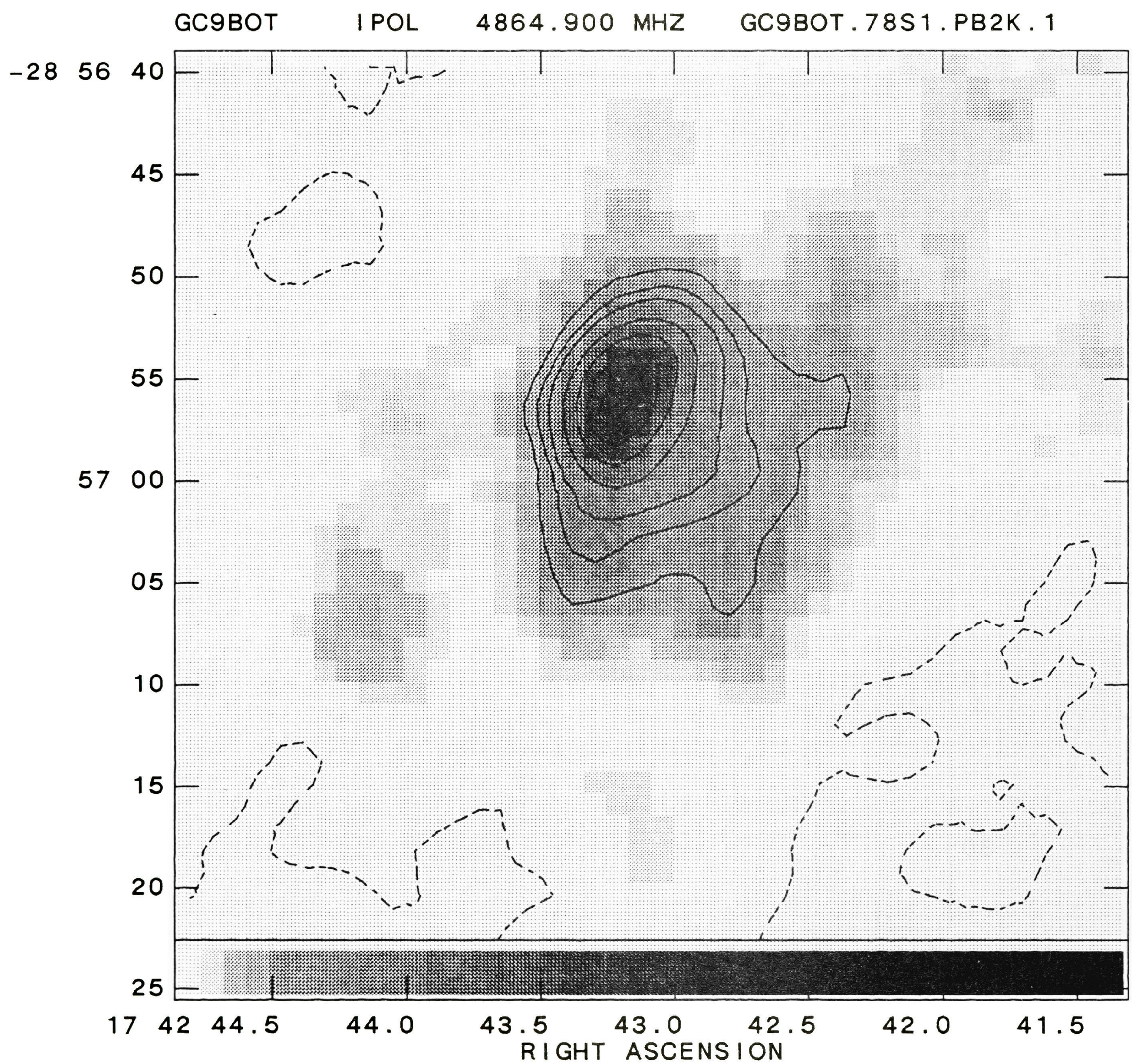

GC9BOT IPOL 4864.900 MHZ GC9BOT.78S1.PB2K.1
-28 56 40
45
50
55
57 00
05
10
15
20
25
DECLINATION
17 42 44.5
44.0
43.5
43.0
42.5
42.0
41.5
RIGHT ASCENSION

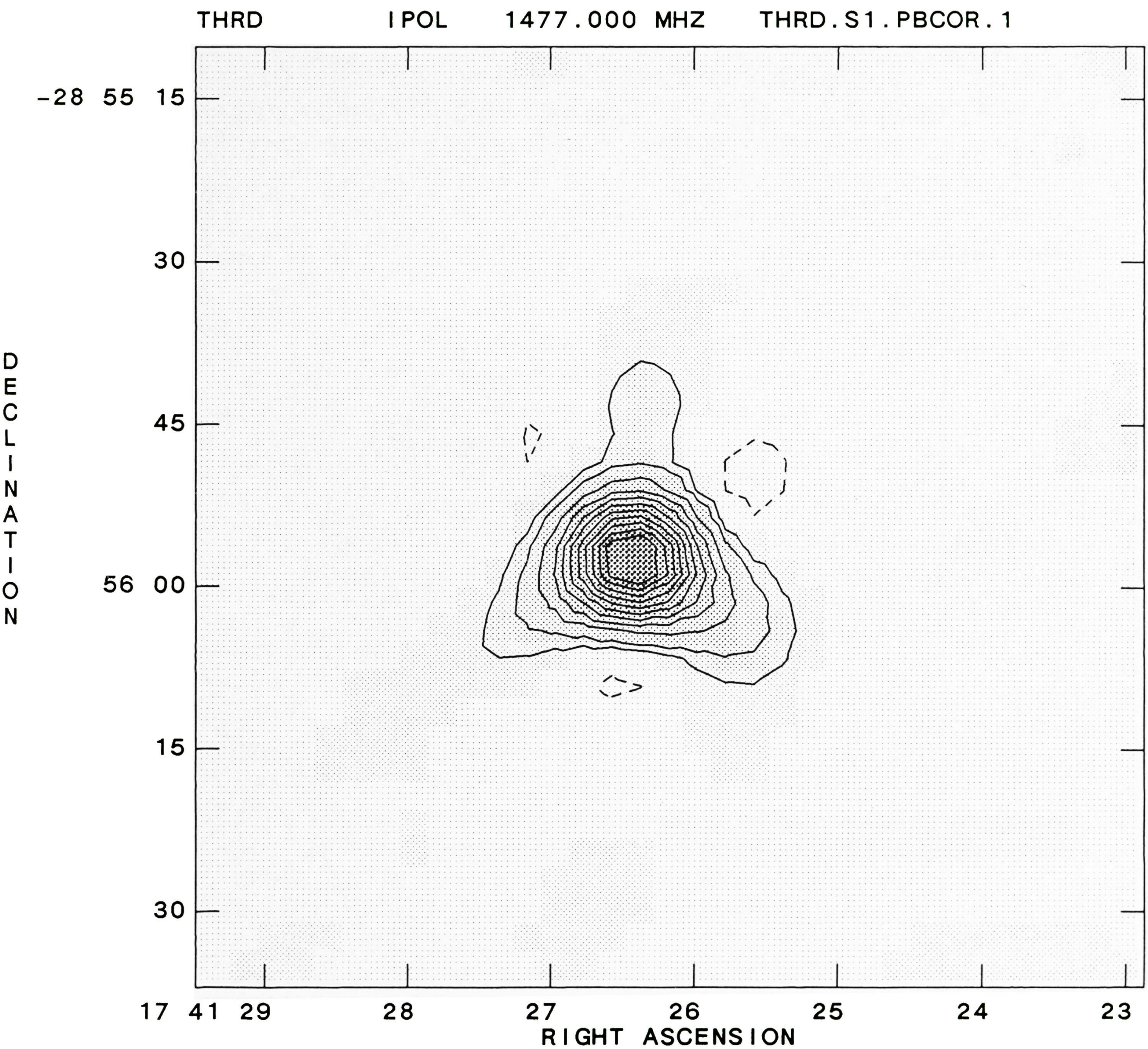

Figure 12(g): The contours of total intensity at 20 cm with intervals of -5, 5, 10, 20, ... 100, 120, 140, ..., 200, 250, 300, 350, mJy/beam area (see figure 12 a for further details).

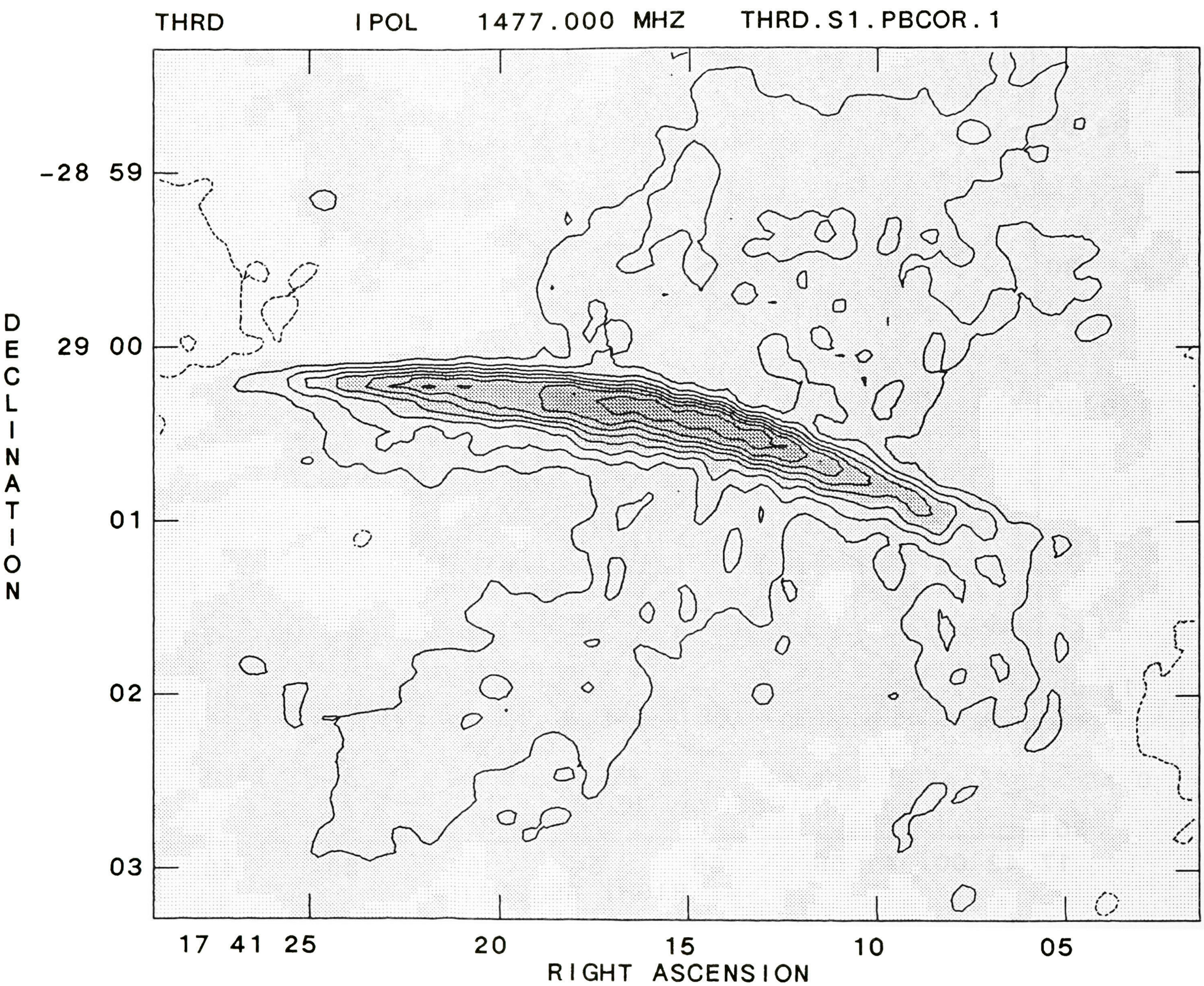

Figure 12(h): The same of (g) except that the intervals are -3, 3, 6, ..., 21, 25, 30, mJy/beam area.

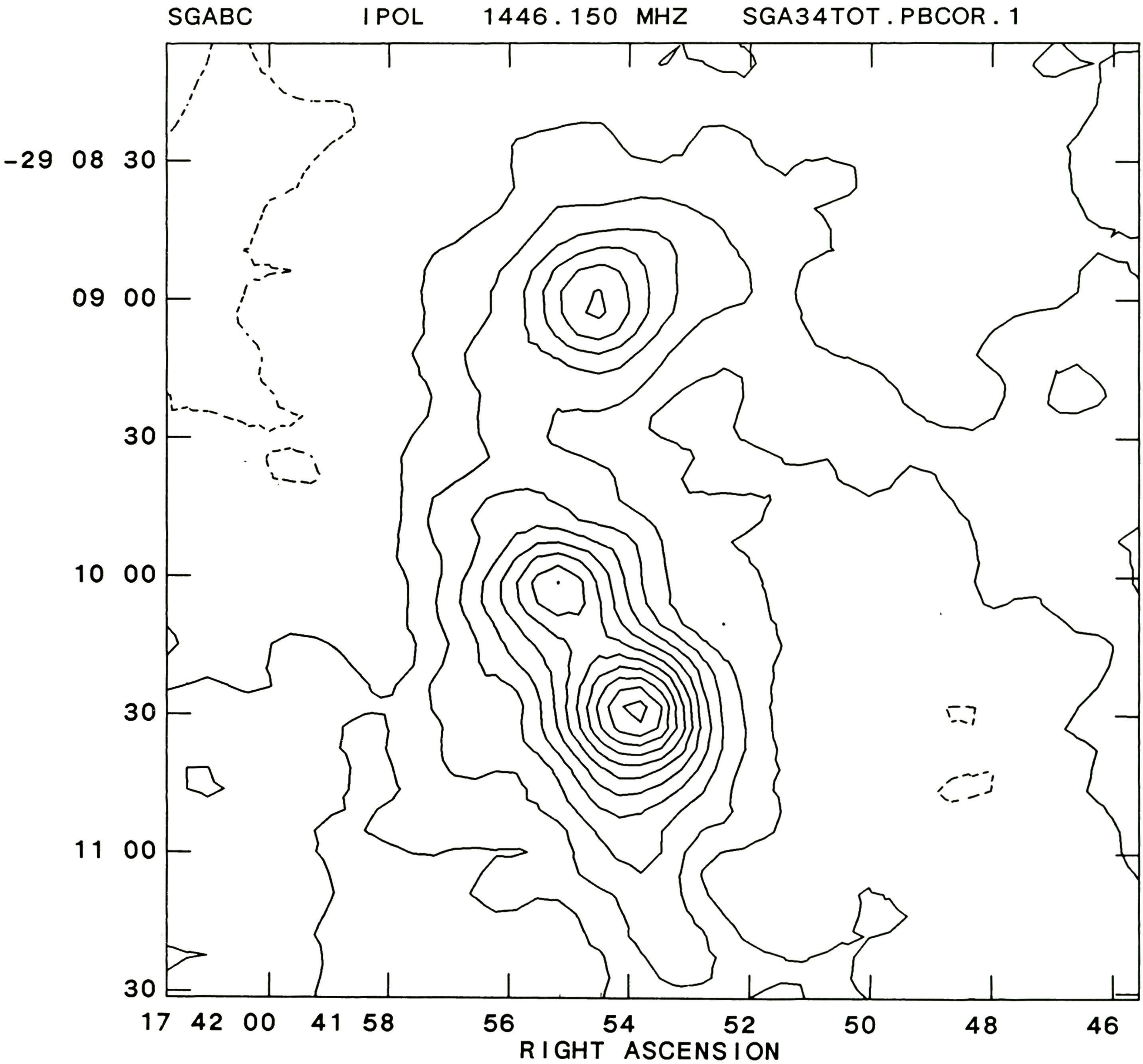

Figure 12(i): The contours of the total intensity at 20 cm with intervals of -10, 10, 20, 30, 40, ..., 120 mJy/beam area. This is based on the observations that were carried out using the hybrid B/C[1] and the C/D[2] arrays. (The Sgr A Halo field in Table 1 of chapter 2). FWHM = 15.5"×12.6". The rms noise level is 7 mJy/beam area.

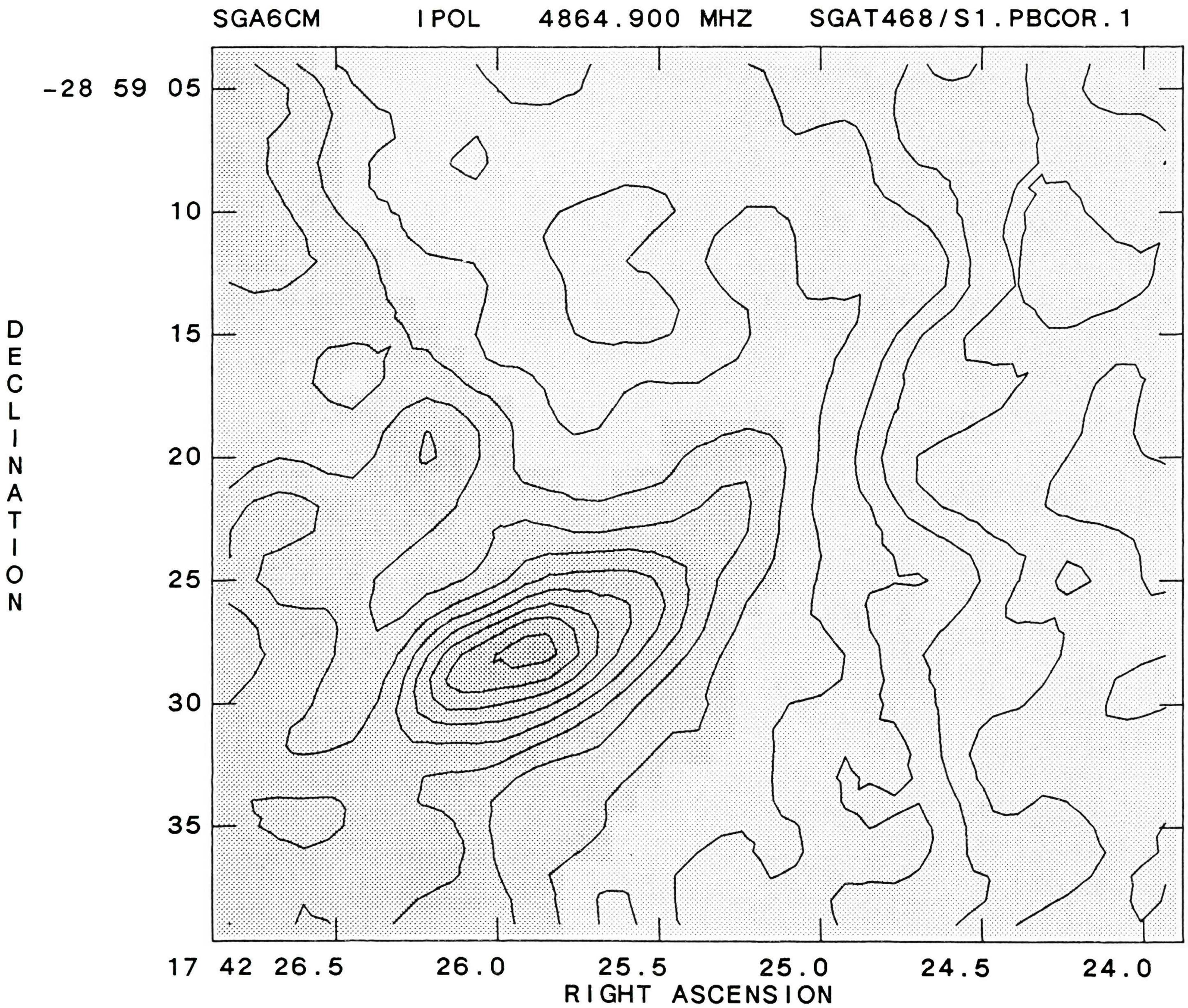

Figure 12(j): The contours of total intensity at 6 cm with intervals of -1, -0.5, 0.5, 1, 3, 5, 7, 8, 11, 14, 18, 23, 28, 33, 38, 43, 48, 53, 58, 68, 78, 88 mJy/beam area (see figures 5-7 for further details).

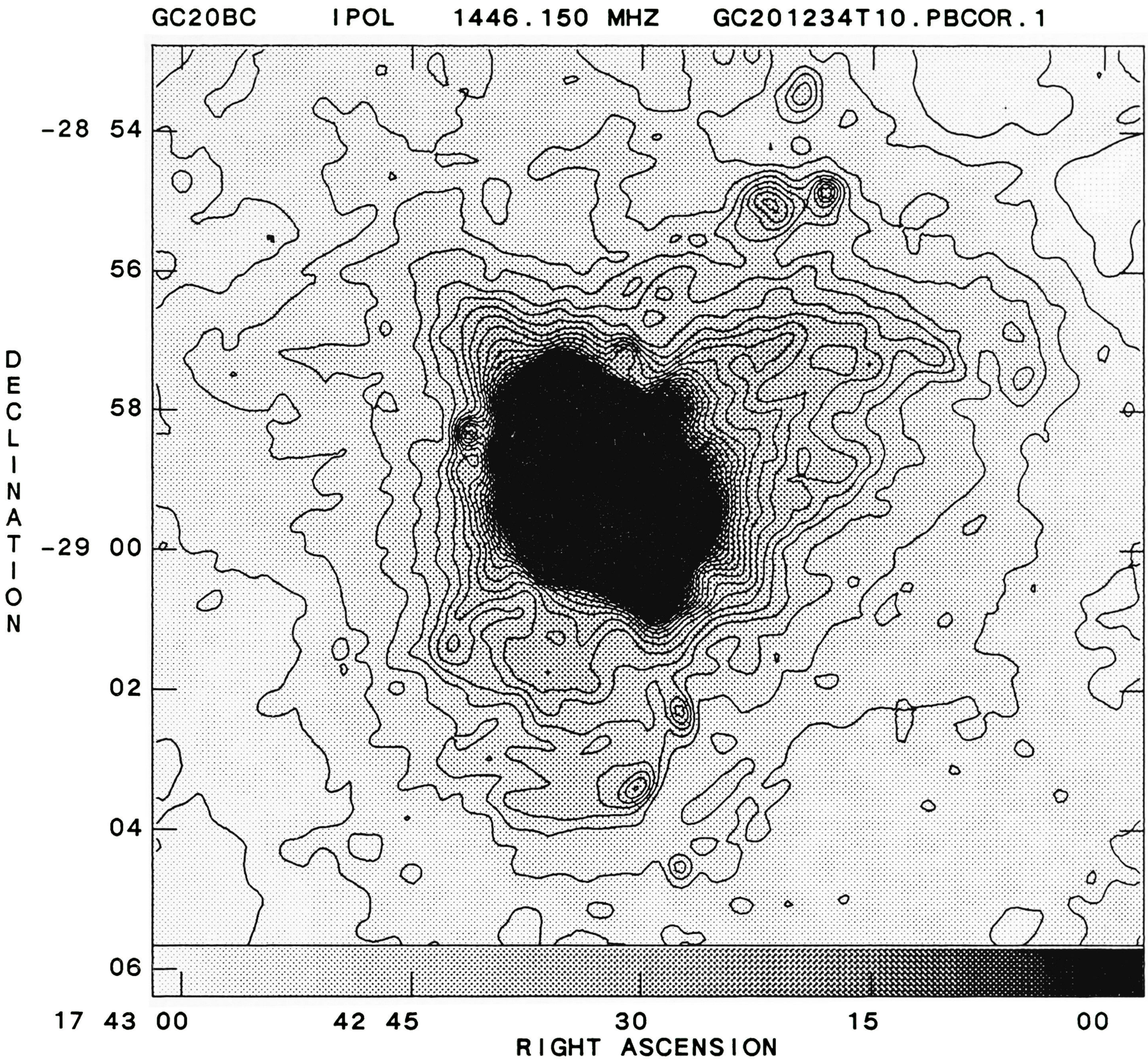

Figure 12(k): The contours of the total intensity at 20 cm with intervals of −20, 20, 50, 100, 150, ..., 1000 mJy/beam area. This map is based on the data set which is tapered at 10 kλ (GC20 field) and has a resolution of 17.1"×16.4" (P.A. = 35°). The rms residual flux is 14.6 mJy/beam area.

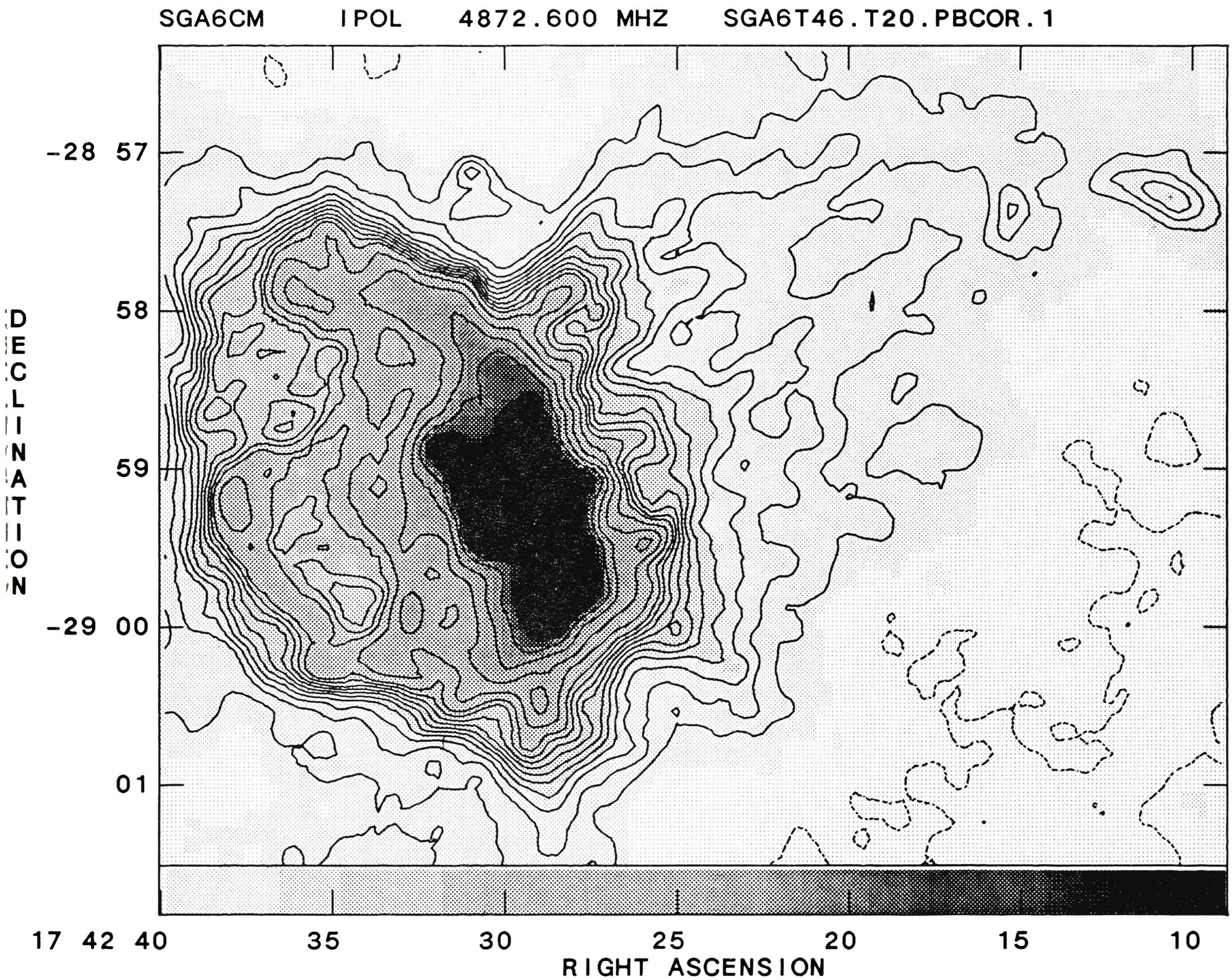

Figure 12(1): The contours of the total intensity at 6 cm with intervals of -10, 10, 20, 30, 40, 50, 60, 70, 80, 90, 100, 120, 150, 190, 240, 300, 370, mJy/beam area. This map is based on the data set used for constructing figures 5-7 except that the data is tapered at 20 kλ. FWHM = 7.7×7.2 mJy/beam area.

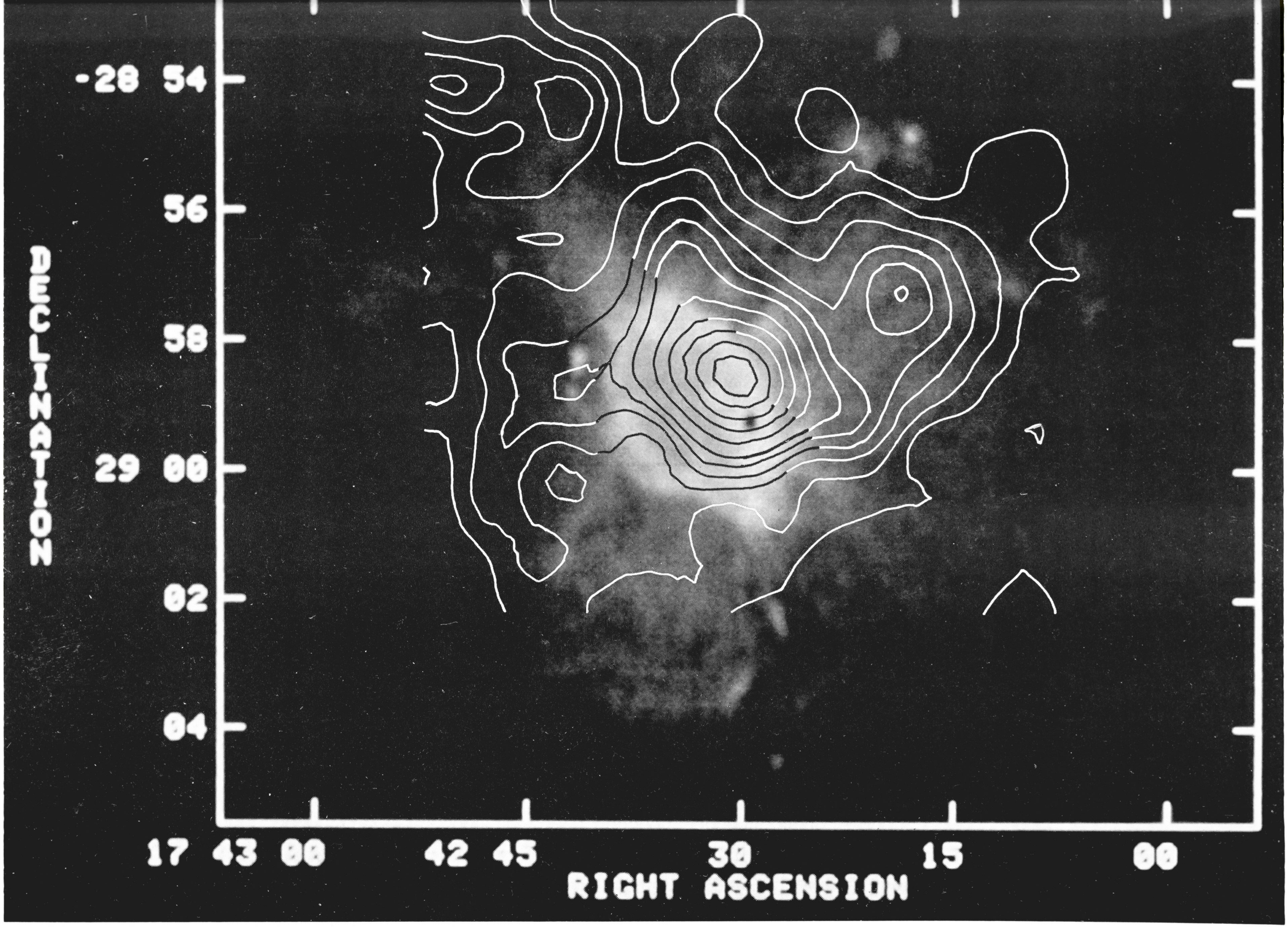

Figure 13: This 20-cm figure is identical to figure 2-4 except that the X-ray map (Watson et al. 1981) with a 1' resolution is superimposed on this figure.

# Chapter 7

## A LOW-ENERGY JET EMANATING FROM THE GALACTIC NUCLEUS?[5]

"because scientists must be men, must be fallible,
and yet as men must be willing and as a
society must be organized to correct their errors.
William Blake said that 'to be an Error
and to be Cast out is a part of God's design.'
It is certainly part of the design of science."
J. Bronowski

## I. INTRODUCTION

Understanding the dynamics and energetics of gas in galactic nuclei is imperative if we are to unravel the mysteries of their central energy sources. Our own Galaxy demonstrates the complexity of the nuclear region even for a relatively quiescent example of a galactic nucleus.

We report in this chapter a new low-frequency radio feature that may signify a new and important aspect of the activity in our galactic nucleus. This feature is approximately perpendicular to the galactic plane, and appears to join, or originate at, the nucleus. It extends ~ 30 pc toward negative latitudes but there is no evidence for a positive-latitude counterpart. Its shape is suggestive of a collimated ejection from the galactic nucleus or from the region near the nucleus.

[5] This chapter is the product of a collaboration between Dr. Morris (UCLA), Drs. Slee and Nelson (both from C.S.I.R.O., Division of Radiophysics).

Recent radio interferometric observations of the galactic center region demonstrate that the inner 20 pc of the Galaxy -- known to radioastronomers as the Sgr A complex (see chapter 6) -- consists of emission features having a wide range of scale sizes. On the small scale, Sgr A West, which consists of ionized, thermally emitting gas, occupies the inner 2 pc of the Galaxy (Ekers et al. 1975, 1983; Brown and Johnston 1983; Lo and Claussen 1984). Apparently buried within Sgr A West is a nonthermal compact radio source having a diameter $\leq 3 \times 10^{14}$ cm (Lo et al. 1981; Brown and Lo 1982; Lo et al. 1985).

On a scale of approximately 10 pc, one finds Sgr A East, which is a nonthermal shell source surrounding Sgr A West (Ekers et al. 1983). Several investigators have argued that Sgr A East is a supernova remnant (Downes and Maxwell 1965; Jones 1974; Ekers et al. 1975; Gopal-Krishna and Swarup 1976; Goss et al. 1983). Surrounding the Sgr A East shell, one finds a 20-pc halo of relatively weak emission, some of which takes the form of radio protrusions, apparently associated with Sgr A West (chapter 6). Most of the halo, however, is apparently an extension of Sgr A East and may therefore be nonthermal. Finally, on the largest scale (approximately 1° or 180 pc), a large radio "lobe" extends as much as one degree toward positive latitudes (Sofue and Handa 1984).

The new large scale feature we describe in this paper was identified from 160 and 327-MHz maps made with the Culgoora Circular Array; like the radio lobe seen at positive latitudes, it is asymmetrically situated with respect to the galactic plane. Its location and orientation suggest a link to small-scale structure in Sgr A West and possibly to the elongated, compact nonthermal source. A hint of this structure was present in a previous 160-MHz map of Sgr A (Dulk and Slee 1974; Slee 1977), but the large field of view of the present observations was required to reveal it fully. Another 160-MHz emission source in the field of view associated with the Galactic Center Arc is discussed elsewhere (chapter 4). Details of the 160-MHz and 327-MHz observations are also presented there. Caution should be exercised in accepting the details of the brightness distribution in the two outer contours of the 327-MHz map due to the high sidelobe level. The 160-MHz map is reliable to the 10% level of peak brightness.

## II. Results

### a) The 160-MHz Map

The most interesting result -- the recognition of a large-scale feature aligned roughly perpendicular to the galactic plane -- is shown in Figure 1, in which contours of 160 MHz emission are superimposed upon a 1.4 GHz radiograph made with the Very Large Array. The elongated 160 MHz feature, which appears to be linked to Sgr A, has dimensions exceeding 10' × 5'. The outer contours of this structure narrow slightly with increasing negative latitude until they disappear beyond the edge of the observed field of view at a distance of 10' (~30 pc) from the galactic nucleus. The addition of this elongated feature embellishes our view of this region by adding a new large-scale component to the phenomenologically rich collection of radio structures constituting the Sgr A complex.

The total flux density of the elongated 160-MHz radio source is ~118 Jy. It appears to consist of two unresolved sources within the 5'-wide envelope of an extended ridge surrounding Sgr A East, and Sgr A West which includes the nonthermal compact source. The superposition of the 160-MHz map and the 1.4 GHz map confirms the earlier report by Dulk and Slee (1974) that the brightest component of the 160-MHz map does not coincide with the position of Sgr A West (marked in figure 1 by a black spot). Rather, the 160-MHz intensity peak, which has a brightness temperature $T_b \sim 6.35 \times 10^4$ °K, is located close to the northern tip of the Sgr A East shell, at ($\alpha$ = $17^h42^m36.7^s$, $\delta$ = -28°58'17").

The extended 160-MHz ridge appears to bend at a position ($\alpha \sim 17^h42^m40^s$, $\delta \sim$ -29°01') roughly coinciding with a narrow protrusion from Sgr A visible in the 1.4 GHz map. The second compact component along the ridge can also be seen in figure 1 at $\alpha$ = $17^h42^m58^s$, $\delta$ = -29°03'40". Neither this second peak nor the extended ridge have been seen at frequencies other than 160-MHz. We have confidence in the reality of the new structure seen at 160-MHz for three reasons: first, the peak coincident with Sgr A East, known from previous observations, is present in the map at about the expected intensity; second, the field of view which is shown in Figure 1 is a selected portion of a map having a 30' field of view which includes both the Sgr A complex and the Arc. The 160-MHz map of the Arc, which is discussed in chapter 4 agrees well with the results obtained at 80

MHz by LaRosa and Kassim (1985); third, the extension is also evident on the 160-MHz maps of Dulk and Slee (1974) and Slee (1977). It is probable, however, that the feature is considerably smoothed in the latter maps by the averaging of 10 sets of data taken on different nights having varying degrees of ionospheric refraction.

b) The 327-MHz map

Figure 2 presents the 327-MHz map of the Sgr A region made with a resolution (FWHM) of 56" × 56". The peak brightness temperature, ~ $7.5 \times 10^4$ °K, occurs at a position coincident with that at 160-MHz. The extended ridge and the weaker of the two emission peaks appearing in the 160-MHz map are absent in the 327-MHz map. The total flux emitted from the region south of $\delta = -29°\ 02'$ is ~38 Jy at 160-MHz, but emission here is absent at 327-MHz with a limit of ~1.5 Jy/beam area. We note that the ridge seen at 160-MHz clearly has a steep spectrum ($\alpha < -0.85$; $F_\nu \propto \nu^{+\alpha}$), since it is not visible in the 408-MHz map made by Little (1974) nor on the 843-MHz map of Mills and Drinkwater (1984). The 408-MHz map does, however, show a steeper fall-off of surface brightness at the lowest contour levels to the northwest of Sgr A than along the 160-MHz emission ridge, indicating, perhaps, that this ridge is present, but unresolved.

The total flux seen to the north of $-29°\ 02'$ is ~228 and 80 Jy at 327 and 160 MHz, respectively, giving an apparent spectral index for the Sgr A East component of +1.46. If the northwest extension in the 327-MHz map is excluded, the total flux in this region is ~201 Jy and thus $\alpha$ is ~1.29. This high positive spectral index strongly suggests that the northern area of Sgr A East is strongly absorbed by an intervening HII region as first suggested by Dulk and Slee (1974); the southern extended ridge with its steep negative spectral index is probably not significantly absorbed.

Table 1 lists the surface brightnesses of the two peaks seen in the 160-MHz map plus the brightest component of the Sgr A complex, i.e. Sgr A West, at 160, 327, and 1440 MHz. Because of the high noise level seen in the 327-MHz map (see the details of the 327-MHz observations in chapter 4), the surface brightness of the second weak peak at 327 MHz is quite uncertain and its spectral index should be much lower than the upper limit indicated in Table 1. (Figure 5 shows contours of the total intensity at 20 cm with the same resolution as that of the 160-MHz map seen in figure 1). High-resolution observation at 327 MHz, which will be possible in the near future with the VLA, would be extremely useful to determine the detailed structure of this ridge.

## III. DISCUSSION

### a) The Location of Sgr A East

Because the ridge seen at 160-MHz has a high brightness temperature and a steep spectrum (being prominent at 160 MHz and absent at 327, 408, and 843-MHz), we suggest that the character of the new features presented in Figures 1 and 2 is nonthermal. We argue that these features are most readily understood if Sgr A East lies behind the nucleus, and if the extended steep-spectrum emission is linked to the nucleus. That is, the peaks associated with Sgr A East and the ridge are hypothesized to be two independent, but superimposed sources.

A picture in which Sgr A East is located on the far side of the Sgr A West complex was suggested by Güsten and Downes (1980) and Güsten and Henkel (1983) and received recent observational support by our VLA observations discussed in chapter 6. The low-frequency observations reported here also supply clues having a bearing on the suggested geometry. Indeed, if the Sgr A East shell emits nonthermal radiation uniformly around its perimeter as it appears to do in high-frequency maps, then it appears that the low-frequency flux from Sgr A East is reduced substantially in the vicinity of Sgr A West, presumably as a result of free-free absorption by Sgr A West. The peak flux in the meter-wavelength maps is located at the northeastern extremity of the Sgr A East shell (cf., Fig. 1), where the surface brightness there is relatively low at 20 cm. Sgr A West, on the other hand, lies near the southwestern perimeter of this shell. At the electron density ($10^3$ - $10^4$ cm$^{-3}$) and electron temperature (5000 - 8000°K) suggested by H 110α and [Ne II] line data (van Gorkom et al. 1983; Lacy et al. 1980), Sgr A West should have a substantial opacity (> 1) at meter wavelengths, whereas in the extended medium surrounding Sgr A West and in the region between the high-density ionized clumps in Sgr A West, in which the temperature is similar, but the electron density is lower (< $10^2$ cm$^{-3}$, see Watson et al. 1980), the free-free opacity is substantially reduced to < 0.9 at 160

MHz. Indeed, the above argument is justified when the 327-MHz map is superimposed on the 6-cm radiograph of Sgr A East and West which is shown in figure 3. At the location of the "3-arm spiral" (i.e. Sgr A West), the brightness of the 327-MHz emission is reduced by a factor of $\geq 4$. Based on equation 1 of chapter 4, emission measure toward Sgr A, East, has to be $\geq 1.66\ T_e^{1.33}$, where $T_e$ is the electron temperature.

Is the ridge of 160 MHz emission linked to Sgr A East? Such a possibility appears unlikely: The Sgr A East shell is elongated along the galactic plane -- perpendicular to the 160 MHz ridge -- and the surface brightness of the radio halo surrounding the shell decreases monotonically outward from the shell (see chapter 6). There is no obvious way to simultaneously produce a collimated radio feature such as is seen in the 160-MHz map and an elongated structure having an orientation perpendicular to the direction of that collimation.

b) Geometry of the Ridge

The appearance of a one-sided nonthermal radio feature emanating from the nucleus and running roughly along the rotation axis of the Galaxy resembles a scaled-down version of the anomalous radio features seen at higher frequencies toward the central regions of several edge-on spiral galaxies (de Bruyn 1978; Hummel et al. 1982; Duric et al. 1983). The characteristics of radio emission from the nuclei of galaxies showing elongated vertical structures are summarized by Hummel, van Gorkom and Kotanyi (1983). Here, we draw attention to the analogy between the structures seen in the nucleus of our Galaxy and in other spiral galaxies (see Hummel et al. 1983 and the references cited therein). There are several interesting points of comparison:

1) The features seen in NGC 6500, NGC 4438, and NGC 2992 appear to be one-sided, similar to what is seen in Figure 1 at 160 MHz. The asymmetry of the 160-MHz ridge of emission in our Galaxy with respect to the galactic plane recalls the asymmetry of the Ω-shaped radio lobe studied by Sofue and Handa (1984), which is also asymmetric about the galactic plane, but with the opposite asymmetry -- it is present almost exclusively at positive latitudes. We speculate that the 160-MHz ridge and the Ω-shaped radio lobe might be manifestations of one-sided nuclear activity occurring in two different episodes. We speculate that it is also possible that the "central thread" of radio emission which has a jet-like appearance and which is observed at 1.44 GHz (see chapter 5) accounts for the counterpart to the 160-MHz ridge. Figure 4 shows clearly that these two features namely, the central thread and the 160-MHz ridge are oppositely directed from each other. The problems with interpreting the thread as the counterpart to the 160-MHz ridge is that the central thread shows a discontinuity in its orientation at $\alpha \sim 17^h42^m15^s$, $\delta \sim -28°55'$ as it is followed toward the nucleus. Furthermore, this interpretation can certainly not account for other similarly-looking threads. However, if we assume that these two features (the central thread and the ridge) are associated and that both result from a nuclear activity, we speculate that the relativistic electrons associated with the 160-MHz ridge have lost much of their high energy electrons as a result of interaction with the cool gas possibly associated with the 20-km $s^{-1}$ molecular cloud (Güsten et al. 1981; Ho et al. 1985). Indeed, distribution of molecular gas in the region near the Sgr A conmplex is maximized on the negative-latitude side of the galactic plane and thus, this hypothesized interaction could broaden the shape and steepen the spectrum of the 160-MHz ridge, whereas the central thread has not been affected by the ram pressure which the 160-MHz ridge has experienced. Alternatively, the asymmetry of the 160-MHz ridge of emission might be caused by the presence of more ionized gas at positive than at negative latitudes, which thus preferentially suppresses the positive latitude counterpart to the observed ridge. More sensitive observations at 160 MHz toward positive latitudes would be helpful in constraining the cause of the asymmetry.

2) NGC 3079 and NGC 6500 show compact core sources. In particular, NGC 6500 shows a milliarcsecond radio core (~ 5 pc) which is elongated along the galaxy's minor, or rotation axis (Jones, Sramek and Terzian 1981). The nonthermal compact radio source located at the center of our Galaxy has a diameter of only $10^{14}$ - $10^{15}$ cm or less (Lo et al. 1981), but is elongated parallel to the rotation axis of the Galaxy (Lo et al. 1985). Also, there is a 1 pc bar of radio emission coincident with Sgr A West which also aligns along the Galaxy's rotation axis (Lacy et al. 1980; van Gorkom et al. 1983). As is believed to be the case in other galaxies, the colinear radio structures seen on widely different scales in our own galactic nucleus are likely to be manifestations of a single phenomenon.

3) The extended ridge of emission in most of these galaxies, including ours, is observed to align with the galaxy's minor, or projected rotational axis (Duric et al. 1983). Furthermore, the alignment with that axis improves with increasing distance from the nucleus. The apparent misalignment near the nucleus of our Galaxy, however, is very possibly due to the superposition of two unrelated sources, as discussed in § III.a.

4) All five galaxies listed by Hummel et al. (1983), including our own Galaxy (Watson et al. 1981), appear to have an X-ray source in their nucleus, except NGC 6500. However, many of them show Seyfert-like emission lines from the nucleus, whereas our Galaxy is unobservable in this regard, and all of them have a substantially larger radio luminosity than our Galaxy.

5) The linear dimension of the elongated, vertical radio features now known in external spiral galaxies is ~20 to 200 times larger than that seen in Figure 1. However, the ratio of the angular scale of the long axis of the vertical ridge of emission to that of the compact nuclear radio source is roughly the same in both external galaxies and our own.

6) A major difference between the 160-MHz ridge observed in Sgr A and that detected in the nuclei of edge-on spiral galaxies is the spectral index of the ridge and the frequency at which the ridge manifests itself.

These comparisons between the radio characteristics of the nucleus of our Galaxy and those of several edge-on spirals having radio-bright nuclei suggest circumstantially that the 160 MHz ridge is associated with the nucleus, but is a lower-energy version of phenomena seen elsewhere.

In view of the probable inhomogeneities in the medium lying between the Earth and the galactic center, the possibility must be considered that the 160-MHz ridge corresponds to a cleft in the distribution of intervening optical depth at this frequency. Then, although the 160-MHz background from the galactic center region might intrinsically be fairly uniform, we would be viewing it only through this cleft. With the present data base, this possibility cannot be ruled out, but it does imply both a remarkably improbable coincidence of location and orientation and the existence of a diffuse HII region which has not been detected toward the galactic center at high frequencies.

c) Emission Mechanism and Origin

Because the observed spectrum from the 160-MHz ridge is nonthermal and because the observed emission from the possibly analogous, elongated feature in NGC 3079 (Duric et al. 1983) is highly polarized (up to 50% at 6 cm), we argue that the radiation emitted from the 160-MHz ridge connected to Sgr A is produced by the synchrotron mechanism. The equipartition magnetic field in the 160-MHz ridge is then $\sim 8 \times 10^{-6}$ gauss and the synchrotron lifetime at this frequency is $\sim 10^{8}$ years. Since the free expansion-time for a cloud of relativistic particles having the dimensions of the 160-MHz ridge is only ~100 years, one might conclude that the initial energy distribution of electrons injected into this source is quite steep.

However, the ambient gas pressure in the nucleus of the Galaxy might be large enough to confine the collimated ridge. The low-thrust jet might have been affected severely by the rotating gas associated with the galactic disk. The electron density, $n_e \sim 16$ cm$^{-3}$, and electron temperature, $T_e \sim 10^4$ °K, estimated from the total 5 GHz flux emitted from the inner 61'×20' ($\theta_\ell \times \theta_b$) of the Galaxy (Mezger et al. 1974) indicate that the thermal pressure of the surrounding gas could be an order of magnitude greater than the internal pressure of relativistic particles -- assumed to be in equipartition with the magnetic field pressure.

We suggest that the elongated ridge of emission is possibly produced by the injection of relativistic electrons from the non-thermal compact source directed along the rotation axis of the Galaxy. The elongated ridge of emission might then be a jet, related to the much larger scale jets occurring in much more active galaxies than our own. If the steep spectrum can be attributed to age and if the expansion of relativistic particles is severely affected by the ambient gas pressure in the Sgr A environment, then this structure might be a "fossil jet". Alternatively, this radio ridge could be a manifestation of a continuous low energy jet emanating from the nucleus. Higher angular resolution observations of this feature at low frequencies would be fruitful in clarifying the nature of this ridge of radio emission.

# IV. SUMMARY

A narrow ridge of low frequency radio emission is found to lie along the rotation axis of the Galaxy and to extend ~ 30 pc from the nucleus toward negative latitudes. No positive-latitude counterpart is evident at 160-MHz. This radio feature is observed only at 160 MHz using the Culgoora Circular Array and the constraints on its spectral index indicate that it is nonthermal. We argue that it results from the continuous, or frequent ejection of mildly relativistic particles from the galactic nucleus. It is reminiscent of radio structures seen at higher frequencies in the nuclei of a number of spiral galaxies having Seyfert-like properties. Hypotheses which might account for such a structure are discussed.

Table 1

Spectra of Three Positions along the 160 MHz Ridge

| Source | $\alpha(1950)$ | $\delta(1950)$ | Peak Brightness Temperature (K) | | | Spectal Index ($\alpha$) | | |
|---|---|---|---|---|---|---|---|---|
| | | | 160 MHz | 327 MHz | 1.44 GHz | 0.16–0.327 | 0.16–1.44 | 0.327–1.44 |
| First Peak | $17^h42^m36\overset{s}{.}7$ | −28°58'17" | $6.35\times10^4$ | $7.5\times10^4$ | $4.2\times10^3$ | +2.3 | +0.76 | +0.05 |
| Second Peak | $17^h42^m58\overset{s}{.}0$ | −29°03'40" | $1.9\times10^4$ | $< 7.9\times10^3$ | < 140 | < +0.78 | < −0.23 | — |
| Sgr A West | $17^h42^m29\overset{s}{.}0$ | −28°59'22" | $2.5\times10^4$ | $1.5\times10^4$ | $1.4\times10^4$ | 1.28 | 1.73 | 1.95 |

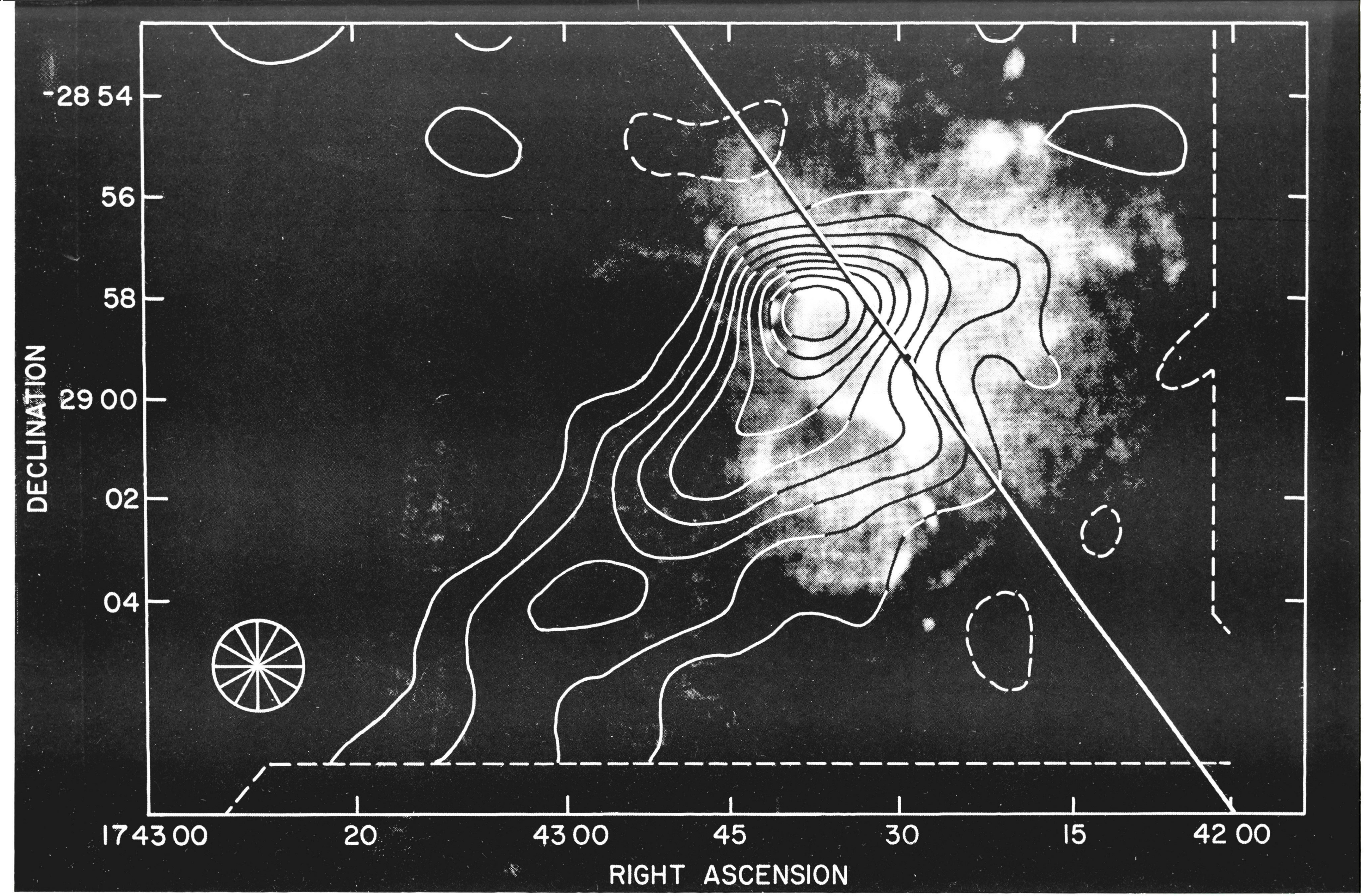

Figure 1: The 160-MHz map of Sgr A, made with a resolution of 1.9' x 1.9', is superimposed upon the 1.4 GHz VLA map, which has a beam (FWHM) of 5" x 9" ($\alpha \times \delta$). The 160-MHz contour levels are 10%, 20%, ... 90% of the peak brightness, which corresponds to 12.0 Jy/beam area. The solid black and white line shows the orientation of the galactic plane. The black spot indicates the position of galactic center.

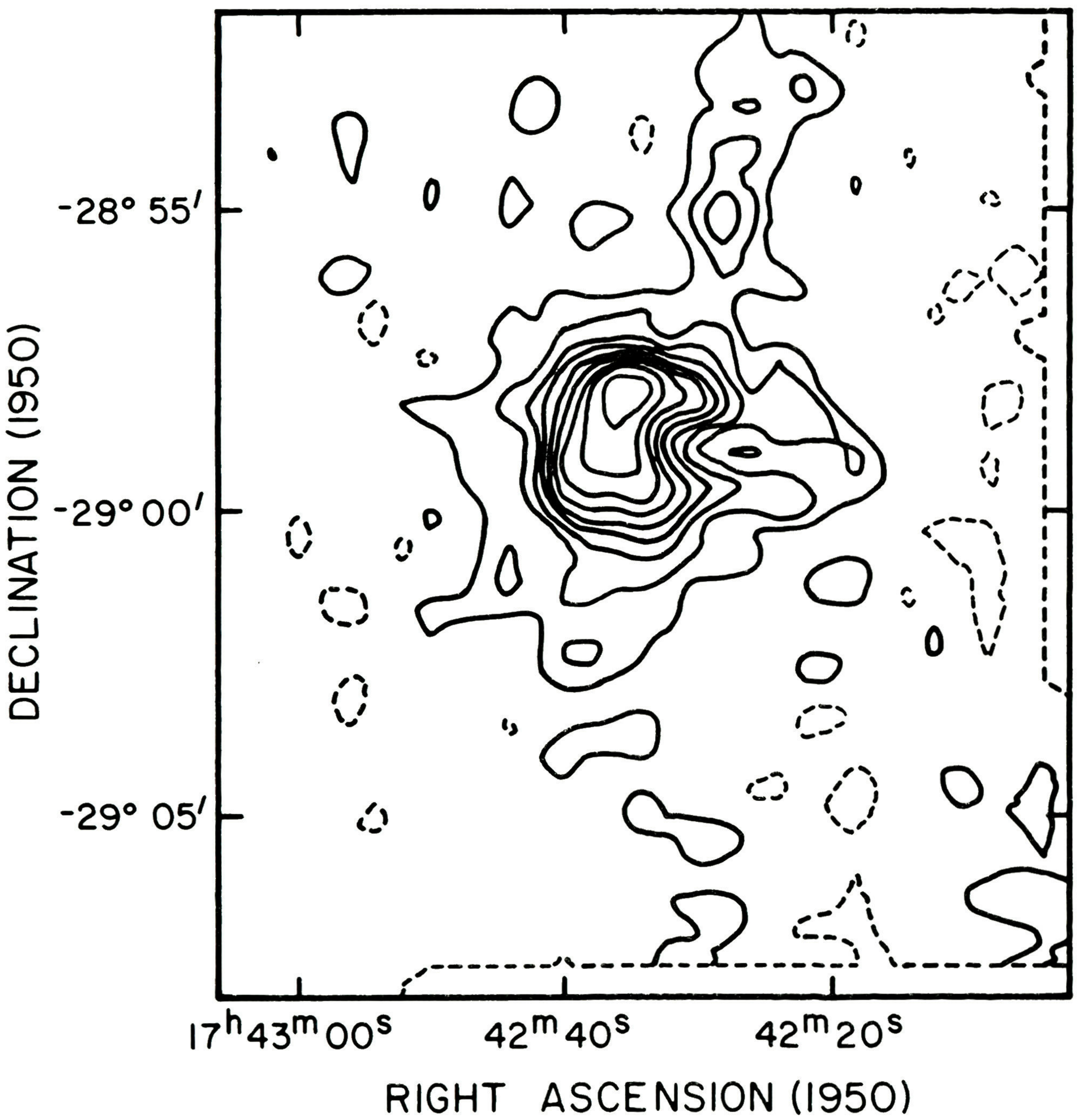

Figure 2: The 327-MHz map of Sgr A, made with a resolution (FWHM) of 56" x 56". The contour levels are 10%, 20%, ... 90% of the peak brightness, which corresponds to 14.2 Jy/beam.

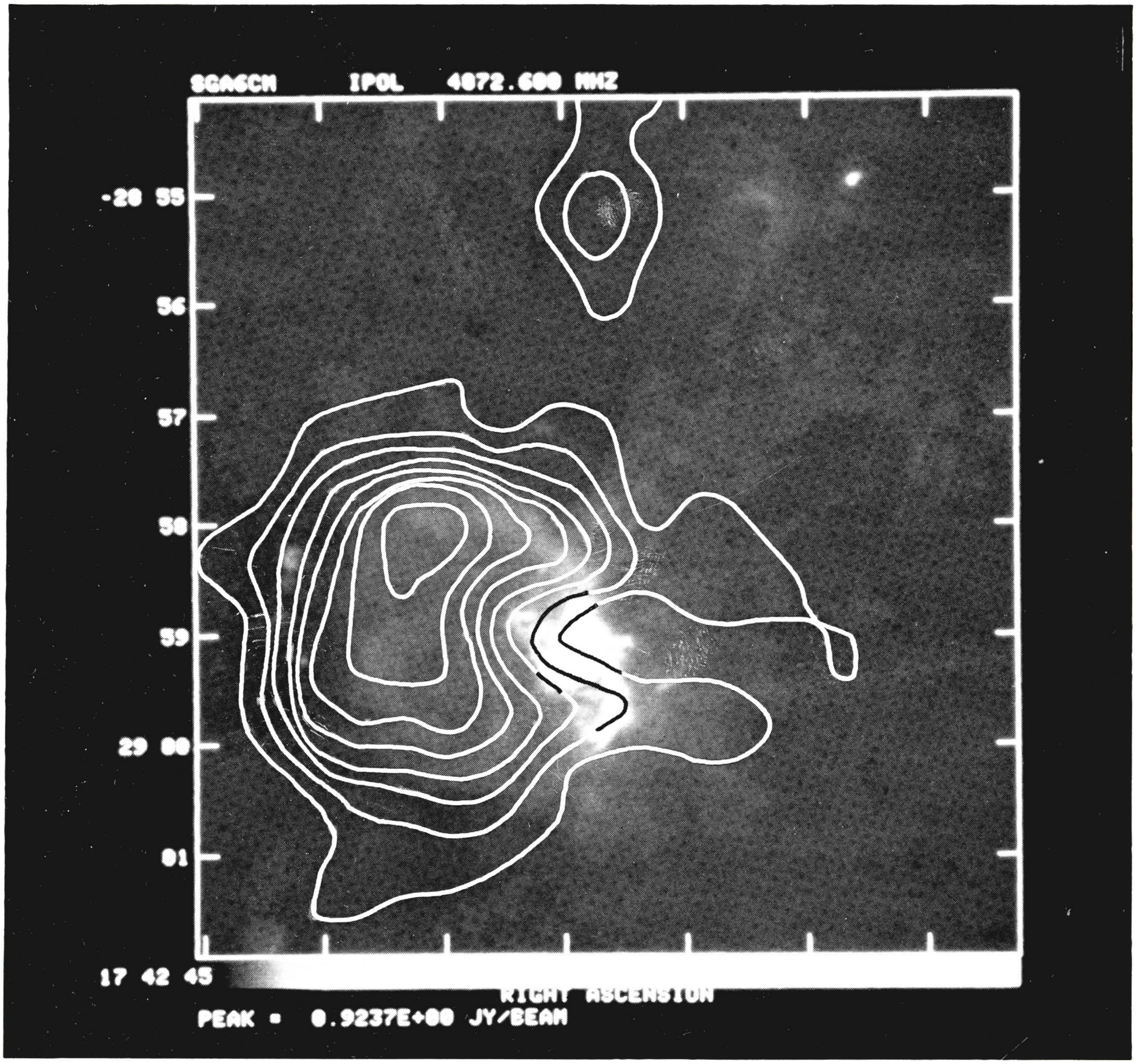

Figure 3: The 327-MHz map is superimposed on the 5-GHz image of Sgr A East and West which is identical to figure 5 of chapter 5 except that the transfer function is different.

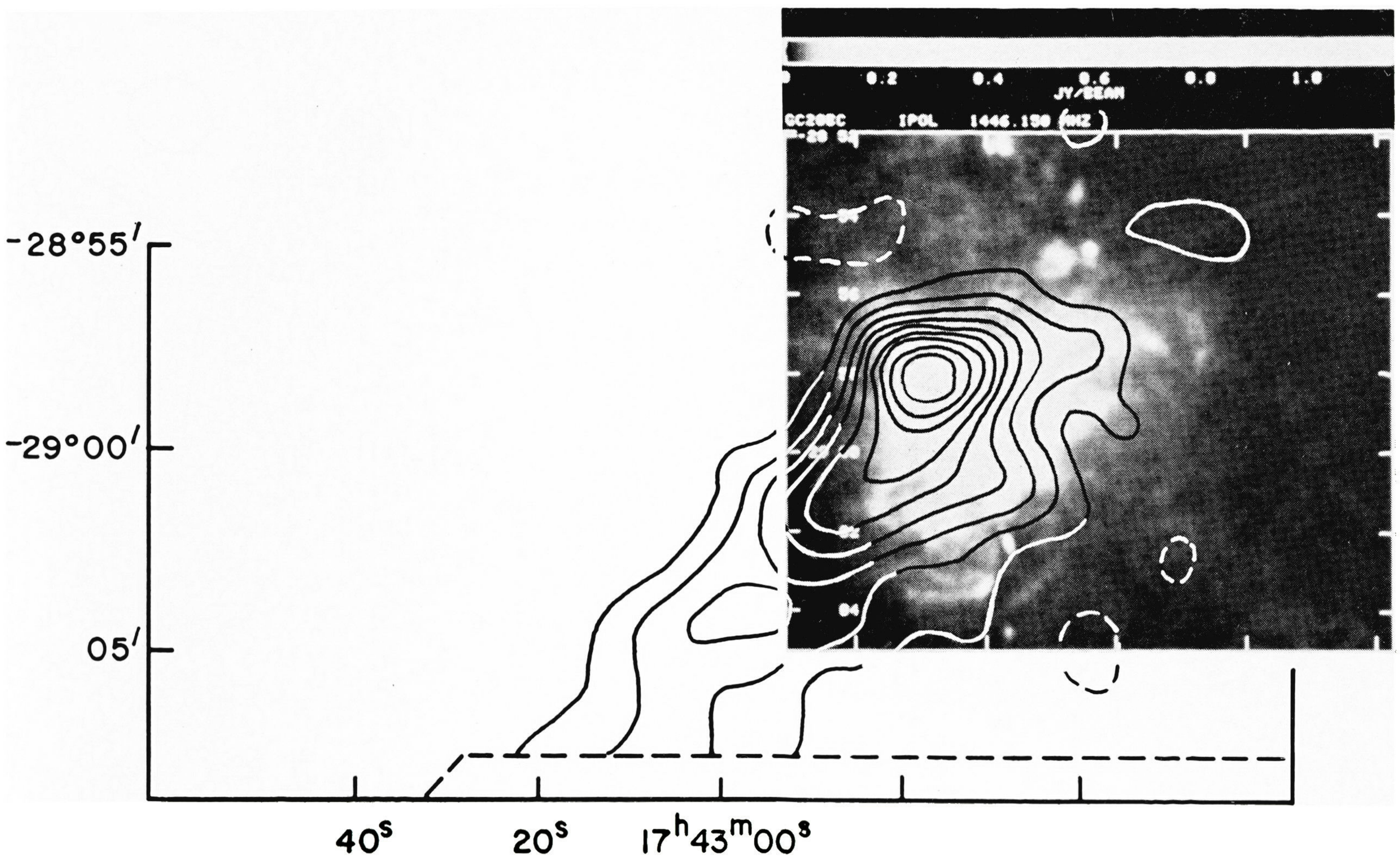

Figure 4: This figure is identical to figure 1 except that a different transfer function is chosen for the 20-cm image in order to show the central thread.

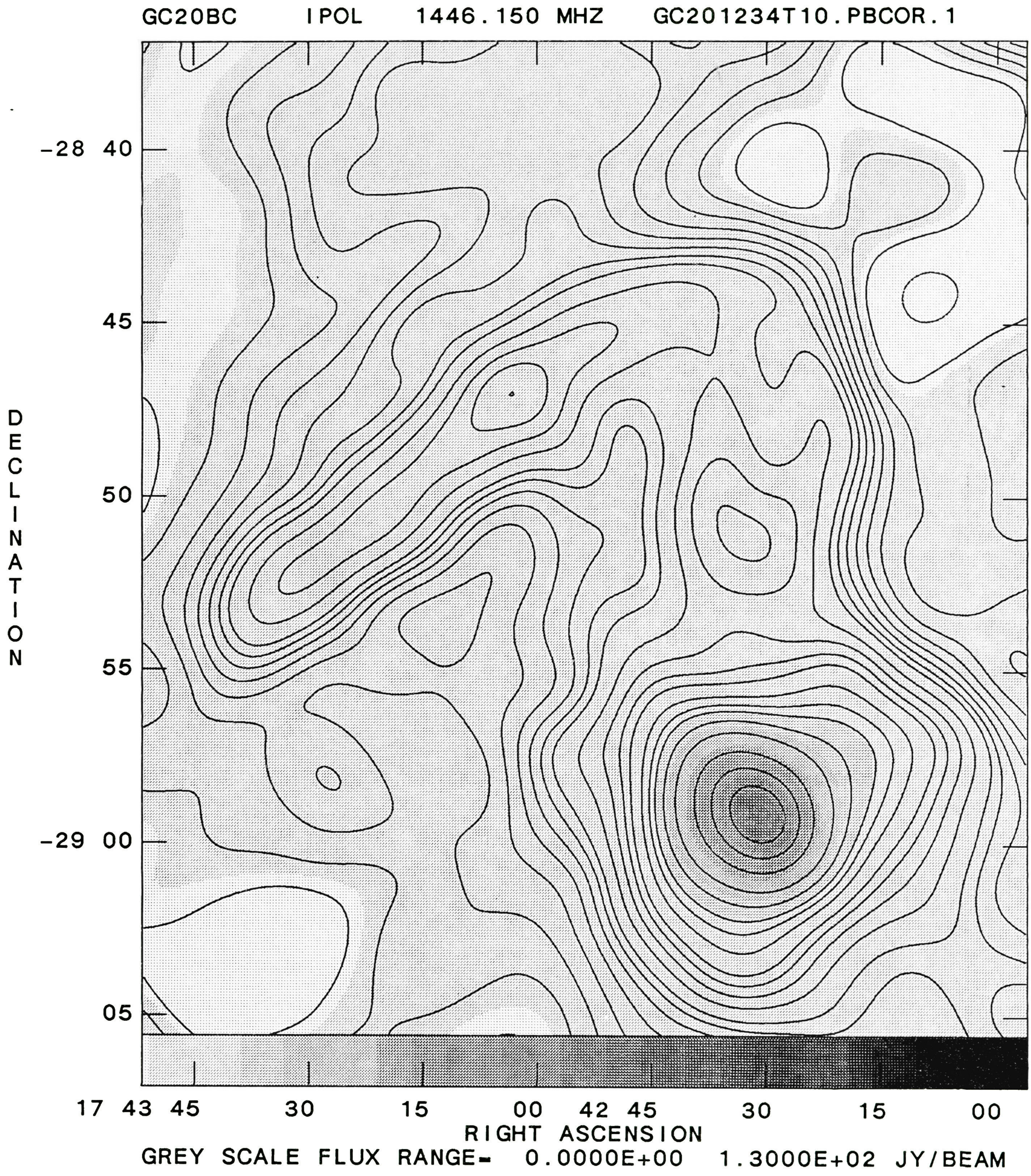

Figure 5: The 1.4 GHz contour map with a resolution of 1.9' (i.e. FWHM) is shown with intervals of -.5, .5, 1, 1.5, 2, 2.5, 3, 3.5, 4, 5, 6, 7, 8, 10, 12, 14, 18, 25, 40, 55, 70, 85, 100 Jy/beam area.

# Chapter 8

## A SYMMETRICAL LARGE-SCALE POLARIZATION STRUCTURE NEAR THE ARC[6]

"God has asked me to tell you the meaning of what you have seen. The Arc of hot gases is being formed by Space Brothers bringing high energy from other parts of the Galaxy ..., it is time to bring the earth off its axis"

In His Service, Edna Chappel

### I. INTRODUCTION

The last several chapters have dealt exclusively with radio-interferometric observations of the Arc and the Sgr A complex. Here we present a 1°× 1° radio image of the galactic center region based on single-dish observations. These observations, which were made with the 100-m Bonn telescope, signify a new aspect of large-scale radio structure seen near the galactic center. This new structure consists of two polarized lobes of radio emission which are aligned with the linear portion of the Arc, and which are symmetrically placed with respect to the galactic plane. (The results of recent observations of these lobes using the VLA are reported in chapter 10).

Past single-dish observations have indicated a number of large scale ridges rising away from the galactic plane up to $b = \pm 0.5$ (Altenhoff et al. 1978; Whiteoak and Gardner 1973; Haynes et al. 1978). A more detailed study of this region by Sofue and Handa

[6]This chapter is the product of a collaboration with Dr. Seiradakis (University of Thessaloniki), Dr. Lasenby (MRAL, Cambridge), and Drs. Wielebinski and Klein (both at MPIfR).

(1984) showed that two of these radio ridges form an Ω-shaped radio loop lying at the positive-latitude side of the galactic plane. The two feet of the loop, which are projected out of the plane at $\ell$ = 0.18° and $\ell$ = 359.4°, appear to merge with the linear portion of the Arc and Sgr C, respectively. (The Arc and Sgr C are located on opposite sides of the galactic center.) Subsequent study by Inoue et al. (1984) and Tsuboi et al. (1985) showed that a portion of the large-scale loop, located at positive latitudes is highly polarized and has a large Faraday rotation.

We were motivated to determine, in the first place, if the linear portion of the Arc is part of a broader envelope of radio emission which extends continuously to almost a degree above and below the galactic plane (see figure 4 in chapter 1 made by Altenhoff et al. 1978). Furthermore, we hoped to determine the physical relationship between the remarkable loop structure, recognized by Sofue and Handa (1984), and the southern ridge seen in the radio survey of Altenhoff et al. (1978) at $\ell$ > 0.0.

## II. OBSERVATION

The observations were performed using the 100 m radiotelescope at Effelsberg, near Bonn, between July and August 1983 (4.75 GHz data) and in July 1984 (10.7 GHz data). Both receivers (parametric amplifiers with $T_{sys}$ = 70 K at 4.75 GHz and FET amplifiers with $T_{sys}$ = 100 K at 10.7 GHz) were used in single horn mode and had a band-width of 500 MHz. The half power beamwidths were 2.4 and 1.2 arcmin, respectively, for the two frequencies. The weather was excellent throughout the observations, enabling unswitched total power information to be used. The Stokes parameters I, U and Q were recorded simultaneously. Several calibration sources were observed at different elevations, including 3C 286, 3C 48, 3C 147 and 3C 409. These data were used for estimating the receiver characteristics and atmospheric extinction, the latter being particularly important at the higher frequency (10.7 GHz).

At 4.75 GHz we obtained three coverages scanned in RA and two in Dec, covering a $40' \times 40'$ field, but with the last RA coverage being extended to $1° \times 1°$. The galactic center is visible from Effelsberg for about 1.5 hours daily, therefore for the higher resolution 10.7 GHz map the region was observed in 8 independent strips, with appropriate overlap for the off-line processing and combination into one map. Because of atmospheric extinction, only scans in RA (corresponding roughly to azimuth) were made. The region covered was $1° \times 1\overset{\circ}{.}1$, the northern extension being made to include all of Sgr B2. At both frequencies the maps (made using the NOD2 procedure of Haslam [1974]) were absolutely calibrated using both calibration source information and existing lower resolution data covering a larger region. This was important for setting the edges of the maps to the correct absolute values. At 4.75 GHz we consulted the maps of Haynes _et al._ (1978) and Altenhoff _et al._ (1978) and at 10.7 GHz the edge calibration was obtained using the map of Sofue _et al._ (1984).

## III. RESULTS

Figure 1 is a radiograph of the total intensity of 10.7 GHz emission. Most of the features seen in this map are identical to those features brought out with the VLA observations and are described in chapter 3. It should be mentioned, however, that single-dish observations do not suffer from problems associated with radio interferometers (i.e. lack of short [u,v] spacings). Therefore, the flux density measurements of large-scale features are much more accurate using single-dish observations than with radio interferometers measurements, especially for extended radio sources lying along the galactic plane.

The two arched filaments reaching from the Sgr A complex to the NW end of the linear filaments, as shown in figure 1, are matched by a symmetrical pair of filaments reaching to their SE end. The NW arched filaments have a peak flux of 2.15 Jy/beam whereas the SE arched filaments, the so called "counter-arch" in chapter 3 (§II.3f), have a peak flux of 0.6 Jy/beam. Another feature linking the Sgr A complex to the sickle-shaped feature (G0.18-0.04) can be seen in this figure. Overall there is a remarkable symmetry in the features seen at positive and negative latitudes, although not in absolute surface brightness.

The most interesting result is shown in Figure 2, which is a radiograph of the intensity of polarized emission at 10.7 GHz. A core/lobe (A,B,C) structure is visible at a P.A. of 120°, parallel to the Arc. The two lobes (B,C) resembling a plume-like geometry lie symmetrically with respect to the true galactic plane. The lobe B lies to the north of the region where the linear and arched filaments meet (see figure 1 in chapter 3). The core coincides with G0.16-0.04 whose high-resolution polarization properties at 6 cm were described in detail in chapter 3. The lobe C which is located at negative latitude is the newest component added to the rich collection of radio features viewed in the galactic center region. The core A and the lobes B and C have peak percentage polarizations of 22, 32 and 30%, respectively, in a 1.2' beam. The two other patches of polarization visible correspond to Sgr A and a region close to the peak of Sgr B2. The percentage values here, 1 and 3% respectively, represent the level of instrumental polarization for Sgr A, but appear to indicate some intrinsic polarization SE of Sgr B2. In figure 2 the core A and NW lobe correspond to the sources A and B of Inoue et al. (1984). New detailed mapping of the inner 1° × 1° by Tsubsoi et al. (1986) exhibits the same polarization morphology described in this chapter.

The average rotation measures for the A and B components, as measured by Inoue et al. (1984), are -1660 and +800 rad $m^{-2}$ respectively. If the symmetrical lobes have similar rotation measures, the intrinsic magnetic field directions are found to lie +30, +14° and -25° relative to the 120° P.A. of the linear portion of the Arc for the components A, B, and C respectively. Alignment of the magnetic field with the direction of the core/lobe structure is best seen in a recent map made by Tsuboi et al. (1986).

The extension of the linear portion of the Arc can be seen in a low contrast version of the 4.75 GHz map in figure 3. The north-western extension of the Arc in this figure can also be seen in 10.5 GHz Nobeyama data filtered to remove large-scale background structure (Fig. 8 of Sofue 1985) and it appears to indicate smooth continuation to the large scale ($\sim 1° \times 1°$) $\Omega$-shaped loop which extends to northern galactic latitude $b = 0\overset{\circ}{.}5$.

## III. DISCUSSION

The core/lobe polarization structure seen in figure 2 is reminiscent of the total intensity picture of a classical double radio source. The 160-MHz map of the Arc shows also that G0.16-0.04 (i.e. the core) is the only source seen along the Arc (see chapter 4). Radio recombination line observations by Pauls et al. (1976) and Pauls and Mezger (1980) show a lack of emission from G0.16-0.04. In contrast to the polarizations asymmetries seen in the Arc, the total intensity pictures at high frequencies (see chapters 1 & 3) show a continuous source of emission which links core A to lobe B. The model which was invoked to account for the 160-MHz distribution is used here to account for the structures seen in polarized and total intensity maps (i.e Figures 1 & 2). That is, a non-uniform screen of thermal gas, threaded by the ambient magnetic field, attenuates the nonthermal emission from the Arc at low frequencies and depolarizes the underlying synchrotron radiation. This non-uniformity in the thermal gas distribution is suggested to be fueled by high molecular column density (40 km $s^{-1}$ cloud) seen in the region between core A and lobe B (see chapters 4 & 9). However, it would be difficult to generate the complete cocoon of ionized gas fueled only by the 40 km $s^{-1}$ molecular cloud since a steep gradient of molecular and dust column density density is seen in the region between sources A and C at negative galactic latitudes (see figures 4 and 5 of chapter 4). Alternatively, it is possible that the cocoon of ionized gas is fed along the arched filaments from the vicinity of the galactic nucleus as part of a supply of matter to the envelope surrounding the regions between the core and its 2 lobes. Such an outflow model from the galactic center or its vicinity is discussed further in chapters 9 and 10. We argue further in chapter 10 that the non-uniform thermal gas lies very near the non-thermal features.

One of the main motivations for single-dish observations olf the Arc was to seek evidence which would indicate that the filamentary Arc has true extensions at both positive and negative latitudes. Figure 4 shows the contours of total intensity at 6 cm superimposed on a high-resolution VLA image at 20 cm. The NW and SE extensions of the filaments which compose the linear portion of the Arc strongly suggests that the Arc extends $\pm 0.5°$ away from the galactic plane. Figure 3 shows the NW at the location where the linear filaments are crossed by the arched filaments and where the brightness of the filaments is weakened suddenly by at least a factor of 10. Curiously, the NW extension of the filaments at positive latitude shows a strong linear polarization immediately to the north of the arched filaments whereas the polarized emission from the symmetrical counterpart of lobe B (i.e. lobe C) is not positioned at the location where the "counter-arch" filaments cross the filaments. In other words, the polarized emission is not seen to be symmetrical with respect to the arch and counter-arch.

The linear portion of the Arc appears to lie along a surface of constant angular velocity based on the mass distribution given by Oort (1977) and consists of a number of filaments which are continuous over 40 pc. In contrast, the two lobes seen in figure 2 do not lie along a surface of constant angular velocity and do not show any filamentary structure in single-dish maps. Thus, if they are associated physically with the Arc, the differential rotation of the Galaxy would distort their alignment with the Arc in several rotation periods. An extension from a filamentary structure in the Arc to a diffuse structure with a weaker surface brightness than those seen in the linear filaments, at least by a factor of 10, is not understood if the two structures are physically linked as we argue here. Further discussion of the diffuse polarized components and the filaments are found in chapter 10.

The apparent absence of a loop structure at negative galactic latitudes compable to that seen by Sofue and Handa (1984) and the nevertheless remarkably symmetric structure of the polarized components A, B and C in polarization and of the Arc and its extensions suggest that caution is required in identifying the Sofue lobe directly as a continuation of the Arc. There also exist a problem with reconciling the apparent one-dimensional geometry of the filaments in the Arc, as was argued in chapter 3, with the implied two dimensional geometry of the Ω-shaped feature, as argued by Sofue and Handa (1984), who interpreted it a limb-brightened cylinder.

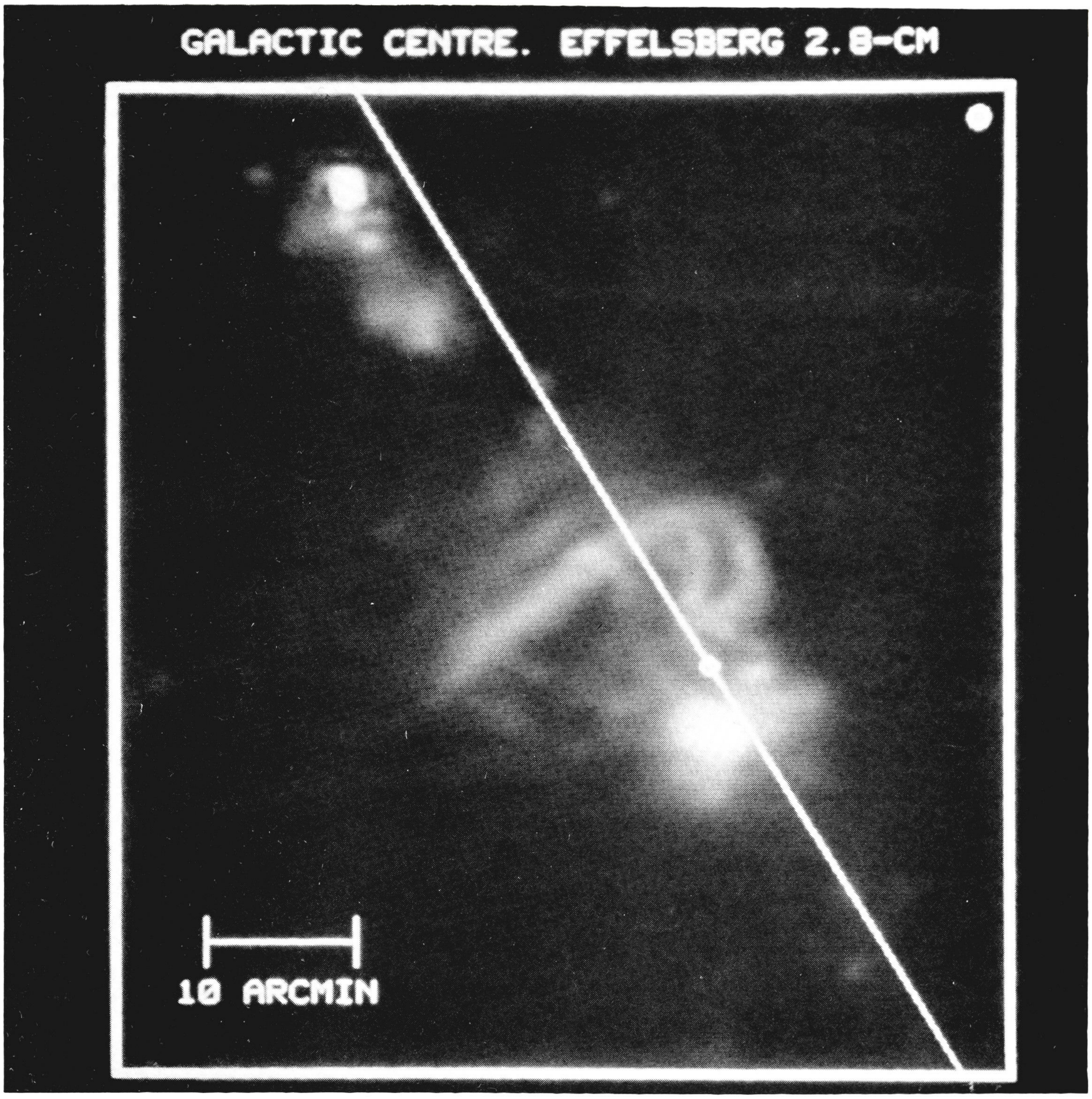

Figure 1: Total intensity radiograph (logarithmic scale) of a 60' × 68' region around the galactic center at 10.7 GHz. The coordinates of the bottom left corner ar RA $17^h45^m28\overset{s}{.}4$, Dec -29°23'00". The half-power beam is shown as a filled circle at the upper right corner. The position of zero galactic coordinates ($\ell$ = b = 0) is marked by a circle on the diagonal line which indicates the b = 0 plane. The peak flux density of Sgr A is 26.8 Jy and Sgr B2, 17.6 Jy.

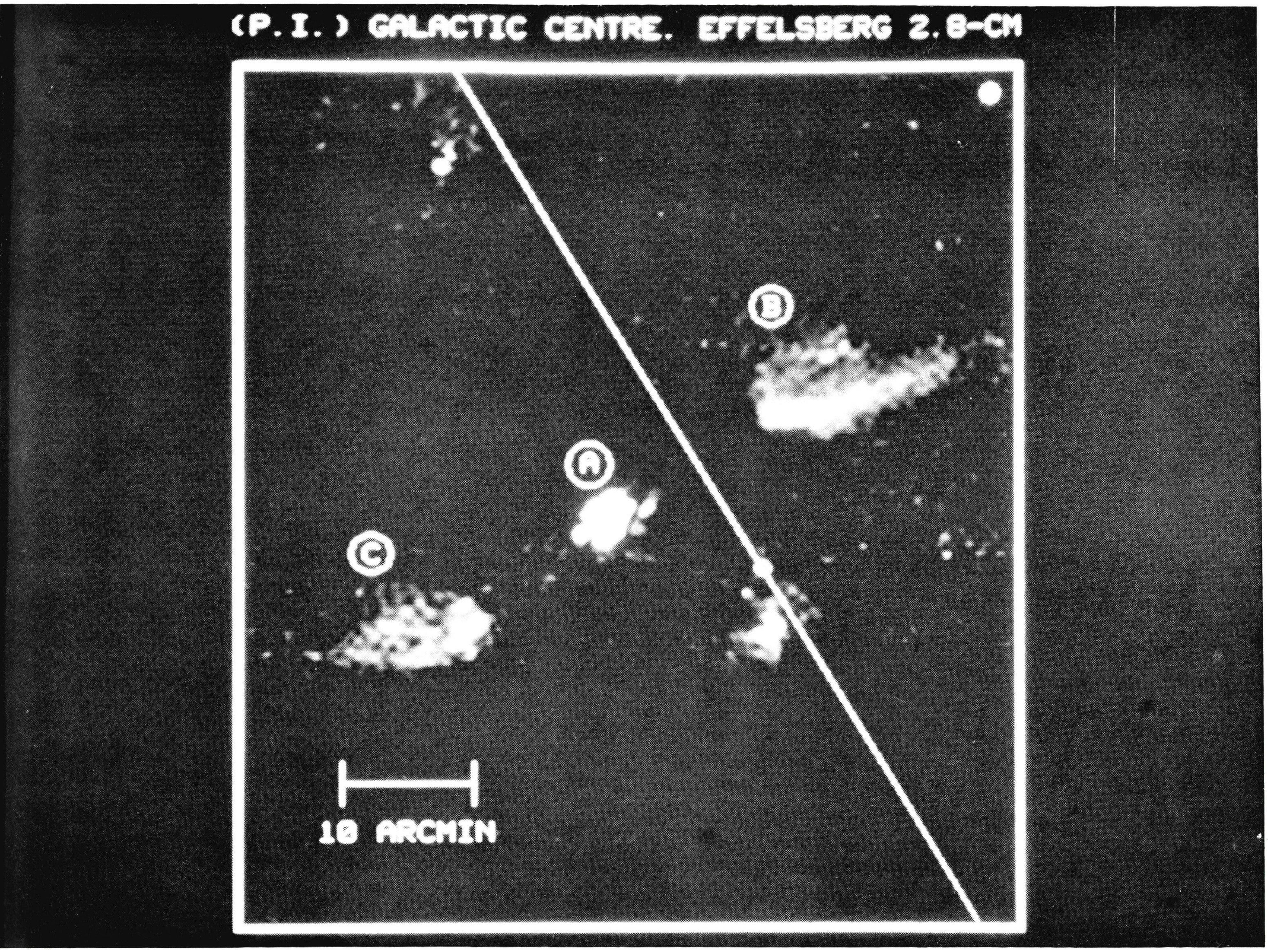

Figure 2: Polarized intensity radiograph (linear scale) of the same region as in Figure 1 at 10.7 GHz. The peak polarized flux density of components A, B and C is 690, 310 and 140 mJy respectively while that for Sgr A is 260 mJy. For further details see Figure 1.

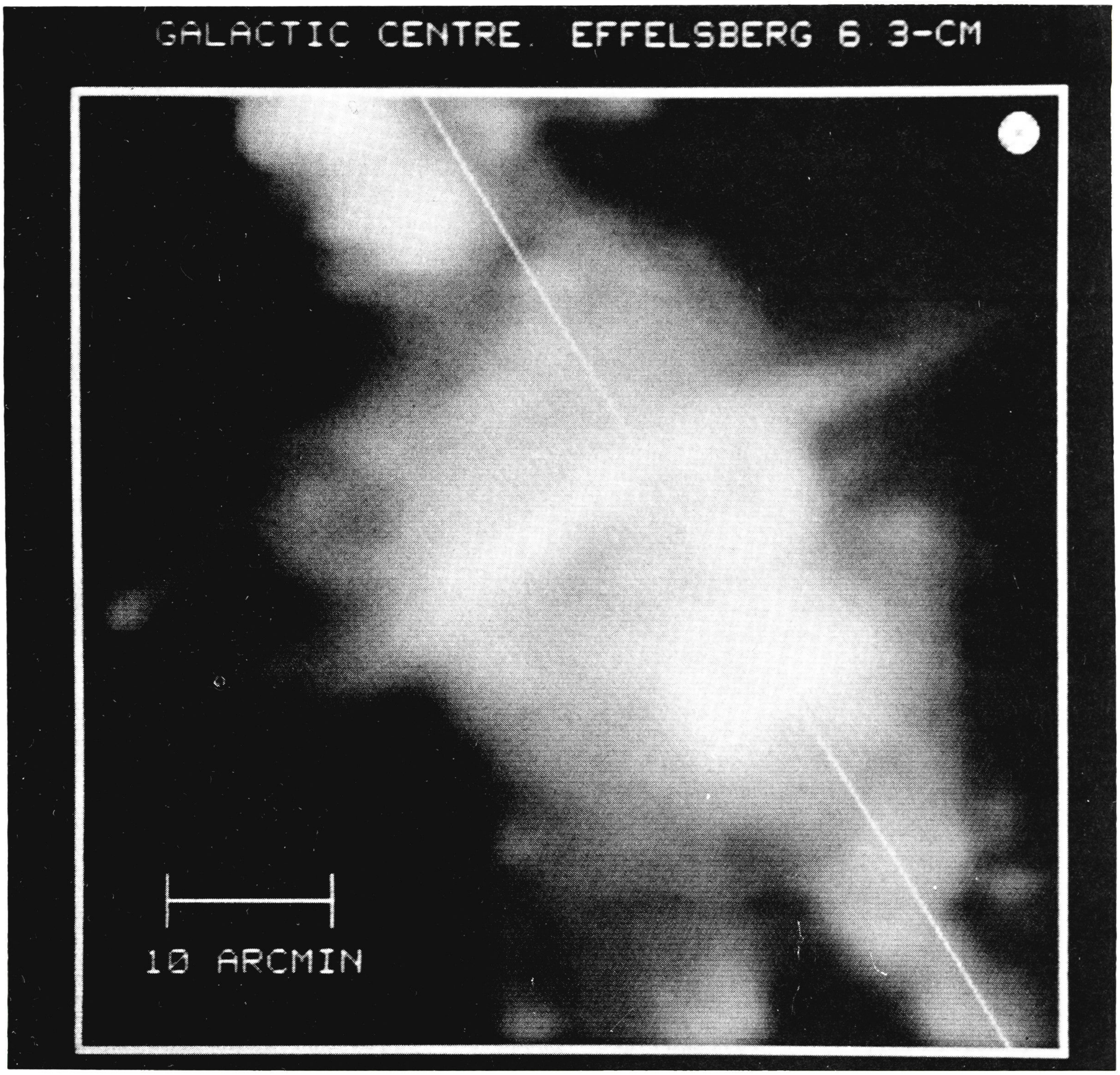

Figure 3: Low contrast (histogram equalized-linear scale) radiograph of total intensity of a 60' × 60' region around the galactic center at 4.75 GHz. The bottom left corner of the map coincides with the bottom left corner of the map in Figure 1. The peak flux density for Sgr A is 58.5 Jy. The peak flux of the NW arched filaments is 4.5 Jy. For further details see Figure 1.

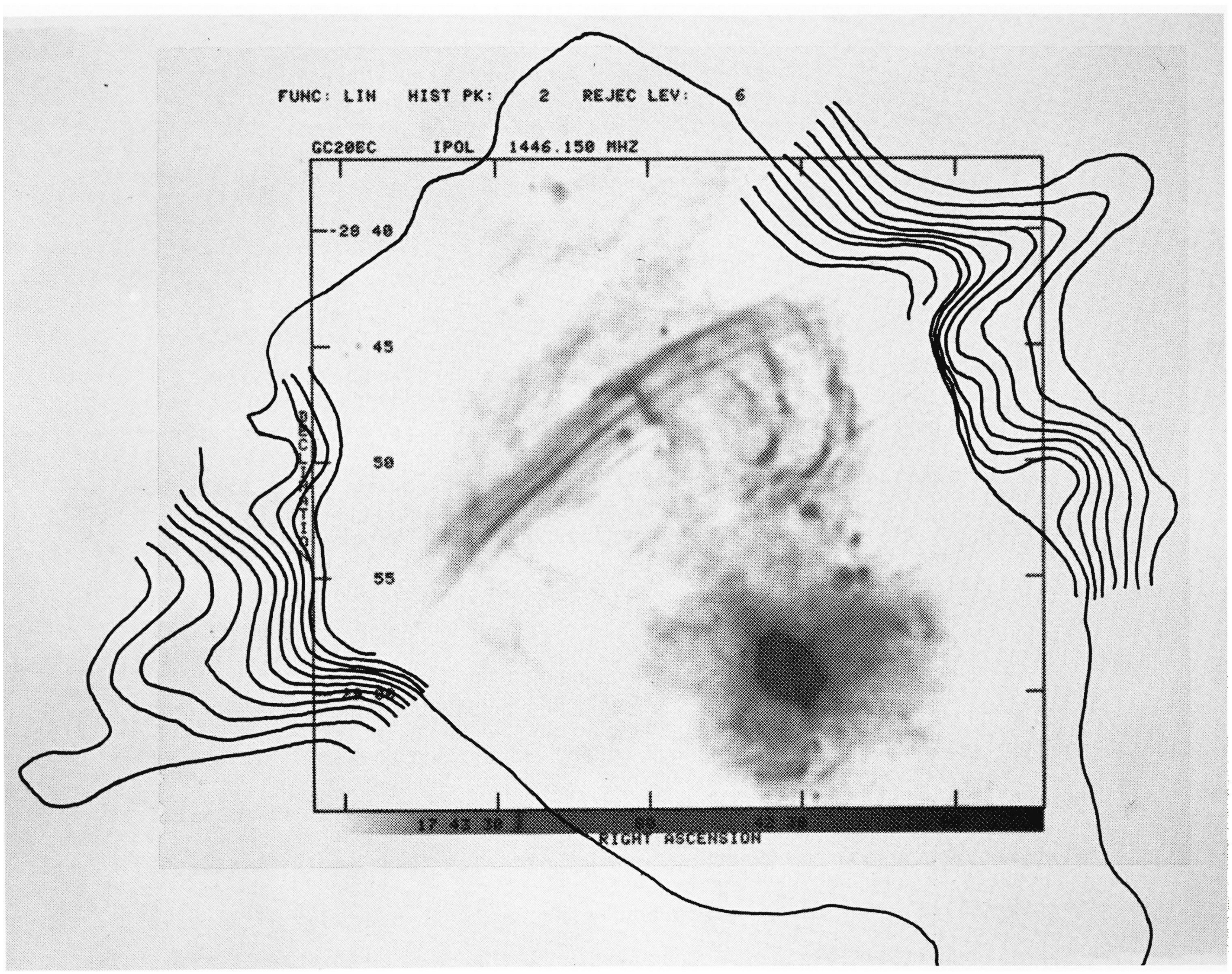

Figure 4: The 6-cm radio contours of the image shown in figure 3 is superimposed on the 1.4 GHz image of the filamentary Arc with a resolution 8.1"× 7.1".

# Chapter 9

# RECOMBINATION LINE EMISSION FROM THE GALACTIC CENTER ARC

"Into the Universe, and Why not knowing Nor Whence,
like Water Willy-nilly flowing; And out of it,
as Wind along the Waste, I know not Whither,
willy-nilly, blowing."
Omar Khayyam

## I. Introduction

Multiconfiguration radio continuum studies of the galactic center region using the VLA reveal that the radio Arc ($\ell$ = 0°2) consists of a number of radio features, the most important of which are: 1) A network of linear filaments which appear to be coherent over a scale of 40 pc, and which are oriented perpendicular to the galactic plane. 2) A system of arched filaments (G0.1+0.08) which meet the linear filaments at positive latitudes and show a high degree of non-uniformity in their appearance. 3) Finally, a non-filamentary structure (G0.18-0.04) which appears to cross the linear filaments along the galactic plane. Based on its appearance, we argued in chapter 3 that this structure (i.e. the sickle-shaped feature) is interacting with the linear filaments. We hypothesized in chapter 4 that two radiation mechanisms operate in the radio Arc: nonthermal emission occurs in the system of linear filaments and thermal emission arises in the arched filaments and in a halo which is centered on the linear filaments and which includes G0.18-0.04. In presenting the results of hydrogen recombination line emission in this chapter, we are concentrating entirely on the thermal component of radio emission from the Arc.

Hydrogen recombination line emission has previously been observed at several places along the Arc with single-dish telescopes. A search for the H85α line was first carried out by Pauls et al. (1976) with a beamwidth of 3' at 10.5 GHz. They found that many lines show non-gaussian profiles and possibly have multiple velocity components. However, no lines were seen in the southern portion of the Arc at G0.16-0.15. In other studies, Gardner and Whiteoak (1977) observed the Arc at 4.875 GHz (H110α) with a lower spatial resolution (4.5') than that of Pauls et al. (1976) but with a greater velocity coverage. They concluded that the linear emission is displaced to the outside of the Arc with peaks at three positions along the Arc and that a velocity pattern was evident. Subsequently, Pauls and Mezger (1980) sampled the entire region of the Arc with a 2!6 beam at 5 GHz (H109α). Their line intensity map, as shown in figure 6 in chapter 1, illustrates two regions of significant line emission: G0.18-0.04 and G0.10+0.08. These authors argued that the line emission at G0.18-0.04 may signal the location of a site of star formation associated with the 40 km $s^{-1}$ molecular cloud (Fukui et al. 1977; Güsten et al. 1981; Brown and Liszt 1984).

We report here the first aperture synthesis observations of radio recombination line emission (H110α) from these two regions (The results of the sickle-shaped structure will be very brief and preliminary). High resolution radio continuum images of these two regions will be shown in order to identify the structural details of the thermal component of the Arc.

## II. Observations

Full description of the techniques which were employed in making the line observation is given in chapter 2. Here, we describe the construction of some of the figures which will be discussed in this chapter. The radiograph shown in figure 1 is our highest resolution image of the continuum emission from the arched filaments at 6 cm. The visibility data corresponding to this figure is based on combining the continuum channel of the line observations using the C/D array and the continuum observations with the B, B/C[1] and C/D[1] arrays (see Arc No. 4 in Table 1 of chapter 2) having the same phase center.

The radiograph and the contour presentation of the total line emission from the arched filaments having spatial resolution, i.e., FWHM, of $22''.47 \times 11''.3$ are exhibited in figures 2 and 3, respectively [note that a galactic (equatorial) coordinate system is used for figure 2(3)]. These maps are constructed by integrating the line flux exceeding a chosen threshold level over all channels. The radiograph (the contour presentation) shown in figure 4(5) is produced by subtracting the continuum channels from all channels in order to display the line emission arising from the arched filaments in each of the central 16(21) consecutive velocity channel centered on 32-channel spectrum (see chapter 2 §II). A map of the intensity-weighted velocity centroid of the region shown in figure 3 is presented in figure 6. The spectra presented in figure 8 (1 to 26) correspond to positions shown with crosses on figure 7, which is the continuum map based on the continuum channel; this map has spatial resolution identical to that of figure 3. Figures 9 and 10 show the high and low resolution images of G0.18-0.04, respectively. Figure 10 is based on the continuum channel of the line observations (FWHM = 22"× 12"). The spectra presented in figures 12 (1-6) correspond to positions (1-6) marked on figure 10. Figure 11 presents a map of the intensity-weighted velocity centroid of the region shown in figure 10.

## III. Results

The results and discussions of each of the fields which depict the arched filaments and G0.18-0.04 are presented separately in the following two sections.

### The Arched Filaments (G0.1+0.08)

The radiograph of continuum emission from the arched filaments in figure 1 shows at least four sets of filaments. These were designated as two western and two eastern filaments (W1, W2, E1 and E2) in figure 10 of chapter 3. (We use these designations throughout this chapter.) The most important result is that the arched filaments, as seen in figures 2 and 3, have thermal characteristics. This supports the earlier suggestion made in chapter 3 that these filaments have different radiation characteristics than those in the linear portion of the Arc. The most noteworthy kinematic information derived from the recombination line is described next.

1) Almost all of the line emission arises at negative velocities (see

figures 4 and 5). This is in agreement with earlier single-dish observations (Pauls and Mezger 1980; Pauls et al. 1976). The negative radial velocities seen in this region ($\ell \simeq 0\overset{\circ}{.}1$) are forbidden in the sense of galactic rotation.

2) The velocities corresponding to the southern portion of the arched filaments are generally more negative than their counterparts to the north. This can best be seen in figure 4 at velocities between -68 to -38 km $s^{-1}$. Another presentation of this kinematic structure can be seen in figure 6. The dashed lines and a thin solid line correspond to radial velocities $\leq$ -40 km $s^{-1}$ and the dotted and thick dashed lines correspond to radial velocities $\geq$ -25 km $s^{-1}$.

3) The rough outline of the velocity structure seen in figure 6 indicates that the more negative-velocity emitting regions are generally located to the West (W1 and W2). The intensity-weighted radial velocities seen in the western and eastern filaments are ~ -32 ± 6 km $s^{-1}$ and -20 ± 6 km $s^{-1}$, respectively.

4) Many of the spectra presented in figure 8 (1-26) show multiple velocity components. For example, figures 8-2 and 8-3 indicate that there are at least 2 velocity components overlapping with each other. (The line surface brightness and the radial velocity of the components recognized in each spectrum are listed in Table 1.)

5) A number of velocity profiles were examined along the westernmost

portion of the arched filaments (W2). The profiles corresponding to the southern (spectra 1 to 6) and northern (spectra 9 and 10) portion of W2 in figure 7 appear to show different characteristics: (a) the line emission from the northern portion is stronger than its southern counterpart. This trend is also present in the continuum image (figure 1). The ratio of line to continuum brightness temperature is ~5 to 10% with much uncertainty. [This is because, for an extended structure, the lack of short spacings affects the line and the continuum maps differently.] (b) the velocity gradient ($\Delta V$) of the northern portion tends to be greater in its absolute value than that of the southern portion. This can be seen in two adjacent spectra 2 and 3 and 8 and 9 whose velocity components are easily recognized. The northern and southern portions show $|\Delta V|$ ~ 11 and 4.5 km $s^{-1}$ $arcmin^{-1}$. (c) Although a proper gaussian shaped profile has not been fitted to the velocity profiles seen in figure 8. We note a larger velocity width ($\Delta V_{FWHM}$) in the northern portion than that of the southern portion of W2. Spectra 4, 5, 6, and 7 with their single-component profile have a typical width of 12-14 km $s^{-1}$ whereas spectrum 10 with its two lines separated by 15.8 km $s^{-1}$ shows a width of ~18-20 km $s^{-1}$, which is a typical width for HII regions in which the electron temperature is ~$10^4$ K. (d) We note a velocity jump of ~15 km $s^{-1}$ between spectra 1-4 and 5-6.

6) Positions 11 and 12 are located at a junction where the two western filaments (W1 and W2) meet. Their profile consist of at least two velocity components whose most negative component (-29 km $s^{-1}$) appears to be associated with W1. This inference is based on the trends in velocity outline and on the continuous structure of the arched filaments seen in figures 6 and 1, respectively.

7) The brightest western filament (W1) shows a clear velocity increase as it is followed northward. The profiles corresponding to positions 13 and 14 are dominated by a single-velocity component with a velocity gradient ~ $-3.8$ km $s^{-1}$ $arcmin^{-1}$. Figure 1 and figure 4 of chapter 5 show W1 to consist of at least two components running southward parallel to each other. The brightness and profile shapes of 13 and 14 are very different from those seen along the southern portion of W2. However, the central velocity and the velocity gradient of the southern portion of both W1 and W2 are similar to each other. We note in figure 1 that diffuse and weak emission is associated with the southern portion of W1 in a number of channels between -62 and to -80 km $s^{-1}$. This emission is seen only on the blue side of the central velocities corresponding to profiles 13 and 14.

8) The most intense lines from the arched filaments are located along a peanut-shaped feature which lies at the junction of the two eastern filaments (E1 and E2). The velocity profiles of two positions along this feature, 15 and 16, show at least two velocity components. These two components have the largest velocity gradients in this region and, perhaps, the most curious kinematic property. The velocity component with $V_{LSR}$ = $-42.7$ km $s^{-1}$ in position 15 shows a VV

= 26 km $s^{-1}$ $arcmin^{-1}$ whereas the component with $V_{LSR}$ = -10.3 km $s^{-1}$ shows $\nabla V$ = -20 km $s^{-1}$ $arcmin^{-1}$ as both of these velocity components merge when the eastern filament is followed northward.

9) Positions 17, 18, and 19 coincide with the peaks seen along the northern portion of the eastern filaments (E1 and E2). Comparison of profiles 17 and 18 reveals that the velocity of peak emission decreases (becomes more negative) as the eastern filament (E2) is continued northward. This trend in velocity gradient is similar to one of the two velocity components seen further south in positions 15 and 16 having a negative velocity gradient (-20 km $s^{-1}$ $arcmin^{-1}$). Indeed, because of weak signal to noise in the region discussed here and because of the overlap of numerous features, it is not readily obvious which of the northeastern filaments (E1, E2, or both) are associated with the southern counterpart.

10) Positions 20 to 26 correspond to a region of the G0.1+0.08 complex where the arched filaments lose their coherence and break up into a number of hot spots, most of which are joined to each other in projection by an extended envelope of emission. Positions 20 and 21 represent the most compact and intense sources in this region. Position 22 is at a location where much of the high negative velocity emission can be clearly seen, i.e. ~ -72 km $s^{-1}$ (see figure 4). Its profile shows emission centered at 0 km $s^{-1}$. Positions 24 and 25 shows characteristics similar to positions 20 and 21. Source 24 is located at a position where a weak thread-like structure was seen to emerge or end (see chapter 5). Positions 25 and 26 show a weak emission with a very low signal to noise ratio at large negative velocities. The compact source at position 25 is either a background (foreground) source or has a central velocity whose frequency is outside the bandwidth used here or has a high electron temperature.

Table 1

List of Thermal Sources in G0.1+0.08

| Source Name | Right Ascension (1950) h | m | s | Declination (1950) ° | ′ | ″ | Peak Surface Brightness Continuum mJy/beam | Line ($V_{LSR}$) mJy/beam (km $s^{-1}$) |
|---|---|---|---|---|---|---|---|---|
| 1 | 17 | 42 | 27.0 | -28 | 51 | 19 | 80 | 4.6* (-51.2), 5.7 (-42.6) |
| 2 | 17 | 42 | 24.7 | -28 | 50 | 55 | 60 | 2.9* (-65), 5.8* (-51.2), 6.3 (-44) |
| 3 | 17 | 42 | 22.9 | -28 | 50 | 10 | 60 | 5.1* (-55.1), 6(-39.7) |
| 4 | 17 | 42 | 21.3 | -28 | 50 | 01 | 40 | 4.1 (-43.4) |
| 5 | 17 | 42 | 20.6 | -28 | 49 | 40 | 40 | 4.3 (-27.6) |
| 6 | 17 | 42 | 20.6 | -28 | 49 | 30 | 60 | 2.6* (-38), 6.4 (-25) |
| 7 | 17 | 42 | 20.3 | -28 | 49 | 10 | 80 | 8.2 (-31.5), 7.5* (-21.7) |
| 8 | 17 | 42 | 20.6 | -28 | 48 | 43 | 120 | 13.4 (-25.6) |
| 9 | 17 | 42 | 20.1 | -28 | 48 | 22 | 100 | 10.9 (-21.7) |
| 10 | 17 | 42 | 20.4 | -28 | 47 | 43 | 80 | 5.3 (-26,3), 5.4 (-10.5) |
| 11 | 17 | 42 | 22.7 | -28 | 46 | 52 | 100 | 6.8 (-19.5), 5(-7.3) |
| 12 | 17 | 42 | 24.3 | -28 | 46 | 52 | 100 | 7.3 (-29.3), 7.1 (-23.1) |
| 13 | 17 | 42 | 25.4 | -28 | 51 | 37 | 180 | 16.95 (-39) |
| 14 | 17 | 42 | 28.8 | -28 | 51 | 40 | 180 | 13.8 (-44) |
| 15 | 17 | 42 | 36.8 | -28 | 51 | 49 | 240 | 10.8 (-42.7), 11.2 (-10.3) |
| 16 | 17 | 42 | 35.5 | -28 | 49 | 13 | 260 | 5.6 (-86.5), 18.8 (-24.4), 6.3 (-8.5) |
| 17 | 17 | 42 | 33.2 | -28 | 47 | 49 | 100 | 7.1 (+1.8) |
| 18 | 17 | 42 | 34.3 | -28 | 47 | 10 | 100 | 4.4* (-36.5), 6.3 (-17), 5.9* (-2.4) |
| 19 | 17 | 42 | 39.8 | -28 | 47 | 28 | 40 | 2.8 (-50), 4.1 (-17.6) |
| 20 | 17 | 42 | 34.8 | -28 | 50 | 49 | 160 | 10.8* (-33), 11.7 (-25.6), 7.8* (-14.6) |
| 21 | 17 | 42 | 27.9 | -28 | 52 | 13 | 240 | 17.3 (-47.5), 11.2* (-33), 7.3 (-19.5) |
| 22 | 17 | 42 | 34.6 | -28 | 51 | 52 | 100 | 3.3 (-68.2), 3.7 (-6) |
| 23 | 17 | 42 | 21.8 | -28 | 55 | 01 | 80 | 3.7* (-52.5), 6.8 (-41.5), 4.1* (-25) |
| 24 | 17 | 42 | 19.5 | -28 | 53 | 31 | 60 | 5.9 (-47), 4.6* (-37.8) |
| 25 | 17 | 42 | 18.1 | -28 | 55 | 10 | 120 | — |
| 26 | 17 | 42 | 40.3 | -28 | 51 | 28 | 80 | — |

*These components may not be significant because of their high noise level.

B) The Sickle-shaped Feature (G0.18-0.04)

Structural details of the high-resolution image of G0.18-0.04, which is shown in figure 9, were fully described in chapter 3 [see also figures 9 (a-b) in chapter 3]. Here, we briefly point out the main features which comprise this small portion of the Arc as seen in figures 9 and 10: A sickle-shaped feature crosses the network of parallel filaments oriented perpendicular to the galactic plane. The southeastern extension of the sickle feature continues along the galactic plane but with a much smaller surface brightness (i.e. a factor of ~2 to 6) than that where the filaments and the sickle cross each other. In addition to the sickle, recombination line emission was also detected from a bright spot (G0.15-0.05) which is evident to the south of G0.18-0.04 in figure 9 and which has a pistol-like structure in the high-resolution map and an isolated compact source (source 01 in Table 1 of chapter 3 and source 5 in figure 10).

The most interesting result has been the recognition that almost all of the extended recombination line emission arised in sickle-shaped structure; none can be seen from the filaments. The kinematic information which is extracted from figures 11, 12, and 13 follow next.

1) All the line emission arising from the inner portion of figure 11 has positive radial velocities, predominantly between 30 and 40 km $s^{-1}$. This result is consistent with earlier single-dish results (see §I).

2) The spectra seen in figure 12 have typical line widths (FWHM) between 12 and 15 km $s^{-1}$.

3) Figure 13 shows an increase in radial velocity in the direction of increasing right ascension. The velocity trend with respect to position 4 is best seen by comparing figures 10 and 11: the radial velocity is seen to increase somewhat continuously in the direction to the south and east of this position with $\nabla V \approx 20$ km $s^{-1}$ $arcmin^{-1}$. The line emission from the region to the northwest of position 4 is weak and broken in its appearance, thus a clear velocity trend can not be identified in that direction. We note, however, that the velocity of the line emission merges with that at position 5 whose peak emission is at 44 km $s^{-1}$ (see Table 2).

4) The highest velocity components are shown with vertical lines in figure 11 and can be identified in two regions: one is along the southern tail of the sickle (positions 2 and 3) which shows $V_{LSR} \sim 105$ km $s^{-1}$ (see Table 2). The largest velocity gradient can be noted between positions 1 and 4. The other is situated on the western edge of G0.15-.05 (position 6) with a central radial velocity greater than 110 km $s^{-1}$, i.e. beyond the edge of the velocity range which we employed ($-80 < V_{LSR} < +110$ km $s^{-1}$).

Table 2

List of Thermal Sources in G0.18-0.04

| Source Name | Right Ascension (1950) h | m | s | Declination (1950) ° | ' | " | Peak Surface Brightness Line ($V_{LSR}$) mJy/beam (km $s^{-1}$) |
|---|---|---|---|---|---|---|---|
| 1 | 17 | 43 | 00 | -28 | 47 | 27 | 8.3 (39) |
| 2 | 17 | 42 | 56.2 | -28 | 48 | 33 | 3.3 (13.9), 2.5 (101.5) |
| 3 | 17 | 42 | 55.3 | -28 | 49 | 15 | 2.2 (30), 3.9 (104) |
| 4 | 17 | 43 | 03.1 | -28 | 46 | 33 | 7.6 (8), 2.2 (107) |
| 5 | 17 | 42 | 56.7 | -28 | 44 | 27 | 6.8 (44.1) |
| 6 | 17 | 43 | 04 | -28 | 48 | 45 | 7.1 (>110) |

## IV. Discussion

A) G0.1+0.08

Kinematic information described in previous sections suggests that the complex associated with the arched filaments consists of a number of ionized features with central velocities usually having continuous gradients in a north-south direction. Furthermore, some of these features, such as the western filament (W1), show coherent structure over 5 to 6 arcminutes. These characteristics argue in favor of a flow of ionized gas along some of the more organized structures such as W1.

The origin of the flow of ionized gas in such a fashion is unknown. The ionized gas associated with Sgr A West shows strong gradients as a result of orbital or infall motion relative to the nucleus of the Galaxy where a massive object may reside (van Gorkom et al. 1983; Serabyn and Lacy 1984; Brown and Liszt 1984). But the ionized flow seen along the arched filaments has very different kinematical thermodynamical properties from Sgr A West. An object of $\sim 2\times 10^6$ $M_\odot$ located ~ 10 pc from the center of W1 is needed for the gas to rotate about this object if such a situation is to account for the observed velocities. This also doesn't account for W2, E1, E2 or for the global arrangement. There is no indication - based on past observations - that such a massive object exists near the Arc, thus its existence is very unlikely.

It has been suggested that material is being fed along the arched filaments from the vicinity of the galactic nucleus (Chapter 8; Seiradakis et al. 1985) or from the nucleus itself. One drawback of these suggestions is the direction in which the arched filaments are bent toward the galactic plane and not toward the nucleus. Furthermore, we do not see any source of activity near the nucleus which hints that such an organized outflow has been operating there. It is possible, however, that the ionized material flows in the direction toward the galactic plane at $\ell \simeq 0.1°$. Such an assessment is justified on the basis of the following argument: The high resolution 20-cm continuum image of the Arc and Sgr A (see figure 1 of chapter 3) plus the 6-cm radiograph shown in figure 1 indicate that the ionized gas in the region where the arched filaments and the Sgr A complex meet is more non-uniform and disorganized in its appearance than other parts of the complex associated with the arched filaments. Furthermore, the velocity profiles associated with this region (spectra 20, 21, 22) show a number of velocity components with no obvious sign of continuous flow of ionized gas. Thus, one might expect that the ionized material flows from the organized portion of the arched filaments (north of W1) toward a region with much non-uniformity in its structure.

The non-uniform portion of the arched filaments is a region where much of the molecular gas and dust appear to be concentrated. The molecular distribution of $^{13}CO$ which was measured by Bally et al. (1986) shows that, at negative velocities, much of the molecular material is projected to the south (i.e., closer to the galactic center along the galactic plane) of the arched filaments (see figure 14), although a ridge appears to follow the eastern arched filaments. The radial velocity of the molecular gas at positive latitudes agrees well with that of the recombination line emission at positions where they coincide. We note that an $NH_3$ condensation, M0.1-0.01, which is mapped by Güsten et al. (1981), appears to be near the location where the two eastern filaments (E1 and E2) join (see figure 15). The molecular component is suggested to be bound to the 50 km $s^{-1}$ cloud complex seen in figure 15. Thus, the arched filaments and M0.1-0.01 are probably not associated. However, the $^{13}CO$ distribution, seen in figure 14 suggests an association with the positive-latitude molecular ridge because they both have negative radial velocities (i.e. opposite to the sense of galactic rotation) and because the molecular gas peaks generally at the location where the continuum emission is weak and non-uniform in its distribution.

Further support for this association is evident in figure 16 where the 55 μm dust distribution, measured by Dent et al. (1982), is superimposed on the 20-cm emission. The bright infrared component G0.074 +0.04 appears to lie at an emission minimum at 20-cm wavelength. This infrared source which is also seen at 125 μ (see figure 5 of chapter 4) is also peaked at the location where the 20-cm emission is minimized. We also note an infrared ridge at both 55 and 125 μm along the arched filaments. Thus, a hint of anticorrelation between the dust and the ionized and molecular gas is evident in G0.07+0.04. Furthermore, it is possible that the ionized gas in the arched filaments has molecular and dust counterparts. Indeed, a high-resolution picture of the infrared and molecular line distribution of the arched filaments would clarify whether any material occupied the inner region between the western and eastern arched filaments.

A continuous flow of ionized gas in the fashion described in previous section does not generally resemble the environments which are ever associated with star forming regions. Further arguments against star formation hypothesis are as follows: Based on a change in the surface brightness of the arched filaments as they cross the northern thread, we suggested that a physical interaction between these two features is taking place. This interaction is also supported by occurence of a change in the kinematic properties of W2 at the point where it crosses the thread (see §II.A5). Indeed, a situation in which magnetic structures such as the thread (see chapters 5 and 10) and the thermal features interacting with each other, does not occur in the star forming regions. Additional complexities such as the interaction of the arched and nonthermal linear filaments (see chapter 10), and the forbidden radial velocities of the molecular and ionized components, count against the star formation hypothesis.

Since the northern thread appears to be coherent as it crosses the western arched filaments (see chapter 5) then, if they are interacting, we estimate that the magnetic field has to be > $\left(\frac{n_e}{4\times 10^2}\right)^{1/2} \left(\frac{V}{30 \text{ km s}^{-1}}\right) 2.6\times 10^{-4}$ Gauss based on equating the magnetic field and kinetic energy densities. The number density of electrons, $n_e$, is estimated by assuming that the plasma has an electron temperature, $T_e$, ~ $10^4$ °K under LTE condition. The magnetic energy density in the thread corresponding to the above derived field strength is also similar to the thermal energy density of the ionized plasma in the filaments. It is possible that dissipation of magnetic energy might play a role in the ionization of the observed structures.

The above LTE assumption might not be valid because of the particular kinematics of the ionized gas and the importance of the magnetic field in this region. Thus, it is plausible that the ionized features are rapidly evolving structures. However, the derived number density based on LTE assumptions is not greatly different from the estimates that are made based on the rotation measure (chapter 10) and the optical depth at 160 MHz (chapter 4).

One possibility for the origin of the arched filaments is that the -30 km $s^{-1}$ molecular cloud has plunged through the linear filaments and has dragged the field lines associated with the filaments with it toward the galactic plane. The coupling between the neutral gas in the cloud (the peak brightness of the $^{13}CO$ can be seen to the south of the arched filaments at G0.04+0.04 and G0.04-0.02) and the magnetic lines of force are made by the fractional ionization in the molecular cloud. We speculate that the northern thread was part of a system of linear filaments before the molecular cloud crossed a portion of the linear filaments and dragged away the magnetic lines of force which were once associated with the linear portion of the Arc. Radio images of the northern thread show clearly that the northern thread approached a group of linear filaments away from the galactic plane (see figure 1 of chapter 10) and thus supports that the northern thread, though isolated, might have been associated with the system of filamentary structure constituting the Arc. It is not clear, however, that how the magnetic lines of force associated with the Arc can be pushed southward and yet form a gentle curvature seen along the northern thread, assuming that the suggested hypothesis is at work.

Based on earlier argument that the source of ionization is not likely to be stellar, we suggest that the collisional ionization might be relevant in ionization the gas. It is necessary to have a high electron temperature, $T_e$, $10^5$-$10^6$ °K to ionize the gas. On the other hand the electron temperature has to be ≈ $10^4$ °K to allow observable radio recombination lines to form. A wide range of electron temperature, therefore, is needed if such a process is at work. It is possible that much of the ionization and heating is done by relativistic electrons whose synchrotron emission was associated with the linear filaments before the impact of the neutral gas upon the filaments. It is necessary that the heating rate by the relativistic electrons be similar to the total recombination line cooling $\sim 4.4\times 10^{-11}$ recombination $s^{-1}$, assuming that $T_e \sim 10^4$ °K and $n_e \sim 500$ $cm^{-3}$.

Another source of heating ($T < 10^4$ °K) might be provided by the reduction of magnetic flux through the exchange of momentum in collisions between ionized and neutral gas (i.e. ambipolar diffusion). The diffusion time scale for supersonic drift is inversely proportional to $B^{-2}$ and thus is shortened considerably if the magnetic field strength is large throughout the cloud (Spitzer 1978; Elmegreen 1981). The magnetic diffusion heating, which is proportional to $B^4$ can be much larger than typical values found in the clouds in other parts of the Galaxy unless a hydromagnetic shock is generated when the drift velocity, which is proportional to $B^2$, is enhanced substantially. If such a mechanism is applicable, one then expects to find shocked material residing at the edge of the filament boundaries. High-resolution spatial distribution of molecular gas along the filaments might be very useful for further understanding of the ionization process that take place in this puzzling region.

(B) G0.18-0.04

The lack of any recombination line emission from the bright southern linear filaments provides additional evidence in support of the idea that they are non-thermal (see chapter 3 for the polarization structure of the southern linear filaments at 2 cm). However, the filament extending to the northwestern from position 4 in G0.18-0.04 (see figures 9 and 10) coincides with weak line emission having $V_{LSR} \approx 30$ km $s^{-1}$. The northwestern continuation of this filamentary structure ($\alpha \sim 17^h42^m30^s$, $\delta \sim -28°42'$) is linked to the region where the northern tip of the western arched filaments merges with the northern filaments (see the 6 and 20-cm images of this region in figures 3 and 2 of chapters 10 and 3, respectively). Because of the widening and brightening of the northwestern filaments as they approach toward more negative latitudes where the sickle feature lies, because of the brightening of the knots which compose the upper portion of the sickle in the high-resolution image displayed in figure 9 and because of the line emission from a portion of the northwest filaments, whose extensions toward more positive latitudes are highly polarized (see figure 5 of chapter 10), we argue that not only are the linear filaments located in the outskirts of the sickle-shaped feature but also these two systems of thermal and non-thermal features are interacting with each other.

The plausibility of such an interaction was also discussed in chapter 3 based on both a gradual decrease in the brightness and a subtle discontinuity in the linearity of the filaments as they protrude through the sickle feature and extend toward more positive latitudes. A slightly distorted nature of the southern filaments could be argued for the possibility that the sickle is generated by the impact of the relativistic gas in the linear filaments upon ambient gas along the galactic plane.

The 50 km $s^{-1}$ molecular has been suggested to be the parent cloud from which G0.18-0.04 is produced (Pauls and Mezger 1980). The inference is based on similar radial velocities seen in both the molcular and ionized gas, i.e. 40-50 km $s^{-1}$. Anticorrelation in the appearance of the molecular distribution of $NH_3$ (Güsten et al. 1980; Armstrong and Barrett 1984; see figure 10), and HCN (Fukui et al. 1977) and that of the ionized gas in the sickle also support an association between the 50 km $s^{-1}$ molecular cloud and the sickle feature, though their relative geometry is not yet clear. The source(s) of ionization were previously suggested to be OB stars in the 50 km $s^{-1}$ cloud (Pauls ans Mezger 1980), however, the ionized and molecular components have very different velocity gradients (the fluctuation in the 50 km $s^{-1}$ velocity of the molecular cloud is very small [Güsten et al. 1981]). It is unlikely that OB stars as the source of ionization could account for the differences (although not impossible) noted here. Furthermore, the complex kinematics with a wide range of $V_{LSR}$ and structural details of G0.18-0.04 plus the interaction between it and the linear filaments can not be identified with typical HII regions powered solely by OB stars; however, the association between the sickle feature and the 50 km $s^{-1}$ molecular cloud is very likely.

It is possible that the relativistic electrons association with the linear filaments deposit a portion of their energy as they press through the sickle feature. Although the brightness of the southern linear filaments weakens northwestward and thus supports the above suggestion, the northern linear filaments brighten substantially as they slip through the upper portion of the sickle feature (see figure 9) in the direction from NW to SE. The slowing down time of relativistic particles as they pass through a medium of particle density of 500 $cm^{-3}$ and a kinetic temperature of $10^4$ °K is $\sim 4 \times 10^7$ sec (see Spitzer (1978) equation 2-4). The energy that is deposited by the high energy particles along the filaments with $n_e \sim 10^{-4}$ $cm^{-3}$ - based on equipartition of energy - is roughly equal to the energy lost by cooling of plasma along the sickle feature (i.e. $2.5 \times 10^{-17}$ erg $s^{-1}$ $cm^{-3}$). It is then possible, that the source of ionization along the southern (northern) portion of the sickle feature can be attributed to the collision of high energy particles from the southern (northern) filaments. On the other hand, it is plausible that the source of relativistic electrons along the linear filametns are supplemented by the ionized gas associated with G0.18-0.04 that illuminates the field lines. This hypothesis might be relevant since a number of linear filaments - unlike the southern and northern filaments whose brightnesses decrease gradually - either end exactly at or emerge from relatively bright portions of the sickle feature. Indeed, the mechanism by which the electrons along the linear filaments are accelerated remain a mystery.

Since most of the linear filaments maintain their coherence as they pass through the sickle,the field strength is estimated to be > $(4\pi n_e mv^2)^{1/2}$ where v, thermal speed, is considered to be similar to $V_{LSR}$, m is the mass of the proton and $n_e$ is estimated to be ~500 $cm^{-3}$, is the number density of electrons. Thus, the field strength of $\gtrsim 3\text{-}5 \times 10^{-4}$ is needed if the magnetic pressure is similar to the thermal pressure. This estimate of magnetic field strength, however, is incorrect if the field lines slip through the sickle feature and are shielded from any form of interactions with the ionized gas.

(c) Common Characteristics of G0.1+0.08 and G0.18-0.04

The general morphology of ionized gas in G0.18-0.04 and G0.1+0.08 shows that they are, unlike the linear filaments, broad and diffuse in their appearance. Furthermore, they both have a similar curvature, i.e., concave with respect to the galactic plane. Because both G0.18-0.04 and G0.1+0.08 are an integral part of the Arc and because both of these features are associated with +50 and -30 km $s^{-1}$ molecular components, we infer that these two molecular components are physically coupled. If so, the 50 km $s^{-1}$ molecular cloud, which has a very uniform velocity, at $\ell > 0\overset{\circ}{.}0$ shows the highest velocity gradients away from the galactic plane at $b > 0\overset{\circ}{.}0$. One can deduce from the evidence that the non-thermal and thermal features are associated and the thermal features have molecular counterparts that the large magnetic field strengths found along the linear filaments pervades throughout the molecular clouds. If indeed the mean magnetic field strength is as much as $10^{-4}$ Gauss in the galactic center region, two main problems can be resolved in this region: one is associated with the large kinetic temperature of the molecular clouds in the galactic center region (Morris et al. 1973; Güsten et al. 1981, 1985) including the 50 km $s^{-1}$ cloud. Heating by ambipolar diffusion should be the dominant heating mechanism based on earlier discussion section A. The other problem is associated with the lack of star formations in the galactic center region. The magnetic field acts as a stabilizer against contraction of the clouds if B $> \sqrt{4\pi G M \rho_{H_2}/R}$ (Fleck 1980) where M, $\rho_{H_2}$, R are the mass, hydrogen molecule density and the size radius of the cloud. For a giant molecular cloud such as the 50 km $s^{-1}$ cloud, we require B $> 1.4\times10^{-4}$ Gauss if $M = 4\times10^5\ M_\odot$, $n_{H_2} = 500\ cm^{-3}$ and R = 20 pc (Fleck 1980). It is then possible that the lack of significant $H_2O$ masers and HII regions associated with the giant molecular clouds are explained by a large magnetic field in the galactic center region. The geometry of such a field is discussed in chapters 6 and 10.

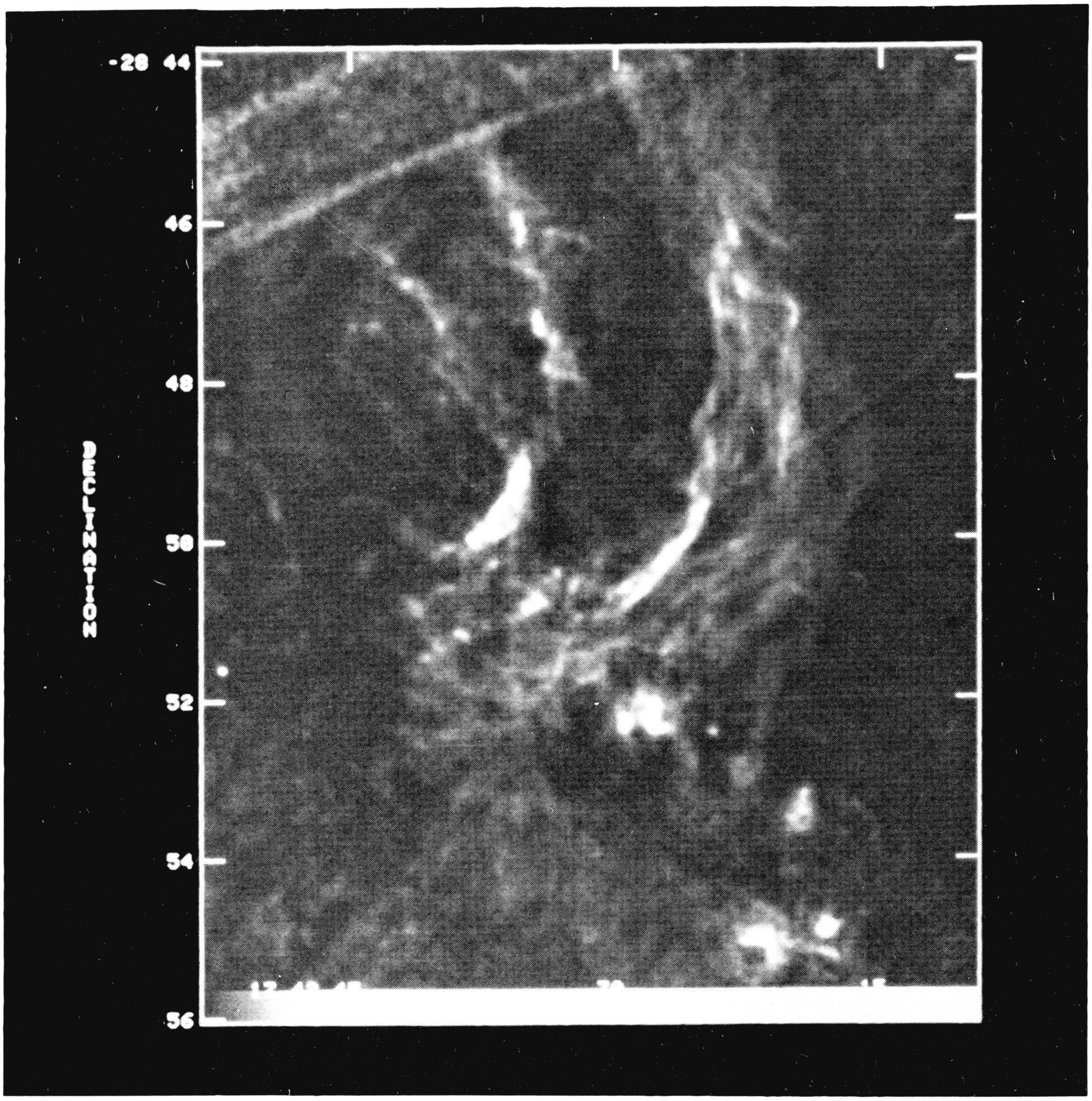

Figure 1: The image is identical to figure 10 of chapter 3 except that the visibility data is not tapered. FWHM = 3.8"×3.1".

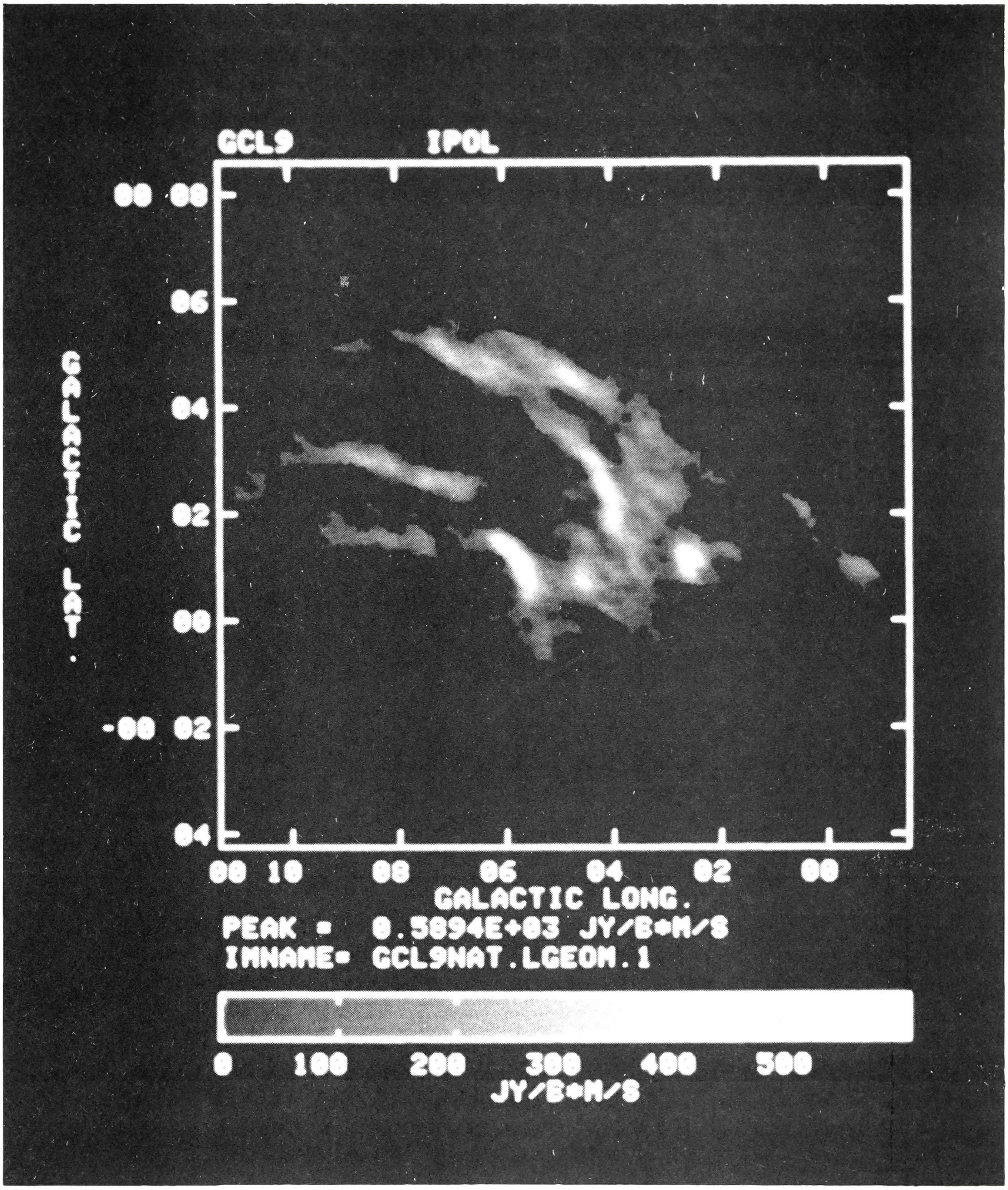

Figure 2: This radiograph shows the total emission from the arched filaments. FWHM = 22.5"×11.3". Natural weighting was applied to the visibility data before it was Fourier transformed (see chapter 2).

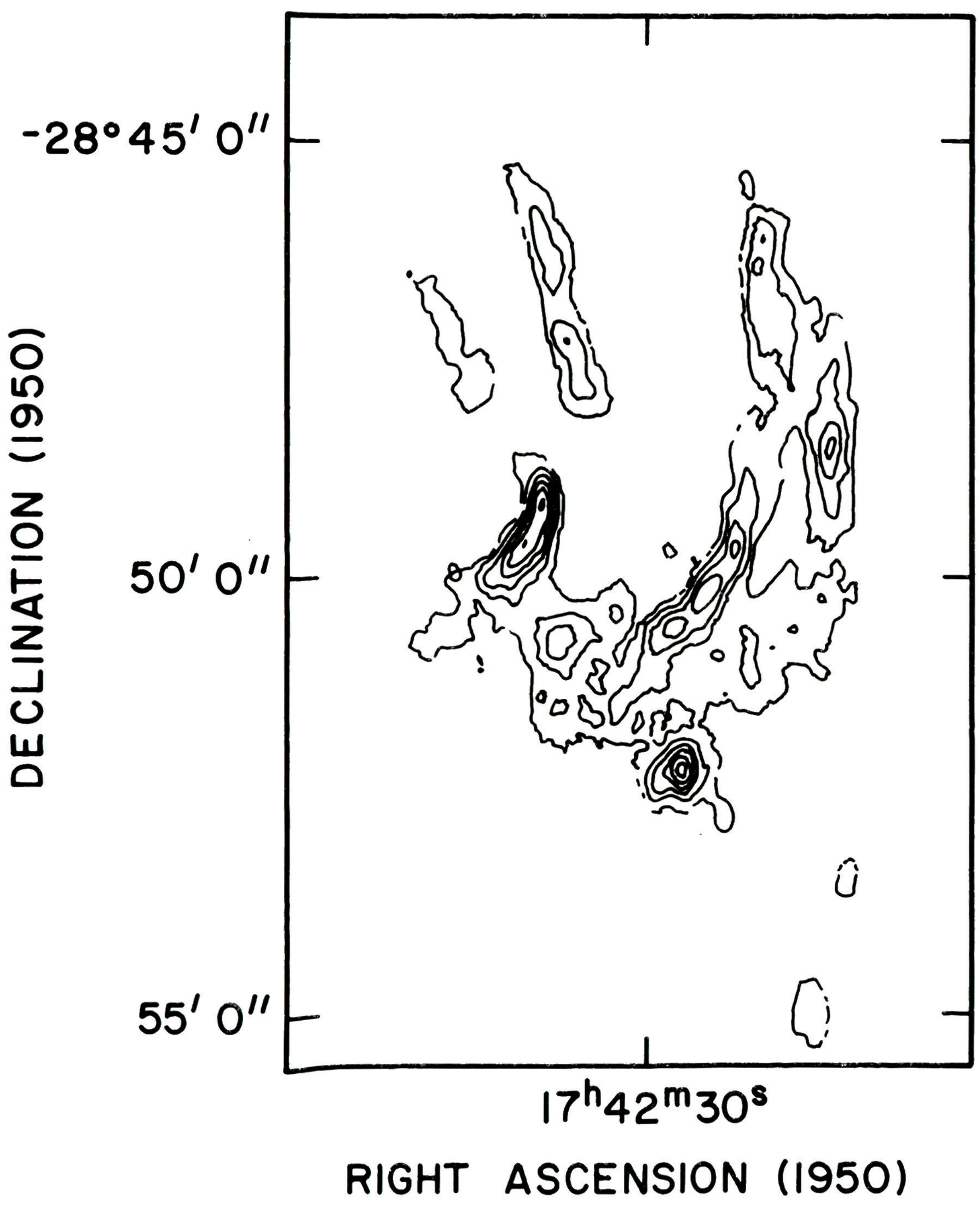

Figure 3: Contours of the velocity weighted line emission are shown with intervals of 0.5, 1.5, 2.5, 3.5, 4.5, 5.5 mJy/beam km $s^{-1}$.

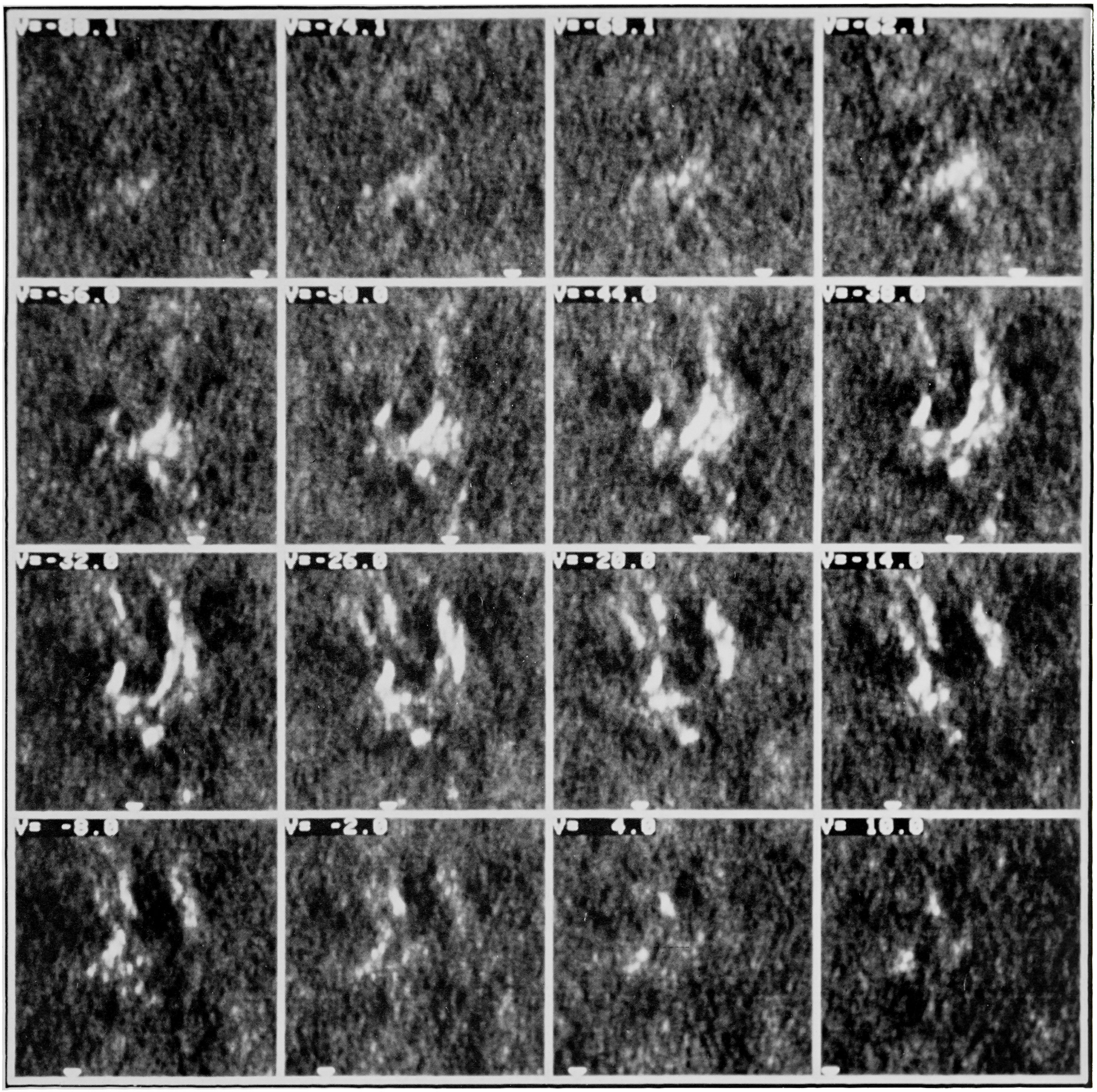

Figure 4: The radiograph shows 16 consecutive velocity channels separated by 6 km $s^{-1}$. The radial velocity with respect to local standard of rest ($V_{LSR}$) is indicated in the upper left corner of each channel in km $s^{-1}$.

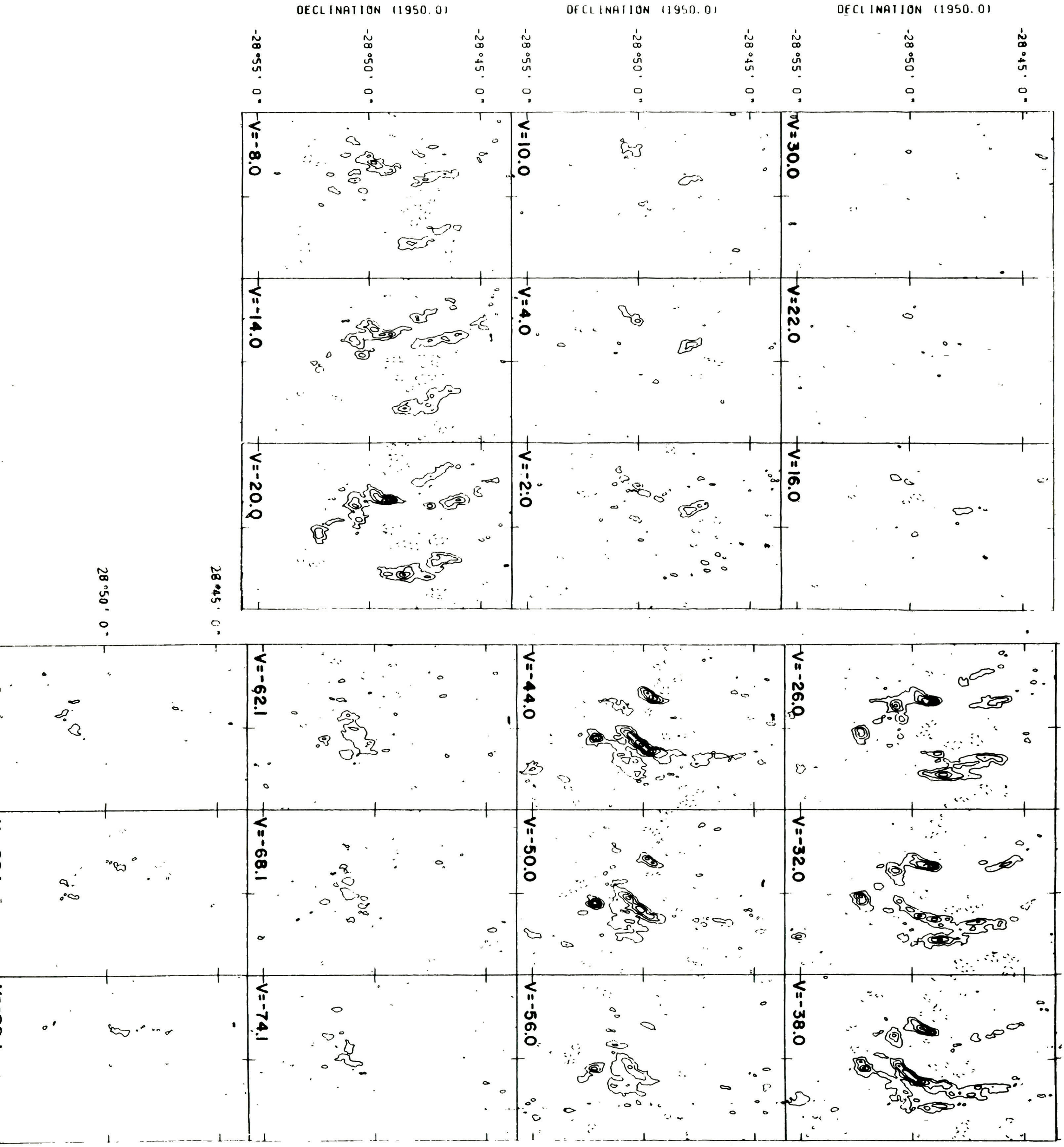

Figure 5: Contours of line emission for a given channel is represented with intervals of -6, -3, 3, 6, ..., 18 mJy/beam area. The $V_{LSR}$ is indicated in the bottom left corner of each channel map.

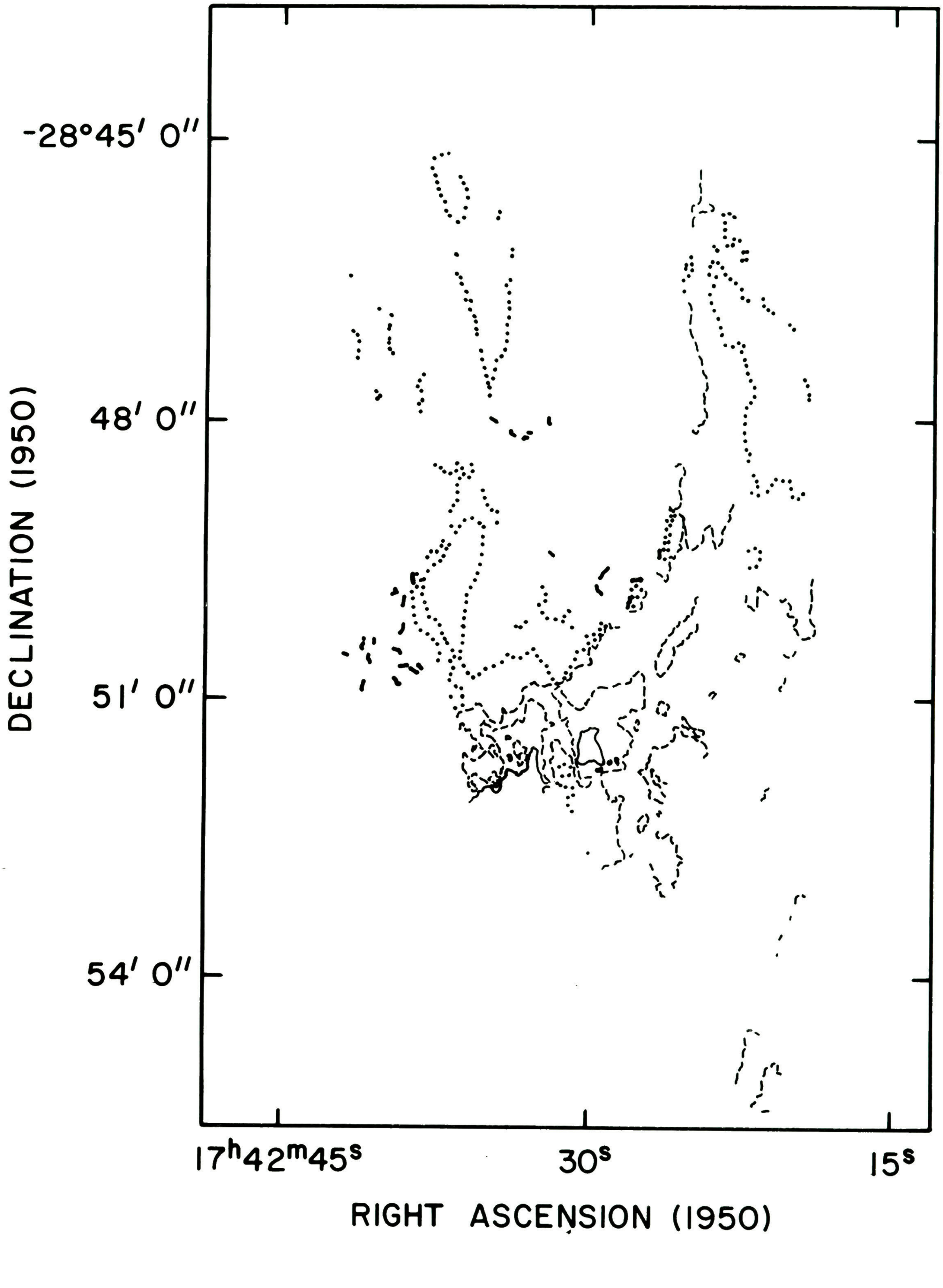

Figure 6: The kinematic structure is represented with dotted, thin-dashed, thin-solid and thick-dashed lines which correspond to $-10 < V_{LSR} < -25$, $-55 \leqslant V_{LSR} \leqslant -40$, $-85 \leqslant V_{LSR} \leqslant -70$, $20 \leqslant 50$ km s$^{-1}$, respectively.

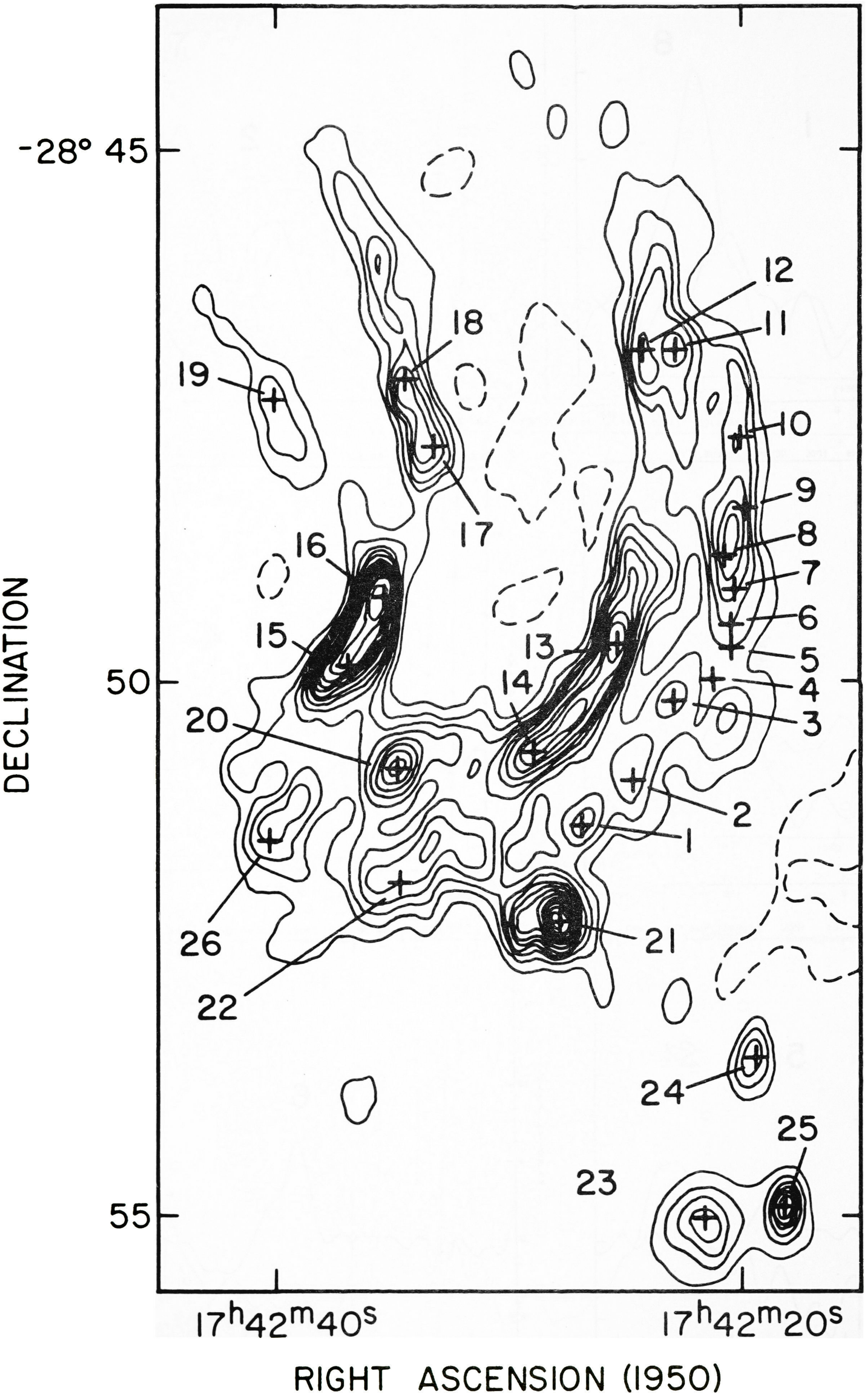

Figures 7 and 8: Contours of the total continuum emission (i.e. the continuum channel) is shown with intervals of -40, -20, 20, 40, 260 mJy/beam area. FWHM = 22.4" x 11.3". Figure 8 shows the spectra of the positions noted in figure 7.

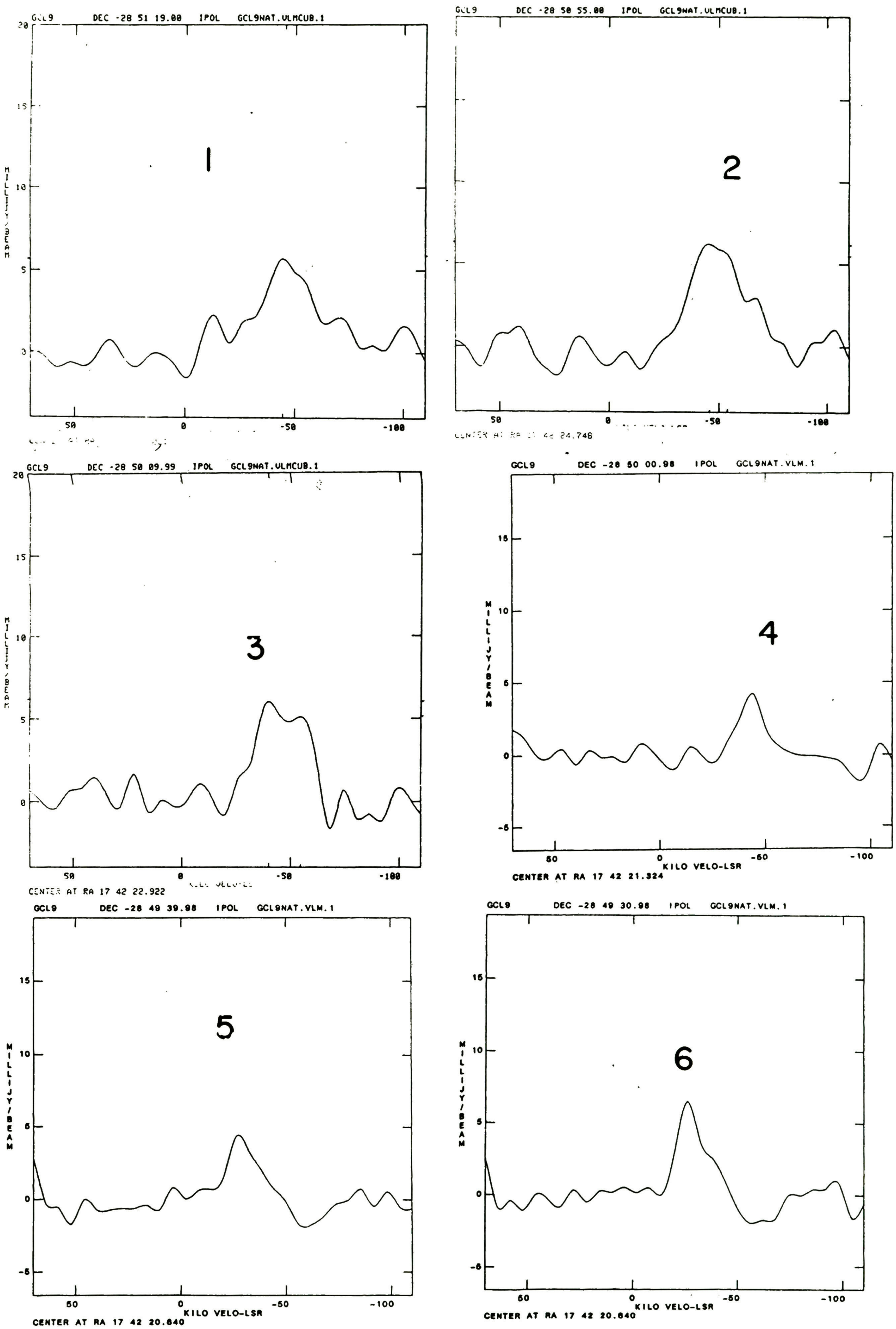
GCL9 DEC -28 51 19.00 IPOL GCL9NAT.ULMCUB.1
1
MILLIJY/BEAM
GCL9 DEC -28 50 55.00 IPOL GCL9NAT.ULMCUB.1
2
CENTER AT RA 17 42 24.746
GCL9 DEC -28 50 09.99 IPOL GCL9NAT.ULMCUB.1
3
MILLIJY/BEAM
CENTER AT RA 17 42 22.922
GCL9 DEC -28 50 00.98 IPOL GCL9NAT.VLM.1
4
MILLIJY/BEAM
KILO VELO-LSR
CENTER AT RA 17 42 21.324
GCL9 DEC -28 49 39.98 IPOL GCL9NAT.VLM.1
5
MILLIJY/BEAM
KILO VELO-LSR
CENTER AT RA 17 42 20.640
GCL9 DEC -28 49 30.98 IPOL GCL9NAT.VLM.1
6
MILLIJY/BEAM
KILO VELO-LSR
CENTER AT RA 17 42 20.640

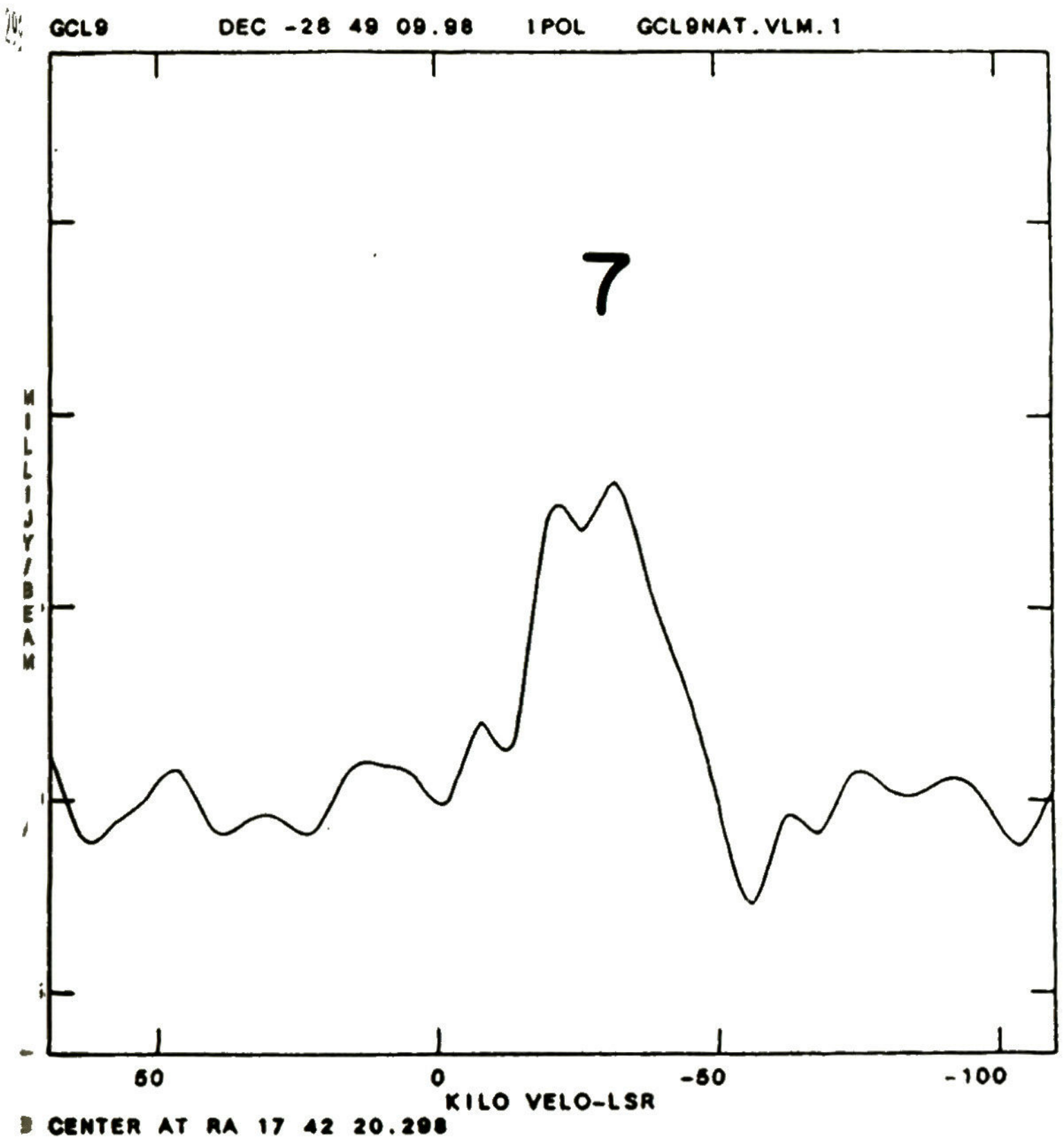
GCL9 DEC -28 49 09.98 IPOL GCL9NAT.VLM.1
7
MILLIJY/BEAM
50
0
-50
-100
KILO VELO-LSR
CENTER AT RA 17 42 20.298

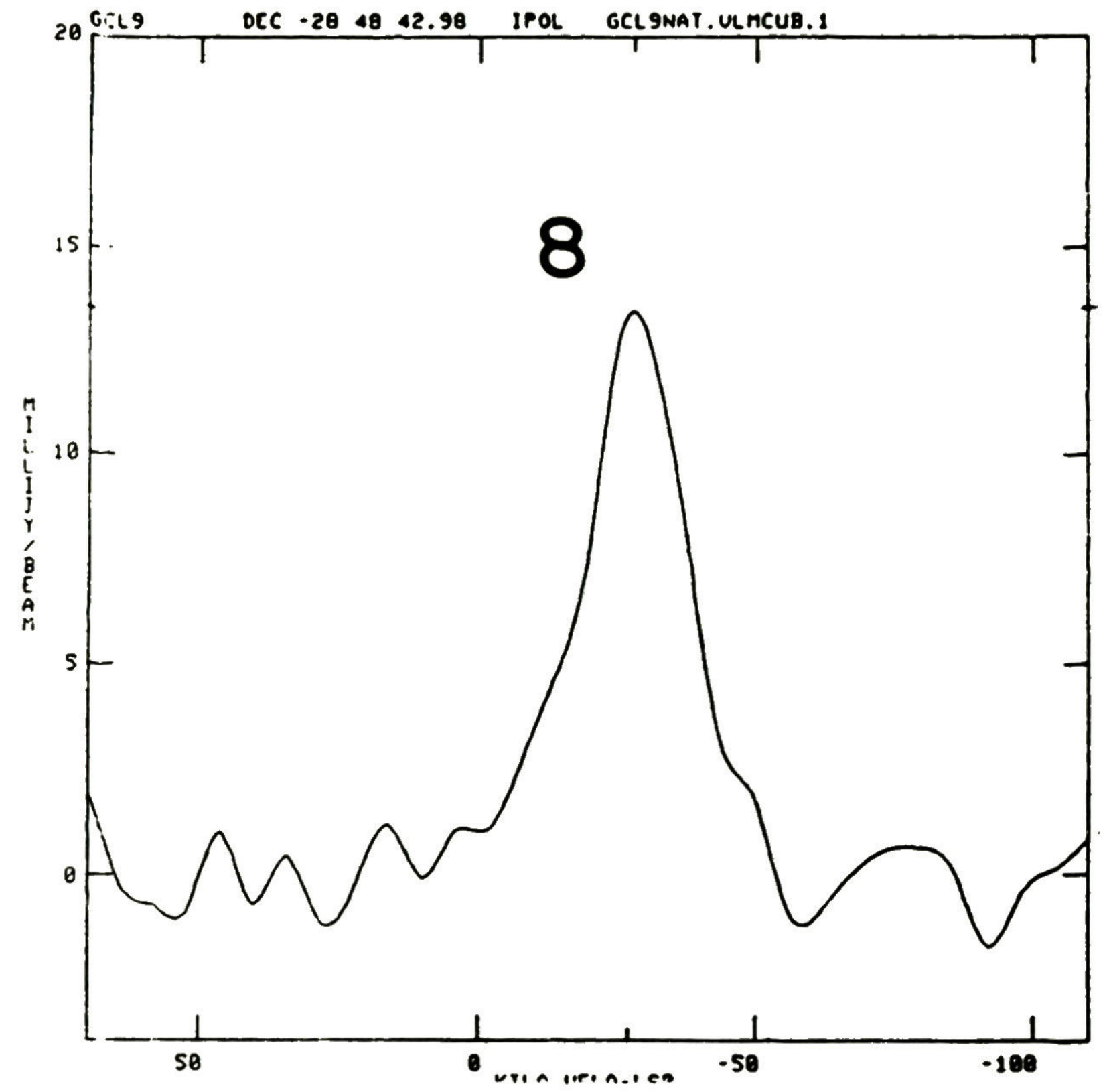
GCL9 DEC -28 48 42.98 IPOL GCL9NAT.VLMCUB.1
8
20
15
10
5
0
MILLIJY/BEAM
50
0
-50
-100

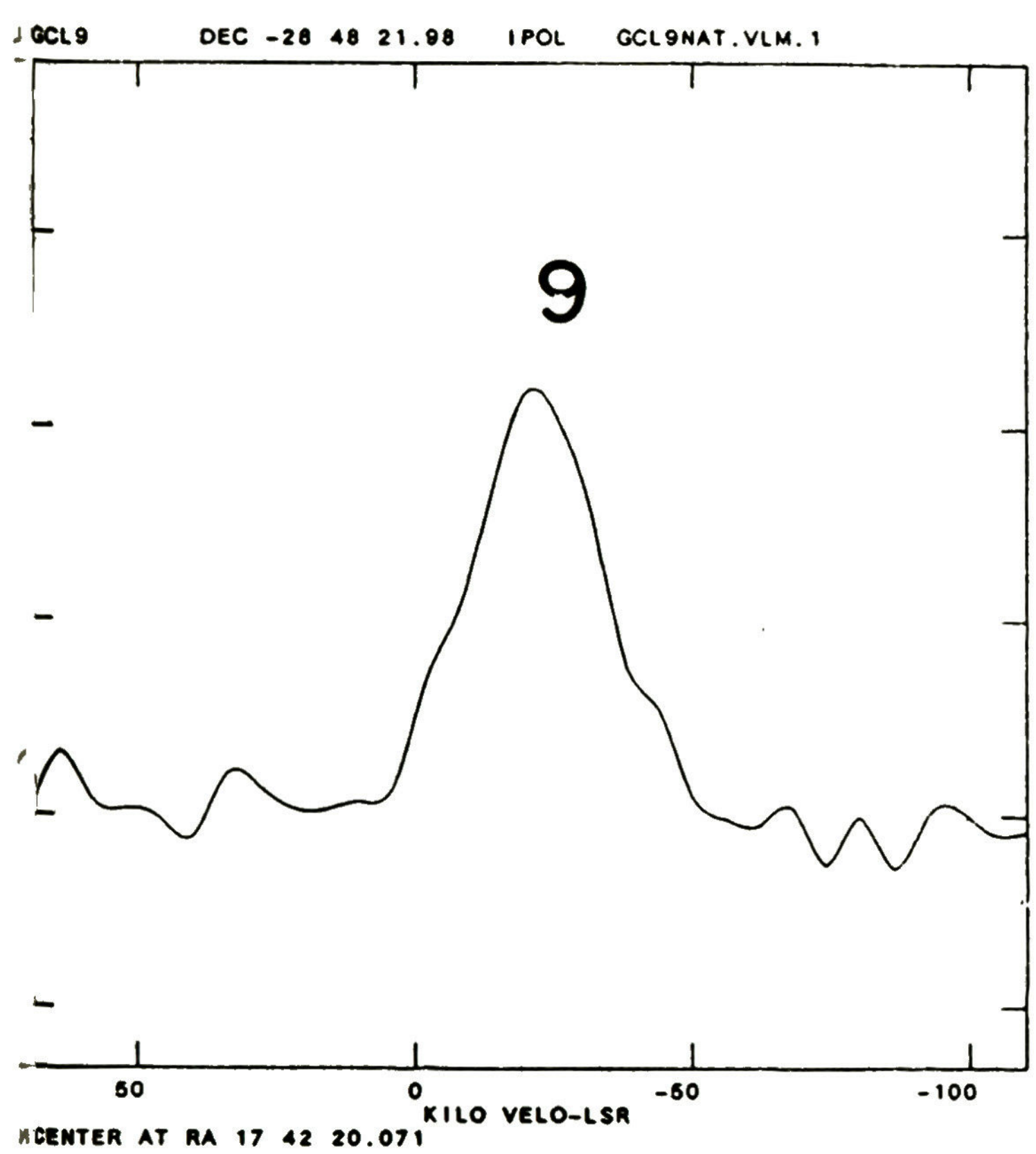
GCL9 DEC -28 48 21.98 IPOL GCL9NAT.VLM.1
9
50
0
-50
-100
KILO VELO-LSR
CENTER AT RA 17 42 20.071

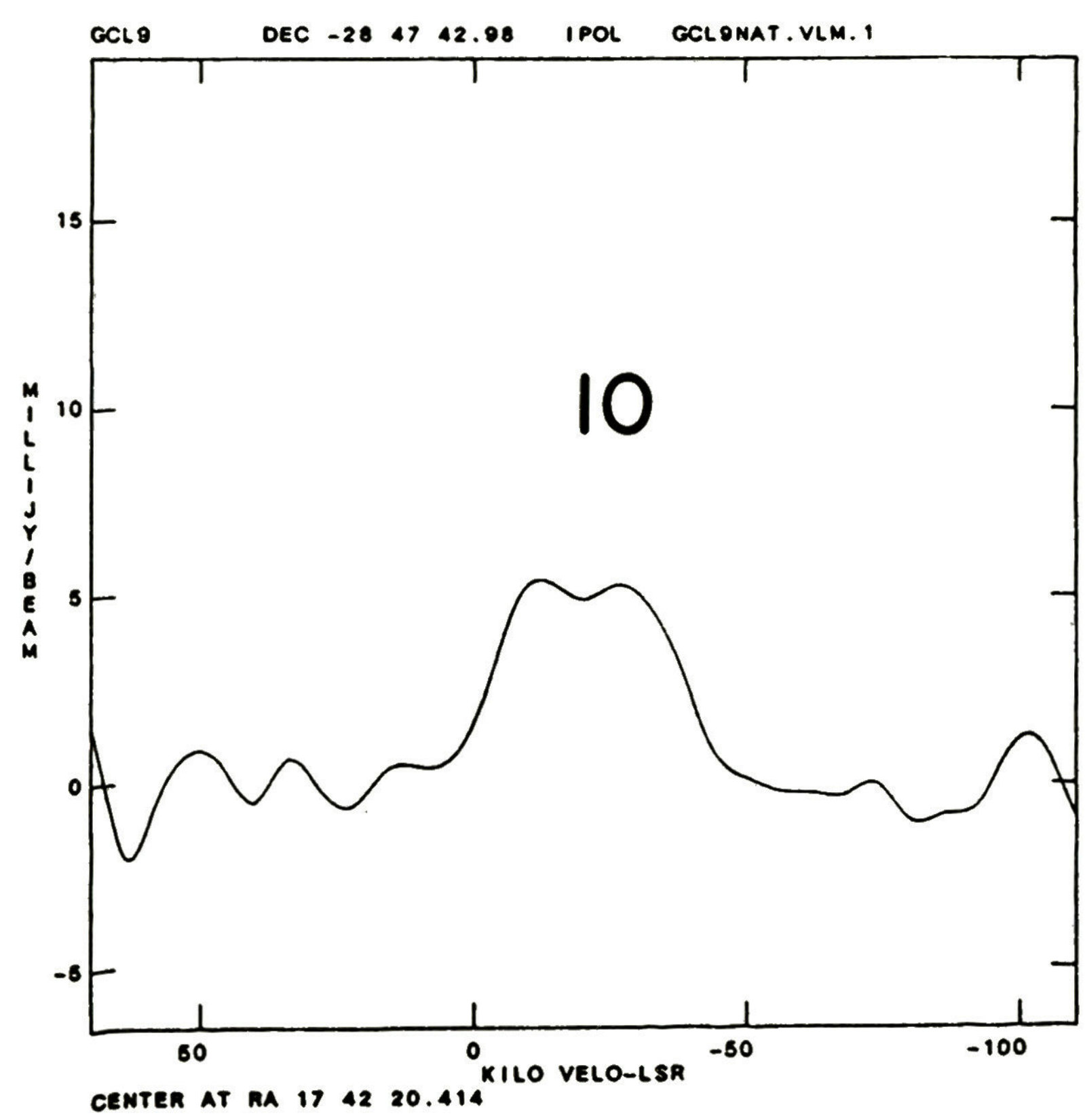
GCL9 DEC -28 47 42.98 IPOL GCL9NAT.VLM.1
10
15
10
5
0
-5
MILLIJY/BEAM
50
0
-50
-100
KILO VELO-LSR
CENTER AT RA 17 42 20.414

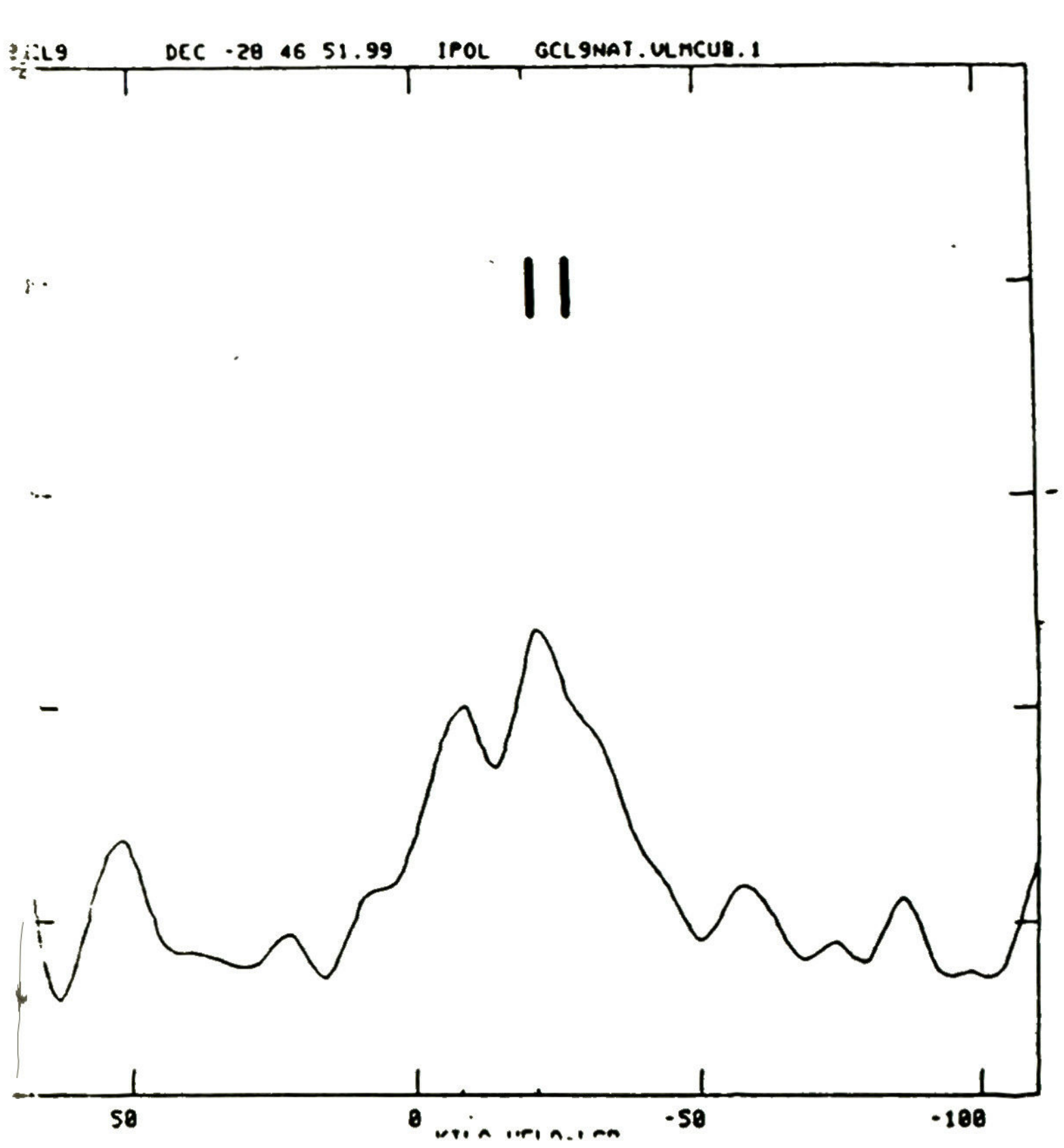
DEC -28 46 51.99 IPOL GCL9NAT.VLMCUB.1
11
50
0
-50
-100

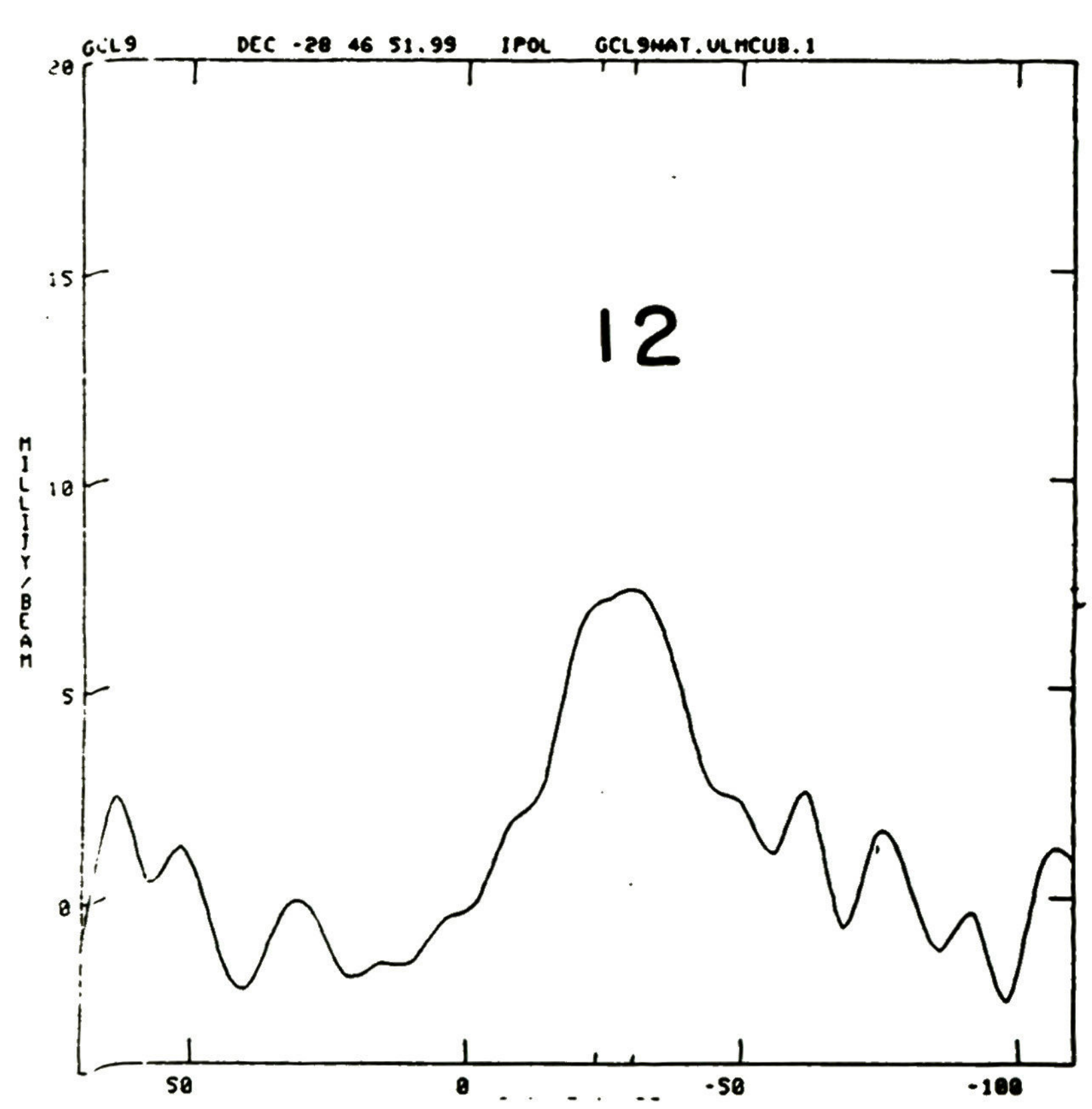
GCL9 DEC -28 46 51.99 IPOL GCL9NAT.VLMCUB.1
12
20
15
10
5
0
MILLIJY/BEAM
50
0
-50
-100

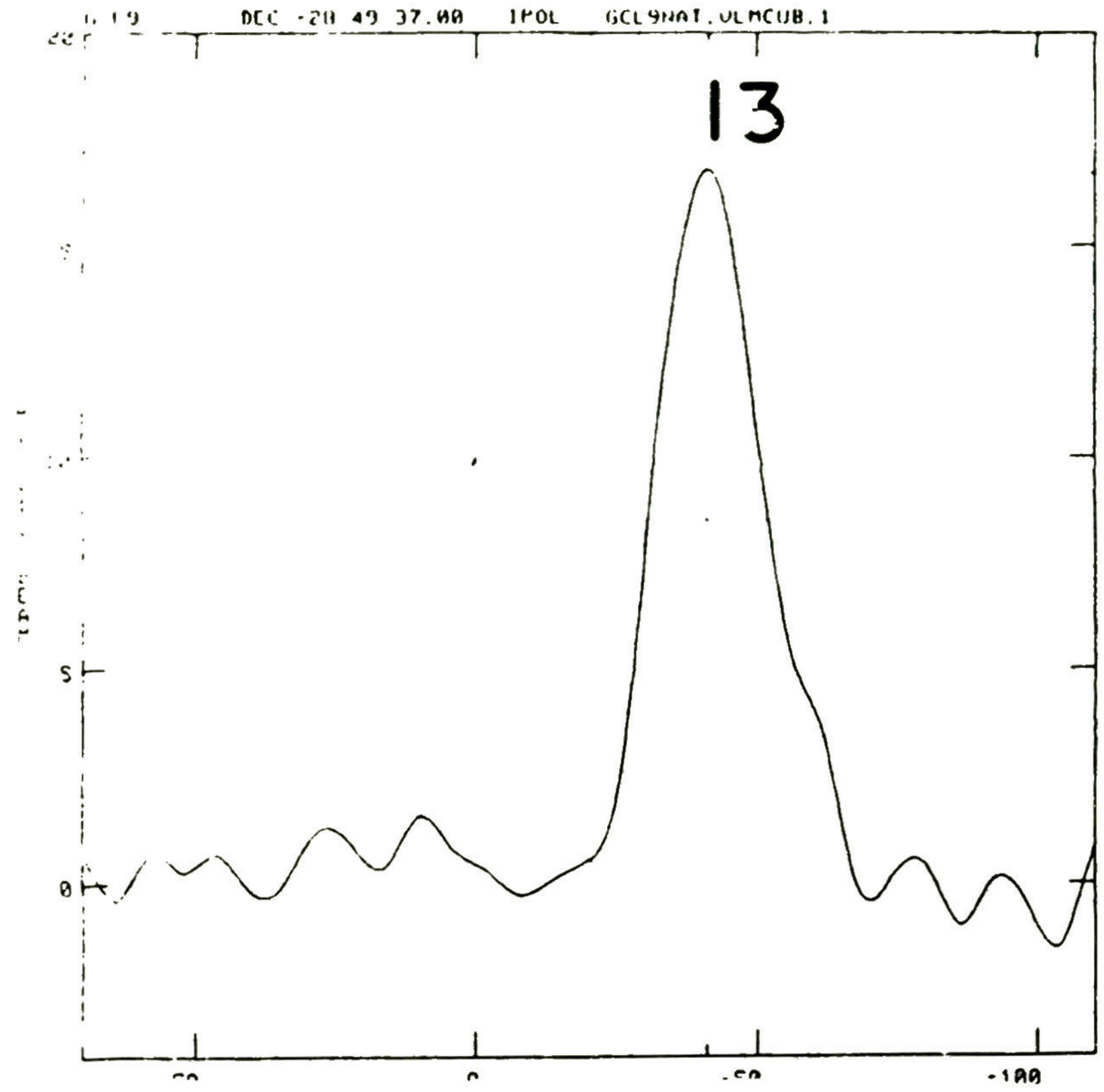
DEC -28 49 37.00 IPOL GCL9NAT.VLMCUB.1
13
5
0
50
0
-50
-100

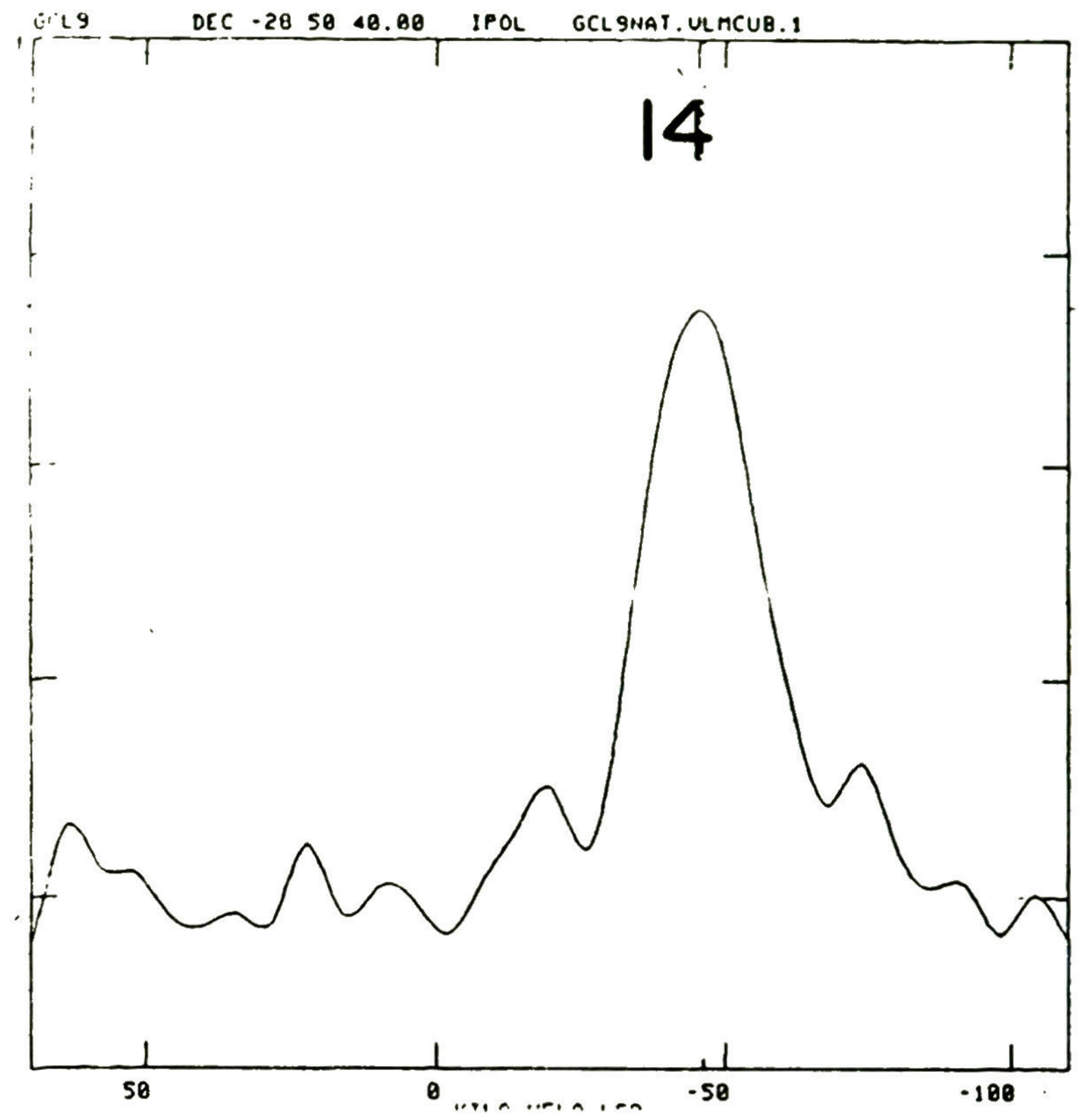
GCL9 DEC -28 50 40.00 IPOL GCL9NAT.VLMCUB.1
14
50
0
-50
-100

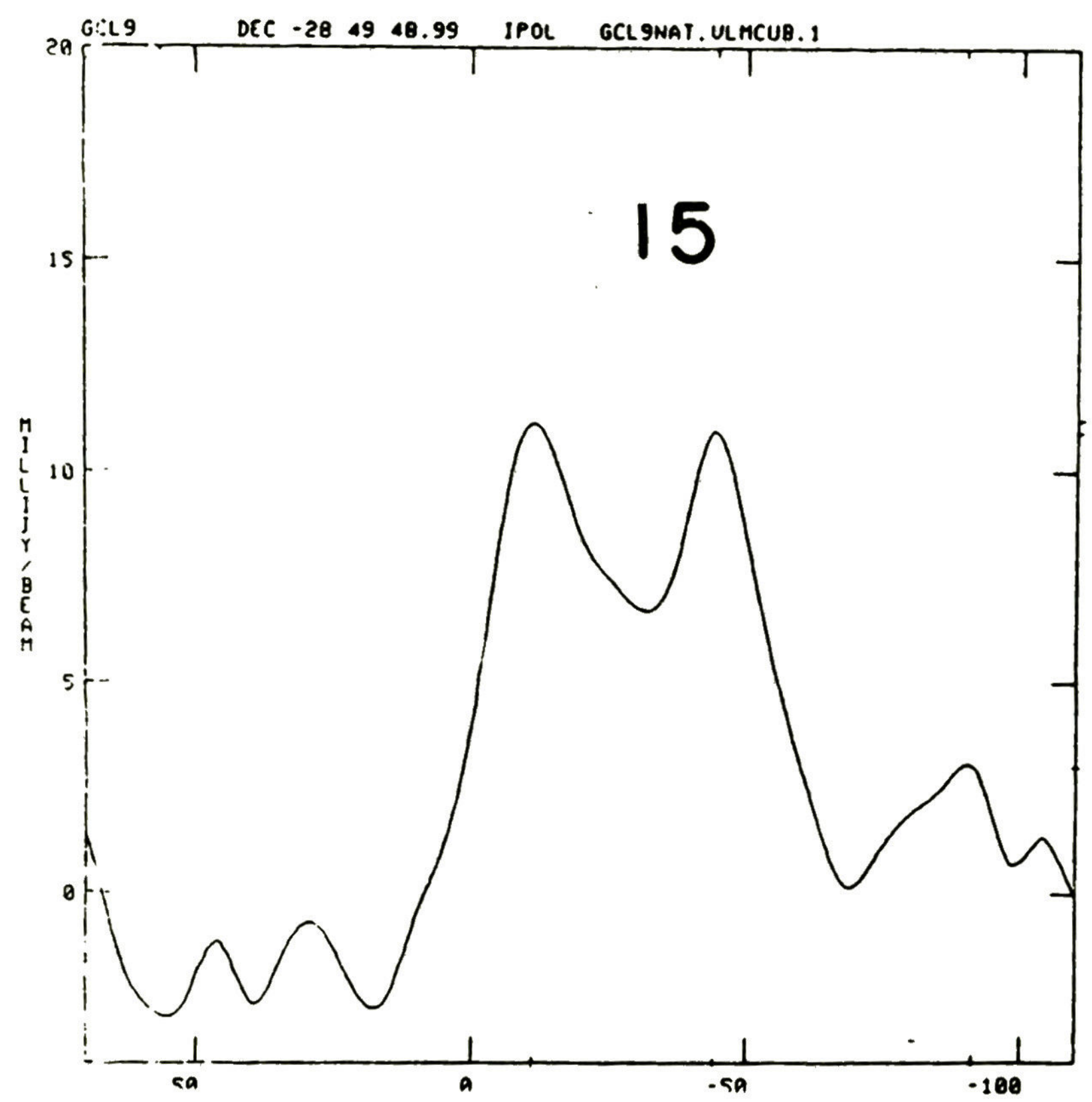
GCL9 DEC -28 49 48.99 IPOL GCL9NAT.VLMCUB.1
15
20
15
10
5
0
MILLIJY/BEAM
50
0
-50
-100

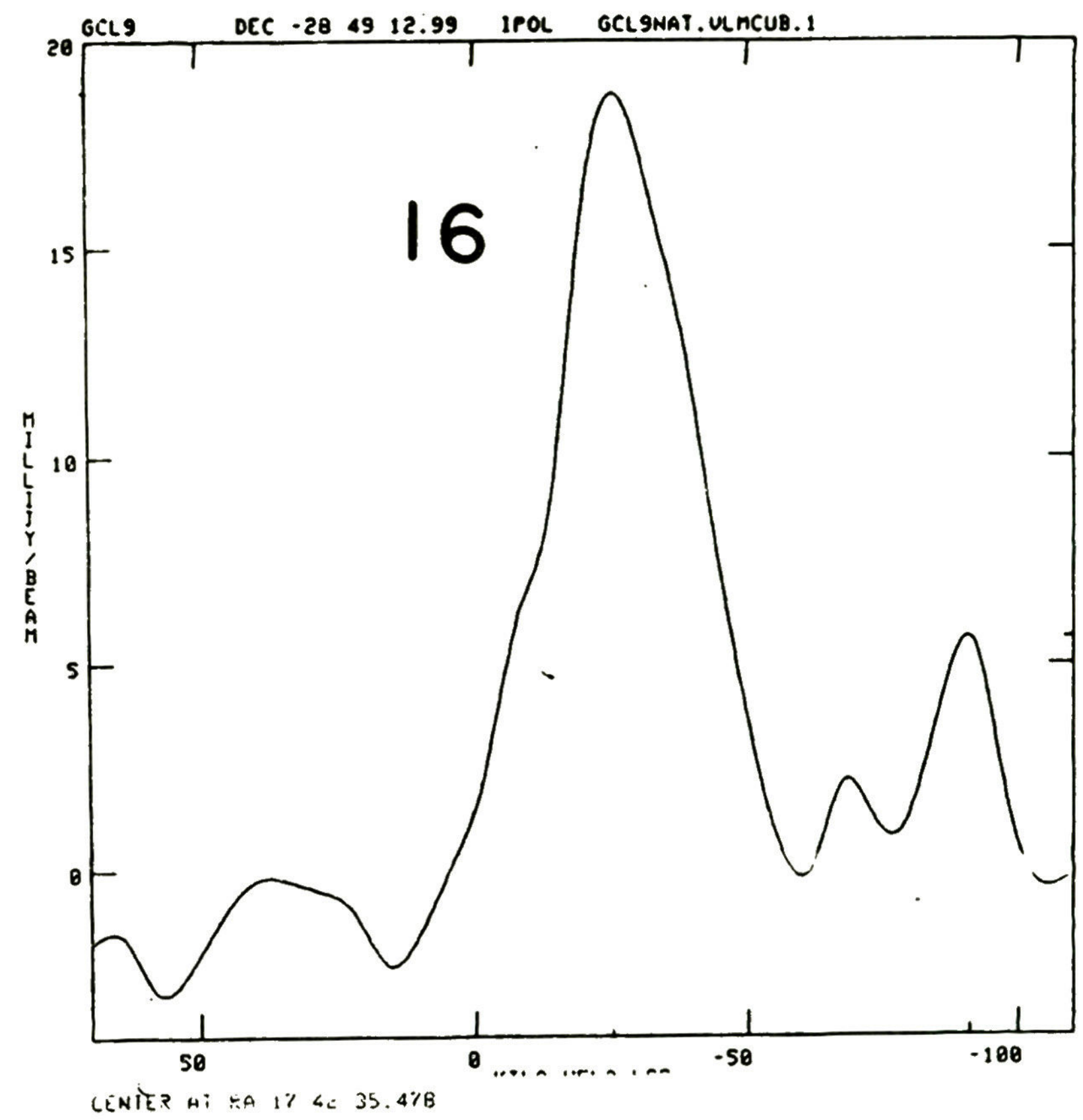
GCL9 DEC -28 49 12.99 IPOL GCL9NAT.VLMCUB.1
16
20
15
10
5
0
MILLIJY/BEAM
50
0
-50
-100
CENTER AT RA 17 42 35.478

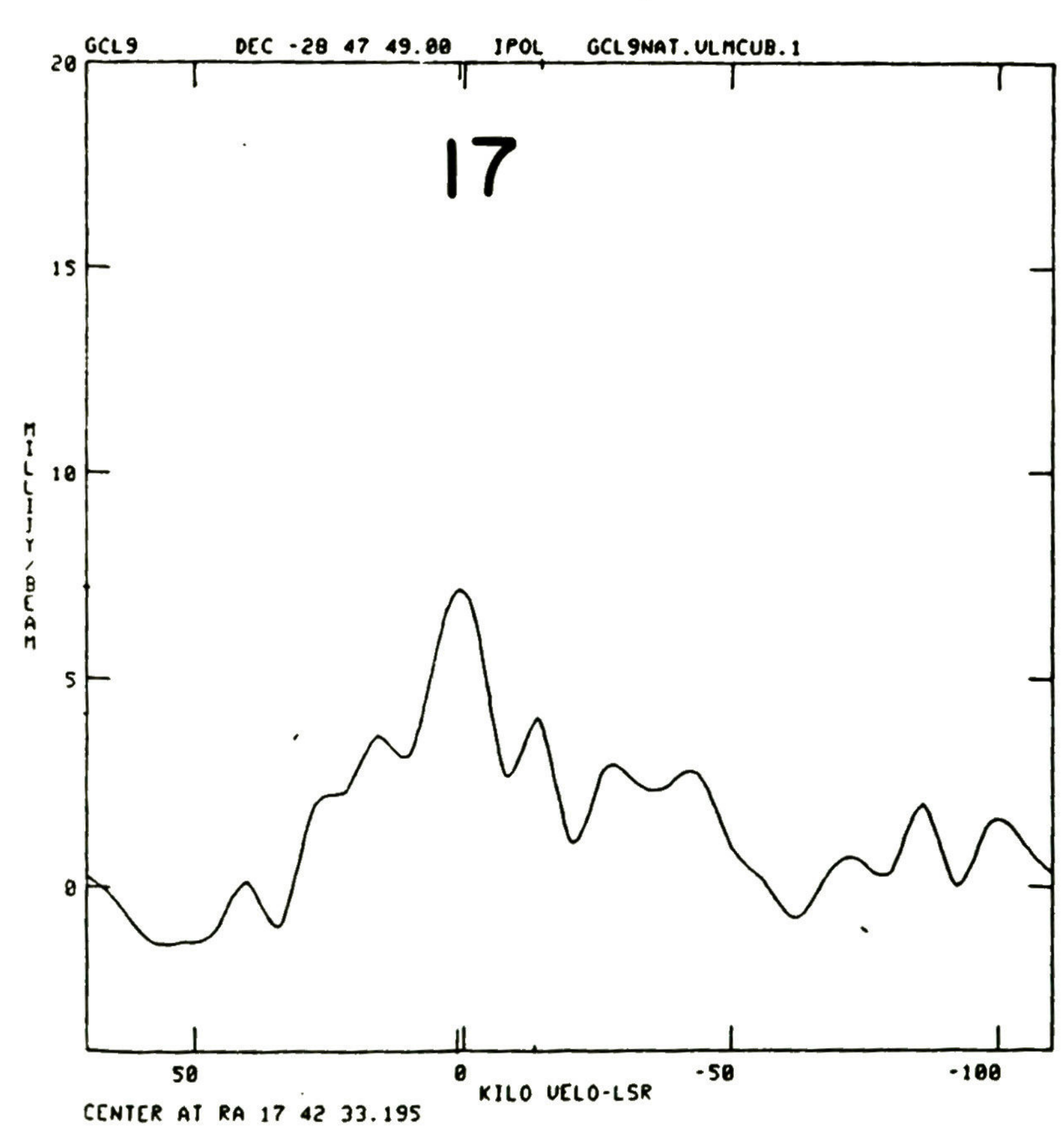
GCL9 DEC -28 47 49.00 IPOL GCL9NAT.VLMCUB.1
17
20
15
10
5
0
MILLIJY/BEAM
50
0
-50
-100
KILO VELO-LSR
CENTER AT RA 17 42 33.195

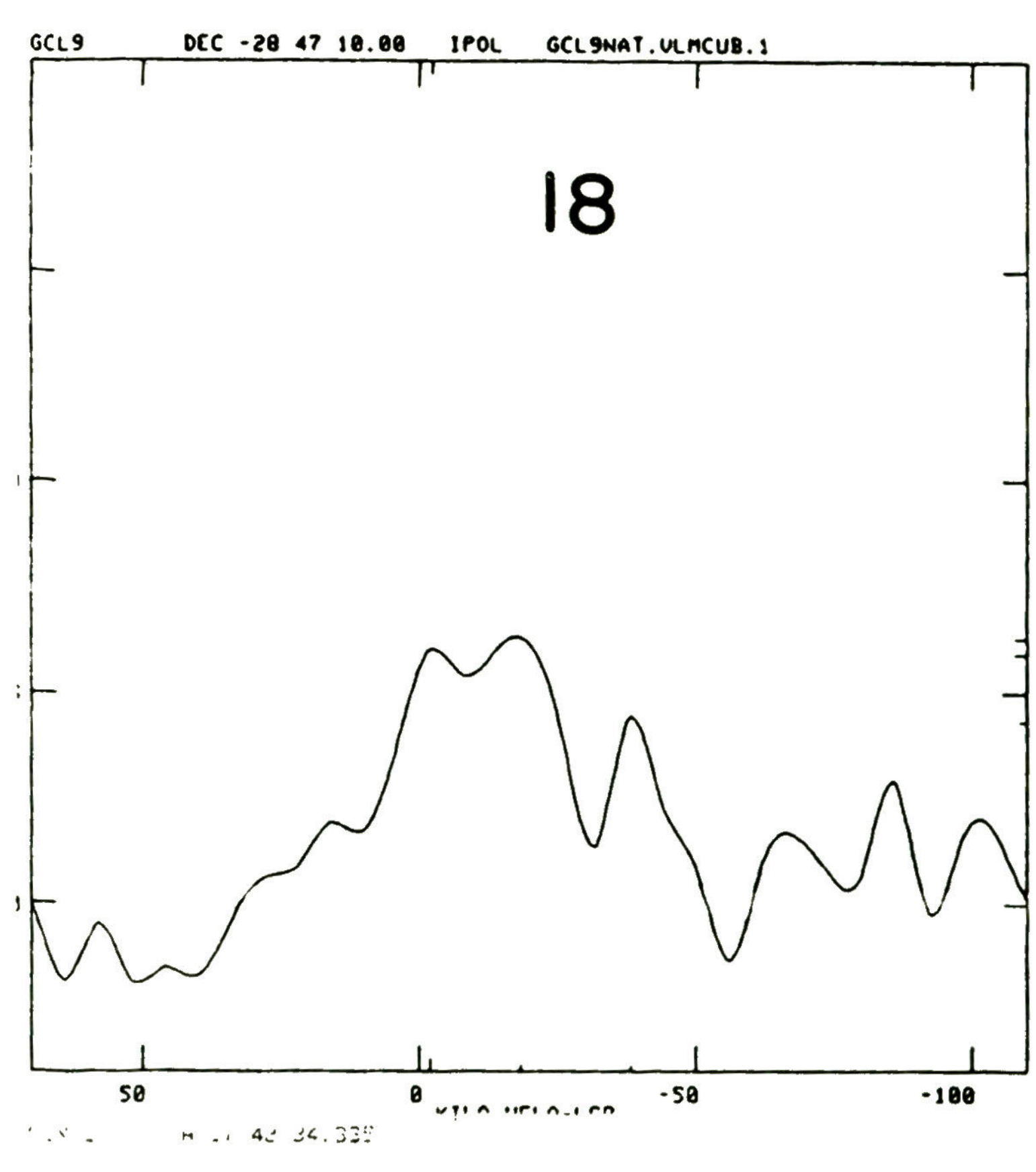
GCL9 DEC -28 47 10.00 IPOL GCL9NAT.VLMCUB.1
18
50
0
-50
-100

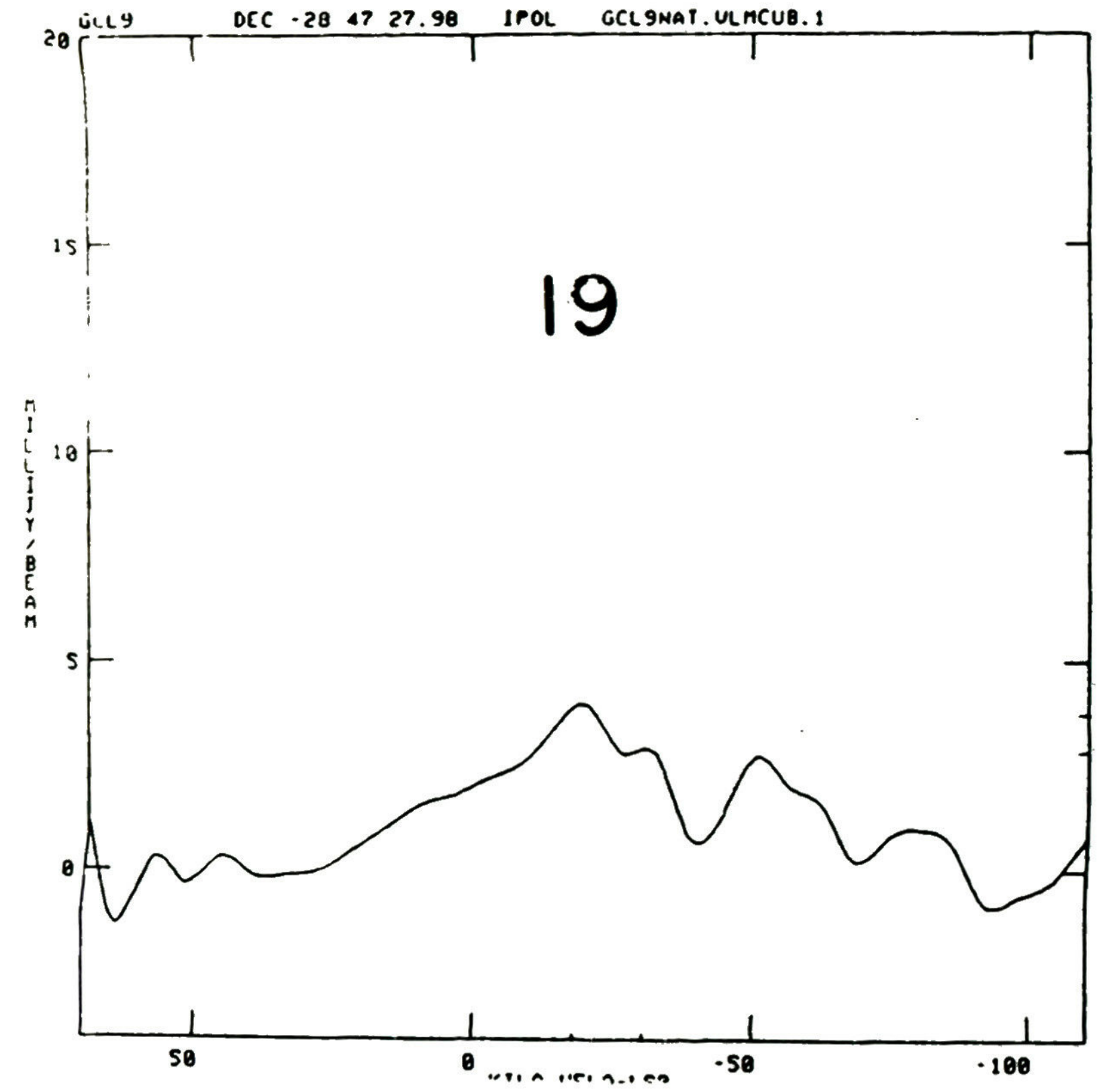
GCL9 DEC -28 47 27.98 IPOL GCL9NAT.ULMCUB.1
19
MILLIJY/BEAM
20
15
10
5
0
50
0
-50
-100

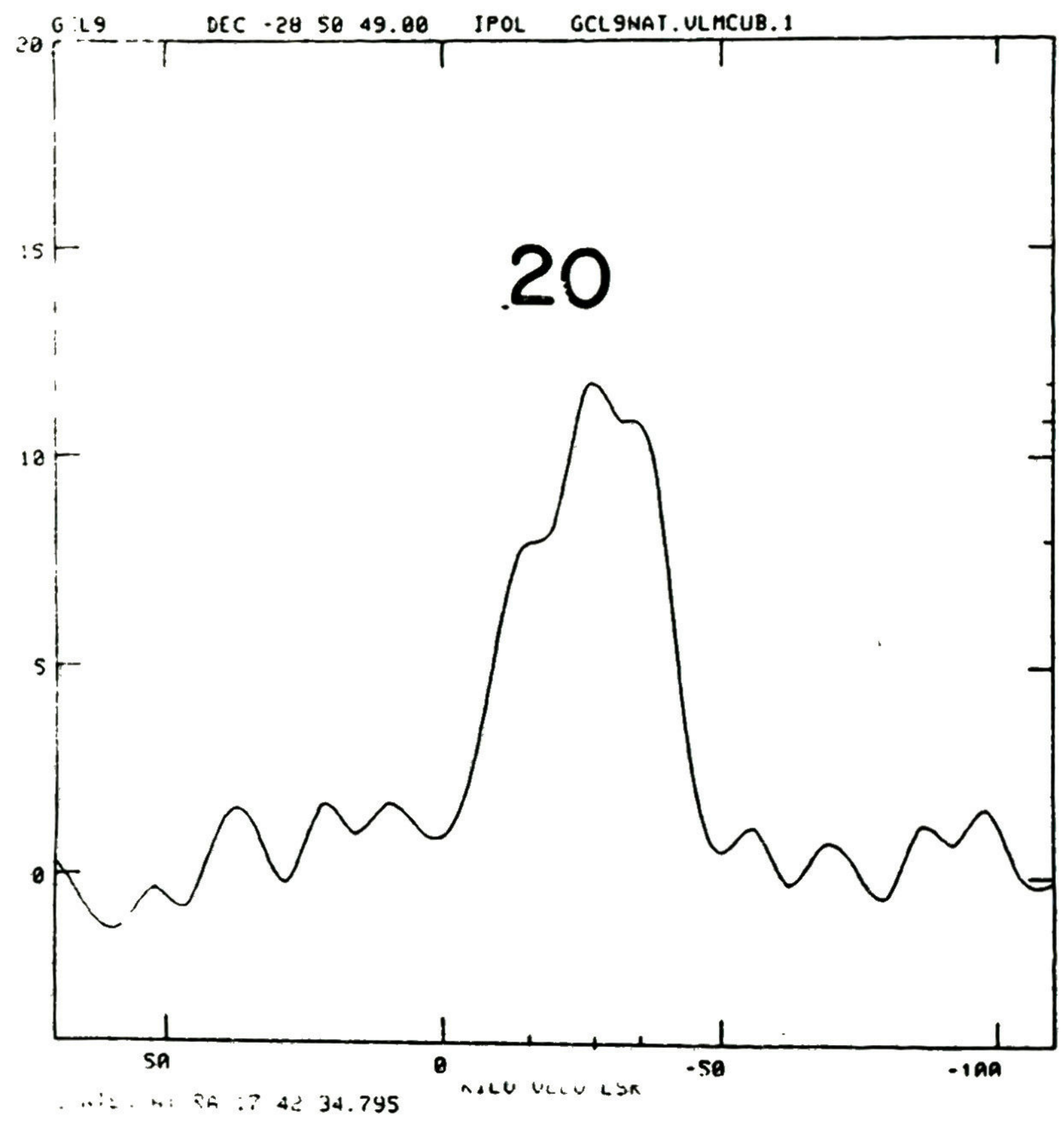
DEC -28 50 49.00 IPOL GCL9NAT.ULMCUB.1
20
20
15
10
5
0
50
0
-50
-100
KILO VELO-LSR
RA 17 42 34.795

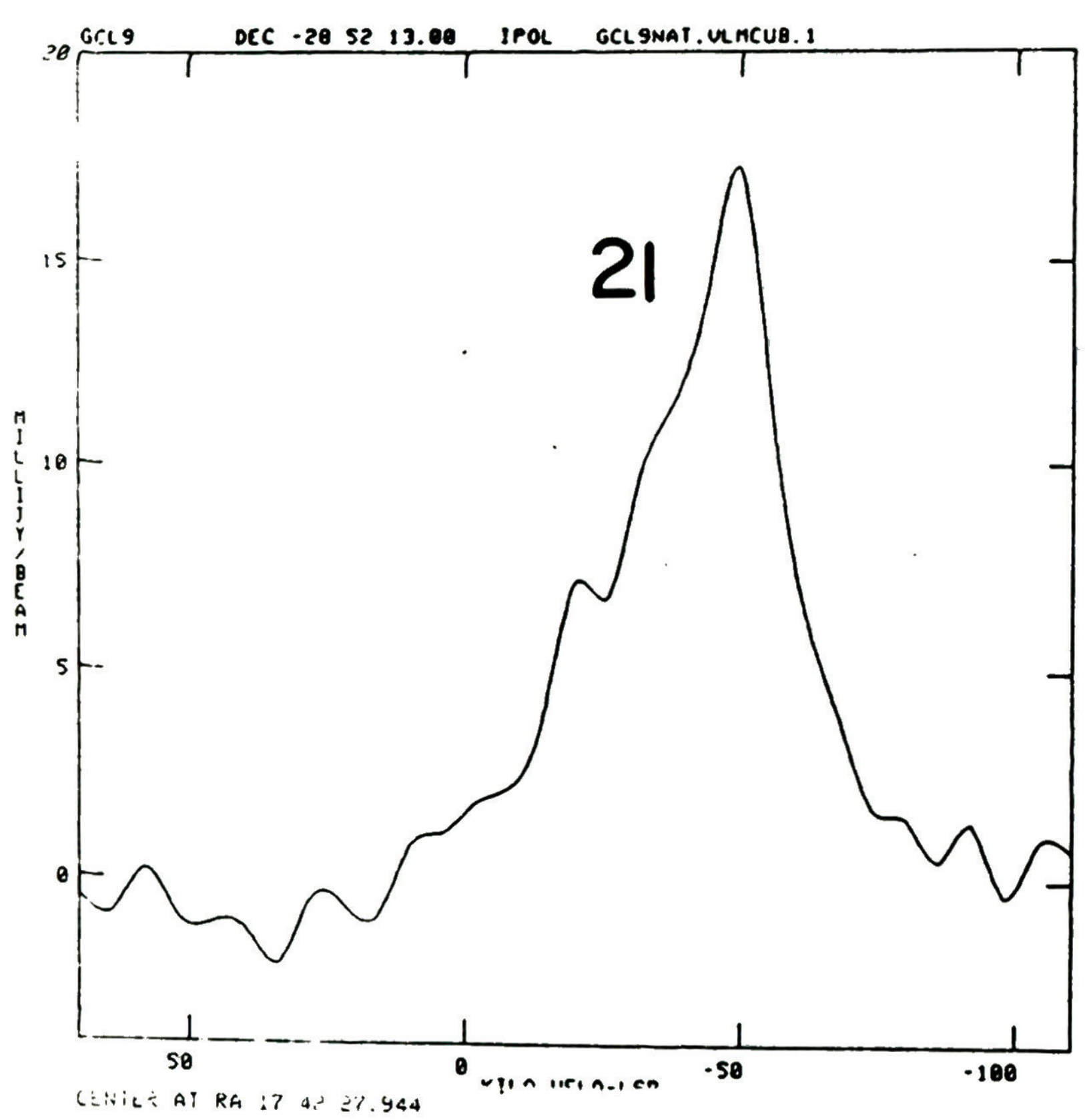
GCL9 DEC -28 52 13.00 IPOL GCL9NAT.ULMCUB.1
21
MILLIJY/BEAM
20
15
10
5
0
50
0
-50
-100
CENTER AT RA 17 42 27.944

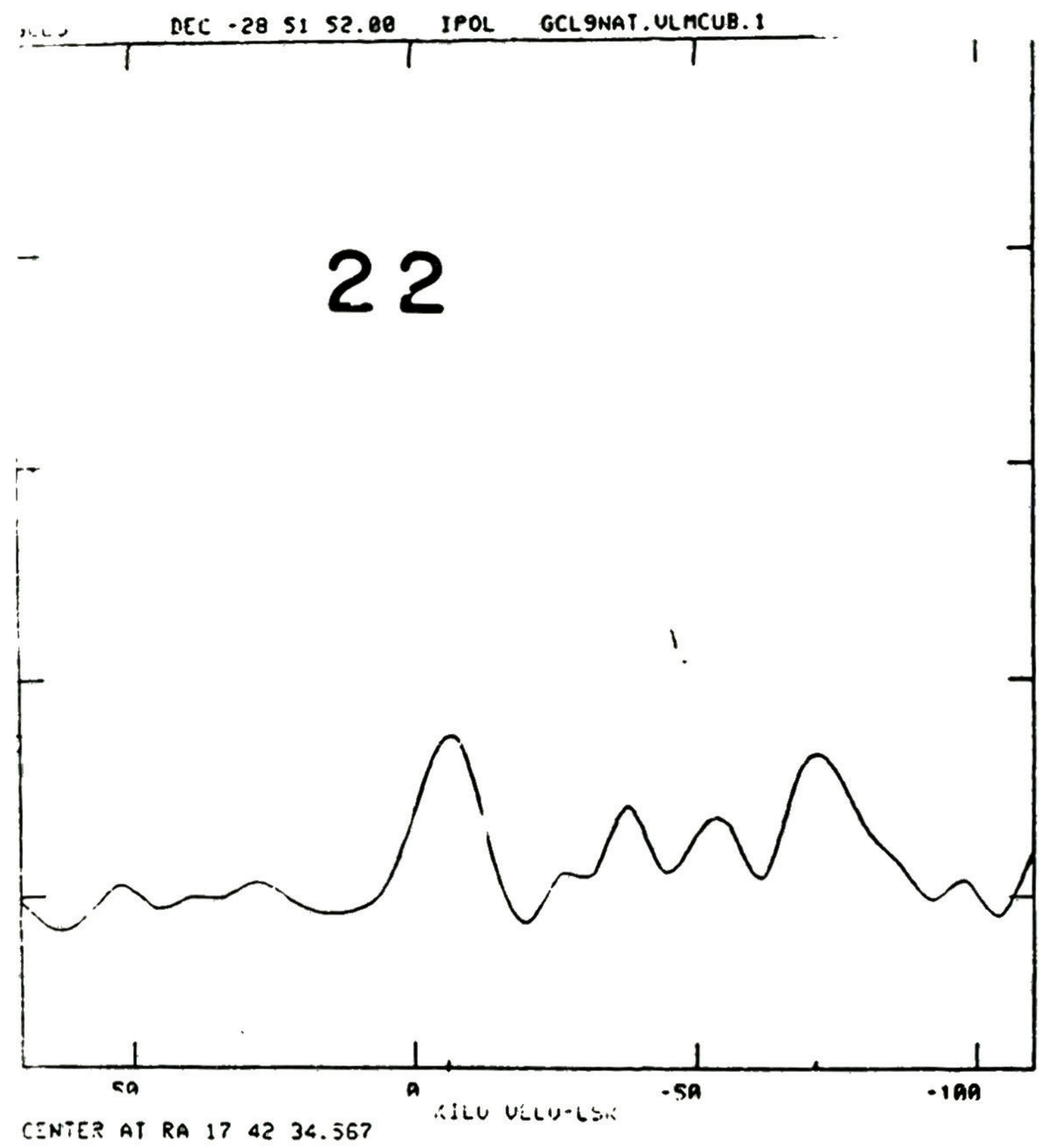
DEC -28 51 52.00 IPOL GCL9NAT.ULMCUB.1
22
50
0
-50
-100
KILO VELO-LSR
CENTER AT RA 17 42 34.567

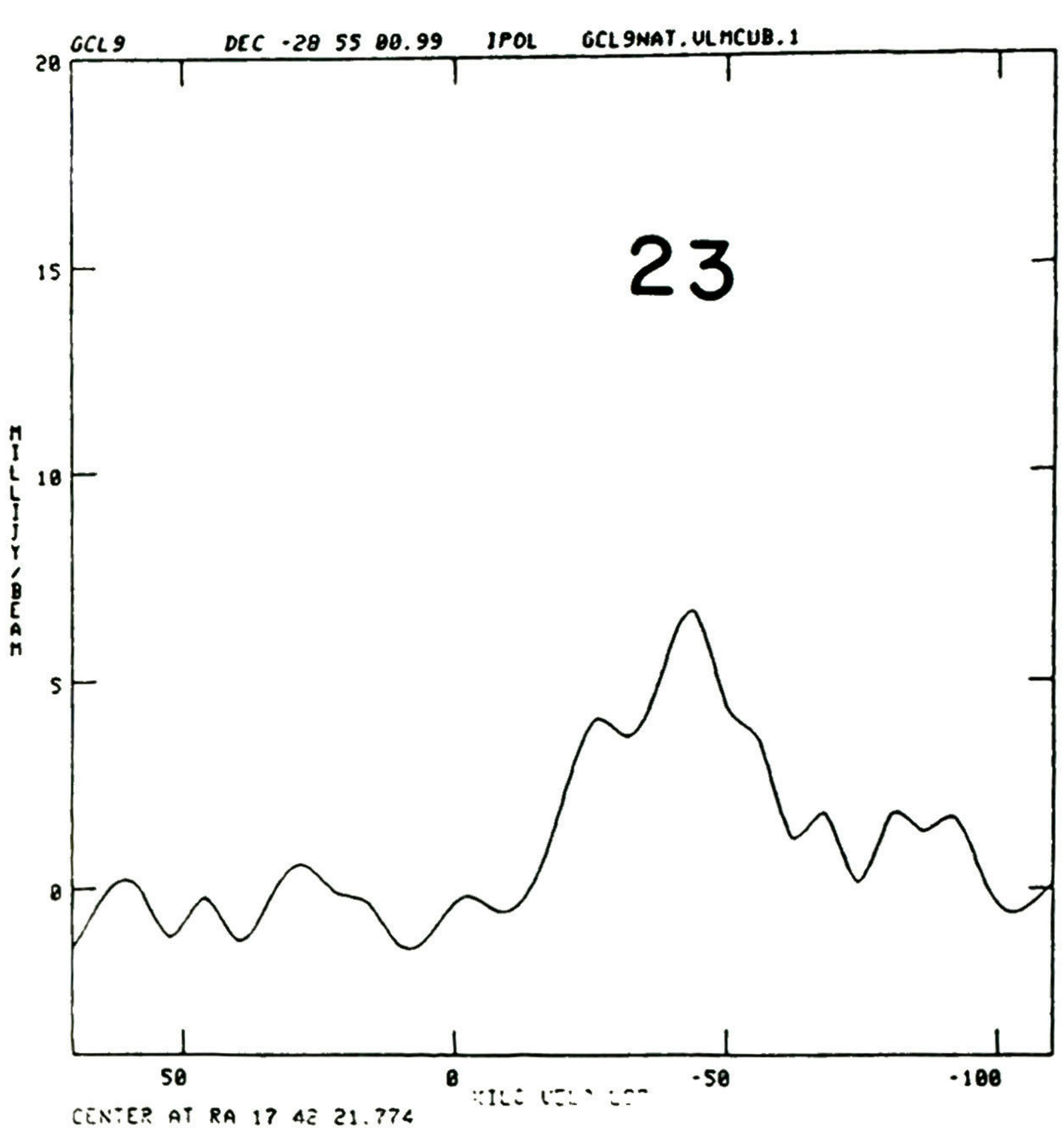
GCL9 DEC -28 55 00.99 IPOL GCL9NAT.ULMCUB.1
23
MILLIJY/BEAM
20
15
10
5
0
50
0
-50
-100
CENTER AT RA 17 42 21.774

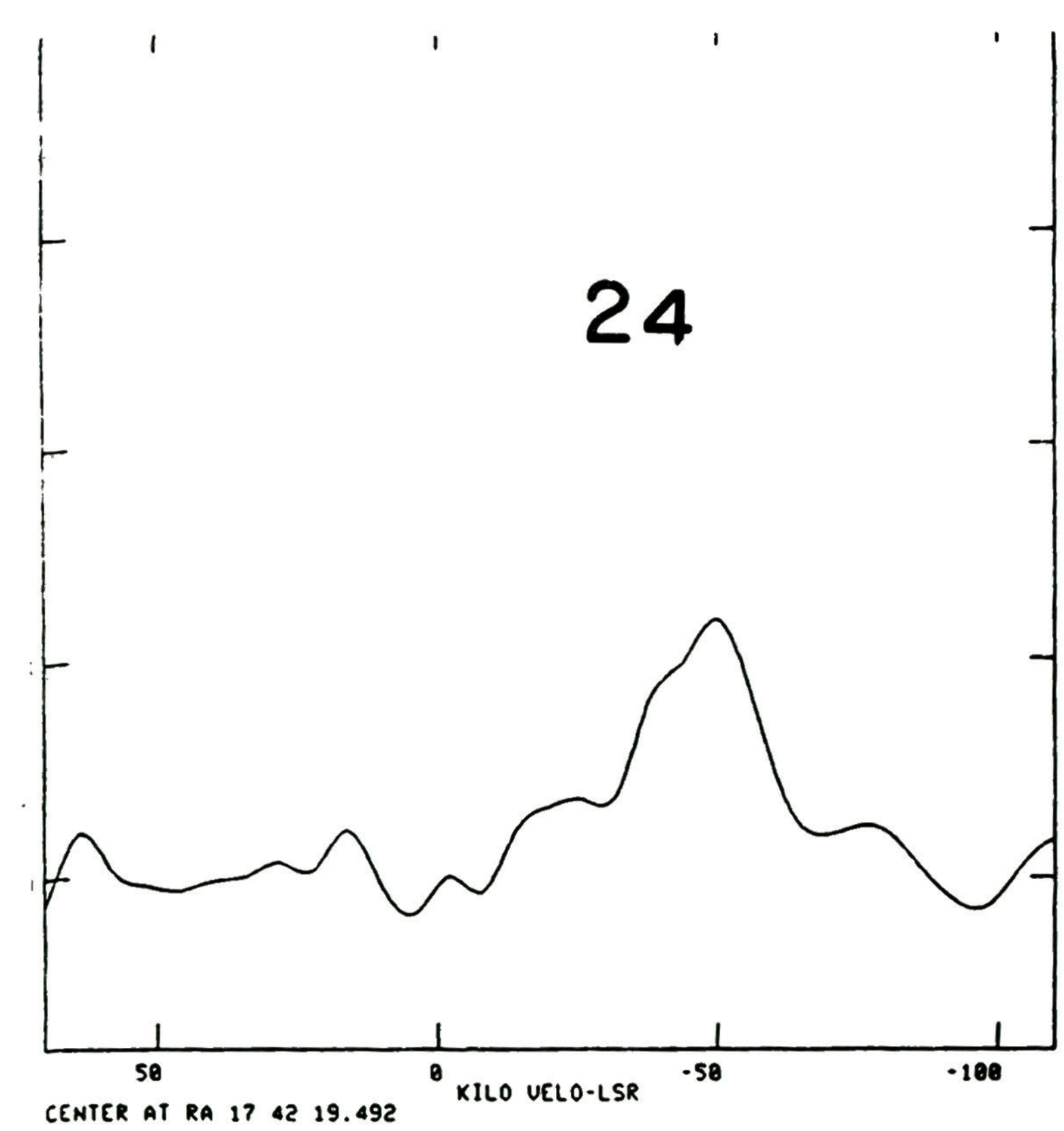
24
50
0
-50
-100
KILO VELO-LSR
CENTER AT RA 17 42 19.492

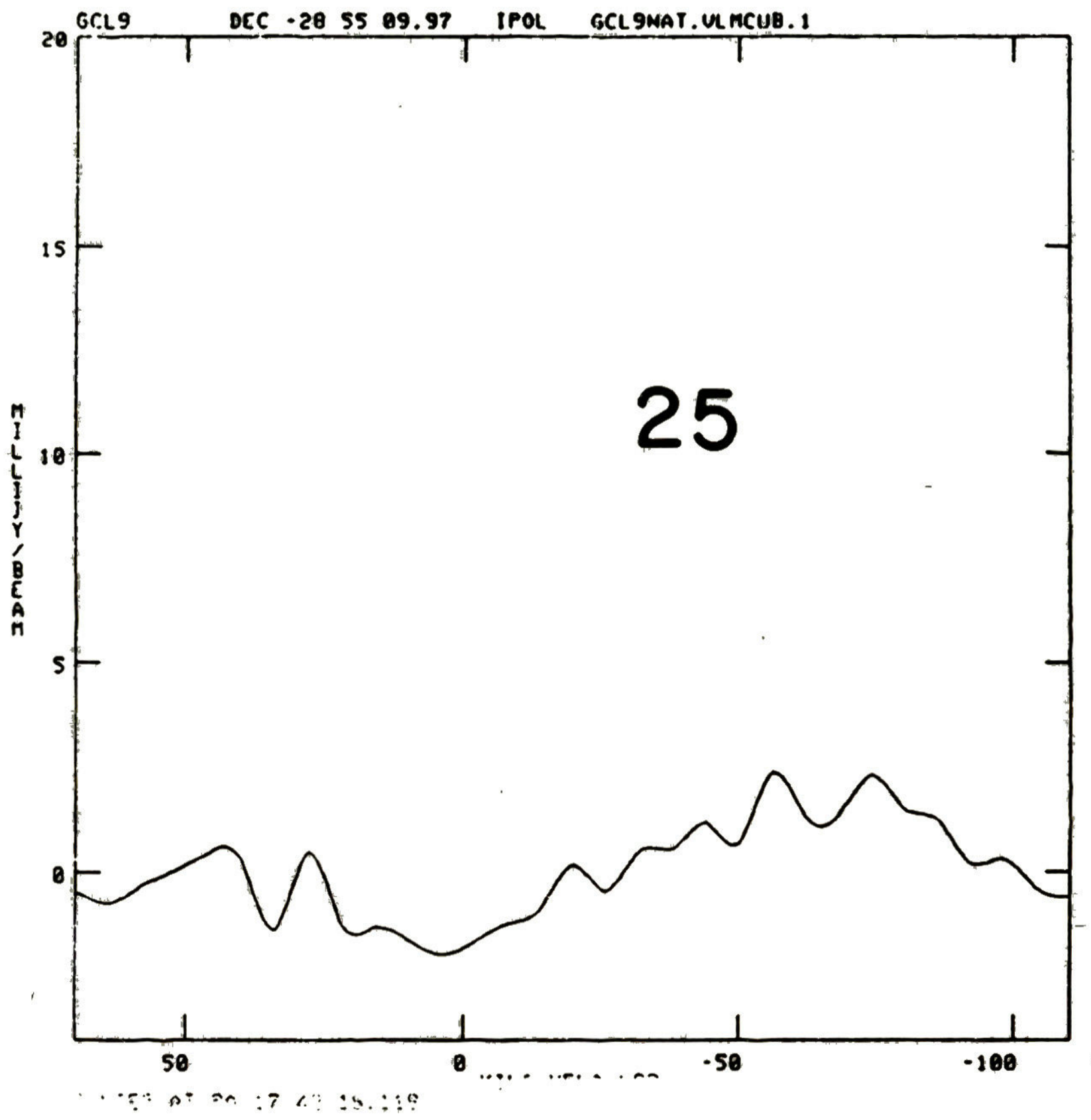
GCL9 DEC -28 55 09.97 IPOL GCL9NAT.ULMCUB.1
25
MILLIJY/BEAM
20
15
10
5
0
50
0
-50
-100

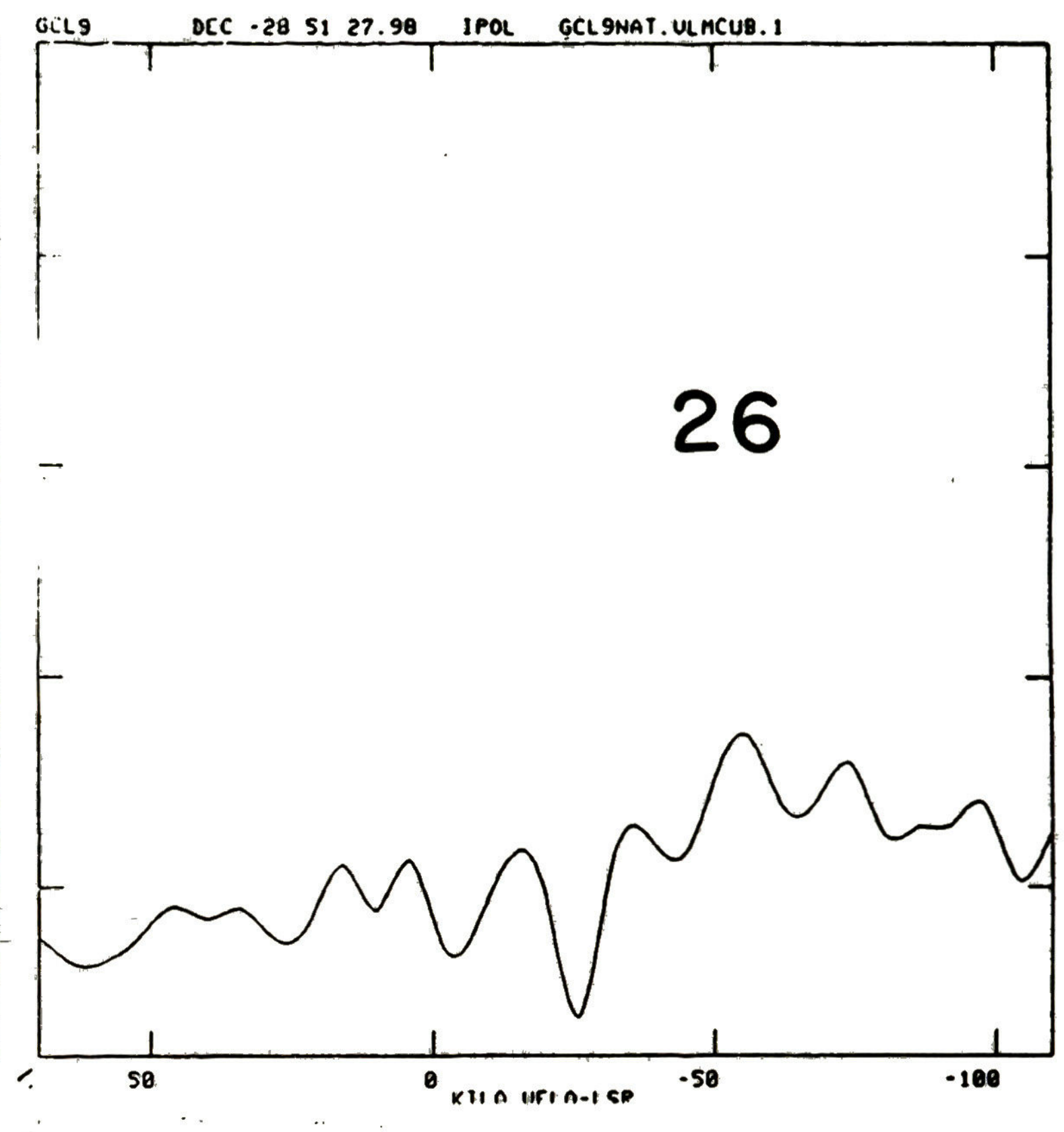
GCL9 DEC -28 51 27.98 IPOL GCL9NAT.ULMCUB.1
26
50
0
-50
-100
KILO VELO-LSR

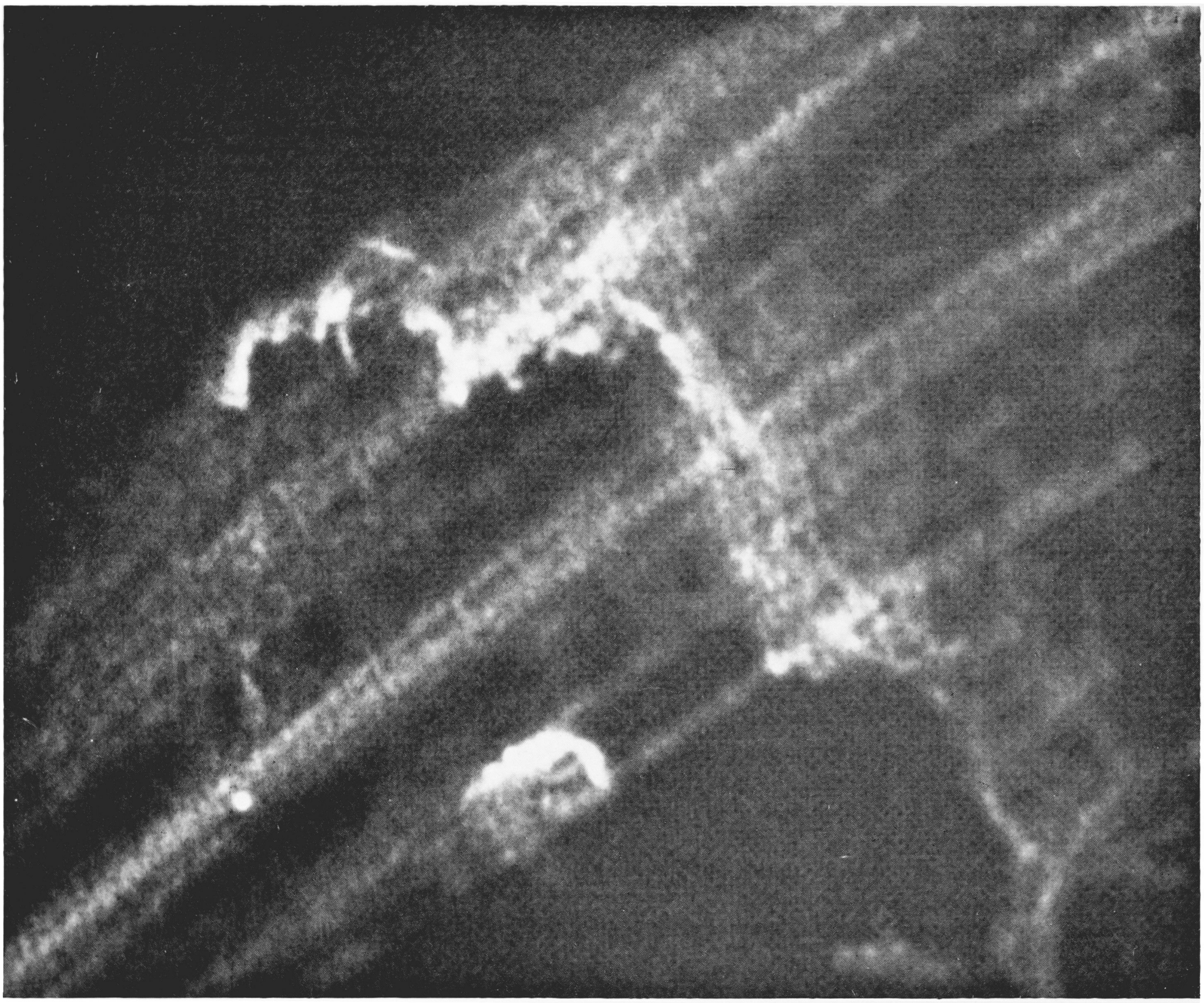

Figure 9: This radiograph is identical to figure 9 of chapter 3 except that CLEAN deconvolution algorithm is used in making this image.

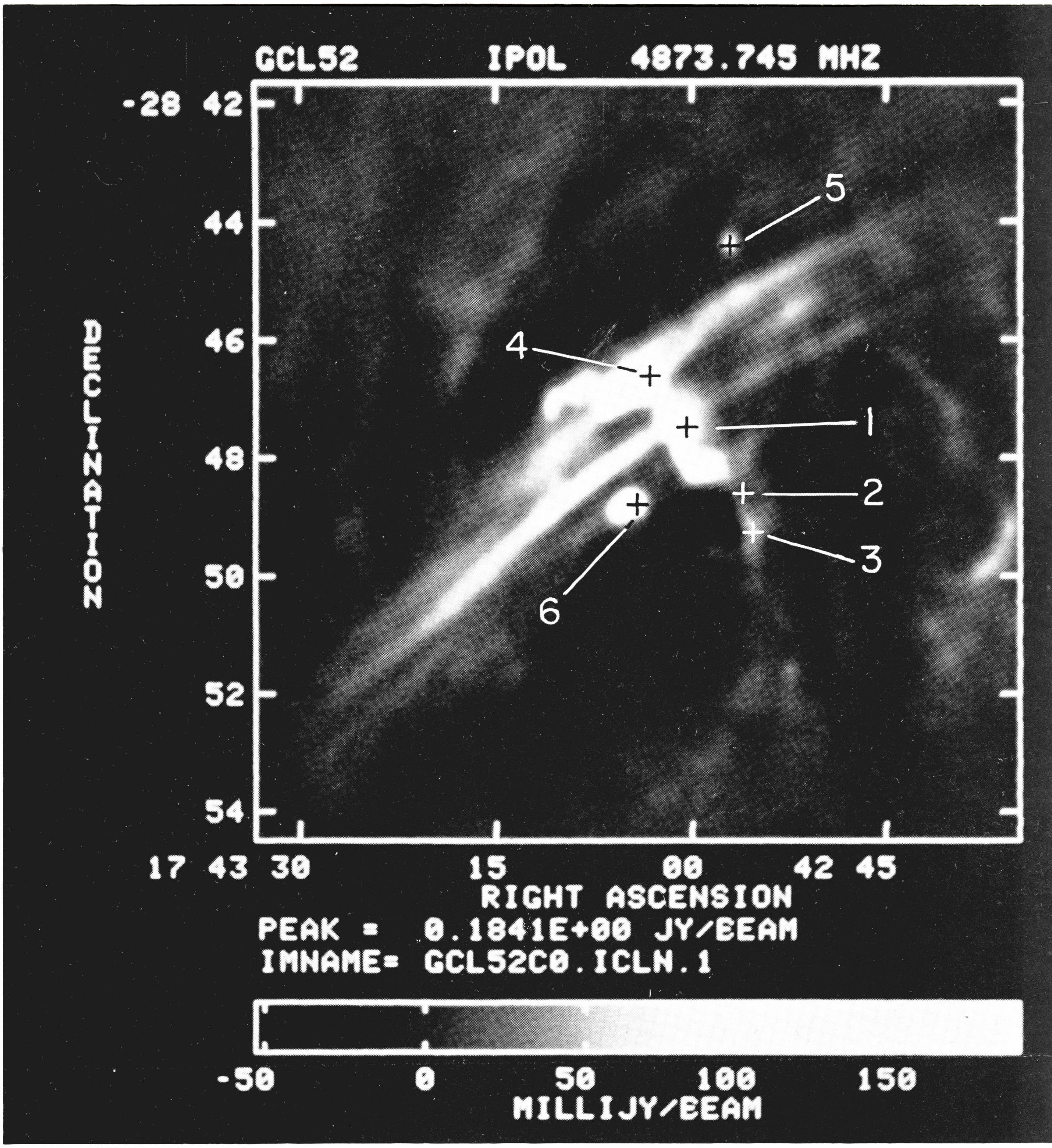

Figures 10 and 12: This radiograph shows the continuum emission from G0.18-0.04 with a resolution similar to that of figure 2. The image was also based on naturally weighted visibility data. The spectra shown in figure 12 corresponds to positions seen in figure 10.

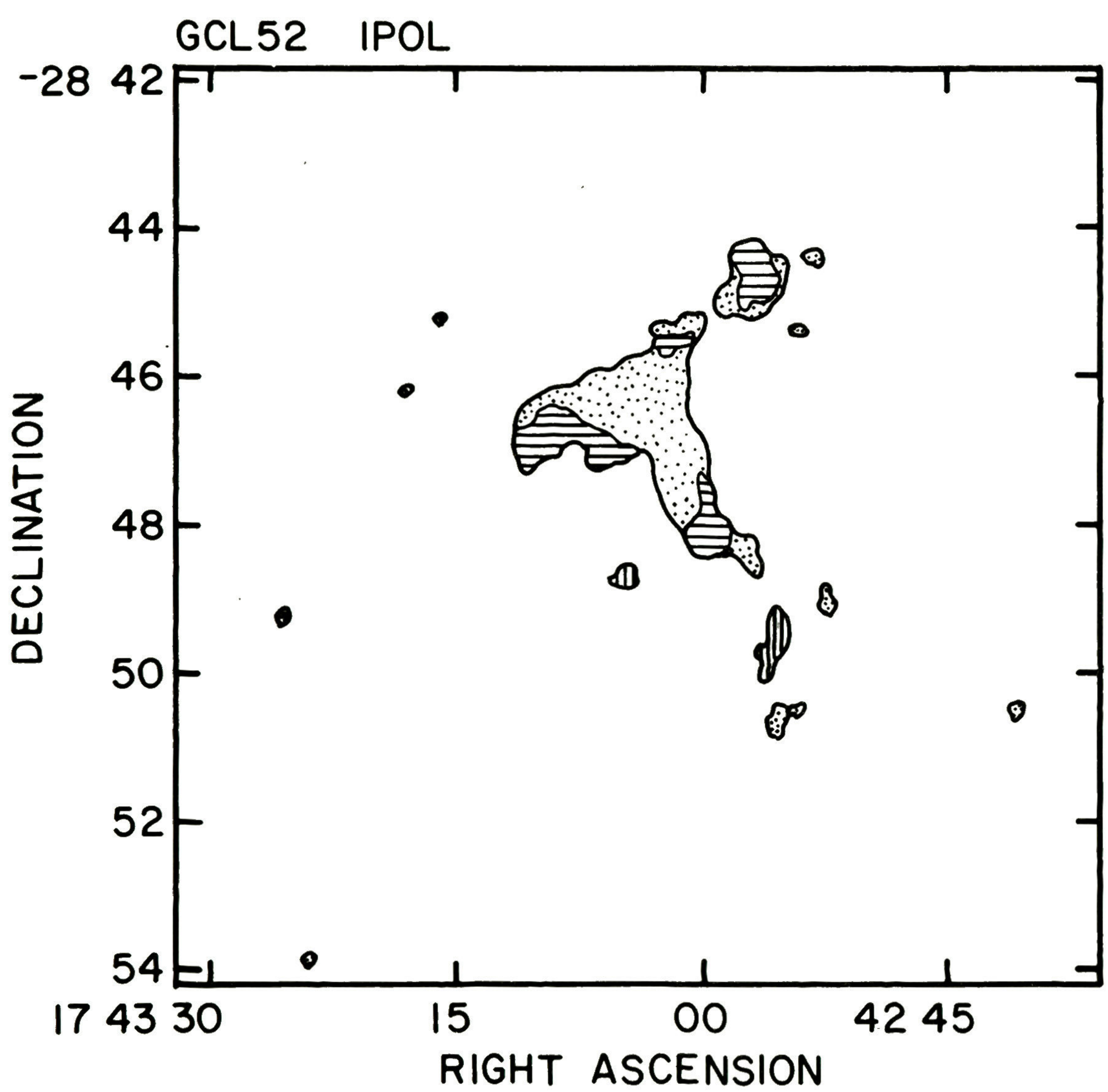

Figure 11: The intensity-weighted velocity structure of the region shown in figure 9. The horizontal lines, the vertical lines, the dotted region represent line emission with 35 $\leqslant$ $V_{LSR} \leqslant 65$, $65 < V_{LSR} < 110$ and $5 < V_{LSR} < 35$ km $s^{-1}$, respectively.

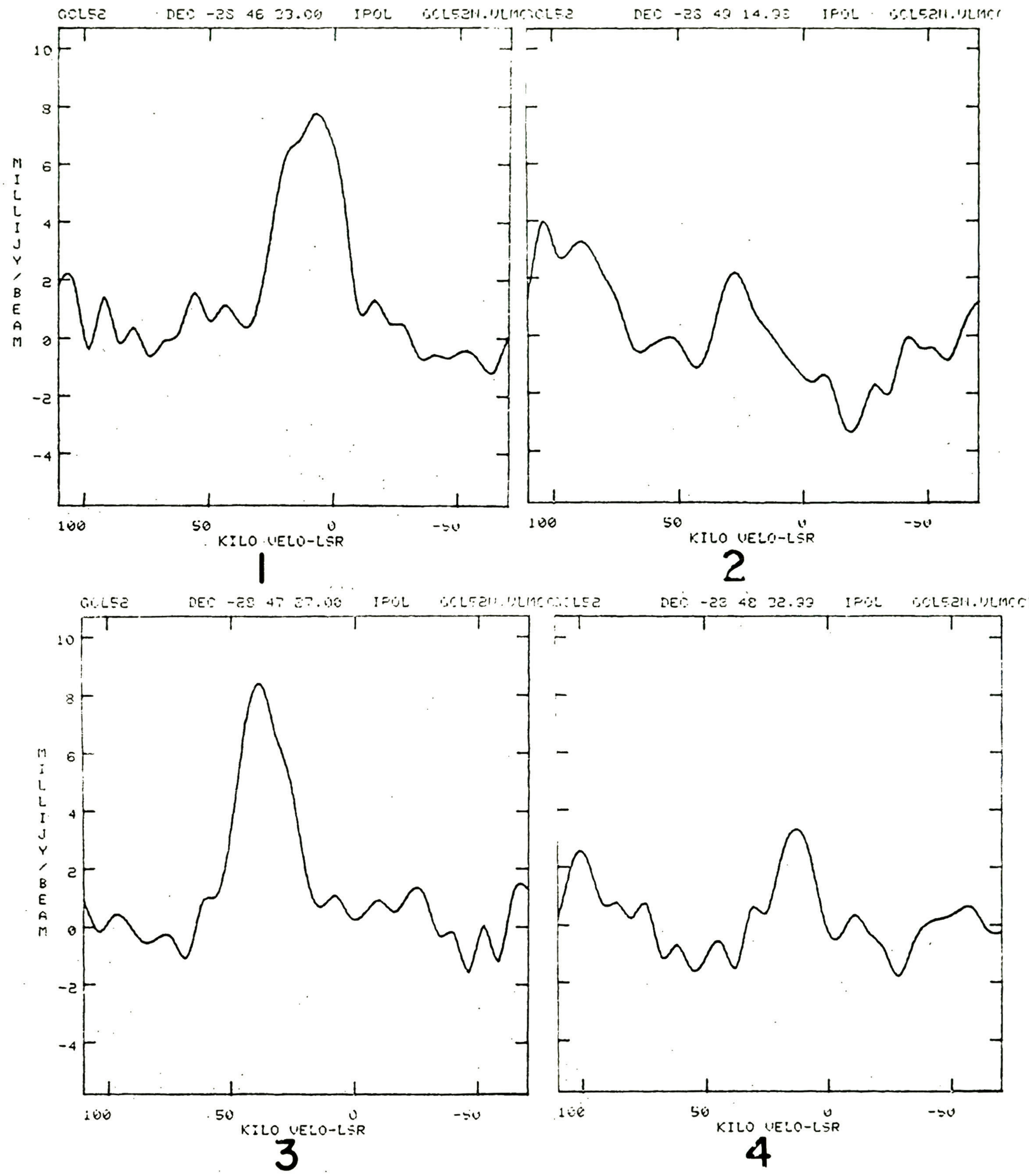

Figure 12.

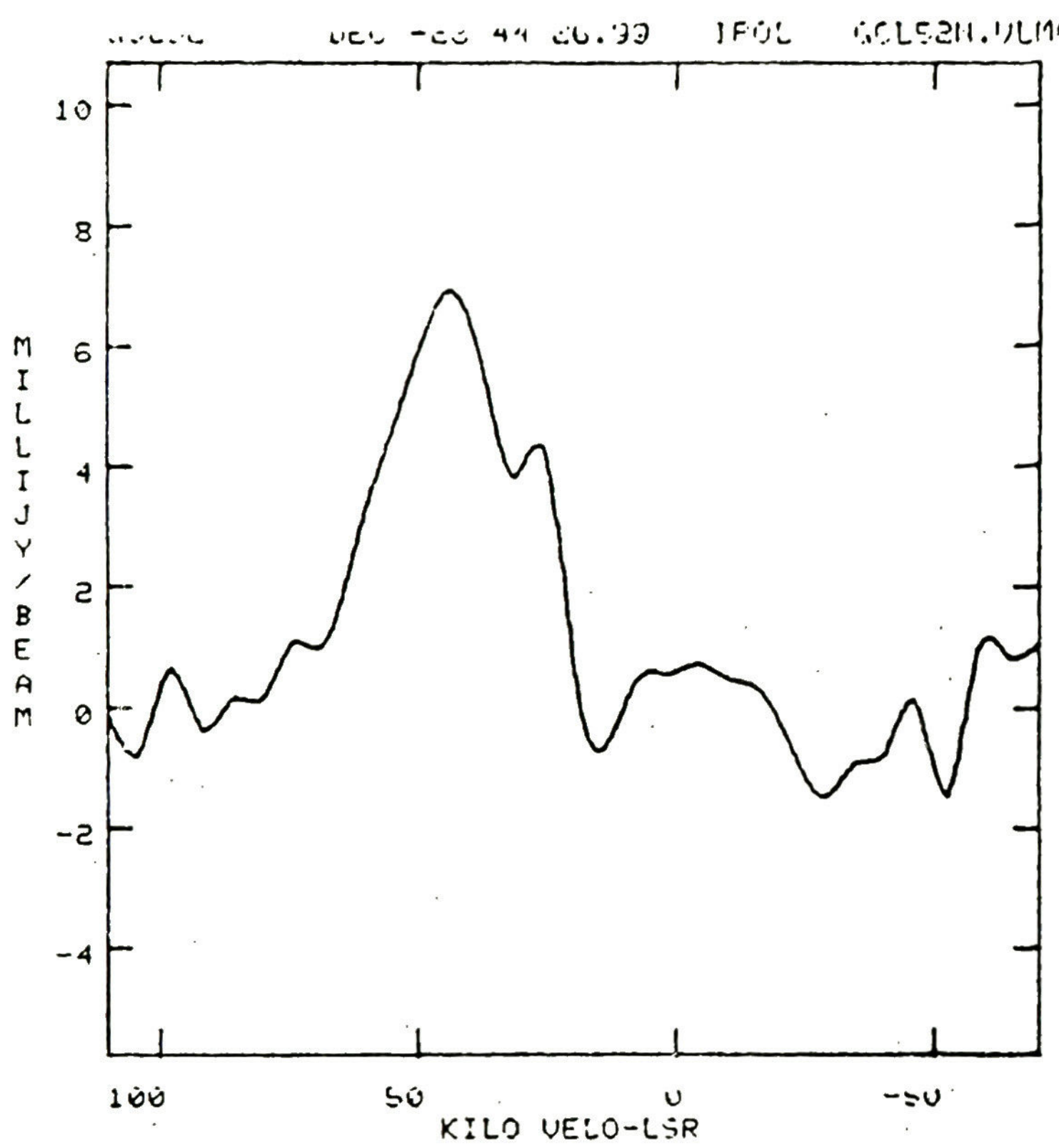
IPOL
MILLIJY/BEAM
10
8
6
4
2
0
-2
-4
100
50
0
-50
KILO VELO-LSR

5

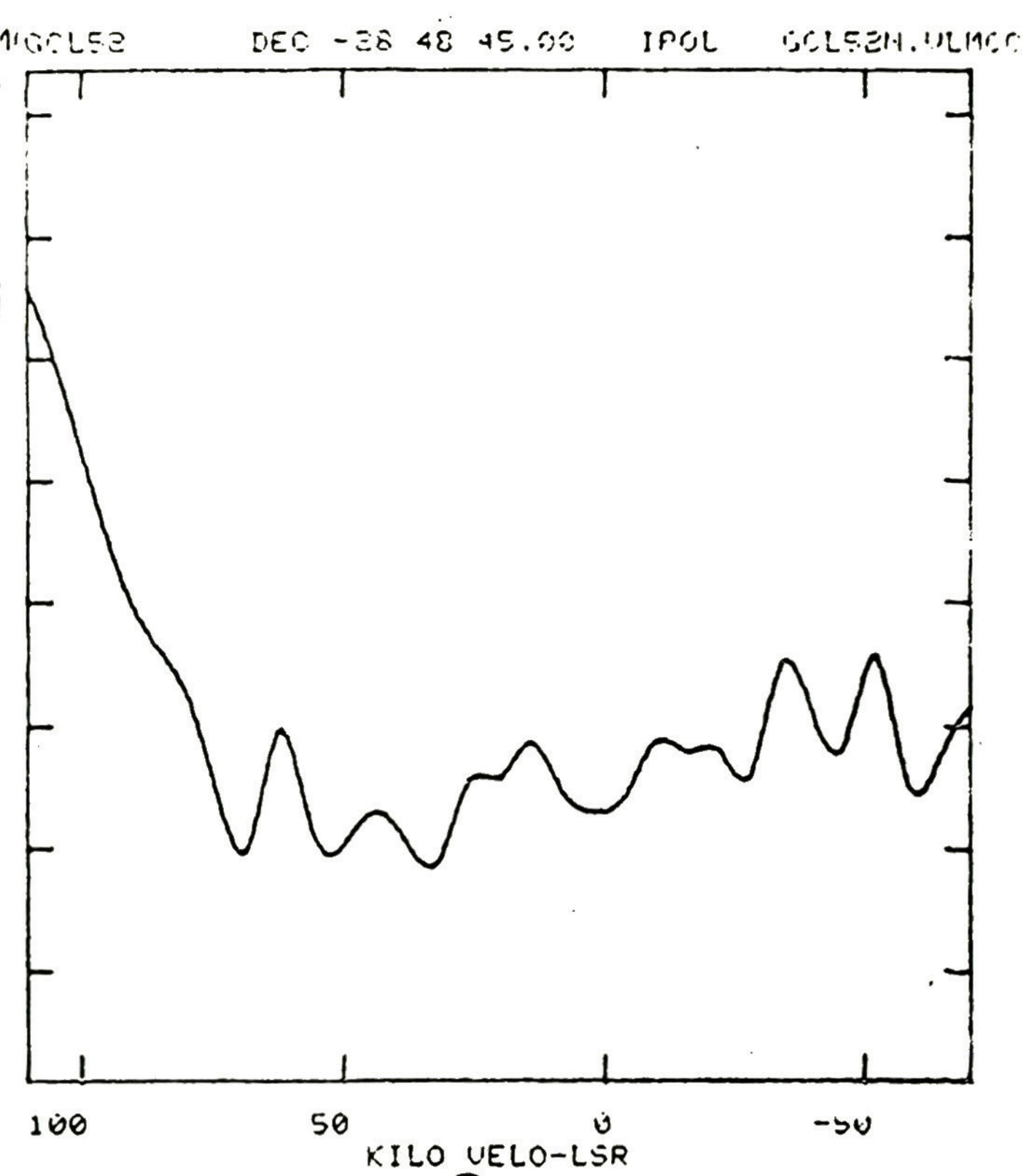
GCL52
DEC -28 48 45.00
IPOL
100
50
0
-50
KILO VELO-LSR

6

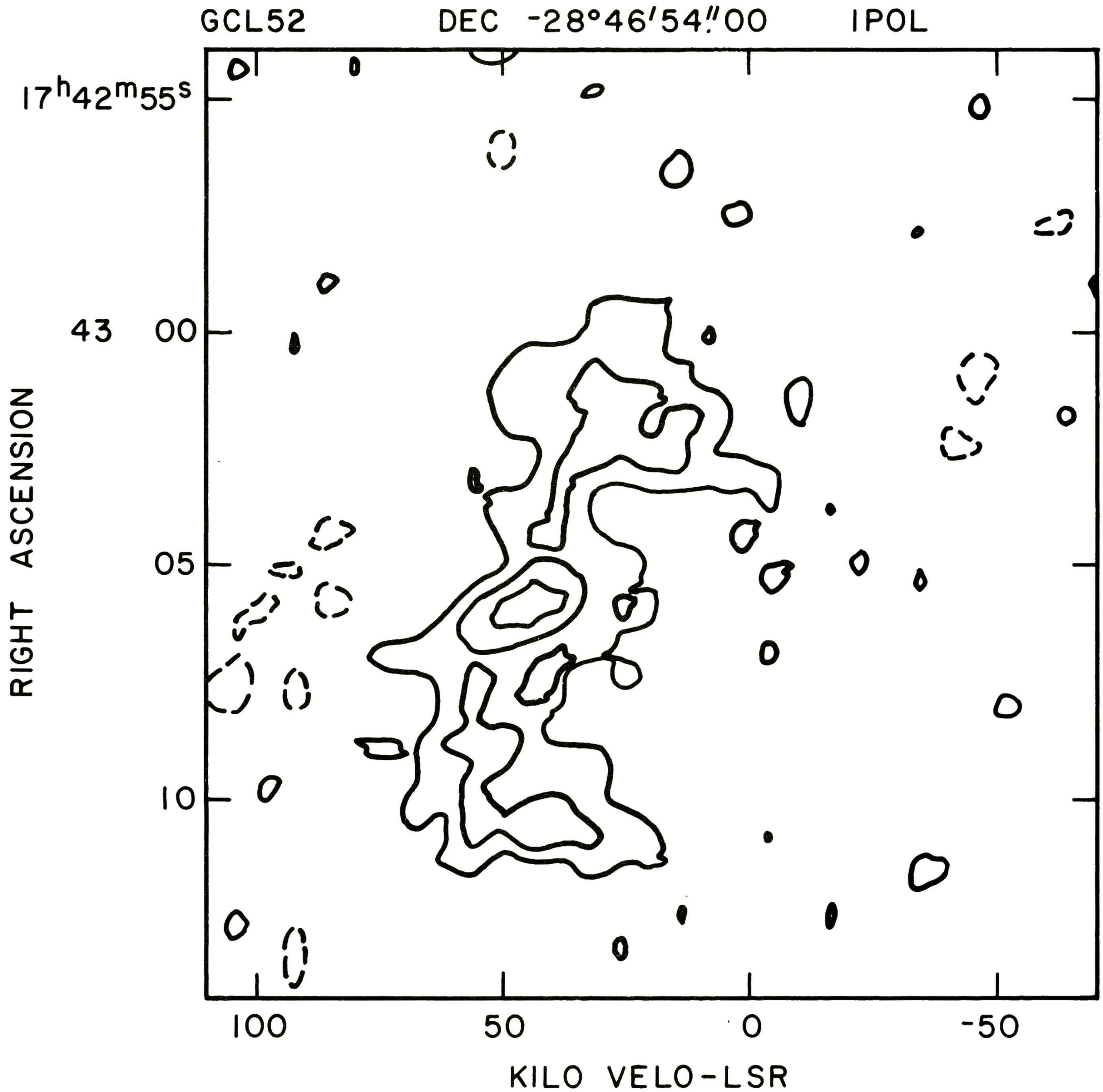

Figure 13: A cut is made along the right ascension at δ = -28°46'54". Contours of the line emission have intervals of -6, -4, -2, 2, 4, 6, 8, 10, 12 mJy/beam area.

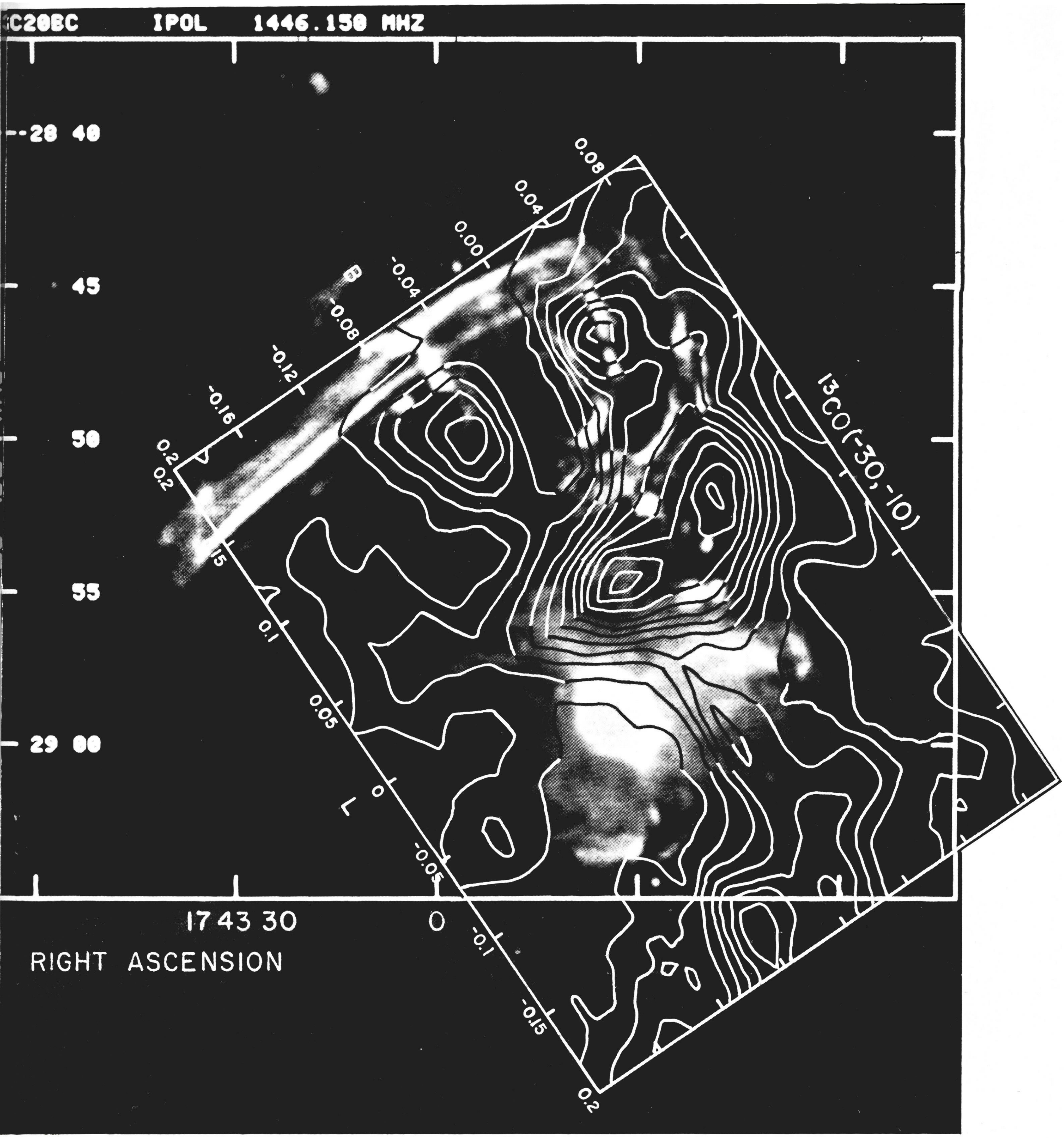

Figure 14: Contours of $^{13}CO$ distribution (Bally et al. 1986) having $-30 < V_{LSR} < -10$ km $s^{-1}$ is superimposed on the 20 cm radiograph.

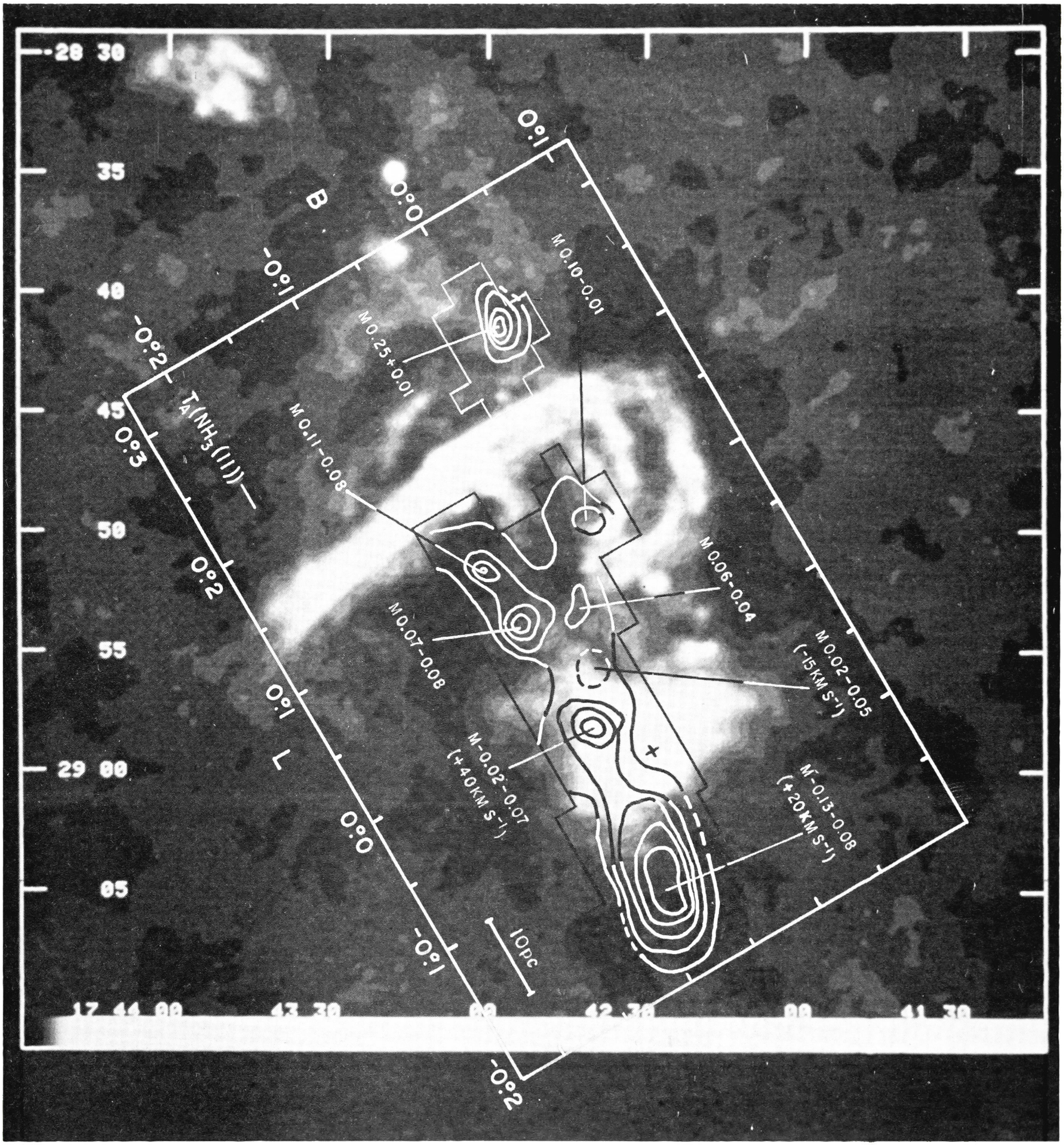

Figure 15: Contours of $NH_3$ distribution (Güsten et al. 1981) is superimposed on the 20 cm radiograph.

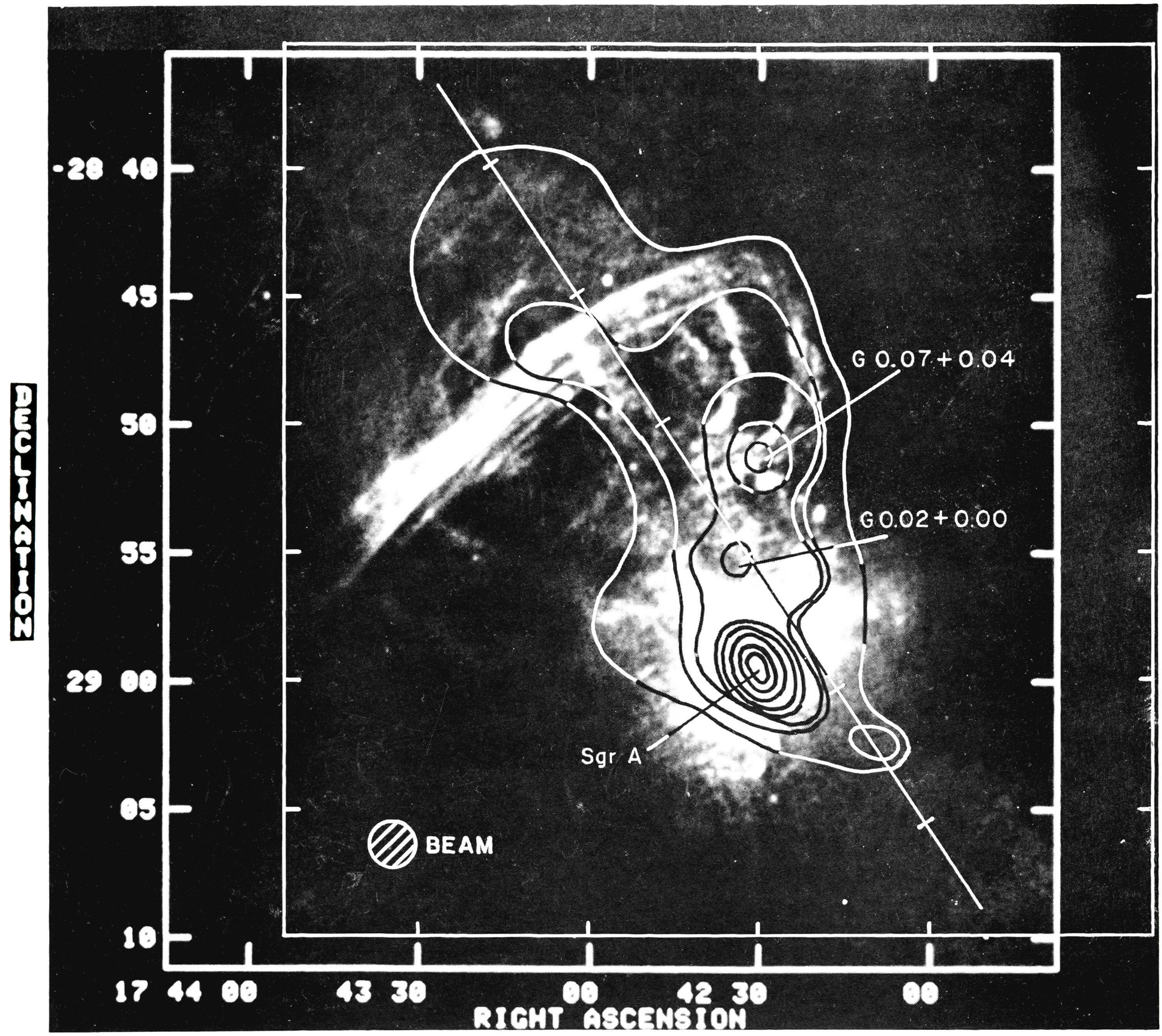

Figure 16: The 55 μm dust distribution (Dent et al.) is superimposed on the 20-cm image.

# CHAPTER 10

# VLA OBSERVATIONS OF THE POLARIZED LOBES NEAR THE ARC

"a bird made diaphanous suddenly,
a spider Webbing the sky,
and then gone."
Pablo Neruda

## I. Introduction

Recent observations of the inner one degree of the galactic center region using single-dish telescopes have revealed three polarized sources at 3 cm: a relatively compact one is located in the Arc is identified with G0.16-0.15, while the other more extended structures are situated along the northwest and southeast extensions of the Arc on opposite sides of the galactic plane (see chapter 8 and the references cited therein). These three polarized components resemble a core/lobe structure which is reminiscent of classical double radio source (Seiradakis _et al._ 1985). However, the total intensity counterpart of the polarized components shows a continuous system of filaments (i.e. the Arc) which occupies the region between the polarized lobes (see chapters 3 and 8). A model in which a non-uniform distribution of thermal gas surrounds the filamentary Arc is invoked in order to account for the distributions of both the total and polarized emission from the Arc and its extensions (lobes B and C in chapter 8). This thermal plasma is hypothesized to originate either by means of an outflow from the galactic center (Seiradakis _et_ al. 1985) or/and by means of an interaction between the 50 km $s^{-1}$ molecular cloud and the Arc (see chapters 4 and 8; Seiradakis et al. 1985). More recent low-frequency observations (i.e. at 120 and 160 MHz) show three peaks of emission roughly coincident with the polarized regions (Kassim, La Rosa and Erickson 1986; chapter 4) and thus reveal the regions in which absorption of low-frequency emission is minimized.

VLA results of the central polarized component lying along the Arc (source A in chapter 8) were fully described in chapter 3. There, we find that this source has a very large Faraday rotation (~ -3000 rad $m^{-2}$) and has an elongated and filamentary structure at 6 and 2 cm, respectively. Here we concentrate exclusively on the results of the total and polarized emission from the out-of-plane extensions of the Arc (i.e. the NE and SW Arc lobes) based on the VLA data. These results show clearly that both extensions of the Arc have filamentary characteristics, though deviate somewhat from the linear filaments which compose the Arc. Furthermore, a number of sharp absorption features which depolarize emission from the northwestern extension of the Arc are revealed.

## II. OBSERVATIONS

Radio continuum observations of the northern and southern lobes were carried out using the VLA at both 6 and 20 cm (Arc no. 6 to 11 in Table 1 of chapter 2). We used both the hybrid B/C and C/D arrays for the fields centered on the eastern edge of the NE lobe, whereas we employed only the C/D array for the western edge of the SE lobe. Full description of the continuum observations can be found in chapter 2.

All the figures presented in the following section except the polarization maps - figures 8, 9, 13 and 14 - are based on combined data from the two sets IF's used throughout the observations (c.f., chapter 2). Figures 8 and 9 show contours of polarized intensity in the northern lobe at 4.72 and 4.87 GHz, respectively. The line segments superimposed on both maps represent the orientation of electric vectors, and their lengths are proportional to the intensity of polarized emission. Figures 13 and 14 show contours of polarized intensity in the southern lobe at 4.87 GHz. The line segments superimposed on both maps correspond to the difference in position angles of the electric vectors at 4.87 and 4.72 GHz (i.e. the 4.72 GHz map is subtracted from 4.87-GHz map in this chapter; the reverse of this operation is carried out in Faraday rotation maps shown in chapter 3). The length of the line segments is proportional to the polarized intensity at 4.87 GHz. None of the figures shown in this chapter are corrected for the response of the primary beam. The scale sizes of structures appearing in the 6-cm maps of polarized and total intensity are rather different. The missing short interferometer spacings that we could not sample in our observations prevent us from properly imaging the more extended structures seen in the total intensity maps. Consequently, we are able to present reliable maps of percentage polarization and we do not attempt to do so.

# III. RESULTS

"I know that you believe you understand
what you think I wrote,
But, I am not sure you realize that
what you read is not what I meant."
Anonymous

## (A) NW Arc Lobe

The 20-cm radiograph shown in figure 1 depicts both the arched and the linear filaments which compose the Arc, the radio threads and the Sgr A complex (see chapters 3, 5, 6 and 9). The newest radio structure in this region is a continuous extension of the linear filaments as they protrude through their intersection with the arched filaments. This structure, which continues up to b = 0.4°, appears in single-dish maps as a ridge of emission (see figures 3 and 4 in chapter 8). Another large-scale feature is the northern radio thread (see chapter 5) which rises from the southern portion of the arched filaments and appears to merge with the column of radio structures which constitute the northwestern Arc lobe. At the junction between these two structures, the emission becomes diffuse, and resembles a plume.

A discrete source which is embedded within the NW Arc lobe (G0.17+0.15) appears to be bounded on its eastern and western sides by two sets of large-scale filaments, one of which is an extension of the northern linear filaments depicted in figure 2 of chapter 3. The other is a prominent distorted filament which lies to the west of the discrete source and shows a gentle change in its curvature a few arcminutes north of the discrete source toward more positive latitude at $\ell \simeq 0.15°$, $b \simeq 0.2°$. The south eastern extension of the distorted filament into the network of filamentary structures associated with the linear portion of the Arc appears incoherent and disjoint as is placed in the region between the northern and southern linear filaments (see figure 4 in chapter 8 and figures 1 and 2 in chapter 3). These linear filaments are seen in the 6-cm images of figures 2 and 3 as they cross the arched filaments. We note that the distorted filamentary extension of the linear filaments - as seen in figure 1 - is not brought out well in figure 2 whereas the extension of the southern linear filament, which is apparent in figure 1, can be identified clearly in figure 2 (this discrepancy occurs partly because the 20 and 6-cm images preferentially emphasize the large and small-scale structures, respectively.)

We note from figures 2 and 3 that the discrete source G0.17+0.15 is resolved into an oval-shaped ring structure with an enhancement at its center (see contour map in figure 6). Comparison of the 6 and 20-cm maps of this discrete source reveals that it has a flat spectrum. It is not clear whether this enhancement and a bright curved feature to its south can be interpreted in terms of filaments interacting with the ring structure.

Another important feature is the divergence of the northern and southern linear filaments as they traverse the discontinuity in brightness at the intersection with the arched filaments (fig. 2). These two sets of filaments were noted in chapter 3 to diverge gradually from each other in the linear portion of the Arc but not to the extent seen in figure 2. The strikingly linear northern filament is distorted exactly at the location where the brightness of the western arched filament is reduced substantially ($\alpha \sim 17^h42^m32^s$, $\delta \sim -28°$ $43'$. Another curious feature is the extension of the southern linear filament, the appearance of which is very similar to that of the northern thread. These two structures have widths of ~ 0.25 pc and appear to cross the southern and northern ends of the westernmost arched filaments whose brightnesses drop substantially if continued further to the south (north) of the thread (southern linear filament). Further high-sensitivity measurements of this region should indicate if the northern thread and the southern linear filament join each other. (Interestingly, but perhaps fortuitously, they appear to form two segments of a projected circle.)

A close up view of figure 2 is shown in figure 3 in order to identify a long and narrow radio shadow (width of ~ 25") which runs tangentially along the southern linear filament. Indeed, a sharp discontinuity of ~25" width can be recognized in the brightness distribution of the eastern arched filaments as they cross both the southern linear filament and its adjacent shadow counterpart. This radio shadow appears in many of our 6 and 20-cm radio pictures with different phase centers and can not be accounted for by a "negative bowl" (see chapter 2) which is attributed to a lack of short (u,v) spacings and which usually suppresses emission from the extended sources. This "negative bowl" is visible in the vicinity of the arched filaments in figures 2 and 3 but, unlike the shadow feature which appears only along the northern side of the southern linear filament and not near any other linear structures, it is patchy and extended in its appearance.

The polarized emission from the features shown in figures 2 and 3 is presented in figures 4 and 5. Because of the small size of the primary beam at 6 cm, the scale of the polarized region seen in these figures constitute only a small fraction of a much larger structure brought out by single-dish measurements (chapter 8; Tsuboi et al. 1985; Seiradakis et al. 1985). The region shown in figure 4 is centered on the most intense portion of the polarized structure displayed in figure 2 of chapter 8 (i.e. the eastern edge of lobe B). The polarized region which appears like a comet emerges suddenly at a location northwest of the junction between the system of arched and linear filaments and is coincident with the brightness discontinuity along the linear filaments. This polarized structure consists of at least three components: a fine linear structure to the south, a large and bright comet-like structure separated by a gap from the linear component and a layer of weak and diffuse emission lying to the north of the bright component.

The extension of the southern linear filaments which is identified by a fine linear structure is seen to be linearly polarized. This can best be noted in figure 7 where a linear structure ($\alpha \sim 17^h42^m14^s$, $\delta \sim -28°43'20''$) lies along the southern linear filament. This is the finest scale ($\sim 1.8' \times 10''$) in which polarized emission has been manifested at 6 cm. We also note an enhancement of an extremely weak polarized emission from the northern thread. The directions of the electric field vectors in the northern lobe, shown in figure 8, change rapidly along the southern filament. The extension of the northern linear filament is the most prominent polarized feature in the extended portion of the polarized region (7'×1.5'). The polarized emission from the northern filament and peaks at $\alpha \sim 17^h42^m31^s$, $\delta \sim -28°42'20^s$ and its brightness decreases gradually to the west. This is also the location where the width of the northern filament is maximized and the orientation of the electric vectors differ most (i.e. by ~ 50°) from other portions of this polarized filament. The direction of the electric vectors along the northern filament are typically 35° different from the position angle of the northern filament. The easternmost part of the polarized region resembles a limb-brightened cylinder (see figure 4). This effect is especially noticeable toward the northern edge of this cylinder where the northern filament lies.

A number of depolarizing features can also be recognized in figure 4, some of which appear to be well defined and organized: one is a sharp zig zag-edge structure outlining the southern portion of the cometary structure. The other is a gap seen between the large-scale component and the southern filament. Irregular depolarizing features can be seen throughout the bright portion of the polarized region, the most prominent of which is the shell-like source seen in the total intensity image. This source causes the western extension of the polarized emission from the northern filament to be reduced substantially. In fact, we note an interesting reversal in the appearance of the northern filament in the polarization map (fig. 4). Toward the East, it appears as a fine polarized structure, but at the position of G0.17+0.15, it becomes a fine depolarized filament from that point westward.

The polarized structure to the south of the G0.17+0.15 has a spiky appearance and is associated with the distorted filament seen in figure 1. The orientation of the electric field vectors for the distorted filament as shown in figure 8 are roughly perpendicular to the direction of the filament, however, there is a considerable gradient in the direction of electric vectors from south to north throughout the polarized region of the NW Arc lobe.

The weak layer of diffuse emission situated to the north of the northern filament is also cut off along a line nearly perpendicular to the direction of linear filaments at $\alpha = 17^h42^m33^s$, $\delta = -28°41'$ suggesting that the depolarization medium surrounding the vertical filament is rather broad and has a well-defined edge that is similarly broad.

Measurements of the Faraday rotation, which is illustrated in terms of p.a. of the line segments, indicate that the rotation measure is ~ +1450 rad $m^{-2}$ near the region where the polarized emission is maximized. We note that the rotation measure along the cometary structure decreases smoothly in a direction from east to west by at least an order of magnitude. The maximum rotation measure seen here is in full agreement with single-dish measurements made at 3 cm (Tsuboi <u>et al.</u> 1985).

(B) SE Arc Lobe

The 20-cm image shown in figure 10 displays the southern end of the filamentary Arc and its extension toward negative latitudes. A number of curved structures, the so-called counter-arched filaments (see chapters 3 and 8) appear to cross the southern segment of the Arc at $\alpha \sim 17^h43^m30^s$, $\delta \sim -28°58'$. A very long and straight filament (~ 25') traverses the field as it appears to link a weak compact source at $\alpha \sim 17^h45^m32^s$, $\delta \sim -29°07'$ to the network of filaments system associated with the Arc. This collimated structure has a width of ~ 30"; its surface brightness is weak and uniform except at $\alpha \sim 17^h44^m30$, $\delta \sim -29°$. Comparison of this feature with the extension of the northern filament toward positive latitudes suggests that the two features may be related; both features diverge from the bright, linear portion of the Arc as if they are released at the ends of the vertical portion of the Arc from a restraint that binds together all the filaments. However, the collimated feature seen here - unlike its northwestern counterpart (see figure 1) - does not merge continuously with the easternmost portion of the northern filament and does not show any signs of distortion or bending in its coherent appearance. Indeed, the characteristics associated with this structure are similar to those which were identified with the central radio thread (see chapter 5).

We note that the surface brightnesses of the southern and northern linear filaments decrease gradually as they run toward increasing negative latitudes. This characteristic of the filaments and their widening from each other can also be noted in the 6-cm image of the the immediate vicinity of the southeastern portion of the Arc shown in figure 11.

The polarized emission from the region shown in figure 11 is seen in figure 12, which reveals a non-uniform distribution of linear polarization along the extension of the northern and southern filaments. The northwestern blob of polarized emission corresponds to the core structure in figure 2 of chapter 8 and is fully discussed in chapter 3. The southeastern portion of polarized emission coincides with the western edge of the polarized lobe seen in single dish maps of lobe C (chapter 8). Indeed, the polarization structure seen toward the southeast of figure 12, although patchy in its distribution, can be identified exactly with the extension of two sets of filaments - unlike the NW Arc lobe at positive latitudes - is non-uniform but - like the NW Arc lobe - has a positive rotation measure, i.e. ~ +300 - 3000 rad $m^{-2}$. The direction of the electric vectors and the Faraday rotations in the region to the northwest of figures 13 and 14 agree well with the measurements which were based on observations with a different phase center and which were described in chapter 3.

## IV. DISCUSSION

Much insight about the relative placement of the linear and arched filaments is obtained by studying the brightness distributions of the arched filaments as they cross the elongated shadow feature (see figure 3). We argue that the arched filaments pass both in front of and behind the southern linear filaments, assuming that the shadow feature is associated with the southern filaments and is a coherent structure which runs parallel to the southern linear filaments. This inference is based on the sharp reductions in the surface brightnesses of the eastern and westernmost arched filaments as they cross the southern linear filaments and continue northward. On the other hand, the eastern portion of the system of western arched filaments (W1 in figure 10 of chapter 3) shows an enhancement in its brightness as it crosses the southern linear filament (see figure 7).

Absorption of background emission from the arched filaments by the shadow structure can account for the surface-brightness variation along the thermal arched filaments. Both the temperature and density of the absorption feature can be estimated roughly by assuming that 1) the thermal arched filaments are in pressure equilibrium with the elongated absorption feature and 2) radio recombination line emission from the arched filaments are in LTE [indeed none of these assumptions can be justified properly (see chapter 9)]. Based on the reduction of the surface brightness along the filaments, we find that the optical depth toward the absorption feature has to be $\geq 2$ (see equation 1 in chapter 4). Using the thermal nature of the arched filaments with electron temperature ($T_e$) of ~ 10000 °K and electron density ($n_e$) of ~500 $cm^{-3}$ (see chapter 9), we find that $T_e$ and $n_e$ in the absorption feature with a thickness of 0.5 pc - with much uncertainties - have to be ~ 2000 $cm^{-3}$ and 2500 °K, respectively.

The nature of such a cool and dense thermal plasma collimated as it is precisely along one side of an elongated non-thermal stucture is not understood. However, this geometry indicates that the non-thermal and thermal plasma are interacting with each other in the region where the arched and linear filaments are crossing each other. This suggestion is strengthened by the appearance of the arched filaments as they cross the northern linear filaments (see figure 3). Indeed, the brightness and the linear structure of the northern linear filaments are changed exactly at the location where the western arched filaments merge smoothly with the northern linear filaments ($\alpha \sim 17^h42^m32^s$, $\delta \sim -28°43'$), although some diffuse and weak emission can also be noted to the north of this position. Further pictorial evidence that the arched and linear filaments are interacting is provided by the substantial decrease in the surface brightness of the southern and northern linear filaments as they are continued north westward where the polarized emission emerges.

The collimated absorption feature which accompanies the southern linear filament possibly accounts for the gap seen between the large-scale comet-like component of the polarized emission and the southern linear filament (see figure 4). This coherent absorbing feature could then cause depolarization of non-thermal emission from the region which separates the southern filament from the rest of the polarization complex. Indeed, if fluctuations in the magnetic field amount to $> 10^{-6}$ Gauss over the half-parsec of the filaments, then with a number density of $\sim 2000$ $cm^{-3}$ along the absorbing feature, emission from the linear filaments would be readily polarized (Burn 1966).

We note that most of the depolarizing features seen in the NW Arc lobe are sharp and collimated; the best examples of such structures are the absorption features seen along extensions of the northern filaments and the sharp boundary seen to the south of the comet-like polarization complex (see figure 4). These all suggest that the depolarizing features are located in the outskirts of the Arc.

It is plausible that magnetic field plays a strong role in shaping thermal plasma in this region. If so, the strength of magnetic field (B) has to be $> 1.4\times 10^{-4}$ Gauss using $B^2/8\pi > nkT$ where k is the Boltzmann constant. Indeed, such a large magnetic field strength is needed to keep the northern linear filament (see figure 2) relatively straight as it interacts with the arched filaments at $\alpha \sim 17^h42^m30^s$, $\delta \sim -28°43'$.

Another aspect of the depolarizing features can be seen at the eastern edge of the diffuse layer of polarized emission (i.e. figure 4). The cut off seen at this boundary reveals the cool and diffuse portion of thermal plasma associated with the arched filaments (see lower right panel of figure 7). Indeed, the polarized emission associated with this layer might originate from lateral diffusion of relativistic particles away from the northern filament. This suggestion is based on the intensity gradient seen along the layer of polarized emission from the region to the north of the northern filaments (see figure 4). The time it takes for relativistic electrons to travel across this diffuse layer is estimated to be $\sim$100 years. Although this time scale is much less than the synchrotron lifetime at 6 cm with $B \sim 10^{-4}$, it signifies that the complex associated with the Arc is very young and is a rapidly evolving structure.

Comparisons of the northwest and southeast extensions of the Arc show a number of structural details which appear somewhat differently in the region above (b > 0) and below the galactic plane (b < 0).

i) For one thing, the northwestern extensions of the filaments are continuous but get distorted clearly at two different locations: one is the region where the northern filament bends slightly as it crosses the arched filaments and the other is where the southern and central filaments change their curvature as they graze over the oval-shaped source. However, the striking long feature seen along the southeastern extension of the filaments is not continuous with respect to the filaments and appears to be extremely linear.

ii) The region above the galactic plane shows at least two thread-like structures which do not seem to be associated with the Arc. But no such isolated thread-like features, which appear to have magnetic structures, are recognized below the galactic plane.

iii) At the junction of the arched and linear filaments (b > 0), a number of sharp and elongated absorbing features were recognized. Furthermore, we argued that a physical interaction between thermal and non-thermal filaments are plausible. The counter-arched filaments cross the linear filaments below the galactic plane but their appearance suggests no interaction. Figure 15, which shows the distribution of thermal emission from dust (Odenwald and Fazio 1985), supports the above picture where much of the thermal structures are presumably interacting with the nonthermal components - and where the interacting features are located toward positive latitudes.

iv) Recent studies by Sofue _et al._ (1984) show a large-scale Ω-shaped

structure above the galactic plane. A physical link between the NW Arc lobe and this large-scale structure is not yet established, although very suggestive. A counterpart to this large-scale feature has not been seen at negative latitudes (see chapter 8).

v) Low-frequency observations at 120 MHz show a steep-spectrum structure lying roughly along the extension of the arched filaments in the direction away from the galactic center but above the galactic plane (LaRosa and Kassim 1985).

It should be pointed out that the VLA images of the region to the northwest of the Arc and the Sgr A complex have a better sensitivity and higher spatial resolution than those of the region at negative latitudes, so the above VLA comparisons might be observationally biased. Further high-resolution observations of the southwestern extension of the Arc could be very fruitful in clarifying the nature of such spatial asymmetry. It is also important to find contributions of thermal emission in both regions in order to examine if the large-scale magnetic field lines (see chapter 5; Uchida et al. 1985) in the region above the galactic plane are manifested because of the high density of thermal gas which is being accelerated along the field lines.

A recent theoretical study by Uchida, Shibata and Sofue (1985) suggests that the radio lobes could have originated from a distortion of poloidal magnetic field lines by a coupling between the rotation and contraction of orbiting gas. The fact that the radio lobes are associated physically with the filamentary Arc contradicts the above modelling. This is because the linear filaments are well organized, straight and show no sign of radially or azimuthally-directed field lines at the galactic plane where the model predicts otherwise. In fact, the linear filaments are distorted only slightly as they cross the galactic plane (see figure 9 of chapter 3), and in the direction away from the galactic plane. Furthermore, the geometry of the arched and counter-arched filaments are contrary to those predicted by the model of Uchida et al. (1985). It should be pointed out, however, that the geometry of magnetic field that Uchida et al. assume would explain numerous features observed near the galactic center (see chapter 6).

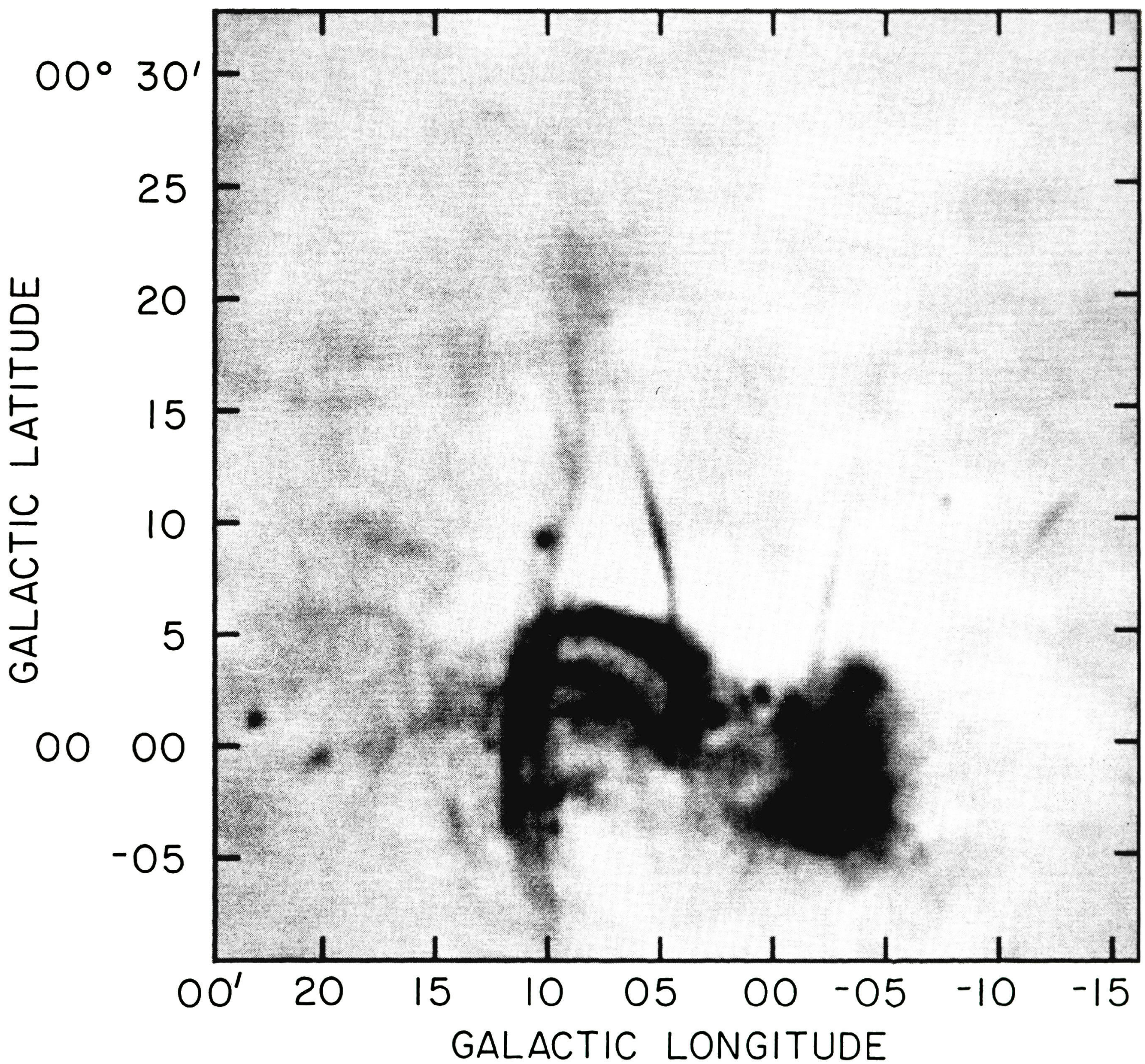

Figure 1: The designated field corresponding to this figure is Arc No. 10 (see Table 1 in chapter 2). This map which is based on a self-calibrated data set has a gaussian CLEAN beam (FWHM) of 30"×30".

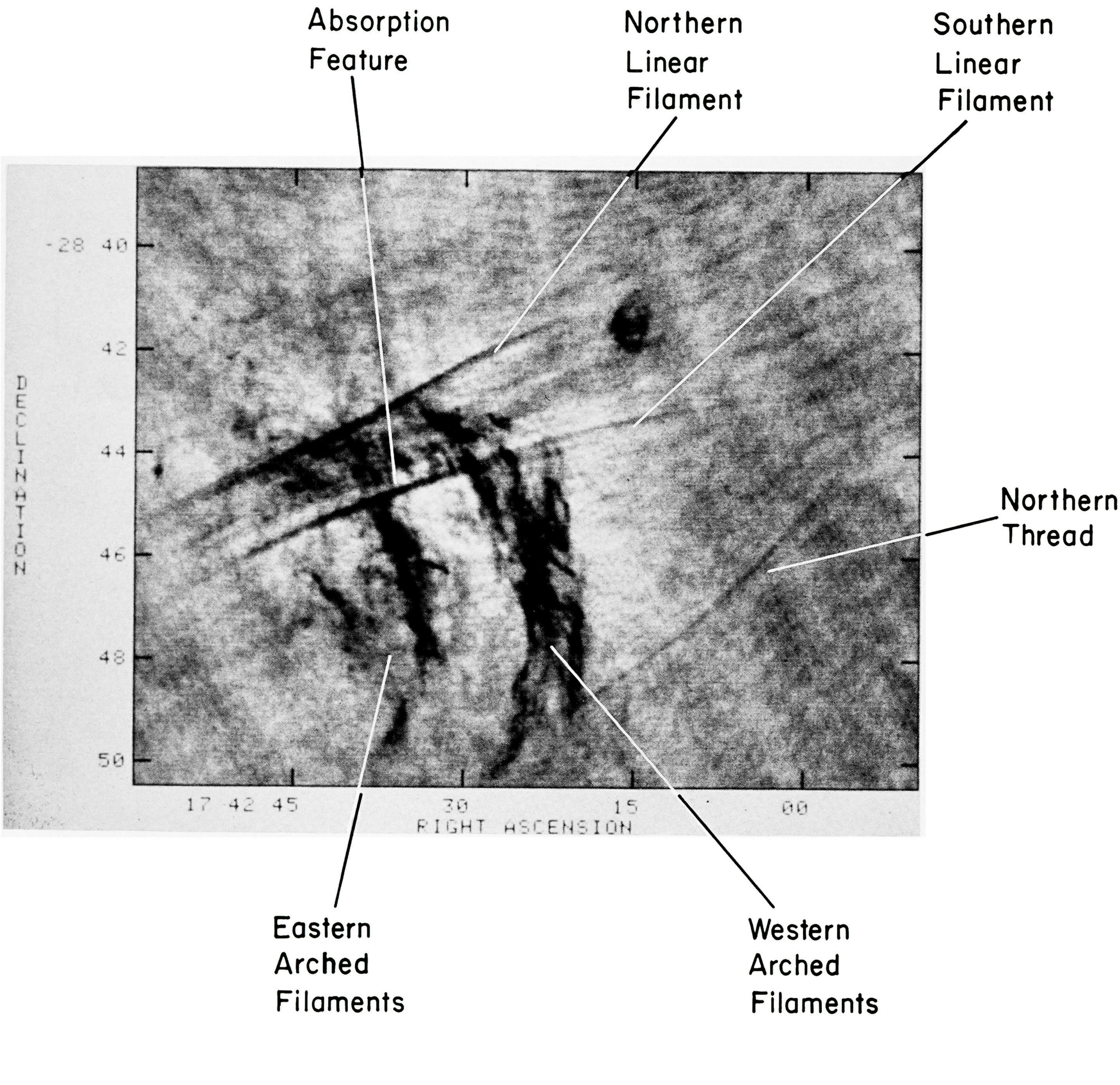

Figure 2 and 3: The designated field is Arc No. 6. FWHM = 3.65"×2.8" (P.A. = −90°). The noise level is 0.14 mJy/beam area. λ = 6cm.

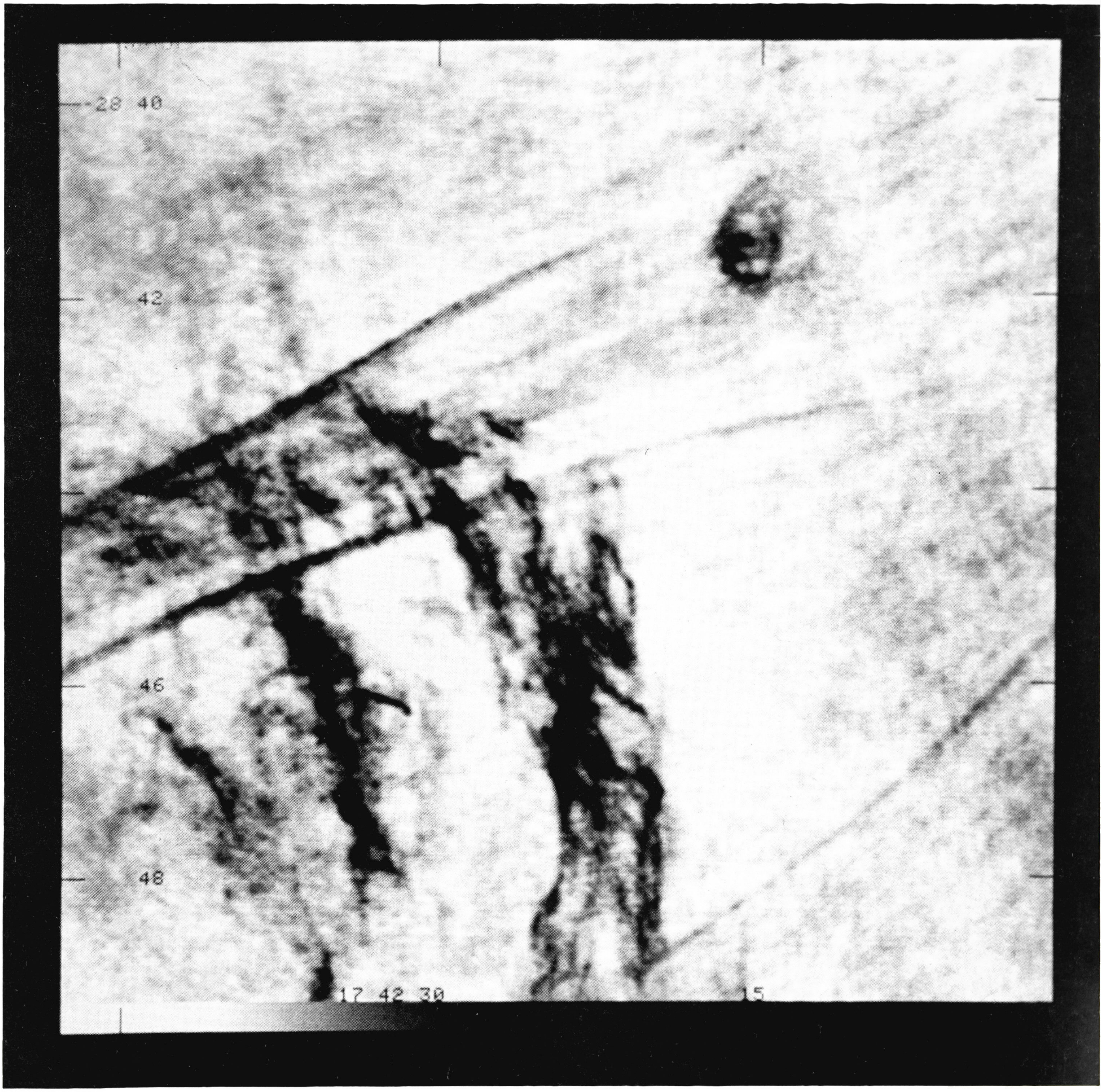
-28 40
42
46
48
17 42 30
15

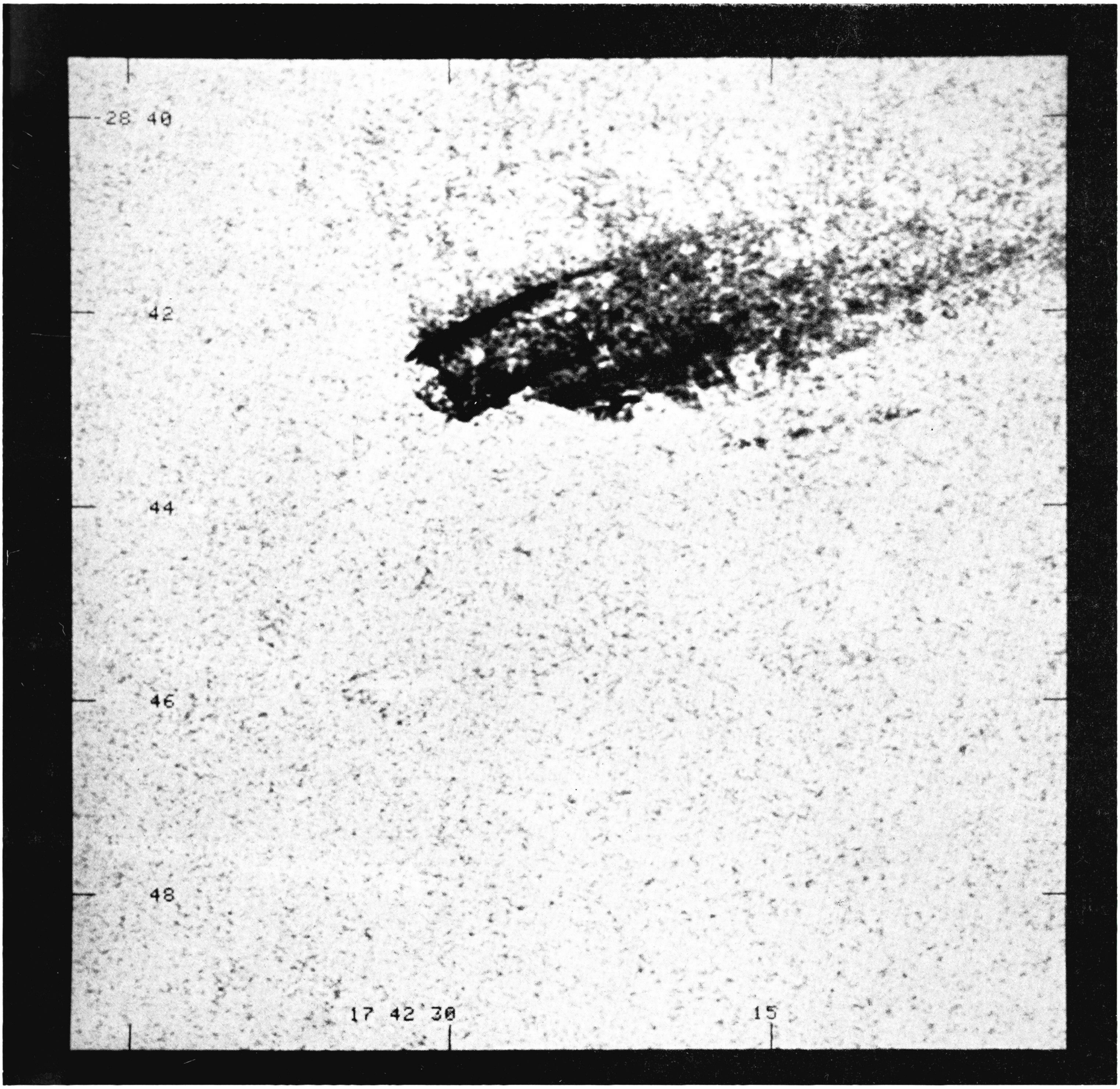

Figure 4: The polarized intensity of the region shown in this figure has identical scale and spatial resolution as that of figure 3. The noise level is 43 μJy/beam area.

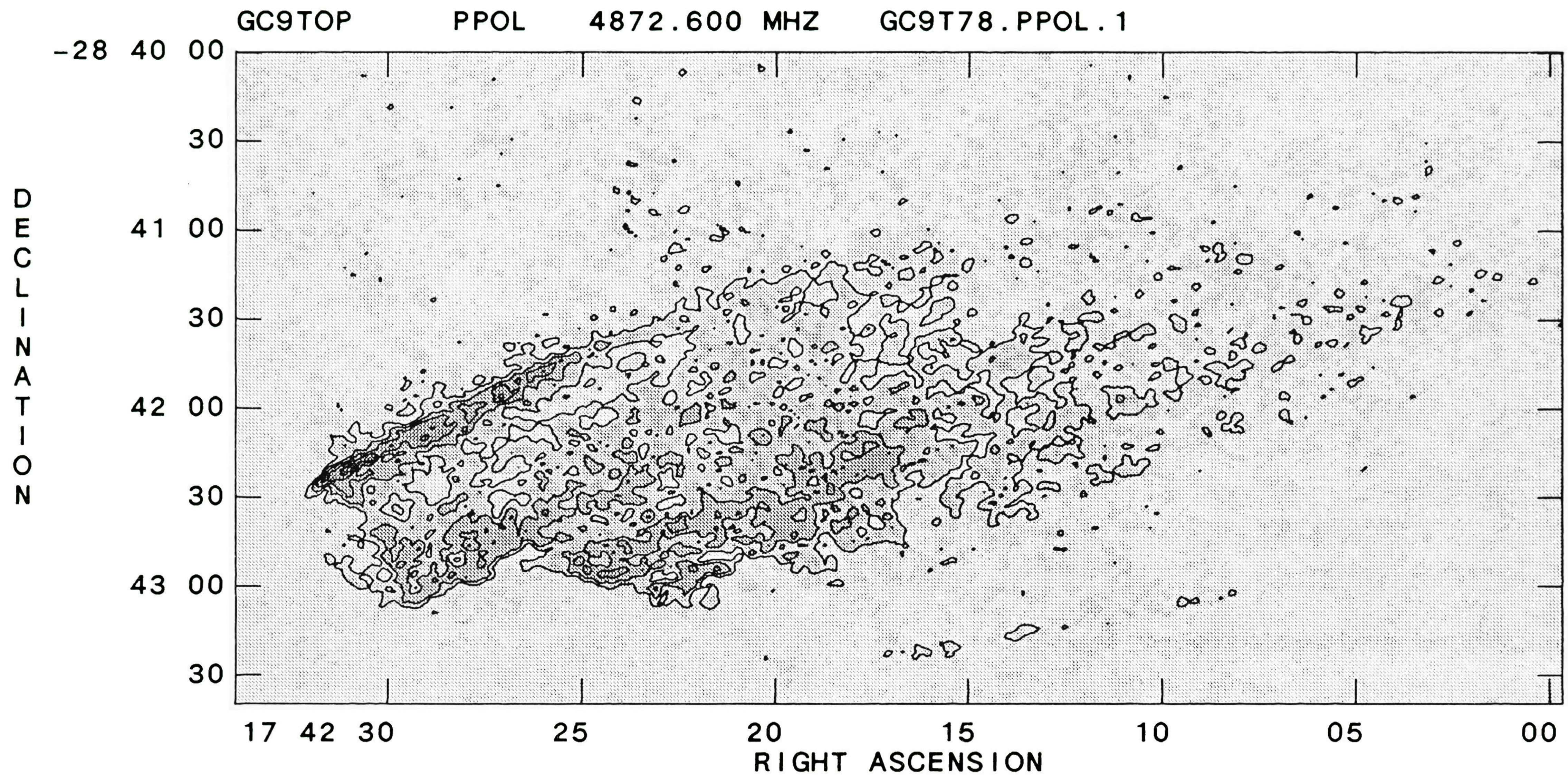

Figure 5: Contours of the polarized intensity are shown with intervals of 0.3, 0.5, 0.7, 0.9, 1.1, 1.5, 2.3, 3, 3.7, 4.5, 5.3 mJy/beam area. This map is based on the same data set as that of figure 2 and is not corrected for the response of the primary beam.

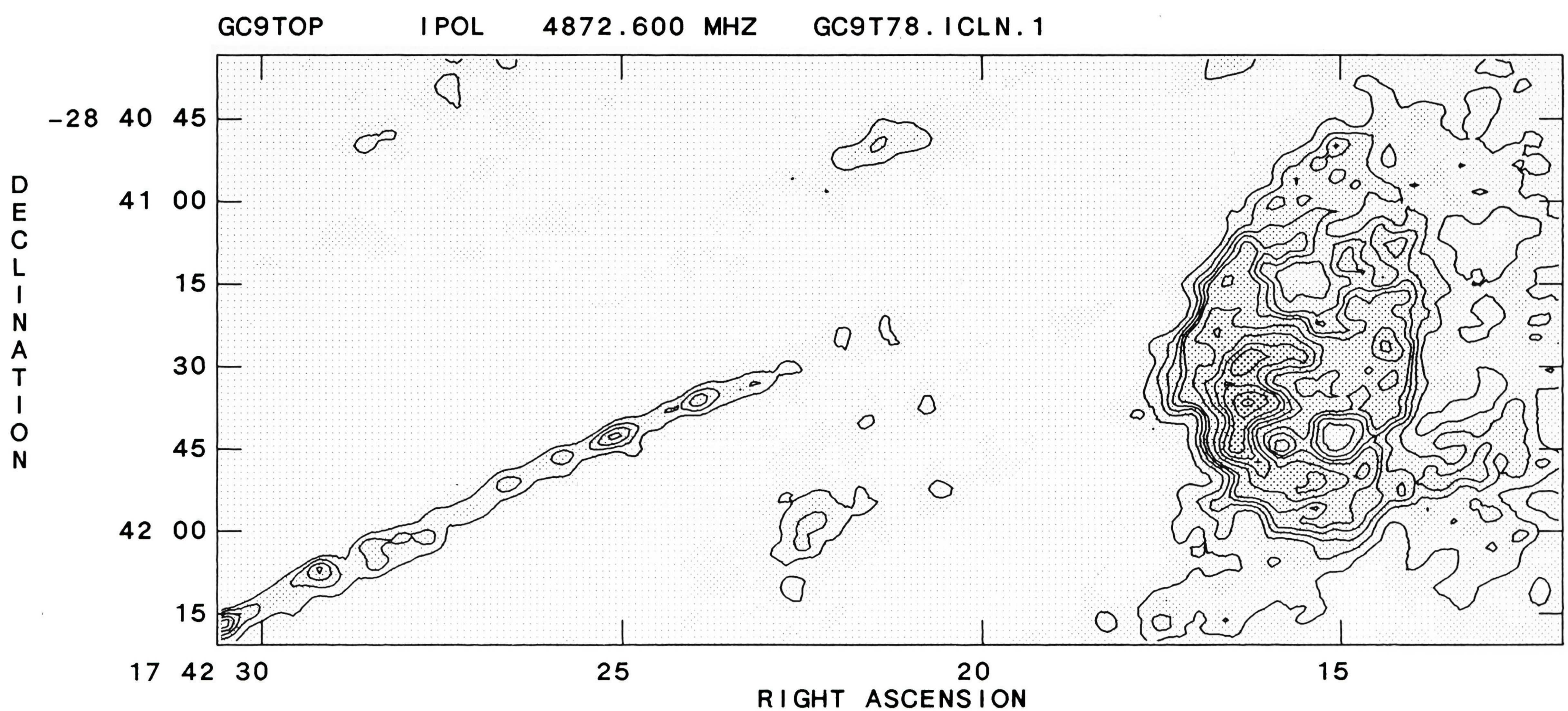

Figure 6: Contours of the total intensity with intervals of 0.2, 0.4, 0.6, 0.8, 1, 1.5, 2, 2.5, 3, 4, 5, 6 are shown in this figure (see figure 2 for more details).

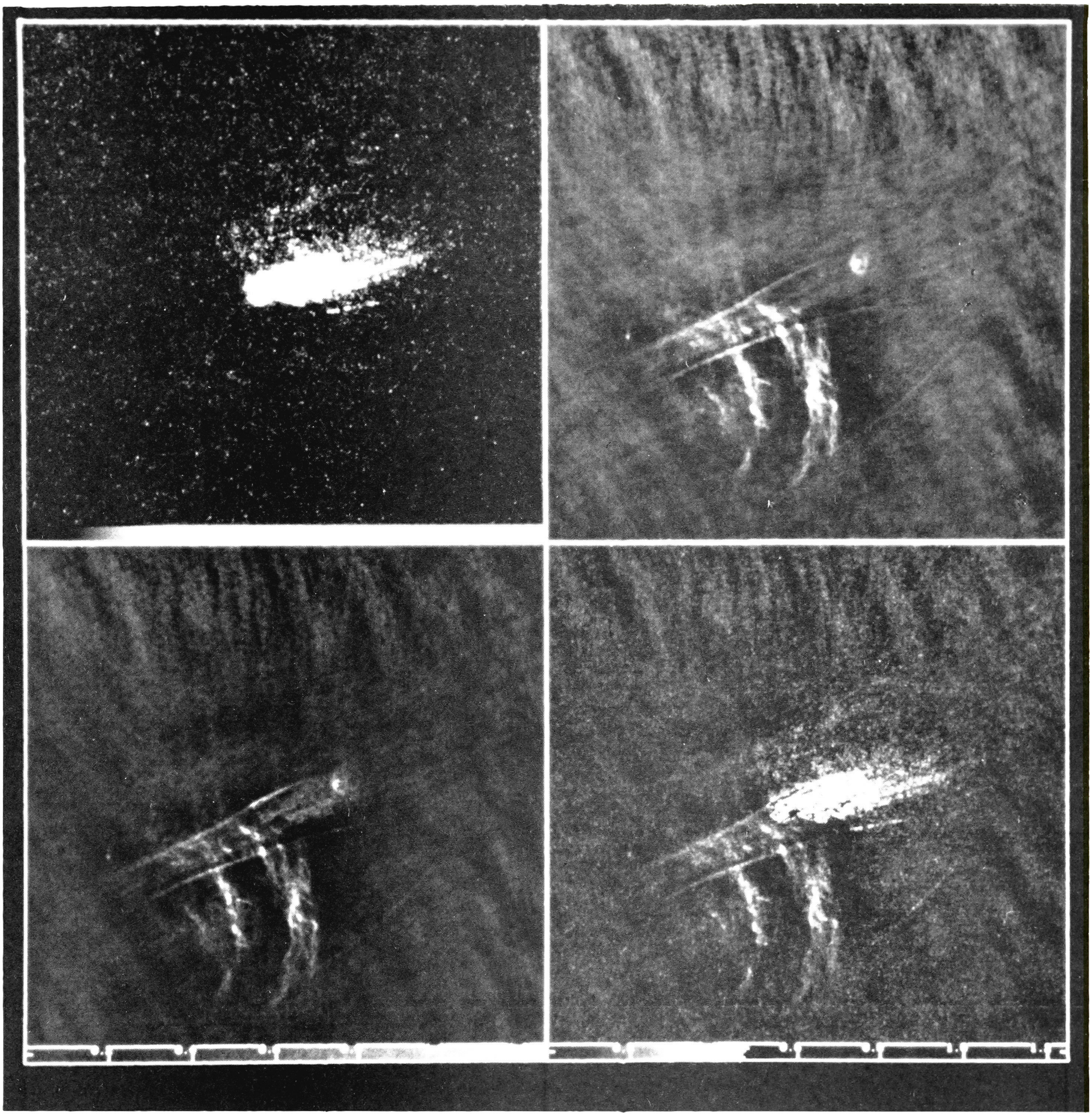

Figure 7: Juxtaposition of the polarization and total intensity images are shown in the top two boxes. The bottom two boxes show the polarization image - with two different contrast levels - superimposed onto the total intensity image (see figures 2 and 4 for more details).

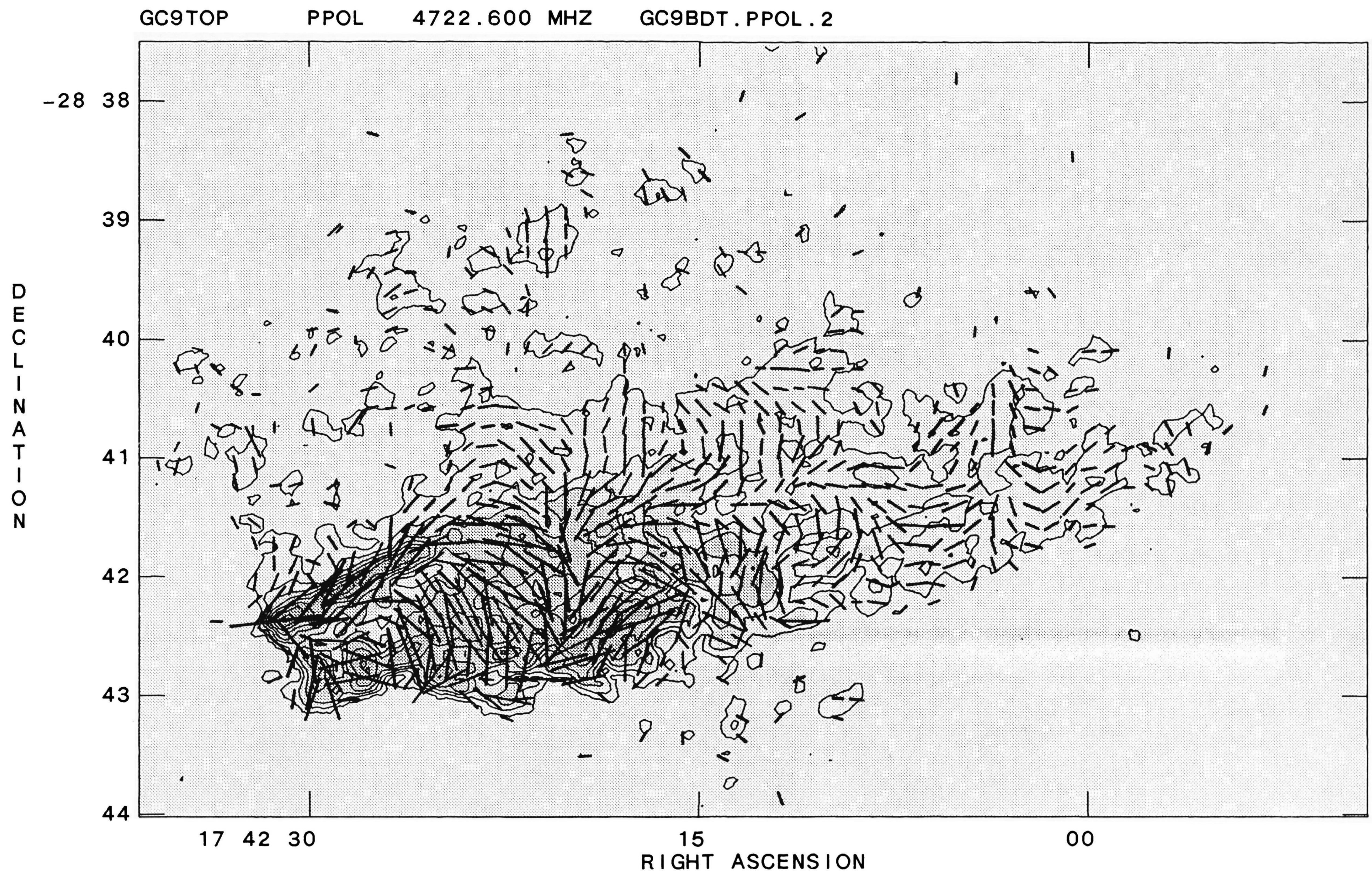

Figure 8: Contours of the polarized intensity are shown with intervals of 0.5, 1, 1.5, 2, 2.5, ... 5 mJy/beam area.

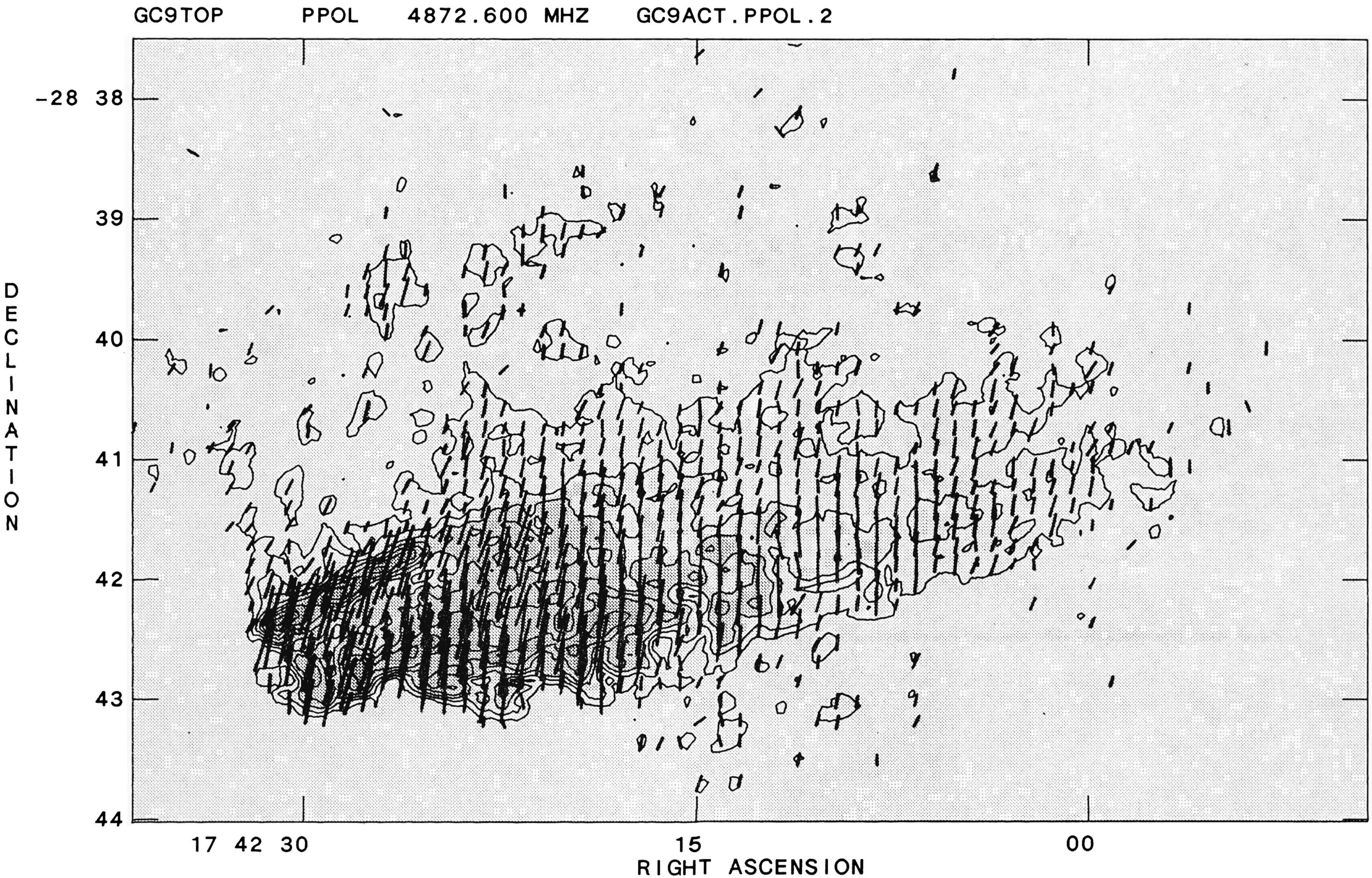

Figure 9: The same as figure 8 except that the direction of the electric vectors are a measure of Faraday rotation (see text).

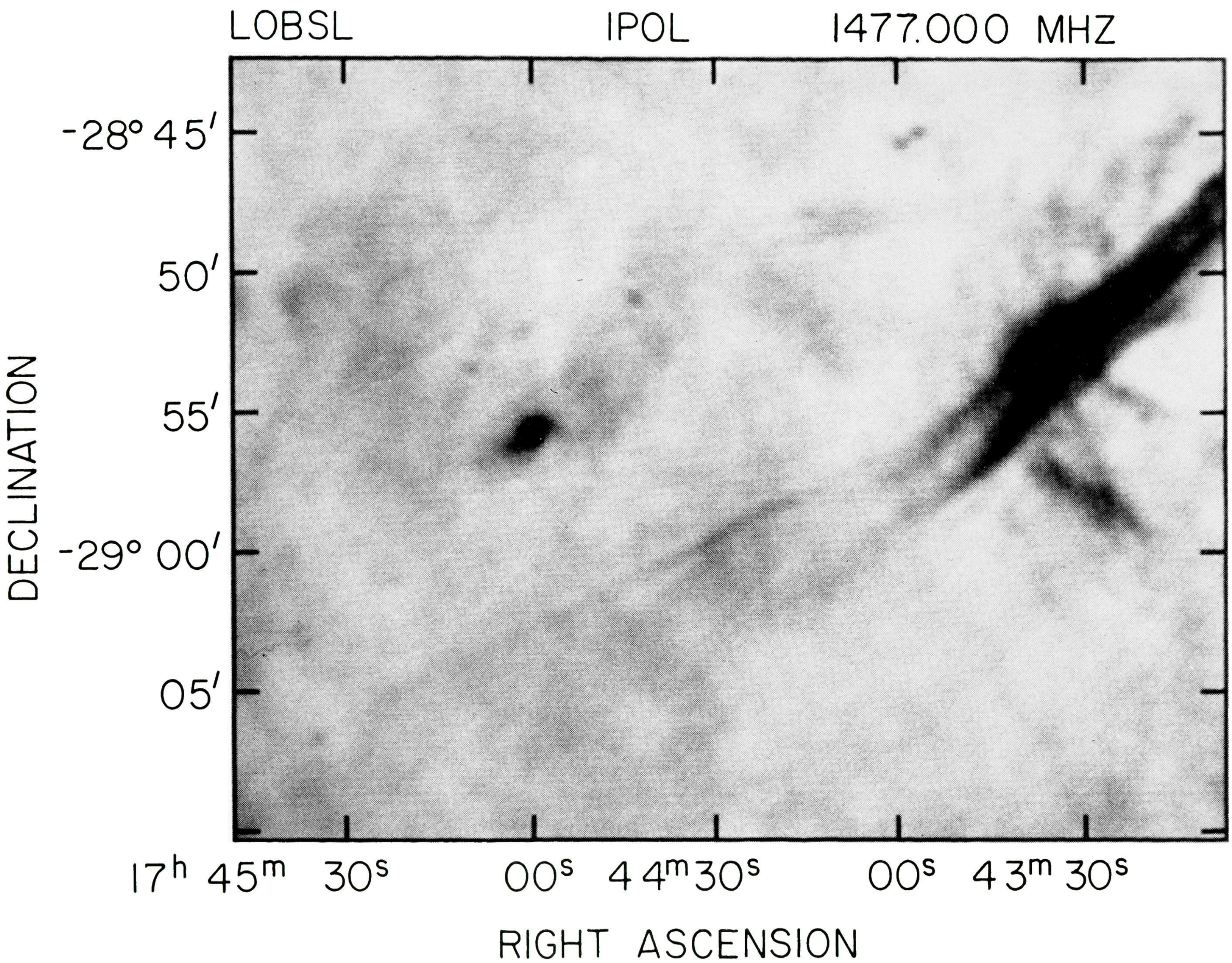

Figure 10: The designated field corresponding to this figure is Arc No. 11. The CLEAN beam = 30.2" x 27.2" (P.A. = 89°). The peak residual flux is 10.9 mJy/beam area.

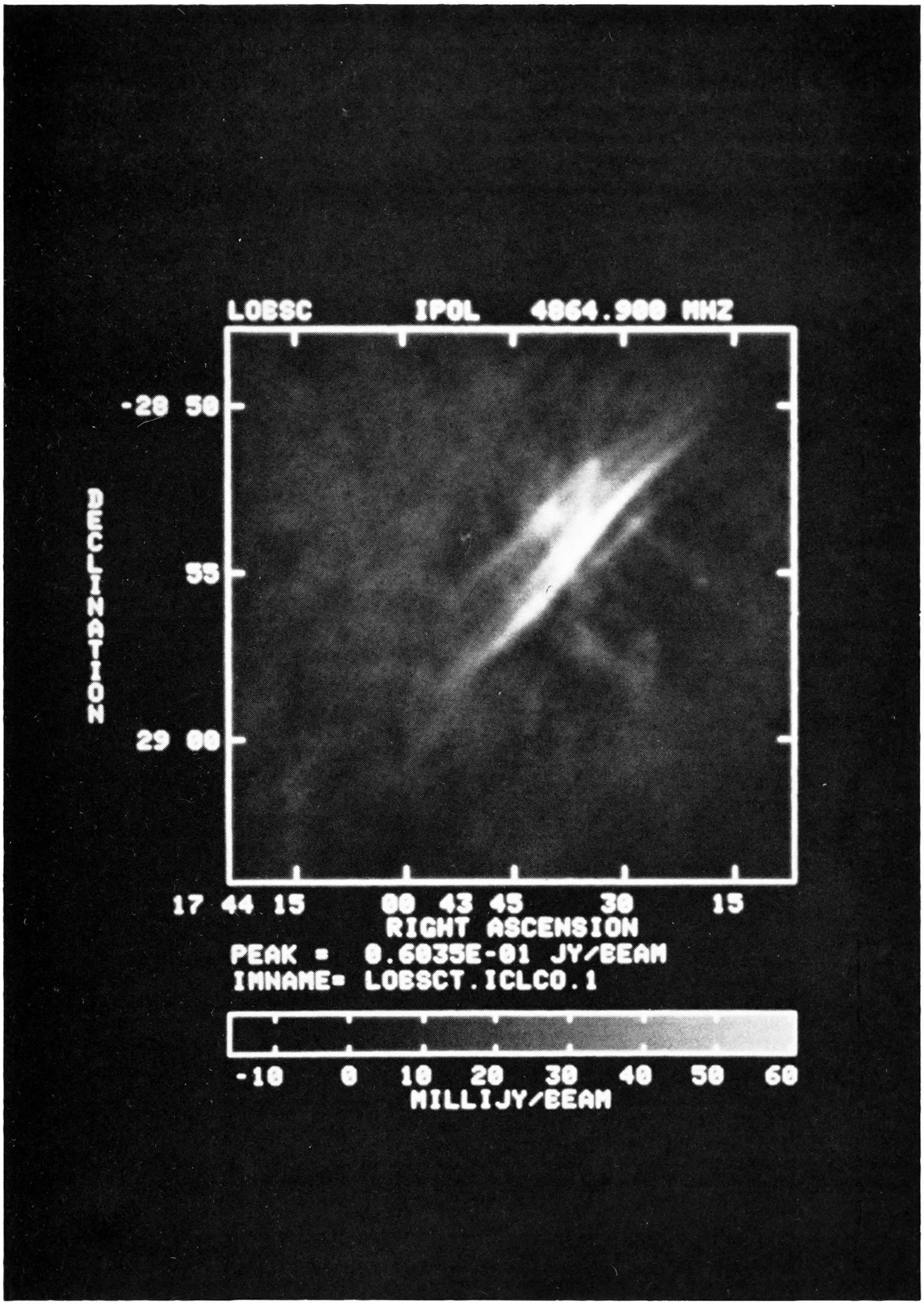

Figure 11 and 12: The polarization and total intensity images of the same region are shown in these two figures. These two maps are convolved with a gaussian beam of 15"×15". The designated field for these figures is Arc No. 8.

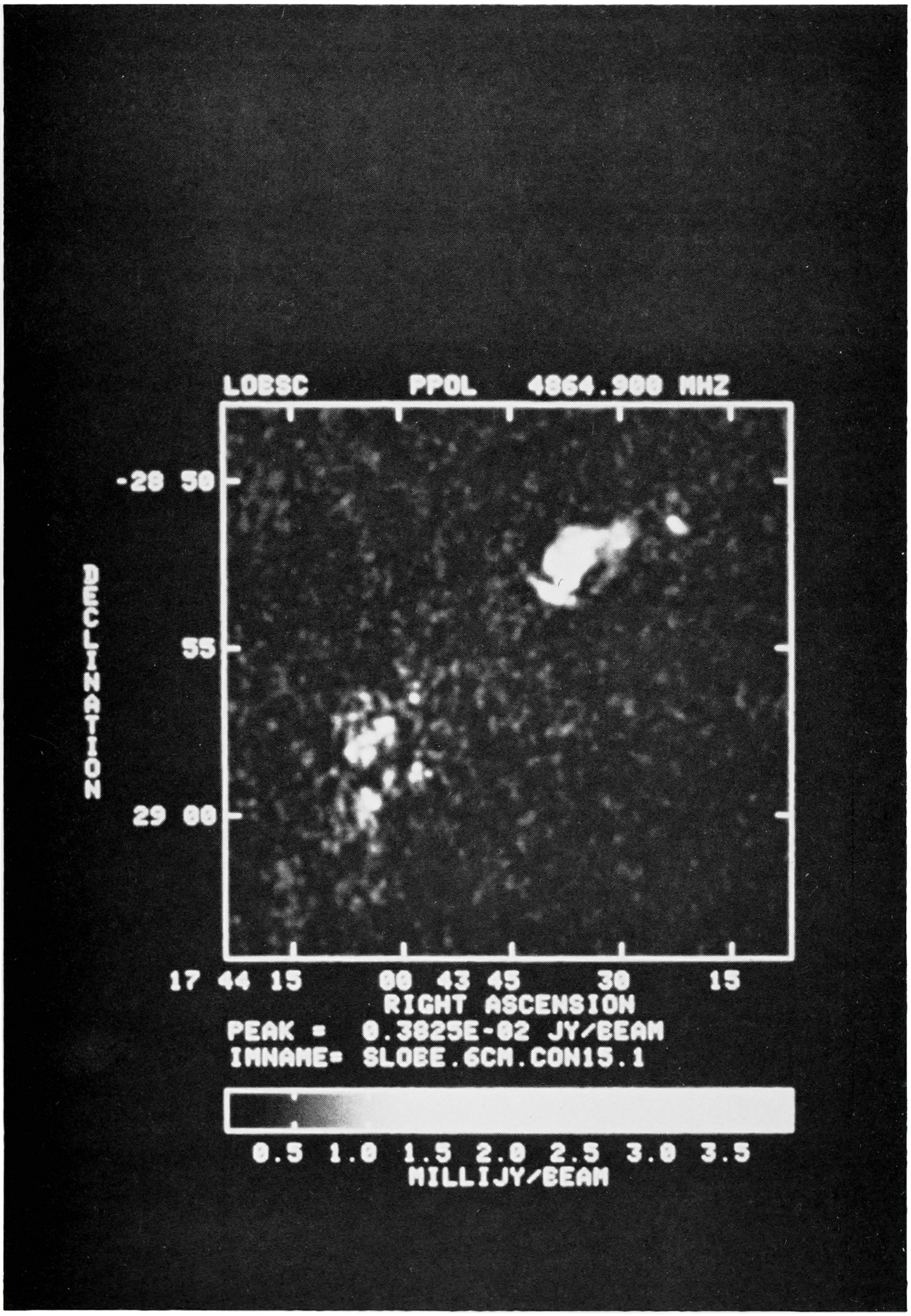
LOBSC PPOL 4864.900 MHZ
-28 50
55
29 00
DECLINATION
17 44 15 00 43 45 30 15
RIGHT ASCENSION
PEAK = 0.3825E-02 JY/BEAM
IMNAME= SLOBE.6CM.CON15.1
0.5 1.0 1.5 2.0 2.5 3.0 3.5
MILLIJY/BEAM

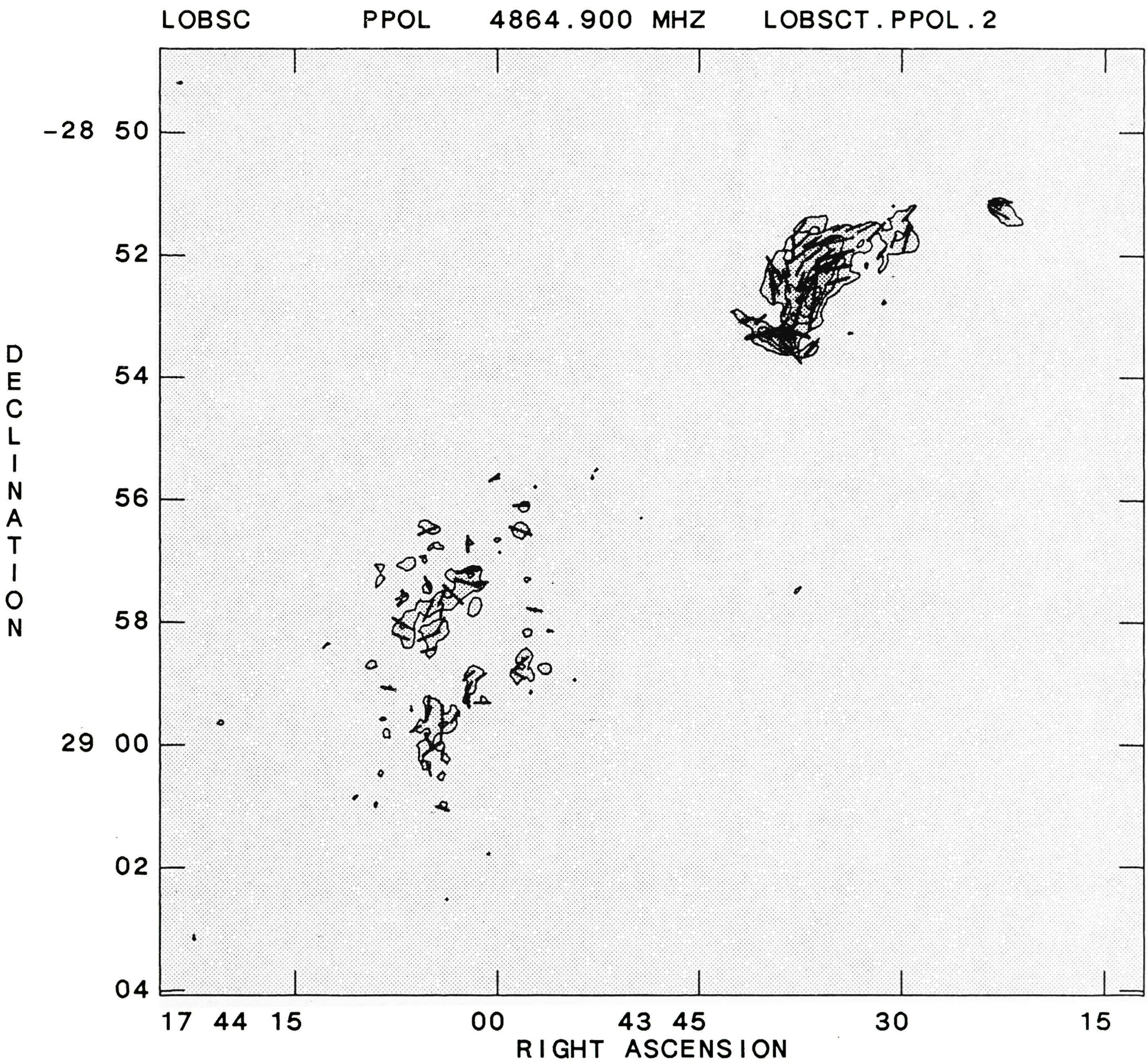

Figure 13: Contours of the polarized intensity are shown with intervals of 0.5, 1, 1.5, ..., 5 mJy/beam area. The line segments show distribution of electric field vectors. FWHM = 11.8" x 10.7" (P.A. = -75°).

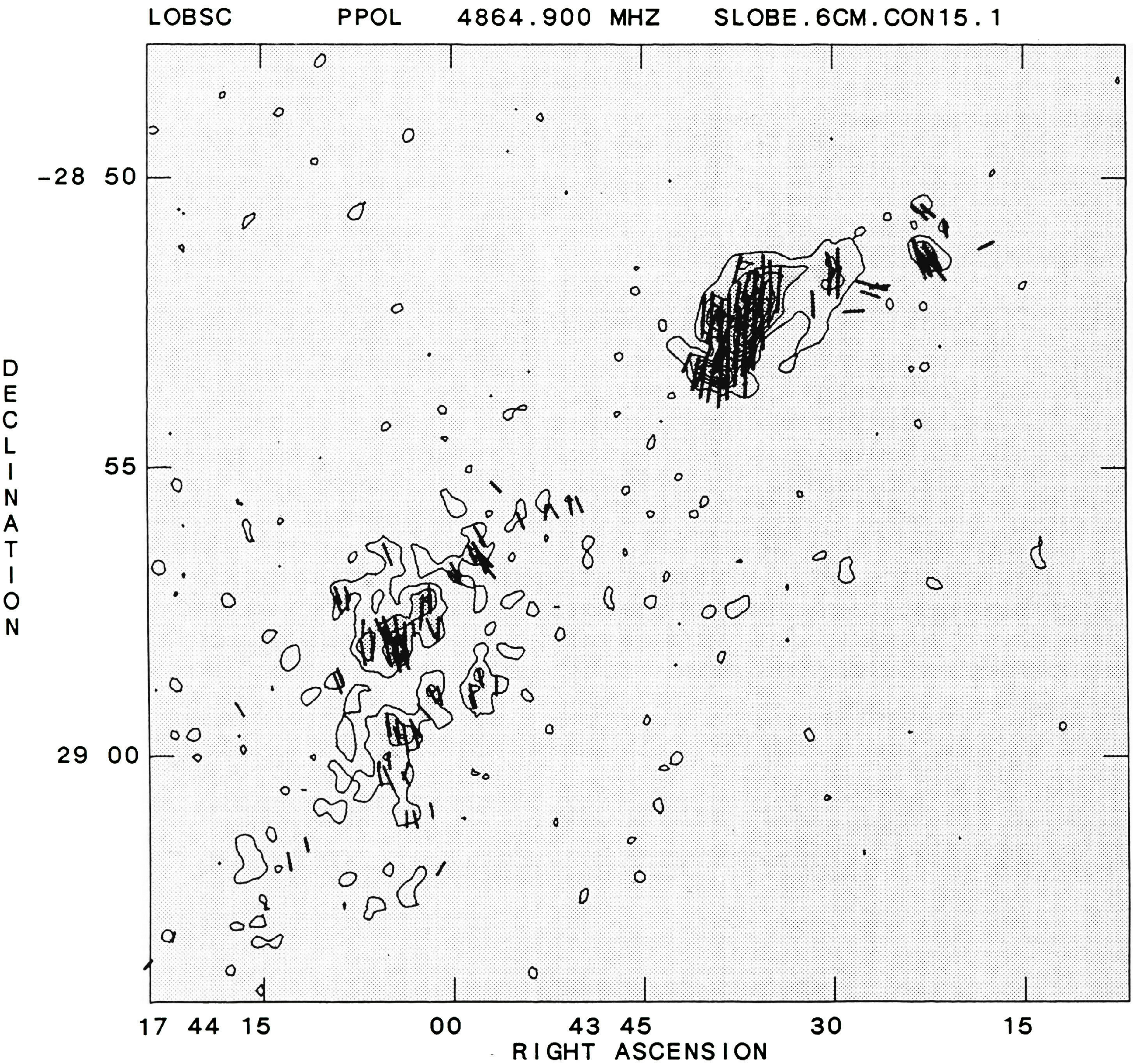

Figure 14: The same contour levels as those of figure 13 are shown in this figure. The resolution is 15" x 15". The line segments are a measure of Faraday rotation (see text).

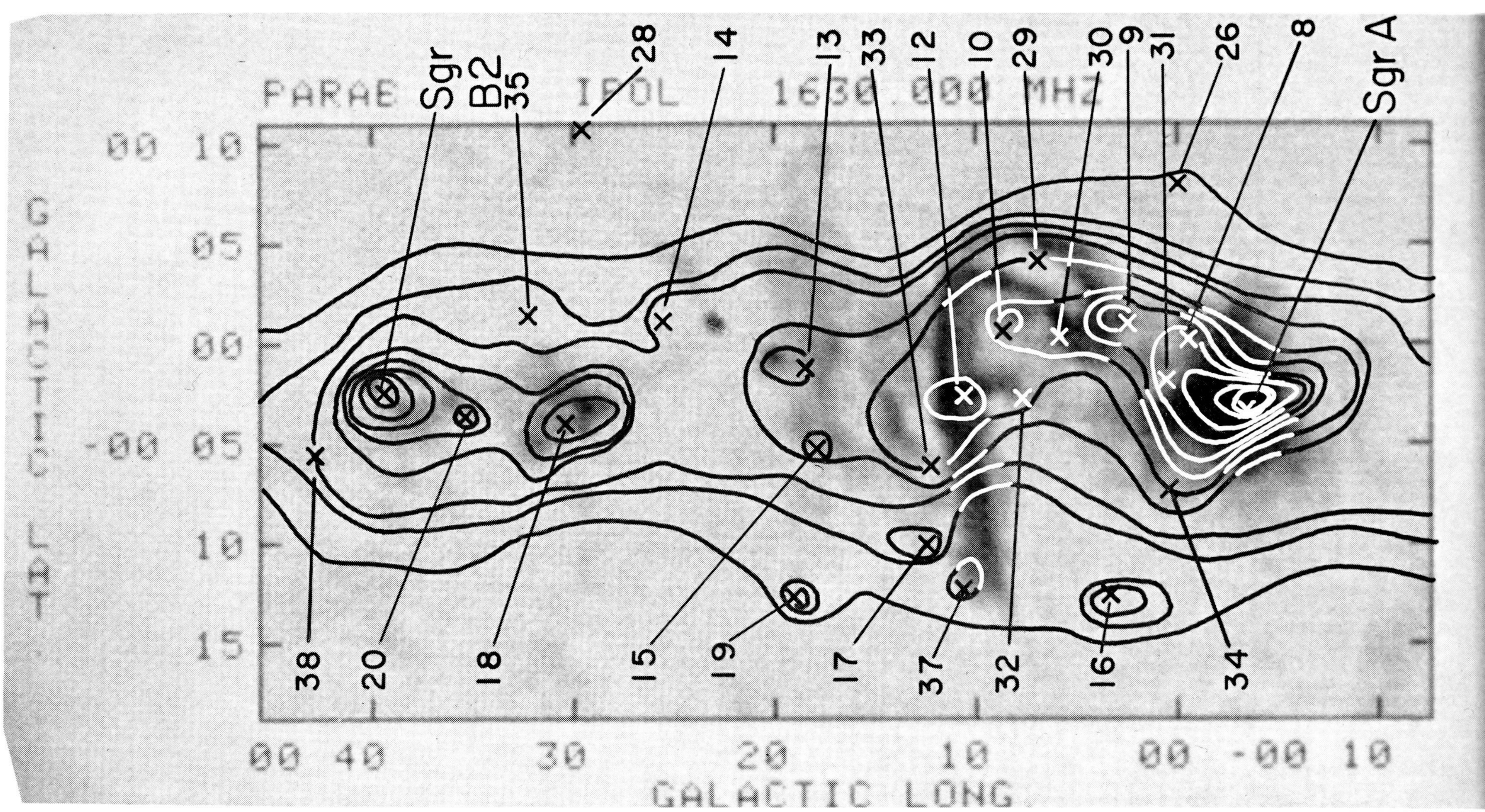

Figure 15: Contours of the far-infrared emission between 40-250 $\mu$m are superimposed on the 20-cm map of the inner 1°×0.5° of the galactic center. The infrared map is based on observations made by Odenwald and Fazio (1985).

# Epilogue

"This is Ze end," she said
"No, this is the end," he joked
pointing at the tail

A scene in Last Tango in Paris

It has been about two and a half years since VLA observations showed that the Arc is a coherent structure and that it has a filamentary nature. Working on the unique environment of the galactic center has been tremendously overwhelming but has also been frustrating at times. This is because the new features were brought out slowly and piecemeal throughout this period and thus the ideas on the nature of these features - if any - had to be adjusted, shifted, shuffled or thrown out, accordingly. I say this because I feel that a more coherent understanding will be made in the near future unless researchers in the field assume that all the facts are out and thus there is no point of observing further.

As William Saroyan said on his deathbed,
"What next?"

Russel Baker

Future observations, some of which are under progress, are listed next and it is hoped that the body of useful information continued to grow as it has.

- 6 and 2-cm observations of the threads and long filaments in the southern lobe should indicate the direction of the magnetic field lines in these enigmatic structures. They are least affected by thermal plasma and it is hoped that their spectral indeces should reveal something about their evolution with respect to time.

- High-resolution observations of threads extended over a few years would allow us to determine if the long threads are moving relativistically. It is speculated that either the threads look similar to the so-called relativatistically moving "cosmic strings" or the "cosmic strings" are expected to look similar to radio threads [Ekers and Field (private communication), Paczynsky 1986]. It is expected that this observation would provide a constraint on future theoretical modelling of this object.

- High-resolution recombination line observations of the arched filaments and their surrounding regions should give useful information on the kinematics of the ionized flow.

- A complete high-resolution molecular line observations of the Arc should be made in order to establish the relationship between the +50 and -30 km $s^{-1}$ molecular clouds. High-resolution infrared and molecular observations should reveal whether the distribution of the ionized gas in this region is anticorrelated or correlated with those of the dust and molecules.

- A high degree of linear polarization from the Arc and its associates should be sufficient to give rise to circular polarization measurements which are not affected by very high Faraday rotations that have been seen toward the Arc

- High-resolution observations of the $\Omega$-shaped feature are needed to establish directly if this feature is physically associated with the Arc.

- Radio recombination line observations of the counter-arched filaments and the Arch should be very useful in separating the thermal from non-thermal components. Indeed, the nature of the counter-arched filaments is not known.

- Extensive surveys of radio continuum observations of the inner region of the Galaxy and the galactic plane using the VLA would be useful to verify if the features that are observed toward the galactic center are unique.

"When I wrote this poem my God and I
knew what it meant
now, it is only God
who knows what it means."

Anonymous

References

Allen, D.A., Hyland, A.R. and Jones, T.J., 1983, M.N.R.A.S., **204**, 1145.

Altenhoff, W.J., Downes, D., Goad, L., Maxwell, A., Rinehart, R. 1970, Astr. Astrophys. (Suppl.) **1**, 319.

Altenhoff, W.J., Downes, D., Pauls, T., and Schraml, J. 1979, Astr. Astrophy. Supp. **35**, 23.

Alvarez, J.A., Furniss, I., Jennings, R.E., King, K.J. and Moorwood, A.F.M. 1974, in HII Regions and the Galactic Center, ed. A.F. Moorwood (Proceedings of the Eighth ESCAB Symposium, ESTEC, Noordwijk, Netherlands), p. 69.

Audouze, J. 1978, in IAU Coll. No. 45, Chemical and Dynamical Evolution of our Galaxy, ed. E. Basinska-Grzesik and M. Mayor, p. 79.

Audouze, J., Lequeux, J., Mansou, J.-L. and Puget, J.-L. 1979, Astron. & Astrophys., **80**, 276.

Balick, B. and Brown, R.L. 1974, Ap.J. **194**, 265.

Balick, B. and Sanders, R.H. 1974, Ap.J. **92**, 325.

Bally, J., Stark, T., Wilson, R., Henkel, C. 1986, in preparation.

Bania, T.M. 1977, Ap.J., **216**, 381.

Becklin, E.E., Gatley, I. and Werner, I. 1982, Ap.J., **258**, 135.

Becklin, E.E., Mathews, K., Neugebauer, G. and Willner, S.P. 1978, Ap.J., **220**, 831.

Becklin, E.E. and Neugebauer, G. 1975, Ap.J. (Letters) **200**, L71.

Bernstein, I.B. and Kulsrud, R.M. 1965, Ap.J., **142**, 479.

Bieging, J., Downes, D., Wilson, T.L., Martin, A.H.M., Güsten, R. 1980, Astr. Ap. Suppl., **42**, 163.

Bignell, R.C. 1982, lecture No. 6 in Proceedings of the NRAO-VLA Workshop, eds. A.R. Thompson and L.R. D'Addario.

Biraud, F., Lequeux, J. and LeRoux, E. 1960, Observatory **80**, 116.

Blitz, L., Bloeman, J.B.G.M., Hermsen, W. and Bania, T.M. 1985, Astron. Astrophys., **143**, 267.

Brezgunov, V.N., Dagkensmansky, R.D., and Udal'tsov, V.A. 1971, Ap.J. (Letters) **9**, 117.

Bridle, A.H., Perly, R.A. 1984, Ann. Rev. Astron. Astrophys., **22**, 319.

Broten, N.W., Cooper, B.F.C., Gardner, F.F., Minnet, H.C. et al. 1965, Australia J. Phys. **18**, 85.

Brown, R.L. and Johnston, K.J. 1983, Ap.J. (Letters) **268**, 185.

Brown, R.L. and Lo, K.Y. 1982, Ap.J., **253**, 108.

Brown, R.L., Lo, K.Y., Johnston, K.J. 1978, Astron.J. **83**, 1594.

Brown, R.L. and Liszt, H.S. 1984, Ann. Rev. Astr. Ap., **22**, 223.

Burn, B.J. 1966, M.N.R.A.S., **133**, 67.

Capps, R.N. and Knacke, R.F. 1976, Ap.J., **210**, 76.

Castor, J., McCray, R. and Weaver, R. 1975, Ap.J. (Letters), **200**, L107.

Chandrasekhar, S. 1956, Proc. Nat. Acad. Sci., **42**, 1.

Chandrasekhar, S. and Kendall, P.C. 1957, Ap.J., **126**, 257.

Clark, B.G. 1980, Astron. Astrophys., **89**, 377.

Cohen, R.J. and Few, R.W. 1976, MNRAS, **176**, 495.

Cohen, R.J. and Dent, W.R.F., 1983, in Surveys of the Southern Galaxy, eds. W.B. Burton and F.P. Israel.

Cooper, B.F.C. and Price, R.M. 1964, The Galaxy and the Magellanic Clouds, eds. F.J. Kerr and A.W. Rodgers, p. 168.

Cornwell, T.J. and Evans, K.F. 1985, Astron. Astrophys., **143**, 77.

Davis, L. and Greenstein, J. 1951, Ap.J., **114**, 206.

de Bruyn, A.G. 1978, in Structure and Properties of Nearby Galaxies, eds. E.M. Berkhuijsen and R. Wielebinski, p. 205.

Dent, W.A., et al. 1982, AIP Conference Proceeding, No. 83, "The Galactic Center".

Downes, D., Goss, W.M., Schwarz, U.J., and Wooterloot, J.G.A. 1978, Ast. & Ap. Suppl., **35**, 1.

Downes, D. 1974, Proceeding of 8th ESLAB Symp., "HII Reg. Galactic Center", p. 247.

Downes, D. and Martin, A.H.M. 1971, Nature, **233**, 112.

Downes, D. and Maxwell, A. 1966, Ap.J., **146**, 653.

Downes, D., Maxwell, A., and Meeks. M.L. 1965, Nature, **208**, 1189.

Downes, D., Wilson, T.L., Biegin, J., Wink, J. 1980, Astron. Astrophys. Suppl., **40**, 379.

Drake, F. 1959, A.J., **64**, 329.

Dulk, G.A. 1970, Astrophys. Lett., **7**, 137.

Dulk, G.A. and Slee, O.B. 1974, Nature, **248**, 33.

Duric, N., Seaquist, E.R., Crane, P.C., Bignel, R.C., and Davis, L.E. 1983, Ap.J. (Letters), **273**, L11.

Ekers, R.D., Goss, W.M., Schwarz, U.J., Downes, D., and Rogstad, D.H. 1975, Astron. Astrophys., **43**, 159.

Ekers, R.D. and Lynden-Bell, D. 1971, Ap.J. (Lett.), **9**, 189.

Ekers, R.D., van Gorkom, J.H., Schwarz, U.J., Goss, W.M. 1983, Astron. Astrophys., **122**, 143.

Elmegreen, B.G. 1981, in The Formation of Planetary Systems, ed. A. Brahic (Touleuse: Capadues Editions).

Elmegreen, B.G., Genzel, R., Moran, J.M., Reid, M.J., and Walker, R.C. 1980, Ap.J., **241**, 1007.

Fleck, Jr., R.C. 1980, Ap.J., **242**, 1019.

Forrest, W.J., Pipher, J.L., and Stein, W.A. 1986, Ap.J. (Letters), **301**, L49.

Fomalont, E.B. and Wright, C.H. 1974, in Galactic and Extragalactic Radio Astronomy, eds. G.L. Verschuur and K.I. Kellerman (New York: Springer), p. 25.

Fukui, Y. et al. 1977, Publ. Astron. Soc. Japan, **29**, 643.

Gardner, F.F. and Whiteoak, J.B. 1977, Proc. Astron. Soc. Australia, **3**, 150.

Gardner, F.F. and Whiteoak, J.B. 1962, Phys. Rev. Letters, **9**, 197.

Gatley, I. 1983, in Galactic and Extragalactic Infrared Spectroscopy, eds. M.F. Kessler and J.P. Phillips (Paris: E.S.A), p. 347.

Gatley, I. 1984, in Galactic and Extragalactic Infrared Spectroscopy, M.F. Kessler and J.P. Phillips (eds.), p. 351.

Gatley, I., Becklin, E.E., Werner, M.W., and Wynn-Williams, C.G., 1977, Ap.J., **216**, 277.

Gatley, I., Hyland, A.R., Jones, T.J., Beattie, D.H., and Lee, T.J. 1984, M.N.R.A.S., **210**, 565.

Geballe, T.R., Krisciunas, K., Lee, T.J., Gatley, I., Wade, R., Duncan, W.D., Garden, R., and Becklin, E.E. 1984, Ap.J. **284**, 118.

Genzel, R. and Downes, D. 1980, Astron. Astrophys., **87**, 6.

Genzel, R., Downes, D. and Bieging, J. 1976, M.N.R.A.S., **177**, 101.

Genzel, R., Watson, D.M., Townes, C.H., Dinerstein, H.L., Hollenback, D., Lester, D.F., Werner, M. and Storey, J.W.V. 1984, Ap.J., **276**, 551.

Gopal-Krishna and Swarup, G., 1976, Astrophys. Letters, **17**, 45.

Gopal-Krishna, Swarup, G., Sarma, N.V.G., and Joshi, M.N. 1972, Nature, **239**, 91.

Gordon, M.A. 1974, in Galactic Radio Astronomy, eds. F.J. Kerr and S.C. Simonson, III, p. 477.

Goss, W.M., Schwarz, U.J., Ekers, R.D., and van Gorkom, J.H. 1983, in IAU Symposium No. 101, Supernova Remnants and Their X-ray Emission, eds. J. Danzinger and P. Gorenstein, p. 65.

Goss, W.M., Schwarz, U.J., van Gorkom, J.H., Ekers, R.D. 1985, M.N.R.A.S., **215**, 69P.

Güsten, R. 1982, in The Galactic Center, ed. G.R. Riegler and R.D. Blandford, p. 9.

Güsten, R., and Downes, D. 1980, Astron. Astrophys., **87**, 6.

Güsten, R. and Henkel, C. 1983, Astron. Astrophys., **126**, 136.

Güsten, R., Walmsley, C.M., Pauls, T. 1981, Astron. Astrophys., **103**, 197.

Hall, D.N.B., Kleinmann, S.G., and Scoville, N.Z. 1982, Ap.J. (Letters), **260**, L53.

Hall, J.S. 1949, Science, **109**, 166.

Haslam, C.G.T. 1974, Astr. Astrophys. Suppl., **15**, 333.

Haynes, R.F., Caswell, J.L. and Simons, L.W.J. 1978, Aust. J. Phys. Suppl., **45**, 1.

Heiles, C. 1976, Ann. Rev. Astron. Astrophys., **14**, 1.

Henry, J.P., Depoy, D.L., Becklin, E.E., 1984, Ap.J. (Letters), **285**, L27.

Hilderbrandt, R.H., Whitcomb, S.E., Winston, R., Stiening, R.F., Harper, D.A., Mosley, S.H. 1978, Ap.J. (Lett.), **219**, L101.

Hiltner, W.A. 1949, Ap.J., **109**, 471.

Hiltner, W.A. 1951, Ap.J., **114**, 241.

Hiltner, W.A. 1956, Ap.J. Suppl., **2**, 389.

Hjellming, R.M. 1982, An Introduction to the NRAO Very Large Array.

Ho, P.T.P., Jackson, J.M., Barret, A.H. and Armstrong, J.T. 1985, Ap.J., **288**, 575.

Hogbom, J.A. 1974, Astron. Astrophys. Suppl., **15**, 417.

Hollinger, J.P. 1965, Ap.J., **142**, 609.

Hummel, E., van Gorkom, J.H., and Kotanyi, C.G. 1983, Ap.J. (Letters), **267**, L5.

Inoue, M., Takahashi, T., Tabaram H., Kato, Tsuboi, M. 1984, Publ. Astr. Soc. Japan, **36**, 633.

Isaacman, R. 1981, Astron. Astrophys. Suppl., **43**, 405.

Jones, T.W. 1974, Astron. Astrophys. , **30**, 37.

Jones, D.L., Sramek, R.A. and Terzian, Y. 1981, Ap.J., **246**, 28.

Jordon, S. 1981, The Sun as a Star, NASA monograph series.

Kaifu, N., Morris, M. Palmer, P. and Zuckerman, B. 1975, Ap.J., **201**, 98.

Kapitzky, J.E. and Dent, W.A. 1974, Ap.J., **188**, 27.

Kaplan, S.A. 1959, A. Astrophys., **36**, 778.

Kassim, N., LaRosa, T.N., Erickson, W.C. 1986, in press.

Kerr, F.J. 1966, (see Downes and Maxwell 1966)

Kerr, F.J. and Sandqvist, A. 1968, Ap.J. (Lett.), **2**, 195.

Knacke, R.F. and Capps, R.W. 1977, Ap.J., **216**, 271.

Knight, F.K., Johnson, W.N., Kurfess, J.D. and Strickman, M.S. 1983,

Kulsrud, R.M., Bernstein, I.B., and Kruskal, M. 1965, Ap.J., **142**, 491.

Lacy, J.H., Baars, F., Townes, C.H., Geballe, T.R. 1979, Ap.J. (Letters), **227**, L17.

Lacy, J.H., Townes, C.H., Geballe, T.R., Hollenbach, D.J. 1980, Ap.J., **262**, 120.

LaRosa, T.N. and Kassim, N. 1985, Ap.J. (Letters),, **299**,

Lacy, J.H., Townes, C.H., and Hollenbach, D.J. 1982, Ap.J., **262**, 120.

Lewtas, J., Lasenby, A., Yusef-Zadeh, F. 1986, in preparation.

Liszt, H. 1985, Ap.J. (Letters), **293**, L65.

Liszt, H. 1985, Ap.J. (Lett.), **293**, L65.

Liszt, H., Burton, W.B. 1978, Ap.J., **226**, 790.

Little, A.G. 1974, in Galactic Radio Astronomy, eds. F.J. Kerr and S.C. Simonson III, (Dordrecht: Reidel), p. 491.

Lo, K.Y. 1985, preprint.

Lo, K.Y., Backer, D.C., Ekers, R.D., Kellerman, K.I., Reid, M., Moran, J.M. 1985, Nature, **315**, 124.

Lo, K.Y., Claussen, M.J. 1984, Nature, **306**, 647.

Lo, K.Y., Cohen, M.H., Readhead, A.C.S. and Backer, D.C. 1981, Ap.J., **249**, 504.

Loose, H.H., Krügel, E. and Tutukov, A. 1982, Astron. Astrophys., **105**, 342.

Lynden-Bell, D. and Rees, M.J. 1971, M.N.R.A.S., **152**, 461.

Maihara, T., Noguchi, K., Okuda, H., Sato, S., Oishi, M. 1977, Publ. Astron. Soc. Japan, **29**, 415.

Martin, A.H.M. and Downes, D. 1972, Ap.J. (Lett)., **11**, 219.

Maxwell, A. and Taylor, J.H. 1968, Ap.J. (Lett.), **2**, 191.

Mayer, C.H., Hollinger, J.P. and Allen, P.J. 1963, Ap.J., **137**, 1309.

Mayer, C.H., McCullough, T.P. and Sloanaker, R.M. 1964, Ap.J., **139**, 258.

Mezger, P.G., Churchwell, E.B., and Pauls, T.A. 1974, in Stars and the Milky Way System, ed. L.N. Mavridis, p. 140.

Mezger, P.G. and Hoglund, B. 1967, Ap.J., **147**, 490.

Mezger, P.G., Pauls, T. 1979, in The Large-Scale Structure of the Galaxy, IAU Symposium 84, ed. W. Burton, p. 357.

Mezger, P.G., Schraml, J. and Terzian, Y. 1967, Ap.J., **150**, 807.

Mezger, P.G. and Wink, J.E. 1986, Astron. Astrophys., **157**, 252.

Mills, B.Y. and Drinkwater, M.J. 1984, J. Astrophys. Astron., **5**, 43.

Moffatt, H.K. 1978, Magnetic Field Generation in Electrically Conducting Fluids, Cambridge Univ. Press.

Morfill, G.E., O'C Dury, L. Aschenbach, B. 1984, Nature, **317**, 358.

Morris, M. Polish, N., Zuckerman, B., Kaifu, N. 1983, Ap.J., **88**, 1228.

Morris, M., Yusef-Zadeh, F. 1985, A.J., **90**, 2511.

Morris, M., Yusef-Zadeh, F., Chance, D. 1984 in Birth and Infancy of Stars, R. Lucas, A. Omont, and R. Sora, eds.

Napier, P.J., Thompson, R., Ekers, R.D. 1983, Proceedings of the IEEE, **11**, 1295.

Odenwald, S.F. and Fazio, G.G. 1984, Ap.J., **283**, 601.

Oort, J.H. 1977, Ann. Rev. Astron. Astrophys., **15**, 296.

Paczynsky, B. 1986, Nature, **319**,

Parker, E.N. 1955, Ap.J., **122**, 293.

Parker, E.N. 1970, Ap.J., **162**, 665.

Parker, E.N. 1971a, Ap.J., **163**, 255.

Parker, E.N. 1971b, Ap.J., **163**, 279.

Parker, E.N. 1971c, Ap.J., **166**, 295.

Parker, E.N. 1976, in The Physics of Non-Thermal Radio Sources, ed. G. Setti, p. 169.

Pauls, T. 1979, in Radio Recombination Lines, ed. P.A. Shaver, p. 159.

Pauls, T., Downes, D., Mezger, P.G. and Churchwell, E. 1976, Astron. and Astrophys., **46**, 407.

Pauls, T. and Mezger, P.G. 1980, Astr. Astrophys., **85**, 26.

Pauls, T.A., Mezger, P.G. and Churchwell, E.B. 1974, Astr. Ap., **34**, 327.

Piddington, J.H. and Minnett, H.C. 1951, Australian J. Sc. Res., **4**, 459.

Pik'el'ner, S.B. 1961, Fundamentals of Cosmic Electrodynamics, NASA Technical Translation.

Priest, E.R. 1984, Solar Magnetohydrodynamics, (Dordrecht: Reidel).

Quinn, P.J., Sussman, G.J. 1985, Ap.J., **288**, 377.

Rieke, G.H. 1981, in Infrared Astronomy, eds. C.G. Wynn-Williams and D.P. Cruishank, p. 317

Rieke, G.H., Low, F.J. 1973, Ap.J., **184**, 415.

Rimer, N. and Jen, N.C. 1973, Ap.J., **184**, 887.

Rodriguez, L.F. and Chaisson, E.J. 1979, Ap.J., **228**, 734.

Ruzmaikin, A.A., Sokolov, D.D. and Shukurov, A.M. 1979, Magnetohydrodynamics, **16**, 15.

Sanders, R.H. and Lowinger, T. 1972, Astron. J., **77**, 292.

Sanders, R.H. and Wrixon, G.T. 1973, Astr. Astrophys., **26**, 365.

Sanders, R.H. 1979, IAU Symposium No. 84., p. 383.

Sandqvist, Aa, 1974, Astron. Astrophys., **33**, 413.

Sandqvist, Aa, 1971, in Lunar Occultations of the Galactic Center Region in HI, OH and CH2O Lines, University of Maryland (Ph.D. Thesis).

Schluter, A. 1957, Z. Naturforsch., **12a**, 855.

Schmidt, R. 1980, Ph.D. Thesis, Bonn University.

Schwab, F. 1981, VLA Scientific Memo No. 36.

Schwab, F. 1981, Robust Solutions for Antenna Gains, VLA Scientific Memo, No. 136.

Scoville, N.Z. 1972, Ap.J. (Letters), **175**, L127.

Seiradakis, J.H., Lasenby, A., Yusef-Zadeh, F., Wielebinski, R. and Klein, U. 1985, Nature, **317**, 697.

Serabyn, E. 1984, Ph.D. Thesis, University of California, Berkeley.

Serabyn, E. and Lacy, J.H. 1985, Ap.J., **293**, 445.

Slee, B. 1977, Aust. J. Phys., Astrophys. Suppl., **43**.

Sofue, Y. and Handa, T. 1984, Nature, **310**, 568.

Sofue, Y. 1985, Publ. Astr. Soc. Japan, in press.

Soward, A.M. 1978, Astron. Nachr., 299, 25.

Spitzer, L., Jr. 1978, Physical Processes in the Interstellar Medium (New York: John Wiley and Sons).

Steenbeck, M., Krause, F. and Radler, K.H. 1966, Zh. Naturforsch., **21a**, 369.

Steinberg, J.L. and Lequeux, J. 1963, in Radio Astronomy, (New York: McGraw-Hill), p. 193.

Steward, G.C., Fabian, A.C., Seward, F.D. 1983, in Supernova Remnants and their X-ray Emission, IAU Symposium No. 101, p. 59.

Stier, M.T., Dwek, E., Silverberg, R.F., Hauser, M.G., Cheung, L., Kelsall, T., and Gezari, D.Y. 1982, The Galactic Center, eds. G.R. Riegler and R.D. Blandford, (New York: American Institute of Physics), p. 77.

Stix, M. 1975, Astron. Astrophys., **42**, 85.

Storey, J.W.V., Allen, D.A. 1983, M.N.R.A.S., **204**, 1153.

Thompson, A.R. and D'Addario, L.R. 1982, Proc. NRAO Workshop, Synthesis Mapping.

Thompson, A.R., Clark, B.G., Wade, C.M. and Napier, P.J. 1980, Astrophys. J. Suppl., **44**, 151.

Thompson, A.R., Riddle, A.C., and Lang, K.R. 1969, Astrophys. Lett., **3**, 49.

Townes, C.H., Lacy, J.H., Geballe, T.R., and Hollenbach, D.J. 1983, Nature, **301**, 661.

Tsuboi, M., Inoue, M., Handa, T., Tabara, H., and Kato, T. 1985, Publ. Astr. Soc. Japan,

Tsuboi, M., Inoue, M., Handa, T., Tabara, H. et al. 1986, in press.

Uchida, Y., Shibata, K., Sofue, Y. 1985, Nature, **317**, 699.

Vallee, J.P. and Simard-Normandin, M. 1985, Astron. and Astrophys., **243**, 274.

Vainshtein, S.I. and Ruzmaikin, A.A. 1972, Soviet Astronomy A.J., **5**, 714.

van Gorkom, J.H., Schwarz, U.J., Bregman, J.D. 1984, in The Milky Way Galaxy, IAU Symposium 106, eds. H. van Woerden, W.B. Burton, and R.J. Allen.

van den Bergh, S. 1983, Ap.J., **265**, 719.

van den Bergh, S. 1971, Ap.J., **165**, 259.

Verschuur, G.L. 1974, in Galactic and Extragalactic Radio Astronomy, eds. G.L. Verschuur and Kellermann, p. 179.

Watson, D.M., Storey, J.W.V., Townes, C.H., Haller, E.E. 1980, Ap.J. (Letters), **241**, L43.

Watson, M.G., Willingale, R., Grindlay, J.E., and Hertz, P. 1981, Ap.J., **250**, 142.

White, M.P. 1978, Astron. Nachr., **299**. 209.

White, R.L. 1985, NASA Conf. Publ., NASA CP-2358, p. 136.

Whiteoak, J.B. and Gardner, F.F. 1973, Astrophys. Lett., **13**, 205.

Wollman, E.R., Geballe, T.R., Lacy, J.H., Townes, C.H. and Rank, D.M. 1977, Ap.J. (Lett.), **218**, L103.

Yusef-Zadeh, F., Morris, M., and Chance, D. 1984, Nature, **310**, 557.